\documentclass[12pt]{article}
\usepackage{amsmath}
\usepackage{amssymb}
\usepackage{bm}
\usepackage{graphicx}
\usepackage{tikz}
\usepackage{hyperref}
\usepackage{lmodern}
\def\baselinestretch{1.1}

\numberwithin{equation}{section}

\def\bC{\mathbb{C}}

\def\bR{\mathbb{R}}
\def\bZ{\mathbb{Z}}

\def\cH{\mathcal{H}}
\def\cM{\mathcal{M}}
\def\cN{\mathcal{N}}
\def\cG{\mathcal{G}}

\def\CP{\mathbb{CP}}
\def\fg{{\displaystyle\mathfrak{g}}}
\def\SU{\mathrm{SU}}

\def\SO{\mathrm{SO}}
\def\Sp{\mathrm{Sp}}
\def\U{\mathrm{U}}

\def\tr{\mathrm{tr}}
\def\vev#1{\langle#1\rangle}
\def\ket#1{|#1\rangle}
\def\bra#1{\langle#1|}
\def\vec#1{\bm{#1}}
\def\bo{J}

\def\node#1#2{\overset{#1}{\underset{#2}{\circ}}}
\def\Node#1#2{\overset{#1}{\underset{#2}{\bullet}}}
\def\ver#1#2{\overset{{\llap{$\scriptstyle#1$}\displaystyle\circ{\rlap{$\scriptstyle#2$}}}}{\scriptstyle\vert}}
\def\Ver#1#2{\overset{{\llap{$\scriptstyle#1$}\displaystyle\bullet{\rlap{$\scriptstyle#2$}}}}{\scriptstyle\vert}}

\usetikzlibrary{arrows}
\tikzstyle{every picture}+=[remember picture]
\tikzstyle{na} = [baseline=-.5ex]

\DeclareMathSizes{6}{.8}{.8}{.8}

\newcommand{\be}{\begin{equation}}
\newcommand{\ee}{\end{equation}}
\newcommand{\beq}{\begin{equation}}
\newcommand{\eeq}{\end{equation}}
\newcommand{\ba}{\begin{array}}
\newcommand{\ea}{\end{array}}
\newcommand{\bi}{\begin{itemize}}
\newcommand{\ei}{\end{itemize}}
\def\bea#1\eea{\allowdisplaybreaks \begin{eqnarray}#1\end{eqnarray}}
\newcommand{\ben}{\begin{enumerate}}
\newcommand{\een}{\end{enumerate}}
\newcommand{\bean}{\begin{eqnarray*}}
\newcommand{\eean}{\end{eqnarray*}}
\newcommand{\eref}[1]{(\ref{#1})}

\newcommand{\fref}[1]{Figure~\ref{#1}}

\newcommand{\BC}{\mathbb{C}}

\newcommand{\BZ}{\mathbb{Z}}

\newcommand{\comment}[1]{}

\newcommand{\p}{\partial}
\newcommand{\al}{\bm{\alpha}}
\newcommand{\e}{\bm{\epsilon}}

\begin{document}

\begin{titlepage}

\begin{flushright}
IPMU-11-0191\\
CALT-68-2855 \\
MPP-2011-130
\end{flushright}
\vskip 3cm

\begin{center}
{\Large \bfseries
The ABCDEFG of Instantons and W-algebras
}

\vskip 1.2cm

Christoph A. Keller$^\sharp$,
Noppadol Mekareeya$^\flat$,
Jaewon Song$^\sharp$, 
and Yuji Tachikawa$^\natural$

\bigskip
\bigskip

\begin{tabular}{ll}
$^\sharp$ & California Institute of Technology,  Pasadena, CA 91125, USA \\[.5em]
$^\flat$ & Max-Planck-Institut f\"ur Physik (Werner-Heisenberg-Institut),\\
&F\"ohringer Ring 6, 80805 M\"unchen, Deutschland \\[.5em]
$^\natural$ & IPMU, University of Tokyo,  Kashiwa, Chiba 277-8583, Japan
\end{tabular}

\vskip 1.5cm

\textbf{Abstract}
\end{center}

\medskip

For arbitrary gauge groups, we check at the one-instanton level 
that the Nekrasov partition function of pure $\cN=2$ super Yang-Mills
is equal to the norm of a certain coherent state of the 
corresponding W-algebra. For non-simply-laced gauge groups,
we confirm in particular that the coherent state is in the twisted sector
of a simply-laced W-algebra.

\bigskip
\vfill
\end{titlepage}

\setcounter{tocdepth}{2}
\tableofcontents

\section{Introduction}
It has been almost twenty years since it was realized that the quantum dynamics of 4d $\cN=2$ gauge theory is encoded in the classical geometry of a two-dimensional Riemann surface $\Sigma$, called the Seiberg-Witten curve \cite{Seiberg:1994rs,Seiberg:1994aj}.  The curve $\Sigma$ was originally introduced as an auxiliary construct, but it was later recognized \cite{Klemm:1996bj,Witten:1997sc,Gaiotto:2009we,Gaiotto:2009hg} that $\Sigma$ is a branched covering of another two-dimensional surface $C$ on which a 6d $\cN=(2,0)$ theory is compactified to obtain the 4d $\cN=2$ theory. 

From this point of view, it is not surprising that the 2d quantum dynamics on $\Sigma$ or $C$ have some bearing on the 4d gauge dynamics. Indeed the role of 2d free bosons in this setup has long been recognized: see e.g.~\cite{Klemm:2002pa,Dijkgraaf:2002yn,Fuji:2004vf} for the identification of the gravitational factor on the $u$-plane  \cite{Witten:1995gf,Moore:1997pc} as the one-loop determinant of a free boson on $\Sigma$, and \cite{Aganagic:2003db,Nekrasov:2003rj,Dijkgraaf:2007sw} for a more general analysis.  
More recently, it was noticed that 2d Toda field theories with W-symmetry arise naturally on $C$  by considering the partition function on $S^4$ of the gauge theory \cite{Alday:2009aq,Wyllard:2009hg}. This correspondence has been studied thoroughly for  gauge groups $\SU(N)$\footnote{The authors apologize for not citing papers on $\SU(N)$.}
but not much work has been done for gauge groups of other types, see e.g.~\cite{Bonelli:2009zp,Itoyama:2009sc,Hollands:2010xa,Hollands:2011zc}. 

The aim of this paper is to rectify the situation, by studying $\cN=2$ pure gauge theory for arbitrary $G$ as an extension of \cite{Gaiotto:2009ma,Marshakov:2009gn,Taki:2009zd}. 
The main tools are the Hilbert series of the moduli space for arbitrary gauge group $G$ found and discussed e.g.~in \cite{VinbergPopov,Garfinkle,Benvenuti:2010pq}, the free-field realization of W-algebras \cite{Bershadsky:1989mf,Feigin:1990pn,Bouwknegt:1992wg}, and K.~Thielemans' Mathematica package \texttt{OPEdefs.m} \cite{Thielemans:1991uw}.

\paragraph{Martinec-Warner solution of $\cN=2$ pure theory:}

\begin{table}
\[
\begin{array}{l||r|l||l||r|l|l }
G&h^\vee &  w_i   &
	G&h^\vee &  w_i  & \Gamma^{(r)} \\
\hline
A_{n-1} & n   & 2,3,\ldots,n &
	B_n & 2n-1   & 2,4,\ldots,2n & A_{2n-1}^{(2)}\\ 
D_n & 2n-2  & 2,4,\ldots,2n-2 ; n &
	C_n & n+1   & 2,4,\ldots,2n  & D_{n+1}^{(2)}\\ 
E_6 & 12 & 2,5,6,8,9,12 &
	F_4 & 9 & 2, 6, 8, 12 & E_{6}^{(2)} \\
E_7 & 18 & 2,6,8,10,12,14,18 &
	G_2 & 4 & 2,6 & D_4^{(3)} \\
E_8 & 30 &  2,8,12,14,18,20,24,30 &&&&
\end{array}
\]
\caption{The dual Coxeter number $h^\vee$ and the dimensions $w_i$ of the Casimir invariants for 
all simple Lie groups $G$.  Recall $A_{n-1}=\SU(n)$, $B_n=\SO(2n+1)$, $C_n=\Sp(n)$, $D_n=\SO(2n)$. For non-simply-laced $G$, the Langlands dual $(G^{(1)})^\vee=\Gamma^{(r)}$ of the affine  $G$ algebra is also shown. \label{gaugedata}}
\end{table}

The Seiberg-Witten curve of pure $\cN=2$ gauge theory for arbitrary $G$ 
was constructed in \cite{Martinec:1995by} as the spectral curve
of a Toda lattice. (See also \cite{Klemm:1994qs,Gorsky:1995zq,Itoyama:1995nv,Lerche:1996an}.)
From a modern perspective  \cite{Klemm:1996bj,Gaiotto:2009we,Gaiotto:2009hg},
their construction reads as follows.  Let us first consider the case when $G$ is simply-laced. 
Take  6d $\cN=(2,0)$ theory of type $G$, and compactify this theory on $C=\CP^1$ parametrized by $z$, with two codimension 2 defects at $z=0$ and $z=\infty$. 
The 6d theory has world-volume fields $\phi^{(w_i)}(z)$ on $C$, transforming as degree $w_i$ multi-differentials, where $w_i$ is the degree of the Casimir invariants of $G$ given in Table~\ref{gaugedata}.
We then set \begin{align}
\phi^{(w_i)}(z)&=u^{(w_i)}\left({dz}/z\right)^{w_i}, & (w_i \neq h^\vee); \label{foo1}\\
\phi^{(h^\vee)}(z)&=\left(\Lambda^{h^\vee}z+ u^{(h^\vee)} +\frac{\Lambda^{h^\vee}}z\right)\left(dz/z\right)^{h^\vee}&  (w_i=h^\vee).\label{foo2}
\end{align} Here $u^{(w_i)}$ is the vev of the dimension $w_i$ Coulomb branch operator, and $\Lambda$ is the holomorphic dynamical scale of the gauge theory.
From this data one can then construct the Seiberg-Witten curve $\Sigma$ \cite{Hollowood:1997pp}, or equivalently the fibration of the ALE space of type $G$ \cite{Lerche:1996an,Diaconescu:2006ry}. In the following we label the degrees $w_i$ so that $h^\vee=w_n$.

\begin{figure}
\def\circn#1{\tikz[na]\node(#1){$\circ$};}
\def\seg{\,\tikz[na] \draw(0,0)--(1em,0) ;}
\def\ud{{\color{red}\updownarrow}}
\[
\begin{array}{|l|c@{}c@{}c@{}c@{}c@{\hskip-.3em}c@{}c||l|c@{}c@{}c@{}c@{}c@{\hskip-.3em}c@{}c|}
\hline
&\circ &  \seg & \circ & \cdots  &\circn{A}  & 
	&&&&&&&& &\circn{D} \\
\Gamma= A_{2n- 1}&\ud&&\ud && \ud & &\circn{B} & 
	\Gamma= D_{n+1} & \circ & \seg  &\circ &  \cdots  & \circn{E} & &\ud \\
&\circ & \seg  & \circ  & \cdots  & \circn{C} &
	&&&&&&&& &\circn{F}\\
G= B_n & \circ & \seg  &\circ &  \cdots  & \circ & \Rightarrow  & \circn{}  &
	G=C_n & \circ & \seg  &\circ & \cdots  & \circ & \Leftarrow  & \circn{}  \\[3pt]
\hline
&&&&&\circn{G}&\seg &\circn{}  &
	 &&& & & \circn{J} && \\
\Gamma=E_6& \circ &\seg & \circn{H} && \ud &&\ud &
	 \Gamma=D_4 &&& \circn{K} & \seg  & \circn{M}&&  \\
&&&&&\circn{I}&\seg &\circn{}  &
	 && &&&\circn{L} & &\\
G=F_4 & \circ & \seg  &\circ  &\Leftarrow & \circ&\seg  & \circn{}  &
	G= G_2 && &\circ & \Lleftarrow & \circ &&\\[3pt]
\hline
\end{array}
\]

\begin{tikzpicture}[overlay,thin]
        \draw (A) --  (B) --(C);
        \draw (D) --  (E) --(F);
        \draw (G) --  (H) --(I);
        \draw (J) --  (K) --(L);
        \draw[arrows={}-{angle 90},red] (J)  ..  controls +(.5,-.3) ..  (M);
        \draw[arrows={}-{angle 90},red] (M)  ..  controls +(.5,-.3) ..  (L);
        \draw[arrows={}-{angle 90},red] (L)  ..  controls  +(1,0)  and +(1,.1) ..  (J);
\end{tikzpicture}

\caption{The relation between a non-simply-laced Lie algebra $G$, its associated simply-laced algebra $\Gamma$, and the outer automorphism used to fold $\Gamma$ to obtain $G$.  \label{outer}}
\end{figure}

When $G$ is non simply-laced, we take a pair $(\Gamma,\bZ_r)$ such that the twisted affine Lie algebra $\Gamma^{(r)}$ is the Langlands dual to $G^{(1)}$, the untwisted affine algebra of $G$. In other words, the Dynkin diagram of $G$  is obtained by folding the Dynkin diagram of $\Gamma$ as in Fig.~\ref{outer}.\footnote{
$\Gamma$ is called the associated simply-laced algebra of $G$ \cite{Slodowy}.
$G$ is also known as the orbit Lie algebra of the pair $(\Gamma,\bZ_r)$, see \cite{Fuchs:1995zr}.
Note that $G$ is \emph{not} the $\bZ_r$-invariant part of $\Gamma$ in general, as explained in Appendix~\ref{nonsimplydata}.}
For example, when $G=G_2$, $\Gamma=\SO(8)$ and $r=3$. 
We then put 6d theory of type $\Gamma$ on $C=\CP^1$, with a twist by $\bZ_r$ around two singularities at $z=0,\infty$.  The fields of the 6d theory are divided into two sets,   $\phi^{(\hat w_i)}(z)$ which are invariant under $\bZ_r$ action, and $\phi^{(\tilde w_i)}(z)$ which transform nontrivially under $\bZ_r$. 
We then take \begin{align}
\phi^{(\hat w_i)}(z)&=u^{(\hat w_i)}\left({dz}/z\right)^{\hat w_i},  \label{fooX}\\
\phi^{(\tilde w_i)}(z)&=0, &  (\tilde w_i \ne h^\vee),\\
\phi^{(\tilde w_i)}(z)&=\left(\Lambda^{\tilde w_i}z^{1/r}+ \frac{\Lambda^{\tilde w_i}}{z^{1/r}}\right)\left(dz/z\right)^{\tilde w_i}&  (\tilde w_i = h^\vee)\label{fooZ}.
\end{align} Here, $\Lambda$ is the dynamical scale and $u^{(\hat w_i)}$ is the vev of the degree-$\hat w_i$ Coulomb branch operator; note that the degrees of Casimirs of $G$ are exactly the degrees of Casimirs of $\Gamma$ invariant under $\bZ_r$. Note also that the dual Coxeter number of $G$ are exactly the highest degree of Casimirs of $\Gamma$ not invariant under $\bZ_r$.

\begin{figure}
{\[
\begin{array}{lr@{}c@{}c}
\text{6d $\Gamma$ theory}&&\vcenter{\hbox{\includegraphics[scale=.5]{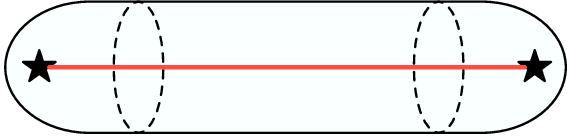}}} \\
 &&\downarrow\\
\text{5d $G$ theory}&&\vcenter{\hbox{\includegraphics[scale=.5]{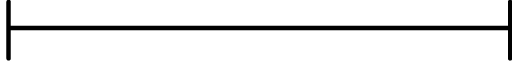}}} \\[10pt]
\text{2d W-algebra} &\bra{\underline{\cG}}& \Lambda ^{2h^\vee r L_0} &\ket{\underline{\cG}}
\end{array}
\]}
\caption{ Top: the Seiberg-Witten solution of pure $\cN=2$ super Yang-Mills theory with gauge group $G$ in terms of 6d $\cN=(2,0)$ theory of type $\Gamma$ on $C=\CP^1$ with the $\bZ_r$ twist line from $z=0$ to $z=\infty$. Middle: the $S^1$ reduction to the 5d maximally supersymmetric Yang-Mills theory with gauge group $G$ on a segment, with a suitable half-BPS boundary condition on both ends. Bottom: In the 2d description, the coherent state $\bra{\underline{\cG}}$ is produced by the BPS boundary condition. It is then propagated along the horizontal direction and annihilated by $\ket{\underline{\cG}}$.   \label{curve}}
\end{figure}

The construction is summarized in Fig.~\ref{curve}: the 6d $\cN=(2,0)$ theory of type $\Gamma$ on a circle with $\bZ_r$ twist gives maximally supersymmetric 5d Yang-Mills theory of gauge group $G$. To obtain pure $\cN=2$ Yang-Mills theory, we need to put the 5d Yang-Mills theory on a segment with an appropriate half-BPS boundary condition on both ends. The boundary condition then becomes the prescribed singularity of the worldvolume fields $\phi^{(w_i)}(z)$. When $G$ is classical, the 5d Yang-Mills theory can be realized on coincident D4-branes, possibly on top of an O4-plane. Then the BPS boundary condition comes from ending them on an NS5-brane \cite{Witten:1997sc,Evans:1997hk,Landsteiner:1997vd,Brandhuber:1997cc}.

\paragraph{Correspondence with W-algebra:}
The essence of the 2d-4d correspondence is that the correlator of the 2d Toda theory equals the Nekrasov partition function of the 4d theory with $\Omega$ deformation parameters $\epsilon_{1,2}$, and that the vevs of its W-currents $W^{(w_i)}(z)$ become the world-volume fields $\phi^{(w_i)}$ determining the Seiberg-Witten curve:  
\begin{equation}
\lim_{\epsilon_{1,2}\to 0} \vev{W^{(w_i)}(z)}(dz)^{w_i} \to \phi^{(w_i)}(z)\ .\label{babar}
\end{equation} 
In particular, the instanton contribution to Nekrasov's partition function should equal the conformal block of the W-algebra.  With the help of \eqref{babar} we translate the conditions on the singularities of $\phi^{(w_i)}(z)$ at $z=0,\infty$ into  conditions on the state $\ket \cG$, which we call the Gaiotto-Whittaker state,  inserted at $z=0,\infty$.
This state turns out to be a certain coherent state in the Verma module of the untwisted sector of the $W(G)$-algebra when $G$ is simply-laced, and in the $\bZ_r$-twisted sector of the $W(\Gamma)$-algebra when $G$ is non-simply-laced. We come back to the details in Sec.~\ref{coherent}. The most important relation is  \begin{equation}
W^{(h^\vee)}_{1/r}\ket{\cG} = 
\Lambda^{h^\vee}\ket{\cG}.\label{GW}
\end{equation}
Under a suitable identification of parameters we should then have the equality 
\begin{equation}\label{mainEq}
Z_\text{inst}(\vec a; \epsilon_{1,2}) = \vev{\cG|\cG}=\vev{\underline{\cG}|\Lambda^{2h^\vee r L_0}|\underline{\cG}}\ .
\end{equation}
On the right hand side, $\ket{\underline{\cG}}$ is the coherent state \eqref{GW} defined by setting $\Lambda=1$. The relation can be understood as in Fig.~\ref{curve}: the boundary condition creates the state $\bra{\underline{\cG}}$, which is then propagated by the distance $\propto\log\Lambda$, then is annihilated by $\ket{\underline{\cG}}$. Indeed, $\log\Lambda$ is the UV coupling, and is proportional to the length of the fifth direction. 

This relation was first considered for $\SU(2)$ in \cite{Gaiotto:2009ma}, and its calculation was later streamlined in \cite{Marshakov:2009gn}. The check for $\SU(3)$ was performed in \cite{Taki:2009zd} using the known explicit commutation relation of the $W_3$-algebra.
In this paper we will check (\ref{mainEq})  at the one-instanton level uniformly for all $G$. We will use the free-field realization of the W-algebra, without explicitly writing down the complicated commutation relation of the modes of its modes.
Before proceeding, it is to be noted that relation \eqref{mainEq} when we have a full surface operator has already been rigorously proved for all groups to all order in \cite{Braverman:2004vv,Braverman:2004cr}.

\bigskip

The remainder of this paper is organized as follows. In Sec.~\ref{counting} we review how we obtain the instanton contribution to Nekrasov's partition function, and evaluate the one-instanton term for all $G$. In Sec.~\ref{freeW}, we review the construction of the W-algebras in terms of free bosons.   In Sec.~\ref{comparison}, we identify the coherent state $\ket \cG$ from the behavior of the worldvolume fields $\phi^{(w_i)}(z)$, and compare its norm to the instanton contribution to Nekrasov's partition function. 
Most of the calculations are relegated to the appendices.

\section{Instanton calculation}\label{counting}
\subsection{Generalities}
Nekrasov's partition function was introduced in \cite{Nekrasov:2002qd}, as a culmination of a long series of works e.g.~\cite{Finnell:1995dr,Ito:1996qj,Dorey:2002ik} on the instanton calculation of the nonperturbative effects in $\cN=2$ gauge theory. We follow the presentation of \cite{Nekrasov:2004vw} in the following. 
Consider the partition function of  the 5d supersymmetric field theory with the same matter content on the spacetime of the form $\bC^2\times \bR$ parameterized by $(z_1,z_2,x_5)$, with the identification \begin{equation}
(z_1,z_2,x_5)\sim (z_1 e^{\epsilon_1 \beta}, z_2 e^{\epsilon_2 \beta}, x_5+\beta),
\end{equation} together with an appropriate $\SU(2)_R$ symmetry rotation to preserve supersymmetry. 
The vev of the 4d scalar field can be included either via the vev of the 5d scalar field, or via the Wilson line of the gauge field around $x^5$. 
Here we use the latter. Then the partition function is  \begin{equation}
Z_{5d}=\tr (-1)^F\exp \left[ i\beta H + \beta(\epsilon_1 J_1 + \epsilon_2 J_2 + a_i  H^i )\right]
\end{equation} where $H$ is the Hamiltonian, $J_{1,2}$  are the rotations of $z_{1,2}$-planes corrected with an appropriate amount of the $\SU(2)$ R-symmetry to commute with the supercharge, and $H^i$ are the generators of the Cartan of the gauge group.
Here the trace is taken in the field theory Hilbert space.\footnote{Physically, it would be more natural to take $\epsilon_{1,2}$ and $a_i$ to be purely imaginary, but supersymmetry guarantees that $Z$ is holomorphic with respect to them. For convenience we  regard $\epsilon_{1,2}$ as real and $a_i$ as purely imaginary.}

Using the localization, the partition function  can be written as the product of the one-loop contribution and the instanton contribution. The contribution $Z_k$ from $k$-instanton configurations is  \begin{equation}
Z_{k,5d}=\tr_{\cH_{k,\text{BPS}}} \exp \beta(\epsilon_1 J_1 + \epsilon_2 J_2 + a_i  H^i )
\end{equation} where $\cH_{k,\text{BPS}}$ is the BPS subspace of the Hilbert space of the supersymmetric quantum mechanics from the $k$-instanton configurations.
For the pure gauge theory, this is just the supersymmetric sigma model whose target space is the $k$-instanton moduli space $\cM_{G,k}$ of gauge group $G$. Then the BPS subspace $\cH_{k,\text{BPS}}$ is the space of holomorphic functions on $\cM_{G,k}$. Therefore  $Z_{k,5d}$ is just the character of the holomorphic functions on  $\cM_{G,k}$ under the action of the spacetime rotation $\U(1)^2$ and the gauge rotation $G$.
This quantity is also known as the Hilbert series. 

In the $\beta \to 0$ limit, $Z_{k,5d}$ is known to behave as \begin{equation}
Z_{k,5d} \sim \beta^{-2kh^\vee} Z_{k}(\vec a;\epsilon_{1,2})
\end{equation} and then the 4d instanton partition function is given by \begin{equation}
Z_\text{inst}=\sum_k \Lambda^{2kh^\vee} Z_{k}(\vec a;\epsilon_{1,2}).
\end{equation} Then the instanton part of the prepotential is given by \begin{equation}
F_\text{inst}=\lim_{\epsilon_1,\epsilon_2\to 0} \epsilon_1\epsilon_2 \log Z_\text{inst}.\label{pre}
\end{equation}

\subsection{One-instanton contribution}\label{s:1instanton}
Let us calculate the one-instanton contribution for arbitrary $G$. When $G$ is classical, we can use the ADHM description of the instanton moduli space to obtain the contribution \cite{Nekrasov:2004vw,Marino:2004cn,Hollands:2010xa}. Here, we use a more direct approach. 

The one-instanton configuration of arbitrary gauge group $G$ is obtained by embedding an $\SU(2)$ BPST instanton into $G$ via a map $\SU(2) \hookrightarrow G$ associated to the long root of $G$  \cite{Bernard:1977nr,Vainshtein:1981wh,Kronheimer,Br,KS1}. 
Therefore the one-instanton moduli space has a decomposition \begin{equation}
\cM_{G,1}=\bC^2 \times \tilde\cM_{G,1}
\end{equation} where the factor $\bC^2$ stands for the center of the instanton, and $\tilde\cM_{G,1}$ stands for the size and the gauge direction of the instanton.  $\cM_{G,1}$ is a hyperk\"ahler cone of real dimension $4(h^\vee-1)$.

Holomorphic functions on $\bC^2$ are just polynomials of the coordinates $z_1$ and $z_2$, and the character under the rotations $\U(1)^2\subset \SO(4)$  is just $(1-e^{\beta\epsilon_1})^{-1}(1-e^{\beta\epsilon_2})^{-1}$.
The space of holomorphic functions on $\tilde \cM_{G,1}$ was known to mathematicians e.g.~\cite{VinbergPopov,Garfinkle} using the fact that $\tilde\cM_{G,1}$ is the orbit of the highest weight vector in $\fg_\bC$ under $G_\bC$. The same space was also studied from a physical point of view by \cite{Benvenuti:2010pq}. The conclusion is that as the representation of $\U(1)^2 \times G$, the space of the holomorphic functions on $\cM_{G,1}$ is decomposed as 
\begin{equation}
\bigoplus_m (T_1T_2)^{\otimes m} \otimes V(-m\vec\alpha_0)
\end{equation} 
where $T_i$ is the 1-dimensional representation of $\U(1)$ with character $e^{\beta\epsilon_i}$, $V(\vec w)$ is the irreducible representation of $G$ of the highest weight $\vec w$, and $-\vec\alpha_0$ is the highest root of the root system of $G$. 
The factor $(T_1T_2)^{\otimes m}$ arises from the fact that  the radial direction is  generated by $e^{(\epsilon_1+\epsilon_2)/2}$  under the same $\U(1)^2\subset \SO(4)$.

Using the Weyl character formula, the character of this representation can be expressed as a summation over Weyl group elements, which can then be simplified as a summation over long roots, as we explain in Appendix~\ref{1instHilbert}. 
The end result is that the 4d 1-instanton contribution, including the contribution from the centre of mass, is given by 
\begin{equation}\label{Z1instanton}
Z_{k=1}=-\frac{1}{\epsilon_1\epsilon_2}\sum_{\vec\gamma\in\Delta_l}  \frac{1} {(\epsilon_1+\epsilon_2+\vec\gamma\cdot \vec a) (\vec\gamma\cdot\vec a ) \prod_{\vec\gamma^\vee\cdot\vec\alpha=1,\ \vec\alpha\in\Delta } (\vec\alpha\cdot \vec a)}\ ,
\end{equation} 
where $\Delta$ and $\Delta_l $ are the sets of the roots and the long roots, respectively. 
The one-instanton contribution to the prepotential via \eqref{pre} reproduces the instanton calculation by Ito and Sasakura \cite{Ito:1996hy}.
Explicit results for individual $G$ will be discussed in Sec.~\ref{explicitcomparison}.

\section{Free field realization of W-algebras}\label{freeW}

We will construct our W-algebras of type $\Gamma=A_n,D_n$ and $E_n$ from free fields
using the quantum version \cite{Bershadsky:1989mf,Feigin:1990pn} of the Drinfeld-Sokolov reduction \cite{Drinfeld:1984qv}. 
For our purpose this boils down to the following steps:
We introduce the free bosons $\vec\varphi$ living
in the weight space of the semisimple Lie group $\Gamma$ of rank $n$.
We normalize the OPE of the free bosons so that $J^k = i\partial \varphi^i$ satisfies  \begin{equation}
J^k(z)J^l(w)\sim \frac{\delta^{kl}}{(z-w)^2}.
\end{equation}
We assume $\Gamma$ is simply-laced, and normalize the roots to have squared length $2$.
The W-algebra is then given by the centralizer of the
screening charges $Q^\pm_i$ defined as follows: 
For each simple root $\vec \alpha_i$ there are
charges
\begin{equation}
Q^\pm_i = \oint \tilde s^\pm_i dz\ ,\qquad \tilde s^\pm_i = 
\exp(b^{\pm 1} \vec\alpha_i\cdot \vec\varphi)\ . \label{screening}
\end{equation}
Then we find the operators constructed from the bosons $\vec\varphi$
 which commute with all the screening charges. 
It is known that there are $n$ independent generators with weights $w_i$, tabulated in Table.~\ref{gaugedata}.
The weight 2 operator is the energy-momentum tensor given by 
\begin{equation}
T(z)=W^{(2)}(z)=-\frac12 (\partial\vec\varphi\cdot\partial\vec\varphi)(z) + Q \vec\rho \cdot \partial^2 \vec\varphi(z)\ ,
\end{equation} 
where $\vec\rho$ is the Weyl vector and $Q=b+1/b$.

Our computations were all done in Mathematica using the package \texttt{OPEdefs.m} developed by K. Thielemans \cite{Thielemans:1991uw}. For more details on W-algebras, the readers should refer to the review \cite{Bouwknegt:1992wg}.
In the following, all composite operators are understood to be OPE-normal ordered.

\subsection{Simply laced W-algebras}
\subsubsection{$A_n$}
For $A_n$ we make use of the quantum Miura transform \cite{Fateev:1987zh,Lukyanov:1987xg}.
Let  $\vec e_i, i=1,\ldots n+1$ be the weights of the fundamental $n+1$ dimensional representation of $A_n$, as in Appendix~\ref{data}.
We then construct a set of generators $U^{(k)}(z)$ from
\begin{equation}
R^{(n+1)}(z)= - \sum_{k=0}^{n+1} U^{(k)}(z)(Q \partial)^{n+1-k}
= (Q \partial - \vec e_1\cdot \partial \vec\varphi(z))\cdots
(Q \partial - \vec e_{n+1}\cdot \partial \vec\varphi(z))\ . \label{Miura-A}
\end{equation}
One can show that the singular part of 
the OPE of $R^{(n+1)}$ with $\tilde s^\pm_i$ is
a total derivative, which means that
the $U^{(k)}(z)$ are indeed in the centralizer
of the screening charges. Since $U^{(1)}(z)$ vanishes,
the remaining set of generators has the
correct dimensions $w_i$. One can
also show that they are independent, from which
we conclude that we have a full set of generators.
Note however that these generators are certainly
not unique, as we can always add suitable products
and derivatives of lower order generators.

\subsubsection{$D_n$}\label{s:WDn}
For $D_n$ we introduce $\vec e_i, i=1,\ldots n$ such
that $\pm\vec e_i$ form the weights of the fundamental $2n$ dimensional representation.
We repeat the construction of 
$R^{(n)}(z)$ as
\begin{equation}
R^{(n)}(z)= - \sum_{k=0}^{n} V^{(k)}(z)(Q \partial)^{n-k}\\
= (Q \partial - \vec e_1\cdot i\partial \vec\varphi(z))\cdots
(Q \partial - \vec e_{n}\cdot i\partial \vec\varphi(z))\ . \label{Miura-D}
\end{equation}
In this case however it turns out that only $V^{(n)}(z)$ commutes
with the screening charges. To obtain the rest of the generators,
we can take the OPE of $V^{(n)}$ with itself,
\begin{equation} \label{eq:miuraD}
V^{(n)}(z)V^{(n)}(w) = \frac{a_n}{(z-w)^{2n}} + \sum_{k=1}^{n-1}
\frac{a_{n-k}}{(z-w)^{2(n-k)}}\left(U^{(2k)}(z)+U^{(2k)}(w)\right)\ ,
\end{equation}
where we choose the normalization 
$a_k= \prod_{j=1}^{k-1}(1-2j(2j+1)Q^2)$.
Once again one can show that $V^{(n)}$ and the $U^{(2k)}$ are independent,
which gives us a set of $n$ generators of the correct weights
\cite{Lukyanov:1989gg,Lukyanov:1990tf}.

\subsubsection{$E_n$}\label{WE6preliminary}
For W-algebras of type $E_n$, the concise Miura transforms such as \eqref{Miura-A} for type $A_n$ 
and \eqref{Miura-D} for type $D_n$ are not known. 
Thus, one is forced to construct the commutants of the screening operators \eqref{screening} directly.
Note that an operator $O(z)$ commute with the screening charge $Q_i$ if and only if 
it has the form
\begin{equation}
O(z)=\sum_a X_a(T_j)(z) Y_a(\partial\varphi^1,\ldots,\partial\varphi^{i-1},\partial\varphi^{i+1},\ldots,\partial\varphi^n)(z)\label{decomp-foo}
\end{equation}  where
\begin{itemize}
\item $X_a$ and $Y_a$ stand for normal-ordered polynomials, possibly with derivatives, constructed from their respective arguments,
\item $T_i$ is the energy momentum tensor for the boson $\varphi_i=\vec\alpha_i\cdot\vec\varphi$ along the root $\vec\alpha_i$, i.e.~$T_i(z)=- \partial\varphi_i\partial\varphi_i(z) + Q\partial^2\varphi_i/2$ with central charge $1+6Q^2$,
\item and $\varphi^i=\vec w^i \cdot\vec\varphi$ where $\vec w^i$ are the fundamental weights, so that $Y_a$ are constructed from bosons perpendicular to the root $\vec\alpha_i$.
\end{itemize}
Therefore,  $O(z)$ is in the W-algebra of type $E_n$ if and only if $O(z)$ has the decomposition \eqref{decomp-foo} for each simple root $\vec\alpha_j$.
Details of the construction of the $W(E_6)$ algebra are presented in Appendix~\ref{WE6}.
The authors did not attempt to construct W-algebras of type $E_{7,8}$.

\subsection{Twisted sectors of the simply-laced W-algebras}
We need to consider the sectors of the simply-laced algebras twisted by their outer automorphisms to compare with the instanton partition function for non-simply-laced gauge groups.\footnote{The W-algebras for non-simply-laced groups  $B_n,C_n,G_2,F_4$ can also be determined via Drinfeld-Sokolov reduction, see e.g. \cite{Ito:1995ny}, but that is not what we use. }

The $\bZ_r$ outer automorphism acts on the simple roots as shown in Fig.~\ref{outer}.
This induces an action  on the free bosons $\vec\varphi$.
Since this is also a symmetry of the W-algebra, 
and we can consider a $\bZ_r$-twisted state.
In practice  we pick new linear combinations of bosons
$\tilde \varphi_j$ which are eigenstates of the $\mathbb{Z}_r$ action:
\begin{equation}\label{twistBoson}
\p \tilde \varphi_j \mapsto e^{{2\pi i k_j}/{r}}\p\tilde\varphi_j\ , \qquad k_j=0,\ldots r-1\ .
\end{equation}
Their modes are therefore in $\mathbb{Z}+{k_j}/{r}$. 
The set of generators $W^{(w_i)}$ therefore
decompose into generators $W^{(\hat w_i)}$ with integer modes, and generators $W^{(\tilde w_i)}$
with non-integer modes. The former correspond to the invariants of the non-simply-laced gauge group that we want to construct, and the states of the lowest level in the twisted Verma module is generated by $W^{(\tilde w_i)}_{-1/r}$.

The actions on the W-generators are given explicitly as follows:
\begin{itemize}
\item The $\bZ_2$ action on  $W(A_n)$ maps $\tilde U^{(k)}\to (-1)^k \tilde U^{(k)}$ where $\tilde U^{(k)}$ is a suitable redefinition of $U^{(k)}$  defined in \eqref{Miura-A}. For example, in the case of $W(A_5)$, 
\begin{align}
 \tilde{U}^{(2)} &= U^{(2)},   &
 \tilde{U}^{(3)} &= U^{(3)} - 2 Q \p \tilde{U}^{(2)}, \nonumber \\
 \tilde{U}^{(4)} &= U^{(4)} - \frac{3}{2} Q \p \tilde{U}^{(3)} , &
 \tilde{U}^{(5)} &= U^{(5)} - Q \p \tilde{U}^{(4)} + Q^3 \p^3 \tilde{U^{(2)}}, \nonumber \\
 \tilde{U}^{(6)} &= U^{(6)} - \frac{1}{2} Q \p \tilde{U}^{(5)} + \frac{1}{4} Q^3 \p^3 \tilde{U}^{(3)}. \label{WA5}
\end{align}
The redefined currents are determined by requiring W-generators to have definite $Z_2$ eigenvalues.
\item The $\bZ_2$ action on $W(D_n)$ maps $V^{(n)} \to - V^{(n)}$, $U^{(2k)}\to U^{(2k)}$ where $V^{(n)}$ and $U^{(2k)}$ are defined in \eqref{Miura-D}, \eqref{eq:miuraD}.
\item The $\bZ_2$ action on $W(E_6)$ maps $W^{(n)} \to (-1)^n W^{(n)}$, where $W^{(n)}$ is defined in Appendix~\ref{WE6}.
\item The $\bZ_3$ action on $W(D_4)$ is given is induced from the $\bZ_3$ action on the four free bosons \eqref{twistBoson}. Explicitly, we first define $W^{(2,4,6)}$ and $\tilde W^{(4)}$ by 
\begin{align}
 W^{(2)} &= U^{(2)}, &
 W^{(4)} &= U^{(4)} + \frac{1}{4} \p^2 W^{(2)} - \frac{1}{2} U^{(4)} W^{(2)} + \frac{9Q^2(1-4 Q^2)}{2 - 12 Q^2} \p^2 W^{(2)}, \nonumber \\
 \tilde{W}^{(4)} &= \sqrt{3} V^{(4)}, &
 W^{(6)} &= U^{(6)} + \frac{1}{2} \p^2 W^{(4)} - \frac{1}{3} W^{(4)} W^{(2)} + \frac{5}{6} Q^2 \p^2 W^{(4)} \ .
\end{align}
where $V^{(4)}$ and $U^{(2,4,6)}$ are defined in \eqref{Miura-D}, \eqref{eq:miuraD}.
Then $W^{(2,6)}$ is invariant and $W^{(4)} + i \tilde W^{(4)} \to e^{2\pi i/3}(W^{(4)} + i \tilde W^{(4)}) $  under the $\bZ_3$ action.
\end{itemize}

\subsection{Basic properties of the Verma module} \label{basicVerma}
Before continuing, we recall two  basic features of the untwisted and twisted Verma module.
The first is their Weyl invariance.  
The zero modes of the W-generators are given in terms of the zero modes
$\vec\bo_0$ of the free bosons $\vec\varphi$. If we define $\vec a$ by \begin{equation}
\vec a=\vec \bo_0-Q\vec\rho \label{shiftedzero}
\end{equation} then the zero modes of W-generators are Weyl-invariant polynomials of $\vec a$. 
For twisted sectors, the twisted bosons do not have zero modes. Correspondingly, $\vec a$ and $\vec\bo_0$ are invariant under the twist; $\vec\rho$ is automatically invariant.

The second is the Kac determinant at the lowest level, which we  detail  in Appendix~\ref{KacDeterminant}. Here we just quote the result, which is given by
\begin{equation}
\text{(Kac determinant at level $-1/r$)} \propto \prod_{\vec\gamma\in\Delta_{l}} (\vec\gamma\cdot \vec a+ Q)\label{Kac2}
\end{equation} 
where 
 \begin{equation}
\Delta_l =\left\{ 
\begin{array}{ll}
\Delta, & (r=1) \\
\{ \vec\alpha + o(\vec\alpha)  \}, & (r=2) \\
\{ \vec\alpha + o(\vec\alpha) +o^2(\vec\alpha) \}.  &  (r=3)
\end{array}\right.
\end{equation} 
Here   $\vec\alpha$ runs over roots with $\vec\alpha\neq o(\vec\alpha)$.
As explained in Appendix~\ref{nonsimplydata}, 
$\Delta_l$  can be identified with the set of long roots of $G$, which is the S-dual of 
 the $\bZ_r$ invariant subgroup of $\Gamma$.

\section{Instantons and coherent states of W-algebras}\label{comparison}

\subsection{Identification of the coherent state}\label{coherent}
Under the correspondence of Nekrasov's partition functions and conformal blocks of W-algebras, the key relation is that the vev of the W-currents should become the fields $\phi^{(w_i)}(z)$ in the limit $\epsilon_{1,2}\ll a$: \begin{equation}
\lim_{\epsilon_{1,2}\to 0} \vev{W^{(w_i)}(z)}(dz)^{w_i} \to \phi^{(w_i)}(z).
\end{equation}
The fields $\phi^{(w_i)}(z)$ have two singularities $z=0,\infty$, which means that there is a state $\bra \cG$ at $z=\infty$ and a state $\ket \cG$ at $z=0$. The behavior of $\phi^{(w_i)}(z)$ at $z=\infty$ has the same form as the behavior at $z=0$ by the map $w=1/z$. Therefore the state $\bra \cG$ is a conjugate of the state $\ket \cG$.

When $G$ is simply-laced, the conditions \eqref{foo1} and \eqref{foo2} imply that $\ket \cG$ is in the Verma module of $W(G)$-algebra generated from the highest weight state $\ket{\vec w}$ with \begin{equation}
W^{(w_i)}_0 \ket{\vec w}=w^{(w_i)} \ket{\vec w} \label{zeromode1}
\end{equation}  where the eigenvalues $w^{(w_i)}$ should equal the vev $u^{(w_i)}$ up to some quantization error involving $\epsilon_{1,2}$.
The condition \eqref{foo2} then tells us that
\begin{equation}\label{Gstaterelations}
W^{(w_i)}_\ell \ket \cG=0  \quad \text{for $\ell >0$ unless} \quad
W^{(h^\vee)}_1\ket \cG=\Lambda^{h^\vee} \ket \cG.
\end{equation}

When $G$ is non-simply-laced, the conditions \eqref{fooX} through \eqref{fooZ} imply that $\ket \cG$ is in the Verma module generated by the $\bZ_r$-twisted vacuum $\ket{\vec w}$ of $W(\Gamma)$-algebra determined by \begin{equation}
W^{(\hat w_i)}_0 \ket{\vec w}=w^{(\hat w_i)} \ket{\vec w}\label{zeromode2}
\end{equation}  where the eigenvalues $w^{(\hat w_i)}$ should equal the vev $u^{(\hat w_i)}$ up to the quantization error involving $\epsilon_{1,2}$.
The condition \eqref{fooZ} then says \begin{equation} \label{eq:GWvector}
W^{(w_i)}_\ell \ket \cG=0  \quad \text{for $\ell>0$ unless} \quad
W^{(h^\vee)}_{1/r}\ket \cG=\Lambda^{h^\vee} \ket \cG.
\end{equation}
Then we should have the relation \begin{equation}
Z(\vec a,\epsilon_{1,2})=\vev{\cG|\cG}.
\end{equation}
Both sides are power series; the $k$-instanton contribution on the left hand side corresponds to the level-$k/r$ contribution on the right hand side.

Note that the norm of $\ket \cG$ does not change if we change
the definition of $W^{(h^\vee)}$ by adding products and
derivatives of lower degree generators, since $W^{(h^\vee)}_{-\ell}$
only changes by negative modes of lower generators,
which annihilate $\ket \cG$ anyway.

\subsection{Coherent state at the lowest level}
Here, we discuss the procedure to compute the norm of the Gaiotto-Whittaker vector $\vev{\cG|\cG}$ using the free-bosons representation of the W-algebras. The method is the same as in \cite{Marshakov:2009gn,Taki:2009zd}.
We stick to the untwisted representations of the W-algebras for the moment. 
Let us  expand  $\ket \cG$ in terms of levels of descendants
\be
 \ket{\cG} = \ket{\vec{w}} + \Lambda^{h^\vee} \ket{\cG_1} + (\Lambda^{h^\vee})^2 \ket{\cG_2} + \cdots 
\ee
so that $\ket{\cG_\ell}$ has conformal weight $\ell$. 
The condition \eqref{eq:GWvector} is now
\be \label{eq:GWvec}
 W^{(w_i)}_\ell \ket {\cG_\ell} =0  \quad \text{for $n>0$ except for} \qquad
 W^{(h^\vee)}_1 \ket{\cG_\ell} = \ket{\cG_{\ell-1}} 
\ee
with $\ket{\cG_0} = \ket{\vec{w}}$. In what follows, we will compute $\vev{\cG_1|\cG_1}$ and compare against the 1-instanton computations. Expressed in terms of descendants at level 1 $\ket{\cG_1}$ is 
\be
 \ket{\cG_1} = \sum_i A_i W^{(w_i)}_{-1} \ket{\vec{w}} . 
\ee
Now, use \eqref{eq:GWvec} to get
\begin{alignat}{3}
 0 &= \bra{\vec{w}} W^{(w_i)}_1 \ket{\cG_1} &= \sum_j A_j \bra{\vec{w}} W^{(w_i)}_1 W^{(w_j)}_{-1} \ket{\vec{w}}  &= \sum_j K^{(1)}_{ij} A_j &\ \, \text{for $w_i  \neq h^\vee$}, \\
 1 = \vev{\vec{w}|\vec{w}} &= \bra{\vec{w}} W^{(h^\vee)}_{1} \ket{\cG_1} & =  \sum_j A_j \bra{\vec{w}} W^{(h^\vee)}_{1} W^{(w_j)}_{-1} \ket{\vec{w}}  &= \sum_j K^{(1)}_{n j}A_j
\end{alignat}
where $K^{(\ell)}$ is the Kac-Shapovalov matrix at level $\ell$. We can solve for $A$ so that $A_i = (K^{(1)})^{-1}_{i n}$, and the norm is given by
\be
 \vev{\cG_1|\cG_1} = \sum_{i, j} A_i K^{(1)}_{ij} A_j = (K^{(1)})^{-1}_{nn} . \label{eq:norm}
\ee
It can be easily generalized to arbitrary level to get $\vev{\cG_\ell|\cG_\ell} = (K^{(n)})^{-1}_{n^\ell,n^\ell }$ where the index $n^\ell$ means we pick the element corresponds to $(W^{(w_n)}_{-1})^\ell \ket{\vec{w}}$. In order to get the norm of the Gaiotto states for the $\bZ_r$ twisted sector, we simply look for the descendants of level-$1/r$, and take the corresponding element of the Kac-Shapovalov matrix. 

To evaluate the Kac-Shapovalov matrix write the $W$ generators in terms of free bosons and expand as
\be
 J^k(z) = i \p \varphi^k (z) = \sum_{m \in \bZ} \bo^k_m z^{-m-1}
\ee
with the usual commutation relation
\be
 [\bo^k_m, \bo^l_n] = m \delta^{k, l} \delta_{m+n, 0} . 
\ee
Then the W-algebra vacuum $\ket{\vec w}$ is  represented by 
the free-boson vacuum $\ket{\vec a}$ where  $\vec a=\vec \bo_0-Q\vec\rho$ is the shifted zero mode of the free bosons. Note that the bra $\bra{\vec w}$ then corresponds to $\bra{-\vec a}$ due to the background charge $Q\vec\rho$ and the shift. 

Let us first discuss the simply-laced case. 
As we are dealing with normal ordered products, the descendant states at level one can
be expressed as
\begin{align}
 W^{(w_i)}_{-1} \ket{\vec{a}} &= \sum_j M_{ij}(\vec a) \bo^j_{-1} \ket{\vec{a}} , &
\bra{-\vec{a}}  W^{(w_i)}_{1}  &= \sum_j \bra{-\vec a} \bo^i_1 M_{ij}(-\vec a)  .
\end{align}
The coefficient $M_{ij}(\vec a)$ is a polynomial in $\vec a$ and $Q$.  The Kac-Shapovalov matrix is given by \begin{equation}
K_{ij}(\vec a)=\sum_k M_{ik}(-\vec a) M_{kj}(\vec a).
\end{equation}

The twisted case is slightly more involved. Again we know that
the lowest descendant states can be written as
\be
 W^{(\tilde w_i)}_{-1/r} \ket{\vec{a}} = \sum_j \tilde M_{ij}(\vec a) \bo^j_{-1/r} \ket{\vec{a}} , 
\ee
where $\tilde M_{ij}(\vec a)$ is again a polynomial in the zero modes of the
untwisted bosons. To compute it, we need to find modes of the form $:(J^j)^m:_{-1/r}$,
which can be found from the original prescription of OPE-normal ordering, that is by
subtracting the singular part of the correlator of the $m$ bosons.
For instance, to obtain the constant $C_{2m+1}$ of the state
$:(J^j)^{2m+1}:_{-1/2}|\vec a\rangle = C_{2m+1} \bo_{-1/2}|\vec a\rangle$
we extract the regular part of the correlator
\be
 \lim_{z_i \to z_1}  \bra{-\vec{a}} \bo^j_{1/2} J^j(z_{2m+1}) J^j(z_{2m}) \cdots J^j(z_1) \ket{\vec{a}}\Bigm|_{reg} 
 = \frac{C_{2m+1}}{2z_1^{2m+1/2} }\ , \label{regularizedtwisted}
\ee
and similarly for the zero modes of even powers of $J$.

\subsection{Comparison}\label{explicitcomparison}

After all these preparations, now we can compare the norm of the Gaiotto-Whittaker vector and the one-instanton partition function. 
Obviously, the norm of the Gaiotto-Whittaker vector can have poles only at the zero of the Kac determinant,  \eqref{Kac2}, i.e.~when $\vec\beta\cdot\vec a_\text{boson}+Q=0$
for a long root $\beta$.
We also have the formula of the one-instanton expression in the gauge theory side, \eqref{Z1instanton},
which has apparent poles when $\vec\beta\cdot\vec a_\text{gauge}=\epsilon_1+\epsilon_2$ for a long root $\beta$, 
or when $\vec\gamma\cdot\vec a_\text{gauge}=0$ for an arbitrary root $\gamma$.
 In order for them to have any chance of agreement, we need to identify \begin{equation}
\vec a_\text{boson}=\frac{\vec a_\text{gauge}}{\sqrt{\epsilon_1\epsilon_2}},\qquad 
Q=\frac{\epsilon_1+\epsilon_2}{\sqrt{\epsilon_1\epsilon_2}}.
\end{equation}

Using the procedure outlined above, we have checked the agreement between the norm of the coherent state and instanton partition function at 1-instanton level  
\begin{itemize}
\item for simply-laced algebras $A_{1, 2, 3, 4, 5, 6}$, $D_{ 4}$, and $E_6$,
\item and for non-simply-laced algebras $B_{2, 3}$, $C_n$, $F_4$ and $G_2$.
\end{itemize}
In general, the agreement comes with a multiplicative ambiguity due to the normalization of W-currents. It can be easily absorbed into the redefinition of the expansion parameter $\Lambda$.\footnote{When the underlying gauge theory is conformal, one may encounter much more intricate map between expansion parameters. See \cite{Hollands:2010xa, Hollands:2011zc} for the details. }
For higher rank algebras, such as $A_5, A_6$, $F_4$ and $E_6$, due to the computational complexity, we checked the agreement by plugging in several set of test numbers for the zero modes and $Q$ parameter instead of leaving it as a symbolic expression. 
Let us now discuss the cases $A_n$, $D_n$, $B_n$, $C_n$, $G_2$ and $F_4$ in this order.

\subsubsection{$A_n$}  

The W-algebra calculation leads to the following explicit form of the Gaiotto-Whittaker vector at level one:  \begin{equation}
\ket{\cG_1} = \sum_i v_i(\vec a) \bo_{i,-1} \ket{\vec a}
\end{equation} where \begin{equation}
v_i(\vec a) = \sum_j C_{ij}(\vec a) w_j(\vec a)  , \quad 
w_i(\vec a)= \frac1{\prod_{x<i}(a_x-a_i) \prod_{i<x }(Q-a_i+a_x) } 
\label{triangular1}
\end{equation} where \begin{align}
C_{ij}(\vec a) &= 0, & (i<j) \\
C_{ij}(\vec a) &= 1, & (i=j) \\
C_{ij}(\vec a) &= (-1)^{i-j}Q\frac{\prod_{j<k< i} (Q-a_j+a_k)}{\prod_{j<k\le i} (a_j-a_k)  }.& (i>j)
\end{align}
Here $a_i=\vec e_i\cdot \vec a$.
We checked the validity for small $n$; we believe it is true in general.

The corresponding bra is given by
 \begin{equation}
\bra{\cG_1} = \sum_i \bra{-\vec a}\bo_{i,1} v_i(-\vec a).
\end{equation} 
Now, it can be checked  that \begin{equation}
v_i(-\vec a) = \sum_j\tilde w_j(\vec a) (C^{-1}(\vec a))_{ji}, \ \text{with}\ 
\tilde w_i(\vec a) = \frac1{\prod_{x<i}(Q-a_x+a_i) \prod_{i<x }(a_i-a_x) }. \label{unstable}
\end{equation}
Therefore, \begin{equation}
\vev{\cG_1|\cG_1} = \sum_i \prod_{i\ne j} \frac{1}{ (a_i-a_j)(Q -a_i+a_j)} \label{total}
\end{equation}
which is indeed the one-instanton contribution of Nekrasov's partition function \cite{Flume:2002az,Bruzzo:2002xf, Nakajima:2003pg, Taki:2009zd} calculated from the geometry of the one-instanton moduli space.
For example,
\bea
Z_{\SU(2),1}  &=& - \frac{2}{4a^2 -Q^2 }~, \\
Z_{\SU(3),1}  &=& \frac{6 \left(a_1^2+a_1 a_2+a_2^2- Q^2 \right)}{[(2a_1+a_2)^2 - Q^2][(a_1+2a_2)^2 - Q^2][(a_1-a_2)^2 - Q^2]}~. 
\eea

Two comments are in order. 
First, note that the uniform formula \eqref{Z1instanton} instead gives the following form \begin{equation}
Z_{k=1}=\sum_{i\ne j}\frac{1}{(a_i-a_j)(Q-a_j-a_i)}\prod_{k\ne i}\frac{1}{a_i-a_k}. \label{bosh}
\end{equation} The agreement of \eqref{total} and \eqref{bosh} are not easy to see, but they are equal nonetheless.
Second, recall that each summand in the formula \eqref{total} comes from the contribution of a fixed point in the resolution of the one-instanton moduli space. Then the relation \eqref{triangular1} means that the free boson basis $\bo^i_{-1}\ket{\vec a}$ is an upper-triangular redefinition of the basis formed by the fixed points, explicitly confirming the results of Maulik and Okounkov \cite{MO}.

\subsubsection{$D_n$}
The $W(D_n)$ contains W-generators of dimension $2, 4, \cdots, 2n-2$ and additionally of dimension $n$. 
The Nekrasov partition function for $D_n = SO(2n)$ can be obtained easily by evaluating the contour integral expression in \cite{Nekrasov:2004vw}. We have, 
\be
 Z_{SO(2n), 1} = - \sum_{i=1}^n 
 \left[ \frac{(a_i \pm Q) (2a_i \pm Q)}{\prod_{j \neq i} (a_i^2 - a_j^2) \left( (a_i \pm Q)^2 - a_j^2 \right)} \right] , 
\ee
where $\pm$ means we sum over both signs and $a_i=\vec e_i\cdot \vec a$. 
We have checked the agreement up to $D_4$. 

\subsubsection{$B_n$}
To get the coherent state for $B_n$, we start from $W(A_{2n-1})$. We need to evaluate the $n$-dimensional Kac-Shapovalov matrix, since all the odd-dimensional W-currents are twisted by $\bZ_2$ automorphism. 
The $\bZ_2$ action maps  $\varphi_i $ to $ - \varphi_{2n +1 - i}$. Then the eigenstates are
\be \label{eq:bntwbo}
 \tilde{\varphi}^+_i = \varphi_i - \varphi_{2n+1-i} ~~\textrm{and}~~ \tilde{\varphi}^-_i =  \varphi_i + \varphi_{2n+1-i} \ .
\ee
Then, the twisted W-currents can be written as
\be
 \tilde{U}^{(m)}_{-1/2} \ket {\vec a} = \sum_i B_{m, i} (\vec a)  \bo_{-1/2}^{-, i} \ket {\vec a} , 
\ee
where $B_{m, i}$ is a function of zero modes of the untwisted free bosons.  
Now, we evaluate the Kac-Shapovalov matrix to obtain the norm of the coherent state. For example, when $n=3$, 
\be
 K_{B_3} = \left(
 \begin{array}{cc}
  \vev{-\vec{a}| \tilde{U}^{(3)}_{1/2} \tilde{U}^{(3)}_{-1/2} | \vec{a}} &   \vev{-\vec{a}| \tilde{U}^{(3)}_{1/2} \tilde{U}^{(5)}_{-1/2} | \vec{a}} \\
    \vev{-\vec{a}| \tilde{U}^{(5)}_{1/2} \tilde{U}^{(3)}_{-1/2} | \vec{a}} &   \vev{-\vec{a}| \tilde{U}^{(5)}_{1/2} \tilde{U}^{(5)}_{-1/2} | \vec{a}}
  \end{array}
 \right) 
\ee
and take the inverse of $K_{B_3}$ and read off the $(2, 2)$ component of it. 

On the instanton side, we have,
\be
 Z_{SO(2n+1), 1} = \sum_{i=1}^n 
 \left[ \frac{(2a_i \pm Q)}{a_i \prod_{j \neq i} (a_i^2 - a_j^2) \left( (a_i \pm Q)^2 - a_j^2 \right)} \right] , 
\ee
where $\pm$ means that we sum over both signs and $a_i=\vec \epsilon_i \cdot \vec a/\sqrt{2}$. 
We find that they agree with the norm of the corresponding coherent states up to numeric constants.

\subsubsection{$C_n$} 
The Gaiotto-Whittaker vector corresponds to $C_n$ can be obtained from the $W(D_{n+1})$-algebra. We can just follow the same procedure as $B_n$ case, but in this case we can do much easily. There is only one $W(D_n)$-generator that is not invariant under $\bZ_2$ which is $V^{(n)}$. In terms of free bosons, it only shifts the sign of one of the bosons.  Therefore, the Kac-Shapovalov matrix is just a number, we can simply use \eqref{Kac2}. We get the norm of the Gaiotto-Whittaker vector to be
\be
 \vev{\cG_1 | \cG_1} \propto \frac{1}{\prod_{i=1}^n (Q^2 - 4 a_i^2)} . 
\ee

It is known that the moduli space (neglecting the centre of mass contribution) of one $\Sp(n)$ instanton is $\BC^{2n}/\BZ_2$, and the corresponding Hilbert series is given by (3.32) of \cite{Benvenuti:2010pq}, whose $\beta\to 0$ limit can be easily taken. Or, equivalently, note that $\U(1)^2\times \Sp(n)$ acts on $\bC^{2n}$ with the eigenvalues $(\epsilon_1+\epsilon_2)/2\pm a_i$. Therefore the integral is just 
\bea
Z_{\Sp(n) ,1} = \frac12 \frac1{\prod_{i=1}^n (Q^2/4 - a_i^2)}
\eea where $a_i= \vec e_i \cdot \vec a$ and the factor $1/2$ comes from the orbifolding.
This expression can, of course, also be obtained from Formula \eref{Z1instcentre}. We see that they agree completely up to a multiplicative constant. 

\subsubsection{$G_2$}
There is only one $W$-generator that has a $-{1}/{3}$ mode: $W^{(4, \frac{2}{3})} = W^{(4)} + i \tilde{W}^{(4)}$. Therefore, the Kac-Shapovalov matrix is one-dimensional, which is given by
\be
 \vev{\vec a | W^{(4, \frac{2}{3})}_{1/3}  W^{(4, \frac{2}{3})}_{-1/3} | \vec a } 
 \propto  \frac{1}{4} \left(Q^2 - 6 a_2^2\right) \left[ 4 Q^4 -12 Q^2 \left(3 a_1^2+a_2^2\right) + 9 \left(3 a_1^2 - a_2^2 \right)^2 \right] . 
\ee
We can use the formula \eqref{Kac2} for this example as well. The norm of Gaiotto-Whittaker state is given as the inverse of the the above expression. 

The Hilbert series was given in  (5.48) and (5.49) of \cite{Benvenuti:2010pq}.  Taking the limit $\beta \rightarrow 0$, we obtain
\be
Z_{G_2,1}  = \frac{72}{\left(Q^2-6 a_2^2\right) \left[4 Q^4 - 12 Q^2 \left(3 a_1^2+a_2^2\right) +9 \left(3 a_1^2 - a_2^2\right)^2 \right]}
\ee
in our variables.\footnote{Here $\vec a_\text{ours} = \sqrt{3} \vec a_\text{theirs}$} Here, $a_i=\vec\epsilon_i\cdot\vec a$.
This expression can also be obtained from Formula \eref{Z1instcentre}. See the section \ref{nonsimplydata} for $G_2$ for explicit expression for the roots and their basis. We see that the Hilbert series result completely agrees with the norm of Gaiotto-Whittaker state up to a multiplicative constant.

\subsubsection{$F_4$}
The Kac-Shapovalov matrix we need to compute is 
\be
 K_{F_4} = \left(
 \begin{array}{cc}
  \vev{-\vec{a}| W^{(5)}_{1/2} W^{(5)}_{-1/2} | \vec{a}} &   \vev{-\vec{a}| W^{(5)}_{1/2} W^{(9)}_{-1/2} | \vec{a}} \\
    \vev{-\vec{a}| W^{(9)}_{1/2} W^{(5)}_{-1/2} | \vec{a}} &   \vev{-\vec{a}| W^{(9)}_{1/2} W^{(9)}_{-1/2} | \vec{a}}
  \end{array}
 \right).
\ee

Honest computation of this matrix is too time consuming for a desktop computer of 2011, due to the complication in the evaluation of the normal ordering of twisted bosons in the expressions involving $W^{(9)}$ as in \eqref{regularizedtwisted}.
Thankfully, the Kac determinant is known in closed form in \eqref{Kac2}. 
Therefore we find \begin{equation}
\vev{\cG_1|\cG_1} =  \vev{-\vec{a}| W^{(5)}_{1/2} W^{(5)}_{-1/2} | \vec{a}}  / (\text{Kac determinant}),
\end{equation} which is fairly straightforward to compute. It was checked that it agrees with the instanton expression \eqref{Z1instanton}.
The explicit results for $E_6$ and $F_4$ are too lengthy to put here. They can be found in the supplementary files.

\section*{Acknowledgements}
The authors thank the initial wavefunction of the universe, which arranged so that NM and JS visited IPMU on the same week, without which this paper would not have been written.
They also thank Kris Thielemans, who provided them the Mathematica package \texttt{OPEdefs.m}, without which the calculation was not possible. 
They also thank IPMU mathematicians Alexey Bondal and Scott Carnahan for their help in simplifying the Hilbert series into the closed form. 
They also thank Amihay Hanany, Lotte Hollands and Michael Kay for useful discussions.
They also thank Nick Dorey for his advice on the correct usage of the English definite article \emph{the} used in the title of the paper.
NM and JS also thank the hospitality of IPMU during their visit. 

The work of CAK is supported by a John A. McCone Postdoctoral Fellowship. 
The work of NM is supported by a research grant of the Max Planck Society.
This work is in addition supported in part by the DOE grant DE-FG03-92-ER40701. 
The  work of YT is  supported in part by World Premier International Research Center Initiative
(WPI Initiative),  MEXT, Japan through the Institute for the Physics and Mathematics
of the Universe, the University of Tokyo.

\appendix

\section{Roots of simple Lie algebras}\label{data}

Here we list  the roots for all Lie algebras, emphasising  how to embed the root space of a non-simply-laced algebra $G$ to that of a simply-laced algebra $\Gamma$. First let us present the simply-laced ones in detail.  Note that the simple roots are named as  in \fref{untwisted}.  

\subsection{Simply-laced algebras}

\paragraph{Roots of $A_n = \SU(n+1)$.}  
We let  $\{ \vec e_i : 1 \leq i  \leq n+1 \}$ be an orthonormal basis. The positive roots are
\bea
\Delta^+  = \{\vec e_i -\vec e_j:   1 \leq i < j \leq n+1 \}~.
\eea
Note that the span of roots is only $n$-dimensional.
The simple roots are \begin{equation}
\vec \alpha_1=\vec e_1-\vec e_2, \quad \vec \alpha_2=\vec e_2-\vec e_3, \quad \ldots, \quad \vec \alpha_n=\vec e_n-\vec e_{n+1}~. \label{Ansimpleroots}
\end{equation}

\paragraph{Roots of $D_n = \SO(2n)$.} The positive roots are
\bea
\Delta^+ = \{ \vec e_i \pm \vec e_j :  1 \leq i < j \leq n \}~.
\eea
The simple roots are 
\begin{equation}
\vec\alpha_1=\vec e_1-\vec e_2, \quad \vec\alpha_2=\vec e_2-\vec e_3, \quad \ldots, \quad  \vec \alpha_{n-1}=\vec e_{n-1}-\vec e_n,\ 
\vec\alpha_n=\vec e_{n-1}+\vec e_n. \label{Dnsimpleroots}
\end{equation}

\paragraph{Roots of $E_6$.} The 36 positive roots are
\bea
\Delta^+ &=& \{\vec e_i + \vec e_j , \vec e_i - \vec e_j \}_{i < j \leq 5}  
 \cup \left \{ \frac{1}{2} ( \pm \vec e_1 \pm \vec e_2 \pm \vec e_3 \pm \vec e_4 \pm \vec e_5 +\sqrt{3}\vec e_6) \right \}_{\text{\# minus signs even}}~.
\eea 
The simple roots are 
\begin{align}
 \vec\alpha_1&= \frac{1}{2} \left(\vec e_1 - \vec e_2 - \vec e_3 - \vec e_4 -\vec e_5 + \sqrt3 e_6\right)~, \nonumber \\
 \vec\alpha_i &= \vec e_i - \vec e_{i-1}  \ (i=2,3,4,5), \quad\vec \alpha_6 = \vec e_1+ \vec e_2~.  \label{E6simpleroots}
\end{align}

\paragraph{Roots of $E_7$.} The 63 positive roots are
\bea
\Delta^+ &=& \{\vec e_i + \vec e_j, \vec e_i - \vec e_j \}_{i < j \leq 6} \cup \{\sqrt{2} \vec e_7 \}  
 \cup \left \{ \frac{1}{2} ( \pm \vec e_1 \pm \ldots \pm \vec e_6 +\sqrt{2}\vec e_7) \right \}_{\text{\# minus signs odd}}\hspace*{-5em},\hspace*{5em}
\eea

\paragraph{Roots of $E_8$.} The 120 positive roots are
\bea
\Delta^+ &=&\{\vec e_i + \vec e_j \}_{i < j \leq 8}  \cup \{\vec e_i - \vec e_j \}_{i < j \leq 8} 
\cup \left \{ \frac{1}{2} ( \pm \vec e_1 \pm \ldots \pm \vec e_7 + \vec e_8) \right \}_{\text{\# minus signs even}}~\hspace*{-5em}.\hspace*{5em}
\eea

\begin{figure}
\[
\begin{array}{r@{\quad}l@{\qquad}@{\quad}l}
A_n & \node{}{\alpha_1} - \node{}{\alpha_2} - \cdots - \node{}{\alpha_{n -1}}-\node{}{\alpha_n } &
\begin{array}{c}
\raisebox{-12pt}{\rotatebox{30}{$-\!\!-\!\!-$}}\Node{}{1}\raisebox{0pt}{\rotatebox{-30}{$-\!\!-\!\!-$}} \\[-7pt]
\node{}{1}-\node{}{1}-\cdots-\node{}{1} 
\end{array}\\
B_n  & \node{}{\alpha_1} - \node{}{\alpha_2} - \cdots - \node{}{\alpha_{n -1}} \Rightarrow\node{}{\alpha_n } \tikz[na]\node(B){};&
 \node{}{1}-\node{\Ver{}{1}}{2}-\node{}{2}-\cdots-\node{}{2}\Rightarrow\node{}{1}  \ \tikz[na]\node(B1){};\\
C_n  & \node{}{\alpha_1} - \node{}{\alpha_2} - \cdots - \node{}{\alpha_{n -1}} \Leftarrow\node{}{\alpha_n } \tikz[na]\node(C){};&
\Node{}{1}\Rightarrow \node{}{1}-\cdots-\node{}{1}\Leftarrow\node{}{1} \ \tikz[na]\node(C1){};\\
D_n & \node{}{\alpha_1} - \node{}{\alpha_2} - \cdots - \node{\ver{}{\alpha_{n }}}{\alpha_{n -2}} -\node{}{\alpha_{n -1}} &
\node{}{1}-\node{\Ver{}{1}}{2}-\node{}{2}-\cdots-\node{\ver{}{1}}{2}-\node{}{1} \\
E_6& \node{}{\alpha_1} - \node{}{\alpha_2} - \node{\ver{}{\alpha_{6}}}{\alpha_{3}} -\node{}{\alpha_{4}} -\node{}{\alpha_5} &
\node{}{1}-\node{}{2}-\node{\overset{\Ver{}{1}}{\ver{}{2}}}{3}-\node{}{2}-\node{}{1} \\
E_7& \node{}{\alpha_1} - \node{}{\alpha_2} - \node{\ver{}{\alpha_{7}}}{\alpha_{3}} -\node{}{\alpha_{4}} -\node{}{\alpha_5} -\node{}{\alpha_6} &
 \Node{}{1}-\node{}{2}-\node{}{3}-\node{\ver{}{2}}{4}-\node{}{3}-\node{}{2}-\node{}{1}\\
E_8& \node{}{\alpha_1} - \node{}{\alpha_2} - \node{}{\alpha_{3}} -\node{}{\alpha_{4}} -\node{\ver{}{\alpha_{8}}}{\alpha_5} -\node{}{\alpha_6}-\node{}{\alpha_7} &
 \Node{}{1}-\node{}{2}-\node{}{3}-\node{}{4}-\node{}{5}-\node{\ver{}{3}}{6}-\node{}{4}-\node{}{2} \\
F_4 & \node{}{\alpha_1} - \node{}{\alpha_2} \Rightarrow \node{}{\alpha_3} -\node{}{\alpha_4}&
\Node{}{1}-\node{}{2}-\node{}{3}\Rightarrow\node{}{2}-\node{}{1} \quad\tikz[na]\node(F41){};\\
G_2 & \node{}{\alpha_1} \Rrightarrow \node{}{\alpha_2} &
 \Node{}{1}-\node{}{2}\Rrightarrow\node{}{1} \quad\tikz[na]\node(G21){};
\end{array}
\]
\caption{Dynkin diagrams of simple Lie algebras, our labeling of the simple roots, and the comarks. The extended node is shown by a black blob. \label{untwisted} }

\end{figure}

\subsection{Non-simply-laced algebras}\label{nonsimplydata}
Let $\Gamma$ be a simply-laced algebra which is not $ A_{2n}$,
and let $o$ be a symmetry of the Dynkin diagram.  Note that $o$ can be viewed as an outer-automorphism acting on the Lie algebra of type $\Gamma$.
It can be checked that  $\alpha$ and $o(\alpha)$ are perpendicular
when $\alpha\neq o(\alpha)$.
Let us take
\begin{alignat}{3}
\Delta_s &= \{\vec\alpha &&: \vec\alpha\in\Delta, \ \vec\alpha=o(\vec\alpha)\}~,  \label{Ds} \\
\Delta_l &= \{\vec\alpha + o(\vec\alpha) &&: \vec\alpha\in\Delta, \ \vec\alpha\neq o(\vec\alpha)\}  &\text{when $r=2$}~, \label{Dl2} \\
\Delta_l &= \{\vec\alpha + o(\vec\alpha) + o^2(\vec\alpha) &&: \vec\alpha\in\Delta, \ \vec\alpha\neq o(\vec\alpha)\}  &\text{when $r=3$}~. \label{Dl3}
\end{alignat}
Then, $\Delta'=\Delta_s \sqcup \Delta_l$ is a non-simply-laced root system, with $\Delta_s$ and $\Delta_l$ short and long roots respectively.  In this  normalization,  the short root has length $\sqrt2$.   
Possible outer-automorphisms of simply-laced algebras that can be used are depicted in \fref{outer}. Every non-simply-laced algebra arises in this manner.

Note that one can also take the averages $\frac{1}{2}[\vec\alpha + o(\vec\alpha)]$ and $\frac{1}{3}[\vec\alpha + o(\vec\alpha) + o^2(\vec\alpha)]$ in \eref{Dl2} and \eref{Dl3} respectively.  A root system obtained this way corresponds to the subalgebra of $\Gamma$ invariant under the $\bZ_r$ action, and is Langlands dual to the one obtained 
via \eqref{Ds}--\eqref{Dl3}.
It is natural in this latter convention for the long roots to have length $\sqrt2$.  
Even when the resulting root system is the same in two ways of folding, as in $G=G_2$ and $\Gamma=\SO(8)$, the embedding of the root space of $G$ into the root space of $\Gamma$ is  different in the two cases.  
In this paper, we adhere to the former convention of folding.  This convention is more natural in the context of singularity theory, see e.g.~\cite{Slodowy,Arnold}.  Let us now go over the root systems of the non-simply-laced algebras one by one.

\paragraph{Roots of $B_n = \SO(2n+1)$.}   The short positive roots and the long positive roots are
\begin{align}
\Delta^+_s &= \{\vec\epsilon_i,  :  1 \leq i \leq n \},~  &
\Delta^+_l &= \{ \vec\epsilon_i \pm \vec\epsilon_j :  1 \leq i < j \leq n \}~. \label{posrootBn}
\end{align}
where $\vec\epsilon_i\cdot\vec\epsilon_j=2 \delta_{ij}$ so that the lengths of the short roots are $\sqrt{2}$.

This root system comes from a $\BZ_2$ outer-automorphism $o$ of  the root system of $A_{2n-1}=\SU(2n)$ with the action mapping the simple root $\al_i $  to $\al_{2n-i} $. 
The short and long positive roots of $B_n$ are then respectively
\begin{align}
 \Delta_s^+ &= \{ -\vec e_i + \vec e_{2n+1-i} : i = 1, \ldots,n \}~, \label{DseBn}\\
 \Delta_l^+ &= \{ ( -\vec e_i + \vec e_{2n+1-i} ) \pm ( -\vec e_j + \vec e_{2n+1-j} ) : 1 \leq i < j \leq n   \} . \label{DleBn}
\end{align}
Setting 
\bea
\vec\epsilon_i = -\vec e_i + \vec e_{2n+1-i}, \quad  i = 1, \ldots,n~,
\eea
the positive roots \eref{DseBn} and \eref{DleBn} become \eref{posrootBn} as required.

\paragraph{Roots of $C_n = \Sp(n)$.}  
The short positive roots and the long positive roots are
\begin{align}
\Delta^+_s &= \{ \vec e_i \pm \vec e_j :  1 \leq i < j \leq n \} ,&
\Delta^+_l &= \{  2 \vec e_i :  1 \leq i \leq n \}~, \label{posrootCn}
\end{align}
where $\vec e_i\cdot\vec e_j= \delta_{ij}$ so that the lengths of the short roots are $\sqrt{2}$.

This root system is obtained by applying a $\BZ_2$ outer-automorphism to the root system of $D_{n+1} = \SO(2n+2)$, which acts as follows:
\bea
o: \al_{n-1}~\mapsto~\al_{n}, \quad \al_{n} ~\mapsto~\al_{n-1}, \quad \al_i ~\mapsto~\al_i ~~ \text{for $i \neq n, n-1$}~.
\eea
From \eref{Dnsimpleroots}, it is clear that $o$ maps $\vec e_{n}$ to $-\vec e_{n}$ and leaves other $e_i$ invariant.  
Following the procedure, we easily get \eqref{posrootCn}.

\paragraph{Roots of $F_4$.} The short positive roots are
\bea
\Delta^+_s = \{\vec\epsilon_i \}_{i \leq 4} \cup \left \{ \frac{1}{2} ( \vec\epsilon_1 \pm \vec\epsilon_2 \pm \vec\epsilon_3 \pm \vec\epsilon_4) \right \}~, \label{DpsF4}
\eea
and the long positive roots are 
\bea
\Delta^+_l =  \{\vec\epsilon_i + \vec\epsilon_j \}_{i < j \leq 4}  \cup \{\vec\epsilon_i - \vec\epsilon_j \}_{i < j \leq 4}~, \label{DlsF4}
\eea  
where $\vec\epsilon_i\cdot \vec\epsilon_j= 2 \delta_{ij}$ so that the length of the short roots is $\sqrt{2}$. 

This root system can be obtained by applying a $\BZ_2$ outer-automorphism to the root system of $E_6$
which maps $(\al_1, \al_2, \al_3, \al_4, \al_5, \al_6)$ to $(\al_5, \al_4, \al_3, \al_2, \al_1, \al_6)$. 
Therefore the simple roots of $F_4$ are given by
\bea
\widehat{\vec\alpha}_1=\al_1+\al_5,\quad 
\widehat{\vec\alpha}_2=\al_2+\al_4,\quad 
\widehat{\vec\alpha}_3=\al_3,\quad 
\widehat{\vec\alpha}_4=\al_6.
\label{newalpha1}
\eea
We then introduce new basis $\e_{1,2,3,4}$ via 
\bea
\widehat{\al_1} = \e_2 -\e_3~, \quad \widehat{\al_2} = \e_3 -\e_4~, \quad  \widehat{\al_3} = \e_4~, \quad \widehat{\al_4} = \frac{1}{2}(\e_1 - \e_2 - \e_3 - \e_4)~.
\eea
These simple roots give rise to the positive roots listed in \eref{DpsF4} and \eref{DlsF4}.

\paragraph{Roots of $G_2$.} The short positive roots and long positive roots are
\bea
 \Delta^+_s = \left \{\sqrt{2} \vec \e_1,~ \pm \frac{1}{\sqrt{2}} \vec \e_1 + \sqrt{\frac{3}{2}}  \vec \e_2 \right\}~, \qquad  \Delta^+_l = \left \{ \pm \frac{3}{\sqrt{2}} \vec \e_1 + \sqrt{\frac{3}{2}} \vec \e_2 , ~ \sqrt{6} \vec \e_2  \right \} , \label{G2roots}
\eea
where $\vec\epsilon_i\cdot \vec\epsilon_j=  \delta_{ij}$ so that the length of the short roots is $\sqrt2$. 

This root system is obtained by applying a $\BZ_3$ outer-automorphism to the root system of $D_4$, which maps the set of simple roots $(\al_1, \al_2, \al_3, \al_4)$ to $(\al_3, \al_2, \al_4, \al_1)$.  
The simple roots of $G_2$ is then \begin{equation}
\widehat\alpha_1=\al_1+\al_3+\al_4,\quad
\widehat\alpha_2=\al_2.
\end{equation}
We then let \begin{equation}
\vec\e_1= \frac{1}{\sqrt{2}}(\vec e_1-\vec e_3), \quad
\vec \e_2= \frac{1}{\sqrt{6}}(\vec e_1+2\vec e_2-\vec e_3),
\end{equation} resulting in \eqref{G2roots}.

\section{Hilbert series of the one-instanton moduli space}\label{1instHilbert}

\begin{table}
\[
\begin{array}{l|rrr||l|rrr}
 & |W|& |\Delta_l|  &G_0   &
 	 & |W|& |\Delta_l|  &G_0  \\
\hline 
A_n  & (n+1)! & n^2+n & A_{n-2} &
	B_n  & 2^n n! & 2n(n-1) & A_1\times B_{n-2}\\
D_n & 2^{n-1} n!  & 2n(n-1) & A_1\times D_{n-2} &
	C_n & 2^n n! & 2n &  C_{n-2} \\
 E_6  & 72\cdot 6! & 72& A_5&
	 F_4   & 1152 & 24 & C_3\\
 E_7   & 72\cdot 8!  & 126& D_6 &
	 G_2   & 12 & 6 & A_1\\
 E_8 & 192 \cdot 10! & 240 & E_7  &&&&
 \end{array}
\]

\caption{Additional data of groups, required for the analysis in Appendix~\ref{1instHilbert}.   \label{t:groupdata}}
\end{table}

In this appendix we derive expression (\ref{Z1instanton}) by computing the
Hilbert series of the one-instanton moduli space.
Let $V(\vec w)$ be the highest weight representation of $G$ of highest weight $\vec w$.
As is conventional, we denote the highest root by $-\vec\alpha_0$, 
so that $V(-\vec\alpha_0)$ is the adjoint representation.
As explained in Sec.~\ref{s:1instanton}, the holomorphic function on the centred 1-instanton moduli space $\tilde \cM_{G,1}$ has the irreducible decomposition
\begin{equation}
V=\bigoplus_{m=0}^\infty V(-m\vec\alpha_0)\otimes T^{\otimes m} \ ,
\end{equation} 
under the action of $\U(1)\times G$,
where $T$ is a one-dimensional representation of $\U(1)$. The character, or equivalently the Hilbert series is then 
\begin{equation}
Z=\tr_V e^{\vec\phi} e^\mu
\end{equation} 
where $\vec\phi$ is an element of the Cartan subalgebra of $G$, and $e^\mu$ is the $\U(1)$ action.
We will abbreviate $e^{\vec\alpha\cdot \vec\phi}$ as $e^{\vec\alpha}$.
The Weyl character formula then gives 
\begin{align}
Z&= \sum_{m=0}^\infty e^{m\mu} \frac{\sum_{w\in W} s(w) e^{-wm\vec\alpha_0+w\vec\rho}}{\prod_{\vec\alpha\in \Delta^+} (e^{\vec\alpha/2}-e^{-\vec\alpha/2})}
= \frac{\sum_{w\in W} s(w) e^{w\vec\rho}/(1-e^{\mu-w\vec\alpha_0}) }{\prod_{\vec\alpha\in \Delta^+} (e^{\vec\alpha/2}-e^{-\vec\alpha/2})} \nonumber \\
&= \sum_{\vec\gamma\in\Delta_l}  \frac1{(1-e^{\mu+\vec\gamma})\prod_{\vec\alpha\in \Delta^+} (e^{\vec\alpha/2}-e^{-\vec\alpha/2})}
\sum_{\substack{w\in W\\ -w\vec\alpha_0=\vec\gamma}} s(w) e^{w\vec\rho} .\label{foo}
\end{align} 
Here, $\Delta$ is the set of roots, $\Delta_l$ is the set of long roots, $\Delta^+$ is the set of positive roots, $\vec\rho$ is the Weyl vector, and $s(w)$ is the sign of an element $w$ of the Weyl group $W$.

Now, consider a subgroup  $G_0$ of $G$, whose Dynkin diagram  is formed by the nodes of the Dynkin diagram of $G$ which is \emph{not} connected to the extended node of the affine Dynkin diagram, see Figure~\ref{untwisted} and Table~\ref{t:groupdata}. 
The Weyl group of $G_0$ fixes $-\vec\alpha_0$ by construction. 
Furthermore, $|W(G)|= |\Delta_l|\cdot |W(G_0)|$. Therefore, $W(G_0)$ is exactly the subgroup of $W(G)$ which fixes $-\vec\alpha_0$.

The difference  $\vec\rho(G)-\vec\rho(G_0)$ of the Weyl vectors of $G$ and $G_0$ 
is perpendicular  to all $\vec\alpha_i$ of $G_0$, and therefore is proportional to $-\vec\alpha_0$.
Therefore there is a constant $c$ such that
\begin{equation}
\vec\rho(G)=-c\vec\alpha_0 + \vec\rho(G_0).
\end{equation}  
To fix $c$, take the inner product with respect to $-\vec\alpha_0$,
using the expansion $-\vec\alpha_0^\vee=\sum n_i^\vee \alpha_i^\vee$ where $n_i^\vee$ are the comarks shown in \fref{untwisted}.
 We find \begin{equation}
2c=-\vec\alpha_0^\vee \cdot\vec\rho(G) = \sum n_i^\vee \vec\alpha_i^\vee\cdot\vec\rho(G)=\sum n_i^\vee = h^\vee-1.
\end{equation} 
Here we used that the sum of the comarks is  $h^\vee-1$.

For a  long root $\vec\gamma$, we pick a Weyl group element $w_{\vec\gamma}$ such that
$-w_{\vec\gamma}\vec\alpha_0=\vec\gamma$. The set of $w$ such that $-w\vec\alpha_0=\vec\gamma$ is then simply 
$w_{\vec\gamma} W(G_0)$. Therefore, 
\begin{multline}
\sum_{-w\vec\alpha_0=\vec\gamma} s(w) e^{w\rho} =
\sum_{w\in W(G_0)} s(w_{\vec\gamma} w) e^{w_{\vec\gamma} w\rho}  
=e^{-(h^\vee-1)w_{\vec\gamma} \vec\alpha_0/2 }s(w_{\vec\gamma})\sum_{w\in W(G_0)} s(w) e^{w_{\vec\gamma} w\rho(G_0)} \\
=e^{(h^\vee-1)\vec\gamma/2 } s(w_{\vec\gamma}) \prod_{\vec\alpha\in \Delta^+(G_0)} (e^{w_{\vec\gamma}\vec\alpha/2}-e^{-w_{\vec\gamma}\vec\alpha/2})\ .
\end{multline} 
Plugging it in to \eqref{foo}, we have 
\begin{align}
Z&=\sum_{\vec\gamma\in\Delta_l}  \frac{e^{(h^\vee-1)\vec\gamma/2 } s(w_{\vec\gamma}) \prod_{\vec\alpha\in \Delta^+(G_0)} (e^{w_{\vec\gamma}\vec\alpha/2}-e^{-w_{\vec\gamma}\vec\alpha/2}) }{(1-e^{\mu+\vec\gamma})\prod_{\vec\alpha\in \Delta^+} (e^{\vec\alpha/2}-e^{-\vec\alpha/2})} \nonumber \\
&=\sum_{\vec\gamma\in\Delta_l}  \frac{e^{(h^\vee-1)\vec\gamma/2 } s(w_{\vec\gamma}) \prod_{\vec\alpha\in \Delta^+(G_0)} (e^{w_{\vec\gamma}\vec\alpha/2}-e^{-w_{\vec\gamma}\vec\alpha/2}) }{(1-e^{\mu+\vec\gamma}) s(w_{\vec\gamma})\prod_{\vec\alpha\in \Delta^+} (e^{w_{\vec\gamma}\vec\alpha/2}-e^{-w_{\vec\gamma}\vec\alpha/2})} \nonumber \\
&=\sum_{\vec\gamma\in\Delta_l}  \frac{e^{(h^\vee-1)\vec\gamma/2 }  }{(1-e^{\mu+\vec\gamma})\prod_{\vec\alpha\in \Delta^+ \setminus \Delta^+(G_0) } (e^{w_{\vec\gamma}\vec\alpha/2}-e^{-w_{\vec\gamma}\vec\alpha/2})}\ . \label{bar}
\end{align}
Recall that the inner product $-\vec\alpha_0^\vee \cdot \vec\alpha$  is 
\begin{itemize}
\item 2 if and only if $\vec\alpha=-\vec\alpha_0$, 
\item 1 if and only if $\vec\alpha\in \Delta^+\setminus \Delta^+(G_0)$ and $\vec\alpha\neq -\vec\alpha_0$,
\end{itemize} 
so that \eqref{bar} can be further written as 
\begin{align}
&=\sum_{\vec\gamma\in\Delta_l}  \frac{e^{(h^\vee-1)\vec\gamma/2 }  }{(1-e^{\mu+\vec\gamma}) (e^{-w_{\vec\gamma}\vec\alpha_0/2}-e^{w_{\vec\gamma}\vec\alpha_0/2} ) \prod_{-\vec\alpha_0^\vee\cdot\vec\alpha=1 } (e^{w_{\vec\gamma}\vec\alpha/2}-e^{-w_{\vec\gamma}\vec\alpha/2})} \nonumber \\
&=\sum_{\vec\gamma\in\Delta_l}  \frac{e^{(h^\vee-1)\vec\gamma/2 }  }{(1-e^{\mu+\vec\gamma}) (e^{\vec\gamma/2}-e^{-\vec\gamma/2} ) \prod_{\vec\gamma^\vee\cdot\vec\alpha=1 } (e^{\vec\alpha/2}-e^{-\vec\alpha/2})}\ . \label{baz}
\end{align}  
Now the elements $w_{\vec \gamma}$ is gone.

Note that $|\{ \vec\alpha \ |\  \vec\gamma^\vee\cdot\vec\alpha=1\}|$ is $2h^\vee-4$, so that \eqref{baz} has $2h^\vee-2$ terms in the denominator, which is the expected number for the centered instanton moduli space. 
To get the explicit expressions, we let 
\bea
e^{\vec\alpha}\equiv e^{\vec\alpha\cdot\phi} = e^{\beta\vec\alpha\cdot\vec a}, \qquad e^\mu=e^{\beta(\epsilon_1+\epsilon_2)}.
\eea
In the papers on Hilbert series, e.g.~\cite{Benvenuti:2010pq}, the variables $x_i=e^{a_i}$ were used instead.
The 4d version is obtained by taking the $\beta\to 0$ limit, giving 
\begin{equation}
\beta^{2h^\vee-2} Z\to \sum_{\vec\gamma\in\Delta_l}  \frac{-1} {(\epsilon_1+\epsilon_2+\vec\gamma\cdot\vec a) (\vec\gamma\cdot \vec a ) \prod_{\vec\gamma^\vee\cdot\vec\alpha=1 } (\vec\alpha\cdot \vec a)}. \label{Z1instcentre}
\end{equation}
Together with the contribution of $\mathbb{C}^2$ we indeed get (\ref{Z1instanton}).

\section{Kac determinant at the lowest level}\label{KacDeterminant}
Here we consider the Kac determinant at the lowest level. 
The result for the untwisted case is well known, see e.g.~\cite{Mizoguchi:1988vk,Bouwknegt:1992wg}. The twisted case also follows straightforwardly by modifying the derivation for the untwisted case. For example, the Kac determinant for the Drinfeld-Sokolov reduction with respect to the minimal nilpotent was determined in \cite{KacWakimoto}. Here we need to perform it with respect to the principal nilpotent.

First, consider the untwisted Verma module.
Let the zero modes of the free bosons be $\vec\bo_0$.
Pick a simple root $\vec\alpha_i$, and 
let $\varphi_i=\vec\alpha_i \cdot \vec\varphi$. 
As explained in Sec.~\ref{WE6preliminary}, elements in the W-algebra can be expanded in terms of the energy-momentum tensor $T_i$ for $\varphi_i$ and the free bosons perpendicular to $\varphi_i$. 
At level one, $(T_i)_{-1} = (\vec\alpha_i\cdot\vec\bo_0) \bo_{i,-1}$,
and similarly, the $(-1)$-mode of any operator constructed out of $T_i$ comes with a factor of $\vec\alpha_i\cdot\vec\bo_0$.
Thus,  there is a null state in the W-algebra Verma module when $\vec\alpha_i\cdot\vec\bo_0=0$. 
Therefore the Kac determinant at level one is divisible by $\vec\alpha_i\cdot\vec\bo_0=\vec \alpha_i\cdot \vec a+Q$. Due to the shifted Weyl invariance, the Kac determinant is divisible by $\vec\alpha\cdot \vec a + Q$ for all roots $\alpha$. 
The Kac determinant has degree $2\sum (w_i-1)$ in $\vec a$, which equals the number of roots.
It follows that 
\begin{equation}
\text{(Kac determinant at level 1)} \propto \prod_{\vec\alpha\in\Delta } (\vec\alpha\cdot \vec a + Q)\ .
\end{equation} 

Next, consider the states at level $-1/r$ of the $\bZ_r$-twisted Verma module.
Denote by $o$ the $\bZ_r$ action, under which the zero mode $\vec\bo_0$ is invariant.
We regard it as in the Cartan of $G$, the S-dual of the $\bZ_r$-invariant subalgebra  of $\Gamma$. 
Recall that the long simple roots $\gamma_i$ of $G$ are given by $\gamma_i=\vec\alpha_i+\vec\alpha_{o(i)}$ when $r=2$ and similarly for $r=3$, \eqref{Dl2}, \eqref{Dl3}.

Take a simple root such that $\vec\alpha_i\ne \vec\alpha_{o(i)}$.  
Assuming $\Gamma\neq A_{2n}$, 
we have $\vec\alpha_i \cdot\vec\alpha_{o(i)}=0$. 
Therefore the operators in the W-algebra can be written in terms of the energy-momentum tensors $T_i$, $T_{o(i)}$ (and $T_{o^2(i)}$ if $r=3$) for the free bosons $\varphi_i$,  $\varphi_{o(i)}$ (and  $\varphi_{o^2(i)}$) and free bosons perpendicular to $\varphi_i$,  $\varphi_{o(i)}$ (and  $\varphi_{o^2(i)}$).
Now, the level $-(1/r)$ states arise from $T_i - T_{o(i)}$ when $r=2$,
and from $T_i + e^{2\pi i/3} T_{o(i)} + e^{4\pi i/3} T_{o^2(i)}$ when $r=3$.
Recall that $T_i \propto -(\partial \varphi_i\partial \varphi_i) +  Q \partial^2\varphi_i/2$.
Therefore the $(-1/r)$-mode always arises with the combination \begin{equation}
(\vec\alpha_i\cdot\vec \bo_0-Q(1-1/r) ) \bo_{i,-1/r}.
\end{equation} 
It follows the Kac determinant has a zero when $\vec\alpha_i\cdot\vec \bo_0=Q(1-1/r)$ for a non-invariant simple root $\vec\alpha_i$, or in other words 
 $\vec\gamma_i\cdot \vec a+ Q=0$
 for a long root $\vec\gamma_i$ and the shifted zero mode $\vec a$. 
 From Weyl invariance, the Kac determinant has a factor $\vec\gamma\cdot\vec a+Q$ for each long root $\gamma$. 
The Kac determinant has degree $2\sum_i (w_i-1)$ where the sum is over the degrees $w_i$ of $W(\Gamma)$-generators not invariant under $\bZ_r$. This equals the number of long roots of $G$.
We thus conclude that
\begin{equation}
\text{(Kac determinant at level $1/r$)} \propto \prod_{\vec\gamma\in\Delta_l  } (\vec\gamma\cdot \vec a + Q).\label{twistedKac}
\end{equation}

\section{Construction of the W-algebra of type $E_6$}\label{WE6}
\subsection{Detailed procedure}
We provide some more details on the construction of the $W(E_6)$-algebra.
We first construct the two lowest generators $W^{(2)}$ and $W^{(5)}$ 
utilizing $W(A_5)$ subalgebra. The higher generators $W^{(6,8,9,12)}$
we then obtain from suitable OPEs of $W^{(5)}$. 

Let $J^{1,\ldots,6}=i\partial\varphi^{1,\ldots,6}$ be six orthonormal free bosons. 
Take the $A_2\times A_2\times A_1$ subalgebra of $E_6$, corresponding to the nodes of the Dynkin diagram except the central node. We can then choose the bosons in such a way that $J^{1,2}$, $J^{3,4}$ span the Cartan of the two $A_2$, and $J_{6}$ spans the Cartan of the $A_1$ respectively. 
Note that the $\bZ_2$ outer automorphism exchanges the two $A_2$ subalgebras.

Consider now the $A_5$ subalgebra of $E_6$. It contains $A_2\times A_2$, but not the $A_1$ described above. Construct the generators $U_{2,\ldots,5}(z)$ of the $W(A_5)$ algebra via \eqref{Miura-A}, and obtain the $\bZ_2$ eigenstates \eqref{WA5}.
We introduce the  boson 
\begin{equation}
K(z)=({\sqrt3J^5+J^6})/{2\sqrt{2}} \ ,
\end{equation} 
which is perpendicular to the Cartan of $A_5$. The most general ansatz for $W^{(2)}$ and $W^{(5)}$ compatible with $\bZ_2$ is then
\begin{align}
W^{(2)}&=U^{(2)} + c_1 K^2 + c_2  \partial K,\\
W^{(5)}&=\tilde U^{(5)}+  c_3 \tilde U^{(3)} U^{(2)} + c_4 \tilde U^{(3)} K^2 + c_5 \partial^2\tilde U^{(3)} + c_6 \partial\tilde U^{(3)} K + c_7\tilde U^{(3)} \partial K\ . 
\label{W2W5}
\end{align} 
The constants $c_{1,\ldots,7}$ can be determined by the method explained 
in section~\ref{WE6preliminary}:  we  expand (\ref{W2W5}) in the $J^i$,
and require that $J^6$ only appears in the form of its energy momentum tensor 
\begin{equation}
T(z)=-(J^6){}^2/2 + (Q/\sqrt{2}) \partial J^6\label{T6}\ .
\end{equation}  
This fixes the constants to be
\begin{equation}
c_1=1,\ 
c_2=-11Q,\ 
c_3=0,\ 
c_4=1,\ 
c_5=\frac{3}2Q^2,\ 
c_6=-3Q,\ 
c_7=2Q.
\end{equation}

The higher generators can now be obtained by repeatedly taking the OPEs of $W^{(5)}$. Let $[O_1O_2]_p(w)$ be the coefficient of $(z-w)^{-p}$ of the OPE $O_1(z)O_2(w)$. We let 
\begin{align}
W^{(6)}&=(1+12Q^2)^{-1}[W^{(5)}W^{(5)}]_4,&
W^{(8)}&=[W^{(5)}W^{(5)}]_2,\\
W^{(9)}&=[W^{(5)}W^{(6)}]_2,& 
W^{(12)}&=[W^{(6)}W^{(8)}]_2.\label{OPEWE6}
\end{align} 
These OPEs a priori are not guaranteed to generate independent fields, but could contain just products of  lower lying fields, possibly with derivatives.
Independence can be confirmed as follows. Let $P^{(d)}$ be the terms in $W^{(d)}$ which do not contain derivatives, regarded as polynomials in $J^i$.  It then suffices to check that they are algebraically independent, for which we can compute the Jacobian $\det(\partial P^{(d)}/\partial J^i)$ and check that it is non-zero. 
Note that W-generators constructed in this manner automatically have definite parity under $\bZ_2$.

Unfortunately these OPE calculations, when expressed in terms of six free bosons, are too much for a laptop computer of 2011. It is thus necessary to use the $A_2\times A_2\times A_1$ subalgebra to organize the computation. 
The generator of $W(A_1)$ is $T(z)$ in \eqref{T6}.
Let $U^{(2,3)}$ be the generators of the first $W(A_2)$, and 
$\tilde U^{(2,3)}$ be those of the second, as defined in \eqref{Miura-A}.
Then, we can rewrite $W^{(2)}$ and $W^{(5)}$ determined in \eqref{W2W5} in terms of $U^{(2,3)}$,  
$\tilde U^{(2,3)}$, $J_5$, and $T$.
Now the OPEs in \eqref{OPEWE6} can be and were performed using the known OPEs of $W(A_2)\times W(A_2)\times W(A_1)$ in about three hours on a 2GHz machine with 3Gbytes of memory;
the independence of the resulting generators can be and was checked.
The denominator $(1+12Q^2)$ for $W^{(6)}$ in \eqref{OPEWE6} is inserted because the OPE turns out to be divisible by this factor. 

\subsection{Explicit generators}
In the following, we list the generators thus constructed; we set $J=J_5$, $A_i=U^{(i)}+\tilde U^{(i)}$ and $B_i=U^{(i)}-\tilde U^{(i)}$. With explicit generators, it is easy to check that the level-1 contribution to the norm of the coherent state as calculated by \eqref{eq:norm} is equal to the 1-instanton result \eqref{Z1instanton}, once when the zero modes are substituted with random numbers.

\begingroup

\def\sqrt#1{\surd(#1)}
\def\baselinestretch{.1}
\lineskiplimit1pt
\lineskip.5pt
\mathsurround=0pt
\scriptspace=.1pt
\def\generatorhead{\subsubsection}
 \tiny

$\  $ 
 \generatorhead{$W_{2}$} \noindent $W_{2} = $%
$T - 7\sqrt{3/2}Q$%
$\partial J + $%
$A_2 + 
    $%
$J{}^{2}/2.
\  $ 
 \generatorhead{$W_{5}$} \noindent $W_{5} = (-Q/12 - 5Q{}^{3})$%
$\partial{}^{3} B_{2} + (1/4 + 3Q{}^{2})$%
$\partial{}^{2} B_{3} + 
    2\sqrt{6}Q{}^{2}$%
$J \partial{}^{2} B_{2} - 
    \sqrt{6}Q$%
$J \partial B_{3} + 3\sqrt{3/2}Q{}^{2}
     $%
$\partial J \partial B_{2} - \sqrt{3/2}Q$%
$\partial J B_3 + 
    \sqrt{2/3}Q{}^{2}$%
$\partial{}^{2} J B_2 + (Q$%
$A_2 \partial B_{2})/
     2 - $%
$A_2 B_3/2 - (Q$%
$\partial A_{2} B_2)/2 + 
    $%
$A_3 B_2/2 + 2Q$%
$B_2 \partial T + 
    Q$%
$\partial B_{2} T + $%
$B_3 T - 
    (3Q$%
$J{}^{2}\partial B_{2})/2 + 
    $%
$J{}^{2}B_3/2 - 
    \sqrt{2/3}$%
$J B_2 T - 
    Q$%
$\partial J J B_2 + 
    $%
$J{}^{3}B_2/(3\sqrt{6}).
\  $

 \generatorhead{$W_{6}$} 

 \noindent $W_{6} = (7 Q{}^{2}/3 + 70Q{}^{4})$%
$\partial{}^{4} T + 
    ((-7Q)/(9\sqrt{6}) - 245 Q{}^{3}/(3\sqrt{6}) - 1043\sqrt{2/3}Q{}^{5})
     $%
$\partial{}^{5} J + (13 Q{}^{2}/6 + 130Q{}^{4})$%
$\partial{}^{4} A_{2} + 
    (Q - 28Q{}^{3})$%
$\partial{}^{3} A_{3} - 16Q{}^{2}$%
$\partial T \partial T + 
    (-10/9 - 68 Q{}^{2}/3)$%
$\partial{}^{2} T T + 
    ((-2\sqrt{2/3}Q)/3 + 20\sqrt{6}Q{}^{3})$%
$J \partial{}^{3} T + 
    ((-2\sqrt{2/3}Q)/3 - 56\sqrt{2/3}Q{}^{3})$%
$J \partial{}^{3} A_{2} + 
    16\sqrt{6}Q{}^{2}$%
$J \partial{}^{2} A_{3} + 
    ((11\sqrt{2/3}Q)/3 + 370\sqrt{2/3}Q{}^{3})$%
$\partial J \partial{}^{2} T + 
    ((11\sqrt{2/3}Q)/3 + 232\sqrt{2/3}Q{}^{3})$%
$\partial J \partial{}^{2} A_{2} + 
    14\sqrt{6}Q{}^{2}$%
$\partial J \partial A_{3} + 
    (4\sqrt{6}Q + 536\sqrt{2/3}Q{}^{3})$%
$\partial{}^{2} J \partial T + 
    (-35Q{}^{2} - 4286 Q{}^{4}/3)$%
$\partial{}^{2} J \partial{}^{2} J + 
    (4\sqrt{6}Q + 458\sqrt{2/3}Q{}^{3})$%
$\partial{}^{2} J \partial A_{2} + 
    14\sqrt{2/3}Q{}^{2}$%
$\partial{}^{2} J A_3 + 
    ((23\sqrt{2/3}Q)/3 + 302\sqrt{2/3}Q{}^{3})$%
$\partial{}^{3} J T + 
    ((-125Q{}^{2})/3 - 1670Q{}^{4})$%
$\partial{}^{3} J \partial J + 
    ((23\sqrt{2/3}Q)/3 + 286\sqrt{2/3}Q{}^{3})$%
$\partial{}^{3} J A_2 + 
    (5/54 + 50 Q{}^{2}/9 + 90Q{}^{4})$%
$\partial{}^{4} J J + 
    (2/9 - 44 Q{}^{2}/3)$%
$A_2 \partial{}^{2} T + 
    (-4/3 - 52Q{}^{2})$%
$\partial A_{2} \partial T + (-1/3 - 30Q{}^{2})
     $%
$\partial A_{2} \partial A_{2} + 4Q$%
$\partial A_{2} A_3 + 
    (2/9 - 128 Q{}^{2}/3)$%
$\partial{}^{2} A_{2} T + (-1/9 - 104 Q{}^{2}/3)
     $%
$\partial{}^{2} A_{2} A_2 + 28Q$%
$A_3 \partial T - 
    4$%
$A_3 A_3 + 28Q$%
$\partial A_{3} T + 
    4Q$%
$B_2 \partial B_{3} + (1/3 + 20Q{}^{2})
     $%
$\partial B_{2} \partial B_{2} - 12Q$%
$\partial B_{2} B_3 + 
    (-1/3 + 8Q{}^{2})$%
$\partial{}^{2} B_{2} B_2 + 
    6$%
$B_3 B_3 - 8\sqrt{2/3}Q$%
$J \partial T 
      T + (1/9 - 130 Q{}^{2}/3)$%
$J{}^{2}\partial{}^{2} T + 
    (1/9 + 32 Q{}^{2}/3)$%
$J{}^{2}\partial{}^{2} A_{2} - 
    18Q$%
$J{}^{2}\partial A_{3} - 
    4\sqrt{6}Q$%
$J A_2 \partial T + 
    8\sqrt{2/3}Q$%
$J \partial A_{2} T + 
    8\sqrt{2/3}Q$%
$J \partial A_{2} A_2 - 
    20\sqrt{2/3}$%
$J A_3 T - 
    16\sqrt{2/3}Q$%
$J \partial B_{2} B_2 - 
    4\sqrt{6}Q$%
$\partial J T{}^{2}+ 
    (-4/3 - 96Q{}^{2})$%
$\partial J J \partial T + 
    (-4/3 - 12Q{}^{2})$%
$\partial J J \partial A_{2} - 
    16Q$%
$\partial J J A_3 + (2/9 + 124 Q{}^{2}/3)
     $%
$(\partial J){}^{2} T + 
    ((5\sqrt{2/3}Q)/3 + 20\sqrt{2/3}Q{}^{3})$%
$(\partial J){}^{3} + (2/9 + 172 Q{}^{2}/3)$%
$(\partial J){}^{2} A_2 - 
    28\sqrt{2/3}Q$%
$\partial J A_2 T - 
    3\sqrt{6}Q$%
$\partial J A_2{}^{2}- 
    \sqrt{6}Q$%
$\partial J B_2{}^{2}+ 
    (-10/9 - 124 Q{}^{2}/3)$%
$\partial{}^{2} J J T + 
    (-10/9 - 56 Q{}^{2}/3)$%
$\partial{}^{2} J J A_2 + 
    ((65\sqrt{2/3}Q)/3 + 806\sqrt{2/3}Q{}^{3})$%
$\partial{}^{2} J \partial J 
      J + (25 Q/(3\sqrt{6}) + 155\sqrt{2/3}Q{}^{3})
     $%
$\partial{}^{3} J J{}^{2}+ (8$%
$A_2 T{}^{2})/
     3 - (2$%
$A_2{}^{2}T)/3 + 
    (2$%
$A_2{}^{3})/3 - 
    (2$%
$A_2 B_2{}^{2})/3 + 
    (10$%
$B_2{}^{2}T)/3 + 
    (4$%
$J{}^{2}T{}^{2})/3 + 
    (28\sqrt{2/3}Q$%
$J{}^{3}\partial T)/3 - 
    (20\sqrt{2/3}Q$%
$J{}^{3}\partial A_{2})/3 + 
    (10\sqrt{2/3}$%
$J{}^{3}A_3)/3 + 
    (8$%
$J{}^{2}A_2 T)/3 - 
    $%
$J{}^{2}A_2{}^{2}/3 + 
    (5$%
$J{}^{2}B_2{}^{2})/3 - 
    6\sqrt{6}Q$%
$\partial J J{}^{2}A_2 + 
    (-5/9 - 4 Q{}^{2}/3)$%
$(\partial J){}^{2} J{}^{2}+ 
    (-5/9 - 146 Q{}^{2}/9)$%
$\partial{}^{2} J J{}^{3}- 
    (4$%
$J{}^{4}T)/9 + 
    (2$%
$J{}^{4}A_2)/3 - 
    (7\sqrt{2/3}Q$%
$\partial J J{}^{4})/
     3 + $%
$J{}^{6}/
     27.
\  $ 
 \generatorhead{$W_{8}$}
  \noindent $W_{8} = (7 Q{}^{2}/90 - 7 Q{}^{4}/5 - 28Q{}^{6})$%
$\partial{}^{6} T + 
    (-Q/(54\sqrt{6}) - 11 Q{}^{3}/(6\sqrt{6}) - 175 Q{}^{5}/(3\sqrt{6}) - 
      78\sqrt{6}Q{}^{7})$%
$\partial{}^{7} J + (1/1080 + 13 Q{}^{2}/180 + 104 Q{}^{4}/15 + 
      70Q{}^{6})$%
$\partial{}^{6} A_{2} + (-Q/60 - 5 Q{}^{3}/2 - 42Q{}^{5})$%
$\partial{}^{5} A_{3} + 
    (-5/54 - Q{}^{2} + 4 Q{}^{4}/3)$%
$\partial{}^{4} T T + 
    (-(\sqrt{2/3}Q)/15 + (41\sqrt{2/3}Q{}^{3})/5 + 36\sqrt{6}Q{}^{5})
     $%
$J \partial{}^{5} T + (-(\sqrt{2/3}Q)/15 - (43\sqrt{2/3}Q{}^{3})/5 - 
      28\sqrt{6}Q{}^{5})$%
$J \partial{}^{5} A_{2} + 
    (2\sqrt{6}Q{}^{2} + 22\sqrt{6}Q{}^{4})$%
$J \partial{}^{4} A_{3} + 
    (-Q/(18\sqrt{6}) + 71 Q{}^{3}/\sqrt{6} + 430\sqrt{2/3}Q{}^{5})
     $%
$\partial J \partial{}^{4} T + (-Q/(18\sqrt{6}) - 35 Q{}^{3}/(2\sqrt{6}) - 
      80\sqrt{2/3}Q{}^{5})$%
$\partial J \partial{}^{4} A_{2} + 
    (16\sqrt{2/3}Q{}^{2} + 52\sqrt{6}Q{}^{4})$%
$\partial J \partial{}^{3} A_{3} + 
    ((\sqrt{2/3}Q)/3 + 74\sqrt{2/3}Q{}^{3} + 280\sqrt{6}Q{}^{5})
     $%
$\partial{}^{2} J \partial{}^{3} T + ((\sqrt{2/3}Q)/3 + (20\sqrt{2/3}Q{}^{3})/3 + 
      20\sqrt{6}Q{}^{5})$%
$\partial{}^{2} J \partial{}^{3} A_{2} + 
    (43 Q{}^{2}/\sqrt{6} + 72\sqrt{6}Q{}^{4})$%
$\partial{}^{2} J \partial{}^{2} A_{3} + 
    ((2\sqrt{2/3}Q)/3 + 26\sqrt{6}Q{}^{3} + 280\sqrt{6}Q{}^{5})
     $%
$\partial{}^{3} J \partial{}^{2} T + ((-35Q{}^{2})/9 - 560 Q{}^{4}/3 - 1680Q{}^{6})
     $%
$\partial{}^{3} J \partial{}^{3} J + ((2\sqrt{2/3}Q)/3 + 65 Q{}^{3}/\sqrt{6} + 
      105\sqrt{6}Q{}^{5})$%
$\partial{}^{3} J \partial{}^{2} A_{2} + 
    (23 Q{}^{2}/\sqrt{6} + 35\sqrt{6}Q{}^{4})$%
$\partial{}^{3} J \partial A_{3} + 
    (\sqrt{2/3}Q + 18\sqrt{6}Q{}^{3} + 168\sqrt{6}Q{}^{5})$%
$\partial{}^{4} J \partial T + 
    ((-35Q{}^{2})/6 - 833 Q{}^{4}/3 - 2492Q{}^{6})$%
$\partial{}^{4} J \partial{}^{2} J + 
    (\sqrt{2/3}Q + 37\sqrt{2/3}Q{}^{3} + 105\sqrt{6}Q{}^{5})
     $%
$\partial{}^{4} J \partial A_{2} + (5 Q{}^{2}/\sqrt{6} + 7\sqrt{6}Q{}^{4})
     $%
$\partial{}^{4} J A_3 + (67 Q/(90\sqrt{6}) + 
      173 Q{}^{3}/(5\sqrt{6}) + 154\sqrt{2/3}Q{}^{5})$%
$\partial{}^{5} J T + 
    ((-77Q{}^{2})/36 - 595 Q{}^{4}/6 - 882Q{}^{6})$%
$\partial{}^{5} J \partial J + 
    (67 Q/(90\sqrt{6}) + (221\sqrt{2/3}Q{}^{3})/15 + 42\sqrt{6}Q{}^{5})
     $%
$\partial{}^{5} J A_2 + (1/324 + Q{}^{2}/9 + 14 Q{}^{4}/9 + 8Q{}^{6})
     $%
$\partial{}^{6} J J + (1/54 + Q{}^{2} + 52 Q{}^{4}/3)
     $%
$A_2 \partial{}^{4} T + (-Q/3 + Q{}^{3})$%
$A_2 \partial{}^{3} A_{3} + 
    ((-4Q{}^{2})/3 + 14Q{}^{4})$%
$\partial A_{2} \partial{}^{3} T + 
    3Q{}^{3}$%
$\partial A_{2} \partial{}^{2} A_{3} + (-5Q{}^{2} - 42Q{}^{4})
     $%
$\partial{}^{2} A_{2} \partial{}^{2} T + ((-9Q{}^{2})/2 - 63Q{}^{4})
     $%
$\partial{}^{2} A_{2} \partial{}^{2} A_{2} + (3 Q/2 + 21Q{}^{3})
     $%
$\partial{}^{2} A_{2} \partial A_{3} + ((-22Q{}^{2})/3 - 80Q{}^{4})
     $%
$\partial{}^{3} A_{2} \partial T + ((-19Q{}^{2})/3 - 82Q{}^{4})
     $%
$\partial{}^{3} A_{2} \partial A_{2} + (Q/3 + 9Q{}^{3})$%
$\partial{}^{3} A_{2} A_3 + 
    (1/54 - 17 Q{}^{2}/6 - 116 Q{}^{4}/3)$%
$\partial{}^{4} A_{2} T + 
    (-1/108 - 25 Q{}^{2}/12 - 83 Q{}^{4}/3)$%
$\partial{}^{4} A_{2} A_2 + 
    (4Q + 42Q{}^{3})$%
$A_3 \partial{}^{3} T + (9Q + 126Q{}^{3})
     $%
$\partial A_{3} \partial{}^{2} T + (-1 - 21Q{}^{2})$%
$\partial A_{3} \partial A_{3} + 
    (9Q + 132Q{}^{3})$%
$\partial{}^{2} A_{3} \partial T + (-1/2 - 18Q{}^{2})
     $%
$\partial{}^{2} A_{3} A_3 + (2Q + 48Q{}^{3})$%
$\partial{}^{3} A_{3} T + 
    9Q{}^{3}$%
$B_2 \partial{}^{3} B_{3} + (-2Q - 3Q{}^{3})
     $%
$\partial B_{2} \partial{}^{2} B_{3} + (9 Q{}^{2}/2 + 63Q{}^{4})
     $%
$\partial{}^{2} B_{2} \partial{}^{2} B_{2} + ((-3Q)/2 - 21Q{}^{3})
     $%
$\partial{}^{2} B_{2} \partial B_{3} + (-1/9 + 20 Q{}^{2}/3 + 82Q{}^{4})
     $%
$\partial{}^{3} B_{2} \partial B_{2} + ((-5Q)/3 - 19Q{}^{3})
     $%
$\partial{}^{3} B_{2} B_3 + (-1/36 + 7 Q{}^{2}/12 + 11Q{}^{4})
     $%
$\partial{}^{4} B_{2} B_2 + (1 + 21Q{}^{2})$%
$\partial B_{3} \partial B_{3} + 
    (2 + 24Q{}^{2})$%
$\partial{}^{2} B_{3} B_3 + 
    (-2\sqrt{2/3}Q - 8\sqrt{6}Q{}^{3})$%
$J \partial{}^{2} T \partial T + 
    (-2\sqrt{2/3}Q - 8\sqrt{6}Q{}^{3})$%
$J \partial{}^{3} T T + 
    (1/108 - 31 Q{}^{2}/6 - 190 Q{}^{4}/3)$%
$J{}^{2}\partial{}^{4} T + 
    (1/108 + 35 Q{}^{2}/12 + 74 Q{}^{4}/3)$%
$J{}^{2}\partial{}^{4} A_{2} + 
    ((-7Q)/3 - 16Q{}^{3})$%
$J{}^{2}\partial{}^{3} A_{3} + 
    ((-4\sqrt{2/3}Q)/3 - 14\sqrt{6}Q{}^{3})$%
$J A_2 \partial{}^{3} T + 
    (-(\sqrt{2/3}Q) - 22\sqrt{6}Q{}^{3})$%
$J \partial A_{2} \partial{}^{2} T + 
    (\sqrt{2/3}Q - 10\sqrt{6}Q{}^{3})$%
$J \partial{}^{2} A_{2} \partial T + 
    (\sqrt{6}Q + 16\sqrt{6}Q{}^{3})$%
$J \partial{}^{2} A_{2} \partial A_{2} - 
    2\sqrt{6}Q{}^{2}$%
$J \partial{}^{2} A_{2} A_3 + 
    ((5\sqrt{2/3}Q)/3 + 4\sqrt{6}Q{}^{3})$%
$J \partial{}^{3} A_{2} T + 
    ((4\sqrt{2/3}Q)/3 + 20\sqrt{2/3}Q{}^{3})$%
$J \partial{}^{3} A_{2} 
      A_2 + (-2\sqrt{2/3} - 8\sqrt{6}Q{}^{2})$%
$J A_3 
      \partial{}^{2} T + (-4\sqrt{2/3} - 18\sqrt{6}Q{}^{2})$%
$J \partial A_{3} 
      \partial T + 4\sqrt{6}Q$%
$J \partial A_{3} A_3 + 
    (-2\sqrt{2/3} - 12\sqrt{6}Q{}^{2})$%
$J \partial{}^{2} A_{3} T - 
    2\sqrt{6}Q{}^{2}$%
$J B_2 \partial{}^{2} B_{3} - 
    4\sqrt{6}Q{}^{2}$%
$J \partial B_{2} \partial B_{3} + 
    (-5\sqrt{2/3}Q - 16\sqrt{6}Q{}^{3})$%
$J \partial{}^{2} B_{2} \partial B_{2} - 
    2\sqrt{6}Q{}^{2}$%
$J \partial{}^{2} B_{2} B_3 + 
    ((-7\sqrt{2/3}Q)/3 - 20\sqrt{2/3}Q{}^{3})$%
$J \partial{}^{3} B_{2} 
      B_2 - 2\sqrt{6}Q$%
$J \partial B_{3} B_3 + 
    (-4\sqrt{2/3}Q - 16\sqrt{6}Q{}^{3})$%
$\partial J \partial T \partial T + 
    (-2\sqrt{6}Q - 24\sqrt{6}Q{}^{3})$%
$\partial J \partial{}^{2} T T + 
    ((-70Q{}^{2})/3 - 280Q{}^{4})$%
$\partial J J \partial{}^{3} T + 
    (11Q{}^{2} + 100Q{}^{4})$%
$\partial J J \partial{}^{3} A_{2} + 
    (-9Q - 60Q{}^{3})$%
$\partial J J \partial{}^{2} A_{3} + 
    (-10Q{}^{2} - 120Q{}^{4})$%
$(\partial J){}^{2} \partial{}^{2} T + 
    (20Q{}^{2} + 219Q{}^{4})$%
$(\partial J){}^{2} \partial{}^{2} A_{2} + 
    (-7Q - 51Q{}^{3})$%
$(\partial J){}^{2} \partial A_{3} + 
    (-5\sqrt{2/3}Q - 32\sqrt{6}Q{}^{3})$%
$\partial J A_2 \partial{}^{2} T - 
    \sqrt{3/2}Q{}^{2}$%
$\partial J A_2 \partial A_{3} + 
    (-8\sqrt{2/3}Q - 46\sqrt{6}Q{}^{3})$%
$\partial J \partial A_{2} \partial T + 
    (-(Q/\sqrt{6}) - 3\sqrt{6}Q{}^{3})$%
$\partial J \partial A_{2} \partial A_{2} + 
    5\sqrt{3/2}Q{}^{2}$%
$\partial J \partial A_{2} A_3 + 
    (-(\sqrt{2/3}Q) - 18\sqrt{6}Q{}^{3})$%
$\partial J \partial{}^{2} A_{2} T + 
    (-(Q/\sqrt{6}) - 5\sqrt{3/2}Q{}^{3})$%
$\partial J \partial{}^{2} A_{2} A_2 + 
    (-4\sqrt{2/3} - 4\sqrt{6}Q{}^{2})$%
$\partial J A_3 \partial T - 
    3\sqrt{3/2}Q$%
$\partial J A_3 A_3 + 
    (-4\sqrt{2/3} - 6\sqrt{6}Q{}^{2})$%
$\partial J \partial A_{3} T - 
    \sqrt{3/2}Q{}^{2}$%
$\partial J B_2 \partial B_{3} + 
    ((-7Q)/\sqrt{6} - 7\sqrt{6}Q{}^{3})$%
$\partial J \partial B_{2} \partial B_{2} - 
    15\sqrt{3/2}Q{}^{2}$%
$\partial J \partial B_{2} B_3 + 
    (-3\sqrt{3/2}Q - 25\sqrt{3/2}Q{}^{3})$%
$\partial J \partial{}^{2} B_{2} 
      B_2 + 5\sqrt{3/2}Q$%
$\partial J B_3 B_3 + 
    (-8\sqrt{2/3}Q - 32\sqrt{6}Q{}^{3})$%
$\partial{}^{2} J \partial T T + 
    ((-92Q{}^{2})/3 - 368Q{}^{4})$%
$\partial{}^{2} J J \partial{}^{2} T + 
    (15Q{}^{2} + 148Q{}^{4})$%
$\partial{}^{2} J J \partial{}^{2} A_{2} + 
    (-12Q - 92Q{}^{3})$%
$\partial{}^{2} J J \partial A_{3} + 
    (8 Q{}^{2}/3 + 32Q{}^{4})$%
$\partial{}^{2} J \partial J \partial T + 
    (51Q{}^{2} + 562Q{}^{4})$%
$\partial{}^{2} J \partial J \partial A_{2} + 
    (-7Q - 54Q{}^{3})$%
$\partial{}^{2} J \partial J A_3 + 
    (1/18 + 37 Q{}^{2}/3 + 140Q{}^{4})$%
$\partial{}^{2} J \partial{}^{2} J T + 
    (5 Q/(6\sqrt{6}) - 121 Q{}^{3}/\sqrt{6} - 262\sqrt{6}Q{}^{5})
     $%
$\partial{}^{2} J \partial{}^{2} J \partial J + (1/18 + 21Q{}^{2} + 704 Q{}^{4}/3)
     $%
$\partial{}^{2} J \partial{}^{2} J A_2 + 
    (-10\sqrt{2/3}Q - 122\sqrt{2/3}Q{}^{3})$%
$\partial{}^{2} J A_2 \partial T - 
    (Q{}^{2}$%
$\partial{}^{2} J A_2 A_3)/\sqrt{6} + 
    (-5\sqrt{2/3}Q - 26\sqrt{6}Q{}^{3})$%
$\partial{}^{2} J \partial A_{2} T + 
    (-(\sqrt{6}Q) - 67 Q{}^{3}/\sqrt{6})$%
$\partial{}^{2} J \partial A_{2} A_2 + 
    (-5\sqrt{2/3} - 10\sqrt{6}Q{}^{2})$%
$\partial{}^{2} J A_3 T - 
    (5Q{}^{2}$%
$\partial{}^{2} J B_2 B_3)/\sqrt{6} + 
    (-5\sqrt{2/3}Q - 83 Q{}^{3}/\sqrt{6})$%
$\partial{}^{2} J \partial B_{2} 
      B_2 + (-2\sqrt{2/3}Q - 8\sqrt{6}Q{}^{3})
     $%
$\partial{}^{3} J T{}^{2}+ (-2/9 - 18Q{}^{2} - 184Q{}^{4})
     $%
$\partial{}^{3} J J \partial T + (-2/9 + 11 Q{}^{2}/3 + 48Q{}^{4})
     $%
$\partial{}^{3} J J \partial A_{2} + ((-10Q)/3 - 22Q{}^{3})
     $%
$\partial{}^{3} J J A_3 + (2/27 + 14Q{}^{2} + 472 Q{}^{4}/3)
     $%
$\partial{}^{3} J \partial J T + 
    ((5\sqrt{2/3}Q)/9 - 20\sqrt{2/3}Q{}^{3} - 320\sqrt{2/3}Q{}^{5})
     $%
$\partial{}^{3} J (\partial J){}^{2} + (2/27 + 25Q{}^{2} + 826 Q{}^{4}/3)
     $%
$\partial{}^{3} J \partial J A_2 + 
    ((10\sqrt{2/3}Q)/3 + (382\sqrt{2/3}Q{}^{3})/3 + 1048\sqrt{2/3}Q{}^{5})
     $%
$\partial{}^{3} J \partial{}^{2} J J + 
    ((-14\sqrt{2/3}Q)/3 - 52\sqrt{2/3}Q{}^{3})$%
$\partial{}^{3} J A_2 
      T + ((-4\sqrt{2/3}Q)/3 - 5\sqrt{6}Q{}^{3})
     $%
$\partial{}^{3} J A_2{}^{2}+ 
    ((-2\sqrt{2/3}Q)/3 - 5\sqrt{2/3}Q{}^{3})$%
$\partial{}^{3} J B_2{}^{2}+ (-5/54 - 13 Q{}^{2}/3 - 116 Q{}^{4}/3)
     $%
$\partial{}^{4} J J T + (-5/54 - Q{}^{2}/3 + 10 Q{}^{4}/3)
     $%
$\partial{}^{4} J J A_2 + 
    (65 Q/(18\sqrt{6}) + 143 Q{}^{3}/\sqrt{6} + 598\sqrt{2/3}Q{}^{5})
     $%
$\partial{}^{4} J \partial J J + 
    (17 Q/(36\sqrt{6}) + 115 Q{}^{3}/(6\sqrt{6}) + 27\sqrt{6}Q{}^{5})
     $%
$\partial{}^{5} J J{}^{2}+ 
    4Q{}^{2}$%
$A_2 \partial T \partial T + (4/3 + 4Q{}^{2})
     $%
$A_2 \partial{}^{2} T T + (-1/3 - 4Q{}^{2})
     $%
$A_2{}^{2}\partial{}^{2} T - 
    Q$%
$A_2 A_3 \partial T + $%
$A_2 A_3 
     A_3 - Q$%
$A_2 \partial A_{3} T - 
    Q$%
$A_2 B_2 \partial B_{3} + 
    (-1/3 - 6Q{}^{2})$%
$A_2 \partial B_{2} \partial B_{2} + 
    4Q$%
$A_2 \partial B_{2} B_3 - 
    3Q{}^{2}$%
$A_2 \partial{}^{2} B_{2} B_2 - 
    2$%
$A_2 B_3 B_3 + 
    (4/3 + 16Q{}^{2})$%
$\partial A_{2} \partial T T - 
    7Q{}^{2}$%
$\partial A_{2} A_2 \partial T - 
    Q$%
$\partial A_{2} A_2 A_3 + 
    (-1/3 - 6Q{}^{2})$%
$\partial A_{2} \partial A_{2} T + 
    (1/3 + 8Q{}^{2})$%
$\partial A_{2} \partial A_{2} A_2 + 
    9Q$%
$\partial A_{2} A_3 T - 
    Q$%
$\partial A_{2} B_2 B_3 - 
    2Q{}^{2}$%
$\partial A_{2} \partial B_{2} B_2 + 
    6Q{}^{2}$%
$\partial{}^{2} A_{2} T{}^{2}+ 
    (-1/3 - Q{}^{2})$%
$\partial{}^{2} A_{2} A_2 T + 
    (1/6 + 5Q{}^{2})$%
$\partial{}^{2} A_{2} A_2{}^{2}+ 
    (-1/6 - 2Q{}^{2})$%
$\partial{}^{2} A_{2} B_2{}^{2}- 
    12Q$%
$A_3 \partial T T - 
    3$%
$A_3 A_3 T + $%
$A_3 B_2 
     B_3 - 2Q$%
$A_3 \partial B_{2} B_2 - 
    6Q$%
$\partial A_{3} T{}^{2}+ 
    Q$%
$\partial A_{3} B_2{}^{2}+ 
    (1/3 + 4Q{}^{2})$%
$B_2{}^{2}\partial{}^{2} T - 
    5Q$%
$B_2 B_3 \partial T - 
    9Q$%
$B_2 \partial B_{3} T + (4/3 + 17Q{}^{2})
     $%
$\partial B_{2} B_2 \partial T + (1/3 + 10Q{}^{2})
     $%
$\partial B_{2} \partial B_{2} T - 
    7Q$%
$\partial B_{2} B_3 T + 
    (1 + 15Q{}^{2})$%
$\partial{}^{2} B_{2} B_2 T + 
    $%
$B_3 B_3 T + (2/3 + 8Q{}^{2})
     $%
$J{}^{2}\partial{}^{2} T T + 
    ((5\sqrt{2/3}Q)/3 + 20\sqrt{2/3}Q{}^{3})$%
$J{}^{3}\partial{}^{3} T + ((-13Q)/(9\sqrt{6}) - 2\sqrt{2/3}Q{}^{3})
     $%
$J{}^{3}\partial{}^{3} A_{2} + 
    (\sqrt{2/3}/3 - 2\sqrt{6}Q{}^{2})$%
$J{}^{3}\partial{}^{2} A_{3} + 18Q{}^{2}$%
$J{}^{2}A_2 \partial{}^{2} T - 
    (Q$%
$J{}^{2}A_2 \partial A_{3})/2 + 
    (2/3 + 28Q{}^{2})$%
$J{}^{2}\partial A_{2} \partial T + 
    (-1/6 - 3Q{}^{2})$%
$J{}^{2}\partial A_{2} \partial A_{2} + 
    (Q$%
$J{}^{2}\partial A_{2} A_3)/2 + 
    10Q{}^{2}$%
$J{}^{2}\partial{}^{2} A_{2} T + 
    (-1/6 - 9 Q{}^{2}/2)$%
$J{}^{2}\partial{}^{2} A_{2} A_2 - 
    6Q$%
$J{}^{2}A_3 \partial T - 
    (3$%
$J{}^{2}A_3 A_3)/2 - 
    2Q$%
$J{}^{2}\partial A_{3} T + 
    (7Q$%
$J{}^{2}B_2 \partial B_{3})/2 + 
    (1/6 - 3Q{}^{2})$%
$J{}^{2}\partial B_{2} \partial B_{2} + 
    (9Q$%
$J{}^{2}\partial B_{2} B_3)/2 + 
    (1/2 - Q{}^{2}/2)$%
$J{}^{2}\partial{}^{2} B_{2} B_2 + 
    $%
$J{}^{2}B_3 B_3/2 + 
    8\sqrt{2/3}Q$%
$J A_2 \partial T T + 
    \sqrt{6}Q$%
$J A_2{}^{2}\partial T - 
    \sqrt{2/3}$%
$J A_2 A_3 T + 
    2\sqrt{2/3}Q$%
$J A_2 \partial B_{2} B_2 + 
    \sqrt{2/3}Q$%
$J \partial A_{2} A_2 T - 
    2\sqrt{2/3}Q$%
$J \partial A_{2} A_2{}^{2}+ 
    2\sqrt{6}$%
$J A_3 T{}^{2}+ 
    \sqrt{6}Q$%
$J B_2{}^{2}\partial T + 
    7\sqrt{2/3}$%
$J B_2 B_3 T + 
    \sqrt{2/3}Q$%
$J \partial B_{2} B_2 T + 
    (4/3 + 16Q{}^{2})$%
$\partial J J \partial T T + 
    (2\sqrt{6}Q + 24\sqrt{6}Q{}^{3})$%
$\partial J J{}^{2}\partial{}^{2} T + ((-11Q)/\sqrt{6} - 13\sqrt{6}Q{}^{3})$%
$\partial J J{}^{2}\partial{}^{2} A_{2} + (2\sqrt{2/3} - 7\sqrt{6}Q{}^{2})
     $%
$\partial J J{}^{2}\partial A_{3} + 
    (4/3 + 28Q{}^{2})$%
$\partial J J A_2 \partial T + 
    (4/3 + 36Q{}^{2})$%
$\partial J J \partial A_{2} T - 
    4Q{}^{2}$%
$\partial J J \partial A_{2} A_2 - 
    24Q$%
$\partial J J A_3 T + 
    4Q$%
$\partial J J B_2 B_3 + 
    (4/3 - 4Q{}^{2})$%
$\partial J J \partial B_{2} B_2 + 
    (4\sqrt{2/3}Q + 16\sqrt{6}Q{}^{3})$%
$(\partial J){}^{2} J 
      \partial T + (-13\sqrt{2/3}Q - 38\sqrt{6}Q{}^{3})$%
$(\partial J){}^{2} 
      J \partial A_{2} + (2\sqrt{2/3} - \sqrt{6}Q{}^{2})
     $%
$(\partial J){}^{2} J A_3 + 
    (2\sqrt{2/3}Q + 8\sqrt{6}Q{}^{3})$%
$(\partial J){}^{3} 
      T + (10 Q{}^{2}/3 + 40Q{}^{4})$%
$(\partial J){}^{4} + ((-5\sqrt{2/3}Q)/3 - 13\sqrt{2/3}Q{}^{3})
     $%
$(\partial J){}^{3} A_2 - 
    4Q{}^{2}$%
$(\partial J){}^{2} A_2 T + 
    (-1/3 - 11 Q{}^{2}/2)$%
$(\partial J){}^{2} A_2{}^{2}+ 
    (1/3 + 3 Q{}^{2}/2)$%
$(\partial J){}^{2} B_2{}^{2}+ 
    2\sqrt{6}Q$%
$\partial J A_2 T{}^{2}+ 
    (Q$%
$\partial J A_2{}^{3})/\sqrt{6} - 
    (Q$%
$\partial J A_2 B_2{}^{2})/\sqrt{6} + 
    2\sqrt{6}Q$%
$\partial J B_2{}^{2}T + 
    (2/3 + 8Q{}^{2})$%
$\partial{}^{2} J J T{}^{2}+ 
    (2\sqrt{6}Q + 24\sqrt{6}Q{}^{3})$%
$\partial{}^{2} J J{}^{2}\partial T + ((-19Q)/\sqrt{6} - 83\sqrt{2/3}Q{}^{3})
     $%
$\partial{}^{2} J J{}^{2}\partial A_{2} + 
    (5/\sqrt{6} + 3\sqrt{6}Q{}^{2})$%
$\partial{}^{2} J J{}^{2}A_3 + (4/3 + 44 Q{}^{2}/3)$%
$\partial{}^{2} J J A_2 
      T + (-1/6 - 3Q{}^{2})$%
$\partial{}^{2} J J A_2{}^{2}+ (5/6 + 5Q{}^{2})$%
$\partial{}^{2} J J B_2{}^{2}+ (-16\sqrt{2/3}Q - 54\sqrt{6}Q{}^{3})
     $%
$\partial{}^{2} J \partial J J A_2 + 
    (101 Q{}^{2}/3 + 404Q{}^{4})$%
$\partial{}^{2} J (\partial J){}^{2} J + 
    (-5/36 + 17 Q{}^{2}/6 + 54Q{}^{4})$%
$\partial{}^{2} J \partial{}^{2} J J{}^{2}+ ((\sqrt{2/3}Q)/3 + 4\sqrt{2/3}Q{}^{3})
     $%
$\partial{}^{3} J J{}^{2}T + 
    ((-10\sqrt{2/3}Q)/3 - 32\sqrt{2/3}Q{}^{3})$%
$\partial{}^{3} J J{}^{2}A_2 + (-5/27 + 8 Q{}^{2}/3 + 176 Q{}^{4}/3)
     $%
$\partial{}^{3} J \partial J J{}^{2}+ 
    (-5/108 - 23 Q{}^{2}/18 - 26 Q{}^{4}/3)$%
$\partial{}^{4} J J{}^{3}- $%
$A_2{}^{2}T{}^{2}/3 + 
    $%
$A_2{}^{3}T/3 - 
    $%
$A_2{}^{4}/12 + 
    $%
$A_2{}^{2}B_2{}^{2}/6 - 
    $%
$A_2 B_2{}^{2}T/3 - 
    $%
$B_2{}^{2}T{}^{2}- 
    $%
$B_2{}^{4}/12 + 
    (-1/9 - 4 Q{}^{2}/3)$%
$J{}^{4}\partial{}^{2} T - (11Q{}^{2}$%
$J{}^{4}\partial{}^{2} A_{2})/6 + (19Q$%
$J{}^{4}\partial A_{3})/6 - (16\sqrt{2/3}Q$%
$J{}^{3}A_2 \partial T)/3 + $%
$J{}^{3}A_2 A_3/(3\sqrt{6}) - 4\sqrt{2/3}Q
     $%
$J{}^{3}\partial A_{2} T + 
    (7Q$%
$J{}^{3}\partial A_{2} A_2)/
     (3\sqrt{6}) + 2\sqrt{2/3}$%
$J{}^{3}A_3 T - (7$%
$J{}^{3}B_2 B_3)/(3\sqrt{6}) + 
    (23Q$%
$J{}^{3}\partial B_{2} B_2)/
     (3\sqrt{6}) - (4$%
$J{}^{2}A_2 T{}^{2})/
     3 - $%
$J{}^{2}A_2{}^{2}T/3 + 
    $%
$J{}^{2}A_2{}^{3}/6 - 
    $%
$J{}^{2}A_2 B_2{}^{2}/6 - 
    $%
$J{}^{2}B_2{}^{2}T + 
    (-2/9 - 8 Q{}^{2}/3)$%
$\partial J J{}^{3}\partial T + (2/3 - 10 Q{}^{2}/3)$%
$\partial J J{}^{3}\partial A_{2} + 8Q$%
$\partial J J{}^{3}A_3 - 2\sqrt{2/3}Q$%
$\partial J J{}^{2}A_2 T + \sqrt{2/3}Q$%
$\partial J J{}^{2}A_2{}^{2}+ 4\sqrt{2/3}Q
     $%
$\partial J J{}^{2}B_2{}^{2}+ 
    (-2/3 - 8Q{}^{2})$%
$(\partial J){}^{2} J{}^{2}T + 
    (2$%
$(\partial J){}^{2} J{}^{2}A_2)/3 + 
    ((-8\sqrt{2/3}Q)/3 - 32\sqrt{2/3}Q{}^{3})$%
$(\partial J){}^{3} J{}^{2}+ (-4/9 - 16 Q{}^{2}/3)
     $%
$\partial{}^{2} J J{}^{3}T + 
    (2/3 + 38 Q{}^{2}/9)$%
$\partial{}^{2} J J{}^{3}A_2 + ((-13\sqrt{2/3}Q)/3 - 52\sqrt{2/3}Q{}^{3})
     $%
$\partial{}^{2} J \partial J J{}^{3}+ 
    ((-4\sqrt{2/3}Q)/9 - (16\sqrt{2/3}Q{}^{3})/3)$%
$\partial{}^{3} J J{}^{4}+ 
    (2\sqrt{2/3}Q$%
$J{}^{5}\partial A_{2})/3 - $%
$J{}^{5}A_3/\sqrt{6} + 
    (4$%
$J{}^{4}A_2 T)/
     9 - $%
$J{}^{4}A_2{}^{2}/12 - $%
$J{}^{4}B_2{}^{2}/4 + \sqrt{3/2}Q$%
$\partial J J{}^{4}A_2 + 
    (1/9 + 4 Q{}^{2}/3)$%
$(\partial J){}^{2} J{}^{4}+ (1/18 + 2 Q{}^{2}/3)$%
$\partial{}^{2} J J{}^{5}- 
    $%
$J{}^{6}A_2/27.
\  $ 
 \generatorhead{$W_{9}$} \noindent $W_{9} = (12959 Q/45360 + 9517 Q{}^{3}/756 + 104687 Q{}^{5}/126 + 10966Q{}^{7})
     $%
$\partial{}^{7} B_{2} + (-323/540 - 1847 Q{}^{2}/60 - 28349 Q{}^{4}/30 - 6594Q{}^{6})
     $%
$\partial{}^{6} B_{3} + ((2\sqrt{2/3}Q{}^{2})/45 - (12044\sqrt{2/3}Q{}^{4})/15 - 
      4396\sqrt{6}Q{}^{6})$%
$J \partial{}^{6} B_{2} + 
    ((-271\sqrt{2/3}Q)/45 + (6598\sqrt{2/3}Q{}^{3})/15 + 2222\sqrt{6}Q{}^{5})
     $%
$J \partial{}^{5} B_{3} + ((-1309Q{}^{2})/(60\sqrt{6}) - 
      71633 Q{}^{4}/(15\sqrt{6}) - 40525\sqrt{2/3}Q{}^{6})
     $%
$\partial J \partial{}^{5} B_{2} + ((-437Q)/(6\sqrt{6}) + 95\sqrt{3/2}Q{}^{3} + 
      5525\sqrt{6}Q{}^{5})$%
$\partial J \partial{}^{4} B_{3} + 
    ((-4235Q{}^{2})/(27\sqrt{6}) - (59366\sqrt{2/3}Q{}^{4})/9 - 
      92258\sqrt{2/3}Q{}^{6})$%
$\partial{}^{2} J \partial{}^{4} B_{2} + 
    ((-172\sqrt{2/3}Q)/9 + (4672\sqrt{2/3}Q{}^{3})/3 + 11944\sqrt{6}Q{}^{5})
     $%
$\partial{}^{2} J \partial{}^{3} B_{3} + ((-4174\sqrt{2/3}Q{}^{2})/27 - 
      160525 Q{}^{4}/(9\sqrt{6}) - 108760\sqrt{2/3}Q{}^{6})
     $%
$\partial{}^{3} J \partial{}^{3} B_{2} + (631 Q/(9\sqrt{6}) + 
      17681 Q{}^{3}/(3\sqrt{6}) + 13044\sqrt{6}Q{}^{5})
     $%
$\partial{}^{3} J \partial{}^{2} B_{3} + (-131\sqrt{2/3}Q{}^{2} - 6073\sqrt{2/3}Q{}^{4} - 
      21896\sqrt{6}Q{}^{6})$%
$\partial{}^{4} J \partial{}^{2} B_{2} + 
    (901 Q/(9\sqrt{6}) + (6064\sqrt{2/3}Q{}^{3})/3 + 6594\sqrt{6}Q{}^{5})
     $%
$\partial{}^{4} J \partial B_{3} + ((-22697Q{}^{2})/(270\sqrt{6}) - 
      (16250\sqrt{2/3}Q{}^{4})/9 - 18599\sqrt{2/3}Q{}^{6})
     $%
$\partial{}^{5} J \partial B_{2} + (611 Q/(30\sqrt{6}) + 
      (1832\sqrt{2/3}Q{}^{3})/5 + 1099\sqrt{6}Q{}^{5})$%
$\partial{}^{5} J B_3 + 
    ((-1429Q{}^{2})/(135\sqrt{6}) - (9829\sqrt{2/3}Q{}^{4})/45 - 
      2186\sqrt{2/3}Q{}^{6})$%
$\partial{}^{6} J B_2 + 
    ((-23Q)/72 - 8884 Q{}^{3}/45 - 9601 Q{}^{5}/3)$%
$A_2 \partial{}^{5} B_{2} + 
    (74/27 + 4187 Q{}^{2}/18 + 2379Q{}^{4})$%
$A_2 \partial{}^{4} B_{3} + 
    (25 Q/36 - 1123 Q{}^{3}/3 - 5985Q{}^{5})$%
$\partial A_{2} \partial{}^{4} B_{2} + 
    (113/27 + 4189 Q{}^{2}/9 + 6064Q{}^{4})$%
$\partial A_{2} \partial{}^{3} B_{3} + 
    (25 Q/18 - 2062 Q{}^{3}/9 - 9650 Q{}^{5}/3)$%
$\partial{}^{2} A_{2} \partial{}^{3} B_{2} + 
    (19/3 + 1487 Q{}^{2}/3 + 7306Q{}^{4})$%
$\partial{}^{2} A_{2} \partial{}^{2} B_{3} + 
    (71 Q/18 + 131Q{}^{3} + 2510Q{}^{5})$%
$\partial{}^{3} A_{2} \partial{}^{2} B_{2} + 
    (121/27 + 2780 Q{}^{2}/9 + 4300Q{}^{4})$%
$\partial{}^{3} A_{2} \partial B_{3} + 
    (463 Q/108 + 4213 Q{}^{3}/18 + 3405Q{}^{5})$%
$\partial{}^{4} A_{2} \partial B_{2} + 
    (79/36 + 283 Q{}^{2}/3 + 1075Q{}^{4})$%
$\partial{}^{4} A_{2} B_3 + 
    (1699 Q/540 + 8947 Q{}^{3}/90 + 1111Q{}^{5})$%
$\partial{}^{5} A_{2} B_2 + 
    (-167/108 - 1217 Q{}^{2}/18 - 1111Q{}^{4})$%
$A_3 \partial{}^{4} B_{2} + 
    27Q$%
$A_3 \partial{}^{3} B_{3} + (-32/27 - 2029 Q{}^{2}/9 - 4480Q{}^{4})
     $%
$\partial A_{3} \partial{}^{3} B_{2} - 18Q$%
$\partial A_{3} \partial{}^{2} B_{3} + 
    (-59/18 - 1217 Q{}^{2}/3 - 6882Q{}^{4})$%
$\partial{}^{2} A_{3} \partial{}^{2} B_{2} + 
    (9Q + 108Q{}^{3})$%
$\partial{}^{2} A_{3} \partial B_{3} + 
    (-59/27 - 2380 Q{}^{2}/9 - 4516Q{}^{4})$%
$\partial{}^{3} A_{3} \partial B_{2} + 
    (9Q + 36Q{}^{3})$%
$\partial{}^{3} A_{3} B_3 + 
    (71/216 - 541 Q{}^{2}/9 - 1111Q{}^{4})$%
$\partial{}^{4} A_{3} B_2 + 
    ((-1159Q)/135 - 17894 Q{}^{3}/45 - 4372Q{}^{5})$%
$B_2 \partial{}^{5} T + 
    ((-1205Q)/54 - 13358 Q{}^{3}/9 - 19710Q{}^{5})$%
$\partial B_{2} \partial{}^{4} T + 
    ((-80Q)/3 - 7096 Q{}^{3}/3 - 35336Q{}^{5})$%
$\partial{}^{2} B_{2} \partial{}^{3} T + 
    ((-292Q)/27 - 1849Q{}^{3} - 95984 Q{}^{5}/3)$%
$\partial{}^{3} B_{2} \partial{}^{2} T + 
    ((-65Q)/9 - 2674 Q{}^{3}/3 - 15948Q{}^{5})$%
$\partial{}^{4} B_{2} \partial T + 
    ((-2857Q)/540 - 3791 Q{}^{3}/15 - 12934 Q{}^{5}/3)$%
$\partial{}^{5} B_{2} T + 
    (-79/18 - 638 Q{}^{2}/3 - 2150Q{}^{4})$%
$B_3 \partial{}^{4} T + 
    (-242/27 - 5884 Q{}^{2}/9 - 8456Q{}^{4})$%
$\partial B_{3} \partial{}^{3} T + 
    (-11 - 2299 Q{}^{2}/3 - 12152Q{}^{4})$%
$\partial{}^{2} B_{3} \partial{}^{2} T + 
    (98/27 - 3518 Q{}^{2}/9 - 6944Q{}^{4})$%
$\partial{}^{3} B_{3} \partial T + 
    (505/54 + 295 Q{}^{2}/9 - 882Q{}^{4})$%
$\partial{}^{4} B_{3} T + 
    ((-1469Q)/1080 + 15019 Q{}^{3}/90 + 6865 Q{}^{5}/3)
     $%
$J{}^{2}\partial{}^{5} B_{2} + 
    (505/108 - 785 Q{}^{2}/18 - 585Q{}^{4})$%
$J{}^{2}\partial{}^{4} B_{3} + 
    ((392\sqrt{2/3}Q{}^{2})/3 + 2536\sqrt{2/3}Q{}^{4})$%
$J A_2 
      \partial{}^{4} B_{2} + ((-232\sqrt{2/3}Q)/3 - 1340\sqrt{2/3}Q{}^{3})
     $%
$J A_2 \partial{}^{3} B_{3} + 
    (62\sqrt{6}Q{}^{2} + 1056\sqrt{6}Q{}^{4})$%
$J \partial A_{2} 
      \partial{}^{3} B_{2} + (-32\sqrt{6}Q - 588\sqrt{6}Q{}^{3})
     $%
$J \partial A_{2} \partial{}^{2} B_{3} + 
    ((112\sqrt{2/3}Q{}^{2})/3 + 848\sqrt{2/3}Q{}^{4})$%
$J \partial{}^{2} A_{2} 
      \partial{}^{2} B_{2} + ((-56\sqrt{2/3}Q)/3 - 640\sqrt{2/3}Q{}^{3})
     $%
$J \partial{}^{2} A_{2} \partial B_{3} + 
    (-12\sqrt{6}Q{}^{2} - 144\sqrt{6}Q{}^{4})$%
$J \partial{}^{3} A_{2} 
      \partial B_{2} + (2\sqrt{6}Q + 24\sqrt{6}Q{}^{3})$%
$J \partial{}^{3} A_{2} 
      B_3 - 24\sqrt{6}Q{}^{4}$%
$J \partial{}^{4} A_{2} B_2 + 
    (-8\sqrt{6}Q + 24\sqrt{6}Q{}^{3})$%
$J A_3 \partial{}^{3} B_{2} + 
    144\sqrt{6}Q{}^{3}$%
$J \partial A_{3} \partial{}^{2} B_{2} - 
    36\sqrt{6}Q{}^{2}$%
$J \partial A_{3} \partial B_{3} + 
    (6\sqrt{6}Q + 144\sqrt{6}Q{}^{3})$%
$J \partial{}^{2} A_{3} \partial B_{2} - 
    36\sqrt{6}Q{}^{2}$%
$J \partial{}^{2} A_{3} B_3 + 
    (-2\sqrt{6}Q + 24\sqrt{6}Q{}^{3})$%
$J \partial{}^{3} A_{3} B_2 + 
    (79/(9\sqrt{6}) + (566\sqrt{2/3}Q{}^{2})/3 + 2150\sqrt{2/3}Q{}^{4})
     $%
$J B_2 \partial{}^{4} T + 
    ((242\sqrt{2/3})/27 + (5776\sqrt{2/3}Q{}^{2})/9 + 8888\sqrt{2/3}Q{}^{4})
     $%
$J \partial B_{2} \partial{}^{3} T + 
    ((59\sqrt{2/3})/9 + (2654\sqrt{2/3}Q{}^{2})/3 + 15044\sqrt{2/3}Q{}^{4})
     $%
$J \partial{}^{2} B_{2} \partial{}^{2} T + 
    ((118\sqrt{2/3})/27 + (5597\sqrt{2/3}Q{}^{2})/9 + 12308\sqrt{2/3}Q{}^{4})
     $%
$J \partial{}^{3} B_{2} \partial T + 
    (5/(27\sqrt{6}) + (1961\sqrt{2/3}Q{}^{2})/9 + 1598\sqrt{6}Q{}^{4})
     $%
$J \partial{}^{4} B_{2} T + (8\sqrt{6}Q - 72\sqrt{6}Q{}^{3})
     $%
$J B_3 \partial{}^{3} T + 
    ((-2\sqrt{2/3}Q)/3 - 1396\sqrt{2/3}Q{}^{3})$%
$J \partial B_{3} 
      \partial{}^{2} T + (13\sqrt{6}Q - 732\sqrt{6}Q{}^{3})$%
$J \partial{}^{2} B_{3} 
      \partial T + ((-43\sqrt{2/3}Q)/3 - 1448\sqrt{2/3}Q{}^{3})
     $%
$J \partial{}^{3} B_{3} T + 
    (569 Q/54 + 7301 Q{}^{3}/9 + 8338Q{}^{5})$%
$\partial J J 
      \partial{}^{4} B_{2} + (98/27 - 185 Q{}^{2}/9 + 1248Q{}^{4})
     $%
$\partial J J \partial{}^{3} B_{3} + 
    (662 Q/27 + 2549 Q{}^{3}/3 + 6790 Q{}^{5}/3)$%
$(\partial J){}^{2} 
      \partial{}^{3} B_{2} + (-11 + 8Q{}^{2} + 7158Q{}^{4})$%
$(\partial J){}^{2} 
      \partial{}^{2} B_{3} + (54\sqrt{6}Q{}^{2} + 1239\sqrt{6}Q{}^{4})
     $%
$\partial J A_2 \partial{}^{3} B_{2} + 
    (20\sqrt{2/3}Q - 149\sqrt{6}Q{}^{3})$%
$\partial J A_2 
      \partial{}^{2} B_{3} + ((440\sqrt{2/3}Q{}^{2})/3 + 3181\sqrt{2/3}Q{}^{4})
     $%
$\partial J \partial A_{2} \partial{}^{2} B_{2} + 
    ((272\sqrt{2/3}Q)/3 + 1936\sqrt{2/3}Q{}^{3})$%
$\partial J \partial A_{2} 
      \partial B_{3} + ((436\sqrt{2/3}Q{}^{2})/3 + 2873\sqrt{2/3}Q{}^{4})
     $%
$\partial J \partial{}^{2} A_{2} \partial B_{2} + (24\sqrt{6}Q + 621\sqrt{6}Q{}^{3})
     $%
$\partial J \partial{}^{2} A_{2} B_3 + 
    ((313\sqrt{2/3}Q{}^{2})/3 + 2057\sqrt{2/3}Q{}^{4})$%
$\partial J \partial{}^{3} A_{2} 
      B_2 + ((-298\sqrt{2/3}Q)/3 - 2021\sqrt{2/3}Q{}^{3})
     $%
$\partial J A_3 \partial{}^{2} B_{2} - 18\sqrt{6}Q{}^{2}
     $%
$\partial J A_3 \partial B_{3} + 
    ((-596\sqrt{2/3}Q)/3 - 4096\sqrt{2/3}Q{}^{3})$%
$\partial J \partial A_{3} 
      \partial B_{2} - 18\sqrt{6}Q{}^{2}$%
$\partial J \partial A_{3} B_3 + 
    ((-298\sqrt{2/3}Q)/3 - 2075\sqrt{2/3}Q{}^{3})$%
$\partial J \partial{}^{2} A_{3} 
      B_2 + ((242\sqrt{2/3})/27 + (1804\sqrt{2/3}Q{}^{2})/9 - 
      20\sqrt{6}Q{}^{4})$%
$\partial J B_2 \partial{}^{3} T + 
    ((118\sqrt{2/3})/9 + (1756\sqrt{2/3}Q{}^{2})/3 + 5308\sqrt{2/3}Q{}^{4})
     $%
$\partial J \partial B_{2} \partial{}^{2} T + 
    ((118\sqrt{2/3})/9 + 364\sqrt{6}Q{}^{2} + 14780\sqrt{2/3}Q{}^{4})
     $%
$\partial J \partial{}^{2} B_{2} \partial T + 
    ((118\sqrt{2/3})/27 + (5135\sqrt{2/3}Q{}^{2})/9 + 3468\sqrt{6}Q{}^{4})
     $%
$\partial J \partial{}^{3} B_{2} T + 
    (-172\sqrt{2/3}Q - 1652\sqrt{6}Q{}^{3})$%
$\partial J B_3 \partial{}^{2} T + 
    ((-1624\sqrt{2/3}Q)/3 - 11864\sqrt{2/3}Q{}^{3})$%
$\partial J \partial B_{3} 
      \partial T + (-233\sqrt{2/3}Q - 2404\sqrt{6}Q{}^{3})
     $%
$\partial J \partial{}^{2} B_{3} T + 
    (1157 Q/27 + 2344Q{}^{3} + 73540 Q{}^{5}/3)$%
$\partial{}^{2} J J 
      \partial{}^{3} B_{2} + (-71/9 - 1106 Q{}^{2}/3 - 1532Q{}^{4})
     $%
$\partial{}^{2} J J \partial{}^{2} B_{3} + 
    (965 Q/9 + 37334 Q{}^{3}/9 + 94972 Q{}^{5}/3)$%
$\partial{}^{2} J \partial J 
      \partial{}^{2} B_{2} + (-242/9 - 620 Q{}^{2}/3 + 8008Q{}^{4})
     $%
$\partial{}^{2} J \partial J \partial B_{3} + 
    (1303 Q/18 + 8398 Q{}^{3}/3 + 25182Q{}^{5})$%
$\partial{}^{2} J \partial{}^{2} J 
      \partial B_{2} + (-91/6 - 338Q{}^{2} - 954Q{}^{4})$%
$\partial{}^{2} J \partial{}^{2} J 
      B_3 + ((1984\sqrt{2/3}Q{}^{2})/9 + (11450\sqrt{2/3}Q{}^{4})/3)
     $%
$\partial{}^{2} J A_2 \partial{}^{2} B_{2} + 
    ((-32\sqrt{2/3}Q)/3 + 272\sqrt{2/3}Q{}^{3})$%
$\partial{}^{2} J A_2 
      \partial B_{3} + (152\sqrt{2/3}Q{}^{2} + 946\sqrt{6}Q{}^{4})
     $%
$\partial{}^{2} J \partial A_{2} \partial B_{2} + (28\sqrt{6}Q + 666\sqrt{6}Q{}^{3})
     $%
$\partial{}^{2} J \partial A_{2} B_3 + 
    ((1298\sqrt{2/3}Q{}^{2})/9 + (8902\sqrt{2/3}Q{}^{4})/3)
     $%
$\partial{}^{2} J \partial{}^{2} A_{2} B_2 + 
    (-44\sqrt{6}Q - 918\sqrt{6}Q{}^{3})$%
$\partial{}^{2} J A_3 \partial B_{2} - 
    6\sqrt{6}Q{}^{2}$%
$\partial{}^{2} J A_3 B_3 + 
    (-44\sqrt{6}Q - 924\sqrt{6}Q{}^{3})$%
$\partial{}^{2} J \partial A_{3} B_2 + 
    ((121\sqrt{2/3})/9 + (4238\sqrt{2/3}Q{}^{2})/9 + (5968\sqrt{2/3}Q{}^{4})/3)
     $%
$\partial{}^{2} J B_2 \partial{}^{2} T + 
    ((242\sqrt{2/3})/9 + (3928\sqrt{2/3}Q{}^{2})/3 + 4068\sqrt{6}Q{}^{4})
     $%
$\partial{}^{2} J \partial B_{2} \partial T + 
    ((121\sqrt{2/3})/9 + (9838\sqrt{2/3}Q{}^{2})/9 + (42860\sqrt{2/3}Q{}^{4})/3)
     $%
$\partial{}^{2} J \partial{}^{2} B_{2} T + (-104\sqrt{6}Q - 2196\sqrt{6}Q{}^{3})
     $%
$\partial{}^{2} J B_3 \partial T + 
    ((-1220\sqrt{2/3}Q)/3 - 8368\sqrt{2/3}Q{}^{3})$%
$\partial{}^{2} J \partial B_{3} 
      T + (1831 Q/27 + 26090 Q{}^{3}/9 + 29956Q{}^{5})
     $%
$\partial{}^{3} J J \partial{}^{2} B_{2} + (-170/9 - 546Q{}^{2} - 4228Q{}^{4})
     $%
$\partial{}^{3} J J \partial B_{3} + 
    (2642 Q/27 + 34472 Q{}^{3}/9 + 105356 Q{}^{5}/3)$%
$\partial{}^{3} J \partial J 
      \partial B_{2} + (-170/9 - 1388 Q{}^{2}/3 - 2084Q{}^{4})
     $%
$\partial{}^{3} J \partial J B_3 + 
    (1177 Q/27 + 1722Q{}^{3} + 49544 Q{}^{5}/3)$%
$\partial{}^{3} J \partial{}^{2} J 
      B_2 + ((683\sqrt{2/3}Q{}^{2})/9 + (4567\sqrt{2/3}Q{}^{4})/3)
     $%
$\partial{}^{3} J A_2 \partial B_{2} + (12\sqrt{6}Q + 177\sqrt{6}Q{}^{3})
     $%
$\partial{}^{3} J A_2 B_3 + 
    ((317\sqrt{2/3}Q{}^{2})/3 + 1681\sqrt{2/3}Q{}^{4})$%
$\partial{}^{3} J \partial A_{2} 
      B_2 + ((-230\sqrt{2/3}Q)/3 - 1165\sqrt{2/3}Q{}^{3})
     $%
$\partial{}^{3} J A_3 B_2 + 
    ((170\sqrt{2/3})/9 + 596\sqrt{2/3}Q{}^{2} + 4564\sqrt{2/3}Q{}^{4})
     $%
$\partial{}^{3} J B_2 \partial T + 
    ((170\sqrt{2/3})/9 + (7802\sqrt{2/3}Q{}^{2})/9 + (27304\sqrt{2/3}Q{}^{4})/3)
     $%
$\partial{}^{3} J \partial B_{2} T + 
    (-194\sqrt{2/3}Q - 1000\sqrt{6}Q{}^{3})$%
$\partial{}^{3} J B_3 T + 
    (2327 Q/54 + 14654 Q{}^{3}/9 + 15266Q{}^{5})$%
$\partial{}^{4} J J 
      \partial B_{2} + (-175/18 - 860 Q{}^{2}/3 - 2222Q{}^{4})
     $%
$\partial{}^{4} J J B_3 + 
    (1277 Q/54 + 8050 Q{}^{3}/9 + 23950 Q{}^{5}/3)$%
$\partial{}^{4} J \partial J 
      B_2 + ((214\sqrt{2/3}Q{}^{2})/9 + (1232\sqrt{2/3}Q{}^{4})/3)
     $%
$\partial{}^{4} J A_2 B_2 + 
    (175/(9\sqrt{6}) + (2848\sqrt{2/3}Q{}^{2})/9 + (7790\sqrt{2/3}Q{}^{4})/3)
     $%
$\partial{}^{4} J B_2 T + 
    (611 Q/90 + 3634 Q{}^{3}/15 + 2174Q{}^{5})$%
$\partial{}^{5} J J 
      B_2 + (199 Q/18 + 772 Q{}^{3}/3)$%
$A_2{}^{2}\partial{}^{3} B_{2} + (-241/18 - 742 Q{}^{2}/3)$%
$A_2{}^{2}\partial{}^{2} B_{3} + (35/9 + 670 Q{}^{2}/3)$%
$A_2 A_3 
      \partial{}^{2} B_{2} + (178/9 + 1394 Q{}^{2}/3)$%
$A_2 \partial A_{3} 
      \partial B_{2} + (62/9 + 670 Q{}^{2}/3)$%
$A_2 \partial{}^{2} A_{3} 
      B_2 + (368 Q/9 + 2464 Q{}^{3}/3)$%
$A_2 B_2 
      \partial{}^{3} T + (78Q + 2106Q{}^{3})$%
$A_2 \partial B_{2} \partial{}^{2} T + 
    (856 Q/9 + 5864 Q{}^{3}/3)$%
$A_2 \partial{}^{2} B_{2} \partial T + 
    (24Q + 726Q{}^{3})$%
$A_2 \partial{}^{3} B_{2} T + 
    (68/3 + 358Q{}^{2})$%
$A_2 B_3 \partial{}^{2} T + 
    (104/9 + 1636 Q{}^{2}/3)$%
$A_2 \partial B_{3} \partial T + 
    (188/9 + 418 Q{}^{2}/3)$%
$A_2 \partial{}^{2} B_{3} T + 
    (152 Q/9 + 1336 Q{}^{3}/3)$%
$\partial A_{2} A_2 \partial{}^{2} B_{2} + 
    (-214/9 - 2114 Q{}^{2}/3)$%
$\partial A_{2} A_2 \partial B_{3} + 
    6Q$%
$\partial A_{2} \partial A_{2} \partial B_{2} + 
    (-10 - 294Q{}^{2})$%
$\partial A_{2} \partial A_{2} B_3 + 
    (4 + 258Q{}^{2})$%
$\partial A_{2} A_3 \partial B_{2} + 
    18Q$%
$\partial A_{2} A_3 B_3 + 
    (10 + 312Q{}^{2})$%
$\partial A_{2} \partial A_{3} B_2 + 
    (494 Q/9 + 3046 Q{}^{3}/3)$%
$\partial A_{2} B_2 \partial{}^{2} T + 
    (48Q + 1548Q{}^{3})$%
$\partial A_{2} \partial B_{2} \partial T + 
    (80 Q/9 + 2002 Q{}^{3}/3)$%
$\partial A_{2} \partial{}^{2} B_{2} T + 
    (16 + 384Q{}^{2})$%
$\partial A_{2} B_3 \partial T + 
    (56/9 + 640 Q{}^{2}/3)$%
$\partial A_{2} \partial B_{3} T + 
    ((-34Q)/9 - 350 Q{}^{3}/3)$%
$\partial{}^{2} A_{2} A_2 \partial B_{2} + 
    (-13 - 348Q{}^{2})$%
$\partial{}^{2} A_{2} A_2 B_3 + 
    ((-172Q)/9 - 1310 Q{}^{3}/3)$%
$\partial{}^{2} A_{2} \partial A_{2} B_2 + 
    (55/9 + 428 Q{}^{2}/3)$%
$\partial{}^{2} A_{2} A_3 B_2 + 
    (76 Q/9 + 776 Q{}^{3}/3)$%
$\partial{}^{2} A_{2} B_2 \partial T + 
    ((-56Q)/3 - 10Q{}^{3})$%
$\partial{}^{2} A_{2} \partial B_{2} T + 
    (-20/3 + 98Q{}^{2})$%
$\partial{}^{2} A_{2} B_3 T + 
    ((-92Q)/9 - 670 Q{}^{3}/3)$%
$\partial{}^{3} A_{2} A_2 B_2 + 
    ((-254Q)/9 - 670 Q{}^{3}/3)$%
$\partial{}^{3} A_{2} B_2 T + 
    36Q$%
$A_3 A_3 \partial B_{2} - 
    18$%
$A_3 A_3 B_3 + (-80/9 + 50 Q{}^{2}/3)
     $%
$A_3 B_2 \partial{}^{2} T + (16 + 168Q{}^{2})
     $%
$A_3 \partial B_{2} \partial T + (224/9 + 886 Q{}^{2}/3)
     $%
$A_3 \partial{}^{2} B_{2} T - 
    36Q$%
$A_3 B_3 \partial T - 
    108Q$%
$A_3 \partial B_{3} T - 
    18Q$%
$\partial A_{3} A_3 B_2 + 
    (16 + 132Q{}^{2})$%
$\partial A_{3} B_2 \partial T + 
    (16/9 + 1016 Q{}^{2}/3)$%
$\partial A_{3} \partial B_{2} T - 
    36Q$%
$\partial A_{3} B_3 T + (116/9 + 562 Q{}^{2}/3)
     $%
$\partial{}^{2} A_{3} B_2 T + (472 Q/9 + 4376 Q{}^{3}/3)
     $%
$B_2 \partial{}^{2} T \partial T + (512 Q/9 + 2680 Q{}^{3}/3)
     $%
$B_2 \partial{}^{3} T T + 
    (15$%
$B_2{}^{2}\partial{}^{2} B_{3})/2 + 
    18Q$%
$B_2 \partial B_{3} B_3 + 
    (72Q + 1656Q{}^{3})$%
$\partial B_{2} \partial T \partial T + 
    (892 Q/9 + 7232 Q{}^{3}/3)$%
$\partial B_{2} \partial{}^{2} T T + 
    (-6 - 72Q{}^{2})$%
$\partial B_{2} B_2 \partial B_{3} - 
    6Q$%
$\partial B_{2} \partial B_{2} \partial B_{2} + 
    (6 + 36Q{}^{2})$%
$\partial B_{2} \partial B_{2} B_3 - 
    54Q$%
$\partial B_{2} B_3 B_3 + (784 Q/9 + 7016 Q{}^{3}/3)
     $%
$\partial{}^{2} B_{2} \partial T T + (3 - 18Q{}^{2})$%
$\partial{}^{2} B_{2} B_2 
      B_3 + (6Q + 108Q{}^{3})$%
$\partial{}^{2} B_{2} \partial B_{2} B_2 + 
    (238 Q/9 + 1460 Q{}^{3}/3)$%
$\partial{}^{3} B_{2} T{}^{2}+ 
    ((-9Q)/2 + 6Q{}^{3})$%
$\partial{}^{3} B_{2} B_2{}^{2}+ 
    (8 + 408Q{}^{2})$%
$B_3 \partial T \partial T + 
    (44 + 480Q{}^{2})$%
$B_3 \partial{}^{2} T T + 
    18$%
$B_3 B_3 B_3 + (536/9 + 3472 Q{}^{2}/3)
     $%
$\partial B_{3} \partial T T + (-56/9 + 1052 Q{}^{2}/3)
     $%
$\partial{}^{2} B_{3} T{}^{2}+ 
    (-5/(162\sqrt{6}) + 367 Q{}^{2}/(27\sqrt{6}) + (2531\sqrt{2/3}Q{}^{4})/3)
     $%
$J{}^{3}\partial{}^{4} B_{2} + 
    (65 Q/(3\sqrt{6}) - 508\sqrt{2/3}Q{}^{3})$%
$J{}^{3}\partial{}^{3} B_{3} + ((-284Q)/9 - 1447 Q{}^{3}/3)
     $%
$J{}^{2}A_2 \partial{}^{3} B_{2} + 
    (94/9 + 425 Q{}^{2}/3)$%
$J{}^{2}A_2 \partial{}^{2} B_{3} + 
    ((-320Q)/9 - 1663 Q{}^{3}/3)$%
$J{}^{2}\partial A_{2} 
      \partial{}^{2} B_{2} + (28/9 + 320 Q{}^{2}/3)$%
$J{}^{2}\partial A_{2} 
      \partial B_{3} + (20 Q/9 - 431 Q{}^{3}/3)$%
$J{}^{2}\partial{}^{2} A_{2} \partial B_{2} + (-10/3 - 23Q{}^{2})$%
$J{}^{2}\partial{}^{2} A_{2} B_3 + ((-19Q)/9 - 47 Q{}^{3}/3)
     $%
$J{}^{2}\partial{}^{3} A_{2} B_2 + 
    (112/9 + 11 Q{}^{2}/3)$%
$J{}^{2}A_3 \partial{}^{2} B_{2} + 
    18Q$%
$J{}^{2}A_3 \partial B_{3} + 
    (8/9 - 140 Q{}^{2}/3)$%
$J{}^{2}\partial A_{3} \partial B_{2} + 
    54Q$%
$J{}^{2}\partial A_{3} B_3 + 
    (58/9 + 65 Q{}^{2}/3)$%
$J{}^{2}\partial{}^{2} A_{3} B_2 + 
    (256 Q/9 + 1340 Q{}^{3}/3)$%
$J{}^{2}B_2 \partial{}^{3} T + 
    (226 Q/9 + 1688 Q{}^{3}/3)$%
$J{}^{2}\partial B_{2} \partial{}^{2} T + 
    ((-184Q)/9 - 668 Q{}^{3}/3)$%
$J{}^{2}\partial{}^{2} B_{2} \partial T + 
    ((-82Q)/9 - 1364 Q{}^{3}/3)$%
$J{}^{2}\partial{}^{3} B_{2} T + 
    (-2 + 528Q{}^{2})$%
$J{}^{2}B_3 \partial{}^{2} T + 
    (268/9 + 3248 Q{}^{2}/3)$%
$J{}^{2}\partial B_{3} \partial T + 
    (-56/9 + 1700 Q{}^{2}/3)$%
$J{}^{2}\partial{}^{2} B_{3} T - 
    16\sqrt{6}Q{}^{2}$%
$J A_2{}^{2}\partial{}^{2} B_{2} + 
    14\sqrt{6}Q$%
$J A_2{}^{2}\partial B_{3} - 
    12\sqrt{6}Q$%
$J A_2 A_3 \partial B_{2} - 
    12\sqrt{6}Q$%
$J A_2 \partial A_{3} B_2 + 
    ((-124\sqrt{2/3})/9 - (1124\sqrt{2/3}Q{}^{2})/3)$%
$J A_2 
      B_2 \partial{}^{2} T + ((-248\sqrt{2/3})/9 - (2194\sqrt{2/3}Q{}^{2})/3)
     $%
$J A_2 \partial B_{2} \partial T + 
    ((-232\sqrt{2/3})/9 - (1574\sqrt{2/3}Q{}^{2})/3)$%
$J A_2 
      \partial{}^{2} B_{2} T - 18\sqrt{6}Q$%
$J A_2 
      B_3 \partial T + 22\sqrt{6}Q$%
$J A_2 
      \partial B_{3} T + 12\sqrt{6}Q{}^{2}$%
$J \partial A_{2} 
      A_2 \partial B_{2} + 12\sqrt{6}Q$%
$J \partial A_{2} 
      A_2 B_3 + 12\sqrt{6}Q{}^{2}$%
$J \partial A_{2} 
      \partial A_{2} B_2 - 12\sqrt{6}Q$%
$J \partial A_{2} 
      A_3 B_2 + (-32\sqrt{2/3} - 202\sqrt{6}Q{}^{2})
     $%
$J \partial A_{2} B_2 \partial T + 
    (-8\sqrt{2/3} - 148\sqrt{6}Q{}^{2})$%
$J \partial A_{2} \partial B_{2} 
      T - 42\sqrt{6}Q$%
$J \partial A_{2} B_3 T + 
    12\sqrt{6}Q{}^{2}$%
$J \partial{}^{2} A_{2} A_2 B_2 + 
    ((52\sqrt{2/3})/9 - (694\sqrt{2/3}Q{}^{2})/3)$%
$J \partial{}^{2} A_{2} 
      B_2 T + 54\sqrt{6}Q$%
$J A_3 
      B_2 \partial T + 6\sqrt{6}Q$%
$J A_3 \partial B_{2} 
      T + 24\sqrt{6}$%
$J A_3 B_3 T + 
    30\sqrt{6}Q$%
$J \partial A_{3} B_2 T + 
    (-8\sqrt{2/3} - 160\sqrt{6}Q{}^{2})$%
$J B_2 \partial T 
      \partial T + (-44\sqrt{2/3} - 232\sqrt{6}Q{}^{2})$%
$J B_2 
      \partial{}^{2} T T + 6\sqrt{6}Q$%
$J B_2{}^{2}\partial B_{3} + ((-536\sqrt{2/3})/9 - (4552\sqrt{2/3}Q{}^{2})/3)
     $%
$J \partial B_{2} \partial T T + 
    12\sqrt{6}Q$%
$J \partial B_{2} B_2 B_3 - 
    24\sqrt{6}Q{}^{2}$%
$J \partial B_{2} \partial B_{2} B_2 + 
    ((-124\sqrt{2/3})/9 - (1592\sqrt{2/3}Q{}^{2})/3)$%
$J \partial{}^{2} B_{2} 
      T{}^{2}- 12\sqrt{6}Q{}^{2}$%
$J \partial{}^{2} B_{2} B_2{}^{2}+ 96\sqrt{6}Q$%
$J B_3 \partial T T + 
    44\sqrt{6}Q$%
$J \partial B_{3} T{}^{2}+ 
    ((-59\sqrt{2/3})/27 + 4765 Q{}^{2}/(9\sqrt{6}) + 2750\sqrt{6}Q{}^{4})
     $%
$\partial J J{}^{2}\partial{}^{3} B_{2} + 
    ((-335Q)/\sqrt{6} - 1538\sqrt{6}Q{}^{3})$%
$\partial J J{}^{2}\partial{}^{2} B_{3} + ((-832Q)/9 - 5402 Q{}^{3}/3)$%
$\partial J J 
      A_2 \partial{}^{2} B_{2} + (104/9 + 790 Q{}^{2}/3)
     $%
$\partial J J A_2 \partial B_{3} + 
    (-104Q - 1848Q{}^{3})$%
$\partial J J \partial A_{2} \partial B_{2} + 
    (16 + 294Q{}^{2})$%
$\partial J J \partial A_{2} B_3 + 
    ((-200Q)/9 - 874 Q{}^{3}/3)$%
$\partial J J \partial{}^{2} A_{2} 
      B_2 + (16 + 78Q{}^{2})$%
$\partial J J A_3 
      \partial B_{2} + 36Q$%
$\partial J J A_3 B_3 + 
    (16 + 114Q{}^{2})$%
$\partial J J \partial A_{3} B_2 + 
    (1604 Q/9 + 11512 Q{}^{3}/3)$%
$\partial J J B_2 
      \partial{}^{2} T + (2096 Q/9 + 15592 Q{}^{3}/3)$%
$\partial J J 
      \partial B_{2} \partial T + (32Q + 1152Q{}^{3})$%
$\partial J J 
      \partial{}^{2} B_{2} T + (64 + 1392Q{}^{2})$%
$\partial J J 
      B_3 \partial T + (536/9 + 3976 Q{}^{2}/3)$%
$\partial J J 
      \partial B_{3} T + ((-59\sqrt{2/3})/9 + (2264\sqrt{2/3}Q{}^{2})/3 + 
      6800\sqrt{6}Q{}^{4})$%
$(\partial J){}^{2} J \partial{}^{2} B_{2} + 
    (-470\sqrt{2/3}Q - 3130\sqrt{6}Q{}^{3})$%
$(\partial J){}^{2} J 
      \partial B_{3} + ((-118\sqrt{2/3})/27 + (3460\sqrt{2/3}Q{}^{2})/9 + 
      3704\sqrt{6}Q{}^{4})$%
$(\partial J){}^{3} \partial B_{2} + 
    (-172\sqrt{2/3}Q - 1208\sqrt{6}Q{}^{3})$%
$(\partial J){}^{3} 
      B_3 + ((-754Q)/9 - 5102 Q{}^{3}/3)$%
$(\partial J){}^{2} 
      A_2 \partial B_{2} + (68/3 + 484Q{}^{2})$%
$(\partial J){}^{2} 
      A_2 B_3 + ((-154Q)/9 - 662 Q{}^{3}/3)
     $%
$(\partial J){}^{2} \partial A_{2} B_2 + 
    (-80/9 - 832 Q{}^{2}/3)$%
$(\partial J){}^{2} A_3 B_2 + 
    (592 Q/3 + 4376Q{}^{3})$%
$(\partial J){}^{2} B_2 \partial T + 
    (1604 Q/9 + 11692 Q{}^{3}/3)$%
$(\partial J){}^{2} \partial B_{2} 
      T + (-4 - 204Q{}^{2})$%
$(\partial J){}^{2} B_3 T + 
    19\sqrt{6}Q{}^{2}$%
$\partial J A_2{}^{2}\partial B_{2} - 
    24\sqrt{6}Q$%
$\partial J A_2{}^{2}B_3 + 
    28\sqrt{6}Q$%
$\partial J A_2 A_3 B_2 + 
    ((-248\sqrt{2/3})/9 - (1240\sqrt{2/3}Q{}^{2})/3)$%
$\partial J A_2 
      B_2 \partial T + ((-248\sqrt{2/3})/9 - (1654\sqrt{2/3}Q{}^{2})/3)
     $%
$\partial J A_2 \partial B_{2} T + 
    30\sqrt{6}Q$%
$\partial J A_2 B_3 T - 
    22\sqrt{6}Q{}^{2}$%
$\partial J \partial A_{2} A_2 B_2 + 
    (-32\sqrt{2/3} - 206\sqrt{6}Q{}^{2})$%
$\partial J \partial A_{2} B_2 
      T + 46\sqrt{6}Q$%
$\partial J A_3 B_2 T + 
    ((-536\sqrt{2/3})/9 - (3112\sqrt{2/3}Q{}^{2})/3)$%
$\partial J B_2 
      \partial T T + ((-248\sqrt{2/3})/9 - (2176\sqrt{2/3}Q{}^{2})/3)
     $%
$\partial J \partial B_{2} T{}^{2}- 
    9\sqrt{6}Q{}^{2}$%
$\partial J \partial B_{2} B_2{}^{2}+ 
    96\sqrt{6}Q$%
$\partial J B_3 T{}^{2}+ 
    (-121/(9\sqrt{6}) + (1721\sqrt{2/3}Q{}^{2})/9 + (19834\sqrt{2/3}Q{}^{4})/3)
     $%
$\partial{}^{2} J J{}^{2}\partial{}^{2} B_{2} + 
    (-90\sqrt{6}Q - 1488\sqrt{6}Q{}^{3})$%
$\partial{}^{2} J J{}^{2}\partial B_{3} + ((-1060Q)/9 - 5768 Q{}^{3}/3)$%
$\partial{}^{2} J J 
      A_2 \partial B_{2} + (68/3 + 340Q{}^{2})$%
$\partial{}^{2} J J 
      A_2 B_3 + ((-388Q)/9 - 1688 Q{}^{3}/3)
     $%
$\partial{}^{2} J J \partial A_{2} B_2 + 
    (100/9 + 212 Q{}^{2}/3)$%
$\partial{}^{2} J J A_3 B_2 + 
    (1696 Q/9 + 11216 Q{}^{3}/3)$%
$\partial{}^{2} J J B_2 
      \partial T + (208 Q/3 + 1904Q{}^{3})$%
$\partial{}^{2} J J \partial B_{2} 
      T + (56 + 720Q{}^{2})$%
$\partial{}^{2} J J B_3 T + 
    ((-242\sqrt{2/3})/9 + 580\sqrt{2/3}Q{}^{2} + 22544\sqrt{2/3}Q{}^{4})
     $%
$\partial{}^{2} J \partial J J \partial B_{2} + 
    (-628\sqrt{2/3}Q - 3656\sqrt{6}Q{}^{3})$%
$\partial{}^{2} J \partial J J 
      B_3 + ((-121\sqrt{2/3})/9 - (490\sqrt{2/3}Q{}^{2})/9 + 
      (15994\sqrt{2/3}Q{}^{4})/3)$%
$\partial{}^{2} J (\partial J){}^{2} 
      B_2 + ((-184Q)/3 - 1084Q{}^{3})$%
$\partial{}^{2} J \partial J 
      A_2 B_2 + (2728 Q/9 + 16856 Q{}^{3}/3)
     $%
$\partial{}^{2} J \partial J B_2 T + 
    (-91/(3\sqrt{6}) - (2842\sqrt{2/3}Q{}^{2})/9 - (1862\sqrt{2/3}Q{}^{4})/3)
     $%
$\partial{}^{2} J \partial{}^{2} J J B_2 - 
    5\sqrt{2/3}Q{}^{2}$%
$\partial{}^{2} J A_2{}^{2}B_2 + 
    ((-304\sqrt{2/3})/9 - (1244\sqrt{2/3}Q{}^{2})/3)$%
$\partial{}^{2} J A_2 
      B_2 T + ((-304\sqrt{2/3})/9 - (1364\sqrt{2/3}Q{}^{2})/3)
     $%
$\partial{}^{2} J B_2 T{}^{2}- 
    \sqrt{6}Q{}^{2}$%
$\partial{}^{2} J B_2{}^{3}+ 
    ((-85\sqrt{2/3})/9 - (905\sqrt{2/3}Q{}^{2})/9 + (1832\sqrt{2/3}Q{}^{4})/3)
     $%
$\partial{}^{3} J J{}^{2}\partial B_{2} + 
    (-109\sqrt{2/3}Q - 476\sqrt{6}Q{}^{3})$%
$\partial{}^{3} J J{}^{2}B_3 + ((-244Q)/9 - 1196 Q{}^{3}/3)$%
$\partial{}^{3} J J 
      A_2 B_2 + (748 Q/9 + 3536 Q{}^{3}/3)
     $%
$\partial{}^{3} J J B_2 T + 
    ((-170\sqrt{2/3})/9 - (1274\sqrt{2/3}Q{}^{2})/3 - 1352\sqrt{2/3}Q{}^{4})
     $%
$\partial{}^{3} J \partial J J B_2 + 
    (-175/(18\sqrt{6}) - (1156\sqrt{2/3}Q{}^{2})/9 - (2555\sqrt{2/3}Q{}^{4})/3)
     $%
$\partial{}^{4} J J{}^{2}B_2 - 
    7Q$%
$A_2{}^{3}\partial B_{2} + 
    7$%
$A_2{}^{3}B_3 - 
    7$%
$A_2{}^{3}3 B_2 - 
    10Q$%
$A_2{}^{2}B_2 \partial T + 
    2Q$%
$A_2{}^{2}\partial B_{2} T - 
    12$%
$A_2{}^{2}B_3 T + 
    16$%
$A_2 A_3 B_2 T - 
    80Q$%
$A_2 B_2 \partial T T - 
    3$%
$A_2 B_2{}^{3}3 - 
    44Q$%
$A_2 \partial B_{2} T{}^{2}+ 
    3Q$%
$A_2 \partial B_{2} B_2{}^{2}+ 
    7Q$%
$\partial A_{2} A_2{}^{2}B_2 + 
    20Q$%
$\partial A_{2} A_2 B_2 T + 
    4Q$%
$\partial A_{2} B_2 T{}^{2}- 
    3Q$%
$\partial A_{2} B_2{}^{3}- 
    40$%
$A_3 B_2 T{}^{2}+ 
    3$%
$A_3 B_2{}^{3}- 
    16Q$%
$B_2 \partial T T{}^{2}- 
    6Q$%
$B_2{}^{3}\partial T - 
    12$%
$B_2{}^{3}3 T - 
    8Q$%
$\partial B_{2} T{}^{3}- 
    30Q$%
$\partial B_{2} B_2{}^{2}T - 
    8$%
$B_3 T{}^{3}+ ((-17Q)/2 - 285Q{}^{3})
     $%
$J{}^{4}\partial{}^{3} B_{2} + 
    (-14/9 + 155 Q{}^{2}/3)$%
$J{}^{4}\partial{}^{2} B_{3} + ((116\sqrt{2/3})/27 - (5\sqrt{2/3}Q{}^{2})/9)
     $%
$J{}^{3}A_2 \partial{}^{2} B_{2} - 
    \sqrt{6}Q$%
$J{}^{3}A_2 \partial B_{3} + 
    ((4\sqrt{2/3})/3 - 22\sqrt{2/3}Q{}^{2})$%
$J{}^{3}\partial A_{2} \partial B_{2} - \sqrt{6}Q$%
$J{}^{3}\partial A_{2} B_3 + ((-26\sqrt{2/3})/27 - (301\sqrt{2/3}Q{}^{2})/9)
     $%
$J{}^{3}\partial{}^{2} A_{2} B_2 + 
    23\sqrt{6}Q$%
$J{}^{3}A_3 
      \partial B_{2} - 4\sqrt{6}$%
$J{}^{3}A_3 
      B_3 + 19\sqrt{6}Q$%
$J{}^{3}\partial A_{3} B_2 + ((-266\sqrt{2/3})/27 - 
      (1636\sqrt{2/3}Q{}^{2})/9)$%
$J{}^{3}B_2 
      \partial{}^{2} T + ((-76\sqrt{2/3})/9 - (740\sqrt{2/3}Q{}^{2})/3)
     $%
$J{}^{3}\partial B_{2} \partial T + 
    ((-124\sqrt{2/3})/27 - (1232\sqrt{2/3}Q{}^{2})/9)$%
$J{}^{3}\partial{}^{2} B_{2} T - 40\sqrt{6}Q
     $%
$J{}^{3}B_3 \partial T - 
    28\sqrt{6}Q$%
$J{}^{3}\partial B_{3} T + 
    5Q$%
$J{}^{2}A_2{}^{2}\partial B_{2} - 
    6$%
$J{}^{2}A_2{}^{2}B_3 + 
    8$%
$J{}^{2}A_2 A_3 B_2 - 
    40Q$%
$J{}^{2}A_2 B_2 \partial T - 
    12Q$%
$J{}^{2}A_2 \partial B_{2} T - 
    14Q$%
$J{}^{2}\partial A_{2} A_2 B_2 + 
    4Q$%
$J{}^{2}\partial A_{2} B_2 T - 
    40$%
$J{}^{2}A_3 B_2 T - 
    16Q$%
$J{}^{2}B_2 \partial T T - 
    6$%
$J{}^{2}B_2{}^{3}3 + 
    28Q$%
$J{}^{2}\partial B_{2} T{}^{2}+ 
    21Q$%
$J{}^{2}\partial B_{2} B_2{}^{2}- 
    60$%
$J{}^{2}B_3 T{}^{2}- 
    4\sqrt{2/3}$%
$J A_2{}^{2}B_2 T + 
    40\sqrt{2/3}$%
$J A_2 B_2 T{}^{2}+ 
    8\sqrt{2/3}$%
$J B_2 T{}^{3}+ 
    4\sqrt{6}$%
$J B_2{}^{3}T + 
    ((-3032Q)/27 - 20920 Q{}^{3}/9)$%
$\partial J J{}^{3}\partial{}^{2} B_{2} + (268/9 + 980 Q{}^{2}/3)
     $%
$\partial J J{}^{3}\partial B_{3} + 
    ((124\sqrt{2/3})/9 + (143\sqrt{2/3}Q{}^{2})/3)$%
$\partial J J{}^{2}A_2 \partial B_{2} + 
    9\sqrt{6}Q$%
$\partial J J{}^{2}A_2 B_3 + 
    (16\sqrt{2/3} + 27\sqrt{6}Q{}^{2})$%
$\partial J J{}^{2}\partial A_{2} B_2 + 41\sqrt{6}Q$%
$\partial J J{}^{2}A_3 B_2 + 
    ((-268\sqrt{2/3})/9 - (1556\sqrt{2/3}Q{}^{2})/3)$%
$\partial J J{}^{2}B_2 \partial T + (-32\sqrt{2/3} - 240\sqrt{6}Q{}^{2})
     $%
$\partial J J{}^{2}\partial B_{2} T + 
    24\sqrt{6}Q$%
$\partial J J{}^{2}B_3 T + 
    2Q$%
$\partial J J A_2{}^{2}B_2 - 
    144Q$%
$\partial J J A_2 B_2 T - 
    104Q$%
$\partial J J B_2 T{}^{2}+ 
    6Q$%
$\partial J J B_2{}^{3}+ 
    ((-2306Q)/9 - 15454 Q{}^{3}/3)$%
$(\partial J){}^{2} J{}^{2}\partial B_{2} + (30 + 306Q{}^{2})$%
$(\partial J){}^{2} 
      J{}^{2}B_3 + 
    ((124\sqrt{2/3})/9 + (512\sqrt{2/3}Q{}^{2})/3)$%
$(\partial J){}^{2} 
      J A_2 B_2 + 
    ((-4\sqrt{2/3})/9 + (412\sqrt{2/3}Q{}^{2})/3)$%
$(\partial J){}^{2} 
      J B_2 T + ((-428Q)/3 - 3028Q{}^{3})
     $%
$(\partial J){}^{3} J B_2 + 
    ((-1144Q)/9 - 6128 Q{}^{3}/3)$%
$\partial{}^{2} J J{}^{3}\partial B_{2} + (28 + 272Q{}^{2})$%
$\partial{}^{2} J J{}^{3}B_3 + 
    ((152\sqrt{2/3})/9 + (502\sqrt{2/3}Q{}^{2})/3)$%
$\partial{}^{2} J J{}^{2}A_2 B_2 + 
    ((-200\sqrt{2/3})/9 - (772\sqrt{2/3}Q{}^{2})/3)$%
$\partial{}^{2} J J{}^{2}B_2 T + ((-2404Q)/9 - 14300 Q{}^{3}/3)
     $%
$\partial{}^{2} J \partial J J{}^{2}B_2 + 
    ((-934Q)/27 - 4448 Q{}^{3}/9)$%
$\partial{}^{3} J J{}^{3}B_2 + ((31\sqrt{2/3})/27 + (146\sqrt{2/3}Q{}^{2})/9)
     $%
$J{}^{5}\partial{}^{2} B_{2} + 
    7\sqrt{6}Q$%
$J{}^{5}\partial B_{3} + 21Q$%
$J{}^{4}A_2 \partial B_{2} + 17Q$%
$J{}^{4}\partial A_{2} B_2 - 
    10$%
$J{}^{4}A_3 
      B_2 - 4Q$%
$J{}^{4}B_2 \partial T + 2Q$%
$J{}^{4}\partial B_{2} T + 10$%
$J{}^{4}B_3 T + 
    (2\sqrt{2/3}$%
$J{}^{3}A_2{}^{2}B_2)/3 + (40\sqrt{2/3}$%
$J{}^{3}A_2 B_2 T)/3 + 
    (20\sqrt{2/3}$%
$J{}^{3}B_2 T{}^{2})/3 - 2\sqrt{2/3}$%
$J{}^{3}B_2{}^{3}+ 
    ((110\sqrt{2/3})/9 + (424\sqrt{2/3}Q{}^{2})/3)$%
$\partial J J{}^{4}\partial B_{2} + 
    28\sqrt{6}Q$%
$\partial J J{}^{4}B_3 + (152Q$%
$\partial J J{}^{3}A_2 B_2)/3 - 
    (88Q$%
$\partial J J{}^{3}B_2 T)/
     3 + ((538\sqrt{2/3})/27 + (2906\sqrt{2/3}Q{}^{2})/9)
     $%
$(\partial J){}^{2} J{}^{3}B_2 + 
    ((328\sqrt{2/3})/27 + (1361\sqrt{2/3}Q{}^{2})/9)$%
$\partial{}^{2} J J{}^{4}B_2 + 
    (11Q$%
$J{}^{6}\partial B_{2})/3 - 
    (7$%
$J{}^{6}B_3)/3 - (10\sqrt{2/3}$%
$J{}^{5}A_2 B_2)/3 + 
    (2\sqrt{2/3}$%
$J{}^{5}B_2 T)/3 + (34Q$%
$\partial J J{}^{5}B_2)/3 - 
    (\sqrt{2/3}$%
$J{}^{7}B_2)/3.
\  $ 
 \generatorhead{$W_{12}$} \noindent $W_{12} = (-1/972 - 49571 Q{}^{2}/85050 - 2769829 Q{}^{4}/85050 - 271978 Q{}^{6}/225 - 
      3015164 Q{}^{8}/135 - 415184 Q{}^{10}/3)$%
$\partial{}^{10} T + 
    (1147 Q/(8748\sqrt{6}) + 176479 Q{}^{3}/(8019\sqrt{6}) + 
      7854071 Q{}^{5}/(5346\sqrt{6}) + 41300825 Q{}^{7}/(891\sqrt{6}) + 
      (33314140\sqrt{2/3}Q{}^{9})/99 + 586040\sqrt{6}Q{}^{11})$%
$\partial{}^{11} J + 
    (-53377/2449440 - 2141653 Q{}^{2}/1020600 - 27450461 Q{}^{4}/340200 - 
      6853219 Q{}^{6}/2835 - 36309Q{}^{8} - 579632 Q{}^{10}/3)$%
$\partial{}^{10} A_{2} + 
    (39397 Q/15120 + 388891 Q{}^{3}/7560 + 249469 Q{}^{5}/630 + 
      434768 Q{}^{7}/105 + 22048Q{}^{9})$%
$\partial{}^{9} A_{3} + 
    (185/486 + 949 Q{}^{2}/9 + 192754 Q{}^{4}/27 + 509452 Q{}^{6}/3 + 
      3552304 Q{}^{8}/3)$%
$\partial{}^{4} T \partial{}^{4} T + 
    (200/243 + 3052 Q{}^{2}/15 + 337732 Q{}^{4}/27 + 1435544 Q{}^{6}/5 + 
      1978592Q{}^{8})$%
$\partial{}^{5} T \partial{}^{3} T + 
    (121/243 + 50326 Q{}^{2}/405 + 106168 Q{}^{4}/15 + 7265072 Q{}^{6}/45 + 
      3367616 Q{}^{8}/3)$%
$\partial{}^{6} T \partial{}^{2} T + 
    (4852/5103 + 690484 Q{}^{2}/8505 + 1347844 Q{}^{4}/405 + 62181368 Q{}^{6}/945 + 
      430944Q{}^{8})$%
$\partial{}^{7} T \partial T + (1823/3402 + 25168 Q{}^{2}/945 + 
      159380 Q{}^{4}/189 + 539964 Q{}^{6}/35 + 98608Q{}^{8})$%
$\partial{}^{8} T T + 
    ((5827\sqrt{2/3}Q)/25515 - (20242\sqrt{2/3}Q{}^{3})/567 - 
      (3943636\sqrt{2/3}Q{}^{5})/2835 - (1733884\sqrt{2/3}Q{}^{7})/105 - 
      21424\sqrt{6}Q{}^{9})$%
$J \partial{}^{9} T + 
    (83761 Q/(102060\sqrt{6}) + 94523 Q{}^{3}/(972\sqrt{6}) + 
      (1104512\sqrt{2/3}Q{}^{5})/945 + (11322572\sqrt{2/3}Q{}^{7})/945 + 
      44096\sqrt{2/3}Q{}^{9})$%
$J \partial{}^{9} A_{2} + 
    ((-1685Q{}^{2})/(21\sqrt{6}) - (3011\sqrt{2/3}Q{}^{4})/5 - 
      (15916\sqrt{6}Q{}^{6})/7 - 10704\sqrt{6}Q{}^{8})$%
$J \partial{}^{8} A_{3} + 
    ((-1027Q)/(630\sqrt{6}) - (130642\sqrt{2/3}Q{}^{3})/315 - 
      (1887392\sqrt{2/3}Q{}^{5})/105 - (10039202\sqrt{2/3}Q{}^{7})/35 - 
      517784\sqrt{6}Q{}^{9})$%
$\partial J \partial{}^{8} T + 
    ((-929Q)/(648\sqrt{6}) - 1149073 Q{}^{3}/(7560\sqrt{6}) - 
      4078667 Q{}^{5}/(315\sqrt{6}) - (766813\sqrt{2/3}Q{}^{7})/5 - 
      341380\sqrt{6}Q{}^{9})$%
$\partial J \partial{}^{8} A_{2} + 
    ((-1557\sqrt{3/2}Q{}^{2})/35 - (723691\sqrt{2/3}Q{}^{4})/315 - 
      29143\sqrt{2/3}Q{}^{6} - 126460\sqrt{2/3}Q{}^{8})$%
$\partial J \partial{}^{7} A_{3} + 
    ((-38431\sqrt{2/3}Q)/8505 - (4435066\sqrt{2/3}Q{}^{3})/2835 - 
      (65350442\sqrt{2/3}Q{}^{5})/945 - (361370696\sqrt{2/3}Q{}^{7})/315 - 
      6417920\sqrt{2/3}Q{}^{9})$%
$\partial{}^{2} J \partial{}^{7} T + 
    ((-142381Q)/(17010\sqrt{6}) - 11099461 Q{}^{3}/(11340\sqrt{6}) - 
      60776813 Q{}^{5}/(945\sqrt{6}) - (220712708\sqrt{2/3}Q{}^{7})/315 - 
      1519892\sqrt{6}Q{}^{9})$%
$\partial{}^{2} J \partial{}^{7} A_{2} + 
    ((-743\sqrt{3/2}Q{}^{2})/5 - 1407437 Q{}^{4}/(90\sqrt{6}) - 
      (167993\sqrt{6}Q{}^{6})/5 - 452900\sqrt{2/3}Q{}^{8})
     $%
$\partial{}^{2} J \partial{}^{6} A_{3} + ((-31679Q)/(1215\sqrt{6}) - 
      (1487107\sqrt{2/3}Q{}^{3})/405 - (4400302\sqrt{2/3}Q{}^{5})/27 - 
      (124788356\sqrt{2/3}Q{}^{7})/45 - 15883280\sqrt{2/3}Q{}^{9})
     $%
$\partial{}^{3} J \partial{}^{6} T + ((-15584\sqrt{2/3}Q)/1215 - 
      404131 Q{}^{3}/(135\sqrt{6}) - 8020853 Q{}^{5}/(45\sqrt{6}) - 
      (84788467\sqrt{2/3}Q{}^{7})/45 - 12146848\sqrt{2/3}Q{}^{9})
     $%
$\partial{}^{3} J \partial{}^{6} A_{2} + ((-44611Q{}^{2})/(45\sqrt{6}) - 
      (746129\sqrt{2/3}Q{}^{4})/45 - (598880\sqrt{2/3}Q{}^{6})/3 - 
      295792\sqrt{6}Q{}^{8})$%
$\partial{}^{3} J \partial{}^{5} A_{3} + 
    ((-62629Q)/(1215\sqrt{6}) - 4958897 Q{}^{3}/(405\sqrt{6}) - 
      (36304108\sqrt{2/3}Q{}^{5})/135 - (207346424\sqrt{2/3}Q{}^{7})/45 - 
      8870848\sqrt{6}Q{}^{9})$%
$\partial{}^{4} J \partial{}^{5} T + 
    ((-62629Q)/(1215\sqrt{6}) - 4979551 Q{}^{3}/(810\sqrt{6}) - 
      (22864567\sqrt{2/3}Q{}^{5})/135 - (155534063\sqrt{2/3}Q{}^{7})/45 - 
      7307888\sqrt{6}Q{}^{9})$%
$\partial{}^{4} J \partial{}^{5} A_{2} + 
    ((-91799Q{}^{2})/(54\sqrt{6}) - (206912\sqrt{2/3}Q{}^{4})/9 - 
      (721688\sqrt{2/3}Q{}^{6})/3 - 334544\sqrt{6}Q{}^{8})
     $%
$\partial{}^{4} J \partial{}^{4} A_{3} + ((-38227\sqrt{2/3}Q)/1215 - 
      (929618\sqrt{2/3}Q{}^{3})/135 - (41252806\sqrt{2/3}Q{}^{5})/135 - 
      (26532872\sqrt{2/3}Q{}^{7})/5 - (92767808\sqrt{2/3}Q{}^{9})/3)
     $%
$\partial{}^{5} J \partial{}^{4} T + (-56/243 + 187985 Q{}^{2}/972 + 
      9318533 Q{}^{4}/450 + 549410239 Q{}^{6}/675 + 68696194 Q{}^{8}/5 + 
      79378824Q{}^{10})$%
$\partial{}^{5} J \partial{}^{5} J + 
    ((-38227\sqrt{2/3}Q)/1215 - 2176823 Q{}^{3}/(270\sqrt{6}) - 
      116184061 Q{}^{5}/(270\sqrt{6}) - (12942424\sqrt{2/3}Q{}^{7})/3 - 
      (81474680\sqrt{2/3}Q{}^{9})/3)$%
$\partial{}^{5} J \partial{}^{4} A_{2} + 
    ((-121958\sqrt{2/3}Q{}^{2})/135 - 1990069 Q{}^{4}/(45\sqrt{6}) - 
      (2922478\sqrt{2/3}Q{}^{6})/15 - 646856\sqrt{2/3}Q{}^{8})
     $%
$\partial{}^{5} J \partial{}^{3} A_{3} + ((-45218\sqrt{2/3}Q)/1215 - 
      (267946\sqrt{2/3}Q{}^{3})/45 - (34207852\sqrt{2/3}Q{}^{5})/135 - 
      (21734264\sqrt{2/3}Q{}^{7})/5 - 8397088\sqrt{6}Q{}^{9})
     $%
$\partial{}^{6} J \partial{}^{3} T + (-1225/2916 + 154825 Q{}^{2}/486 + 
      13915397 Q{}^{4}/405 + 182621708 Q{}^{6}/135 + 22828188Q{}^{8} + 
      131803728Q{}^{10})$%
$\partial{}^{6} J \partial{}^{4} J + 
    ((-45218\sqrt{2/3}Q)/1215 - (37291\sqrt{2/3}Q{}^{3})/9 - 
      (8815309\sqrt{2/3}Q{}^{5})/45 - (167261992\sqrt{2/3}Q{}^{7})/45 - 
      22825160\sqrt{2/3}Q{}^{9})$%
$\partial{}^{6} J \partial{}^{3} A_{2} + 
    ((-26749Q{}^{2})/(27\sqrt{6}) - (565324\sqrt{2/3}Q{}^{4})/45 - 
      (1752442\sqrt{2/3}Q{}^{6})/15 - 132920\sqrt{6}Q{}^{8})
     $%
$\partial{}^{6} J \partial{}^{2} A_{3} + ((-368737Q)/(8505\sqrt{6}) - 
      (9232747\sqrt{2/3}Q{}^{3})/2835 - (132330112\sqrt{2/3}Q{}^{5})/945 - 
      (21835732\sqrt{2/3}Q{}^{7})/9 - 4709264\sqrt{6}Q{}^{9})
     $%
$\partial{}^{7} J \partial{}^{2} T + (-616/2187 + 1855667 Q{}^{2}/10206 + 
      168981461 Q{}^{4}/8505 + 444026162 Q{}^{6}/567 + 1382035856 Q{}^{8}/105 + 
      75670816Q{}^{10})$%
$\partial{}^{7} J \partial{}^{3} J + 
    ((-368737Q)/(8505\sqrt{6}) - 14665961 Q{}^{3}/(2835\sqrt{6}) - 
      (112620028\sqrt{2/3}Q{}^{5})/945 - (693397004\sqrt{2/3}Q{}^{7})/315 - 
      4430432\sqrt{6}Q{}^{9})$%
$\partial{}^{7} J \partial{}^{2} A_{2} + 
    ((-2737Q{}^{2})/(9\sqrt{6}) - (1108687\sqrt{2/3}Q{}^{4})/315 - 
      (3199472\sqrt{2/3}Q{}^{6})/105 - 33072\sqrt{6}Q{}^{8})
     $%
$\partial{}^{7} J \partial A_{3} + ((-131333\sqrt{2/3}Q)/8505 - 
      9086237 Q{}^{3}/(2835\sqrt{6}) - 22761289 Q{}^{5}/(189\sqrt{6}) - 
      (304776746\sqrt{2/3}Q{}^{7})/315 - 5388512\sqrt{2/3}Q{}^{9})
     $%
$\partial{}^{8} J \partial T + (-911/6804 + 34541 Q{}^{2}/486 + 
      22444256 Q{}^{4}/2835 + 589903009 Q{}^{6}/1890 + 1638398842 Q{}^{8}/315 + 
      88930336 Q{}^{10}/3)$%
$\partial{}^{8} J \partial{}^{2} J + 
    ((-131333\sqrt{2/3}Q)/8505 - (4092427\sqrt{2/3}Q{}^{3})/2835 - 
      41994227 Q{}^{5}/(378\sqrt{6}) - (289575659\sqrt{2/3}Q{}^{7})/315 - 
      5213648\sqrt{2/3}Q{}^{9})$%
$\partial{}^{8} J \partial A_{2} + 
    ((-77897Q{}^{2})/(1890\sqrt{6}) - (138584\sqrt{2/3}Q{}^{4})/315 - 
      (74579\sqrt{2/3}Q{}^{6})/21 - 11024\sqrt{2/3}Q{}^{8})
     $%
$\partial{}^{8} J A_3 + ((-400279Q)/(51030\sqrt{6}) - 
      6556673 Q{}^{3}/(8505\sqrt{6}) - 11502733 Q{}^{5}/(405\sqrt{6}) - 
      (212202332\sqrt{2/3}Q{}^{7})/945 - (3703352\sqrt{2/3}Q{}^{9})/3)
     $%
$\partial{}^{9} J T + (-202/5103 + 345349 Q{}^{2}/20412 + 
      6505123 Q{}^{4}/3402 + 12156655 Q{}^{6}/162 + 233390032 Q{}^{8}/189 + 
      6956008Q{}^{10})$%
$\partial{}^{9} J \partial J + 
    ((-400279Q)/(51030\sqrt{6}) - 6278329 Q{}^{3}/(8505\sqrt{6}) - 
      155800847 Q{}^{5}/(5670\sqrt{6}) - (207834586\sqrt{2/3}Q{}^{7})/945 - 
      (3654200\sqrt{2/3}Q{}^{9})/3)$%
$\partial{}^{9} J A_2 + 
    (-113/5103 - 18493 Q{}^{2}/8505 - 129568 Q{}^{4}/1215 - 2590996 Q{}^{6}/945 - 
      32282792 Q{}^{8}/945 - 479456 Q{}^{10}/3)$%
$\partial{}^{10} J J + 
    (31/1890 + 4427 Q{}^{2}/405 + 519298 Q{}^{4}/945 + 1323076 Q{}^{6}/105 + 
      88528Q{}^{8})$%
$A_2 \partial{}^{8} T + 
    ((-184Q)/27 - 69379 Q{}^{3}/945 - 23272 Q{}^{5}/21 - 20656 Q{}^{7}/3)
     $%
$A_2 \partial{}^{7} A_{3} + (232/729 + 555706 Q{}^{2}/8505 + 
      6886624 Q{}^{4}/2835 + 49641752 Q{}^{6}/945 + 374880Q{}^{8})
     $%
$\partial A_{2} \partial{}^{7} T + (1481 Q/135 + 599 Q{}^{3}/6 - 26464 Q{}^{5}/15 - 
      12992Q{}^{7})$%
$\partial A_{2} \partial{}^{6} A_{3} + 
    (13/45 + 12521 Q{}^{2}/81 + 831554 Q{}^{4}/135 + 1239628 Q{}^{6}/9 + 
      3018176 Q{}^{8}/3)$%
$\partial{}^{2} A_{2} \partial{}^{6} T + 
    (1291 Q/54 + 6412 Q{}^{3}/45 - 79126 Q{}^{5}/15 - 31976Q{}^{7})
     $%
$\partial{}^{2} A_{2} \partial{}^{5} A_{3} + (236/405 + 124406 Q{}^{2}/405 + 
      1813847 Q{}^{4}/135 + 12557108 Q{}^{6}/45 + 1909712Q{}^{8})
     $%
$\partial{}^{3} A_{2} \partial{}^{5} T + (3961 Q/162 - 1364 Q{}^{3}/3 - 153104 Q{}^{5}/9 - 
      87800Q{}^{7})$%
$\partial{}^{3} A_{2} \partial{}^{4} A_{3} + 
    (19/81 + 18877 Q{}^{2}/54 + 506134 Q{}^{4}/27 + 1167704 Q{}^{6}/3 + 
      7673648 Q{}^{8}/3)$%
$\partial{}^{4} A_{2} \partial{}^{4} T + 
    (17875/972 + 2707 Q{}^{2}/6 + 393748 Q{}^{4}/27 + 818894 Q{}^{6}/3 + 
      4912984 Q{}^{8}/3)$%
$\partial{}^{4} A_{2} \partial{}^{4} A_{2} + 
    (314 Q/81 - 31568 Q{}^{3}/27 - 242366 Q{}^{5}/9 - 418648 Q{}^{7}/3)
     $%
$\partial{}^{4} A_{2} \partial{}^{3} A_{3} + (236/405 + 138976 Q{}^{2}/405 + 
      58475 Q{}^{4}/3 + 17166206 Q{}^{6}/45 + 2353792Q{}^{8})
     $%
$\partial{}^{5} A_{2} \partial{}^{3} T + (30343/972 + 1321037 Q{}^{2}/1620 + 
      252363 Q{}^{4}/10 + 20450564 Q{}^{6}/45 + 2682112Q{}^{8})
     $%
$\partial{}^{5} A_{2} \partial{}^{3} A_{2} + ((-853Q)/45 - 111527 Q{}^{3}/90 - 
      61576 Q{}^{5}/3 - 95688Q{}^{7})$%
$\partial{}^{5} A_{2} \partial{}^{2} A_{3} + 
    (7/45 + 25361 Q{}^{2}/135 + 1671389 Q{}^{4}/135 + 10899866 Q{}^{6}/45 + 
      4368448 Q{}^{8}/3)$%
$\partial{}^{6} A_{2} \partial{}^{2} T + 
    (78329/4860 + 72107 Q{}^{2}/162 + 1276471 Q{}^{4}/90 + 767576 Q{}^{6}/3 + 
      4541152 Q{}^{8}/3)$%
$\partial{}^{6} A_{2} \partial{}^{2} A_{2} + 
    ((-4823Q)/405 - 17227 Q{}^{3}/27 - 80512 Q{}^{5}/9 - 38808Q{}^{7})
     $%
$\partial{}^{6} A_{2} \partial A_{3} + (407/1458 + 3884143 Q{}^{2}/34020 + 
      17333542 Q{}^{4}/2835 + 14290112 Q{}^{6}/135 + 1773896 Q{}^{8}/3)
     $%
$\partial{}^{7} A_{2} \partial T + (626219/102060 + 6716677 Q{}^{2}/34020 + 
      17333686 Q{}^{4}/2835 + 96374954 Q{}^{6}/945 + 1730552 Q{}^{8}/3)
     $%
$\partial{}^{7} A_{2} \partial A_{2} + ((-116Q)/35 - 111997 Q{}^{3}/945 - 
      89830 Q{}^{5}/63 - 5992Q{}^{7})$%
$\partial{}^{7} A_{2} A_3 + 
    (-3503/68040 + 83561 Q{}^{2}/3240 + 282169 Q{}^{4}/189 + 1581758 Q{}^{6}/63 + 
      138392Q{}^{8})$%
$\partial{}^{8} A_{2} T + 
    (3047/2835 + 80743 Q{}^{2}/1890 + 76511 Q{}^{4}/54 + 212642 Q{}^{6}/9 + 
      133040Q{}^{8})$%
$\partial{}^{8} A_{2} A_2 + 
    ((-73744Q)/2835 - 202088 Q{}^{3}/315 - 1951016 Q{}^{5}/315 - 22048Q{}^{7})
     $%
$A_3 \partial{}^{7} T + ((-8143Q)/81 - 29890 Q{}^{3}/9 - 
      1757992 Q{}^{5}/45 - 154336Q{}^{7})$%
$\partial A_{3} \partial{}^{6} T + 
    ((-7784Q)/45 - 365011 Q{}^{3}/45 - 1666724 Q{}^{5}/15 - 468720Q{}^{7})
     $%
$\partial{}^{2} A_{3} \partial{}^{5} T + ((-9262Q)/81 - 287468 Q{}^{3}/27 - 
      1607336 Q{}^{5}/9 - 2461648 Q{}^{7}/3)$%
$\partial{}^{3} A_{3} \partial{}^{4} T + 
    (10174 Q{}^{2}/9 + 26004Q{}^{4} + 137744Q{}^{6})$%
$\partial{}^{3} A_{3} \partial{}^{3} A_{3} + 
    ((-2561Q)/81 - 84464 Q{}^{3}/9 - 1597072 Q{}^{5}/9 - 844480Q{}^{7})
     $%
$\partial{}^{4} A_{3} \partial{}^{3} T + (-92/9 + 1461Q{}^{2} + 37456Q{}^{4} + 202704Q{}^{6})
     $%
$\partial{}^{4} A_{3} \partial{}^{2} A_{3} + ((-743Q)/135 - 221078 Q{}^{3}/45 - 
      314072 Q{}^{5}/3 - 529760Q{}^{7})$%
$\partial{}^{5} A_{3} \partial{}^{2} T + 
    (-194/9 + 1446 Q{}^{2}/5 + 67228 Q{}^{4}/5 + 77616Q{}^{6})
     $%
$\partial{}^{5} A_{3} \partial A_{3} + ((-551Q)/27 - 51857 Q{}^{3}/30 - 
      558094 Q{}^{5}/15 - 191576Q{}^{7})$%
$\partial{}^{6} A_{3} \partial T + 
    (-203/15 - 4153 Q{}^{2}/45 + 1688Q{}^{4} + 11984Q{}^{6})
     $%
$\partial{}^{6} A_{3} A_3 + (3239 Q/945 - 183418 Q{}^{3}/945 - 
      627194 Q{}^{5}/105 - 100984 Q{}^{7}/3)$%
$\partial{}^{7} A_{3} T + 
    (257 Q/42 + 64706 Q{}^{3}/315 + 13026 Q{}^{5}/35 - 5992Q{}^{7})
     $%
$B_2 \partial{}^{7} B_{3} + (181 Q/6 + 9298 Q{}^{3}/15 + 158Q{}^{5} - 
      23688Q{}^{7})$%
$\partial B_{2} \partial{}^{6} B_{3} + 
    (529 Q/15 + 12023 Q{}^{3}/15 + 5906 Q{}^{5}/5 - 27720Q{}^{7})
     $%
$\partial{}^{2} B_{2} \partial{}^{5} B_{3} + (1133 Q/18 + 19382 Q{}^{3}/9 + 18538Q{}^{5} + 
      45640Q{}^{7})$%
$\partial{}^{3} B_{2} \partial{}^{4} B_{3} + 
    (5177/324 + 11101 Q{}^{2}/54 - 60997 Q{}^{4}/18 - 236204 Q{}^{6}/3 - 
      393048Q{}^{8})$%
$\partial{}^{4} B_{2} \partial{}^{4} B_{2} + 
    (1739 Q/18 + 9548 Q{}^{3}/3 + 34650Q{}^{5} + 132072Q{}^{7})
     $%
$\partial{}^{4} B_{2} \partial{}^{3} B_{3} + (1196/45 + 49786 Q{}^{2}/135 - 
      208828 Q{}^{4}/45 - 1771316 Q{}^{6}/15 - 598608Q{}^{8})
     $%
$\partial{}^{5} B_{2} \partial{}^{3} B_{2} + (947 Q/30 + 24379 Q{}^{3}/15 + 
      126442 Q{}^{5}/5 + 120168Q{}^{7})$%
$\partial{}^{5} B_{2} \partial{}^{2} B_{3} + 
    (22619/1620 + 993 Q{}^{2}/5 - 24877 Q{}^{4}/15 - 744676 Q{}^{6}/15 - 
      259392Q{}^{8})$%
$\partial{}^{6} B_{2} \partial{}^{2} B_{2} + 
    ((-79Q)/9 + 2201 Q{}^{3}/45 + 36062 Q{}^{5}/5 + 48776Q{}^{7})
     $%
$\partial{}^{6} B_{2} \partial B_{3} + (3919/1134 + 37409 Q{}^{2}/945 - 
      65704 Q{}^{4}/105 - 199804 Q{}^{6}/15 - 63600Q{}^{8})
     $%
$\partial{}^{7} B_{2} \partial B_{2} + (313 Q/630 + 53 Q{}^{3}/3 + 10330 Q{}^{5}/7 + 
      10392Q{}^{7})$%
$\partial{}^{7} B_{2} B_3 + 
    (-1423/7560 - 15611 Q{}^{2}/3780 - 2039 Q{}^{4}/14 - 166504 Q{}^{6}/105 - 
      5352Q{}^{8})$%
$\partial{}^{8} B_{2} B_2 + 
    (-167/3 - 2344Q{}^{2} - 32772Q{}^{4} - 149904Q{}^{6})$%
$\partial{}^{3} B_{3} \partial{}^{3} B_{3} + 
    (-110/3 - 8692 Q{}^{2}/3 - 47364Q{}^{4} - 221424Q{}^{6})
     $%
$\partial{}^{4} B_{3} \partial{}^{2} B_{3} + (14 - 9688 Q{}^{2}/15 - 17004Q{}^{4} - 
      85008Q{}^{6})$%
$\partial{}^{5} B_{3} \partial B_{3} + 
    (-52/15 - 180Q{}^{2} - 15996 Q{}^{4}/5 - 15120Q{}^{6})
     $%
$\partial{}^{6} B_{3} B_3 + ((-920Q{}^{2})/3 - 15440Q{}^{4} - 141120Q{}^{6})
     $%
$\partial{}^{2} T \partial{}^{2} T \partial{}^{2} T + (-400/27 - 20128 Q{}^{2}/9 - 80704Q{}^{4} - 
      672000Q{}^{6})$%
$\partial{}^{3} T \partial{}^{2} T \partial T + 
    (-3800/243 - 76288 Q{}^{2}/81 - 663232 Q{}^{4}/27 - 1675520 Q{}^{6}/9)
     $%
$\partial{}^{3} T \partial{}^{3} T T + (-2180/243 - 75976 Q{}^{2}/81 - 
      742792 Q{}^{4}/27 - 1895072 Q{}^{6}/9)$%
$\partial{}^{4} T \partial T \partial T + 
    (-6640/243 - 125264 Q{}^{2}/81 - 1060520 Q{}^{4}/27 - 2662816 Q{}^{6}/9)
     $%
$\partial{}^{4} T \partial{}^{2} T T + (-4208/243 - 78472 Q{}^{2}/81 - 
      595864 Q{}^{4}/27 - 1397216 Q{}^{6}/9)$%
$\partial{}^{5} T \partial T T + 
    (-914/243 - 68672 Q{}^{2}/405 - 460688 Q{}^{4}/135 - 207296 Q{}^{6}/9)
     $%
$\partial{}^{6} T T{}^{2}+ ((2078\sqrt{2/3}Q)/81 + 
      (10240\sqrt{2/3}Q{}^{3})/3 + (520072\sqrt{2/3}Q{}^{5})/9 + 82080\sqrt{6}Q{}^{7})
     $%
$J \partial{}^{4} T \partial{}^{3} T + 
    ((514\sqrt{2/3}Q)/15 + (14332\sqrt{2/3}Q{}^{3})/5 + (72984\sqrt{6}Q{}^{5})/5 + 
      57312\sqrt{6}Q{}^{7})$%
$J \partial{}^{5} T \partial{}^{2} T + 
    ((16426\sqrt{2/3}Q)/405 + (92656\sqrt{2/3}Q{}^{3})/45 + 
      (1236536\sqrt{2/3}Q{}^{5})/45 + 103328\sqrt{2/3}Q{}^{7})
     $%
$J \partial{}^{6} T \partial T + ((50182\sqrt{2/3}Q)/2835 + 
      (76156\sqrt{2/3}Q{}^{3})/105 + (2845064\sqrt{2/3}Q{}^{5})/315 + 
      34528\sqrt{2/3}Q{}^{7})$%
$J \partial{}^{7} T T + 
    (31/3780 + 35062 Q{}^{2}/945 + 426998 Q{}^{4}/315 + 1932698 Q{}^{6}/105 + 
      89624Q{}^{8})$%
$J{}^{2}\partial{}^{8} T + 
    (-3503/136080 - 40583 Q{}^{2}/6480 + 315121 Q{}^{4}/1890 + 416287 Q{}^{6}/63 + 
      47788Q{}^{8})$%
$J{}^{2}\partial{}^{8} A_{2} + 
    (1501 Q/126 + 72319 Q{}^{3}/945 + 5573 Q{}^{5}/15 + 2308 Q{}^{7}/3)
     $%
$J{}^{2}\partial{}^{7} A_{3} + 
    ((15238\sqrt{2/3}Q)/2835 + (714904\sqrt{2/3}Q{}^{3})/945 + 
      (3685624\sqrt{2/3}Q{}^{5})/315 + 48992\sqrt{2/3}Q{}^{7})
     $%
$J A_2 \partial{}^{7} T + 
    ((5272\sqrt{2/3}Q{}^{2})/45 + (9206\sqrt{2/3}Q{}^{4})/5 + 8776\sqrt{2/3}Q{}^{6})
     $%
$J A_2 \partial{}^{6} A_{3} + 
    ((379\sqrt{2/3}Q)/135 + (393736\sqrt{2/3}Q{}^{3})/135 + 
      (2195396\sqrt{2/3}Q{}^{5})/45 + 205744\sqrt{2/3}Q{}^{7})
     $%
$J \partial A_{2} \partial{}^{6} T + 
    ((584\sqrt{2/3}Q{}^{2})/3 + (10232\sqrt{2/3}Q{}^{4})/3 + 13240\sqrt{2/3}Q{}^{6})
     $%
$J \partial A_{2} \partial{}^{5} A_{3} + 
    ((-610\sqrt{2/3}Q)/27 + (247234\sqrt{2/3}Q{}^{3})/45 + 
      (1553752\sqrt{2/3}Q{}^{5})/15 + 153216\sqrt{6}Q{}^{7})
     $%
$J \partial{}^{2} A_{2} \partial{}^{5} T + 
    ((988\sqrt{2/3}Q{}^{2})/3 + (18500\sqrt{2/3}Q{}^{4})/3 + 21808\sqrt{2/3}Q{}^{6})
     $%
$J \partial{}^{2} A_{2} \partial{}^{4} A_{3} + 
    ((-887\sqrt{2/3}Q)/27 + (164228\sqrt{2/3}Q{}^{3})/27 + 
      120616\sqrt{2/3}Q{}^{5} + (1658848\sqrt{2/3}Q{}^{7})/3)
     $%
$J \partial{}^{3} A_{2} \partial{}^{4} T + 
    (616\sqrt{2/3}Q{}^{2} + 13232\sqrt{2/3}Q{}^{4} + 18496\sqrt{6}Q{}^{6})
     $%
$J \partial{}^{3} A_{2} \partial{}^{3} A_{3} + 
    ((-3374\sqrt{2/3}Q)/81 + (30068\sqrt{2/3}Q{}^{3})/9 + 
      (720092\sqrt{2/3}Q{}^{5})/9 + 140384\sqrt{6}Q{}^{7})
     $%
$J \partial{}^{4} A_{2} \partial{}^{3} T + 
    ((-18289\sqrt{2/3}Q)/81 - (206587\sqrt{2/3}Q{}^{3})/27 - 
      (967616\sqrt{2/3}Q{}^{5})/9 - (1364096\sqrt{2/3}Q{}^{7})/3)
     $%
$J \partial{}^{4} A_{2} \partial{}^{3} A_{2} + 
    ((5252\sqrt{2/3}Q{}^{2})/9 + 14348\sqrt{2/3}Q{}^{4} + 76576\sqrt{2/3}Q{}^{6})
     $%
$J \partial{}^{4} A_{2} \partial{}^{2} A_{3} + 
    ((-2897\sqrt{2/3}Q)/135 + (2975\sqrt{2/3}Q{}^{3})/9 + 
      (83686\sqrt{2/3}Q{}^{5})/5 + 118256\sqrt{2/3}Q{}^{7})
     $%
$J \partial{}^{5} A_{2} \partial{}^{2} T + 
    ((-19697\sqrt{2/3}Q)/135 - (216976\sqrt{2/3}Q{}^{3})/45 - 
      (954724\sqrt{2/3}Q{}^{5})/15 - 255328\sqrt{2/3}Q{}^{7})
     $%
$J \partial{}^{5} A_{2} \partial{}^{2} A_{2} + 
    ((9916\sqrt{2/3}Q{}^{2})/45 + (66614\sqrt{2/3}Q{}^{4})/15 + 7680\sqrt{6}Q{}^{6})
     $%
$J \partial{}^{5} A_{2} \partial A_{3} + 
    ((-1534\sqrt{2/3}Q)/81 - (38461\sqrt{2/3}Q{}^{3})/45 - 
      (339838\sqrt{2/3}Q{}^{5})/45 - 11984\sqrt{2/3}Q{}^{7})
     $%
$J \partial{}^{6} A_{2} \partial T + 
    ((-6017\sqrt{2/3}Q)/81 - (309421\sqrt{2/3}Q{}^{3})/135 - 
      (1230664\sqrt{2/3}Q{}^{5})/45 - 103600\sqrt{2/3}Q{}^{7})
     $%
$J \partial{}^{6} A_{2} \partial A_{2} + 
    ((1762\sqrt{2/3}Q{}^{2})/15 + (4342\sqrt{2/3}Q{}^{4})/3 + 1904\sqrt{6}Q{}^{6})
     $%
$J \partial{}^{6} A_{2} A_3 + 
    ((-29416\sqrt{2/3}Q)/2835 - 905831 Q{}^{3}/(945\sqrt{6}) - 
      (558644\sqrt{2/3}Q{}^{5})/105 - (54512\sqrt{2/3}Q{}^{7})/3)
     $%
$J \partial{}^{7} A_{2} T + 
    ((-6437\sqrt{2/3}Q)/405 - (515239\sqrt{2/3}Q{}^{3})/945 - 
      (2089652\sqrt{2/3}Q{}^{5})/315 - (77264\sqrt{2/3}Q{}^{7})/3)
     $%
$J \partial{}^{7} A_{2} A_2 + 
    ((9212\sqrt{2/3})/405 + (14972\sqrt{2/3}Q{}^{2})/45 + 
      (122152\sqrt{2/3}Q{}^{4})/45 + 10064\sqrt{2/3}Q{}^{6})
     $%
$J A_3 \partial{}^{6} T + 
    ((2566\sqrt{2/3})/45 + (66067\sqrt{2/3}Q{}^{2})/45 + 
      (250472\sqrt{2/3}Q{}^{4})/15 + 22032\sqrt{6}Q{}^{6})
     $%
$J \partial A_{3} \partial{}^{5} T + 
    ((1408\sqrt{2/3})/27 + (30314\sqrt{2/3}Q{}^{2})/9 + 
      (152972\sqrt{2/3}Q{}^{4})/3 + 220240\sqrt{2/3}Q{}^{6})
     $%
$J \partial{}^{2} A_{3} \partial{}^{4} T + 
    ((2108\sqrt{2/3})/81 + (41168\sqrt{2/3}Q{}^{2})/9 + 
      (661900\sqrt{2/3}Q{}^{4})/9 + 303520\sqrt{2/3}Q{}^{6})
     $%
$J \partial{}^{3} A_{3} \partial{}^{3} T + 
    ((-2744\sqrt{2/3}Q)/3 - 7280\sqrt{6}Q{}^{3} - 39232\sqrt{6}Q{}^{5})
     $%
$J \partial{}^{3} A_{3} \partial{}^{2} A_{3} + 
    ((-104\sqrt{2/3})/9 + 3272\sqrt{2/3}Q{}^{2} + (178100\sqrt{2/3}Q{}^{4})/3 + 
      255952\sqrt{2/3}Q{}^{6})$%
$J \partial{}^{4} A_{3} \partial{}^{2} T + 
    ((-1240\sqrt{2/3}Q)/3 - 3432\sqrt{6}Q{}^{3} - 19136\sqrt{6}Q{}^{5})
     $%
$J \partial{}^{4} A_{3} \partial A_{3} + 
    ((56\sqrt{2/3})/45 + (8482\sqrt{2/3}Q{}^{2})/5 + (438748\sqrt{2/3}Q{}^{4})/15 + 
      124360\sqrt{2/3}Q{}^{6})$%
$J \partial{}^{5} A_{3} \partial T + 
    ((-988\sqrt{2/3}Q)/15 - (3304\sqrt{6}Q{}^{3})/5 - 3808\sqrt{6}Q{}^{5})
     $%
$J \partial{}^{5} A_{3} A_3 + 
    (4\sqrt{6} + (6082\sqrt{2/3}Q{}^{2})/15 + (34018\sqrt{2/3}Q{}^{4})/5 + 
      10056\sqrt{6}Q{}^{6})$%
$J \partial{}^{6} A_{3} T + 
    ((-1034\sqrt{2/3}Q{}^{2})/5 - (2516\sqrt{2/3}Q{}^{4})/5 + 1904\sqrt{6}Q{}^{6})
     $%
$J B_2 \partial{}^{6} B_{3} + 
    ((-4\sqrt{2/3}Q{}^{2})/5 + (26456\sqrt{2/3}Q{}^{4})/5 + 11488\sqrt{6}Q{}^{6})
     $%
$J \partial B_{2} \partial{}^{5} B_{3} + 
    (106\sqrt{6}Q{}^{2} + 9428\sqrt{2/3}Q{}^{4} + 17360\sqrt{6}Q{}^{6})
     $%
$J \partial{}^{2} B_{2} \partial{}^{4} B_{3} + 
    (-96\sqrt{6}Q{}^{2} + 2416\sqrt{2/3}Q{}^{4} + 7744\sqrt{6}Q{}^{6})
     $%
$J \partial{}^{3} B_{2} \partial{}^{3} B_{3} + 
    ((-878\sqrt{2/3}Q)/9 + (28492\sqrt{2/3}Q{}^{3})/9 + 70888\sqrt{2/3}Q{}^{5} + 
      355424\sqrt{2/3}Q{}^{7})$%
$J \partial{}^{4} B_{2} \partial{}^{3} B_{2} + 
    (-6\sqrt{6}Q{}^{2} + 692\sqrt{2/3}Q{}^{4} - 304\sqrt{6}Q{}^{6})
     $%
$J \partial{}^{4} B_{2} \partial{}^{2} B_{3} + 
    ((-893\sqrt{2/3}Q)/15 + (6416\sqrt{2/3}Q{}^{3})/5 + (57808\sqrt{6}Q{}^{5})/5 + 
      61632\sqrt{6}Q{}^{7})$%
$J \partial{}^{5} B_{2} \partial{}^{2} B_{2} + 
    ((1192\sqrt{6}Q{}^{2})/5 + (28376\sqrt{2/3}Q{}^{4})/5 + 1888\sqrt{6}Q{}^{6})
     $%
$J \partial{}^{5} B_{2} \partial B_{3} + 
    ((-256\sqrt{2/3}Q)/45 + (26092\sqrt{2/3}Q{}^{3})/45 + 
      (16736\sqrt{6}Q{}^{5})/5 + 50176\sqrt{2/3}Q{}^{7})$%
$J \partial{}^{6} B_{2} 
      \partial B_{2} + (106\sqrt{6}Q{}^{2} + (16012\sqrt{2/3}Q{}^{4})/5 + 
      1904\sqrt{6}Q{}^{6})$%
$J \partial{}^{6} B_{2} B_3 + 
    ((29\sqrt{2/3}Q)/5 + (99052\sqrt{2/3}Q{}^{3})/315 + 
      (103384\sqrt{2/3}Q{}^{5})/35 + 8800\sqrt{2/3}Q{}^{7})
     $%
$J \partial{}^{7} B_{2} B_2 + 
    (1372\sqrt{2/3}Q + 7512\sqrt{6}Q{}^{3} + 35232\sqrt{6}Q{}^{5})
     $%
$J \partial{}^{3} B_{3} \partial{}^{2} B_{3} + 
    (254\sqrt{2/3}Q + 2660\sqrt{6}Q{}^{3} + 15120\sqrt{6}Q{}^{5})
     $%
$J \partial{}^{4} B_{3} \partial B_{3} + 
    ((-206\sqrt{2/3}Q)/5 + 68\sqrt{6}Q{}^{3} + 1872\sqrt{6}Q{}^{5})
     $%
$J \partial{}^{5} B_{3} B_3 + 
    ((4780\sqrt{2/3}Q)/27 + (471776\sqrt{2/3}Q{}^{3})/27 + 
      (1184048\sqrt{2/3}Q{}^{5})/3 + (7577920\sqrt{2/3}Q{}^{7})/3)
     $%
$\partial J \partial{}^{3} T \partial{}^{3} T + ((24554\sqrt{2/3}Q)/81 + 
      (774016\sqrt{2/3}Q{}^{3})/27 + (5788660\sqrt{2/3}Q{}^{5})/9 + 
      (12341840\sqrt{2/3}Q{}^{7})/3)$%
$\partial J \partial{}^{4} T \partial{}^{2} T + 
    ((27448\sqrt{2/3}Q)/81 + (543596\sqrt{2/3}Q{}^{3})/27 + 
      (3394292\sqrt{2/3}Q{}^{5})/9 + (6636304\sqrt{2/3}Q{}^{7})/3)
     $%
$\partial J \partial{}^{5} T \partial T + ((56552\sqrt{2/3}Q)/405 + 
      (955976\sqrt{2/3}Q{}^{3})/135 + (5542792\sqrt{2/3}Q{}^{5})/45 + 
      (2098976\sqrt{2/3}Q{}^{7})/3)$%
$\partial J \partial{}^{6} T T + 
    (232/729 + 2620558 Q{}^{2}/8505 + 31689184 Q{}^{4}/2835 + 
      146696888 Q{}^{6}/945 + 779040Q{}^{8})$%
$\partial J J \partial{}^{7} T + 
    (407/1458 - 360041 Q{}^{2}/34020 + 13397273 Q{}^{4}/5670 + 
      8915108 Q{}^{6}/135 + 1317208 Q{}^{8}/3)$%
$\partial J J 
      \partial{}^{7} A_{2} + (4409 Q/135 + 3233 Q{}^{3}/90 - 20456 Q{}^{5}/3 - 
      39424Q{}^{7})$%
$\partial J J \partial{}^{6} A_{3} + 
    (13/45 + 131351 Q{}^{2}/405 + 895804 Q{}^{4}/135 + 767704 Q{}^{6}/45 - 
      589088 Q{}^{8}/3)$%
$(\partial J){}^{2} \partial{}^{6} T + 
    (7/45 - 31679 Q{}^{2}/162 - 99911 Q{}^{4}/15 - 5357456 Q{}^{6}/45 - 
      2334808 Q{}^{8}/3)$%
$(\partial J){}^{2} \partial{}^{6} A_{2} + 
    (4717 Q/135 - 15463 Q{}^{3}/9 - 118250 Q{}^{5}/3 - 168368Q{}^{7})
     $%
$(\partial J){}^{2} \partial{}^{5} A_{3} + 
    ((9079\sqrt{2/3}Q)/135 + (292054\sqrt{2/3}Q{}^{3})/45 + 
      (5761964\sqrt{2/3}Q{}^{5})/45 + (2235488\sqrt{2/3}Q{}^{7})/3)
     $%
$\partial J A_2 \partial{}^{6} T + 
    (13391 Q{}^{2}/(15\sqrt{6}) + (35332\sqrt{2/3}Q{}^{4})/5 + 8214\sqrt{6}Q{}^{6})
     $%
$\partial J A_2 \partial{}^{5} A_{3} + 
    ((73226\sqrt{2/3}Q)/405 + (3010591\sqrt{2/3}Q{}^{3})/135 + 
      (20149876\sqrt{2/3}Q{}^{5})/45 + (7739812\sqrt{2/3}Q{}^{7})/3)
     $%
$\partial J \partial A_{2} \partial{}^{5} T + 
    (6563 Q{}^{2}/(3\sqrt{6}) + 3669\sqrt{6}Q{}^{4} + 3462\sqrt{6}Q{}^{6})
     $%
$\partial J \partial A_{2} \partial{}^{4} A_{3} + 
    ((11831\sqrt{2/3}Q)/81 + (986206\sqrt{2/3}Q{}^{3})/27 + 
      (7446322\sqrt{2/3}Q{}^{5})/9 + (15079868\sqrt{2/3}Q{}^{7})/3)
     $%
$\partial J \partial{}^{2} A_{2} \partial{}^{4} T + 
    ((8260\sqrt{2/3}Q{}^{2})/9 + 6113\sqrt{2/3}Q{}^{4} - 41872\sqrt{2/3}Q{}^{6})
     $%
$\partial J \partial{}^{2} A_{2} \partial{}^{3} A_{3} + 
    ((23920\sqrt{2/3}Q)/81 + (1202488\sqrt{2/3}Q{}^{3})/27 + 
      (8994932\sqrt{2/3}Q{}^{5})/9 + (18499460\sqrt{2/3}Q{}^{7})/3)
     $%
$\partial J \partial{}^{3} A_{2} \partial{}^{3} T + 
    ((3724\sqrt{2/3}Q)/81 + (189722\sqrt{2/3}Q{}^{3})/27 + 
      (2121944\sqrt{2/3}Q{}^{5})/9 + (5656696\sqrt{2/3}Q{}^{7})/3)
     $%
$\partial J \partial{}^{3} A_{2} \partial{}^{3} A_{2} + 
    (317\sqrt{2/3}Q{}^{2} + 3289\sqrt{2/3}Q{}^{4} - 3548\sqrt{6}Q{}^{6})
     $%
$\partial J \partial{}^{3} A_{2} \partial{}^{2} A_{3} + 
    ((13955\sqrt{2/3}Q)/81 + (807775\sqrt{2/3}Q{}^{3})/27 + 
      (6607411\sqrt{2/3}Q{}^{5})/9 + (14370524\sqrt{2/3}Q{}^{7})/3)
     $%
$\partial J \partial{}^{4} A_{2} \partial{}^{2} T + 
    ((5342\sqrt{2/3}Q)/81 + (300388\sqrt{2/3}Q{}^{3})/27 + 
      6962711 Q{}^{5}/(9\sqrt{6}) + (9341030\sqrt{2/3}Q{}^{7})/3)
     $%
$\partial J \partial{}^{4} A_{2} \partial{}^{2} A_{2} + 
    (59 Q{}^{2}/(3\sqrt{6}) + 308\sqrt{2/3}Q{}^{4} + 2106\sqrt{6}Q{}^{6})
     $%
$\partial J \partial{}^{4} A_{2} \partial A_{3} + 
    ((10210\sqrt{2/3}Q)/81 + (1975783\sqrt{2/3}Q{}^{3})/135 + 
      (15417418\sqrt{2/3}Q{}^{5})/45 + (6657064\sqrt{2/3}Q{}^{7})/3)
     $%
$\partial J \partial{}^{5} A_{2} \partial T + 
    (104471 Q/(810\sqrt{6}) + 1974431 Q{}^{3}/(135\sqrt{6}) + 
      19631687 Q{}^{5}/(45\sqrt{6}) + (4925248\sqrt{2/3}Q{}^{7})/3)
     $%
$\partial J \partial{}^{5} A_{2} \partial A_{2} + 
    (25\sqrt{6}Q{}^{2} + 8087 Q{}^{4}/(15\sqrt{6}) - 422\sqrt{2/3}Q{}^{6})
     $%
$\partial J \partial{}^{5} A_{2} A_3 + 
    ((8686\sqrt{2/3}Q)/405 + (85340\sqrt{2/3}Q{}^{3})/27 + 
      (3739018\sqrt{2/3}Q{}^{5})/45 + 570488\sqrt{2/3}Q{}^{7})
     $%
$\partial J \partial{}^{6} A_{2} T + 
    (41761 Q/(810\sqrt{6}) + (313301\sqrt{2/3}Q{}^{3})/135 + 
      (329611\sqrt{2/3}Q{}^{5})/5 + 162354\sqrt{6}Q{}^{7})
     $%
$\partial J \partial{}^{6} A_{2} A_2 + 
    ((2566\sqrt{2/3})/45 - (7\sqrt{2/3}Q{}^{2})/9 - (39884\sqrt{2/3}Q{}^{4})/5 - 
      36044\sqrt{2/3}Q{}^{6})$%
$\partial J A_3 \partial{}^{5} T + 
    ((2816\sqrt{2/3})/27 - (1330\sqrt{2/3}Q{}^{2})/9 - (71702\sqrt{2/3}Q{}^{4})/3 - 
      121196\sqrt{2/3}Q{}^{6})$%
$\partial J \partial A_{3} \partial{}^{4} T + 
    ((2108\sqrt{2/3})/27 + (8834\sqrt{2/3}Q{}^{2})/9 - (62150\sqrt{2/3}Q{}^{4})/3 - 
      187124\sqrt{2/3}Q{}^{6})$%
$\partial J \partial{}^{2} A_{3} \partial{}^{3} T + 
    ((-610\sqrt{2/3}Q)/3 - 381\sqrt{6}Q{}^{3} - 176\sqrt{6}Q{}^{5})
     $%
$\partial J \partial{}^{2} A_{3} \partial{}^{2} A_{3} + 
    ((16534\sqrt{2/3}Q{}^{2})/9 - 4954\sqrt{2/3}Q{}^{4} - 175660\sqrt{2/3}Q{}^{6})
     $%
$\partial J \partial{}^{3} A_{3} \partial{}^{2} T + 
    ((-4366\sqrt{2/3}Q)/9 - 3766\sqrt{2/3}Q{}^{3} - 7220\sqrt{2/3}Q{}^{5})
     $%
$\partial J \partial{}^{3} A_{3} \partial A_{3} + 
    ((-208\sqrt{2/3})/9 + (17483\sqrt{2/3}Q{}^{2})/9 + 13720\sqrt{2/3}Q{}^{4} - 
      24992\sqrt{2/3}Q{}^{6})$%
$\partial J \partial{}^{4} A_{3} \partial T + 
    ((-5399Q)/(9\sqrt{6}) - 775\sqrt{6}Q{}^{3} - 4820\sqrt{2/3}Q{}^{5})
     $%
$\partial J \partial{}^{4} A_{3} A_3 + 
    ((56\sqrt{2/3})/45 + (43522\sqrt{2/3}Q{}^{2})/45 + (18252\sqrt{6}Q{}^{4})/5 + 
      26080\sqrt{2/3}Q{}^{6})$%
$\partial J \partial{}^{5} A_{3} T + 
    ((-2305\sqrt{2/3}Q{}^{2})/9 - (3431\sqrt{2/3}Q{}^{4})/5 + 8098\sqrt{2/3}Q{}^{6})
     $%
$\partial J B_2 \partial{}^{5} B_{3} + 
    (9967 Q{}^{2}/(9\sqrt{6}) + 5471\sqrt{6}Q{}^{4} + 84814\sqrt{2/3}Q{}^{6})
     $%
$\partial J \partial B_{2} \partial{}^{4} B_{3} + 
    ((6509\sqrt{2/3}Q{}^{2})/3 + 15443\sqrt{6}Q{}^{4} + 74596\sqrt{6}Q{}^{6})
     $%
$\partial J \partial{}^{2} B_{2} \partial{}^{3} B_{3} + 
    (1669 Q/(27\sqrt{6}) + (3119\sqrt{2/3}Q{}^{3})/3 + 
      (50116\sqrt{2/3}Q{}^{5})/3 + 83368\sqrt{2/3}Q{}^{7})
     $%
$\partial J \partial{}^{3} B_{2} \partial{}^{3} B_{2} + 
    ((32822\sqrt{2/3}Q{}^{2})/9 + 63886\sqrt{2/3}Q{}^{4} + 296344\sqrt{2/3}Q{}^{6})
     $%
$\partial J \partial{}^{3} B_{2} \partial{}^{2} B_{3} + 
    ((2486\sqrt{2/3}Q)/27 + (7145\sqrt{2/3}Q{}^{3})/3 + 
      186071 Q{}^{5}/(3\sqrt{6}) + 49702\sqrt{6}Q{}^{7})
     $%
$\partial J \partial{}^{4} B_{2} \partial{}^{2} B_{2} + 
    ((27848\sqrt{2/3}Q{}^{2})/9 + 16420\sqrt{6}Q{}^{4} + 203818\sqrt{2/3}Q{}^{6})
     $%
$\partial J \partial{}^{4} B_{2} \partial B_{3} + 
    (51097 Q/(270\sqrt{6}) + (36344\sqrt{2/3}Q{}^{3})/15 + 
      (369539\sqrt{2/3}Q{}^{5})/15 + 31824\sqrt{6}Q{}^{7})
     $%
$\partial J \partial{}^{5} B_{2} \partial B_{2} + 
    ((15847\sqrt{2/3}Q{}^{2})/15 + (78779\sqrt{2/3}Q{}^{4})/5 + 19638\sqrt{6}Q{}^{6})
     $%
$\partial J \partial{}^{5} B_{2} B_3 + 
    (15143 Q/(270\sqrt{6}) + (49346\sqrt{2/3}Q{}^{3})/45 + 
      103721 Q{}^{5}/(5\sqrt{6}) + 31262\sqrt{2/3}Q{}^{7})
     $%
$\partial J \partial{}^{6} B_{2} B_2 + 
    (-508\sqrt{2/3}Q - 2942\sqrt{6}Q{}^{3} - 15924\sqrt{6}Q{}^{5})
     $%
$\partial J \partial{}^{2} B_{3} \partial{}^{2} B_{3} + 
    ((-1591\sqrt{2/3}Q)/3 - 4556\sqrt{6}Q{}^{3} - 23276\sqrt{6}Q{}^{5})
     $%
$\partial J \partial{}^{3} B_{3} \partial B_{3} + 
    ((-1607Q)/(3\sqrt{6}) - 1950\sqrt{6}Q{}^{3} - 8048\sqrt{6}Q{}^{5})
     $%
$\partial J \partial{}^{4} B_{3} B_3 + 
    ((17632\sqrt{2/3}Q)/27 + (186892\sqrt{2/3}Q{}^{3})/3 + 
      (4318192\sqrt{2/3}Q{}^{5})/3 + 9430400\sqrt{2/3}Q{}^{7})
     $%
$\partial{}^{2} J \partial{}^{3} T \partial{}^{2} T + ((63068\sqrt{2/3}Q)/81 + 
      (1345736\sqrt{2/3}Q{}^{3})/27 + (8925292\sqrt{2/3}Q{}^{5})/9 + 
      (18205744\sqrt{2/3}Q{}^{7})/3)$%
$\partial{}^{2} J \partial{}^{4} T \partial T + 
    ((160178\sqrt{2/3}Q)/405 + (936896\sqrt{2/3}Q{}^{3})/45 + 
      (17085664\sqrt{2/3}Q{}^{5})/45 + 2241536\sqrt{2/3}Q{}^{7})
     $%
$\partial{}^{2} J \partial{}^{5} T T + (1538/1215 + 1059422 Q{}^{2}/1215 + 
      4247884 Q{}^{4}/135 + 59303752 Q{}^{6}/135 + 6662560 Q{}^{8}/3)
     $%
$\partial{}^{2} J J \partial{}^{6} T + 
    (2593/2430 - 20591 Q{}^{2}/405 + 74054 Q{}^{4}/15 + 729024 Q{}^{6}/5 + 
      2928016 Q{}^{8}/3)$%
$\partial{}^{2} J J \partial{}^{6} A_{2} + 
    (15958 Q/135 + 14366 Q{}^{3}/15 - 58196 Q{}^{5}/15 - 9592Q{}^{7})
     $%
$\partial{}^{2} J J \partial{}^{5} A_{3} + 
    (236/135 + 519688 Q{}^{2}/405 + 38740 Q{}^{4}/3 - 14897392 Q{}^{6}/45 - 
      3651072Q{}^{8})$%
$\partial{}^{2} J \partial J \partial{}^{5} T + 
    (236/135 - 575474 Q{}^{2}/405 - 8337412 Q{}^{4}/135 - 1188970Q{}^{6} - 
      7681416Q{}^{8})$%
$\partial{}^{2} J \partial J \partial{}^{5} A_{2} + 
    (6811 Q/27 - 130033 Q{}^{3}/27 - 346048 Q{}^{5}/3 - 1040912 Q{}^{7}/3)
     $%
$\partial{}^{2} J \partial J \partial{}^{4} A_{3} + 
    (163/81 + 81638 Q{}^{2}/81 - 737512 Q{}^{4}/81 - 2549800 Q{}^{6}/3 - 
      64693664 Q{}^{8}/9)$%
$\partial{}^{2} J \partial{}^{2} J \partial{}^{4} T + 
    (163/81 - 315785 Q{}^{2}/162 - 7856950 Q{}^{4}/81 - 17520487 Q{}^{6}/9 - 
      113385464 Q{}^{8}/9)$%
$\partial{}^{2} J \partial{}^{2} J \partial{}^{4} A_{2} + 
    (13366 Q/27 + 1358 Q{}^{3}/3 - 34598Q{}^{5} + 142736Q{}^{7})
     $%
$\partial{}^{2} J \partial{}^{2} J \partial{}^{3} A_{3} + 
    ((18634\sqrt{2/3}Q)/81 + (2657317\sqrt{2/3}Q{}^{3})/135 + 
      (5912956\sqrt{2/3}Q{}^{5})/15 + 788368\sqrt{6}Q{}^{7})
     $%
$\partial{}^{2} J A_2 \partial{}^{5} T + 
    ((32206\sqrt{2/3}Q{}^{2})/27 + 16730\sqrt{2/3}Q{}^{4} + 
      (164138\sqrt{2/3}Q{}^{6})/3)$%
$\partial{}^{2} J A_2 \partial{}^{4} A_{3} + 
    ((39935\sqrt{2/3}Q)/81 + (1474142\sqrt{2/3}Q{}^{3})/27 + 
      (10169746\sqrt{2/3}Q{}^{5})/9 + (20376124\sqrt{2/3}Q{}^{7})/3)
     $%
$\partial{}^{2} J \partial A_{2} \partial{}^{4} T + 
    ((6538\sqrt{2/3}Q{}^{2})/3 + 6226\sqrt{6}Q{}^{4} - 6288\sqrt{6}Q{}^{6})
     $%
$\partial{}^{2} J \partial A_{2} \partial{}^{3} A_{3} + 
    ((12470\sqrt{2/3}Q)/27 + (685340\sqrt{2/3}Q{}^{3})/9 + 577312\sqrt{6}Q{}^{5} + 
      10892728\sqrt{2/3}Q{}^{7})$%
$\partial{}^{2} J \partial{}^{2} A_{2} \partial{}^{3} T + 
    ((7243\sqrt{2/3}Q{}^{2})/9 + (9785\sqrt{2/3}Q{}^{4})/3 - 13088\sqrt{6}Q{}^{6})
     $%
$\partial{}^{2} J \partial{}^{2} A_{2} \partial{}^{2} A_{3} + 
    ((13874\sqrt{2/3}Q)/27 + (1893304\sqrt{2/3}Q{}^{3})/27 + 
      (14719694\sqrt{2/3}Q{}^{5})/9 + 10515340\sqrt{2/3}Q{}^{7})
     $%
$\partial{}^{2} J \partial{}^{3} A_{2} \partial{}^{2} T + 
    ((5581\sqrt{2/3}Q)/27 + 1657241 Q{}^{3}/(27\sqrt{6}) + 
      (8640884\sqrt{2/3}Q{}^{5})/9 + 7394384\sqrt{2/3}Q{}^{7})
     $%
$\partial{}^{2} J \partial{}^{3} A_{2} \partial{}^{2} A_{2} + 
    ((-656\sqrt{2/3}Q{}^{2})/3 - 10330\sqrt{2/3}Q{}^{4} - 56972\sqrt{2/3}Q{}^{6})
     $%
$\partial{}^{2} J \partial{}^{3} A_{2} \partial A_{3} + 
    ((29348\sqrt{2/3}Q)/81 + (381980\sqrt{2/3}Q{}^{3})/9 + 
      976360\sqrt{2/3}Q{}^{5} + (18774424\sqrt{2/3}Q{}^{7})/3)
     $%
$\partial{}^{2} J \partial{}^{4} A_{2} \partial T + 
    ((18893\sqrt{2/3}Q)/81 + (214471\sqrt{2/3}Q{}^{3})/9 + 
      438417\sqrt{3/2}Q{}^{5} + (14399050\sqrt{2/3}Q{}^{7})/3)
     $%
$\partial{}^{2} J \partial{}^{4} A_{2} \partial A_{2} + 
    (8267 Q{}^{2}/(27\sqrt{6}) - (3110\sqrt{2/3}Q{}^{4})/3 - 
      (15886\sqrt{2/3}Q{}^{6})/3)$%
$\partial{}^{2} J \partial{}^{4} A_{2} A_3 + 
    ((8260\sqrt{2/3}Q)/81 + (61519\sqrt{2/3}Q{}^{3})/5 + 
      (13329244\sqrt{2/3}Q{}^{5})/45 + 651712\sqrt{6}Q{}^{7})
     $%
$\partial{}^{2} J \partial{}^{5} A_{2} T + 
    (88243 Q/(405\sqrt{6}) + 5057203 Q{}^{3}/(270\sqrt{6}) + 
      (722297\sqrt{2/3}Q{}^{5})/3 + 1706138\sqrt{2/3}Q{}^{7})
     $%
$\partial{}^{2} J \partial{}^{5} A_{2} A_2 + 
    ((1165\sqrt{2/3})/9 + (6676\sqrt{2/3}Q{}^{2})/27 - (48770\sqrt{2/3}Q{}^{4})/3 - 
      (188524\sqrt{2/3}Q{}^{6})/3)$%
$\partial{}^{2} J A_3 \partial{}^{4} T + 
    ((7024\sqrt{2/3})/27 + (4444\sqrt{2/3}Q{}^{2})/9 - (142012\sqrt{2/3}Q{}^{4})/
       3 - 203504\sqrt{2/3}Q{}^{6})$%
$\partial{}^{2} J \partial A_{3} \partial{}^{3} T + 
    ((1054\sqrt{2/3})/9 + (12904\sqrt{2/3}Q{}^{2})/9 - (112240\sqrt{2/3}Q{}^{4})/
       3 - 67748\sqrt{6}Q{}^{6})$%
$\partial{}^{2} J \partial{}^{2} A_{3} \partial{}^{2} T + 
    ((-278\sqrt{2/3}Q)/3 + 13567\sqrt{2/3}Q{}^{3} + 19912\sqrt{6}Q{}^{5})
     $%
$\partial{}^{2} J \partial{}^{2} A_{3} \partial A_{3} + 
    ((2108\sqrt{2/3})/27 + (31736\sqrt{2/3}Q{}^{2})/9 - 2168\sqrt{2/3}Q{}^{4} - 
      103864\sqrt{2/3}Q{}^{6})$%
$\partial{}^{2} J \partial{}^{3} A_{3} \partial T + 
    ((-1118\sqrt{2/3}Q)/3 + 1027\sqrt{2/3}Q{}^{3} + 4132\sqrt{6}Q{}^{5})
     $%
$\partial{}^{2} J \partial{}^{3} A_{3} A_3 + 
    ((272\sqrt{2/3})/27 + (52843\sqrt{2/3}Q{}^{2})/27 + 
      (49484\sqrt{2/3}Q{}^{4})/3 + (146216\sqrt{2/3}Q{}^{6})/3)
     $%
$\partial{}^{2} J \partial{}^{4} A_{3} T + 
    ((-1997Q{}^{2})/(3\sqrt{6}) - 499\sqrt{6}Q{}^{4} + 4274\sqrt{6}Q{}^{6})
     $%
$\partial{}^{2} J B_2 \partial{}^{4} B_{3} + 
    ((6242\sqrt{2/3}Q{}^{2})/3 + 34496\sqrt{2/3}Q{}^{4} + 56948\sqrt{6}Q{}^{6})
     $%
$\partial{}^{2} J \partial B_{2} \partial{}^{3} B_{3} + 
    ((15967\sqrt{2/3}Q{}^{2})/3 + 32089\sqrt{6}Q{}^{4} + 146744\sqrt{6}Q{}^{6})
     $%
$\partial{}^{2} J \partial{}^{2} B_{2} \partial{}^{2} B_{3} + 
    ((8270\sqrt{2/3}Q)/27 + (17084\sqrt{2/3}Q{}^{3})/3 + 18232\sqrt{6}Q{}^{5} + 
      337336\sqrt{2/3}Q{}^{7})$%
$\partial{}^{2} J \partial{}^{3} B_{2} \partial{}^{2} B_{2} + 
    ((54692\sqrt{2/3}Q{}^{2})/9 + (306833\sqrt{2/3}Q{}^{4})/3 + 
      429632\sqrt{2/3}Q{}^{6})$%
$\partial{}^{2} J \partial{}^{3} B_{2} \partial B_{3} + 
    ((7678\sqrt{2/3}Q)/27 + 117299 Q{}^{3}/(9\sqrt{6}) + 
      400109 Q{}^{5}/(3\sqrt{6}) + 100026\sqrt{6}Q{}^{7})
     $%
$\partial{}^{2} J \partial{}^{4} B_{2} \partial B_{2} + 
    (42235 Q{}^{2}/(9\sqrt{6}) + (107051\sqrt{2/3}Q{}^{4})/3 + 
      139262\sqrt{2/3}Q{}^{6})$%
$\partial{}^{2} J \partial{}^{4} B_{2} B_3 + 
    (27083 Q/(135\sqrt{6}) + 561089 Q{}^{3}/(90\sqrt{6}) + 
      182717 Q{}^{5}/(3\sqrt{6}) + 106538\sqrt{2/3}Q{}^{7})
     $%
$\partial{}^{2} J \partial{}^{5} B_{2} B_2 + 
    ((-5824\sqrt{2/3}Q)/3 - 15046\sqrt{6}Q{}^{3} - 65324\sqrt{6}Q{}^{5})
     $%
$\partial{}^{2} J \partial{}^{2} B_{3} \partial B_{3} + 
    ((-2600\sqrt{2/3}Q)/3 - 5446\sqrt{6}Q{}^{3} - 22648\sqrt{6}Q{}^{5})
     $%
$\partial{}^{2} J \partial{}^{3} B_{3} B_3 + 
    ((10840\sqrt{2/3}Q)/27 + (337616\sqrt{2/3}Q{}^{3})/9 + 
      (2661584\sqrt{2/3}Q{}^{5})/3 + 5938240\sqrt{2/3}Q{}^{7})
     $%
$\partial{}^{3} J \partial{}^{2} T \partial{}^{2} T + ((92000\sqrt{2/3}Q)/81 + 
      (663424\sqrt{2/3}Q{}^{3})/9 + (13566928\sqrt{2/3}Q{}^{5})/9 + 
      9437120\sqrt{2/3}Q{}^{7})$%
$\partial{}^{3} J \partial{}^{3} T \partial T + 
    ((54868\sqrt{2/3}Q)/81 + (108868\sqrt{2/3}Q{}^{3})/3 + 
      (6165848\sqrt{2/3}Q{}^{5})/9 + 4165984\sqrt{2/3}Q{}^{7})
     $%
$\partial{}^{3} J \partial{}^{4} T T + (11144/3645 + 1987486 Q{}^{2}/1215 + 
      4627376 Q{}^{4}/81 + 21162952 Q{}^{6}/27 + 3942560Q{}^{8})
     $%
$\partial{}^{3} J J \partial{}^{5} T + 
    (11144/3645 - 146162 Q{}^{2}/1215 + 618254 Q{}^{4}/81 + 31891208 Q{}^{6}/135 + 
      1563216Q{}^{8})$%
$\partial{}^{3} J J \partial{}^{5} A_{2} + 
    (7216 Q/27 + 82300 Q{}^{3}/27 + 94988 Q{}^{5}/9 + 97456Q{}^{7})
     $%
$\partial{}^{3} J J \partial{}^{4} A_{3} + 
    (1954/729 + 396278 Q{}^{2}/243 - 770356 Q{}^{4}/81 - 32683376 Q{}^{6}/27 - 
      31182080 Q{}^{8}/3)$%
$\partial{}^{3} J \partial J \partial{}^{4} T + 
    (1954/729 - 679981 Q{}^{2}/243 - 10863310 Q{}^{4}/81 - 72018005 Q{}^{6}/27 - 
      51915620 Q{}^{8}/3)$%
$\partial{}^{3} J \partial J \partial{}^{4} A_{2} + 
    (6916 Q/9 + 10138 Q{}^{3}/9 - 66214Q{}^{5} + 50084Q{}^{7})
     $%
$\partial{}^{3} J \partial J \partial{}^{3} A_{3} + 
    (932/81 + 142624 Q{}^{2}/81 - 2477828 Q{}^{4}/27 - 33779872 Q{}^{6}/9 - 
      87062080 Q{}^{8}/3)$%
$\partial{}^{3} J \partial{}^{2} J \partial{}^{3} T + 
    (932/81 - 526726 Q{}^{2}/81 - 8850412 Q{}^{4}/27 - 58567162 Q{}^{6}/9 - 
      125439380 Q{}^{8}/3)$%
$\partial{}^{3} J \partial{}^{2} J \partial{}^{3} A_{2} + 
    (15482 Q/9 + 413102 Q{}^{3}/27 + 244336 Q{}^{5}/9 + 441324Q{}^{7})
     $%
$\partial{}^{3} J \partial{}^{2} J \partial{}^{2} A_{3} + 
    (6 + 27220 Q{}^{2}/81 - 795796 Q{}^{4}/9 - 26059240 Q{}^{6}/9 - 21556640Q{}^{8})
     $%
$\partial{}^{3} J \partial{}^{3} J \partial{}^{2} T + 
    ((-40475\sqrt{2/3}Q)/243 - (664610\sqrt{2/3}Q{}^{3})/81 - 
      (4294756\sqrt{2/3}Q{}^{5})/27 - (11131504\sqrt{2/3}Q{}^{7})/9 - 
      887040\sqrt{6}Q{}^{9})$%
$\partial{}^{3} J \partial{}^{3} J \partial{}^{3} J + 
    (6 - 313210 Q{}^{2}/81 - 626336 Q{}^{4}/3 - 38254196 Q{}^{6}/9 - 27521760Q{}^{8})
     $%
$\partial{}^{3} J \partial{}^{3} J \partial{}^{2} A_{2} + 
    (7990 Q/9 + 320768 Q{}^{3}/27 + 600280 Q{}^{5}/9 + 321440Q{}^{7})
     $%
$\partial{}^{3} J \partial{}^{3} J \partial A_{3} + 
    ((34840\sqrt{2/3}Q)/81 + (310538\sqrt{2/3}Q{}^{3})/9 + 
      (2111072\sqrt{2/3}Q{}^{5})/3 + (12985424\sqrt{2/3}Q{}^{7})/3)
     $%
$\partial{}^{3} J A_2 \partial{}^{4} T + 
    ((46279\sqrt{2/3}Q{}^{2})/27 + (66982\sqrt{2/3}Q{}^{4})/3 + 
      (186524\sqrt{2/3}Q{}^{6})/3)$%
$\partial{}^{3} J A_2 \partial{}^{3} A_{3} + 
    ((26644\sqrt{2/3}Q)/27 + (2231948\sqrt{2/3}Q{}^{3})/27 + 
      (15067276\sqrt{2/3}Q{}^{5})/9 + 3403568\sqrt{6}Q{}^{7})
     $%
$\partial{}^{3} J \partial A_{2} \partial{}^{3} T + 
    (1723\sqrt{2/3}Q{}^{2} + (57700\sqrt{2/3}Q{}^{4})/3 + 41852\sqrt{2/3}Q{}^{6})
     $%
$\partial{}^{3} J \partial A_{2} \partial{}^{2} A_{3} + 
    ((16076\sqrt{2/3}Q)/27 + (766982\sqrt{2/3}Q{}^{3})/9 + 
      (5963824\sqrt{2/3}Q{}^{5})/3 + 12862528\sqrt{2/3}Q{}^{7})
     $%
$\partial{}^{3} J \partial{}^{2} A_{2} \partial{}^{2} T + 
    ((1165\sqrt{2/3}Q)/9 + (63794\sqrt{2/3}Q{}^{3})/3 + 650606\sqrt{2/3}Q{}^{5} + 
      1634584\sqrt{6}Q{}^{7})$%
$\partial{}^{3} J \partial{}^{2} A_{2} \partial{}^{2} A_{2} + 
    ((-3413\sqrt{2/3}Q{}^{2})/9 - 14948\sqrt{2/3}Q{}^{4} - 81424\sqrt{2/3}Q{}^{6})
     $%
$\partial{}^{3} J \partial{}^{2} A_{2} \partial A_{3} + 
    ((63868\sqrt{2/3}Q)/81 + (1938548\sqrt{2/3}Q{}^{3})/27 + 
      1534834\sqrt{2/3}Q{}^{5} + 9574340\sqrt{2/3}Q{}^{7})
     $%
$\partial{}^{3} J \partial{}^{3} A_{2} \partial T + 
    ((35924\sqrt{2/3}Q)/81 + (126934\sqrt{2/3}Q{}^{3})/3 + 
      (9831994\sqrt{2/3}Q{}^{5})/9 + 2568796\sqrt{6}Q{}^{7})
     $%
$\partial{}^{3} J \partial{}^{3} A_{2} \partial A_{2} + 
    ((-376\sqrt{2/3}Q{}^{2})/3 - 7495\sqrt{2/3}Q{}^{4} - 34256\sqrt{2/3}Q{}^{6})
     $%
$\partial{}^{3} J \partial{}^{3} A_{2} A_3 + 
    ((16624\sqrt{2/3}Q)/81 + (681185\sqrt{2/3}Q{}^{3})/27 + 
      (1773271\sqrt{2/3}Q{}^{5})/3 + 3836884\sqrt{2/3}Q{}^{7})
     $%
$\partial{}^{3} J \partial{}^{4} A_{2} T + 
    ((6314\sqrt{2/3}Q)/27 + 19706\sqrt{2/3}Q{}^{3} + (4408678\sqrt{2/3}Q{}^{5})/
       9 + (10220324\sqrt{2/3}Q{}^{7})/3)$%
$\partial{}^{3} J \partial{}^{4} A_{2} 
      A_2 + ((2024\sqrt{2/3})/9 + (1004\sqrt{2/3}Q{}^{2})/3 - 
      (86024\sqrt{2/3}Q{}^{4})/3 - 79408\sqrt{2/3}Q{}^{6})
     $%
$\partial{}^{3} J A_3 \partial{}^{3} T + 
    ((8032\sqrt{2/3})/27 + (5158\sqrt{2/3}Q{}^{2})/9 - (180724\sqrt{2/3}Q{}^{4})/
       3 - 180544\sqrt{2/3}Q{}^{6})$%
$\partial{}^{3} J \partial A_{3} \partial{}^{2} T + 
    ((4814\sqrt{2/3}Q)/9 + (50212\sqrt{2/3}Q{}^{3})/3 + 21448\sqrt{6}Q{}^{5})
     $%
$\partial{}^{3} J \partial A_{3} \partial A_{3} + 
    ((8032\sqrt{2/3})/27 + 2414\sqrt{2/3}Q{}^{2} - 11592\sqrt{6}Q{}^{4} - 
      77236\sqrt{2/3}Q{}^{6})$%
$\partial{}^{3} J \partial{}^{2} A_{3} \partial T + 
    ((728\sqrt{2/3}Q)/9 + (37288\sqrt{2/3}Q{}^{3})/3 + 18320\sqrt{6}Q{}^{5})
     $%
$\partial{}^{3} J \partial{}^{2} A_{3} A_3 + 
    ((8032\sqrt{2/3})/81 + (78940\sqrt{2/3}Q{}^{2})/27 + 
      (87434\sqrt{2/3}Q{}^{4})/9 + (80972\sqrt{2/3}Q{}^{6})/3)
     $%
$\partial{}^{3} J \partial{}^{3} A_{3} T + 
    ((-25\sqrt{2/3}Q{}^{2})/9 - (5216\sqrt{2/3}Q{}^{4})/3 + 1904\sqrt{6}Q{}^{6})
     $%
$\partial{}^{3} J B_2 \partial{}^{3} B_{3} + 
    ((10285\sqrt{2/3}Q{}^{2})/3 + 16880\sqrt{6}Q{}^{4} + 70160\sqrt{6}Q{}^{6})
     $%
$\partial{}^{3} J \partial B_{2} \partial{}^{2} B_{3} + 
    ((7345\sqrt{2/3}Q)/27 + (31820\sqrt{2/3}Q{}^{3})/9 + 22946\sqrt{2/3}Q{}^{5} + 
      66696\sqrt{6}Q{}^{7})$%
$\partial{}^{3} J \partial{}^{2} B_{2} \partial{}^{2} B_{2} + 
    (2209\sqrt{6}Q{}^{2} + 106660\sqrt{2/3}Q{}^{4} + 136528\sqrt{6}Q{}^{6})
     $%
$\partial{}^{3} J \partial{}^{2} B_{2} \partial B_{3} + 
    ((14896\sqrt{2/3}Q)/27 + (249550\sqrt{2/3}Q{}^{3})/27 + 
      (680576\sqrt{2/3}Q{}^{5})/9 + 135104\sqrt{6}Q{}^{7})
     $%
$\partial{}^{3} J \partial{}^{3} B_{2} \partial B_{2} + 
    ((30085\sqrt{2/3}Q{}^{2})/9 + 52864\sqrt{2/3}Q{}^{4} + 194816\sqrt{2/3}Q{}^{6})
     $%
$\partial{}^{3} J \partial{}^{3} B_{2} B_3 + 
    ((5746\sqrt{2/3}Q)/27 + (142273\sqrt{2/3}Q{}^{3})/27 + 
      (437612\sqrt{2/3}Q{}^{5})/9 + 185848\sqrt{2/3}Q{}^{7})
     $%
$\partial{}^{3} J \partial{}^{4} B_{2} B_2 + 
    ((-14158\sqrt{2/3}Q)/9 - (95144\sqrt{2/3}Q{}^{3})/3 - 39368\sqrt{6}Q{}^{5})
     $%
$\partial{}^{3} J \partial B_{3} \partial B_{3} + 
    ((-16030\sqrt{2/3}Q)/9 - (97292\sqrt{2/3}Q{}^{3})/3 - 39568\sqrt{6}Q{}^{5})
     $%
$\partial{}^{3} J \partial{}^{2} B_{3} B_3 + 
    ((28120\sqrt{2/3}Q)/27 + (616310\sqrt{2/3}Q{}^{3})/9 + 
      (4306028\sqrt{2/3}Q{}^{5})/3 + 9162832\sqrt{2/3}Q{}^{7})
     $%
$\partial{}^{4} J \partial{}^{2} T \partial T + 
    ((2434\sqrt{2/3}Q)/3 + (387488\sqrt{2/3}Q{}^{3})/9 + 
      (2453996\sqrt{2/3}Q{}^{5})/3 + 1672720\sqrt{6}Q{}^{7})
     $%
$\partial{}^{4} J \partial{}^{3} T T + (1024/243 + 166376 Q{}^{2}/81 + 
      1873540 Q{}^{4}/27 + 8307088 Q{}^{6}/9 + 13637632 Q{}^{8}/3)
     $%
$\partial{}^{4} J J \partial{}^{4} T + 
    (1024/243 - 26041 Q{}^{2}/162 + 238634 Q{}^{4}/27 + 2591200 Q{}^{6}/9 + 
      5769232 Q{}^{8}/3)$%
$\partial{}^{4} J J \partial{}^{4} A_{2} + 
    (37925 Q/81 + 144562 Q{}^{3}/27 + 130528 Q{}^{5}/9 + 294224 Q{}^{7}/3)
     $%
$\partial{}^{4} J J \partial{}^{3} A_{3} + 
    (1426/243 + 93446 Q{}^{2}/81 - 1386448 Q{}^{4}/27 - 19595612 Q{}^{6}/9 - 
      16861264Q{}^{8})$%
$\partial{}^{4} J \partial J \partial{}^{3} T + 
    (1426/243 - 305789 Q{}^{2}/81 - 4946408 Q{}^{4}/27 - 10715888 Q{}^{6}/3 - 
      22850444Q{}^{8})$%
$\partial{}^{4} J \partial J \partial{}^{3} A_{2} + 
    (28036 Q/27 + 78272 Q{}^{3}/9 - 51176 Q{}^{5}/3 + 15500Q{}^{7})
     $%
$\partial{}^{4} J \partial J \partial{}^{2} A_{3} + 
    (629/81 + 31780 Q{}^{2}/81 - 3997102 Q{}^{4}/27 - 42720844 Q{}^{6}/9 - 
      105378896 Q{}^{8}/3)$%
$\partial{}^{4} J \partial{}^{2} J \partial{}^{2} T + 
    (629/81 - 56897 Q{}^{2}/9 - 9105343 Q{}^{4}/27 - 20422658 Q{}^{6}/3 - 
      131631752 Q{}^{8}/3)$%
$\partial{}^{4} J \partial{}^{2} J \partial{}^{2} A_{2} + 
    (38674 Q/27 + 522854 Q{}^{3}/27 + 861160 Q{}^{5}/9 + 391944Q{}^{7})
     $%
$\partial{}^{4} J \partial{}^{2} J \partial A_{3} + 
    (13700/729 - 589244 Q{}^{2}/243 - 20233922 Q{}^{4}/81 - 159496012 Q{}^{6}/27 - 
      39495568Q{}^{8})$%
$\partial{}^{4} J \partial{}^{3} J \partial T + 
    ((-390235Q)/(243\sqrt{6}) - (3134116\sqrt{2/3}Q{}^{3})/81 - 
      (18258529\sqrt{2/3}Q{}^{5})/27 - (33377294\sqrt{2/3}Q{}^{7})/9 + 
      888216\sqrt{6}Q{}^{9})$%
$\partial{}^{4} J \partial{}^{3} J \partial{}^{2} J + 
    (13700/729 - 1699808 Q{}^{2}/243 - 30172979 Q{}^{4}/81 - 193712458 Q{}^{6}/27 - 
      44621808Q{}^{8})$%
$\partial{}^{4} J \partial{}^{3} J \partial A_{2} + 
    (15238 Q/27 + 73348 Q{}^{3}/9 + 135436 Q{}^{5}/3 + 167328Q{}^{7})
     $%
$\partial{}^{4} J \partial{}^{3} J A_3 + 
    (409/162 - 34828 Q{}^{2}/27 - 2375338 Q{}^{4}/27 - 5544752 Q{}^{6}/3 - 
      35375648 Q{}^{8}/3)$%
$\partial{}^{4} J \partial{}^{4} J T + 
    ((-235385Q)/(486\sqrt{6}) - (941755\sqrt{2/3}Q{}^{3})/81 - 
      (4793551\sqrt{2/3}Q{}^{5})/27 - (2492504\sqrt{2/3}Q{}^{7})/9 + 
      7173040\sqrt{2/3}Q{}^{9})$%
$\partial{}^{4} J \partial{}^{4} J \partial J + 
    (409/162 - 149323 Q{}^{2}/81 - 307499 Q{}^{4}/3 - 1990854Q{}^{6} - 
      36996512 Q{}^{8}/3)$%
$\partial{}^{4} J \partial{}^{4} J A_2 + 
    ((17198\sqrt{2/3}Q)/27 + (1128434\sqrt{2/3}Q{}^{3})/27 + 
      (7485262\sqrt{2/3}Q{}^{5})/9 + 1713488\sqrt{6}Q{}^{7})
     $%
$\partial{}^{4} J A_2 \partial{}^{3} T + 
    (6679 Q{}^{2}/(3\sqrt{6}) + (54089\sqrt{2/3}Q{}^{4})/3 + 71476\sqrt{2/3}Q{}^{6})
     $%
$\partial{}^{4} J A_2 \partial{}^{2} A_{3} + 
    ((26314\sqrt{2/3}Q)/27 + (677921\sqrt{2/3}Q{}^{3})/9 + 
      (4651976\sqrt{2/3}Q{}^{5})/3 + 9641296\sqrt{2/3}Q{}^{7})
     $%
$\partial{}^{4} J \partial A_{2} \partial{}^{2} T + 
    (127\sqrt{2/3}Q{}^{2} - 1753\sqrt{2/3}Q{}^{4} - 5096\sqrt{6}Q{}^{6})
     $%
$\partial{}^{4} J \partial A_{2} \partial A_{3} + 
    ((2336\sqrt{2/3}Q)/3 + (641791\sqrt{2/3}Q{}^{3})/9 + 
      (4545860\sqrt{2/3}Q{}^{5})/3 + 3149496\sqrt{6}Q{}^{7})
     $%
$\partial{}^{4} J \partial{}^{2} A_{2} \partial T + 
    ((13199\sqrt{2/3}Q)/27 + (417221\sqrt{2/3}Q{}^{3})/9 + 
      (3474629\sqrt{2/3}Q{}^{5})/3 + 2669632\sqrt{6}Q{}^{7})
     $%
$\partial{}^{4} J \partial{}^{2} A_{2} \partial A_{2} + 
    ((-6911Q{}^{2})/(9\sqrt{6}) - 3124\sqrt{6}Q{}^{4} - 39704\sqrt{2/3}Q{}^{6})
     $%
$\partial{}^{4} J \partial{}^{2} A_{2} A_3 + 
    ((29533\sqrt{2/3}Q)/81 + (922838\sqrt{2/3}Q{}^{3})/27 + 
      (2246512\sqrt{2/3}Q{}^{5})/3 + 4749028\sqrt{2/3}Q{}^{7})
     $%
$\partial{}^{4} J \partial{}^{3} A_{2} T + 
    ((9542\sqrt{2/3}Q)/27 + 488981 Q{}^{3}/(9\sqrt{6}) + 
      (1906271\sqrt{2/3}Q{}^{5})/3 + 1432276\sqrt{6}Q{}^{7})
     $%
$\partial{}^{4} J \partial{}^{3} A_{2} A_2 + 
    ((5290\sqrt{2/3})/27 - (1060\sqrt{2/3}Q{}^{2})/9 - (113266\sqrt{2/3}Q{}^{4})/
       3 - 120512\sqrt{2/3}Q{}^{6})$%
$\partial{}^{4} J A_3 \partial{}^{2} T + 
    ((10580\sqrt{2/3})/27 + (1924\sqrt{2/3}Q{}^{2})/9 - 21986\sqrt{6}Q{}^{4} - 
      65352\sqrt{6}Q{}^{6})$%
$\partial{}^{4} J \partial A_{3} \partial T + 
    ((9752\sqrt{2/3}Q)/9 + (62713\sqrt{2/3}Q{}^{3})/3 + 24192\sqrt{6}Q{}^{5})
     $%
$\partial{}^{4} J \partial A_{3} A_3 + 
    ((5290\sqrt{2/3})/27 + (12346\sqrt{2/3}Q{}^{2})/9 - 14960\sqrt{2/3}Q{}^{4} - 
      28484\sqrt{2/3}Q{}^{6})$%
$\partial{}^{4} J \partial{}^{2} A_{3} T + 
    ((-1177Q{}^{2})/(3\sqrt{6}) - 1269\sqrt{6}Q{}^{4} - 1688\sqrt{6}Q{}^{6})
     $%
$\partial{}^{4} J B_2 \partial{}^{2} B_{3} + 
    ((30377\sqrt{2/3}Q{}^{2})/9 + (152107\sqrt{2/3}Q{}^{4})/3 + 60144\sqrt{6}Q{}^{6})
     $%
$\partial{}^{4} J \partial B_{2} \partial B_{3} + 
    ((14921\sqrt{2/3}Q)/27 + (58942\sqrt{2/3}Q{}^{3})/9 + 8161\sqrt{6}Q{}^{5} + 
      61880\sqrt{6}Q{}^{7})$%
$\partial{}^{4} J \partial{}^{2} B_{2} \partial B_{2} + 
    (62269 Q{}^{2}/(9\sqrt{6}) + (158728\sqrt{2/3}Q{}^{4})/3 + 59976\sqrt{6}Q{}^{6})
     $%
$\partial{}^{4} J \partial{}^{2} B_{2} B_3 + 
    ((25727\sqrt{2/3}Q)/81 + (18796\sqrt{2/3}Q{}^{3})/3 + 
      (419578\sqrt{2/3}Q{}^{5})/9 + 55528\sqrt{6}Q{}^{7})
     $%
$\partial{}^{4} J \partial{}^{3} B_{2} B_2 + 
    ((-20558\sqrt{2/3}Q)/9 - (113383\sqrt{2/3}Q{}^{3})/3 - 42000\sqrt{6}Q{}^{5})
     $%
$\partial{}^{4} J \partial B_{3} B_3 + 
    ((163094\sqrt{2/3}Q)/405 + (3194596\sqrt{2/3}Q{}^{3})/135 + 
      (4150940\sqrt{2/3}Q{}^{5})/9 + (8468656\sqrt{2/3}Q{}^{7})/3)
     $%
$\partial{}^{5} J \partial T \partial T + ((249568\sqrt{2/3}Q)/405 + 
      (4476106\sqrt{2/3}Q{}^{3})/135 + (28631216\sqrt{2/3}Q{}^{5})/45 + 
      (11775904\sqrt{2/3}Q{}^{7})/3)$%
$\partial{}^{5} J \partial{}^{2} T T + 
    (532/81 + 2206186 Q{}^{2}/1215 + 7832776 Q{}^{4}/135 + 98711096 Q{}^{6}/135 + 
      10262560 Q{}^{8}/3)$%
$\partial{}^{5} J J \partial{}^{3} T + 
    (532/81 - 6061 Q{}^{2}/81 + 1115668 Q{}^{4}/135 + 3926792 Q{}^{6}/15 + 
      5335760 Q{}^{8}/3)$%
$\partial{}^{5} J J \partial{}^{3} A_{2} + 
    (3991 Q/9 + 264118 Q{}^{3}/45 + 65332 Q{}^{5}/5 - 29024Q{}^{7})
     $%
$\partial{}^{5} J J \partial{}^{2} A_{3} + 
    (302/81 + 21826 Q{}^{2}/81 - 3045058 Q{}^{4}/45 - 99064856 Q{}^{6}/45 - 
      16284800Q{}^{8})$%
$\partial{}^{5} J \partial J \partial{}^{2} T + 
    (302/81 - 1202902 Q{}^{2}/405 - 6966751 Q{}^{4}/45 - 137426042 Q{}^{6}/45 - 
      19413492Q{}^{8})$%
$\partial{}^{5} J \partial J \partial{}^{2} A_{2} + 
    (18770 Q/27 + 1266976 Q{}^{3}/135 + 30036Q{}^{5} + 47740 Q{}^{7}/3)
     $%
$\partial{}^{5} J \partial J \partial A_{3} + 
    (15812/1215 - 70372 Q{}^{2}/45 - 1479740 Q{}^{4}/9 - 58205108 Q{}^{6}/15 - 
      77581168 Q{}^{8}/3)$%
$\partial{}^{5} J \partial{}^{2} J \partial T + 
    ((-21478\sqrt{2/3}Q)/81 - (344387\sqrt{2/3}Q{}^{3})/27 - 
      (5702104\sqrt{2/3}Q{}^{5})/27 - 905440\sqrt{2/3}Q{}^{7} + 
      (9010624\sqrt{2/3}Q{}^{9})/3)$%
$\partial{}^{5} J \partial{}^{2} J \partial{}^{2} J + 
    (15812/1215 - 67751 Q{}^{2}/15 - 32779588 Q{}^{4}/135 - 209463232 Q{}^{6}/45 - 
      28766612Q{}^{8})$%
$\partial{}^{5} J \partial{}^{2} J \partial A_{2} + 
    (49463 Q/135 + 715882 Q{}^{3}/135 + 376636 Q{}^{5}/15 + 173740 Q{}^{7}/3)
     $%
$\partial{}^{5} J \partial{}^{2} J A_3 + 
    (1702/405 - 512480 Q{}^{2}/243 - 58263826 Q{}^{4}/405 - 406937372 Q{}^{6}/135 - 
      172688656 Q{}^{8}/9)$%
$\partial{}^{5} J \partial{}^{3} J T + 
    ((-96925\sqrt{2/3}Q)/243 - (7881037\sqrt{2/3}Q{}^{3})/405 - 
      (42066637\sqrt{2/3}Q{}^{5})/135 - (40720586\sqrt{2/3}Q{}^{7})/45 + 
      8657432\sqrt{2/3}Q{}^{9})$%
$\partial{}^{5} J \partial{}^{3} J \partial J + 
    (1702/405 - 1218017 Q{}^{2}/405 - 13557725 Q{}^{4}/81 - 145915454 Q{}^{6}/45 - 
      180266464 Q{}^{8}/9)$%
$\partial{}^{5} J \partial{}^{3} J A_2 + 
    ((-40453Q)/(81\sqrt{6}) - (8078566\sqrt{2/3}Q{}^{3})/405 - 
      (88728224\sqrt{2/3}Q{}^{5})/135 - (88814752\sqrt{2/3}Q{}^{7})/9 - 
      (159200384\sqrt{2/3}Q{}^{9})/3)$%
$\partial{}^{5} J \partial{}^{4} J J + 
    ((198139\sqrt{2/3}Q)/405 + (4229969\sqrt{2/3}Q{}^{3})/135 + 
      (9454142\sqrt{2/3}Q{}^{5})/15 + 3941168\sqrt{2/3}Q{}^{7})
     $%
$\partial{}^{5} J A_2 \partial{}^{2} T + 
    (111811 Q{}^{2}/(135\sqrt{6}) + (94949\sqrt{2/3}Q{}^{4})/15 + 
      (72296\sqrt{2/3}Q{}^{6})/3)$%
$\partial{}^{5} J A_2 \partial A_{3} + 
    ((74840\sqrt{2/3}Q)/81 + (788578\sqrt{2/3}Q{}^{3})/15 + 
      979982\sqrt{2/3}Q{}^{5} + (17551492\sqrt{2/3}Q{}^{7})/3)
     $%
$\partial{}^{5} J \partial A_{2} \partial T + 
    (209419 Q/(405\sqrt{6}) + (2509733\sqrt{2/3}Q{}^{3})/135 + 
      36575263 Q{}^{5}/(45\sqrt{6}) + 877940\sqrt{6}Q{}^{7})
     $%
$\partial{}^{5} J \partial A_{2} \partial A_{2} + 
    ((-3229Q{}^{2})/(5\sqrt{6}) - (11654\sqrt{6}Q{}^{4})/5 - 11536\sqrt{6}Q{}^{6})
     $%
$\partial{}^{5} J \partial A_{2} A_3 + 
    ((120499\sqrt{2/3}Q)/405 + (3770971\sqrt{2/3}Q{}^{3})/135 + 
      (27609578\sqrt{2/3}Q{}^{5})/45 + (11739028\sqrt{2/3}Q{}^{7})/3)
     $%
$\partial{}^{5} J \partial{}^{2} A_{2} T + 
    (230027 Q/(405\sqrt{6}) + 6234433 Q{}^{3}/(135\sqrt{6}) + 
      (8052602\sqrt{2/3}Q{}^{5})/15 + 3595816\sqrt{2/3}Q{}^{7})
     $%
$\partial{}^{5} J \partial{}^{2} A_{2} A_2 + 
    ((5176\sqrt{2/3})/27 + (1540\sqrt{2/3}Q{}^{2})/3 - (96754\sqrt{2/3}Q{}^{4})/3 - 
      61124\sqrt{6}Q{}^{6})$%
$\partial{}^{5} J A_3 \partial T + 
    (42373 Q/(45\sqrt{6}) + 86893 Q{}^{3}/(5\sqrt{6}) + 38108\sqrt{2/3}Q{}^{5})
     $%
$\partial{}^{5} J A_3 A_3 + 
    ((5176\sqrt{2/3})/27 + (125696\sqrt{2/3}Q{}^{2})/135 - 
      (395108\sqrt{2/3}Q{}^{4})/15 - (484036\sqrt{2/3}Q{}^{6})/3)
     $%
$\partial{}^{5} J \partial A_{3} T + 
    ((-11963Q{}^{2})/(45\sqrt{6}) - (13169\sqrt{2/3}Q{}^{4})/3 - 
      22988\sqrt{2/3}Q{}^{6})$%
$\partial{}^{5} J B_2 \partial B_{3} + 
    (38923 Q/(135\sqrt{6}) + (2179\sqrt{6}Q{}^{3})/5 - 
      192811 Q{}^{5}/(15\sqrt{6}) - 52976\sqrt{2/3}Q{}^{7})
     $%
$\partial{}^{5} J \partial B_{2} \partial B_{2} + 
    (168391 Q{}^{2}/(45\sqrt{6}) + (472054\sqrt{2/3}Q{}^{4})/15 + 
      44100\sqrt{6}Q{}^{6})$%
$\partial{}^{5} J \partial B_{2} B_3 + 
    (64313 Q/(135\sqrt{6}) + 133663 Q{}^{3}/(15\sqrt{6}) + 
      (82750\sqrt{2/3}Q{}^{5})/3 + 64204\sqrt{2/3}Q{}^{7})
     $%
$\partial{}^{5} J \partial{}^{2} B_{2} B_2 + 
    ((-67763Q)/(45\sqrt{6}) - 402289 Q{}^{3}/(15\sqrt{6}) - 19292\sqrt{6}Q{}^{5})
     $%
$\partial{}^{5} J B_3 B_3 + 
    ((137672\sqrt{2/3}Q)/405 + (2448604\sqrt{2/3}Q{}^{3})/135 + 
      (15118564\sqrt{2/3}Q{}^{5})/45 + (6021520\sqrt{2/3}Q{}^{7})/3)
     $%
$\partial{}^{6} J \partial T T + (14183/3645 + 1201636 Q{}^{2}/1215 + 
      13200124 Q{}^{4}/405 + 56344232 Q{}^{6}/135 + 5829920 Q{}^{8}/3)
     $%
$\partial{}^{6} J J \partial{}^{2} T + 
    (14183/3645 - 3325 Q{}^{2}/243 + 1723768 Q{}^{4}/405 + 18155108 Q{}^{6}/135 + 
      2820224 Q{}^{8}/3)$%
$\partial{}^{6} J J \partial{}^{2} A_{2} + 
    (31486 Q/135 + 196478 Q{}^{3}/45 + 461624 Q{}^{5}/15 + 86896Q{}^{7})
     $%
$\partial{}^{6} J J \partial A_{3} + 
    (22342/3645 - 549208 Q{}^{2}/1215 - 23249908 Q{}^{4}/405 - 
      187551236 Q{}^{6}/135 - 9312784Q{}^{8})$%
$\partial{}^{6} J \partial J \partial T + 
    (22342/3645 - 1996213 Q{}^{2}/1215 - 36211228 Q{}^{4}/405 - 
      229698602 Q{}^{6}/135 - 31292516 Q{}^{8}/3)$%
$\partial{}^{6} J \partial J 
      \partial A_{2} + (20891 Q/135 + 350374 Q{}^{3}/135 + 241168 Q{}^{5}/15 + 
      128356 Q{}^{7}/3)$%
$\partial{}^{6} J \partial J A_3 + 
    (11732/3645 - 1264424 Q{}^{2}/1215 - 10035616 Q{}^{4}/135 - 
      42430952 Q{}^{6}/27 - 90167200 Q{}^{8}/9)$%
$\partial{}^{6} J \partial{}^{2} J 
      T + ((-18475\sqrt{2/3}Q)/81 - (323330\sqrt{2/3}Q{}^{3})/27 - 
      (30487228\sqrt{2/3}Q{}^{5})/135 - (22160804\sqrt{2/3}Q{}^{7})/15 - 
      (3517360\sqrt{2/3}Q{}^{9})/3)$%
$\partial{}^{6} J \partial{}^{2} J \partial J + 
    (11732/3645 - 1861268 Q{}^{2}/1215 - 2360426 Q{}^{4}/27 - 
      229473658 Q{}^{6}/135 - 94387024 Q{}^{8}/9)$%
$\partial{}^{6} J \partial{}^{2} J 
      A_2 + ((-9377Q)/(27\sqrt{6}) - (5683226\sqrt{2/3}Q{}^{3})/405 - 
      (1399094\sqrt{2/3}Q{}^{5})/3 - (316575244\sqrt{2/3}Q{}^{7})/45 - 
      12637072\sqrt{6}Q{}^{9})$%
$\partial{}^{6} J \partial{}^{3} J J + 
    ((16928\sqrt{2/3}Q)/45 + (2555938\sqrt{2/3}Q{}^{3})/135 + 
      (3073474\sqrt{2/3}Q{}^{5})/9 + 2022784\sqrt{2/3}Q{}^{7})
     $%
$\partial{}^{6} J A_2 \partial T + 
    (18229 Q{}^{2}/(135\sqrt{6}) + (4799\sqrt{2/3}Q{}^{4})/5 + 
      (10328\sqrt{2/3}Q{}^{6})/3)$%
$\partial{}^{6} J A_2 A_3 + 
    ((40189\sqrt{2/3}Q)/135 + (458434\sqrt{2/3}Q{}^{3})/27 + 
      (14560858\sqrt{2/3}Q{}^{5})/45 + 1962692\sqrt{2/3}Q{}^{7})
     $%
$\partial{}^{6} J \partial A_{2} T + 
    ((97783\sqrt{2/3}Q)/405 + 789185 Q{}^{3}/(27\sqrt{6}) + 
      (13366949\sqrt{2/3}Q{}^{5})/45 + (5606504\sqrt{2/3}Q{}^{7})/3)
     $%
$\partial{}^{6} J \partial A_{2} A_2 + 
    ((2119\sqrt{2/3})/27 + (37072\sqrt{2/3}Q{}^{2})/27 + 
      (124456\sqrt{2/3}Q{}^{4})/15 + (56180\sqrt{2/3}Q{}^{6})/3)
     $%
$\partial{}^{6} J A_3 T + 
    ((221\sqrt{3/2}Q{}^{2})/5 + (3739\sqrt{2/3}Q{}^{4})/5 + 788\sqrt{6}Q{}^{6})
     $%
$\partial{}^{6} J B_2 B_3 + 
    ((39889\sqrt{2/3}Q)/405 + 181763 Q{}^{3}/(45\sqrt{6}) + 
      (685697\sqrt{2/3}Q{}^{5})/45 + 44276\sqrt{2/3}Q{}^{7})
     $%
$\partial{}^{6} J \partial B_{2} B_2 + 
    ((141539\sqrt{2/3}Q)/2835 + (504376\sqrt{2/3}Q{}^{3})/189 + 
      (15517996\sqrt{2/3}Q{}^{5})/315 + (879440\sqrt{2/3}Q{}^{7})/3)
     $%
$\partial{}^{7} J T{}^{2}+ (2914/729 + 3401326 Q{}^{2}/8505 + 
      32670508 Q{}^{4}/2835 + 134254136 Q{}^{6}/945 + 653536Q{}^{8})
     $%
$\partial{}^{7} J J \partial T + 
    (2914/729 + 1051117 Q{}^{2}/8505 + 11895262 Q{}^{4}/2835 + 
      9768824 Q{}^{6}/135 + 414064Q{}^{8})$%
$\partial{}^{7} J J \partial A_{2} + 
    (2330 Q/63 + 66086 Q{}^{3}/105 + 28416 Q{}^{5}/7 + 10704Q{}^{7})
     $%
$\partial{}^{7} J J A_3 + 
    (7166/5103 - 2524588 Q{}^{2}/8505 - 9057844 Q{}^{4}/405 - 
      448594304 Q{}^{6}/945 - 3017920Q{}^{8})$%
$\partial{}^{7} J \partial J T + 
    ((-143257Q)/(1701\sqrt{6}) - 14218111 Q{}^{3}/(2835\sqrt{6}) - 
      (11169892\sqrt{2/3}Q{}^{5})/189 - (195524722\sqrt{2/3}Q{}^{7})/315 - 
      2398120\sqrt{2/3}Q{}^{9})$%
$\partial{}^{7} J (\partial J){}^{2} + 
    (7166/5103 - 3877297 Q{}^{2}/8505 - 15008702 Q{}^{4}/567 - 
      485025554 Q{}^{6}/945 - 9456512 Q{}^{8}/3)$%
$\partial{}^{7} J \partial J 
      A_2 + ((-15655\sqrt{2/3}Q)/189 - (6475391\sqrt{2/3}Q{}^{3})/945 - 
      (24346402\sqrt{2/3}Q{}^{5})/105 - (370977508\sqrt{2/3}Q{}^{7})/105 - 
      6377072\sqrt{6}Q{}^{9})$%
$\partial{}^{7} J \partial{}^{2} J J + 
    ((43004\sqrt{2/3}Q)/405 + (5088662\sqrt{2/3}Q{}^{3})/945 + 
      (30895586\sqrt{2/3}Q{}^{5})/315 + 583232\sqrt{2/3}Q{}^{7})
     $%
$\partial{}^{7} J A_2 T + 
    ((39938\sqrt{2/3}Q)/945 + (11944\sqrt{2/3}Q{}^{3})/5 + 
      (14545687\sqrt{2/3}Q{}^{5})/315 + (846628\sqrt{2/3}Q{}^{7})/3)
     $%
$\partial{}^{7} J A_2{}^{2}+ 
    ((4345\sqrt{2/3}Q)/567 + (15374\sqrt{2/3}Q{}^{3})/105 + 
      (46549\sqrt{2/3}Q{}^{5})/45 + 2836\sqrt{2/3}Q{}^{7})
     $%
$\partial{}^{7} J B_2{}^{2}+ 
    (10511/10206 + 135572 Q{}^{2}/1701 + 1257908 Q{}^{4}/567 + 750100 Q{}^{6}/27 + 
      390832 Q{}^{8}/3)$%
$\partial{}^{8} J J T + 
    (10511/10206 + 114473 Q{}^{2}/2430 + 3934423 Q{}^{4}/2835 + 
      19064462 Q{}^{6}/945 + 314576 Q{}^{8}/3)$%
$\partial{}^{8} J J 
      A_2 + ((-58673Q)/(1134\sqrt{6}) - (702421\sqrt{2/3}Q{}^{3})/315 - 
      48972317 Q{}^{5}/(315\sqrt{6}) - (6017396\sqrt{2/3}Q{}^{7})/5 - 
      6564920\sqrt{2/3}Q{}^{9})$%
$\partial{}^{8} J \partial J J + 
    ((-91933Q)/(20412\sqrt{6}) - 1370735 Q{}^{3}/(3402\sqrt{6}) - 
      16329163 Q{}^{5}/(1134\sqrt{6}) - (21322894\sqrt{2/3}Q{}^{7})/189 - 
      (1855468\sqrt{2/3}Q{}^{9})/3)$%
$\partial{}^{9} J J{}^{2}+ 
    (-760/81 - 70744 Q{}^{2}/81 - 217808 Q{}^{4}/9 - 1668800 Q{}^{6}/9)
     $%
$A_2 \partial{}^{3} T \partial{}^{3} T + (-3628/243 - 113972 Q{}^{2}/81 - 
      1027952 Q{}^{4}/27 - 2627968 Q{}^{6}/9)$%
$A_2 \partial{}^{4} T \partial{}^{2} T + 
    (-33904/1215 - 450512 Q{}^{2}/405 - 2996024 Q{}^{4}/135 - 1384640 Q{}^{6}/9)
     $%
$A_2 \partial{}^{5} T \partial T + (-7384/405 - 179104 Q{}^{2}/405 - 
      20200 Q{}^{4}/3 - 394816 Q{}^{6}/9)$%
$A_2 \partial{}^{6} T T + 
    (248/135 - 4414 Q{}^{2}/405 - 233296 Q{}^{4}/135 - 153608 Q{}^{6}/9)
     $%
$A_2{}^{2}\partial{}^{6} T + ((-23Q)/45 + 52 Q{}^{3}/3 + 676Q{}^{5})
     $%
$A_2{}^{2}\partial{}^{5} A_{3} + 
    (4232 Q/45 + 73078 Q{}^{3}/45 + 20656 Q{}^{5}/3)$%
$A_2 A_3 
      \partial{}^{5} T + (8335 Q/27 + 20872 Q{}^{3}/3 + 103280 Q{}^{5}/3)
     $%
$A_2 \partial A_{3} \partial{}^{4} T + 
    (16676 Q/27 + 13962Q{}^{3} + 210952 Q{}^{5}/3)$%
$A_2 \partial{}^{2} A_{3} 
      \partial{}^{3} T + (-706/9 - 2750Q{}^{2} - 16592Q{}^{4})$%
$A_2 \partial{}^{2} A_{3} 
      \partial{}^{2} A_{3} + (5122 Q/9 + 126712 Q{}^{3}/9 + 228488 Q{}^{5}/3)
     $%
$A_2 \partial{}^{3} A_{3} \partial{}^{2} T + 
    (-2680/27 - 10396 Q{}^{2}/3 - 64328 Q{}^{4}/3)$%
$A_2 \partial{}^{3} A_{3} 
      \partial A_{3} + (2948 Q/9 + 23096 Q{}^{3}/3 + 40900Q{}^{5})
     $%
$A_2 \partial{}^{4} A_{3} \partial T + (-37/27 - 722Q{}^{2} - 14984 Q{}^{4}/3)
     $%
$A_2 \partial{}^{4} A_{3} A_3 + 
    (4189 Q/135 + 23192 Q{}^{3}/15 + 28132 Q{}^{5}/3)
     $%
$A_2 \partial{}^{5} A_{3} T + 
    ((-3362Q)/135 + 394 Q{}^{3}/3 + 7492 Q{}^{5}/3)$%
$A_2 B_2 
      \partial{}^{5} B_{3} + ((-3256Q)/27 - 3472 Q{}^{3}/3 + 292 Q{}^{5}/3)
     $%
$A_2 \partial B_{2} \partial{}^{4} B_{3} + 
    ((-1564Q)/9 - 3622Q{}^{3} - 15368Q{}^{5})$%
$A_2 \partial{}^{2} B_{2} 
      \partial{}^{3} B_{3} + (-742/81 + 14372 Q{}^{2}/27 + 119536 Q{}^{4}/9 + 
      213968 Q{}^{6}/3)$%
$A_2 \partial{}^{3} B_{2} \partial{}^{3} B_{2} + 
    ((-8632Q)/27 - 20068 Q{}^{3}/3 - 108176 Q{}^{5}/3)
     $%
$A_2 \partial{}^{3} B_{2} \partial{}^{2} B_{3} + 
    (-1678/81 + 1814 Q{}^{2}/3 + 166147 Q{}^{4}/9 + 101788Q{}^{6})
     $%
$A_2 \partial{}^{4} B_{2} \partial{}^{2} B_{2} + 
    ((-4264Q)/27 - 15440 Q{}^{3}/3 - 91676 Q{}^{5}/3)$%
$A_2 \partial{}^{4} B_{2} 
      \partial B_{3} + (-14983/810 + 4766 Q{}^{2}/45 + 292526 Q{}^{4}/45 + 
      36768Q{}^{6})$%
$A_2 \partial{}^{5} B_{2} \partial B_{2} + 
    ((-3776Q)/45 - 32812 Q{}^{3}/15 - 11340Q{}^{5})$%
$A_2 \partial{}^{5} B_{2} 
      B_3 + (-1421/270 - 6596 Q{}^{2}/135 + 1217 Q{}^{4}/3 + 3460Q{}^{6})
     $%
$A_2 \partial{}^{6} B_{2} B_2 + (220 + 4628Q{}^{2} + 25368Q{}^{4})
     $%
$A_2 \partial{}^{2} B_{3} \partial{}^{2} B_{3} + (1946/9 + 6392Q{}^{2} + 34696Q{}^{4})
     $%
$A_2 \partial{}^{3} B_{3} \partial B_{3} + (965/9 + 2376Q{}^{2} + 10624Q{}^{4})
     $%
$A_2 \partial{}^{4} B_{3} B_3 + 
    (-2872/81 - 113048 Q{}^{2}/27 - 965936 Q{}^{4}/9 - 2322208 Q{}^{6}/3)
     $%
$\partial A_{2} \partial{}^{3} T \partial{}^{2} T + (-18052/243 - 306992 Q{}^{2}/81 - 
      1965620 Q{}^{4}/27 - 4290568 Q{}^{6}/9)$%
$\partial A_{2} \partial{}^{4} T \partial T + 
    (-68368/1215 - 708728 Q{}^{2}/405 - 3474368 Q{}^{4}/135 - 1407512 Q{}^{6}/9)
     $%
$\partial A_{2} \partial{}^{5} T T + (5054/1215 - 68867 Q{}^{2}/405 - 
      1512326 Q{}^{4}/135 - 996848 Q{}^{6}/9)$%
$\partial A_{2} A_2 
      \partial{}^{5} T + ((-1253Q)/27 + 68Q{}^{3} + 12008 Q{}^{5}/3)
     $%
$\partial A_{2} A_2 \partial{}^{4} A_{3} + 
    (409/243 - 20614 Q{}^{2}/81 - 383272 Q{}^{4}/27 - 1323848 Q{}^{6}/9)
     $%
$\partial A_{2} \partial A_{2} \partial{}^{4} T + 
    ((-1004Q)/27 + 5152 Q{}^{3}/9 + 25216 Q{}^{5}/3)$%
$\partial A_{2} \partial A_{2} 
      \partial{}^{3} A_{3} + (4613 Q/27 + 7136 Q{}^{3}/3 + 21136 Q{}^{5}/3)
     $%
$\partial A_{2} A_3 \partial{}^{4} T + 
    (3400 Q/9 + 19124 Q{}^{3}/3 + 18816Q{}^{5})$%
$\partial A_{2} \partial A_{3} 
      \partial{}^{3} T + (486Q + 8948Q{}^{3} + 27864Q{}^{5})$%
$\partial A_{2} \partial{}^{2} A_{3} 
      \partial{}^{2} T + (-1694/9 - 20366 Q{}^{2}/3 - 38528Q{}^{4})
     $%
$\partial A_{2} \partial{}^{2} A_{3} \partial A_{3} + 
    (8932 Q/27 + 66340 Q{}^{3}/9 + 85744 Q{}^{5}/3)$%
$\partial A_{2} \partial{}^{3} A_{3} 
      \partial T + (-130/3 - 6062 Q{}^{2}/3 - 12328Q{}^{4})
     $%
$\partial A_{2} \partial{}^{3} A_{3} A_3 + 
    (265 Q/27 + 14482 Q{}^{3}/9 + 26452 Q{}^{5}/3)$%
$\partial A_{2} \partial{}^{4} A_{3} 
      T + (1045 Q/9 + 2338Q{}^{3} + 9956Q{}^{5})$%
$\partial A_{2} B_2 
      \partial{}^{4} B_{3} + ((-712Q)/9 + 152 Q{}^{3}/3 + 2216Q{}^{5})
     $%
$\partial A_{2} \partial B_{2} \partial{}^{3} B_{3} + 
    ((-2776Q)/9 - 6610Q{}^{3} - 34496Q{}^{5})$%
$\partial A_{2} \partial{}^{2} B_{2} 
      \partial{}^{2} B_{3} + (-359/81 + 12704 Q{}^{2}/9 + 331324 Q{}^{4}/9 + 
      590464 Q{}^{6}/3)$%
$\partial A_{2} \partial{}^{3} B_{2} \partial{}^{2} B_{2} + 
    ((-10144Q)/27 - 91222 Q{}^{3}/9 - 164800 Q{}^{5}/3)
     $%
$\partial A_{2} \partial{}^{3} B_{2} \partial B_{3} + 
    (-3658/81 - 2036 Q{}^{2}/27 + 13107Q{}^{4} + 83036Q{}^{6})
     $%
$\partial A_{2} \partial{}^{4} B_{2} \partial B_{2} + 
    ((-6241Q)/27 - 47674 Q{}^{3}/9 - 81076 Q{}^{5}/3)$%
$\partial A_{2} \partial{}^{4} B_{2} 
      B_3 + (-2408/81 - 126803 Q{}^{2}/270 - 45641 Q{}^{4}/45 + 
      16844 Q{}^{6}/3)$%
$\partial A_{2} \partial{}^{5} B_{2} B_2 + 
    (3758/9 + 12160Q{}^{2} + 66616Q{}^{4})$%
$\partial A_{2} \partial{}^{2} B_{3} \partial B_{3} + 
    (2338/9 + 5616Q{}^{2} + 26240Q{}^{4})$%
$\partial A_{2} \partial{}^{3} B_{3} B_3 + 
    ((-23788Q{}^{2})/9 - 78280Q{}^{4} - 564032Q{}^{6})$%
$\partial{}^{2} A_{2} \partial{}^{2} T 
      \partial{}^{2} T + (-3232/81 - 155120 Q{}^{2}/27 - 1188512 Q{}^{4}/9 - 
      2610832 Q{}^{6}/3)$%
$\partial{}^{2} A_{2} \partial{}^{3} T \partial T + 
    (-11872/243 - 243956 Q{}^{2}/81 - 1494812 Q{}^{4}/27 - 3126568 Q{}^{6}/9)
     $%
$\partial{}^{2} A_{2} \partial{}^{4} T T + (893/243 - 33161 Q{}^{2}/81 - 
      651116 Q{}^{4}/27 - 2101792 Q{}^{6}/9)$%
$\partial{}^{2} A_{2} A_2 \partial{}^{4} T + 
    ((-172Q)/9 + 10846 Q{}^{3}/9 + 37424 Q{}^{5}/3)$%
$\partial{}^{2} A_{2} A_2 
      \partial{}^{3} A_{3} + (-764/81 - 31336 Q{}^{2}/27 - 441058 Q{}^{4}/9 - 
      1448384 Q{}^{6}/3)$%
$\partial{}^{2} A_{2} \partial A_{2} \partial{}^{3} T + 
    (488 Q/3 + 4342Q{}^{3} + 23232Q{}^{5})$%
$\partial{}^{2} A_{2} \partial A_{2} 
      \partial{}^{2} A_{3} + ((-3592Q{}^{2})/9 - 29392Q{}^{4} - 309680Q{}^{6})
     $%
$\partial{}^{2} A_{2} \partial{}^{2} A_{2} \partial{}^{2} T + 
    (-1326Q{}^{2} - 36548Q{}^{4} - 238224Q{}^{6})$%
$\partial{}^{2} A_{2} \partial{}^{2} A_{2} 
      \partial{}^{2} A_{2} + (682 Q/3 + 12764 Q{}^{3}/3 + 17248Q{}^{5})
     $%
$\partial{}^{2} A_{2} \partial{}^{2} A_{2} \partial A_{3} + (176Q - 66Q{}^{3} - 16016Q{}^{5})
     $%
$\partial{}^{2} A_{2} A_3 \partial{}^{3} T + 
    (146 Q/3 - 11596 Q{}^{3}/3 - 49952Q{}^{5})$%
$\partial{}^{2} A_{2} \partial A_{3} 
      \partial{}^{2} T + (-1448/9 - 3888Q{}^{2} - 17248Q{}^{4})$%
$\partial{}^{2} A_{2} \partial A_{3} 
      \partial A_{3} + (62Q - 10042 Q{}^{3}/3 - 47072Q{}^{5})
     $%
$\partial{}^{2} A_{2} \partial{}^{2} A_{3} \partial T + 
    (-1198/9 - 10124 Q{}^{2}/3 - 15888Q{}^{4})$%
$\partial{}^{2} A_{2} \partial{}^{2} A_{3} 
      A_3 + ((-616Q)/27 - 2126 Q{}^{3}/3 - 23456 Q{}^{5}/3)
     $%
$\partial{}^{2} A_{2} \partial{}^{3} A_{3} T + (228Q + 4432Q{}^{3} + 17136Q{}^{5})
     $%
$\partial{}^{2} A_{2} B_2 \partial{}^{3} B_{3} + 
    ((-1006Q)/9 - 292 Q{}^{3}/3 + 2240Q{}^{5})$%
$\partial{}^{2} A_{2} \partial B_{2} 
      \partial{}^{2} B_{3} + (8/9 + 2002 Q{}^{2}/3 + 18580Q{}^{4} + 97104Q{}^{6})
     $%
$\partial{}^{2} A_{2} \partial{}^{2} B_{2} \partial{}^{2} B_{2} + 
    ((-1046Q)/3 - 21596 Q{}^{3}/3 - 33376Q{}^{5})$%
$\partial{}^{2} A_{2} \partial{}^{2} B_{2} 
      \partial B_{3} + (-5279/81 - 1786 Q{}^{2}/9 + 162448 Q{}^{4}/9 + 
      337792 Q{}^{6}/3)$%
$\partial{}^{2} A_{2} \partial{}^{3} B_{2} \partial B_{2} + 
    ((-6304Q)/27 - 51616 Q{}^{3}/9 - 85360 Q{}^{5}/3)$%
$\partial{}^{2} A_{2} \partial{}^{3} B_{2} 
      B_3 + (-5378/81 - 28430 Q{}^{2}/27 - 20308 Q{}^{4}/9 + 
      29168 Q{}^{6}/3)$%
$\partial{}^{2} A_{2} \partial{}^{4} B_{2} B_2 + 
    (1916/9 + 5480Q{}^{2} + 28000Q{}^{4})$%
$\partial{}^{2} A_{2} \partial B_{3} \partial B_{3} + 
    (2950/9 + 7068Q{}^{2} + 31856Q{}^{4})$%
$\partial{}^{2} A_{2} \partial{}^{2} B_{3} B_3 + 
    (-2872/81 - 165472 Q{}^{2}/27 - 443708 Q{}^{4}/3 - 2887912 Q{}^{6}/3)
     $%
$\partial{}^{3} A_{2} \partial{}^{2} T \partial T + (-15064/243 - 331064 Q{}^{2}/81 - 
      2121752 Q{}^{4}/27 - 4323064 Q{}^{6}/9)$%
$\partial{}^{3} A_{2} \partial{}^{3} T T + 
    (-232/81 - 87926 Q{}^{2}/81 - 356764 Q{}^{4}/9 - 2976904 Q{}^{6}/9)
     $%
$\partial{}^{3} A_{2} A_2 \partial{}^{3} T + (752 Q/9 + 2618Q{}^{3} + 15224Q{}^{5})
     $%
$\partial{}^{3} A_{2} A_2 \partial{}^{2} A_{3} + 
    (-536/81 - 40924 Q{}^{2}/27 - 188660 Q{}^{4}/3 - 555928Q{}^{6})
     $%
$\partial{}^{3} A_{2} \partial A_{2} \partial{}^{2} T + 
    (7696 Q/27 + 51752 Q{}^{3}/9 + 25368Q{}^{5})$%
$\partial{}^{3} A_{2} \partial A_{2} 
      \partial A_{3} + (-766/81 - 55577 Q{}^{2}/27 - 73636Q{}^{4} - 593920Q{}^{6})
     $%
$\partial{}^{3} A_{2} \partial{}^{2} A_{2} \partial T + 
    (-2636/81 - 198311 Q{}^{2}/27 - 176202Q{}^{4} - 3282512 Q{}^{6}/3)
     $%
$\partial{}^{3} A_{2} \partial{}^{2} A_{2} \partial A_{2} + 
    (5828 Q/27 + 32878 Q{}^{3}/9 + 14048Q{}^{5})$%
$\partial{}^{3} A_{2} \partial{}^{2} A_{2} 
      A_3 + (3808/243 - 63364 Q{}^{2}/81 - 732928 Q{}^{4}/27 - 
      1793552 Q{}^{6}/9)$%
$\partial{}^{3} A_{2} \partial{}^{3} A_{2} T + 
    (-1882/81 - 165100 Q{}^{2}/81 - 409372 Q{}^{4}/9 - 2536784 Q{}^{6}/9)
     $%
$\partial{}^{3} A_{2} \partial{}^{3} A_{2} A_2 + 
    (2180 Q/27 - 20966 Q{}^{3}/9 - 28576Q{}^{5})$%
$\partial{}^{3} A_{2} A_3 
      \partial{}^{2} T + ((-2264Q)/27 - 53572 Q{}^{3}/9 - 54088Q{}^{5})
     $%
$\partial{}^{3} A_{2} \partial A_{3} \partial T + 
    (-2132/9 - 10948 Q{}^{2}/3 - 11920Q{}^{4})$%
$\partial{}^{3} A_{2} \partial A_{3} 
      A_3 + (-58Q - 2658Q{}^{3} - 21560Q{}^{5})$%
$\partial{}^{3} A_{2} \partial{}^{2} A_{3} 
      T + (2080 Q/9 + 11150 Q{}^{3}/3 + 14528Q{}^{5})
     $%
$\partial{}^{3} A_{2} B_2 \partial{}^{2} B_{3} + 
    ((-1804Q)/9 - 2432Q{}^{3} - 5344Q{}^{5})$%
$\partial{}^{3} A_{2} \partial B_{2} 
      \partial B_{3} + (-4619/81 - 814 Q{}^{2}/9 + 177464 Q{}^{4}/9 + 115712Q{}^{6})
     $%
$\partial{}^{3} A_{2} \partial{}^{2} B_{2} \partial B_{2} + 
    ((-1958Q)/9 - 12370 Q{}^{3}/3 - 16752Q{}^{5})$%
$\partial{}^{3} A_{2} \partial{}^{2} B_{2} 
      B_3 + (-22853/243 - 34106 Q{}^{2}/27 - 13216 Q{}^{4}/27 + 
      63680 Q{}^{6}/3)$%
$\partial{}^{3} A_{2} \partial{}^{3} B_{2} B_2 + 
    (3172/9 + 17684 Q{}^{2}/3 + 22032Q{}^{4})$%
$\partial{}^{3} A_{2} \partial B_{3} 
      B_3 + (-2540/243 - 182140 Q{}^{2}/81 - 1414498 Q{}^{4}/27 - 
      2923232 Q{}^{6}/9)$%
$\partial{}^{4} A_{2} \partial T \partial T + 
    (-7156/243 - 249566 Q{}^{2}/81 - 1838498 Q{}^{4}/27 - 3828616 Q{}^{6}/9)
     $%
$\partial{}^{4} A_{2} \partial{}^{2} T T + (-286/243 - 95792 Q{}^{2}/81 - 
      1073684 Q{}^{4}/27 - 2764696 Q{}^{6}/9)$%
$\partial{}^{4} A_{2} A_2 
      \partial{}^{2} T + (2408 Q/27 + 6260 Q{}^{3}/3 + 32164 Q{}^{5}/3)
     $%
$\partial{}^{4} A_{2} A_2 \partial A_{3} + 
    (-3706/243 - 171086 Q{}^{2}/81 - 1628441 Q{}^{4}/27 - 3949324 Q{}^{6}/9)
     $%
$\partial{}^{4} A_{2} \partial A_{2} \partial T + 
    (-5612/243 - 205432 Q{}^{2}/81 - 1513903 Q{}^{4}/27 - 3078452 Q{}^{6}/9)
     $%
$\partial{}^{4} A_{2} \partial A_{2} \partial A_{2} + 
    (3241 Q/27 + 21160 Q{}^{3}/9 + 31372 Q{}^{5}/3)$%
$\partial{}^{4} A_{2} \partial A_{2} 
      A_3 + (5708/243 - 101876 Q{}^{2}/81 - 1228265 Q{}^{4}/27 - 
      3049972 Q{}^{6}/9)$%
$\partial{}^{4} A_{2} \partial{}^{2} A_{2} T + 
    (-7477/243 - 249539 Q{}^{2}/81 - 1949087 Q{}^{4}/27 - 4118620 Q{}^{6}/9)
     $%
$\partial{}^{4} A_{2} \partial{}^{2} A_{2} A_2 + 
    ((-4625Q)/27 - 17300 Q{}^{3}/9 - 20228 Q{}^{5}/3)$%
$\partial{}^{4} A_{2} A_3 
      \partial T + (-2033/27 - 1166Q{}^{2} - 13096 Q{}^{4}/3)
     $%
$\partial{}^{4} A_{2} A_3 A_3 + 
    ((-4727Q)/27 - 5096 Q{}^{3}/3 - 6340 Q{}^{5}/3)$%
$\partial{}^{4} A_{2} \partial A_{3} 
      T + (1703 Q/27 + 4564 Q{}^{3}/3 + 24184 Q{}^{5}/3)
     $%
$\partial{}^{4} A_{2} B_2 \partial B_{3} + 
    (-1787/81 - 3053 Q{}^{2}/27 + 56504 Q{}^{4}/9 + 122336 Q{}^{6}/3)
     $%
$\partial{}^{4} A_{2} \partial B_{2} \partial B_{2} + 
    (-102Q - 6344 Q{}^{3}/3 - 10384Q{}^{5})$%
$\partial{}^{4} A_{2} \partial B_{2} 
      B_3 + (-1877/27 - 21041 Q{}^{2}/27 + 6532 Q{}^{4}/3 + 
      83048 Q{}^{6}/3)$%
$\partial{}^{4} A_{2} \partial{}^{2} B_{2} B_2 + 
    (902/9 + 4946 Q{}^{2}/3 + 7040Q{}^{4})$%
$\partial{}^{4} A_{2} B_3 
      B_3 + (-3608/243 - 733358 Q{}^{2}/405 - 1055716 Q{}^{4}/27 - 
      2099456 Q{}^{6}/9)$%
$\partial{}^{5} A_{2} \partial T T + 
    (-7537/1215 - 183029 Q{}^{2}/162 - 3790814 Q{}^{4}/135 - 1668980 Q{}^{6}/9)
     $%
$\partial{}^{5} A_{2} A_2 \partial T + 
    (1364 Q/45 + 24079 Q{}^{3}/45 + 7492 Q{}^{5}/3)$%
$\partial{}^{5} A_{2} A_2 
      A_3 + (27487/2430 - 359117 Q{}^{2}/405 - 3558089 Q{}^{4}/135 - 
      1633904 Q{}^{6}/9)$%
$\partial{}^{5} A_{2} \partial A_{2} T + 
    (-14317/486 - 748508 Q{}^{2}/405 - 5066117 Q{}^{4}/135 - 2045180 Q{}^{6}/9)
     $%
$\partial{}^{5} A_{2} \partial A_{2} A_2 + 
    ((-5842Q)/27 - 35879 Q{}^{3}/15 - 20876 Q{}^{5}/3)
     $%
$\partial{}^{5} A_{2} A_3 T + 
    (337 Q/27 + 1694 Q{}^{3}/9 + 936Q{}^{5})$%
$\partial{}^{5} A_{2} B_2 
      B_3 + (-25013/810 - 45632 Q{}^{2}/135 + 15796 Q{}^{4}/15 + 
      12048Q{}^{6})$%
$\partial{}^{5} A_{2} \partial B_{2} B_2 + 
    (-443/243 - 104584 Q{}^{2}/405 - 260128 Q{}^{4}/45 - 311936 Q{}^{6}/9)
     $%
$\partial{}^{6} A_{2} T{}^{2}+ (5977/2430 - 152008 Q{}^{2}/405 - 
      1258058 Q{}^{4}/135 - 537964 Q{}^{6}/9)$%
$\partial{}^{6} A_{2} A_2 
      T + (-2306/405 - 24889 Q{}^{2}/90 - 727784 Q{}^{4}/135 - 
      294068 Q{}^{6}/9)$%
$\partial{}^{6} A_{2} A_2{}^{2}+ 
    (-9607/2430 - 21847 Q{}^{2}/810 + 30136 Q{}^{4}/135 + 4424 Q{}^{6}/3)
     $%
$\partial{}^{6} A_{2} B_2{}^{2}+ 
    (12448 Q/9 + 30888Q{}^{3} + 139552Q{}^{5})$%
$A_3 \partial{}^{3} T \partial{}^{2} T + 
    (32804 Q/27 + 62396 Q{}^{3}/3 + 258328 Q{}^{5}/3)
     $%
$A_3 \partial{}^{4} T \partial T + (62716 Q/135 + 84712 Q{}^{3}/15 + 
      51976 Q{}^{5}/3)$%
$A_3 \partial{}^{5} T T + 
    (1483/27 + 1276 Q{}^{2}/3 - 136 Q{}^{4}/3)$%
$A_3 A_3 
      \partial{}^{4} T + (-331/3 - 1486Q{}^{2} - 3864Q{}^{4})$%
$A_3 B_2 
      \partial{}^{4} B_{3} + (-182 - 2988Q{}^{2} - 9072Q{}^{4})$%
$A_3 \partial B_{2} 
      \partial{}^{3} B_{3} + (-104/3 - 1416Q{}^{2} - 8544Q{}^{4})$%
$A_3 \partial{}^{2} B_{2} 
      \partial{}^{2} B_{3} + (260 Q/3 + 3488Q{}^{3} + 23424Q{}^{5})
     $%
$A_3 \partial{}^{3} B_{2} \partial{}^{2} B_{2} + (46 + 44Q{}^{2} - 3696Q{}^{4})
     $%
$A_3 \partial{}^{3} B_{2} \partial B_{3} + 
    (514 Q/3 + 3760Q{}^{3} + 18096Q{}^{5})$%
$A_3 \partial{}^{4} B_{2} 
      \partial B_{2} + (5/3 - 134Q{}^{2} - 936Q{}^{4})$%
$A_3 \partial{}^{4} B_{2} 
      B_3 + (1228 Q/15 + 8024 Q{}^{3}/5 + 6672Q{}^{5})
     $%
$A_3 \partial{}^{5} B_{2} B_2 - 
    48Q$%
$A_3 \partial{}^{2} B_{3} \partial B_{3} + 
    144Q$%
$A_3 \partial{}^{3} B_{3} B_3 + 
    (4828 Q/3 + 40664Q{}^{3} + 200256Q{}^{5})$%
$\partial A_{3} \partial{}^{2} T \partial{}^{2} T + 
    (31376 Q/9 + 203672 Q{}^{3}/3 + 294560Q{}^{5})$%
$\partial A_{3} \partial{}^{3} T 
      \partial T + (14380 Q/9 + 26180Q{}^{3} + 106936Q{}^{5})
     $%
$\partial A_{3} \partial{}^{4} T T + (1928/9 + 3896Q{}^{2} + 18368Q{}^{4})
     $%
$\partial A_{3} A_3 \partial{}^{3} T + (1684/9 + 4816Q{}^{2} + 29456Q{}^{4})
     $%
$\partial A_{3} \partial A_{3} \partial{}^{2} T + 
    48Q$%
$\partial A_{3} \partial A_{3} \partial A_{3} + 
    (-500/3 - 2752Q{}^{2} - 9024Q{}^{4})$%
$\partial A_{3} B_2 \partial{}^{3} B_{3} + 
    (-446/3 - 3644Q{}^{2} - 16848Q{}^{4})$%
$\partial A_{3} \partial B_{2} \partial{}^{2} B_{3} + 
    (220 Q/3 + 2464Q{}^{3} + 16128Q{}^{5})$%
$\partial A_{3} \partial{}^{2} B_{2} 
      \partial{}^{2} B_{2} + (-116/3 - 1432Q{}^{2} - 10752Q{}^{4})
     $%
$\partial A_{3} \partial{}^{2} B_{2} \partial B_{3} + (188Q + 5960Q{}^{3} + 31776Q{}^{5})
     $%
$\partial A_{3} \partial{}^{3} B_{2} \partial B_{2} + (-170/3 - 980Q{}^{2} - 3696Q{}^{4})
     $%
$\partial A_{3} \partial{}^{3} B_{2} B_3 + (102Q + 2804Q{}^{3} + 14304Q{}^{5})
     $%
$\partial A_{3} \partial{}^{4} B_{2} B_2 - 
    48Q$%
$\partial A_{3} \partial B_{3} \partial B_{3} - 
    48Q$%
$\partial A_{3} \partial{}^{2} B_{3} B_3 + 
    (10768 Q/3 + 258896 Q{}^{3}/3 + 426808Q{}^{5})$%
$\partial{}^{2} A_{3} \partial{}^{2} T 
      \partial T + (58916 Q/27 + 42796Q{}^{3} + 570904 Q{}^{5}/3)
     $%
$\partial{}^{2} A_{3} \partial{}^{3} T T + (1736/9 + 15256 Q{}^{2}/3 + 30048Q{}^{4})
     $%
$\partial{}^{2} A_{3} A_3 \partial{}^{2} T + 
    (1996/9 + 24862 Q{}^{2}/3 + 53968Q{}^{4})$%
$\partial{}^{2} A_{3} \partial A_{3} \partial T + 
    (-72Q - 576Q{}^{3})$%
$\partial{}^{2} A_{3} \partial A_{3} A_3 + 
    (860/9 + 1590Q{}^{2} + 6688Q{}^{4})$%
$\partial{}^{2} A_{3} \partial{}^{2} A_{3} T + 
    (-90 - 1724Q{}^{2} - 8304Q{}^{4})$%
$\partial{}^{2} A_{3} B_2 \partial{}^{2} B_{3} + 
    (-202/3 - 1796Q{}^{2} - 10704Q{}^{4})$%
$\partial{}^{2} A_{3} \partial B_{2} \partial B_{3} + 
    (470 Q/3 + 2444Q{}^{3} + 9936Q{}^{5})$%
$\partial{}^{2} A_{3} \partial{}^{2} B_{2} 
      \partial B_{2} + (-120 - 1728Q{}^{2} - 5472Q{}^{4})$%
$\partial{}^{2} A_{3} \partial{}^{2} B_{2} 
      B_3 + ((-2Q)/3 + 1092Q{}^{3} + 8304Q{}^{5})
     $%
$\partial{}^{2} A_{3} \partial{}^{3} B_{2} B_2 + (96Q + 288Q{}^{3})
     $%
$\partial{}^{2} A_{3} \partial B_{3} B_3 + 
    (9608 Q/9 + 87484 Q{}^{3}/3 + 156016Q{}^{5})$%
$\partial{}^{3} A_{3} \partial T 
      \partial T + (35180 Q/27 + 35972Q{}^{3} + 575848 Q{}^{5}/3)
     $%
$\partial{}^{3} A_{3} \partial{}^{2} T T + (204 + 8998 Q{}^{2}/3 + 17096Q{}^{4})
     $%
$\partial{}^{3} A_{3} A_3 \partial T + (24Q + 144Q{}^{3})
     $%
$\partial{}^{3} A_{3} A_3 A_3 + 
    (3044/27 + 6548 Q{}^{2}/3 + 23512 Q{}^{4}/3)$%
$\partial{}^{3} A_{3} \partial A_{3} 
      T + (32 - 240Q{}^{2} - 3840Q{}^{4})$%
$\partial{}^{3} A_{3} B_2 
      \partial B_{3} + (68 Q/3 - 1544 Q{}^{3}/3 - 4928Q{}^{5})
     $%
$\partial{}^{3} A_{3} \partial B_{2} \partial B_{2} + (-218/3 - 1108Q{}^{2} - 3216Q{}^{4})
     $%
$\partial{}^{3} A_{3} \partial B_{2} B_3 + 
    (-106Q - 4436 Q{}^{3}/3 - 4496Q{}^{5})$%
$\partial{}^{3} A_{3} \partial{}^{2} B_{2} 
      B_2 + (72Q - 240Q{}^{3})$%
$\partial{}^{3} A_{3} B_3 B_3 + 
    (5594 Q/9 + 164728 Q{}^{3}/9 + 310160 Q{}^{5}/3)$%
$\partial{}^{4} A_{3} \partial T 
      T + (1745/27 + 994Q{}^{2} + 7768 Q{}^{4}/3)$%
$\partial{}^{4} A_{3} A_3 
      T + (4 - 66Q{}^{2} - 936Q{}^{4})$%
$\partial{}^{4} A_{3} B_2 
      B_3 + ((-550Q)/9 - 3608 Q{}^{3}/3 - 5776Q{}^{5})
     $%
$\partial{}^{4} A_{3} \partial B_{2} B_2 + 
    (8926 Q/135 + 32612 Q{}^{3}/15 + 40288 Q{}^{5}/3)
     $%
$\partial{}^{5} A_{3} T{}^{2}+ ((-548Q)/45 - 3172 Q{}^{3}/15 - 952Q{}^{5})
     $%
$\partial{}^{5} A_{3} B_2{}^{2}+ 
    (-4606/1215 - 12298 Q{}^{2}/135 - 18292 Q{}^{4}/27 - 5032 Q{}^{6}/3)
     $%
$B_2{}^{2}\partial{}^{6} T + 
    (2216 Q/27 + 53726 Q{}^{3}/45 + 4728Q{}^{5})$%
$B_2 B_3 
      \partial{}^{5} T + (11735 Q/27 + 7092Q{}^{3} + 90856 Q{}^{5}/3)
     $%
$B_2 \partial B_{3} \partial{}^{4} T + 
    (4256 Q/9 + 30922 Q{}^{3}/3 + 53264Q{}^{5})$%
$B_2 \partial{}^{2} B_{3} 
      \partial{}^{3} T + (146 Q/3 + 6620Q{}^{3} + 51360Q{}^{5})$%
$B_2 \partial{}^{3} B_{3} 
      \partial{}^{2} T + ((-1607Q)/9 + 3286Q{}^{3} + 29492Q{}^{5})
     $%
$B_2 \partial{}^{4} B_{3} \partial T + 
    ((-842Q)/135 + 24194 Q{}^{3}/15 + 30244 Q{}^{5}/3)
     $%
$B_2 \partial{}^{5} B_{3} T + 
    (-2566/135 - 86567 Q{}^{2}/135 - 287362 Q{}^{4}/45 - 21048Q{}^{6})
     $%
$\partial B_{2} B_2 \partial{}^{5} T + 
    (-983/81 - 20534 Q{}^{2}/27 - 92932 Q{}^{4}/9 - 124192 Q{}^{6}/3)
     $%
$\partial B_{2} \partial B_{2} \partial{}^{4} T + 
    (1463 Q/9 + 10340 Q{}^{3}/3 + 16904Q{}^{5})$%
$\partial B_{2} B_3 
      \partial{}^{4} T + (3176 Q/3 + 22356Q{}^{3} + 113120Q{}^{5})
     $%
$\partial B_{2} \partial B_{3} \partial{}^{3} T + 
    (5678 Q/9 + 78020 Q{}^{3}/3 + 169952Q{}^{5})$%
$\partial B_{2} \partial{}^{2} B_{3} 
      \partial{}^{2} T + (2468 Q/9 + 49520 Q{}^{3}/3 + 118568Q{}^{5})
     $%
$\partial B_{2} \partial{}^{3} B_{3} \partial T + 
    ((-1609Q)/27 + 16298 Q{}^{3}/3 + 113980 Q{}^{5}/3)
     $%
$\partial B_{2} \partial{}^{4} B_{3} T + 
    (-611/27 - 31775 Q{}^{2}/27 - 14560Q{}^{4} - 164200 Q{}^{6}/3)
     $%
$\partial{}^{2} B_{2} B_2 \partial{}^{4} T + 
    (-2108/81 - 8984 Q{}^{2}/3 - 441970 Q{}^{4}/9 - 232624Q{}^{6})
     $%
$\partial{}^{2} B_{2} \partial B_{2} \partial{}^{3} T + 
    (-2268Q{}^{2} - 42952Q{}^{4} - 211344Q{}^{6})$%
$\partial{}^{2} B_{2} \partial{}^{2} B_{2} \partial{}^{2} T + 
    (3176 Q/9 + 26386 Q{}^{3}/3 + 48720Q{}^{5})$%
$\partial{}^{2} B_{2} B_3 
      \partial{}^{3} T + (1494Q + 104692 Q{}^{3}/3 + 185696Q{}^{5})
     $%
$\partial{}^{2} B_{2} \partial B_{3} \partial{}^{2} T + 
    (6338 Q/9 + 30490Q{}^{3} + 185680Q{}^{5})$%
$\partial{}^{2} B_{2} \partial{}^{2} B_{3} 
      \partial T + ((-310Q)/9 + 9782Q{}^{3} + 66952Q{}^{5})
     $%
$\partial{}^{2} B_{2} \partial{}^{3} B_{3} T + 
    (-7004/243 - 3632 Q{}^{2}/3 - 448186 Q{}^{4}/27 - 226672 Q{}^{6}/3)
     $%
$\partial{}^{3} B_{2} B_2 \partial{}^{3} T + 
    (-584/81 - 20716 Q{}^{2}/9 - 425240 Q{}^{4}/9 - 755456 Q{}^{6}/3)
     $%
$\partial{}^{3} B_{2} \partial B_{2} \partial{}^{2} T + 
    (-776/81 - 28520 Q{}^{2}/9 - 530984 Q{}^{4}/9 - 851120 Q{}^{6}/3)
     $%
$\partial{}^{3} B_{2} \partial{}^{2} B_{2} \partial T + 
    (701/81 - 11974 Q{}^{2}/27 - 66584 Q{}^{4}/9 - 97456 Q{}^{6}/3)
     $%
$\partial{}^{3} B_{2} \partial{}^{3} B_{2} T + 
    (13694 Q/27 + 108980 Q{}^{3}/9 + 192704 Q{}^{5}/3)
     $%
$\partial{}^{3} B_{2} B_3 \partial{}^{2} T + 
    (35360 Q/27 + 260498 Q{}^{3}/9 + 436544 Q{}^{5}/3)
     $%
$\partial{}^{3} B_{2} \partial B_{3} \partial T + 
    (12428 Q/27 + 35132 Q{}^{3}/3 + 205744 Q{}^{5}/3)
     $%
$\partial{}^{3} B_{2} \partial{}^{2} B_{3} T + 
    (-1466/81 - 15638 Q{}^{2}/27 - 90700 Q{}^{4}/9 - 169840 Q{}^{6}/3)
     $%
$\partial{}^{4} B_{2} B_2 \partial{}^{2} T + 
    (-2140/81 - 41129 Q{}^{2}/27 - 76621 Q{}^{4}/3 - 128956Q{}^{6})
     $%
$\partial{}^{4} B_{2} \partial B_{2} \partial T + 
    (1376/81 - 6046 Q{}^{2}/9 - 114065 Q{}^{4}/9 - 54356Q{}^{6})
     $%
$\partial{}^{4} B_{2} \partial{}^{2} B_{2} T + 
    (15953 Q/27 + 87074 Q{}^{3}/9 + 134060 Q{}^{5}/3)
     $%
$\partial{}^{4} B_{2} B_3 \partial T + 
    (13088 Q/27 + 28600 Q{}^{3}/3 + 134836 Q{}^{5}/3)
     $%
$\partial{}^{4} B_{2} \partial B_{3} T + 
    (-5281/405 - 84443 Q{}^{2}/270 - 227699 Q{}^{4}/45 - 78844 Q{}^{6}/3)
     $%
$\partial{}^{5} B_{2} B_2 \partial T + 
    (-463/810 - 17632 Q{}^{2}/45 - 286666 Q{}^{4}/45 - 26592Q{}^{6})
     $%
$\partial{}^{5} B_{2} \partial B_{2} T + 
    (8446 Q/45 + 49094 Q{}^{3}/15 + 13452Q{}^{5})$%
$\partial{}^{5} B_{2} B_3 
      T + (-565/162 - 428 Q{}^{2}/3 - 32719 Q{}^{4}/15 - 8524Q{}^{6})
     $%
$\partial{}^{6} B_{2} B_2 T + (-211/9 + 932 Q{}^{2}/3 + 3368Q{}^{4})
     $%
$B_3 B_3 \partial{}^{4} T + (-296/9 + 104 Q{}^{2}/3 - 1344Q{}^{4})
     $%
$\partial B_{3} B_3 \partial{}^{3} T + (-868/9 - 1120Q{}^{2} - 7952Q{}^{4})
     $%
$\partial B_{3} \partial B_{3} \partial{}^{2} T + (-20/9 - 744Q{}^{2} - 12064Q{}^{4})
     $%
$\partial{}^{2} B_{3} B_3 \partial{}^{2} T + (632/9 - 1796Q{}^{2} - 21464Q{}^{4})
     $%
$\partial{}^{2} B_{3} \partial B_{3} \partial T + (-512/3 - 268Q{}^{2} - 744Q{}^{4})
     $%
$\partial{}^{2} B_{3} \partial{}^{2} B_{3} T + (-416/9 - 1092Q{}^{2} - 12784Q{}^{4})
     $%
$\partial{}^{3} B_{3} B_3 \partial T + (-382/9 - 856Q{}^{2} - 4376Q{}^{4})
     $%
$\partial{}^{3} B_{3} \partial B_{3} T + (-55/9 - 748Q{}^{2} - 3872Q{}^{4})
     $%
$\partial{}^{4} B_{3} B_3 T + (512 Q{}^{2}/3 + 2048Q{}^{4})
     $%
$\partial T \partial T \partial T \partial T + (400/9 + 2368Q{}^{2} + 22016Q{}^{4})
     $%
$\partial{}^{2} T \partial T \partial T T + (200/9 + 944Q{}^{2} + 8128Q{}^{4})
     $%
$\partial{}^{2} T \partial{}^{2} T T{}^{2}+ (800/27 + 1440Q{}^{2} + 39040 Q{}^{4}/3)
     $%
$\partial{}^{3} T \partial T T{}^{2}+ (40/9 + 592 Q{}^{2}/3 + 1728Q{}^{4})
     $%
$\partial{}^{4} T T{}^{3}+ 
    ((-2000\sqrt{2/3}Q)/9 - 1152\sqrt{6}Q{}^{3} - 9472\sqrt{2/3}Q{}^{5})
     $%
$J \partial{}^{2} T \partial{}^{2} T \partial T + 
    (-256\sqrt{2/3}Q - 4912\sqrt{2/3}Q{}^{3} - 7360\sqrt{6}Q{}^{5})
     $%
$J \partial{}^{3} T \partial T \partial T + 
    ((-9712\sqrt{2/3}Q)/27 - (20624\sqrt{2/3}Q{}^{3})/3 - 
      (92096\sqrt{2/3}Q{}^{5})/3)$%
$J \partial{}^{3} T \partial{}^{2} T T + 
    ((-2528\sqrt{2/3}Q)/9 - (17248\sqrt{2/3}Q{}^{3})/3 - 28544\sqrt{2/3}Q{}^{5})
     $%
$J \partial{}^{4} T \partial T T + 
    ((-7808\sqrt{2/3}Q)/135 - (17584\sqrt{2/3}Q{}^{3})/15 - 
      (17216\sqrt{2/3}Q{}^{5})/3)$%
$J \partial{}^{5} T T{}^{2}+ 
    (-380/81 - 78368 Q{}^{2}/81 - 191920 Q{}^{4}/9 - 1122112 Q{}^{6}/9)
     $%
$J{}^{2}\partial{}^{3} T \partial{}^{3} T + 
    (-1814/243 - 133696 Q{}^{2}/81 - 996364 Q{}^{4}/27 - 1962416 Q{}^{6}/9)
     $%
$J{}^{2}\partial{}^{4} T \partial{}^{2} T + 
    (-16952/1215 - 442396 Q{}^{2}/405 - 2777932 Q{}^{4}/135 - 1023664 Q{}^{6}/9)
     $%
$J{}^{2}\partial{}^{5} T \partial T + 
    (-3692/405 - 186896 Q{}^{2}/405 - 37544 Q{}^{4}/5 - 354656 Q{}^{6}/9)
     $%
$J{}^{2}\partial{}^{6} T T + 
    ((-106003\sqrt{2/3}Q)/8505 - 90\sqrt{6}Q{}^{3} - (1724012\sqrt{2/3}Q{}^{5})/
       945 - (13648\sqrt{2/3}Q{}^{7})/3)$%
$J{}^{3}\partial{}^{7} T + ((1234\sqrt{2/3}Q)/567 - 964729 Q{}^{3}/(5670\sqrt{6}) - 
      (1784042\sqrt{2/3}Q{}^{5})/945 - (24824\sqrt{2/3}Q{}^{7})/3)
     $%
$J{}^{3}\partial{}^{7} A_{2} + 
    (-2\sqrt{2/3} + (16037\sqrt{2/3}Q{}^{2})/135 + (35203\sqrt{2/3}Q{}^{4})/15 + 
      (36740\sqrt{2/3}Q{}^{6})/3)$%
$J{}^{3}\partial{}^{6} A_{3} + (-2374/1215 - 42944 Q{}^{2}/81 - 140084 Q{}^{4}/15 - 
      427040 Q{}^{6}/9)$%
$J{}^{2}A_2 \partial{}^{6} T + 
    ((-1211Q)/30 - 1372 Q{}^{3}/5 - 186Q{}^{5})$%
$J{}^{2}A_2 \partial{}^{5} A_{3} + (-1804/243 - 818482 Q{}^{2}/405 - 
      4733992 Q{}^{4}/135 - 1553260 Q{}^{6}/9)$%
$J{}^{2}\partial A_{2} \partial{}^{5} T + ((-1453Q)/18 - 3511 Q{}^{3}/9 + 2026 Q{}^{5}/3)
     $%
$J{}^{2}\partial A_{2} \partial{}^{4} A_{3} + 
    (-1028/243 - 260050 Q{}^{2}/81 - 1722046 Q{}^{4}/27 - 3019508 Q{}^{6}/9)
     $%
$J{}^{2}\partial{}^{2} A_{2} \partial{}^{4} T + 
    ((-2356Q)/27 - 2663 Q{}^{3}/3 - 1616 Q{}^{5}/3)$%
$J{}^{2}\partial{}^{2} A_{2} \partial{}^{3} A_{3} + (-2636/243 - 305092 Q{}^{2}/81 - 
      2116156 Q{}^{4}/27 - 3903932 Q{}^{6}/9)$%
$J{}^{2}\partial{}^{3} A_{2} \partial{}^{3} T + (1904/243 + 9166 Q{}^{2}/81 - 108560 Q{}^{4}/27 - 
      563848 Q{}^{6}/9)$%
$J{}^{2}\partial{}^{3} A_{2} \partial{}^{3} A_{2} + 
    (-137Q - 9371 Q{}^{3}/3 - 11868Q{}^{5})$%
$J{}^{2}\partial{}^{3} A_{2} 
      \partial{}^{2} A_{3} + (-1028/243 - 189607 Q{}^{2}/81 - 1522345 Q{}^{4}/27 - 
      3098564 Q{}^{6}/9)$%
$J{}^{2}\partial{}^{4} A_{2} \partial{}^{2} T + 
    (2854/243 + 12890 Q{}^{2}/81 - 405689 Q{}^{4}/54 - 934682 Q{}^{6}/9)
     $%
$J{}^{2}\partial{}^{4} A_{2} \partial{}^{2} A_{2} + 
    ((-4399Q)/54 - 2316Q{}^{3} - 42658 Q{}^{5}/3)$%
$J{}^{2}\partial{}^{4} A_{2} \partial A_{3} + (-1804/243 - 400927 Q{}^{2}/405 - 
      3101458 Q{}^{4}/135 - 1279312 Q{}^{6}/9)$%
$J{}^{2}\partial{}^{5} A_{2} \partial T + (27487/4860 - 31709 Q{}^{2}/810 - 
      1868441 Q{}^{4}/270 - 599272 Q{}^{6}/9)$%
$J{}^{2}\partial{}^{5} A_{2} \partial A_{2} + ((-343Q)/27 + 10427 Q{}^{3}/90 + 602 Q{}^{5}/3)
     $%
$J{}^{2}\partial{}^{5} A_{2} A_3 + 
    (-443/243 - 80152 Q{}^{2}/405 - 44728 Q{}^{4}/9 - 293552 Q{}^{6}/9)
     $%
$J{}^{2}\partial{}^{6} A_{2} T + 
    (5977/4860 - 5036 Q{}^{2}/405 - 291301 Q{}^{4}/135 - 182054 Q{}^{6}/9)
     $%
$J{}^{2}\partial{}^{6} A_{2} A_2 + 
    (31358 Q/135 + 38324 Q{}^{3}/15 + 25988 Q{}^{5}/3)
     $%
$J{}^{2}A_3 \partial{}^{5} T + 
    (10766 Q/27 + 17414 Q{}^{3}/3 + 78964 Q{}^{5}/3)$%
$J{}^{2}\partial A_{3} \partial{}^{4} T + (7858 Q/27 + 10022Q{}^{3} + 201740 Q{}^{5}/3)
     $%
$J{}^{2}\partial{}^{2} A_{3} \partial{}^{3} T + 
    (430/9 + 987Q{}^{2} + 5648Q{}^{4})$%
$J{}^{2}\partial{}^{2} A_{3} 
      \partial{}^{2} A_{3} + ((-1906Q)/27 + 21430 Q{}^{3}/3 + 206068 Q{}^{5}/3)
     $%
$J{}^{2}\partial{}^{3} A_{3} \partial{}^{2} T + 
    (1522/27 + 2794 Q{}^{2}/3 + 19820 Q{}^{4}/3)$%
$J{}^{2}\partial{}^{3} A_{3} \partial A_{3} + ((-5129Q)/27 + 18740 Q{}^{3}/9 + 30136Q{}^{5})
     $%
$J{}^{2}\partial{}^{4} A_{3} \partial T + 
    (1745/54 + 241Q{}^{2} + 4460 Q{}^{4}/3)$%
$J{}^{2}\partial{}^{4} A_{3} 
      A_3 + ((-1454Q)/27 - 1444 Q{}^{3}/15 + 14624 Q{}^{5}/3)
     $%
$J{}^{2}\partial{}^{5} A_{3} T + 
    (8147 Q/135 - 4799 Q{}^{3}/15 - 7918 Q{}^{5}/3)$%
$J{}^{2}B_2 \partial{}^{5} B_{3} + (1415 Q/54 - 4907 Q{}^{3}/3 - 24802 Q{}^{5}/3)
     $%
$J{}^{2}\partial B_{2} \partial{}^{4} B_{3} + 
    ((-443Q)/9 - 1093Q{}^{3} - 316Q{}^{5})$%
$J{}^{2}\partial{}^{2} B_{2} 
      \partial{}^{3} B_{3} + (701/162 + 3565 Q{}^{2}/27 + 11924 Q{}^{4}/9 - 2264 Q{}^{6}/3)
     $%
$J{}^{2}\partial{}^{3} B_{2} \partial{}^{3} B_{2} + 
    ((-3434Q)/27 - 674 Q{}^{3}/3 + 32600 Q{}^{5}/3)$%
$J{}^{2}\partial{}^{3} B_{2} \partial{}^{2} B_{3} + (688/81 + 1123 Q{}^{2}/3 + 70063 Q{}^{4}/18 + 
      6934Q{}^{6})$%
$J{}^{2}\partial{}^{4} B_{2} \partial{}^{2} B_{2} + 
    ((-9080Q)/27 - 3460 Q{}^{3}/3 + 21338 Q{}^{5}/3)$%
$J{}^{2}\partial{}^{4} B_{2} \partial B_{3} + (-463/1620 + 208Q{}^{2} + 174811 Q{}^{4}/45 + 
      13456Q{}^{6})$%
$J{}^{2}\partial{}^{5} B_{2} \partial B_{2} + 
    ((-1421Q)/9 - 18653 Q{}^{3}/15 - 954Q{}^{5})$%
$J{}^{2}\partial{}^{5} B_{2} B_3 + (-565/324 - 1186 Q{}^{2}/45 + 3773 Q{}^{4}/6 + 
      3354Q{}^{6})$%
$J{}^{2}\partial{}^{6} B_{2} B_2 + 
    (-256/3 - 1734Q{}^{2} - 10932Q{}^{4})$%
$J{}^{2}\partial{}^{2} B_{3} 
      \partial{}^{2} B_{3} + (-191/9 - 1836Q{}^{2} - 13324Q{}^{4})
     $%
$J{}^{2}\partial{}^{3} B_{3} \partial B_{3} + 
    (-55/18 - 54Q{}^{2} - 1744Q{}^{4})$%
$J{}^{2}\partial{}^{4} B_{3} 
      B_3 + ((-2584\sqrt{2/3}Q)/3 - (198112\sqrt{2/3}Q{}^{3})/9 - 
      (340448\sqrt{2/3}Q{}^{5})/3)$%
$J A_2 \partial{}^{3} T \partial{}^{2} T + 
    ((-20944\sqrt{2/3}Q)/27 - (143812\sqrt{2/3}Q{}^{3})/9 - 
      (224248\sqrt{2/3}Q{}^{5})/3)$%
$J A_2 \partial{}^{4} T \partial T + 
    ((-14896\sqrt{2/3}Q)/45 - (243764\sqrt{2/3}Q{}^{3})/45 - 
      (67880\sqrt{2/3}Q{}^{5})/3)$%
$J A_2 \partial{}^{5} T T + 
    ((-3874\sqrt{2/3}Q)/45 - (34874\sqrt{2/3}Q{}^{3})/15 - 12064\sqrt{2/3}Q{}^{5})
     $%
$J A_2{}^{2}\partial{}^{5} T + 
    (-40\sqrt{2/3}Q{}^{2} - 656\sqrt{2/3}Q{}^{4})$%
$J A_2{}^{2}\partial{}^{4} A_{3} + ((-991\sqrt{2/3})/27 - 
      (1568\sqrt{2/3}Q{}^{2})/3 - (5672\sqrt{2/3}Q{}^{4})/3)
     $%
$J A_2 A_3 \partial{}^{4} T + 
    ((-2044\sqrt{2/3})/27 - (5860\sqrt{2/3}Q{}^{2})/3 - (27080\sqrt{2/3}Q{}^{4})/3)
     $%
$J A_2 \partial A_{3} \partial{}^{3} T + 
    ((-566\sqrt{2/3})/9 - 3500\sqrt{2/3}Q{}^{2} - 22096\sqrt{2/3}Q{}^{4})
     $%
$J A_2 \partial{}^{2} A_{3} \partial{}^{2} T + 
    (392\sqrt{6}Q + 3120\sqrt{6}Q{}^{3})$%
$J A_2 \partial{}^{2} A_{3} 
      \partial A_{3} + ((-3964\sqrt{2/3})/27 - 1064\sqrt{6}Q{}^{2} - 
      (51272\sqrt{2/3}Q{}^{4})/3)$%
$J A_2 \partial{}^{3} A_{3} 
      \partial T + (296\sqrt{2/3}Q + 976\sqrt{6}Q{}^{3})
     $%
$J A_2 \partial{}^{3} A_{3} A_3 + 
    ((-1258\sqrt{2/3})/27 - (3890\sqrt{2/3}Q{}^{2})/3 - (17672\sqrt{2/3}Q{}^{4})/3)
     $%
$J A_2 \partial{}^{4} A_{3} T + 
    ((-352\sqrt{2/3}Q{}^{2})/3 - 488\sqrt{6}Q{}^{4})$%
$J A_2 
      B_2 \partial{}^{4} B_{3} + (-184\sqrt{6}Q{}^{2} - 992\sqrt{6}Q{}^{4})
     $%
$J A_2 \partial B_{2} \partial{}^{3} B_{3} + 
    (640\sqrt{2/3}Q{}^{2} + 1584\sqrt{6}Q{}^{4})$%
$J A_2 
      \partial{}^{2} B_{2} \partial{}^{2} B_{3} + ((-12896\sqrt{2/3}Q)/27 - 
      (31900\sqrt{2/3}Q{}^{3})/3 - (168016\sqrt{2/3}Q{}^{5})/3)
     $%
$J A_2 \partial{}^{3} B_{2} \partial{}^{2} B_{2} + 
    ((3824\sqrt{2/3}Q{}^{2})/3 + 9920\sqrt{2/3}Q{}^{4})$%
$J A_2 
      \partial{}^{3} B_{2} \partial B_{3} + ((-5434\sqrt{2/3}Q)/27 - 
      (13450\sqrt{2/3}Q{}^{3})/3 - (74192\sqrt{2/3}Q{}^{5})/3)
     $%
$J A_2 \partial{}^{4} B_{2} \partial B_{2} + 
    (472\sqrt{2/3}Q{}^{2} + 1464\sqrt{6}Q{}^{4})$%
$J A_2 
      \partial{}^{4} B_{2} B_3 + ((-1214\sqrt{2/3}Q)/27 - 290\sqrt{6}Q{}^{3} - 
      (13024\sqrt{2/3}Q{}^{5})/3)$%
$J A_2 \partial{}^{5} B_{2} 
      B_2 + (-608\sqrt{6}Q - 4600\sqrt{6}Q{}^{3})
     $%
$J A_2 \partial{}^{2} B_{3} \partial B_{3} + 
    (-584\sqrt{2/3}Q - 1480\sqrt{6}Q{}^{3})$%
$J A_2 
      \partial{}^{3} B_{3} B_3 + ((-2056\sqrt{2/3}Q)/3 - 6944\sqrt{6}Q{}^{3} - 
      38240\sqrt{6}Q{}^{5})$%
$J \partial A_{2} \partial{}^{2} T \partial{}^{2} T + 
    ((-37232\sqrt{2/3}Q)/27 - (315664\sqrt{2/3}Q{}^{3})/9 - 61120\sqrt{6}Q{}^{5})
     $%
$J \partial A_{2} \partial{}^{3} T \partial T + 
    ((-16240\sqrt{2/3}Q)/27 - (39748\sqrt{2/3}Q{}^{3})/3 - 
      (194312\sqrt{2/3}Q{}^{5})/3)$%
$J \partial A_{2} \partial{}^{4} T T + 
    ((-3119\sqrt{2/3}Q)/9 - (104008\sqrt{2/3}Q{}^{3})/9 - 
      (187208\sqrt{2/3}Q{}^{5})/3)$%
$J \partial A_{2} A_2 
      \partial{}^{4} T + (-80\sqrt{6}Q{}^{2} - 1344\sqrt{6}Q{}^{4})
     $%
$J \partial A_{2} A_2 \partial{}^{3} A_{3} + 
    ((-8180\sqrt{2/3}Q)/27 - (112984\sqrt{2/3}Q{}^{3})/9 - 71488\sqrt{2/3}Q{}^{5})
     $%
$J \partial A_{2} \partial A_{2} \partial{}^{3} T + 
    (-464\sqrt{2/3}Q{}^{2} - 2240\sqrt{6}Q{}^{4})$%
$J \partial A_{2} 
      \partial A_{2} \partial{}^{2} A_{3} + ((-1588\sqrt{2/3})/9 - 356\sqrt{6}Q{}^{2} + 
      112\sqrt{6}Q{}^{4})$%
$J \partial A_{2} A_3 \partial{}^{3} T + 
    ((-572\sqrt{2/3})/3 - (6460\sqrt{2/3}Q{}^{2})/3 - 3224\sqrt{2/3}Q{}^{4})
     $%
$J \partial A_{2} \partial A_{3} \partial{}^{2} T + 
    ((2144\sqrt{2/3}Q)/3 + 4432\sqrt{2/3}Q{}^{3})$%
$J \partial A_{2} 
      \partial A_{3} \partial A_{3} + ((-572\sqrt{2/3})/3 - 4492\sqrt{2/3}Q{}^{2} - 
      6840\sqrt{6}Q{}^{4})$%
$J \partial A_{2} \partial{}^{2} A_{3} \partial T + 
    ((1712\sqrt{2/3}Q)/3 + 4432\sqrt{2/3}Q{}^{3})$%
$J \partial A_{2} 
      \partial{}^{2} A_{3} A_3 + (-44\sqrt{2/3} - (4916\sqrt{2/3}Q{}^{2})/3 - 
      8224\sqrt{2/3}Q{}^{4})$%
$J \partial A_{2} \partial{}^{3} A_{3} T + 
    ((-3280\sqrt{2/3}Q{}^{2})/3 - 5288\sqrt{2/3}Q{}^{4})$%
$J \partial A_{2} 
      B_2 \partial{}^{3} B_{3} + (-1600\sqrt{2/3}Q{}^{2} - 2216\sqrt{6}Q{}^{4})
     $%
$J \partial A_{2} \partial B_{2} \partial{}^{2} B_{3} + 
    ((-3424\sqrt{2/3}Q)/9 - (22276\sqrt{2/3}Q{}^{3})/3 - 36736\sqrt{2/3}Q{}^{5})
     $%
$J \partial A_{2} \partial{}^{2} B_{2} \partial{}^{2} B_{2} + 
    (736\sqrt{2/3}Q{}^{2} + 2712\sqrt{6}Q{}^{4})$%
$J \partial A_{2} 
      \partial{}^{2} B_{2} \partial B_{3} + ((-9488\sqrt{2/3}Q)/27 - 
      (69944\sqrt{2/3}Q{}^{3})/9 - (125648\sqrt{2/3}Q{}^{5})/3)
     $%
$J \partial A_{2} \partial{}^{3} B_{2} \partial B_{2} + 
    ((2528\sqrt{2/3}Q{}^{2})/3 + 8536\sqrt{2/3}Q{}^{4})$%
$J \partial A_{2} 
      \partial{}^{3} B_{2} B_3 + ((-728\sqrt{2/3}Q)/27 - 
      (7076\sqrt{2/3}Q{}^{3})/9 - (16592\sqrt{2/3}Q{}^{5})/3)
     $%
$J \partial A_{2} \partial{}^{4} B_{2} B_2 + 
    ((-2896\sqrt{2/3}Q)/3 - 7784\sqrt{2/3}Q{}^{3})$%
$J \partial A_{2} 
      \partial B_{3} \partial B_{3} + ((-2896\sqrt{2/3}Q)/3 - 7496\sqrt{2/3}Q{}^{3})
     $%
$J \partial A_{2} \partial{}^{2} B_{3} B_3 + 
    ((-6320\sqrt{2/3}Q)/9 - (90476\sqrt{2/3}Q{}^{3})/3 - 183248\sqrt{2/3}Q{}^{5})
     $%
$J \partial{}^{2} A_{2} \partial{}^{2} T \partial T + 
    ((-10220\sqrt{2/3}Q)/27 - (48404\sqrt{2/3}Q{}^{3})/3 - 
      (299248\sqrt{2/3}Q{}^{5})/3)$%
$J \partial{}^{2} A_{2} \partial{}^{3} T T + 
    ((-12856\sqrt{2/3}Q)/27 - (147244\sqrt{2/3}Q{}^{3})/9 - 
      (280816\sqrt{2/3}Q{}^{5})/3)$%
$J \partial{}^{2} A_{2} A_2 
      \partial{}^{3} T + (-264\sqrt{6}Q{}^{2} - 2888\sqrt{6}Q{}^{4})
     $%
$J \partial{}^{2} A_{2} A_2 \partial{}^{2} A_{3} + 
    ((-4406\sqrt{2/3}Q)/9 - (63824\sqrt{2/3}Q{}^{3})/3 - 117920\sqrt{2/3}Q{}^{5})
     $%
$J \partial{}^{2} A_{2} \partial A_{2} \partial{}^{2} T + 
    ((-640\sqrt{2/3}Q{}^{2})/3 - 2216\sqrt{2/3}Q{}^{4})$%
$J \partial{}^{2} A_{2} 
      \partial A_{2} \partial A_{3} + ((-1114\sqrt{2/3}Q)/9 - 
      (13232\sqrt{2/3}Q{}^{3})/3 - 8272\sqrt{6}Q{}^{5})$%
$J \partial{}^{2} A_{2} 
      \partial{}^{2} A_{2} \partial T + (772\sqrt{2/3}Q + (51580\sqrt{2/3}Q{}^{3})/3 + 
      80960\sqrt{2/3}Q{}^{5})$%
$J \partial{}^{2} A_{2} \partial{}^{2} A_{2} 
      \partial A_{2} + ((-976\sqrt{2/3}Q{}^{2})/3 - 1760\sqrt{2/3}Q{}^{4})
     $%
$J \partial{}^{2} A_{2} \partial{}^{2} A_{2} A_3 + 
    ((-1822\sqrt{2/3})/9 - (4180\sqrt{2/3}Q{}^{2})/3 - 688\sqrt{2/3}Q{}^{4})
     $%
$J \partial{}^{2} A_{2} A_3 \partial{}^{2} T + 
    ((-812\sqrt{2/3})/9 - (4472\sqrt{2/3}Q{}^{2})/3 - 2120\sqrt{2/3}Q{}^{4})
     $%
$J \partial{}^{2} A_{2} \partial A_{3} \partial T + 
    ((1952\sqrt{2/3}Q)/3 + 3232\sqrt{2/3}Q{}^{3})$%
$J \partial{}^{2} A_{2} 
      \partial A_{3} A_3 + ((-382\sqrt{2/3})/9 - 
      (5156\sqrt{2/3}Q{}^{2})/3 - 3992\sqrt{6}Q{}^{4})$%
$J \partial{}^{2} A_{2} 
      \partial{}^{2} A_{3} T + ((-3520\sqrt{2/3}Q{}^{2})/3 - 7040\sqrt{2/3}Q{}^{4})
     $%
$J \partial{}^{2} A_{2} B_2 \partial{}^{2} B_{3} + 
    ((-2576\sqrt{2/3}Q{}^{2})/3 - 5440\sqrt{2/3}Q{}^{4})$%
$J \partial{}^{2} A_{2} 
      \partial B_{2} \partial B_{3} + ((-2068\sqrt{2/3}Q)/9 - 
      (24856\sqrt{2/3}Q{}^{3})/3 - 42400\sqrt{2/3}Q{}^{5})
     $%
$J \partial{}^{2} A_{2} \partial{}^{2} B_{2} \partial B_{2} + 
    ((992\sqrt{2/3}Q{}^{2})/3 + 4384\sqrt{2/3}Q{}^{4})$%
$J \partial{}^{2} A_{2} 
      \partial{}^{2} B_{2} B_3 + ((2836\sqrt{2/3}Q)/27 - 
      (5864\sqrt{2/3}Q{}^{3})/9 - (18080\sqrt{2/3}Q{}^{5})/3)
     $%
$J \partial{}^{2} A_{2} \partial{}^{3} B_{2} B_2 + 
    ((-1744\sqrt{2/3}Q)/3 - 4544\sqrt{2/3}Q{}^{3})$%
$J \partial{}^{2} A_{2} 
      \partial B_{3} B_3 + ((-7208\sqrt{2/3}Q)/27 - 
      (67240\sqrt{2/3}Q{}^{3})/9 - 12472\sqrt{6}Q{}^{5})$%
$J \partial{}^{3} A_{2} 
      \partial T \partial T + ((-3908\sqrt{2/3}Q)/27 - (74680\sqrt{2/3}Q{}^{3})/9 - 
      52712\sqrt{2/3}Q{}^{5})$%
$J \partial{}^{3} A_{2} \partial{}^{2} T T + 
    ((-11470\sqrt{2/3}Q)/27 - (90538\sqrt{2/3}Q{}^{3})/9 - 
      (147880\sqrt{2/3}Q{}^{5})/3)$%
$J \partial{}^{3} A_{2} A_2 
      \partial{}^{2} T + ((-1136\sqrt{2/3}Q{}^{2})/3 - 1104\sqrt{6}Q{}^{4})
     $%
$J \partial{}^{3} A_{2} A_2 \partial A_{3} + 
    ((-1744\sqrt{2/3}Q)/27 - (26608\sqrt{2/3}Q{}^{3})/9 - 
      (42760\sqrt{2/3}Q{}^{5})/3)$%
$J \partial{}^{3} A_{2} \partial A_{2} 
      \partial T + ((18484\sqrt{2/3}Q)/27 + (42832\sqrt{2/3}Q{}^{3})/3 + 
      (202640\sqrt{2/3}Q{}^{5})/3)$%
$J \partial{}^{3} A_{2} \partial A_{2} 
      \partial A_{2} + ((-976\sqrt{2/3}Q{}^{2})/3 - 2408\sqrt{2/3}Q{}^{4})
     $%
$J \partial{}^{3} A_{2} \partial A_{2} A_3 + 
    ((7850\sqrt{2/3}Q)/27 + (59666\sqrt{2/3}Q{}^{3})/9 + 
      (105632\sqrt{2/3}Q{}^{5})/3)$%
$J \partial{}^{3} A_{2} \partial{}^{2} A_{2} 
      T + ((7196\sqrt{2/3}Q)/9 + (153992\sqrt{2/3}Q{}^{3})/9 + 
      (250480\sqrt{2/3}Q{}^{5})/3)$%
$J \partial{}^{3} A_{2} \partial{}^{2} A_{2} 
      A_2 + (-236\sqrt{2/3} - (11924\sqrt{2/3}Q{}^{2})/3 - 
      4184\sqrt{6}Q{}^{4})$%
$J \partial{}^{3} A_{2} A_3 \partial T + 
    (16\sqrt{6}Q + 64\sqrt{6}Q{}^{3})$%
$J \partial{}^{3} A_{2} A_3 
      A_3 + ((-956\sqrt{2/3})/9 - 2876\sqrt{2/3}Q{}^{2} - 
      13552\sqrt{2/3}Q{}^{4})$%
$J \partial{}^{3} A_{2} \partial A_{3} T + 
    ((-1000\sqrt{2/3}Q{}^{2})/3 - 896\sqrt{6}Q{}^{4})$%
$J \partial{}^{3} A_{2} 
      B_2 \partial B_{3} + ((-124\sqrt{2/3}Q)/3 - 3880\sqrt{2/3}Q{}^{3} - 
      6848\sqrt{6}Q{}^{5})$%
$J \partial{}^{3} A_{2} \partial B_{2} \partial B_{2} + 
    (80\sqrt{6}Q{}^{2} + 576\sqrt{6}Q{}^{4})$%
$J \partial{}^{3} A_{2} \partial B_{2} 
      B_3 + ((1112\sqrt{2/3}Q)/9 - (5992\sqrt{2/3}Q{}^{3})/3 - 
      4480\sqrt{6}Q{}^{5})$%
$J \partial{}^{3} A_{2} \partial{}^{2} B_{2} B_2 + 
    (-64\sqrt{2/3}Q - 160\sqrt{6}Q{}^{3})$%
$J \partial{}^{3} A_{2} B_3 
      B_3 + ((2560\sqrt{2/3}Q)/27 - (6784\sqrt{2/3}Q{}^{3})/9 - 
      4072\sqrt{6}Q{}^{5})$%
$J \partial{}^{4} A_{2} \partial T T + 
    ((-730\sqrt{2/3}Q)/9 - (23356\sqrt{2/3}Q{}^{3})/9 - 
      (42968\sqrt{2/3}Q{}^{5})/3)$%
$J \partial{}^{4} A_{2} A_2 
      \partial T + (-64\sqrt{6}Q{}^{2} - 488\sqrt{6}Q{}^{4})
     $%
$J \partial{}^{4} A_{2} A_2 A_3 + 
    ((7615\sqrt{2/3}Q)/27 + (45086\sqrt{2/3}Q{}^{3})/9 + 24104\sqrt{2/3}Q{}^{5})
     $%
$J \partial{}^{4} A_{2} \partial A_{2} T + 
    ((13220\sqrt{2/3}Q)/27 + (91874\sqrt{2/3}Q{}^{3})/9 + 
      (151088\sqrt{2/3}Q{}^{5})/3)$%
$J \partial{}^{4} A_{2} \partial A_{2} 
      A_2 + ((-2623\sqrt{2/3})/27 - 3140\sqrt{2/3}Q{}^{2} - 
      (53168\sqrt{2/3}Q{}^{4})/3)$%
$J \partial{}^{4} A_{2} A_3 
      T + 32\sqrt{6}Q{}^{4}$%
$J \partial{}^{4} A_{2} B_2 
      B_3 + ((3022\sqrt{2/3}Q)/27 - (5336\sqrt{2/3}Q{}^{3})/3 - 
      (48592\sqrt{2/3}Q{}^{5})/3)$%
$J \partial{}^{4} A_{2} \partial B_{2} 
      B_2 + ((5138\sqrt{2/3}Q)/135 + (12682\sqrt{2/3}Q{}^{3})/15 + 
      (7664\sqrt{2/3}Q{}^{5})/3)$%
$J \partial{}^{5} A_{2} T{}^{2}+ 
    ((16583\sqrt{2/3}Q)/135 + (71072\sqrt{2/3}Q{}^{3})/45 + 
      (18280\sqrt{2/3}Q{}^{5})/3)$%
$J \partial{}^{5} A_{2} A_2 
      T + ((2911\sqrt{2/3}Q)/45 + (11572\sqrt{2/3}Q{}^{3})/9 + 
      (19040\sqrt{2/3}Q{}^{5})/3)$%
$J \partial{}^{5} A_{2} A_2{}^{2}+ ((2029\sqrt{2/3}Q)/45 + (1336\sqrt{2/3}Q{}^{3})/5 - 
      32\sqrt{2/3}Q{}^{5})$%
$J \partial{}^{5} A_{2} B_2{}^{2}+ 
    ((-1948\sqrt{2/3})/9 - 2696\sqrt{6}Q{}^{2} - 38240\sqrt{2/3}Q{}^{4})
     $%
$J A_3 \partial{}^{2} T \partial{}^{2} T + 
    (-272\sqrt{6} - 15008\sqrt{2/3}Q{}^{2} - 56176\sqrt{2/3}Q{}^{4})
     $%
$J A_3 \partial{}^{3} T \partial T + 
    ((-16676\sqrt{2/3})/27 - 8368\sqrt{2/3}Q{}^{2} - (86704\sqrt{2/3}Q{}^{4})/3)
     $%
$J A_3 \partial{}^{4} T T + 
    (-56\sqrt{6}Q - 288\sqrt{6}Q{}^{3})$%
$J A_3 A_3 
      \partial{}^{3} T + (688\sqrt{2/3}Q + 1024\sqrt{6}Q{}^{3})
     $%
$J A_3 B_2 \partial{}^{3} B_{3} + 
    (992\sqrt{2/3}Q + 1760\sqrt{6}Q{}^{3})$%
$J A_3 
      \partial B_{2} \partial{}^{2} B_{3} + (-656\sqrt{2/3}Q{}^{2} - 1664\sqrt{6}Q{}^{4})
     $%
$J A_3 \partial{}^{2} B_{2} \partial{}^{2} B_{2} + 
    (256\sqrt{2/3}Q + 832\sqrt{6}Q{}^{3})$%
$J A_3 \partial{}^{2} B_{2} 
      \partial B_{3} + (-1696\sqrt{2/3}Q{}^{2} - 3328\sqrt{6}Q{}^{4})
     $%
$J A_3 \partial{}^{3} B_{2} \partial B_{2} + 
    (32\sqrt{2/3}Q - 32\sqrt{6}Q{}^{3})$%
$J A_3 \partial{}^{3} B_{2} 
      B_3 + (-1000\sqrt{2/3}Q{}^{2} - 1920\sqrt{6}Q{}^{4})
     $%
$J A_3 \partial{}^{4} B_{2} B_2 + 
    ((-11528\sqrt{2/3})/9 - (100580\sqrt{2/3}Q{}^{2})/3 - 159800\sqrt{2/3}Q{}^{4})
     $%
$J \partial A_{3} \partial{}^{2} T \partial T + 
    ((-32752\sqrt{2/3})/27 - (56464\sqrt{2/3}Q{}^{2})/3 - 
      (194312\sqrt{2/3}Q{}^{4})/3)$%
$J \partial A_{3} \partial{}^{3} T T + 
    ((1232\sqrt{2/3}Q)/3 + 1504\sqrt{2/3}Q{}^{3})$%
$J \partial A_{3} 
      A_3 \partial{}^{2} T + ((128\sqrt{2/3}Q)/3 + 1840\sqrt{2/3}Q{}^{3})
     $%
$J \partial A_{3} \partial A_{3} \partial T + 
    192\sqrt{6}Q{}^{2}$%
$J \partial A_{3} \partial A_{3} A_3 + 
    (800\sqrt{2/3}Q + 1664\sqrt{6}Q{}^{3})$%
$J \partial A_{3} 
      B_2 \partial{}^{2} B_{3} + (512\sqrt{2/3}Q + 1472\sqrt{6}Q{}^{3})
     $%
$J \partial A_{3} \partial B_{2} \partial B_{3} + 
    (-1360\sqrt{2/3}Q{}^{2} - 2752\sqrt{6}Q{}^{4})$%
$J \partial A_{3} 
      \partial{}^{2} B_{2} \partial B_{2} - 192\sqrt{6}Q{}^{3}$%
$J \partial A_{3} 
      \partial{}^{2} B_{2} B_3 + (-528\sqrt{6}Q{}^{2} - 3328\sqrt{6}Q{}^{4})
     $%
$J \partial A_{3} \partial{}^{3} B_{2} B_2 - 
    96\sqrt{6}Q{}^{2}$%
$J \partial A_{3} \partial B_{3} B_3 + 
    ((-1600\sqrt{2/3})/3 - 16492\sqrt{2/3}Q{}^{2} - 30328\sqrt{6}Q{}^{4})
     $%
$J \partial{}^{2} A_{3} \partial T \partial T + 
    (-320\sqrt{6} - (66368\sqrt{2/3}Q{}^{2})/3 - 103456\sqrt{2/3}Q{}^{4})
     $%
$J \partial{}^{2} A_{3} \partial{}^{2} T T + 
    ((3800\sqrt{2/3}Q)/3 + 4720\sqrt{2/3}Q{}^{3})$%
$J \partial{}^{2} A_{3} 
      A_3 \partial T + (312\sqrt{6}Q + 3120\sqrt{6}Q{}^{3})
     $%
$J \partial{}^{2} A_{3} \partial A_{3} T + 
    (304\sqrt{2/3}Q + 832\sqrt{6}Q{}^{3})$%
$J \partial{}^{2} A_{3} B_2 
      \partial B_{3} + (128\sqrt{6}Q{}^{2} + 960\sqrt{6}Q{}^{4})
     $%
$J \partial{}^{2} A_{3} \partial B_{2} \partial B_{2} + 
    (-32\sqrt{6}Q - 480\sqrt{6}Q{}^{3})$%
$J \partial{}^{2} A_{3} \partial B_{2} 
      B_3 + (-560\sqrt{2/3}Q{}^{2} - 1280\sqrt{6}Q{}^{4})
     $%
$J \partial{}^{2} A_{3} \partial{}^{2} B_{2} B_2 + 
    96\sqrt{6}Q{}^{2}$%
$J \partial{}^{2} A_{3} B_3 B_3 + 
    ((-11848\sqrt{2/3})/27 - (41780\sqrt{2/3}Q{}^{2})/3 - 
      (222248\sqrt{2/3}Q{}^{4})/3)$%
$J \partial{}^{3} A_{3} \partial T T + 
    (120\sqrt{6}Q + 1200\sqrt{6}Q{}^{3})$%
$J \partial{}^{3} A_{3} A_3 
      T + (32\sqrt{2/3}Q - 32\sqrt{6}Q{}^{3})$%
$J \partial{}^{3} A_{3} 
      B_2 B_3 + (32\sqrt{6}Q{}^{2} + 448\sqrt{6}Q{}^{4})
     $%
$J \partial{}^{3} A_{3} \partial B_{2} B_2 + 
    ((-1168\sqrt{2/3})/27 - (5984\sqrt{2/3}Q{}^{2})/3 - (35504\sqrt{2/3}Q{}^{4})/3)
     $%
$J \partial{}^{4} A_{3} T{}^{2}+ 
    (-8\sqrt{6}Q{}^{2} + 16\sqrt{6}Q{}^{4})$%
$J \partial{}^{4} A_{3} B_2{}^{2}+ ((-182\sqrt{2/3}Q)/9 - (942\sqrt{6}Q{}^{3})/5 - 
      952\sqrt{6}Q{}^{5})$%
$J B_2{}^{2}\partial{}^{5} T + 
    ((-1061\sqrt{2/3})/9 - (5056\sqrt{2/3}Q{}^{2})/3 - 2200\sqrt{6}Q{}^{4})
     $%
$J B_2 B_3 \partial{}^{4} T + 
    ((-2596\sqrt{2/3})/9 - (10852\sqrt{2/3}Q{}^{2})/3 - 4736\sqrt{6}Q{}^{4})
     $%
$J B_2 \partial B_{3} \partial{}^{3} T + 
    ((-874\sqrt{2/3})/3 - (9796\sqrt{2/3}Q{}^{2})/3 - 14384\sqrt{2/3}Q{}^{4})
     $%
$J B_2 \partial{}^{2} B_{3} \partial{}^{2} T + 
    ((-148\sqrt{2/3})/3 - (4600\sqrt{2/3}Q{}^{2})/3 - 6056\sqrt{2/3}Q{}^{4})
     $%
$J B_2 \partial{}^{3} B_{3} \partial T + 
    ((331\sqrt{2/3})/3 - (838\sqrt{2/3}Q{}^{2})/3 - 784\sqrt{6}Q{}^{4})
     $%
$J B_2 \partial{}^{4} B_{3} T + 
    ((-6293\sqrt{2/3}Q)/27 - (19592\sqrt{2/3}Q{}^{3})/3 - 
      (106888\sqrt{2/3}Q{}^{5})/3)$%
$J \partial B_{2} B_2 
      \partial{}^{4} T + ((-1516\sqrt{2/3}Q)/9 - 7288\sqrt{2/3}Q{}^{3} - 
      43712\sqrt{2/3}Q{}^{5})$%
$J \partial B_{2} \partial B_{2} \partial{}^{3} T + 
    ((-2596\sqrt{2/3})/9 - (18452\sqrt{2/3}Q{}^{2})/3 - 9792\sqrt{6}Q{}^{4})
     $%
$J \partial B_{2} B_3 \partial{}^{3} T + 
    ((-1028\sqrt{2/3})/3 - (29144\sqrt{2/3}Q{}^{2})/3 - 57184\sqrt{2/3}Q{}^{4})
     $%
$J \partial B_{2} \partial B_{3} \partial{}^{2} T + 
    (-148\sqrt{2/3} - 2440\sqrt{6}Q{}^{2} - 14824\sqrt{6}Q{}^{4})
     $%
$J \partial B_{2} \partial{}^{2} B_{3} \partial T + 
    (20\sqrt{6} - 2768\sqrt{2/3}Q{}^{2} - 5536\sqrt{6}Q{}^{4})
     $%
$J \partial B_{2} \partial{}^{3} B_{3} T + 
    ((-5692\sqrt{2/3}Q)/9 - (42944\sqrt{2/3}Q{}^{3})/3 - 69152\sqrt{2/3}Q{}^{5})
     $%
$J \partial{}^{2} B_{2} B_2 \partial{}^{3} T + 
    ((-3526\sqrt{2/3}Q)/9 - (68452\sqrt{2/3}Q{}^{3})/3 - 149392\sqrt{2/3}Q{}^{5})
     $%
$J \partial{}^{2} B_{2} \partial B_{2} \partial{}^{2} T + 
    ((-274\sqrt{2/3}Q)/9 - (31816\sqrt{2/3}Q{}^{3})/3 - 79216\sqrt{2/3}Q{}^{5})
     $%
$J \partial{}^{2} B_{2} \partial{}^{2} B_{2} \partial T + 
    (-122\sqrt{6} - (31732\sqrt{2/3}Q{}^{2})/3 - 56240\sqrt{2/3}Q{}^{4})
     $%
$J \partial{}^{2} B_{2} B_3 \partial{}^{2} T + 
    ((-1972\sqrt{2/3})/3 - 17336\sqrt{2/3}Q{}^{2} - 30440\sqrt{6}Q{}^{4})
     $%
$J \partial{}^{2} B_{2} \partial B_{3} \partial T + 
    ((82\sqrt{2/3})/3 - 2108\sqrt{6}Q{}^{2} - 14336\sqrt{6}Q{}^{4})
     $%
$J \partial{}^{2} B_{2} \partial{}^{2} B_{3} T + 
    ((-16838\sqrt{2/3}Q)/27 - (146228\sqrt{2/3}Q{}^{3})/9 - 
      (252368\sqrt{2/3}Q{}^{5})/3)$%
$J \partial{}^{3} B_{2} B_2 
      \partial{}^{2} T + ((-14888\sqrt{2/3}Q)/27 - (183272\sqrt{2/3}Q{}^{3})/9 - 
      (387152\sqrt{2/3}Q{}^{5})/3)$%
$J \partial{}^{3} B_{2} \partial B_{2} 
      \partial T + ((-5822\sqrt{2/3}Q)/27 - (32464\sqrt{2/3}Q{}^{3})/3 - 
      (212800\sqrt{2/3}Q{}^{5})/3)$%
$J \partial{}^{3} B_{2} \partial{}^{2} B_{2} 
      T + ((-1316\sqrt{2/3})/3 - (29080\sqrt{2/3}Q{}^{2})/3 - 
      50984\sqrt{2/3}Q{}^{4})$%
$J \partial{}^{3} B_{2} B_3 \partial T + 
    ((-1036\sqrt{2/3})/3 - (23920\sqrt{2/3}Q{}^{2})/3 - 46432\sqrt{2/3}Q{}^{4})
     $%
$J \partial{}^{3} B_{2} \partial B_{3} T + 
    ((-11582\sqrt{2/3}Q)/27 - (91856\sqrt{2/3}Q{}^{3})/9 - 
      (160736\sqrt{2/3}Q{}^{5})/3)$%
$J \partial{}^{4} B_{2} B_2 
      \partial T + ((-4273\sqrt{2/3}Q)/27 - (19432\sqrt{2/3}Q{}^{3})/3 - 
      (128120\sqrt{2/3}Q{}^{5})/3)$%
$J \partial{}^{4} B_{2} \partial B_{2} 
      T + ((-703\sqrt{2/3})/3 - 1454\sqrt{6}Q{}^{2} - 6640\sqrt{6}Q{}^{4})
     $%
$J \partial{}^{4} B_{2} B_3 T + 
    ((-18994\sqrt{2/3}Q)/135 - 2408\sqrt{2/3}Q{}^{3} - (38728\sqrt{2/3}Q{}^{5})/3)
     $%
$J \partial{}^{5} B_{2} B_2 T + 
    (56\sqrt{2/3}Q + 480\sqrt{6}Q{}^{3})$%
$J B_3 B_3 
      \partial{}^{3} T + ((-592\sqrt{2/3}Q)/3 + 7840\sqrt{2/3}Q{}^{3})
     $%
$J \partial B_{3} B_3 \partial{}^{2} T + 
    ((4304\sqrt{2/3}Q)/3 + 13528\sqrt{2/3}Q{}^{3})$%
$J \partial B_{3} 
      \partial B_{3} \partial T + ((848\sqrt{2/3}Q)/3 + 11512\sqrt{2/3}Q{}^{3})
     $%
$J \partial{}^{2} B_{3} B_3 \partial T + 
    (216\sqrt{6}Q + 4520\sqrt{6}Q{}^{3})$%
$J \partial{}^{2} B_{3} \partial B_{3} 
      T + (856\sqrt{2/3}Q + 1784\sqrt{6}Q{}^{3})
     $%
$J \partial{}^{3} B_{3} B_3 T + 
    ((-11128\sqrt{2/3}Q)/9 - 39568\sqrt{2/3}Q{}^{3} - 296768\sqrt{2/3}Q{}^{5})
     $%
$\partial J \partial{}^{2} T \partial T \partial T + 
    ((-9496\sqrt{2/3}Q)/9 - 32128\sqrt{2/3}Q{}^{3} - 233600\sqrt{2/3}Q{}^{5})
     $%
$\partial J \partial{}^{2} T \partial{}^{2} T T + 
    ((-50752\sqrt{2/3}Q)/27 - 54112\sqrt{2/3}Q{}^{3} - 
      (1136000\sqrt{2/3}Q{}^{5})/3)$%
$\partial J \partial{}^{3} T \partial T T + 
    ((-12364\sqrt{2/3}Q)/27 - (38648\sqrt{2/3}Q{}^{3})/3 - 
      (265952\sqrt{2/3}Q{}^{5})/3)$%
$\partial J \partial{}^{4} T T{}^{2}+ 
    (-2872/81 - 194632 Q{}^{2}/27 - 1387184 Q{}^{4}/9 - 2618432 Q{}^{6}/3)
     $%
$\partial J J \partial{}^{3} T \partial{}^{2} T + 
    (-18052/243 - 470192 Q{}^{2}/81 - 2956712 Q{}^{4}/27 - 5459104 Q{}^{6}/9)
     $%
$\partial J J \partial{}^{4} T \partial T + 
    (-68368/1215 - 1130432 Q{}^{2}/405 - 1232512 Q{}^{4}/27 - 2187776 Q{}^{6}/9)
     $%
$\partial J J \partial{}^{5} T T + 
    ((-37609\sqrt{2/3}Q)/405 - (44732\sqrt{2/3}Q{}^{3})/135 + 
      (1320376\sqrt{2/3}Q{}^{5})/45 + (718048\sqrt{2/3}Q{}^{7})/3)
     $%
$\partial J J{}^{2}\partial{}^{6} T + 
    ((6011\sqrt{2/3}Q)/135 + (11251\sqrt{2/3}Q{}^{3})/27 + 
      (667399\sqrt{2/3}Q{}^{5})/45 + 168932\sqrt{2/3}Q{}^{7})
     $%
$\partial J J{}^{2}\partial{}^{6} A_{2} + 
    ((-28\sqrt{2/3})/45 + (26311\sqrt{2/3}Q{}^{2})/15 + 
      (103766\sqrt{2/3}Q{}^{4})/3 + 166072\sqrt{2/3}Q{}^{6})
     $%
$\partial J J{}^{2}\partial{}^{5} A_{3} + 
    (-3608/243 - 1325612 Q{}^{2}/405 - 7826324 Q{}^{4}/135 - 2702264 Q{}^{6}/9)
     $%
$\partial J J A_2 \partial{}^{5} T + 
    ((-2486Q)/9 - 3094 Q{}^{3}/3 + 8652Q{}^{5})$%
$\partial J J 
      A_2 \partial{}^{4} A_{3} + (-10180/243 - 838052 Q{}^{2}/81 - 
      4895792 Q{}^{4}/27 - 8171488 Q{}^{6}/9)$%
$\partial J J 
      \partial A_{2} \partial{}^{4} T + ((-5384Q)/9 - 11168 Q{}^{3}/9 + 90560 Q{}^{5}/3)
     $%
$\partial J J \partial A_{2} \partial{}^{3} A_{3} + 
    (-2872/81 - 401444 Q{}^{2}/27 - 2635652 Q{}^{4}/9 - 4711744 Q{}^{6}/3)
     $%
$\partial J J \partial{}^{2} A_{2} \partial{}^{3} T + 
    ((-1148Q)/3 + 190Q{}^{3} + 42936Q{}^{5})$%
$\partial J J \partial{}^{2} A_{2} 
      \partial{}^{2} A_{3} + (-2872/81 - 370484 Q{}^{2}/27 - 2698052 Q{}^{4}/9 - 
      5182816 Q{}^{6}/3)$%
$\partial J J \partial{}^{3} A_{2} \partial{}^{2} T + 
    (-766/81 - 19099 Q{}^{2}/27 - 548174 Q{}^{4}/9 - 1897856 Q{}^{6}/3)
     $%
$\partial J J \partial{}^{3} A_{2} \partial{}^{2} A_{2} + 
    ((-11252Q)/27 - 55048 Q{}^{3}/9 - 16568Q{}^{5})$%
$\partial J J 
      \partial{}^{3} A_{2} \partial A_{3} + (-10180/243 - 623012 Q{}^{2}/81 - 
      4559252 Q{}^{4}/27 - 8971288 Q{}^{6}/9)$%
$\partial J J 
      \partial{}^{4} A_{2} \partial T + (-3706/243 - 77639 Q{}^{2}/81 - 1261055 Q{}^{4}/27 - 
      3716788 Q{}^{6}/9)$%
$\partial J J \partial{}^{4} A_{2} \partial A_{2} + 
    ((-1214Q)/9 - 10496 Q{}^{3}/9 - 18532 Q{}^{5}/3)$%
$\partial J J 
      \partial{}^{4} A_{2} A_3 + (-3608/243 - 813986 Q{}^{2}/405 - 
      5950388 Q{}^{4}/135 - 2381984 Q{}^{6}/9)$%
$\partial J J 
      \partial{}^{5} A_{2} T + (-7537/1215 - 328051 Q{}^{2}/810 - 
      2474846 Q{}^{4}/135 - 1374572 Q{}^{6}/9)$%
$\partial J J 
      \partial{}^{5} A_{2} A_2 + (43208 Q/27 + 22644Q{}^{3} + 256360 Q{}^{5}/3)
     $%
$\partial J J A_3 \partial{}^{4} T + 
    (101344 Q/27 + 70552Q{}^{3} + 1028888 Q{}^{5}/3)$%
$\partial J J 
      \partial A_{3} \partial{}^{3} T + (30080 Q/9 + 96528Q{}^{3} + 575512Q{}^{5})
     $%
$\partial J J \partial{}^{2} A_{3} \partial{}^{2} T + 
    (1996/9 + 310 Q{}^{2}/3 + 544Q{}^{4})$%
$\partial J J \partial{}^{2} A_{3} 
      \partial A_{3} + (13256 Q/9 + 180316 Q{}^{3}/3 + 380536Q{}^{5})
     $%
$\partial J J \partial{}^{3} A_{3} \partial T + 
    (204 + 2014 Q{}^{2}/3 + 536Q{}^{4})$%
$\partial J J \partial{}^{3} A_{3} 
      A_3 + (4654 Q/27 + 120064 Q{}^{3}/9 + 279808 Q{}^{5}/3)
     $%
$\partial J J \partial{}^{4} A_{3} T + 
    (568 Q/9 - 5192 Q{}^{3}/3 - 13132Q{}^{5})$%
$\partial J J 
      B_2 \partial{}^{4} B_{3} + ((-760Q)/9 - 17392 Q{}^{3}/3 - 31672Q{}^{5})
     $%
$\partial J J \partial B_{2} \partial{}^{3} B_{3} + 
    ((-11380Q)/9 - 9810Q{}^{3} - 3152Q{}^{5})$%
$\partial J J \partial{}^{2} B_{2} 
      \partial{}^{2} B_{3} + (-776/81 + 134438 Q{}^{2}/27 + 693952 Q{}^{4}/9 + 
      887408 Q{}^{6}/3)$%
$\partial J J \partial{}^{3} B_{2} \partial{}^{2} B_{2} + 
    ((-44020Q)/27 - 83614 Q{}^{3}/9 + 42464 Q{}^{5}/3)
     $%
$\partial J J \partial{}^{3} B_{2} \partial B_{3} + 
    (-2140/81 + 66548 Q{}^{2}/27 + 134951 Q{}^{4}/3 + 559076 Q{}^{6}/3)
     $%
$\partial J J \partial{}^{4} B_{2} \partial B_{2} + 
    ((-20920Q)/27 - 54100 Q{}^{3}/9 - 11380 Q{}^{5}/3)
     $%
$\partial J J \partial{}^{4} B_{2} B_3 + 
    (-5281/405 + 23803 Q{}^{2}/90 + 398233 Q{}^{4}/45 + 43076Q{}^{6})
     $%
$\partial J J \partial{}^{5} B_{2} B_2 + 
    (632/9 - 3948Q{}^{2} - 39968Q{}^{4})$%
$\partial J J \partial{}^{2} B_{3} 
      \partial B_{3} + (-416/9 - 844Q{}^{2} - 8200Q{}^{4})$%
$\partial J J 
      \partial{}^{3} B_{3} B_3 + (956 Q{}^{2}/9 + 29912Q{}^{4} + 343648Q{}^{6})
     $%
$(\partial J){}^{2} \partial{}^{2} T \partial{}^{2} T + 
    (-3232/81 - 23152 Q{}^{2}/27 + 274000 Q{}^{4}/9 + 1259584 Q{}^{6}/3)
     $%
$(\partial J){}^{2} \partial{}^{3} T \partial T + 
    (-11872/243 - 87140 Q{}^{2}/81 + 143368 Q{}^{4}/27 + 1207904 Q{}^{6}/9)
     $%
$(\partial J){}^{2} \partial{}^{4} T T + 
    ((-53914\sqrt{2/3}Q)/405 + (1198252\sqrt{2/3}Q{}^{3})/135 + 
      (12410992\sqrt{2/3}Q{}^{5})/45 + (5404288\sqrt{2/3}Q{}^{7})/3)
     $%
$(\partial J){}^{2} J \partial{}^{5} T + 
    ((20335\sqrt{2/3}Q)/81 + (172607\sqrt{2/3}Q{}^{3})/27 + 
      (7726099\sqrt{2/3}Q{}^{5})/45 + (4381984\sqrt{2/3}Q{}^{7})/3)
     $%
$(\partial J){}^{2} J \partial{}^{5} A_{2} + 
    ((104\sqrt{2/3})/9 + (69971\sqrt{2/3}Q{}^{2})/9 + (422792\sqrt{2/3}Q{}^{4})/3 + 
      617312\sqrt{2/3}Q{}^{6})$%
$(\partial J){}^{2} J \partial{}^{4} A_{3} + 
    ((-142\sqrt{2/3}Q)/9 + (112634\sqrt{2/3}Q{}^{3})/9 + 
      (890252\sqrt{2/3}Q{}^{5})/3 + 577200\sqrt{6}Q{}^{7})
     $%
$(\partial J){}^{3} \partial{}^{4} T + 
    ((1889\sqrt{2/3}Q)/9 + (10532\sqrt{2/3}Q{}^{3})/3 + 125602\sqrt{2/3}Q{}^{5} + 
      1260890\sqrt{2/3}Q{}^{7})$%
$(\partial J){}^{3} 
      \partial{}^{4} A_{2} + ((85858\sqrt{2/3}Q{}^{2})/9 + 157922\sqrt{2/3}Q{}^{4} + 
      624974\sqrt{2/3}Q{}^{6})$%
$(\partial J){}^{3} \partial{}^{3} A_{3} + 
    (-2056/243 - 152144 Q{}^{2}/81 - 293108 Q{}^{4}/27 + 538208 Q{}^{6}/9)
     $%
$(\partial J){}^{2} A_2 \partial{}^{4} T + 
    ((-8638Q)/27 + 2686 Q{}^{3}/9 + 22958Q{}^{5})$%
$(\partial J){}^{2} 
      A_2 \partial{}^{3} A_{3} + (-2872/81 - 52088 Q{}^{2}/9 - 398408 Q{}^{4}/9 + 
      66232Q{}^{6})$%
$(\partial J){}^{2} \partial A_{2} \partial{}^{3} T + 
    ((-5510Q)/9 - 3442 Q{}^{3}/3 + 35742Q{}^{5})$%
$(\partial J){}^{2} 
      \partial A_{2} \partial{}^{2} A_{3} + ((-63776Q{}^{2})/9 - 236840 Q{}^{4}/3 - 62360Q{}^{6})
     $%
$(\partial J){}^{2} \partial{}^{2} A_{2} \partial{}^{2} T + 
    (2019Q{}^{2} + 155782 Q{}^{4}/3 + 343034Q{}^{6})$%
$(\partial J){}^{2} 
      \partial{}^{2} A_{2} \partial{}^{2} A_{2} + ((-5246Q)/9 - 5510Q{}^{3} + 2390Q{}^{5})
     $%
$(\partial J){}^{2} \partial{}^{2} A_{2} \partial A_{3} + 
    (-2872/81 - 143984 Q{}^{2}/27 - 182956 Q{}^{4}/3 - 96596Q{}^{6})
     $%
$(\partial J){}^{2} \partial{}^{3} A_{2} \partial T + 
    (-536/81 + 85540 Q{}^{2}/27 + 239212 Q{}^{4}/3 + 1574368 Q{}^{6}/3)
     $%
$(\partial J){}^{2} \partial{}^{3} A_{2} \partial A_{2} + 
    ((-6928Q)/27 - 30872 Q{}^{3}/9 - 14242Q{}^{5})$%
$(\partial J){}^{2} 
      \partial{}^{3} A_{2} A_3 + (-2056/243 - 102122 Q{}^{2}/81 - 
      213212 Q{}^{4}/27 + 445604 Q{}^{6}/9)$%
$(\partial J){}^{2} \partial{}^{4} A_{2} 
      T + (-286/243 + 110563 Q{}^{2}/81 + 901294 Q{}^{4}/27 + 
      2034398 Q{}^{6}/9)$%
$(\partial J){}^{2} \partial{}^{4} A_{2} A_2 + 
    (76048 Q/27 + 40344Q{}^{3} + 447536 Q{}^{5}/3)$%
$(\partial J){}^{2} 
      A_3 \partial{}^{3} T + (50360 Q/9 + 322648 Q{}^{3}/3 + 493752Q{}^{5})
     $%
$(\partial J){}^{2} \partial A_{3} \partial{}^{2} T + 
    (1684/9 + 3248Q{}^{2} + 26012Q{}^{4})$%
$(\partial J){}^{2} \partial A_{3} 
      \partial A_{3} + (45368 Q/9 + 316468 Q{}^{3}/3 + 496620Q{}^{5})
     $%
$(\partial J){}^{2} \partial{}^{2} A_{3} \partial T + 
    (1736/9 + 7516 Q{}^{2}/3 + 22380Q{}^{4})$%
$(\partial J){}^{2} \partial{}^{2} A_{3} 
      A_3 + (37580 Q/27 + 35196Q{}^{3} + 519556 Q{}^{5}/3)
     $%
$(\partial J){}^{2} \partial{}^{3} A_{3} T + 
    ((-346Q)/3 - 3834Q{}^{3} - 29142Q{}^{5})$%
$(\partial J){}^{2} B_2 
      \partial{}^{3} B_{3} + ((-7966Q)/9 - 23842 Q{}^{3}/3 - 20614Q{}^{5})
     $%
$(\partial J){}^{2} \partial B_{2} \partial{}^{2} B_{3} + 
    (36725 Q{}^{2}/9 + 62658Q{}^{4} + 229570Q{}^{6})$%
$(\partial J){}^{2} 
      \partial{}^{2} B_{2} \partial{}^{2} B_{2} + ((-3274Q)/3 - 722 Q{}^{3}/3 + 61526Q{}^{5})
     $%
$(\partial J){}^{2} \partial{}^{2} B_{2} \partial B_{3} + 
    (-584/81 + 166784 Q{}^{2}/27 + 868072 Q{}^{4}/9 + 1102292 Q{}^{6}/3)
     $%
$(\partial J){}^{2} \partial{}^{3} B_{2} \partial B_{2} + 
    ((-21694Q)/27 - 18082 Q{}^{3}/9 + 97970 Q{}^{5}/3)
     $%
$(\partial J){}^{2} \partial{}^{3} B_{2} B_3 + 
    (-1466/81 + 12289 Q{}^{2}/9 + 244106 Q{}^{4}/9 + 118122Q{}^{6})
     $%
$(\partial J){}^{2} \partial{}^{4} B_{2} B_2 + 
    (-868/9 - 6112Q{}^{2} - 51812Q{}^{4})$%
$(\partial J){}^{2} \partial B_{3} 
      \partial B_{3} + (-20/9 - 5056Q{}^{2} - 46468Q{}^{4})$%
$(\partial J){}^{2} 
      \partial{}^{2} B_{3} B_3 + (-1820\sqrt{2/3}Q - 54400\sqrt{2/3}Q{}^{3} - 
      119656\sqrt{6}Q{}^{5})$%
$\partial J A_2 \partial{}^{2} T \partial{}^{2} T + 
    ((-105136\sqrt{2/3}Q)/27 - (848536\sqrt{2/3}Q{}^{3})/9 - 
      (1692232\sqrt{2/3}Q{}^{5})/3)$%
$\partial J A_2 \partial{}^{3} T \partial T + 
    ((-16820\sqrt{2/3}Q)/9 - 39764\sqrt{2/3}Q{}^{3} - 231112\sqrt{2/3}Q{}^{5})
     $%
$\partial J A_2 \partial{}^{4} T T + 
    (-128\sqrt{6}Q - (40616\sqrt{2/3}Q{}^{3})/3 - 95876\sqrt{2/3}Q{}^{5})
     $%
$\partial J A_2{}^{2}\partial{}^{4} T + 
    ((-937\sqrt{2/3}Q{}^{2})/3 - 594\sqrt{6}Q{}^{4})$%
$\partial J A_2{}^{2}\partial{}^{3} A_{3} + ((-2044\sqrt{2/3})/27 + 
      (2842\sqrt{2/3}Q{}^{2})/3 + (37486\sqrt{2/3}Q{}^{4})/3)
     $%
$\partial J A_2 A_3 \partial{}^{3} T + 
    ((-796\sqrt{2/3})/9 + (6062\sqrt{2/3}Q{}^{2})/3 + 28718\sqrt{2/3}Q{}^{4})
     $%
$\partial J A_2 \partial A_{3} \partial{}^{2} T + 
    (-626\sqrt{2/3}Q - 2240\sqrt{6}Q{}^{3})$%
$\partial J A_2 
      \partial A_{3} \partial A_{3} + ((-1132\sqrt{2/3})/9 + 
      (2246\sqrt{2/3}Q{}^{2})/3 + 21356\sqrt{2/3}Q{}^{4})
     $%
$\partial J A_2 \partial{}^{2} A_{3} \partial T + 
    (-374\sqrt{2/3}Q - 1896\sqrt{6}Q{}^{3})$%
$\partial J A_2 
      \partial{}^{2} A_{3} A_3 + ((-3964\sqrt{2/3})/27 - 
      (3196\sqrt{2/3}Q{}^{2})/3 + (13972\sqrt{2/3}Q{}^{4})/3)
     $%
$\partial J A_2 \partial{}^{3} A_{3} T + 
    (-164\sqrt{2/3}Q{}^{2} + 492\sqrt{6}Q{}^{4})$%
$\partial J A_2 
      B_2 \partial{}^{3} B_{3} + (-1624\sqrt{2/3}Q{}^{2} - 2822\sqrt{6}Q{}^{4})
     $%
$\partial J A_2 \partial B_{2} \partial{}^{2} B_{3} + 
    ((-3082\sqrt{2/3}Q)/9 - 1267\sqrt{6}Q{}^{3} - 6752\sqrt{2/3}Q{}^{5})
     $%
$\partial J A_2 \partial{}^{2} B_{2} \partial{}^{2} B_{2} + 
    (-602\sqrt{6}Q{}^{2} - 4696\sqrt{6}Q{}^{4})$%
$\partial J A_2 
      \partial{}^{2} B_{2} \partial B_{3} + ((-16525\sqrt{2/3}Q)/27 - 
      (18254\sqrt{2/3}Q{}^{3})/3 - (38072\sqrt{2/3}Q{}^{5})/3)
     $%
$\partial J A_2 \partial{}^{3} B_{2} \partial B_{2} + 
    (-1058\sqrt{2/3}Q{}^{2} - 2566\sqrt{6}Q{}^{4})$%
$\partial J A_2 
      \partial{}^{3} B_{2} B_3 + ((-6544\sqrt{2/3}Q)/27 - 
      2567\sqrt{2/3}Q{}^{3} - (22064\sqrt{2/3}Q{}^{5})/3)
     $%
$\partial J A_2 \partial{}^{4} B_{2} B_2 + 
    (242\sqrt{6}Q + 2516\sqrt{6}Q{}^{3})$%
$\partial J A_2 \partial B_{3} 
      \partial B_{3} + (416\sqrt{6}Q + 3084\sqrt{6}Q{}^{3})
     $%
$\partial J A_2 \partial{}^{2} B_{3} B_3 + 
    ((-60080\sqrt{2/3}Q)/9 - (562388\sqrt{2/3}Q{}^{3})/3 - 
      1181744\sqrt{2/3}Q{}^{5})$%
$\partial J \partial A_{2} \partial{}^{2} T \partial T + 
    ((-106568\sqrt{2/3}Q)/27 - (287704\sqrt{2/3}Q{}^{3})/3 - 
      (1736296\sqrt{2/3}Q{}^{5})/3)$%
$\partial J \partial A_{2} \partial{}^{3} T T + 
    ((-46028\sqrt{2/3}Q)/27 - (547498\sqrt{2/3}Q{}^{3})/9 - 
      434842\sqrt{2/3}Q{}^{5})$%
$\partial J \partial A_{2} A_2 \partial{}^{3} T + 
    ((-688\sqrt{2/3}Q{}^{2})/3 - 3518\sqrt{2/3}Q{}^{4})$%
$\partial J \partial A_{2} 
      A_2 \partial{}^{2} A_{3} + (-1154\sqrt{2/3}Q - 49460\sqrt{2/3}Q{}^{3} - 
      126040\sqrt{6}Q{}^{5})$%
$\partial J \partial A_{2} \partial A_{2} \partial{}^{2} T + 
    ((2062\sqrt{2/3}Q{}^{2})/3 + 3980\sqrt{2/3}Q{}^{4})$%
$\partial J \partial A_{2} 
      \partial A_{2} \partial A_{3} + (-76\sqrt{6} + (1178\sqrt{2/3}Q{}^{2})/3 + 
      23890\sqrt{2/3}Q{}^{4})$%
$\partial J \partial A_{2} A_3 \partial{}^{2} T + 
    ((-1144\sqrt{2/3})/3 - (6148\sqrt{2/3}Q{}^{2})/3 + 21292\sqrt{2/3}Q{}^{4})
     $%
$\partial J \partial A_{2} \partial A_{3} \partial T + 
    ((-4426\sqrt{2/3}Q)/3 - 14828\sqrt{2/3}Q{}^{3})$%
$\partial J \partial A_{2} 
      \partial A_{3} A_3 + ((-572\sqrt{2/3})/3 - 1928\sqrt{2/3}Q{}^{2} + 
      1340\sqrt{6}Q{}^{4})$%
$\partial J \partial A_{2} \partial{}^{2} A_{3} T + 
    ((1492\sqrt{2/3}Q{}^{2})/3 + 2804\sqrt{2/3}Q{}^{4})$%
$\partial J \partial A_{2} 
      B_2 \partial{}^{2} B_{3} + (-928\sqrt{6}Q{}^{2} - 6408\sqrt{6}Q{}^{4})
     $%
$\partial J \partial A_{2} \partial B_{2} \partial B_{3} + 
    ((-2111\sqrt{2/3}Q)/3 - 3076\sqrt{2/3}Q{}^{3} + 8794\sqrt{6}Q{}^{5})
     $%
$\partial J \partial A_{2} \partial{}^{2} B_{2} \partial B_{2} + 
    ((-6940\sqrt{2/3}Q{}^{2})/3 - 16268\sqrt{2/3}Q{}^{4})
     $%
$\partial J \partial A_{2} \partial{}^{2} B_{2} B_3 + 
    ((-13604\sqrt{2/3}Q)/27 - (31438\sqrt{2/3}Q{}^{3})/9 + 450\sqrt{6}Q{}^{5})
     $%
$\partial J \partial A_{2} \partial{}^{3} B_{2} B_2 + 
    (774\sqrt{6}Q + 7020\sqrt{6}Q{}^{3})$%
$\partial J \partial A_{2} \partial B_{3} 
      B_3 + ((-22640\sqrt{2/3}Q)/9 - 28040\sqrt{6}Q{}^{3} - 
      566824\sqrt{2/3}Q{}^{5})$%
$\partial J \partial{}^{2} A_{2} \partial T \partial T + 
    ((-27448\sqrt{2/3}Q)/9 - (331216\sqrt{2/3}Q{}^{3})/3 - 
      772744\sqrt{2/3}Q{}^{5})$%
$\partial J \partial{}^{2} A_{2} \partial{}^{2} T T + 
    ((-5062\sqrt{2/3}Q)/3 - (211420\sqrt{2/3}Q{}^{3})/3 - 546074\sqrt{2/3}Q{}^{5})
     $%
$\partial J \partial{}^{2} A_{2} A_2 \partial{}^{2} T + 
    ((1892\sqrt{2/3}Q{}^{2})/3 + 4090\sqrt{2/3}Q{}^{4})$%
$\partial J \partial{}^{2} A_{2} 
      A_2 \partial A_{3} + ((-16664\sqrt{2/3}Q)/9 - 27060\sqrt{6}Q{}^{3} - 
      652396\sqrt{2/3}Q{}^{5})$%
$\partial J \partial{}^{2} A_{2} \partial A_{2} \partial T + 
    ((-3140\sqrt{2/3}Q)/9 - (88201\sqrt{2/3}Q{}^{3})/3 - 97490\sqrt{6}Q{}^{5})
     $%
$\partial J \partial{}^{2} A_{2} \partial A_{2} \partial A_{2} + 
    (2306\sqrt{2/3}Q{}^{2} + 6128\sqrt{6}Q{}^{4})$%
$\partial J \partial{}^{2} A_{2} 
      \partial A_{2} A_3 + ((-776\sqrt{2/3}Q)/3 - 
      (59596\sqrt{2/3}Q{}^{3})/3 - 200180\sqrt{2/3}Q{}^{5})
     $%
$\partial J \partial{}^{2} A_{2} \partial{}^{2} A_{2} T + 
    (-220\sqrt{2/3}Q - (59335\sqrt{2/3}Q{}^{3})/3 - 200762\sqrt{2/3}Q{}^{5})
     $%
$\partial J \partial{}^{2} A_{2} \partial{}^{2} A_{2} A_2 + 
    ((-3644\sqrt{2/3})/9 - (12026\sqrt{2/3}Q{}^{2})/3 + 9304\sqrt{2/3}Q{}^{4})
     $%
$\partial J \partial{}^{2} A_{2} A_3 \partial T + 
    (-377\sqrt{6}Q - 2982\sqrt{6}Q{}^{3})$%
$\partial J \partial{}^{2} A_{2} A_3 
      A_3 + ((-812\sqrt{2/3})/9 - (7756\sqrt{2/3}Q{}^{2})/3 + 
      1172\sqrt{6}Q{}^{4})$%
$\partial J \partial{}^{2} A_{2} \partial A_{3} T + 
    ((896\sqrt{2/3}Q{}^{2})/3 + 4126\sqrt{2/3}Q{}^{4})$%
$\partial J \partial{}^{2} A_{2} 
      B_2 \partial B_{3} + ((-209\sqrt{2/3}Q)/9 + 
      (7073\sqrt{2/3}Q{}^{3})/3 + 10372\sqrt{6}Q{}^{5})$%
$\partial J \partial{}^{2} A_{2} 
      \partial B_{2} \partial B_{2} + (-904\sqrt{6}Q{}^{2} - 7884\sqrt{6}Q{}^{4})
     $%
$\partial J \partial{}^{2} A_{2} \partial B_{2} B_3 + 
    ((-3130\sqrt{2/3}Q)/9 - (1948\sqrt{2/3}Q{}^{3})/3 + 17414\sqrt{2/3}Q{}^{5})
     $%
$\partial J \partial{}^{2} A_{2} \partial{}^{2} B_{2} B_2 + 
    ((4559\sqrt{2/3}Q)/3 + 13156\sqrt{2/3}Q{}^{3})$%
$\partial J \partial{}^{2} A_{2} 
      B_3 B_3 + ((-61448\sqrt{2/3}Q)/27 - 
      (692344\sqrt{2/3}Q{}^{3})/9 - 177920\sqrt{6}Q{}^{5})
     $%
$\partial J \partial{}^{3} A_{2} \partial T T + 
    ((-40108\sqrt{2/3}Q)/27 - (468890\sqrt{2/3}Q{}^{3})/9 - 130448\sqrt{6}Q{}^{5})
     $%
$\partial J \partial{}^{3} A_{2} A_2 \partial T + 
    ((1268\sqrt{2/3}Q{}^{2})/3 + 1176\sqrt{6}Q{}^{4})$%
$\partial J \partial{}^{3} A_{2} 
      A_2 A_3 + ((-4040\sqrt{2/3}Q)/9 - 
      (285448\sqrt{2/3}Q{}^{3})/9 - (900800\sqrt{2/3}Q{}^{5})/3)
     $%
$\partial J \partial{}^{3} A_{2} \partial A_{2} T + 
    ((-16136\sqrt{2/3}Q)/27 - (294692\sqrt{2/3}Q{}^{3})/9 - 
      (897650\sqrt{2/3}Q{}^{5})/3)$%
$\partial J \partial{}^{3} A_{2} \partial A_{2} 
      A_2 + (-236\sqrt{2/3} - 2208\sqrt{6}Q{}^{2} - 28492\sqrt{2/3}Q{}^{4})
     $%
$\partial J \partial{}^{3} A_{2} A_3 T + 
    (44\sqrt{6}Q{}^{2} + 390\sqrt{6}Q{}^{4})$%
$\partial J \partial{}^{3} A_{2} B_2 
      B_3 + ((-1349\sqrt{2/3}Q)/27 + (428\sqrt{2/3}Q{}^{3})/3 + 
      (9980\sqrt{2/3}Q{}^{5})/3)$%
$\partial J \partial{}^{3} A_{2} \partial B_{2} 
      B_2 + ((-2554\sqrt{2/3}Q)/9 - (37318\sqrt{2/3}Q{}^{3})/3 - 
      32236\sqrt{6}Q{}^{5})$%
$\partial J \partial{}^{4} A_{2} T{}^{2}+ 
    ((-9028\sqrt{2/3}Q)/27 - 17816\sqrt{2/3}Q{}^{3} - (454436\sqrt{2/3}Q{}^{5})/3)
     $%
$\partial J \partial{}^{4} A_{2} A_2 T + 
    ((-9577Q)/(27\sqrt{6}) - 15199 Q{}^{3}/\sqrt{6} - (195736\sqrt{2/3}Q{}^{5})/3)
     $%
$\partial J \partial{}^{4} A_{2} A_2{}^{2}+ 
    (2323 Q/(27\sqrt{6}) + 1523 Q{}^{3}/\sqrt{6} + (5068\sqrt{2/3}Q{}^{5})/3)
     $%
$\partial J \partial{}^{4} A_{2} B_2{}^{2}+ 
    ((-11528\sqrt{2/3})/9 - 3548\sqrt{6}Q{}^{2} + 16808\sqrt{2/3}Q{}^{4})
     $%
$\partial J A_3 \partial{}^{2} T \partial T + 
    ((-32752\sqrt{2/3})/27 - 12080\sqrt{2/3}Q{}^{2} - (65000\sqrt{2/3}Q{}^{4})/3)
     $%
$\partial J A_3 \partial{}^{3} T T + 
    (196\sqrt{2/3}Q - 270\sqrt{6}Q{}^{3})$%
$\partial J A_3 A_3 
      \partial{}^{2} T + (-154\sqrt{2/3}Q + 810\sqrt{6}Q{}^{3})
     $%
$\partial J A_3 B_2 \partial{}^{2} B_{3} + 
    (482\sqrt{2/3}Q + 924\sqrt{6}Q{}^{3})$%
$\partial J A_3 \partial B_{2} 
      \partial B_{3} + (-466\sqrt{6}Q{}^{2} - 2332\sqrt{6}Q{}^{4})
     $%
$\partial J A_3 \partial{}^{2} B_{2} \partial B_{2} + 
    (46\sqrt{2/3}Q - 358\sqrt{6}Q{}^{3})$%
$\partial J A_3 \partial{}^{2} B_{2} 
      B_3 + ((-2132\sqrt{2/3}Q{}^{2})/3 - 6724\sqrt{2/3}Q{}^{4})
     $%
$\partial J A_3 \partial{}^{3} B_{2} B_2 - 
    48\sqrt{6}Q{}^{2}$%
$\partial J A_3 \partial B_{3} B_3 + 
    ((-3200\sqrt{2/3})/3 - (35648\sqrt{2/3}Q{}^{2})/3 - 14440\sqrt{2/3}Q{}^{4})
     $%
$\partial J \partial A_{3} \partial T \partial T + 
    (-640\sqrt{6} - (48976\sqrt{2/3}Q{}^{2})/3 + 3376\sqrt{2/3}Q{}^{4})
     $%
$\partial J \partial A_{3} \partial{}^{2} T T + 
    ((11720\sqrt{2/3}Q)/3 + 24124\sqrt{2/3}Q{}^{3})$%
$\partial J \partial A_{3} 
      A_3 \partial T + 24\sqrt{6}Q{}^{2}$%
$\partial J \partial A_{3} 
      A_3 A_3 + (680\sqrt{6}Q + 4552\sqrt{6}Q{}^{3})
     $%
$\partial J \partial A_{3} \partial A_{3} T + 
    (436\sqrt{2/3}Q + 996\sqrt{6}Q{}^{3})$%
$\partial J \partial A_{3} B_2 
      \partial B_{3} + (-322\sqrt{2/3}Q{}^{2} - 564\sqrt{6}Q{}^{4})
     $%
$\partial J \partial A_{3} \partial B_{2} \partial B_{2} + 
    (-334\sqrt{2/3}Q - 980\sqrt{6}Q{}^{3})$%
$\partial J \partial A_{3} 
      \partial B_{2} B_3 + (-550\sqrt{6}Q{}^{2} - 3916\sqrt{6}Q{}^{4})
     $%
$\partial J \partial A_{3} \partial{}^{2} B_{2} B_2 + 
    120\sqrt{6}Q{}^{2}$%
$\partial J \partial A_{3} B_3 B_3 + 
    ((-11848\sqrt{2/3})/9 - 16340\sqrt{2/3}Q{}^{2} - 29408\sqrt{2/3}Q{}^{4})
     $%
$\partial J \partial{}^{2} A_{3} \partial T T + 
    (2048\sqrt{2/3}Q + 4296\sqrt{6}Q{}^{3})$%
$\partial J \partial{}^{2} A_{3} 
      A_3 T + (-26\sqrt{2/3}Q - 406\sqrt{6}Q{}^{3})
     $%
$\partial J \partial{}^{2} A_{3} B_2 B_3 + 
    (-478\sqrt{6}Q{}^{2} - 2748\sqrt{6}Q{}^{4})$%
$\partial J \partial{}^{2} A_{3} 
      \partial B_{2} B_2 + ((-4216\sqrt{2/3})/27 - 
      (10760\sqrt{2/3}Q{}^{2})/3 - (24548\sqrt{2/3}Q{}^{4})/3)
     $%
$\partial J \partial{}^{3} A_{3} T{}^{2}+ 
    ((-1141\sqrt{2/3}Q{}^{2})/3 - 718\sqrt{6}Q{}^{4})$%
$\partial J \partial{}^{3} A_{3} 
      B_2{}^{2}+ ((-7126\sqrt{2/3}Q)/27 - 
      (15700\sqrt{2/3}Q{}^{3})/3 - (72884\sqrt{2/3}Q{}^{5})/3)
     $%
$\partial J B_2{}^{2}\partial{}^{4} T + 
    ((-2596\sqrt{2/3})/9 - (9038\sqrt{2/3}Q{}^{2})/3 - 2938\sqrt{6}Q{}^{4})
     $%
$\partial J B_2 B_3 \partial{}^{3} T + 
    (-380\sqrt{2/3} - (7546\sqrt{2/3}Q{}^{2})/3 + 3622\sqrt{2/3}Q{}^{4})
     $%
$\partial J B_2 \partial B_{3} \partial{}^{2} T + 
    ((-1748\sqrt{2/3})/3 - (11186\sqrt{2/3}Q{}^{2})/3 + 7388\sqrt{2/3}Q{}^{4})
     $%
$\partial J B_2 \partial{}^{2} B_{3} \partial T + 
    ((-148\sqrt{2/3})/3 - 116\sqrt{6}Q{}^{2} + 1548\sqrt{6}Q{}^{4})
     $%
$\partial J B_2 \partial{}^{3} B_{3} T + 
    ((-51668\sqrt{2/3}Q)/27 - 40042\sqrt{2/3}Q{}^{3} - 
      (569950\sqrt{2/3}Q{}^{5})/3)$%
$\partial J \partial B_{2} B_2 
      \partial{}^{3} T + ((-14222\sqrt{2/3}Q)/9 - (129232\sqrt{2/3}Q{}^{3})/3 - 
      77208\sqrt{6}Q{}^{5})$%
$\partial J \partial B_{2} \partial B_{2} \partial{}^{2} T + 
    (-620\sqrt{2/3} - 11326\sqrt{2/3}Q{}^{2} - 19754\sqrt{6}Q{}^{4})
     $%
$\partial J \partial B_{2} B_3 \partial{}^{2} T + 
    ((-2056\sqrt{2/3})/3 - 3708\sqrt{6}Q{}^{2} - 15876\sqrt{6}Q{}^{4})
     $%
$\partial J \partial B_{2} \partial B_{3} \partial T + 
    (-148\sqrt{2/3} - 4316\sqrt{2/3}Q{}^{2} - 6676\sqrt{6}Q{}^{4})
     $%
$\partial J \partial B_{2} \partial{}^{2} B_{3} T + 
    ((-21898\sqrt{2/3}Q)/9 - (173044\sqrt{2/3}Q{}^{3})/3 - 
      297874\sqrt{2/3}Q{}^{5})$%
$\partial J \partial{}^{2} B_{2} B_2 \partial{}^{2} T + 
    ((-11704\sqrt{2/3}Q)/3 - 100120\sqrt{2/3}Q{}^{3} - 178660\sqrt{6}Q{}^{5})
     $%
$\partial J \partial{}^{2} B_{2} \partial B_{2} \partial T + 
    ((-7324\sqrt{2/3}Q)/9 - 8220\sqrt{6}Q{}^{3} - 136748\sqrt{2/3}Q{}^{5})
     $%
$\partial J \partial{}^{2} B_{2} \partial{}^{2} B_{2} T + 
    (-244\sqrt{6} - (45898\sqrt{2/3}Q{}^{2})/3 - 79784\sqrt{2/3}Q{}^{4})
     $%
$\partial J \partial{}^{2} B_{2} B_3 \partial T + 
    ((-1972\sqrt{2/3})/3 - 4412\sqrt{6}Q{}^{2} - 23308\sqrt{6}Q{}^{4})
     $%
$\partial J \partial{}^{2} B_{2} \partial B_{3} T + 
    ((-46616\sqrt{2/3}Q)/27 - (380704\sqrt{2/3}Q{}^{3})/9 - 74864\sqrt{6}Q{}^{5})
     $%
$\partial J \partial{}^{3} B_{2} B_2 \partial T + 
    ((-38608\sqrt{2/3}Q)/27 - 12296\sqrt{6}Q{}^{3} - (602360\sqrt{2/3}Q{}^{5})/3)
     $%
$\partial J \partial{}^{3} B_{2} \partial B_{2} T + 
    ((-1316\sqrt{2/3})/3 - 8068\sqrt{2/3}Q{}^{2} - 13996\sqrt{6}Q{}^{4})
     $%
$\partial J \partial{}^{3} B_{2} B_3 T + 
    ((-12928\sqrt{2/3}Q)/27 - (36148\sqrt{2/3}Q{}^{3})/3 - 
      (196244\sqrt{2/3}Q{}^{5})/3)$%
$\partial J \partial{}^{4} B_{2} B_2 
      T + ((2840\sqrt{2/3}Q)/3 + 12814\sqrt{2/3}Q{}^{3})
     $%
$\partial J B_3 B_3 \partial{}^{2} T + 
    (936\sqrt{6}Q + 9036\sqrt{6}Q{}^{3})$%
$\partial J \partial B_{3} B_3 
      \partial T + (648\sqrt{6}Q + 6056\sqrt{6}Q{}^{3})$%
$\partial J \partial B_{3} 
      \partial B_{3} T + (488\sqrt{6}Q + 4968\sqrt{6}Q{}^{3})
     $%
$\partial J \partial{}^{2} B_{3} B_3 T + 
    (-592\sqrt{2/3}Q - (56656\sqrt{2/3}Q{}^{3})/3 - 141376\sqrt{2/3}Q{}^{5})
     $%
$\partial{}^{2} J \partial T \partial T \partial T + 
    ((-28712\sqrt{2/3}Q)/9 - (281984\sqrt{2/3}Q{}^{3})/3 - 222848\sqrt{6}Q{}^{5})
     $%
$\partial{}^{2} J \partial{}^{2} T \partial T T + 
    ((-8680\sqrt{2/3}Q)/9 - (83480\sqrt{2/3}Q{}^{3})/3 - 195040\sqrt{2/3}Q{}^{5})
     $%
$\partial{}^{2} J \partial{}^{3} T T{}^{2}+ 
    (-736/27 - 148240 Q{}^{2}/27 - 350816 Q{}^{4}/3 - 1979264 Q{}^{6}/3)
     $%
$\partial{}^{2} J J \partial{}^{2} T \partial{}^{2} T + 
    (-10816/81 - 30544 Q{}^{2}/3 - 1649584 Q{}^{4}/9 - 2892224 Q{}^{6}/3)
     $%
$\partial{}^{2} J J \partial{}^{3} T \partial T + 
    (-31048/243 - 495304 Q{}^{2}/81 - 2620112 Q{}^{4}/27 - 4542656 Q{}^{6}/9)
     $%
$\partial{}^{2} J J \partial{}^{4} T T + 
    ((-85241\sqrt{2/3}Q)/405 + (6116\sqrt{2/3}Q{}^{3})/15 + 
      (4316792\sqrt{2/3}Q{}^{5})/45 + 242912\sqrt{6}Q{}^{7})
     $%
$\partial{}^{2} J J{}^{2}\partial{}^{5} T + 
    ((14072\sqrt{2/3}Q)/81 + 1063661 Q{}^{3}/(135\sqrt{6}) + 
      (4217548\sqrt{2/3}Q{}^{5})/45 + 776600\sqrt{2/3}Q{}^{7})
     $%
$\partial{}^{2} J J{}^{2}\partial{}^{5} A_{2} + 
    ((-136\sqrt{2/3})/27 + 232507 Q{}^{2}/(27\sqrt{6}) + 
      (232238\sqrt{2/3}Q{}^{4})/3 + (997300\sqrt{2/3}Q{}^{6})/3)
     $%
$\partial{}^{2} J J{}^{2}\partial{}^{4} A_{3} + 
    (-9064/243 - 181396 Q{}^{2}/27 - 3119852 Q{}^{4}/27 - 1810264 Q{}^{6}/3)
     $%
$\partial{}^{2} J J A_2 \partial{}^{4} T + 
    ((-15676Q)/27 - 27506 Q{}^{3}/9 + 15696Q{}^{5})$%
$\partial{}^{2} J J 
      A_2 \partial{}^{3} A_{3} + (-10528/81 - 497128 Q{}^{2}/27 - 
      904576 Q{}^{4}/3 - 4516640 Q{}^{6}/3)$%
$\partial{}^{2} J J \partial A_{2} 
      \partial{}^{3} T + ((-2996Q)/3 - 16234 Q{}^{3}/9 + 209776 Q{}^{5}/3)
     $%
$\partial{}^{2} J J \partial A_{2} \partial{}^{2} A_{3} + 
    (-1472/27 - 538360 Q{}^{2}/27 - 3530260 Q{}^{4}/9 - 6487216 Q{}^{6}/3)
     $%
$\partial{}^{2} J J \partial{}^{2} A_{2} \partial{}^{2} T + 
    (-209/27 - 164 Q{}^{2}/3 - 275374 Q{}^{4}/9 - 1016768 Q{}^{6}/3)
     $%
$\partial{}^{2} J J \partial{}^{2} A_{2} \partial{}^{2} A_{2} + 
    (-832Q - 72526 Q{}^{3}/9 - 27320 Q{}^{5}/3)$%
$\partial{}^{2} J J 
      \partial{}^{2} A_{2} \partial A_{3} + (-10528/81 - 141976 Q{}^{2}/9 - 
      898760 Q{}^{4}/3 - 4964200 Q{}^{6}/3)$%
$\partial{}^{2} J J \partial{}^{3} A_{2} 
      \partial T + (-944/27 - 13372 Q{}^{2}/9 - 651524 Q{}^{4}/9 - 1934368 Q{}^{6}/3)
     $%
$\partial{}^{2} J J \partial{}^{3} A_{2} \partial A_{2} + 
    ((-8482Q)/27 - 13676 Q{}^{3}/9 - 552Q{}^{5})$%
$\partial{}^{2} J J 
      \partial{}^{3} A_{2} A_3 + (-9064/243 - 127996 Q{}^{2}/27 - 
      2574830 Q{}^{4}/27 - 1636144 Q{}^{6}/3)$%
$\partial{}^{2} J J 
      \partial{}^{4} A_{2} T + (-4075/243 - 58508 Q{}^{2}/81 - 857942 Q{}^{4}/27 - 
      2419600 Q{}^{6}/9)$%
$\partial{}^{2} J J \partial{}^{4} A_{2} A_2 + 
    (10256 Q/3 + 50368Q{}^{3} + 192432Q{}^{5})$%
$\partial{}^{2} J J 
      A_3 \partial{}^{3} T + (16472 Q/3 + 1036844 Q{}^{3}/9 + 1645336 Q{}^{5}/3)
     $%
$\partial{}^{2} J J \partial A_{3} \partial{}^{2} T + 
    (1804/9 - 860 Q{}^{2}/3 + 1792Q{}^{4})$%
$\partial{}^{2} J J \partial A_{3} 
      \partial A_{3} + (14936 Q/3 + 1098620 Q{}^{3}/9 + 1868632 Q{}^{5}/3)
     $%
$\partial{}^{2} J J \partial{}^{2} A_{3} \partial T + 
    (3164/9 + 2516 Q{}^{2}/3 + 3344Q{}^{4})$%
$\partial{}^{2} J J \partial{}^{2} A_{3} 
      A_3 + (28252 Q/27 + 108308 Q{}^{3}/3 + 587504 Q{}^{5}/3)
     $%
$\partial{}^{2} J J \partial{}^{3} A_{3} T + 
    ((-236Q)/3 - 10606 Q{}^{3}/3 - 25032Q{}^{5})$%
$\partial{}^{2} J J 
      B_2 \partial{}^{3} B_{3} + ((-12424Q)/9 - 51710 Q{}^{3}/3 - 70424Q{}^{5})
     $%
$\partial{}^{2} J J \partial B_{2} \partial{}^{2} B_{3} + 
    (-527/27 + 35248 Q{}^{2}/9 + 182614 Q{}^{4}/3 + 225440Q{}^{6})
     $%
$\partial{}^{2} J J \partial{}^{2} B_{2} \partial{}^{2} B_{2} + 
    ((-6448Q)/3 - 55690 Q{}^{3}/3 - 46104Q{}^{5})$%
$\partial{}^{2} J J 
      \partial{}^{2} B_{2} \partial B_{3} + (-4420/81 + 146396 Q{}^{2}/27 + 
      791660 Q{}^{4}/9 + 329744Q{}^{6})$%
$\partial{}^{2} J J \partial{}^{3} B_{2} 
      \partial B_{2} + ((-39208Q)/27 - 114022 Q{}^{3}/9 - 81400 Q{}^{5}/3)
     $%
$\partial{}^{2} J J \partial{}^{3} B_{2} B_3 + 
    (-3349/81 + 5558 Q{}^{2}/9 + 53530 Q{}^{4}/3 + 236560 Q{}^{6}/3)
     $%
$\partial{}^{2} J J \partial{}^{4} B_{2} B_2 + 
    (-172/9 - 3644 Q{}^{2}/3 - 14712Q{}^{4})$%
$\partial{}^{2} J J \partial B_{3} 
      \partial B_{3} + (508/9 + 700 Q{}^{2}/3 - 7400Q{}^{4})
     $%
$\partial{}^{2} J J \partial{}^{2} B_{3} B_3 + 
    (-2872/27 + 32128 Q{}^{2}/27 + 200912Q{}^{4} + 6167360 Q{}^{6}/3)
     $%
$\partial{}^{2} J \partial J \partial{}^{2} T \partial T + 
    (-15064/81 - 161104 Q{}^{2}/81 + 750160 Q{}^{4}/9 + 8687296 Q{}^{6}/9)
     $%
$\partial{}^{2} J \partial J \partial{}^{3} T T + 
    ((-32558\sqrt{2/3}Q)/81 + 42212\sqrt{2/3}Q{}^{3} + (10967144\sqrt{2/3}Q{}^{5})/
       9 + 7849760\sqrt{2/3}Q{}^{7})$%
$\partial{}^{2} J \partial J J 
      \partial{}^{4} T + ((103720\sqrt{2/3}Q)/81 + (1134200\sqrt{2/3}Q{}^{3})/27 + 
      (9262273\sqrt{2/3}Q{}^{5})/9 + (24094016\sqrt{2/3}Q{}^{7})/3)
     $%
$\partial{}^{2} J \partial J J \partial{}^{4} A_{2} + 
    ((-2108\sqrt{2/3})/27 + (289774\sqrt{2/3}Q{}^{2})/9 + 
      (1553302\sqrt{2/3}Q{}^{4})/3 + 645864\sqrt{6}Q{}^{6})
     $%
$\partial{}^{2} J \partial J J \partial{}^{3} A_{3} + 
    ((3428\sqrt{2/3}Q)/27 + (1703720\sqrt{2/3}Q{}^{3})/27 + 
      1379588\sqrt{2/3}Q{}^{5} + (23063824\sqrt{2/3}Q{}^{7})/3)
     $%
$\partial{}^{2} J (\partial J){}^{2} \partial{}^{3} T + 
    ((47930\sqrt{2/3}Q)/27 + (995380\sqrt{2/3}Q{}^{3})/27 + 
      (7502258\sqrt{2/3}Q{}^{5})/9 + 7108346\sqrt{2/3}Q{}^{7})
     $%
$\partial{}^{2} J (\partial J){}^{2} \partial{}^{3} A_{2} + 
    ((-1054\sqrt{2/3})/9 + (386884\sqrt{2/3}Q{}^{2})/9 + 
      (2016740\sqrt{2/3}Q{}^{4})/3 + 806386\sqrt{6}Q{}^{6})
     $%
$\partial{}^{2} J (\partial J){}^{2} \partial{}^{2} A_{3} + 
    (-5272/81 - 279496 Q{}^{2}/81 + 49648Q{}^{4} + 6862264 Q{}^{6}/9)
     $%
$\partial{}^{2} J \partial J A_2 \partial{}^{3} T + 
    (-1038Q - 8686 Q{}^{3}/3 + 52904Q{}^{5})$%
$\partial{}^{2} J \partial J A_2 
      \partial{}^{2} A_{3} + (-2872/27 - 95452 Q{}^{2}/9 + 329636 Q{}^{4}/9 + 
      3748768 Q{}^{6}/3)$%
$\partial{}^{2} J \partial J \partial A_{2} \partial{}^{2} T + 
    (-1740Q - 126436 Q{}^{3}/9 + 62668 Q{}^{5}/3)$%
$\partial{}^{2} J \partial J 
      \partial A_{2} \partial A_{3} + (-2872/27 - 319148 Q{}^{2}/27 + 19096 Q{}^{4}/9 + 
      3192184 Q{}^{6}/3)$%
$\partial{}^{2} J \partial J \partial{}^{2} A_{2} \partial T + 
    (-764/27 + 40574 Q{}^{2}/3 + 1114628 Q{}^{4}/3 + 2593624Q{}^{6})
     $%
$\partial{}^{2} J \partial J \partial{}^{2} A_{2} \partial A_{2} + 
    ((-7874Q)/9 - 11778Q{}^{3} - 44236Q{}^{5})$%
$\partial{}^{2} J \partial J 
      \partial{}^{2} A_{2} A_3 + (-5272/81 - 9400 Q{}^{2}/3 + 473168 Q{}^{4}/9 + 
      2476984 Q{}^{6}/3)$%
$\partial{}^{2} J \partial J \partial{}^{3} A_{2} T + 
    (-232/27 + 197084 Q{}^{2}/27 + 1791346 Q{}^{4}/9 + 4219244 Q{}^{6}/3)
     $%
$\partial{}^{2} J \partial J \partial{}^{3} A_{2} A_2 + 
    (8792Q + 132352Q{}^{3} + 465992Q{}^{5})$%
$\partial{}^{2} J \partial J A_3 
      \partial{}^{2} T + (148288 Q/9 + 2346584 Q{}^{3}/9 + 2899384 Q{}^{5}/3)
     $%
$\partial{}^{2} J \partial J \partial A_{3} \partial T + 
    (1928/3 + 40504 Q{}^{2}/3 + 102284Q{}^{4})$%
$\partial{}^{2} J \partial J 
      \partial A_{3} A_3 + (64244 Q/9 + 375656 Q{}^{3}/3 + 485656Q{}^{5})
     $%
$\partial{}^{2} J \partial J \partial{}^{2} A_{3} T + 
    (-1202Q - 58526 Q{}^{3}/3 - 104168Q{}^{5})$%
$\partial{}^{2} J \partial J 
      B_2 \partial{}^{2} B_{3} + ((-8852Q)/3 - 57988 Q{}^{3}/3 - 10068Q{}^{5})
     $%
$\partial{}^{2} J \partial J \partial B_{2} \partial B_{3} + 
    (-2108/27 + 159994 Q{}^{2}/9 + 768356 Q{}^{4}/3 + 817552Q{}^{6})
     $%
$\partial{}^{2} J \partial J \partial{}^{2} B_{2} \partial B_{2} + 
    ((-5630Q)/3 - 2566 Q{}^{3}/3 + 92500Q{}^{5})$%
$\partial{}^{2} J \partial J 
      \partial{}^{2} B_{2} B_3 + (-7004/81 + 51574 Q{}^{2}/9 + 896452 Q{}^{4}/9 + 
      1116124 Q{}^{6}/3)$%
$\partial{}^{2} J \partial J \partial{}^{3} B_{2} B_2 + 
    (-296/3 - 17576Q{}^{2} - 153972Q{}^{4})$%
$\partial{}^{2} J \partial J \partial B_{3} 
      B_3 + (-3692/81 + 42536 Q{}^{2}/27 + 281192 Q{}^{4}/3 + 
      2457440 Q{}^{6}/3)$%
$\partial{}^{2} J \partial{}^{2} J \partial T \partial T + 
    (-10036/81 + 18784 Q{}^{2}/27 + 1023224 Q{}^{4}/9 + 1050016Q{}^{6})
     $%
$\partial{}^{2} J \partial{}^{2} J \partial{}^{2} T T + 
    ((-1874\sqrt{2/3}Q)/9 + (358504\sqrt{2/3}Q{}^{3})/9 + 
      1068040\sqrt{2/3}Q{}^{5} + 6720608\sqrt{2/3}Q{}^{7})
     $%
$\partial{}^{2} J \partial{}^{2} J J \partial{}^{3} T + 
    (1685\sqrt{2/3}Q + (1510732\sqrt{2/3}Q{}^{3})/27 + (10522976\sqrt{2/3}Q{}^{5})/
       9 + 8388200\sqrt{2/3}Q{}^{7})$%
$\partial{}^{2} J \partial{}^{2} J J 
      \partial{}^{3} A_{2} + ((-1756\sqrt{2/3})/9 + (202516\sqrt{2/3}Q{}^{2})/9 + 
      (1006864\sqrt{2/3}Q{}^{4})/3 + 1021816\sqrt{2/3}Q{}^{6})
     $%
$\partial{}^{2} J \partial{}^{2} J J \partial{}^{2} A_{3} + 
    ((4520\sqrt{2/3}Q)/27 + (650824\sqrt{2/3}Q{}^{3})/9 + 
      (4707476\sqrt{2/3}Q{}^{5})/3 + 2902000\sqrt{6}Q{}^{7})
     $%
$\partial{}^{2} J \partial{}^{2} J \partial J \partial{}^{2} T + 
    ((77591\sqrt{2/3}Q)/27 + 24738\sqrt{6}Q{}^{3} + (4372798\sqrt{2/3}Q{}^{5})/3 + 
      3655036\sqrt{6}Q{}^{7})$%
$\partial{}^{2} J \partial{}^{2} J \partial J \partial{}^{2} A_{2} + 
    ((-3512\sqrt{2/3})/9 + 38734\sqrt{2/3}Q{}^{2} + (1762394\sqrt{2/3}Q{}^{4})/3 + 
      1854124\sqrt{2/3}Q{}^{6})$%
$\partial{}^{2} J \partial{}^{2} J \partial J 
      \partial A_{3} + ((7640\sqrt{2/3}Q)/27 + (230416\sqrt{2/3}Q{}^{3})/9 + 
      500540\sqrt{2/3}Q{}^{5} + 2808784\sqrt{2/3}Q{}^{7})
     $%
$\partial{}^{2} J \partial{}^{2} J \partial{}^{2} J \partial T + 
    (365/54 - 2835Q{}^{2} - 4784194 Q{}^{4}/27 - 10997492 Q{}^{6}/3 - 
      70539920 Q{}^{8}/3)$%
$\partial{}^{2} J \partial{}^{2} J \partial{}^{2} J \partial{}^{2} J + 
    ((48221\sqrt{2/3}Q)/27 + (426610\sqrt{2/3}Q{}^{3})/9 + 
      (6393286\sqrt{2/3}Q{}^{5})/9 + (13274660\sqrt{2/3}Q{}^{7})/3)
     $%
$\partial{}^{2} J \partial{}^{2} J \partial{}^{2} J \partial A_{2} + 
    ((-1493\sqrt{2/3})/9 + (33430\sqrt{2/3}Q{}^{2})/9 + 21814\sqrt{6}Q{}^{4} + 
      169012\sqrt{2/3}Q{}^{6})$%
$\partial{}^{2} J \partial{}^{2} J \partial{}^{2} J A_3 + 
    (-3424/81 + 1604 Q{}^{2}/27 + 2697460 Q{}^{4}/27 + 8664584 Q{}^{6}/9)
     $%
$\partial{}^{2} J \partial{}^{2} J A_2 \partial{}^{2} T + 
    ((-5597Q)/9 - 40892 Q{}^{3}/9 + 22136 Q{}^{5}/3)$%
$\partial{}^{2} J \partial{}^{2} J 
      A_2 \partial A_{3} + (-13996/81 - 28856 Q{}^{2}/27 + 
      4158724 Q{}^{4}/27 + 13352096 Q{}^{6}/9)$%
$\partial{}^{2} J \partial{}^{2} J 
      \partial A_{2} \partial T + (-77/81 + 152132 Q{}^{2}/27 + 4126720 Q{}^{4}/27 + 
      9554018 Q{}^{6}/9)$%
$\partial{}^{2} J \partial{}^{2} J \partial A_{2} \partial A_{2} + 
    ((-1781Q)/3 - 77300 Q{}^{3}/9 - 96628 Q{}^{5}/3)$%
$\partial{}^{2} J \partial{}^{2} J 
      \partial A_{2} A_3 + (-3424/81 - 1640 Q{}^{2}/3 + 112564Q{}^{4} + 
      3533896 Q{}^{6}/3)$%
$\partial{}^{2} J \partial{}^{2} J \partial{}^{2} A_{2} T + 
    (191/81 + 193741 Q{}^{2}/27 + 5623588 Q{}^{4}/27 + 13629356 Q{}^{6}/9)
     $%
$\partial{}^{2} J \partial{}^{2} J \partial{}^{2} A_{2} A_2 + 
    (19996 Q/3 + 230564 Q{}^{3}/3 + 143792Q{}^{5})$%
$\partial{}^{2} J \partial{}^{2} J 
      A_3 \partial T + (1771/9 + 6288Q{}^{2} + 48098Q{}^{4})
     $%
$\partial{}^{2} J \partial{}^{2} J A_3 A_3 + 
    (5772Q + 203260 Q{}^{3}/3 + 117608Q{}^{5})$%
$\partial{}^{2} J \partial{}^{2} J 
      \partial A_{3} T + ((-9013Q)/9 - 132736 Q{}^{3}/9 - 232952 Q{}^{5}/3)
     $%
$\partial{}^{2} J \partial{}^{2} J B_2 \partial B_{3} + 
    (-1205/27 + 11080 Q{}^{2}/3 + 387388 Q{}^{4}/9 + 43834Q{}^{6})
     $%
$\partial{}^{2} J \partial{}^{2} J \partial B_{2} \partial B_{2} + 
    (-871Q + 37184 Q{}^{3}/9 + 278020 Q{}^{5}/3)$%
$\partial{}^{2} J \partial{}^{2} J 
      \partial B_{2} B_3 + (-769/9 + 35339 Q{}^{2}/9 + 555160 Q{}^{4}/9 + 
      169340Q{}^{6})$%
$\partial{}^{2} J \partial{}^{2} J \partial{}^{2} B_{2} B_2 + 
    (-223/3 - 22400 Q{}^{2}/3 - 66302Q{}^{4})$%
$\partial{}^{2} J \partial{}^{2} J B_3 
      B_3 + (-6184\sqrt{2/3}Q - (1386404\sqrt{2/3}Q{}^{3})/9 - 
      (2848480\sqrt{2/3}Q{}^{5})/3)$%
$\partial{}^{2} J A_2 \partial{}^{2} T \partial T + 
    ((-104192\sqrt{2/3}Q)/27 - (748552\sqrt{2/3}Q{}^{3})/9 - 
      (1489912\sqrt{2/3}Q{}^{5})/3)$%
$\partial{}^{2} J A_2 \partial{}^{3} T T + 
    ((-24502\sqrt{2/3}Q)/27 - (259784\sqrt{2/3}Q{}^{3})/9 - 
      (620212\sqrt{2/3}Q{}^{5})/3)$%
$\partial{}^{2} J A_2{}^{2}\partial{}^{3} T + ((-958\sqrt{2/3}Q{}^{2})/3 - 1026\sqrt{6}Q{}^{4})
     $%
$\partial{}^{2} J A_2{}^{2}\partial{}^{2} A_{3} + 
    ((-1022\sqrt{2/3})/9 + (16730\sqrt{2/3}Q{}^{2})/9 + (83098\sqrt{2/3}Q{}^{4})/3)
     $%
$\partial{}^{2} J A_2 A_3 \partial{}^{2} T + 
    ((-2044\sqrt{2/3})/9 + (26468\sqrt{2/3}Q{}^{2})/9 + 
      (149284\sqrt{2/3}Q{}^{4})/3)$%
$\partial{}^{2} J A_2 \partial A_{3} 
      \partial T + ((-4880\sqrt{2/3}Q)/3 - 14548\sqrt{2/3}Q{}^{3})
     $%
$\partial{}^{2} J A_2 \partial A_{3} A_3 + 
    ((-2042\sqrt{2/3})/9 - (284\sqrt{2/3}Q{}^{2})/3 + 18376\sqrt{2/3}Q{}^{4})
     $%
$\partial{}^{2} J A_2 \partial{}^{2} A_{3} T + 
    ((1906\sqrt{2/3}Q{}^{2})/3 + 5296\sqrt{2/3}Q{}^{4})$%
$\partial{}^{2} J A_2 
      B_2 \partial{}^{2} B_{3} + (-800\sqrt{6}Q{}^{2} - 16414\sqrt{2/3}Q{}^{4})
     $%
$\partial{}^{2} J A_2 \partial B_{2} \partial B_{3} + 
    ((-2234\sqrt{2/3}Q)/3 - (13603\sqrt{2/3}Q{}^{3})/3 + 2068\sqrt{6}Q{}^{5})
     $%
$\partial{}^{2} J A_2 \partial{}^{2} B_{2} \partial B_{2} + 
    ((-7274\sqrt{2/3}Q{}^{2})/3 - 17902\sqrt{2/3}Q{}^{4})
     $%
$\partial{}^{2} J A_2 \partial{}^{2} B_{2} B_3 + 
    ((-12614\sqrt{2/3}Q)/27 - (42023\sqrt{2/3}Q{}^{3})/9 - 
      (31664\sqrt{2/3}Q{}^{5})/3)$%
$\partial{}^{2} J A_2 \partial{}^{3} B_{2} 
      B_2 + ((7624\sqrt{2/3}Q)/3 + 20408\sqrt{2/3}Q{}^{3})
     $%
$\partial{}^{2} J A_2 \partial B_{3} B_3 + 
    ((-43544\sqrt{2/3}Q)/9 - (1121488\sqrt{2/3}Q{}^{3})/9 - 
      (2275424\sqrt{2/3}Q{}^{5})/3)$%
$\partial{}^{2} J \partial A_{2} \partial T \partial T + 
    ((-55900\sqrt{2/3}Q)/9 - (470552\sqrt{2/3}Q{}^{3})/3 - 
      978232\sqrt{2/3}Q{}^{5})$%
$\partial{}^{2} J \partial A_{2} \partial{}^{2} T T + 
    ((-24742\sqrt{2/3}Q)/9 - (863936\sqrt{2/3}Q{}^{3})/9 - 
      (2154802\sqrt{2/3}Q{}^{5})/3)$%
$\partial{}^{2} J \partial A_{2} A_2 
      \partial{}^{2} T + ((6440\sqrt{2/3}Q{}^{2})/9 + (12790\sqrt{2/3}Q{}^{4})/3)
     $%
$\partial{}^{2} J \partial A_{2} A_2 \partial A_{3} + 
    ((-15964\sqrt{2/3}Q)/9 - (581348\sqrt{2/3}Q{}^{3})/9 - 
      (1511488\sqrt{2/3}Q{}^{5})/3)$%
$\partial{}^{2} J \partial A_{2} \partial A_{2} 
      \partial T + ((-2888\sqrt{2/3}Q)/9 - (163192\sqrt{2/3}Q{}^{3})/9 - 
      (471266\sqrt{2/3}Q{}^{5})/3)$%
$\partial{}^{2} J \partial A_{2} \partial A_{2} 
      \partial A_{2} + ((15668\sqrt{2/3}Q{}^{2})/9 + (43678\sqrt{2/3}Q{}^{4})/3)
     $%
$\partial{}^{2} J \partial A_{2} \partial A_{2} A_3 + 
    ((-1588\sqrt{2/3})/3 + (3188\sqrt{2/3}Q{}^{2})/9 + (198964\sqrt{2/3}Q{}^{4})/3)
     $%
$\partial{}^{2} J \partial A_{2} A_3 \partial T + 
    ((-5368\sqrt{2/3}Q)/3 - 15644\sqrt{2/3}Q{}^{3})$%
$\partial{}^{2} J \partial A_{2} 
      A_3 A_3 + ((-908\sqrt{2/3})/3 + 104\sqrt{2/3}Q{}^{2} + 
      18596\sqrt{6}Q{}^{4})$%
$\partial{}^{2} J \partial A_{2} \partial A_{3} T + 
    ((11800\sqrt{2/3}Q{}^{2})/9 + (37220\sqrt{2/3}Q{}^{4})/3)
     $%
$\partial{}^{2} J \partial A_{2} B_2 \partial B_{3} + 
    ((-2440\sqrt{2/3}Q)/9 + (33280\sqrt{2/3}Q{}^{3})/9 + 
      (160622\sqrt{2/3}Q{}^{5})/3)$%
$\partial{}^{2} J \partial A_{2} \partial B_{2} 
      \partial B_{2} + ((-44156\sqrt{2/3}Q{}^{2})/9 - (124642\sqrt{2/3}Q{}^{4})/3)
     $%
$\partial{}^{2} J \partial A_{2} \partial B_{2} B_3 + 
    ((-5572\sqrt{2/3}Q)/9 - (24218\sqrt{2/3}Q{}^{3})/9 + 
      (58694\sqrt{2/3}Q{}^{5})/3)$%
$\partial{}^{2} J \partial A_{2} \partial{}^{2} B_{2} 
      B_2 + ((8068\sqrt{2/3}Q)/3 + 22910\sqrt{2/3}Q{}^{3})
     $%
$\partial{}^{2} J \partial A_{2} B_3 B_3 + 
    ((-12292\sqrt{2/3}Q)/3 - (435176\sqrt{2/3}Q{}^{3})/3 - 
      1010872\sqrt{2/3}Q{}^{5})$%
$\partial{}^{2} J \partial{}^{2} A_{2} \partial T T + 
    ((-8072\sqrt{2/3}Q)/3 - (869696\sqrt{2/3}Q{}^{3})/9 - 
      (2233300\sqrt{2/3}Q{}^{5})/3)$%
$\partial{}^{2} J \partial{}^{2} A_{2} A_2 
      \partial T + ((6818\sqrt{2/3}Q{}^{2})/9 + (19192\sqrt{2/3}Q{}^{4})/3)
     $%
$\partial{}^{2} J \partial{}^{2} A_{2} A_2 A_3 + 
    ((-11006\sqrt{2/3}Q)/9 - 22422\sqrt{6}Q{}^{3} - 619168\sqrt{2/3}Q{}^{5})
     $%
$\partial{}^{2} J \partial{}^{2} A_{2} \partial A_{2} T + 
    ((-4018\sqrt{2/3}Q)/3 - (636347\sqrt{2/3}Q{}^{3})/9 - 
      (1847242\sqrt{2/3}Q{}^{5})/3)$%
$\partial{}^{2} J \partial{}^{2} A_{2} \partial A_{2} 
      A_2 + ((-3658\sqrt{2/3})/9 - 3096\sqrt{6}Q{}^{2} - 
      29468\sqrt{2/3}Q{}^{4})$%
$\partial{}^{2} J \partial{}^{2} A_{2} A_3 T + 
    ((2044\sqrt{2/3}Q{}^{2})/9 + (2930\sqrt{2/3}Q{}^{4})/3)
     $%
$\partial{}^{2} J \partial{}^{2} A_{2} B_2 B_3 + 
    ((-500\sqrt{2/3}Q)/3 + (1204\sqrt{2/3}Q{}^{3})/9 + (40124\sqrt{2/3}Q{}^{5})/3)
     $%
$\partial{}^{2} J \partial{}^{2} A_{2} \partial B_{2} B_2 + 
    ((-23272\sqrt{2/3}Q)/27 - (286168\sqrt{2/3}Q{}^{3})/9 - 
      (691540\sqrt{2/3}Q{}^{5})/3)$%
$\partial{}^{2} J \partial{}^{3} A_{2} T{}^{2}+ 
    ((-25718\sqrt{2/3}Q)/27 - (386578\sqrt{2/3}Q{}^{3})/9 - 
      (1067692\sqrt{2/3}Q{}^{5})/3)$%
$\partial{}^{2} J \partial{}^{3} A_{2} A_2 
      T + ((-13618\sqrt{2/3}Q)/27 - 20098\sqrt{2/3}Q{}^{3} - 
      54492\sqrt{6}Q{}^{5})$%
$\partial{}^{2} J \partial{}^{3} A_{2} A_2{}^{2}+ 
    ((2224\sqrt{2/3}Q)/27 + (4847\sqrt{2/3}Q{}^{3})/3 + 2224\sqrt{6}Q{}^{5})
     $%
$\partial{}^{2} J \partial{}^{3} A_{2} B_2{}^{2}+ 
    ((-2824\sqrt{2/3})/3 - (7688\sqrt{2/3}Q{}^{2})/3 + 16384\sqrt{6}Q{}^{4})
     $%
$\partial{}^{2} J A_3 \partial T \partial T + 
    (-1952\sqrt{2/3} - (47828\sqrt{2/3}Q{}^{2})/3 - 5136\sqrt{6}Q{}^{4})
     $%
$\partial{}^{2} J A_3 \partial{}^{2} T T + 
    (640\sqrt{2/3}Q - 740\sqrt{6}Q{}^{3})$%
$\partial{}^{2} J A_3 A_3 
      \partial T - 24\sqrt{6}Q{}^{2}$%
$\partial{}^{2} J A_3 A_3 
      A_3 + ((860\sqrt{2/3}Q)/3 + 2434\sqrt{2/3}Q{}^{3})
     $%
$\partial{}^{2} J A_3 B_2 \partial B_{3} + 
    (-40\sqrt{2/3}Q{}^{2} - 1948\sqrt{2/3}Q{}^{4})$%
$\partial{}^{2} J A_3 
      \partial B_{2} \partial B_{2} + ((-676\sqrt{2/3}Q)/3 - 1070\sqrt{2/3}Q{}^{3})
     $%
$\partial{}^{2} J A_3 \partial B_{2} B_3 + 
    (-380\sqrt{6}Q{}^{2} - 8836\sqrt{2/3}Q{}^{4})$%
$\partial{}^{2} J A_3 
      \partial{}^{2} B_{2} B_2 + 40\sqrt{6}Q{}^{2}$%
$\partial{}^{2} J A_3 
      B_3 B_3 + ((-22384\sqrt{2/3})/9 - 
      (38752\sqrt{2/3}Q{}^{2})/3 + 54824\sqrt{2/3}Q{}^{4})
     $%
$\partial{}^{2} J \partial A_{3} \partial T T + 
    (3424\sqrt{2/3}Q + 4436\sqrt{6}Q{}^{3})$%
$\partial{}^{2} J \partial A_{3} 
      A_3 T + ((-676\sqrt{2/3}Q)/3 - 830\sqrt{2/3}Q{}^{3})
     $%
$\partial{}^{2} J \partial A_{3} B_2 B_3 + 
    ((-5008\sqrt{2/3}Q{}^{2})/3 - 11008\sqrt{2/3}Q{}^{4})
     $%
$\partial{}^{2} J \partial A_{3} \partial B_{2} B_2 + 
    ((-4916\sqrt{2/3})/9 - (14480\sqrt{2/3}Q{}^{2})/3 - 676\sqrt{2/3}Q{}^{4})
     $%
$\partial{}^{2} J \partial{}^{2} A_{3} T{}^{2}+ 
    ((-2372\sqrt{2/3}Q{}^{2})/3 - 1442\sqrt{6}Q{}^{4})$%
$\partial{}^{2} J \partial{}^{2} A_{3} 
      B_2{}^{2}+ (-722\sqrt{2/3}Q - (113944\sqrt{2/3}Q{}^{3})/
       9 - 52996\sqrt{2/3}Q{}^{5})$%
$\partial{}^{2} J B_2{}^{2}\partial{}^{3} T + ((-1298\sqrt{2/3})/3 - (41234\sqrt{2/3}Q{}^{2})/9 - 
      (43990\sqrt{2/3}Q{}^{4})/3)$%
$\partial{}^{2} J B_2 B_3 
      \partial{}^{2} T + ((-2596\sqrt{2/3})/3 - (38780\sqrt{2/3}Q{}^{2})/9 - 
      (4612\sqrt{2/3}Q{}^{4})/3)$%
$\partial{}^{2} J B_2 \partial B_{3} \partial T + 
    ((-958\sqrt{2/3})/3 - 704\sqrt{6}Q{}^{2} - 4048\sqrt{2/3}Q{}^{4})
     $%
$\partial{}^{2} J B_2 \partial{}^{2} B_{3} T + 
    ((-29602\sqrt{2/3}Q)/9 - (616592\sqrt{2/3}Q{}^{3})/9 - 
      (952042\sqrt{2/3}Q{}^{5})/3)$%
$\partial{}^{2} J \partial B_{2} B_2 
      \partial{}^{2} T + ((-23500\sqrt{2/3}Q)/9 - (505340\sqrt{2/3}Q{}^{3})/9 - 
      (739048\sqrt{2/3}Q{}^{5})/3)$%
$\partial{}^{2} J \partial B_{2} \partial B_{2} 
      \partial T + ((-2596\sqrt{2/3})/3 - (133724\sqrt{2/3}Q{}^{2})/9 - 
      (254476\sqrt{2/3}Q{}^{4})/3)$%
$\partial{}^{2} J \partial B_{2} B_3 
      \partial T + ((-1916\sqrt{2/3})/3 - (27776\sqrt{2/3}Q{}^{2})/3 - 
      53636\sqrt{2/3}Q{}^{4})$%
$\partial{}^{2} J \partial B_{2} \partial B_{3} T + 
    ((-30812\sqrt{2/3}Q)/9 - (638108\sqrt{2/3}Q{}^{3})/9 - 
      (946828\sqrt{2/3}Q{}^{5})/3)$%
$\partial{}^{2} J \partial{}^{2} B_{2} B_2 
      \partial T + ((-8522\sqrt{2/3}Q)/3 - 62578\sqrt{2/3}Q{}^{3} - 
      276832\sqrt{2/3}Q{}^{5})$%
$\partial{}^{2} J \partial{}^{2} B_{2} \partial B_{2} T + 
    ((-2318\sqrt{2/3})/3 - 12160\sqrt{2/3}Q{}^{2} - 21116\sqrt{6}Q{}^{4})
     $%
$\partial{}^{2} J \partial{}^{2} B_{2} B_3 T + 
    ((-31342\sqrt{2/3}Q)/27 - (73682\sqrt{2/3}Q{}^{3})/3 - 
      (335684\sqrt{2/3}Q{}^{5})/3)$%
$\partial{}^{2} J \partial{}^{3} B_{2} B_2 
      T + ((6688\sqrt{2/3}Q)/3 + 21932\sqrt{2/3}Q{}^{3})
     $%
$\partial{}^{2} J B_3 B_3 \partial T + 
    ((7744\sqrt{2/3}Q)/3 + 25796\sqrt{2/3}Q{}^{3})$%
$\partial{}^{2} J \partial B_{3} 
      B_3 T + ((-41080\sqrt{2/3}Q)/27 - 
      (138800\sqrt{2/3}Q{}^{3})/3 - (1008320\sqrt{2/3}Q{}^{5})/3)
     $%
$\partial{}^{3} J \partial T \partial T T + 
    ((-31240\sqrt{2/3}Q)/27 - (102328\sqrt{2/3}Q{}^{3})/3 - 
      (728096\sqrt{2/3}Q{}^{5})/3)$%
$\partial{}^{3} J \partial{}^{2} T T{}^{2}+ 
    (-11984/81 - 297824 Q{}^{2}/27 - 1842640 Q{}^{4}/9 - 3372352 Q{}^{6}/3)
     $%
$\partial{}^{3} J J \partial{}^{2} T \partial T + 
    (-14072/81 - 69776 Q{}^{2}/9 - 1068784 Q{}^{4}/9 - 608832Q{}^{6})
     $%
$\partial{}^{3} J J \partial{}^{3} T T + 
    ((-23333\sqrt{2/3}Q)/81 + (86630\sqrt{2/3}Q{}^{3})/27 + 
      (1805336\sqrt{2/3}Q{}^{5})/9 + (4341952\sqrt{2/3}Q{}^{7})/3)
     $%
$\partial{}^{3} J J{}^{2}\partial{}^{4} T + 
    ((3065\sqrt{2/3}Q)/9 + 182827 Q{}^{3}/(9\sqrt{6}) + 
      4091869 Q{}^{5}/(9\sqrt{6}) + (5171038\sqrt{2/3}Q{}^{7})/3)
     $%
$\partial{}^{3} J J{}^{2}\partial{}^{4} A_{2} + 
    ((-4016\sqrt{2/3})/81 + (174260\sqrt{2/3}Q{}^{2})/27 + 
      (934589\sqrt{2/3}Q{}^{4})/9 + (1109158\sqrt{2/3}Q{}^{6})/3)
     $%
$\partial{}^{3} J J{}^{2}\partial{}^{3} A_{3} + 
    (-20584/243 - 675656 Q{}^{2}/81 - 400496 Q{}^{4}/3 - 2081312 Q{}^{6}/3)
     $%
$\partial{}^{3} J J A_2 \partial{}^{3} T + 
    ((-16568Q)/27 - 34402 Q{}^{3}/9 + 18104Q{}^{5})$%
$\partial{}^{3} J J 
      A_2 \partial{}^{2} A_{3} + (-11984/81 - 487928 Q{}^{2}/27 - 
      2643368 Q{}^{4}/9 - 1498080Q{}^{6})$%
$\partial{}^{3} J J \partial A_{2} 
      \partial{}^{2} T + ((-28304Q)/27 - 96868 Q{}^{3}/9 - 20296Q{}^{5})
     $%
$\partial{}^{3} J J \partial A_{2} \partial A_{3} + 
    (-11984/81 - 154888 Q{}^{2}/9 - 2570908 Q{}^{4}/9 - 1450592Q{}^{6})
     $%
$\partial{}^{3} J J \partial{}^{2} A_{2} \partial T + 
    (-3952/81 - 40234 Q{}^{2}/27 - 599402 Q{}^{4}/9 - 1680416 Q{}^{6}/3)
     $%
$\partial{}^{3} J J \partial{}^{2} A_{2} \partial A_{2} + 
    ((-3524Q)/9 - 11066 Q{}^{3}/9 + 17312 Q{}^{5}/3)$%
$\partial{}^{3} J J 
      \partial{}^{2} A_{2} A_3 + (-20584/243 - 607204 Q{}^{2}/81 - 
      386608 Q{}^{4}/3 - 6073360 Q{}^{6}/9)$%
$\partial{}^{3} J J \partial{}^{3} A_{2} 
      T + (-32 - 105650 Q{}^{2}/81 - 130778 Q{}^{4}/3 - 3075280 Q{}^{6}/9)
     $%
$\partial{}^{3} J J \partial{}^{3} A_{2} A_2 + 
    (99376 Q/27 + 61528Q{}^{3} + 696224 Q{}^{5}/3)$%
$\partial{}^{3} J J 
      A_3 \partial{}^{2} T + (177800 Q/27 + 1027420 Q{}^{3}/9 + 433592Q{}^{5})
     $%
$\partial{}^{3} J J \partial A_{3} \partial T + 
    (976/3 - 2828 Q{}^{2}/3 - 6944Q{}^{4})$%
$\partial{}^{3} J J \partial A_{3} 
      A_3 + (76468 Q/27 + 522044 Q{}^{3}/9 + 247224Q{}^{5})
     $%
$\partial{}^{3} J J \partial{}^{2} A_{3} T + 
    ((-14404Q)/27 - 50486 Q{}^{3}/9 - 21888Q{}^{5})$%
$\partial{}^{3} J J 
      B_2 \partial{}^{2} B_{3} + ((-48760Q)/27 - 168992 Q{}^{3}/9 - 61504Q{}^{5})
     $%
$\partial{}^{3} J J \partial B_{2} \partial B_{3} + 
    (-8032/81 + 166882 Q{}^{2}/27 + 293950 Q{}^{4}/3 + 307232Q{}^{6})
     $%
$\partial{}^{3} J J \partial{}^{2} B_{2} \partial B_{2} + 
    ((-12028Q)/9 - 41038 Q{}^{3}/3 - 40384Q{}^{5})$%
$\partial{}^{3} J J 
      \partial{}^{2} B_{2} B_3 + (-18848/243 + 90730 Q{}^{2}/81 + 
      700630 Q{}^{4}/27 + 260128 Q{}^{6}/3)$%
$\partial{}^{3} J J \partial{}^{3} B_{2} 
      B_2 + (2032/27 + 22124 Q{}^{2}/9 + 6752Q{}^{4})
     $%
$\partial{}^{3} J J \partial B_{3} B_3 + 
    (-20192/243 + 115976 Q{}^{2}/81 + 3281408 Q{}^{4}/27 + 9977728 Q{}^{6}/9)
     $%
$\partial{}^{3} J \partial J \partial T \partial T + 
    (-44296/243 + 82504 Q{}^{2}/81 + 4663696 Q{}^{4}/27 + 14499776 Q{}^{6}/9)
     $%
$\partial{}^{3} J \partial J \partial{}^{2} T T + 
    ((-7744\sqrt{2/3}Q)/27 + (580460\sqrt{2/3}Q{}^{3})/9 + 
      (5228344\sqrt{2/3}Q{}^{5})/3 + 11130400\sqrt{2/3}Q{}^{7})
     $%
$\partial{}^{3} J \partial J J \partial{}^{3} T + 
    ((68240\sqrt{2/3}Q)/27 + (764042\sqrt{2/3}Q{}^{3})/9 + 
      (5288410\sqrt{2/3}Q{}^{5})/3 + 12607756\sqrt{2/3}Q{}^{7})
     $%
$\partial{}^{3} J \partial J J \partial{}^{3} A_{2} + 
    ((-8032\sqrt{2/3})/27 + (296008\sqrt{2/3}Q{}^{2})/9 + 503326\sqrt{2/3}Q{}^{4} + 
      530912\sqrt{6}Q{}^{6})$%
$\partial{}^{3} J \partial J J \partial{}^{2} A_{3} + 
    ((6620\sqrt{2/3}Q)/81 + (1451248\sqrt{2/3}Q{}^{3})/27 + 
      (10474516\sqrt{2/3}Q{}^{5})/9 + (19101776\sqrt{2/3}Q{}^{7})/3)
     $%
$\partial{}^{3} J (\partial J){}^{2} \partial{}^{2} T + 
    ((180656\sqrt{2/3}Q)/81 + (1544602\sqrt{2/3}Q{}^{3})/27 + 
      (9350290\sqrt{2/3}Q{}^{5})/9 + (22551368\sqrt{2/3}Q{}^{7})/3)
     $%
$\partial{}^{3} J (\partial J){}^{2} \partial{}^{2} A_{2} + 
    ((-8032\sqrt{2/3})/27 + (248806\sqrt{2/3}Q{}^{2})/9 + 420998\sqrt{2/3}Q{}^{4} + 
      434908\sqrt{6}Q{}^{6})$%
$\partial{}^{3} J (\partial J){}^{2} \partial A_{3} + 
    (-17656/243 + 4852 Q{}^{2}/81 + 4226860 Q{}^{4}/27 + 13607744 Q{}^{6}/9)
     $%
$\partial{}^{3} J \partial J A_2 \partial{}^{2} T + 
    ((-23560Q)/27 - 51412 Q{}^{3}/9 + 51746 Q{}^{5}/3)
     $%
$\partial{}^{3} J \partial J A_2 \partial A_{3} + 
    (-67024/243 - 255824 Q{}^{2}/81 + 5520172 Q{}^{4}/27 + 19066136 Q{}^{6}/9)
     $%
$\partial{}^{3} J \partial J \partial A_{2} \partial T + 
    (-2756/243 + 610706 Q{}^{2}/81 + 5791880 Q{}^{4}/27 + 13755256 Q{}^{6}/9)
     $%
$\partial{}^{3} J \partial J \partial A_{2} \partial A_{2} + 
    ((-23560Q)/27 - 133028 Q{}^{3}/9 - 61998Q{}^{5})$%
$\partial{}^{3} J \partial J 
      \partial A_{2} A_3 + (-17656/243 - 95768 Q{}^{2}/81 + 
      4260328 Q{}^{4}/27 + 15131240 Q{}^{6}/9)$%
$\partial{}^{3} J \partial J 
      \partial{}^{2} A_{2} T + (-220/243 + 803158 Q{}^{2}/81 + 7868194 Q{}^{4}/27 + 
      19192742 Q{}^{6}/9)$%
$\partial{}^{3} J \partial J \partial{}^{2} A_{2} A_2 + 
    (259864 Q/27 + 351980 Q{}^{3}/3 + 798416 Q{}^{5}/3)
     $%
$\partial{}^{3} J \partial J A_3 \partial T + 
    (6328/27 + 23402 Q{}^{2}/3 + 161186 Q{}^{4}/3)$%
$\partial{}^{3} J \partial J 
      A_3 A_3 + (233384 Q/27 + 997856 Q{}^{3}/9 + 
      815216 Q{}^{5}/3)$%
$\partial{}^{3} J \partial J \partial A_{3} T + 
    ((-4208Q)/3 - 173284 Q{}^{3}/9 - 266822 Q{}^{5}/3)
     $%
$\partial{}^{3} J \partial J B_2 \partial B_{3} + 
    (-5812/81 + 49846 Q{}^{2}/9 + 649180 Q{}^{4}/9 + 134968Q{}^{6})
     $%
$\partial{}^{3} J \partial J \partial B_{2} \partial B_{2} + 
    ((-35872Q)/27 + 24832 Q{}^{3}/9 + 94458Q{}^{5})$%
$\partial{}^{3} J \partial J 
      \partial B_{2} B_3 + (-10252/81 + 51302 Q{}^{2}/9 + 833170 Q{}^{4}/9 + 
      266958Q{}^{6})$%
$\partial{}^{3} J \partial J \partial{}^{2} B_{2} B_2 + 
    (-1888/27 - 81470 Q{}^{2}/9 - 72890Q{}^{4})$%
$\partial{}^{3} J \partial J 
      B_3 B_3 + (-49592/243 + 483736 Q{}^{2}/81 + 
      8604848 Q{}^{4}/27 + 23505728 Q{}^{6}/9)$%
$\partial{}^{3} J \partial{}^{2} J \partial T 
      T + ((-6520\sqrt{2/3}Q)/27 + (2537552\sqrt{2/3}Q{}^{3})/27 + 
      841944\sqrt{6}Q{}^{5} + (49077280\sqrt{2/3}Q{}^{7})/3)
     $%
$\partial{}^{3} J \partial{}^{2} J J \partial{}^{2} T + 
    ((110684\sqrt{2/3}Q)/27 + (1349684\sqrt{2/3}Q{}^{3})/9 + 
      (26166686\sqrt{2/3}Q{}^{5})/9 + (58600240\sqrt{2/3}Q{}^{7})/3)
     $%
$\partial{}^{3} J \partial{}^{2} J J \partial{}^{2} A_{2} + 
    ((-2024\sqrt{2/3})/3 + (254176\sqrt{2/3}Q{}^{2})/9 + 
      (1396994\sqrt{2/3}Q{}^{4})/3 + 1478644\sqrt{2/3}Q{}^{6})
     $%
$\partial{}^{3} J \partial{}^{2} J J \partial A_{3} + 
    ((13024\sqrt{2/3}Q)/9 + (3376760\sqrt{2/3}Q{}^{3})/27 + 
      2296544\sqrt{2/3}Q{}^{5} + (36149248\sqrt{2/3}Q{}^{7})/3)
     $%
$\partial{}^{3} J \partial{}^{2} J \partial J \partial T + 
    ((72944\sqrt{2/3}Q)/9 + (5770594\sqrt{2/3}Q{}^{3})/27 + 
      (26315990\sqrt{2/3}Q{}^{5})/9 + 16807484\sqrt{2/3}Q{}^{7})
     $%
$\partial{}^{3} J \partial{}^{2} J \partial J \partial A_{2} + 
    ((-2024\sqrt{2/3})/3 + (470476\sqrt{2/3}Q{}^{2})/27 + 
      (943852\sqrt{2/3}Q{}^{4})/3 + (2884904\sqrt{2/3}Q{}^{6})/3)
     $%
$\partial{}^{3} J \partial{}^{2} J \partial J A_3 + 
    ((80422\sqrt{2/3}Q)/81 + (1354156\sqrt{2/3}Q{}^{3})/27 + 
      (7188176\sqrt{2/3}Q{}^{5})/9 + (12233216\sqrt{2/3}Q{}^{7})/3)
     $%
$\partial{}^{3} J \partial{}^{2} J \partial{}^{2} J T + 
    (12190/243 - 76708 Q{}^{2}/3 - 124323868 Q{}^{4}/81 - 92086696 Q{}^{6}/3 - 
      1732955360 Q{}^{8}/9)$%
$\partial{}^{3} J \partial{}^{2} J \partial{}^{2} J \partial J + 
    ((233812\sqrt{2/3}Q)/81 + (2295616\sqrt{2/3}Q{}^{3})/27 + 
      (3316078\sqrt{2/3}Q{}^{5})/3 + (17025476\sqrt{2/3}Q{}^{7})/3)
     $%
$\partial{}^{3} J \partial{}^{2} J \partial{}^{2} J A_2 + 
    (-49592/243 + 717944 Q{}^{2}/81 + 1102900 Q{}^{4}/3 + 2745312Q{}^{6})
     $%
$\partial{}^{3} J \partial{}^{2} J A_2 \partial T + 
    ((-12830Q)/27 - 11908 Q{}^{3}/3 + 1202Q{}^{5})$%
$\partial{}^{3} J \partial{}^{2} J 
      A_2 A_3 + (-49592/243 + 46876 Q{}^{2}/9 + 
      8874208 Q{}^{4}/27 + 24319168 Q{}^{6}/9)$%
$\partial{}^{3} J \partial{}^{2} J 
      \partial A_{2} T + (5056/243 + 1324706 Q{}^{2}/81 + 
      12056614 Q{}^{4}/27 + 28271894 Q{}^{6}/9)$%
$\partial{}^{3} J \partial{}^{2} J 
      \partial A_{2} A_2 + (230728 Q/27 + 329404 Q{}^{3}/3 + 
      966392 Q{}^{5}/3)$%
$\partial{}^{3} J \partial{}^{2} J A_3 T + 
    ((-19510Q)/27 - 62152 Q{}^{3}/9 - 64690 Q{}^{5}/3)
     $%
$\partial{}^{3} J \partial{}^{2} J B_2 B_3 + 
    (-2024/9 + 104570 Q{}^{2}/27 + 672410 Q{}^{4}/9 + 591842 Q{}^{6}/3)
     $%
$\partial{}^{3} J \partial{}^{2} J \partial B_{2} B_2 + 
    (-2116/81 + 137500 Q{}^{2}/81 + 611792 Q{}^{4}/9 + 4735232 Q{}^{6}/9)
     $%
$\partial{}^{3} J \partial{}^{3} J T{}^{2}+ 
    ((10684\sqrt{2/3}Q)/27 + (1404680\sqrt{2/3}Q{}^{3})/27 + 
      1185688\sqrt{2/3}Q{}^{5} + (22261216\sqrt{2/3}Q{}^{7})/3)
     $%
$\partial{}^{3} J \partial{}^{3} J J \partial T + 
    ((210754\sqrt{2/3}Q)/81 + (2480684\sqrt{2/3}Q{}^{3})/27 + 
      (13575428\sqrt{2/3}Q{}^{5})/9 + (26949280\sqrt{2/3}Q{}^{7})/3)
     $%
$\partial{}^{3} J \partial{}^{3} J J \partial A_{2} + 
    ((-20776\sqrt{2/3})/81 + (12358\sqrt{2/3}Q{}^{2})/9 + 
      (413536\sqrt{2/3}Q{}^{4})/9 + 138592\sqrt{2/3}Q{}^{6})
     $%
$\partial{}^{3} J \partial{}^{3} J J A_3 + 
    ((6116\sqrt{2/3}Q)/9 + (906536\sqrt{2/3}Q{}^{3})/27 + 162616\sqrt{6}Q{}^{5} + 
      (6580768\sqrt{2/3}Q{}^{7})/3)$%
$\partial{}^{3} J \partial{}^{3} J \partial J 
      T + (13330/729 - 2244334 Q{}^{2}/243 - 44144290 Q{}^{4}/81 - 
      288717488 Q{}^{6}/27 - 198537056 Q{}^{8}/3)$%
$\partial{}^{3} J \partial{}^{3} J 
      (\partial J){}^{2} + ((171418\sqrt{2/3}Q)/81 + 
      (541816\sqrt{2/3}Q{}^{3})/9 + 710102\sqrt{2/3}Q{}^{5} + 
      (9824252\sqrt{2/3}Q{}^{7})/3)$%
$\partial{}^{3} J \partial{}^{3} J \partial J 
      A_2 + (47080/729 - 1819700 Q{}^{2}/243 - 41246152 Q{}^{4}/81 - 
      273900616 Q{}^{6}/27 - 62686368Q{}^{8})$%
$\partial{}^{3} J \partial{}^{3} J \partial{}^{2} J 
      J + (-4232/81 + 388924 Q{}^{2}/81 + 4223636 Q{}^{4}/27 + 
      3351784 Q{}^{6}/3)$%
$\partial{}^{3} J \partial{}^{3} J A_2 T + 
    (4040/243 + 268223 Q{}^{2}/81 + 2313868 Q{}^{4}/27 + 5329264 Q{}^{6}/9)
     $%
$\partial{}^{3} J \partial{}^{3} J A_2{}^{2}+ 
    (-10388/243 - 6901 Q{}^{2}/27 + 31364 Q{}^{4}/27 + 1160Q{}^{6})
     $%
$\partial{}^{3} J \partial{}^{3} J B_2{}^{2}+ 
    ((-86560\sqrt{2/3}Q)/27 - (226256\sqrt{2/3}Q{}^{3})/3 - 
      (1343192\sqrt{2/3}Q{}^{5})/3)$%
$\partial{}^{3} J A_2 \partial T \partial T + 
    ((-122860\sqrt{2/3}Q)/27 - (885644\sqrt{2/3}Q{}^{3})/9 - 
      592448\sqrt{2/3}Q{}^{5})$%
$\partial{}^{3} J A_2 \partial{}^{2} T T + 
    ((-25604\sqrt{2/3}Q)/27 - (286738\sqrt{2/3}Q{}^{3})/9 - 
      239696\sqrt{2/3}Q{}^{5})$%
$\partial{}^{3} J A_2{}^{2}\partial{}^{2} T + 
    (-206\sqrt{2/3}Q{}^{2} - 1690\sqrt{2/3}Q{}^{4})$%
$\partial{}^{3} J A_2{}^{2}\partial A_{3} + ((-1720\sqrt{2/3})/9 + 
      (20482\sqrt{2/3}Q{}^{2})/9 + (113606\sqrt{2/3}Q{}^{4})/3)
     $%
$\partial{}^{3} J A_2 A_3 \partial T + 
    ((-3872\sqrt{2/3}Q)/3 - 9602\sqrt{2/3}Q{}^{3})$%
$\partial{}^{3} J A_2 
      A_3 A_3 + ((-1720\sqrt{2/3})/9 + 
      (15956\sqrt{2/3}Q{}^{2})/9 + (104626\sqrt{2/3}Q{}^{4})/3)
     $%
$\partial{}^{3} J A_2 \partial A_{3} T + 
    ((3334\sqrt{2/3}Q{}^{2})/3 + 8782\sqrt{2/3}Q{}^{4})$%
$\partial{}^{3} J A_2 
      B_2 \partial B_{3} + ((-2044\sqrt{2/3}Q)/27 + 
      (23420\sqrt{2/3}Q{}^{3})/9 + (71288\sqrt{2/3}Q{}^{5})/3)
     $%
$\partial{}^{3} J A_2 \partial B_{2} \partial B_{2} + 
    (-1326\sqrt{6}Q{}^{2} - 9194\sqrt{6}Q{}^{4})$%
$\partial{}^{3} J A_2 
      \partial B_{2} B_3 + ((-11308\sqrt{2/3}Q)/27 - 
      (26788\sqrt{2/3}Q{}^{3})/9 - (4006\sqrt{2/3}Q{}^{5})/3)
     $%
$\partial{}^{3} J A_2 \partial{}^{2} B_{2} B_2 + 
    ((6224\sqrt{2/3}Q)/3 + 14686\sqrt{2/3}Q{}^{3})$%
$\partial{}^{3} J A_2 
      B_3 B_3 + ((-165200\sqrt{2/3}Q)/27 - 
      (1345496\sqrt{2/3}Q{}^{3})/9 - (2734352\sqrt{2/3}Q{}^{5})/3)
     $%
$\partial{}^{3} J \partial A_{2} \partial T T + 
    ((-89468\sqrt{2/3}Q)/27 - (861004\sqrt{2/3}Q{}^{3})/9 - 
      696298\sqrt{2/3}Q{}^{5})$%
$\partial{}^{3} J \partial A_{2} A_2 \partial T + 
    ((11438\sqrt{2/3}Q{}^{2})/9 + (26548\sqrt{2/3}Q{}^{4})/3)
     $%
$\partial{}^{3} J \partial A_{2} A_2 A_3 + 
    ((-20854\sqrt{2/3}Q)/27 - (104350\sqrt{2/3}Q{}^{3})/3 - 
      (905648\sqrt{2/3}Q{}^{5})/3)$%
$\partial{}^{3} J \partial A_{2} \partial A_{2} 
      T + ((-28652\sqrt{2/3}Q)/27 - (371722\sqrt{2/3}Q{}^{3})/9 - 
      320338\sqrt{2/3}Q{}^{5})$%
$\partial{}^{3} J \partial A_{2} \partial A_{2} 
      A_2 + ((-11096\sqrt{2/3})/27 - (78412\sqrt{2/3}Q{}^{2})/9 - 
      8194\sqrt{6}Q{}^{4})$%
$\partial{}^{3} J \partial A_{2} A_3 T + 
    ((4168\sqrt{2/3}Q{}^{2})/9 + (7574\sqrt{2/3}Q{}^{4})/3)
     $%
$\partial{}^{3} J \partial A_{2} B_2 B_3 + 
    ((-10384\sqrt{2/3}Q)/27 - 646\sqrt{6}Q{}^{3} + (8338\sqrt{2/3}Q{}^{5})/3)
     $%
$\partial{}^{3} J \partial A_{2} \partial B_{2} B_2 + 
    ((-30296\sqrt{2/3}Q)/27 - (129004\sqrt{2/3}Q{}^{3})/3 - 
      (926968\sqrt{2/3}Q{}^{5})/3)$%
$\partial{}^{3} J \partial{}^{2} A_{2} T{}^{2}+ 
    ((-11846\sqrt{2/3}Q)/9 - (172202\sqrt{2/3}Q{}^{3})/3 - 159046\sqrt{6}Q{}^{5})
     $%
$\partial{}^{3} J \partial{}^{2} A_{2} A_2 T + 
    ((-18521\sqrt{2/3}Q)/27 - (85876\sqrt{2/3}Q{}^{3})/3 - 
      (687046\sqrt{2/3}Q{}^{5})/3)$%
$\partial{}^{3} J \partial{}^{2} A_{2} A_2{}^{2}+ ((1889\sqrt{2/3}Q)/27 + (5704\sqrt{2/3}Q{}^{3})/3 + 
      (26824\sqrt{2/3}Q{}^{5})/3)$%
$\partial{}^{3} J \partial{}^{2} A_{2} B_2{}^{2}+ ((-46672\sqrt{2/3})/27 - (27736\sqrt{2/3}Q{}^{2})/3 + 
      (53560\sqrt{2/3}Q{}^{4})/3)$%
$\partial{}^{3} J A_3 \partial T T + 
    (1360\sqrt{2/3}Q + 898\sqrt{6}Q{}^{3})$%
$\partial{}^{3} J A_3 
      A_3 T + ((-1252\sqrt{2/3}Q)/3 - 1664\sqrt{2/3}Q{}^{3})
     $%
$\partial{}^{3} J A_3 B_2 B_3 + 
    (136\sqrt{2/3}Q{}^{2} - 504\sqrt{6}Q{}^{4})$%
$\partial{}^{3} J A_3 
      \partial B_{2} B_2 + ((-20368\sqrt{2/3})/27 - 
      (6548\sqrt{2/3}Q{}^{2})/3 + (80584\sqrt{2/3}Q{}^{4})/3)
     $%
$\partial{}^{3} J \partial A_{3} T{}^{2}+ 
    ((-2704\sqrt{2/3}Q{}^{2})/3 - 4792\sqrt{2/3}Q{}^{4})$%
$\partial{}^{3} J \partial A_{3} 
      B_2{}^{2}+ ((-25060\sqrt{2/3}Q)/27 - 
      16222\sqrt{2/3}Q{}^{3} - (189488\sqrt{2/3}Q{}^{5})/3)
     $%
$\partial{}^{3} J B_2{}^{2}\partial{}^{2} T + 
    ((-15208\sqrt{2/3})/27 - (35302\sqrt{2/3}Q{}^{2})/9 - 
      (27970\sqrt{2/3}Q{}^{4})/3)$%
$\partial{}^{3} J B_2 B_3 
      \partial T + ((-15208\sqrt{2/3})/27 - (16652\sqrt{2/3}Q{}^{2})/9 + 
      2318\sqrt{2/3}Q{}^{4})$%
$\partial{}^{3} J B_2 \partial B_{3} T + 
    ((-34012\sqrt{2/3}Q)/9 - (586384\sqrt{2/3}Q{}^{3})/9 - 
      (751598\sqrt{2/3}Q{}^{5})/3)$%
$\partial{}^{3} J \partial B_{2} B_2 
      \partial T + ((-4378\sqrt{2/3}Q)/3 - (244490\sqrt{2/3}Q{}^{3})/9 - 
      (284248\sqrt{2/3}Q{}^{5})/3)$%
$\partial{}^{3} J \partial B_{2} \partial B_{2} 
      T + ((-15208\sqrt{2/3})/27 - (61940\sqrt{2/3}Q{}^{2})/9 - 
      37334\sqrt{2/3}Q{}^{4})$%
$\partial{}^{3} J \partial B_{2} B_3 T + 
    ((-15770\sqrt{2/3}Q)/9 - (294482\sqrt{2/3}Q{}^{3})/9 - 
      (366790\sqrt{2/3}Q{}^{5})/3)$%
$\partial{}^{3} J \partial{}^{2} B_{2} B_2 
      T + ((2468\sqrt{2/3}Q)/3 + 10006\sqrt{2/3}Q{}^{3})
     $%
$\partial{}^{3} J B_3 B_3 T + 
    ((-21604\sqrt{2/3}Q)/27 - (74636\sqrt{2/3}Q{}^{3})/3 - 
      (549968\sqrt{2/3}Q{}^{5})/3)$%
$\partial{}^{4} J \partial T T{}^{2}+ 
    (-17120/243 - 291880 Q{}^{2}/81 - 1643728 Q{}^{4}/27 - 3000512 Q{}^{6}/9)
     $%
$\partial{}^{4} J J \partial T \partial T + 
    (-35872/243 - 516608 Q{}^{2}/81 - 2646608 Q{}^{4}/27 - 4616512 Q{}^{6}/9)
     $%
$\partial{}^{4} J J \partial{}^{2} T T + 
    ((-2395\sqrt{2/3}Q)/9 + (54568\sqrt{2/3}Q{}^{3})/9 + 
      (820646\sqrt{2/3}Q{}^{5})/3 + 1949656\sqrt{2/3}Q{}^{7})
     $%
$\partial{}^{4} J J{}^{2}\partial{}^{3} T + 
    (84947 Q/(81\sqrt{6}) + (443605\sqrt{2/3}Q{}^{3})/27 + 
      104828\sqrt{6}Q{}^{5} + (6556246\sqrt{2/3}Q{}^{7})/3)
     $%
$\partial{}^{4} J J{}^{2}\partial{}^{3} A_{2} + 
    ((-2645\sqrt{2/3})/27 + (48331\sqrt{2/3}Q{}^{2})/9 + 
      (279556\sqrt{2/3}Q{}^{4})/3 + 107258\sqrt{6}Q{}^{6})
     $%
$\partial{}^{4} J J{}^{2}\partial{}^{2} A_{3} + 
    (-17476/243 - 492620 Q{}^{2}/81 - 2694776 Q{}^{4}/27 - 4817152 Q{}^{6}/9)
     $%
$\partial{}^{4} J J A_2 \partial{}^{2} T + 
    ((-11518Q)/27 - 49516 Q{}^{3}/9 - 51016 Q{}^{5}/3)
     $%
$\partial{}^{4} J J A_2 \partial A_{3} + 
    (-52636/243 - 987632 Q{}^{2}/81 - 4698404 Q{}^{4}/27 - 7437256 Q{}^{6}/9)
     $%
$\partial{}^{4} J J \partial A_{2} \partial T + 
    (-5849/243 - 97099 Q{}^{2}/81 - 824398 Q{}^{4}/27 - 1834664 Q{}^{6}/9)
     $%
$\partial{}^{4} J J \partial A_{2} \partial A_{2} + 
    ((-8416Q)/27 - 21944 Q{}^{3}/9 - 2176Q{}^{5})$%
$\partial{}^{4} J J 
      \partial A_{2} A_3 + (-17476/243 - 526388 Q{}^{2}/81 - 
      2813432 Q{}^{4}/27 - 4573432 Q{}^{6}/9)$%
$\partial{}^{4} J J 
      \partial{}^{2} A_{2} T + (-6205/243 - 100682 Q{}^{2}/81 - 961778 Q{}^{4}/27 - 
      2327152 Q{}^{6}/9)$%
$\partial{}^{4} J J \partial{}^{2} A_{2} A_2 + 
    (35608 Q/9 + 60336Q{}^{3} + 202968Q{}^{5})$%
$\partial{}^{4} J J 
      A_3 \partial T + (2527/27 + 1328 Q{}^{2}/3 - 1000 Q{}^{4}/3)
     $%
$\partial{}^{4} J J A_3 A_3 + 
    (92536 Q/27 + 484364 Q{}^{3}/9 + 181776Q{}^{5})$%
$\partial{}^{4} J J 
      \partial A_{3} T + ((-14206Q)/27 - 5212Q{}^{3} - 44312 Q{}^{5}/3)
     $%
$\partial{}^{4} J J B_2 \partial B_{3} + 
    (-3757/81 + 6155 Q{}^{2}/3 + 318794 Q{}^{4}/9 + 327584 Q{}^{6}/3)
     $%
$\partial{}^{4} J J \partial B_{2} \partial B_{2} + 
    ((-21416Q)/27 - 86620 Q{}^{3}/9 - 33208Q{}^{5})$%
$\partial{}^{4} J J 
      \partial B_{2} B_3 + (-6823/81 + 10418 Q{}^{2}/9 + 251234 Q{}^{4}/9 + 
      262136 Q{}^{6}/3)$%
$\partial{}^{4} J J \partial{}^{2} B_{2} B_2 + 
    (539/27 + 6208 Q{}^{2}/9 + 3848Q{}^{4})$%
$\partial{}^{4} J J B_3 
      B_3 + (-31708/243 + 225568 Q{}^{2}/81 + 4697656 Q{}^{4}/27 + 
      13152224 Q{}^{6}/9)$%
$\partial{}^{4} J \partial J \partial T T + 
    ((-9958\sqrt{2/3}Q)/81 + (1684654\sqrt{2/3}Q{}^{3})/27 + 
      (15215068\sqrt{2/3}Q{}^{5})/9 + (33268496\sqrt{2/3}Q{}^{7})/3)
     $%
$\partial{}^{4} J \partial J J \partial{}^{2} T + 
    ((200450\sqrt{2/3}Q)/81 + (2489413\sqrt{2/3}Q{}^{3})/27 + 
      (16052518\sqrt{2/3}Q{}^{5})/9 + (35912684\sqrt{2/3}Q{}^{7})/3)
     $%
$\partial{}^{4} J \partial J J \partial{}^{2} A_{2} + 
    ((-10580\sqrt{2/3})/27 + (49552\sqrt{2/3}Q{}^{2})/3 + 
      (866098\sqrt{2/3}Q{}^{4})/3 + 969232\sqrt{2/3}Q{}^{6})
     $%
$\partial{}^{4} J \partial J J \partial A_{3} + 
    ((36194\sqrt{2/3}Q)/81 + (1102108\sqrt{2/3}Q{}^{3})/27 + 
      (6755950\sqrt{2/3}Q{}^{5})/9 + (11706488\sqrt{2/3}Q{}^{7})/3)
     $%
$\partial{}^{4} J (\partial J){}^{2} \partial T + 
    ((198992\sqrt{2/3}Q)/81 + (1781440\sqrt{2/3}Q{}^{3})/27 + 
      (7701721\sqrt{2/3}Q{}^{5})/9 + (13856708\sqrt{2/3}Q{}^{7})/3)
     $%
$\partial{}^{4} J (\partial J){}^{2} \partial A_{2} + 
    ((-5290\sqrt{2/3})/27 + (45181\sqrt{2/3}Q{}^{2})/9 + 
      (290842\sqrt{2/3}Q{}^{4})/3 + 326348\sqrt{2/3}Q{}^{6})
     $%
$\partial{}^{4} J (\partial J){}^{2} A_3 + 
    (-31708/243 + 345796 Q{}^{2}/81 + 5420008 Q{}^{4}/27 + 13921112 Q{}^{6}/9)
     $%
$\partial{}^{4} J \partial J A_2 \partial T + 
    ((-7795Q)/27 - 9686 Q{}^{3}/3 - 23822 Q{}^{5}/3)$%
$\partial{}^{4} J \partial J 
      A_2 A_3 + (-31708/243 + 174988 Q{}^{2}/81 + 
      4738144 Q{}^{4}/27 + 13591904 Q{}^{6}/9)$%
$\partial{}^{4} J \partial J 
      \partial A_{2} T + (32/243 + 685495 Q{}^{2}/81 + 6519436 Q{}^{4}/27 + 
      15662354 Q{}^{6}/9)$%
$\partial{}^{4} J \partial J \partial A_{2} A_2 + 
    (44228 Q/9 + 69020Q{}^{3} + 217552Q{}^{5})$%
$\partial{}^{4} J \partial J 
      A_3 T + ((-3893Q)/9 - 4450Q{}^{3} - 13018Q{}^{5})
     $%
$\partial{}^{4} J \partial J B_2 B_3 + 
    (-10580/81 + 63241 Q{}^{2}/27 + 150656 Q{}^{4}/3 + 469102 Q{}^{6}/3)
     $%
$\partial{}^{4} J \partial J \partial B_{2} B_2 + 
    (-4042/81 + 179252 Q{}^{2}/81 + 905996 Q{}^{4}/9 + 7227856 Q{}^{6}/9)
     $%
$\partial{}^{4} J \partial{}^{2} J T{}^{2}+ 
    ((50036\sqrt{2/3}Q)/81 + (2286568\sqrt{2/3}Q{}^{3})/27 + 
      (17606632\sqrt{2/3}Q{}^{5})/9 + (37043744\sqrt{2/3}Q{}^{7})/3)
     $%
$\partial{}^{4} J \partial{}^{2} J J \partial T + 
    ((338987\sqrt{2/3}Q)/81 + (4025582\sqrt{2/3}Q{}^{3})/27 + 
      (7326674\sqrt{2/3}Q{}^{5})/3 + (43499132\sqrt{2/3}Q{}^{7})/3)
     $%
$\partial{}^{4} J \partial{}^{2} J J \partial A_{2} + 
    ((-3827\sqrt{2/3})/9 + (45874\sqrt{2/3}Q{}^{2})/27 + 
      (224560\sqrt{2/3}Q{}^{4})/3 + (766412\sqrt{2/3}Q{}^{6})/3)
     $%
$\partial{}^{4} J \partial{}^{2} J J A_3 + 
    ((97604\sqrt{2/3}Q)/81 + (1634212\sqrt{2/3}Q{}^{3})/27 + 
      (7965412\sqrt{2/3}Q{}^{5})/9 + (11960912\sqrt{2/3}Q{}^{7})/3)
     $%
$\partial{}^{4} J \partial{}^{2} J \partial J T + 
    (685/27 - 414299 Q{}^{2}/27 - 894840Q{}^{4} - 52161094 Q{}^{6}/3 - 
      106828632Q{}^{8})$%
$\partial{}^{4} J \partial{}^{2} J (\partial J){}^{2} + 
    ((283226\sqrt{2/3}Q)/81 + (2765264\sqrt{2/3}Q{}^{3})/27 + 
      (10845556\sqrt{2/3}Q{}^{5})/9 + (16210348\sqrt{2/3}Q{}^{7})/3)
     $%
$\partial{}^{4} J \partial{}^{2} J \partial J A_2 + 
    (1420/27 - 488128 Q{}^{2}/81 - 33657592 Q{}^{4}/81 - 74893280 Q{}^{6}/9 - 
      463733504 Q{}^{8}/9)$%
$\partial{}^{4} J \partial{}^{2} J \partial{}^{2} J J + 
    (-8084/81 + 59552 Q{}^{2}/9 + 6359348 Q{}^{4}/27 + 15547240 Q{}^{6}/9)
     $%
$\partial{}^{4} J \partial{}^{2} J A_2 T + 
    (3397/162 + 132011 Q{}^{2}/27 + 1175615 Q{}^{4}/9 + 2752640 Q{}^{6}/3)
     $%
$\partial{}^{4} J \partial{}^{2} J A_2{}^{2}+ 
    (-3827/54 - 13217 Q{}^{2}/27 + 1907Q{}^{4} + 18376 Q{}^{6}/3)
     $%
$\partial{}^{4} J \partial{}^{2} J B_2{}^{2}+ 
    ((77516\sqrt{2/3}Q)/81 + (582548\sqrt{2/3}Q{}^{3})/9 + 
      (11623768\sqrt{2/3}Q{}^{5})/9 + 7831264\sqrt{2/3}Q{}^{7})
     $%
$\partial{}^{4} J \partial{}^{3} J J T + 
    ((181366\sqrt{2/3}Q)/81 + (2439224\sqrt{2/3}Q{}^{3})/27 + 
      (13459228\sqrt{2/3}Q{}^{5})/9 + (25899584\sqrt{2/3}Q{}^{7})/3)
     $%
$\partial{}^{4} J \partial{}^{3} J J A_2 + 
    (54680/729 - 2220758 Q{}^{2}/243 - 51041978 Q{}^{4}/81 - 342530404 Q{}^{6}/27 - 
      78841840Q{}^{8})$%
$\partial{}^{4} J \partial{}^{3} J \partial J J + 
    (13865/972 - 4291 Q{}^{2}/81 - 768703 Q{}^{4}/27 - 6283004 Q{}^{6}/9 - 
      13994752 Q{}^{8}/3)$%
$\partial{}^{4} J \partial{}^{4} J J{}^{2}+ 
    ((-10700\sqrt{2/3}Q)/3 - (701660\sqrt{2/3}Q{}^{3})/9 - 
      (1372168\sqrt{2/3}Q{}^{5})/3)$%
$\partial{}^{4} J A_2 \partial T T + 
    ((-23386\sqrt{2/3}Q)/27 - (225907\sqrt{2/3}Q{}^{3})/9 - 
      (550264\sqrt{2/3}Q{}^{5})/3)$%
$\partial{}^{4} J A_2{}^{2}\partial T + ((-128\sqrt{2/3}Q{}^{2})/3 - 338\sqrt{2/3}Q{}^{4})
     $%
$\partial{}^{4} J A_2{}^{3}3 + 
    ((-1241\sqrt{2/3})/9 - (14014\sqrt{2/3}Q{}^{2})/9 - (11606\sqrt{2/3}Q{}^{4})/3)
     $%
$\partial{}^{4} J A_2 A_3 T + 
    ((86\sqrt{2/3}Q{}^{2})/3 + 358\sqrt{2/3}Q{}^{4})$%
$\partial{}^{4} J A_2 
      B_2 B_3 + ((-1118\sqrt{2/3}Q)/27 + 229\sqrt{2/3}Q{}^{3} + 
      (1214\sqrt{2/3}Q{}^{5})/3)$%
$\partial{}^{4} J A_2 \partial B_{2} 
      B_2 + ((-38842\sqrt{2/3}Q)/27 - (105566\sqrt{2/3}Q{}^{3})/3 - 
      (654392\sqrt{2/3}Q{}^{5})/3)$%
$\partial{}^{4} J \partial A_{2} T{}^{2}+ 
    ((-32831\sqrt{2/3}Q)/27 - (126490\sqrt{2/3}Q{}^{3})/3 - 
      (1003570\sqrt{2/3}Q{}^{5})/3)$%
$\partial{}^{4} J \partial A_{2} A_2 
      T + ((-21962\sqrt{2/3}Q)/27 - 49231 Q{}^{3}/\sqrt{6} - 
      (529114\sqrt{2/3}Q{}^{5})/3)$%
$\partial{}^{4} J \partial A_{2} A_2{}^{2}+ ((164\sqrt{2/3}Q)/3 + 25549 Q{}^{3}/(9\sqrt{6}) + 
      (17560\sqrt{2/3}Q{}^{5})/3)$%
$\partial{}^{4} J \partial A_{2} B_2{}^{2}+ ((-12382\sqrt{2/3})/27 - 4400\sqrt{2/3}Q{}^{2} - 
      (30536\sqrt{2/3}Q{}^{4})/3)$%
$\partial{}^{4} J A_3 T{}^{2}+ 
    (-142\sqrt{2/3}Q{}^{2} - 928\sqrt{2/3}Q{}^{4})$%
$\partial{}^{4} J A_3 
      B_2{}^{2}+ ((-27838\sqrt{2/3}Q)/27 - 
      15479\sqrt{2/3}Q{}^{3} - (153632\sqrt{2/3}Q{}^{5})/3)
     $%
$\partial{}^{4} J B_2{}^{2}\partial T + 
    ((-8659\sqrt{2/3})/27 - (28718\sqrt{2/3}Q{}^{2})/9 - 9074\sqrt{2/3}Q{}^{4})
     $%
$\partial{}^{4} J B_2 B_3 T + 
    ((-17029\sqrt{2/3}Q)/9 - (265826\sqrt{2/3}Q{}^{3})/9 - 
      (296170\sqrt{2/3}Q{}^{5})/3)$%
$\partial{}^{4} J \partial B_{2} B_2 
      T + ((-884\sqrt{2/3}Q)/9 - (9508\sqrt{2/3}Q{}^{3})/3 - 
      23888\sqrt{2/3}Q{}^{5})$%
$\partial{}^{5} J T{}^{3}+ 
    (-106372/1215 - 1313488 Q{}^{2}/405 - 1203952 Q{}^{4}/27 - 1974208 Q{}^{6}/9)
     $%
$\partial{}^{5} J J \partial T T + 
    ((-61747\sqrt{2/3}Q)/405 + (31339\sqrt{2/3}Q{}^{3})/5 + 
      (10951444\sqrt{2/3}Q{}^{5})/45 + 1754368\sqrt{2/3}Q{}^{7})
     $%
$\partial{}^{5} J J{}^{2}\partial{}^{2} T + 
    (338077 Q/(405\sqrt{6}) + 1357679 Q{}^{3}/(45\sqrt{6}) + 
      (12344053\sqrt{2/3}Q{}^{5})/45 + 1770142\sqrt{2/3}Q{}^{7})
     $%
$\partial{}^{5} J J{}^{2}\partial{}^{2} A_{2} + 
    ((-2588\sqrt{2/3})/27 + (240076\sqrt{2/3}Q{}^{2})/135 + 
      (645758\sqrt{2/3}Q{}^{4})/15 + (559390\sqrt{2/3}Q{}^{6})/3)
     $%
$\partial{}^{5} J J{}^{2}\partial A_{3} + 
    (-106372/1215 - 1241908 Q{}^{2}/405 - 5266844 Q{}^{4}/135 - 1737592 Q{}^{6}/9)
     $%
$\partial{}^{5} J J A_2 \partial T + 
    ((-12304Q)/135 - 15608 Q{}^{3}/15 - 8776 Q{}^{5}/3)
     $%
$\partial{}^{5} J J A_2 A_3 + 
    (-106372/1215 - 1681712 Q{}^{2}/405 - 7853024 Q{}^{4}/135 - 2464160 Q{}^{6}/9)
     $%
$\partial{}^{5} J J \partial A_{2} T + 
    (-28732/1215 - 489044 Q{}^{2}/405 - 3663758 Q{}^{4}/135 - 1559000 Q{}^{6}/9)
     $%
$\partial{}^{5} J J \partial A_{2} A_2 + 
    (253912 Q/135 + 31664Q{}^{3} + 397720 Q{}^{5}/3)$%
$\partial{}^{5} J J 
      A_3 T + ((-18652Q)/135 - 60596 Q{}^{3}/45 - 3776Q{}^{5})
     $%
$\partial{}^{5} J J B_2 B_3 + 
    (-5176/81 + 15392 Q{}^{2}/45 + 254638 Q{}^{4}/15 + 244096 Q{}^{6}/3)
     $%
$\partial{}^{5} J J \partial B_{2} B_2 + 
    (-10012/405 + 378532 Q{}^{2}/405 + 224412 Q{}^{4}/5 + 3251536 Q{}^{6}/9)
     $%
$\partial{}^{5} J \partial J T{}^{2}+ 
    ((45896\sqrt{2/3}Q)/135 + (414484\sqrt{2/3}Q{}^{3})/9 + 
      (1804196\sqrt{6}Q{}^{5})/5 + 2315312\sqrt{6}Q{}^{7})
     $%
$\partial{}^{5} J \partial J J \partial T + 
    ((18398\sqrt{2/3}Q)/9 + (3404876\sqrt{2/3}Q{}^{3})/45 + 
      (56290876\sqrt{2/3}Q{}^{5})/45 + (22181764\sqrt{2/3}Q{}^{7})/3)
     $%
$\partial{}^{5} J \partial J J \partial A_{2} + 
    ((-5176\sqrt{2/3})/27 + (2858\sqrt{2/3}Q{}^{2})/3 + 
      (631424\sqrt{2/3}Q{}^{4})/15 + 61360\sqrt{6}Q{}^{6})
     $%
$\partial{}^{5} J \partial J J A_3 + 
    ((23588\sqrt{2/3}Q)/81 + (2125624\sqrt{2/3}Q{}^{3})/135 + 
      (11087366\sqrt{2/3}Q{}^{5})/45 + (3577528\sqrt{2/3}Q{}^{7})/3)
     $%
$\partial{}^{5} J (\partial J){}^{2} T + 
    (302/81 - 22384 Q{}^{2}/9 - 19470316 Q{}^{4}/135 - 41661458 Q{}^{6}/15 - 
      50807608 Q{}^{8}/3)$%
$\partial{}^{5} J (\partial J){}^{3} + 
    ((346591\sqrt{2/3}Q)/405 + (3527248\sqrt{2/3}Q{}^{3})/135 + 
      (523607\sqrt{6}Q{}^{5})/5 + (4218260\sqrt{2/3}Q{}^{7})/3)
     $%
$\partial{}^{5} J (\partial J){}^{2} A_2 + 
    (-20024/405 + 230660 Q{}^{2}/81 + 4760756 Q{}^{4}/45 + 7076104 Q{}^{6}/9)
     $%
$\partial{}^{5} J \partial J A_2 T + 
    (976/135 + 279631 Q{}^{2}/135 + 2528773 Q{}^{4}/45 + 1193320 Q{}^{6}/3)
     $%
$\partial{}^{5} J \partial J A_2{}^{2}+ 
    (-2588/81 - 26491 Q{}^{2}/135 + 2137Q{}^{4} + 13040Q{}^{6})
     $%
$\partial{}^{5} J \partial J B_2{}^{2}+ 
    ((86596\sqrt{2/3}Q)/135 + (658892\sqrt{2/3}Q{}^{3})/15 + 
      (13277648\sqrt{2/3}Q{}^{5})/15 + 1801728\sqrt{6}Q{}^{7})
     $%
$\partial{}^{5} J \partial{}^{2} J J T + 
    ((198488\sqrt{2/3}Q)/135 + (301744\sqrt{2/3}Q{}^{3})/5 + 
      (45233338\sqrt{2/3}Q{}^{5})/45 + (17401396\sqrt{2/3}Q{}^{7})/3)
     $%
$\partial{}^{5} J \partial{}^{2} J J A_2 + 
    (11860/243 - 818966 Q{}^{2}/135 - 11381788 Q{}^{4}/27 - 42545408 Q{}^{6}/5 - 
      52900928Q{}^{8})$%
$\partial{}^{5} J \partial{}^{2} J \partial J J + 
    (16957/729 - 2153 Q{}^{2}/27 - 18679909 Q{}^{4}/405 - 5654238 Q{}^{6}/5 - 
      67936952 Q{}^{8}/9)$%
$\partial{}^{5} J \partial{}^{3} J J{}^{2}+ 
    ((-99998\sqrt{2/3}Q)/135 - (746206\sqrt{2/3}Q{}^{3})/45 - 
      99968\sqrt{2/3}Q{}^{5})$%
$\partial{}^{5} J A_2 T{}^{2}+ 
    ((-3548\sqrt{2/3}Q)/15 - (361117\sqrt{2/3}Q{}^{3})/45 - 
      (192248\sqrt{2/3}Q{}^{5})/3)$%
$\partial{}^{5} J A_2{}^{2}T + ((-13717Q)/(45\sqrt{6}) - 74287 Q{}^{3}/(9\sqrt{6}) - 
      (84020\sqrt{2/3}Q{}^{5})/3)$%
$\partial{}^{5} J A_2{}^{3}+ (4877 Q/(45\sqrt{6}) + 6845 Q{}^{3}/(3\sqrt{6}) + 
      (16084\sqrt{2/3}Q{}^{5})/3)$%
$\partial{}^{5} J A_2 B_2{}^{2}+ ((-68066\sqrt{2/3}Q)/135 - (377089\sqrt{2/3}Q{}^{3})/45 - 
      (103928\sqrt{2/3}Q{}^{5})/3)$%
$\partial{}^{5} J B_2{}^{2}T + (-18874/1215 - 231628 Q{}^{2}/405 - 1030088 Q{}^{4}/135 - 
      324448 Q{}^{6}/9)$%
$\partial{}^{6} J J T{}^{2}+ 
    ((-386\sqrt{2/3}Q)/45 + (188726\sqrt{2/3}Q{}^{3})/45 + 
      (1836622\sqrt{2/3}Q{}^{5})/15 + 850552\sqrt{2/3}Q{}^{7})
     $%
$\partial{}^{6} J J{}^{2}\partial T + 
    (228433 Q/(405\sqrt{6}) + (1446191\sqrt{2/3}Q{}^{3})/135 + 
      (292643\sqrt{6}Q{}^{5})/5 + (3048454\sqrt{2/3}Q{}^{7})/3)
     $%
$\partial{}^{6} J J{}^{2}\partial A_{2} + 
    (-2119/(27\sqrt{6}) - (56204\sqrt{2/3}Q{}^{2})/135 - 
      (17696\sqrt{2/3}Q{}^{4})/15 - (2102\sqrt{2/3}Q{}^{6})/3)
     $%
$\partial{}^{6} J J{}^{2}A_3 + 
    (-37748/1215 - 140048 Q{}^{2}/135 - 1803652 Q{}^{4}/135 - 64168Q{}^{6})
     $%
$\partial{}^{6} J J A_2 T + 
    (-5963/2430 - 51293 Q{}^{2}/405 - 418976 Q{}^{4}/135 - 189800 Q{}^{6}/9)
     $%
$\partial{}^{6} J J A_2{}^{2}+ 
    (-2119/162 - 5551 Q{}^{2}/27 - 10456 Q{}^{4}/9 - 7888 Q{}^{6}/3)
     $%
$\partial{}^{6} J J B_2{}^{2}+ 
    ((11636\sqrt{2/3}Q)/45 + (2498012\sqrt{2/3}Q{}^{3})/135 + 
      (5751808\sqrt{2/3}Q{}^{5})/15 + (7151168\sqrt{2/3}Q{}^{7})/3)
     $%
$\partial{}^{6} J \partial J J T + 
    ((48950\sqrt{2/3}Q)/81 + (3508072\sqrt{2/3}Q{}^{3})/135 + 
      (19932664\sqrt{2/3}Q{}^{5})/45 + 2574052\sqrt{2/3}Q{}^{7})
     $%
$\partial{}^{6} J \partial J J A_2 + 
    (2435/243 - 477359 Q{}^{2}/405 - 11404642 Q{}^{4}/135 - 77615764 Q{}^{6}/45 - 
      10777104Q{}^{8})$%
$\partial{}^{6} J (\partial J){}^{2} J + 
    (3224/243 + 938 Q{}^{2}/27 - 1793716 Q{}^{4}/81 - 566948Q{}^{6} - 
      34466672 Q{}^{8}/9)$%
$\partial{}^{6} J \partial{}^{2} J J{}^{2}+ 
    ((74552\sqrt{2/3}Q)/2835 + (214472\sqrt{2/3}Q{}^{3})/105 + 
      (2747576\sqrt{2/3}Q{}^{5})/63 + 91552\sqrt{6}Q{}^{7})
     $%
$\partial{}^{7} J J{}^{2}T + 
    ((26911\sqrt{2/3}Q)/405 + (2754293\sqrt{2/3}Q{}^{3})/945 + 
      (2269963\sqrt{2/3}Q{}^{5})/45 + (885568\sqrt{2/3}Q{}^{7})/3)
     $%
$\partial{}^{7} J J{}^{2}A_2 + 
    (23981/5103 + 474533 Q{}^{2}/8505 - 15177166 Q{}^{4}/2835 - 
      143384492 Q{}^{6}/945 - 3152656 Q{}^{8}/3)$%
$\partial{}^{7} J \partial J 
      J{}^{2}+ (10511/20412 + 270163 Q{}^{2}/8505 + 
      2500996 Q{}^{4}/2835 + 11079326 Q{}^{6}/945 + 520808 Q{}^{8}/9)
     $%
$\partial{}^{8} J J{}^{3}+ 
    (2096/27 + 8896 Q{}^{2}/3 + 70192 Q{}^{4}/3)$%
$A_2 \partial{}^{2} T \partial T 
      \partial T + (3760/27 + 7696 Q{}^{2}/3 + 50192 Q{}^{4}/3)
     $%
$A_2 \partial{}^{2} T \partial{}^{2} T T + 
    (17984/81 + 42208 Q{}^{2}/9 + 266800 Q{}^{4}/9)$%
$A_2 \partial{}^{3} T 
      \partial T T + (680/9 + 11168 Q{}^{2}/9 + 21424 Q{}^{4}/3)
     $%
$A_2 \partial{}^{4} T T{}^{2}+ 
    (166/9 + 2932 Q{}^{2}/3 + 8672Q{}^{4})$%
$A_2{}^{2}\partial{}^{2} T 
      \partial{}^{2} T + (1384/27 + 47312 Q{}^{2}/27 + 127192 Q{}^{4}/9)
     $%
$A_2{}^{2}\partial{}^{3} T \partial T + 
    (494/9 + 6448 Q{}^{2}/9 + 13016 Q{}^{4}/3)$%
$A_2{}^{2}\partial{}^{4} T 
      T + (-271/27 - 208 Q{}^{2}/9 + 2672 Q{}^{4}/3)
     $%
$A_2{}^{3}\partial{}^{4} T + 
    (10Q - 56 Q{}^{3}/3)$%
$A_2{}^{3}\partial{}^{3} A_{3} + 
    ((-668Q)/9 - 676Q{}^{3})$%
$A_2{}^{3}3 \partial{}^{3} T + 
    (-212Q - 2028Q{}^{3})$%
$A_2{}^{2}\partial A_{3} \partial{}^{2} T + 
    (260/3 + 752Q{}^{2})$%
$A_2{}^{2}\partial A_{3} \partial A_{3} + 
    ((-644Q)/3 - 2140Q{}^{3})$%
$A_2{}^{2}\partial{}^{2} A_{3} \partial T + 
    (184/3 + 696Q{}^{2})$%
$A_2{}^{2}\partial{}^{2} A_{3} A_3 + 
    ((-370Q)/9 - 2092 Q{}^{3}/3)$%
$A_2{}^{2}\partial{}^{3} A_{3} 
      T + ((-44Q)/3 - 348Q{}^{3})$%
$A_2{}^{2}B_2 
      \partial{}^{3} B_{3} + (652 Q/3 + 1200Q{}^{3})$%
$A_2{}^{2}\partial B_{2} \partial{}^{2} B_{3} + (-563/27 - 724Q{}^{2} - 14836 Q{}^{4}/3)
     $%
$A_2{}^{2}\partial{}^{2} B_{2} \partial{}^{2} B_{2} + 
    (740 Q/3 + 2628Q{}^{3})$%
$A_2{}^{2}\partial{}^{2} B_{2} 
      \partial B_{3} + (-1232/81 - 2816 Q{}^{2}/3 - 57976 Q{}^{4}/9)
     $%
$A_2{}^{2}\partial{}^{3} B_{2} \partial B_{2} + 
    (264Q + 1992Q{}^{3})$%
$A_2{}^{2}\partial{}^{3} B_{2} B_3 + 
    (718/81 - 692 Q{}^{2}/9 - 8476 Q{}^{4}/9)$%
$A_2{}^{2}\partial{}^{4} B_{2} B_2 + (-440/3 - 1600Q{}^{2})
     $%
$A_2{}^{2}\partial B_{3} \partial B_{3} + 
    (-856/3 - 1952Q{}^{2})$%
$A_2{}^{2}\partial{}^{2} B_{3} B_3 + 
    ((-13456Q)/9 - 41348 Q{}^{3}/3)$%
$A_2 A_3 \partial{}^{2} T 
      \partial T + ((-5924Q)/9 - 3956Q{}^{3})$%
$A_2 A_3 \partial{}^{3} T 
      T + (-856/9 - 356 Q{}^{2}/3)$%
$A_2 A_3 A_3 
      \partial{}^{2} T + (236 + 964Q{}^{2})$%
$A_2 A_3 B_2 
      \partial{}^{2} B_{3} + (84 + 1336Q{}^{2})$%
$A_2 A_3 \partial B_{2} 
      \partial B_{3} + ((-620Q)/3 - 2664Q{}^{3})$%
$A_2 A_3 
      \partial{}^{2} B_{2} \partial B_{2} + (148/3 + 740Q{}^{2})$%
$A_2 A_3 
      \partial{}^{2} B_{2} B_3 + ((-2168Q)/9 - 3928 Q{}^{3}/3)
     $%
$A_2 A_3 \partial{}^{3} B_{2} B_2 + 
    ((-13312Q)/9 - 38840 Q{}^{3}/3)$%
$A_2 \partial A_{3} \partial T 
      \partial T + ((-14420Q)/9 - 44356 Q{}^{3}/3)$%
$A_2 \partial A_{3} 
      \partial{}^{2} T T + (-560/9 - 5512 Q{}^{2}/3)$%
$A_2 \partial A_{3} 
      A_3 \partial T + (-156 - 512Q{}^{2})$%
$A_2 \partial A_{3} 
      \partial A_{3} T + (272/3 + 1336Q{}^{2})$%
$A_2 \partial A_{3} 
      B_2 \partial B_{3} + ((-812Q)/3 - 1528Q{}^{3})
     $%
$A_2 \partial A_{3} \partial B_{2} \partial B_{2} + 
    (740/3 + 1528Q{}^{2})$%
$A_2 \partial A_{3} \partial B_{2} B_3 + 
    (-76Q - 1240Q{}^{3})$%
$A_2 \partial A_{3} \partial{}^{2} B_{2} B_2 + 
    ((-15476Q)/9 - 42520 Q{}^{3}/3)$%
$A_2 \partial{}^{2} A_{3} \partial T 
      T + (-68 - 336Q{}^{2})$%
$A_2 \partial{}^{2} A_{3} A_3 
      T + (244/3 + 740Q{}^{2})$%
$A_2 \partial{}^{2} A_{3} B_2 
      B_3 + (28 Q/3 + 40Q{}^{3})$%
$A_2 \partial{}^{2} A_{3} 
      \partial B_{2} B_2 + ((-3100Q)/9 - 8180 Q{}^{3}/3)
     $%
$A_2 \partial{}^{3} A_{3} T{}^{2}+ 
    (278 Q/9 + 268Q{}^{3})$%
$A_2 \partial{}^{3} A_{3} B_2{}^{2}+ (1237/81 + 1520 Q{}^{2}/9 + 2768 Q{}^{4}/9)
     $%
$A_2 B_2{}^{2}\partial{}^{4} T + 
    (212 Q/3 + 716Q{}^{3})$%
$A_2 B_2 B_3 \partial{}^{3} T + 
    ((-1604Q)/9 - 4564 Q{}^{3}/3)$%
$A_2 B_2 \partial B_{3} 
      \partial{}^{2} T + ((-2924Q)/9 - 9112 Q{}^{3}/3)$%
$A_2 B_2 
      \partial{}^{2} B_{3} \partial T + ((-1528Q)/3 - 2568Q{}^{3})
     $%
$A_2 B_2 \partial{}^{3} B_{3} T + 
    (4936/81 + 8564 Q{}^{2}/9 + 35012 Q{}^{4}/9)$%
$A_2 \partial B_{2} 
      B_2 \partial{}^{3} T + (48 + 6488 Q{}^{2}/9 + 6736 Q{}^{4}/3)
     $%
$A_2 \partial B_{2} \partial B_{2} \partial{}^{2} T + 
    (1124 Q/9 + 5908 Q{}^{3}/3)$%
$A_2 \partial B_{2} B_3 
      \partial{}^{2} T + (-576Q - 4644Q{}^{3})$%
$A_2 \partial B_{2} \partial B_{3} 
      \partial T + ((-1808Q)/3 - 4556Q{}^{3})$%
$A_2 \partial B_{2} 
      \partial{}^{2} B_{3} T + (1520/27 + 4096 Q{}^{2}/3 + 19324 Q{}^{4}/3)
     $%
$A_2 \partial{}^{2} B_{2} B_2 \partial{}^{2} T + 
    (2668/27 + 5398 Q{}^{2}/3 + 23744 Q{}^{4}/3)$%
$A_2 \partial{}^{2} B_{2} 
      \partial B_{2} \partial T + (1604/27 + 1958 Q{}^{2}/3 + 4096 Q{}^{4}/3)
     $%
$A_2 \partial{}^{2} B_{2} \partial{}^{2} B_{2} T + 
    (1676 Q/9 - 116 Q{}^{3}/3)$%
$A_2 \partial{}^{2} B_{2} B_3 
      \partial T + (-548Q - 3344Q{}^{3})$%
$A_2 \partial{}^{2} B_{2} \partial B_{3} 
      T + (7220/81 + 29450 Q{}^{2}/27 + 44936 Q{}^{4}/9)
     $%
$A_2 \partial{}^{3} B_{2} B_2 \partial T + 
    (4694/81 + 7892 Q{}^{2}/9 + 20176 Q{}^{4}/9)$%
$A_2 \partial{}^{3} B_{2} 
      \partial B_{2} T + (-140Q - 652Q{}^{3})$%
$A_2 \partial{}^{3} B_{2} 
      B_3 T + (2792/81 + 6526 Q{}^{2}/9 + 22240 Q{}^{4}/9)
     $%
$A_2 \partial{}^{4} B_{2} B_2 T + 
    (968/9 + 268 Q{}^{2}/3)$%
$A_2 B_3 B_3 \partial{}^{2} T + 
    (592/9 + 6848 Q{}^{2}/3)$%
$A_2 \partial B_{3} B_3 \partial T + 
    (572/3 + 808Q{}^{2})$%
$A_2 \partial B_{3} \partial B_{3} T + 
    (592/3 + 1112Q{}^{2})$%
$A_2 \partial{}^{2} B_{3} B_3 T + 
    (1184/27 + 17120 Q{}^{2}/9 + 43664 Q{}^{4}/3)$%
$\partial A_{2} \partial T \partial T 
      \partial T + (11216/27 + 29024 Q{}^{2}/3 + 188464 Q{}^{4}/3)
     $%
$\partial A_{2} \partial{}^{2} T \partial T T + 
    (12176/81 + 27680 Q{}^{2}/9 + 152608 Q{}^{4}/9)$%
$\partial A_{2} \partial{}^{3} T 
      T{}^{2}+ (248 + 24292 Q{}^{2}/3 + 63100Q{}^{4})
     $%
$\partial A_{2} A_2 \partial{}^{2} T \partial T + 
    (7184/27 + 40988 Q{}^{2}/9 + 83420 Q{}^{4}/3)$%
$\partial A_{2} A_2 
      \partial{}^{3} T T + (-3244/81 - 7384 Q{}^{2}/27 + 35420 Q{}^{4}/9)
     $%
$\partial A_{2} A_2{}^{2}\partial{}^{3} T + 
    ((-44Q)/3 - 236Q{}^{3})$%
$\partial A_{2} A_2{}^{2}\partial{}^{2} A_{3} + (116 Q/9 - 4160 Q{}^{3}/3)$%
$\partial A_{2} A_2 
      A_3 \partial{}^{2} T + (-128Q - 2092Q{}^{3})$%
$\partial A_{2} A_2 
      \partial A_{3} \partial T + (1852/9 + 6584 Q{}^{2}/3)
     $%
$\partial A_{2} A_2 \partial A_{3} A_3 + 
    ((-608Q)/9 - 3628 Q{}^{3}/3)$%
$\partial A_{2} A_2 \partial{}^{2} A_{3} 
      T + ((-3080Q)/9 - 6184 Q{}^{3}/3)$%
$\partial A_{2} A_2 
      B_2 \partial{}^{2} B_{3} + (1496 Q/3 + 3492Q{}^{3})
     $%
$\partial A_{2} A_2 \partial B_{2} \partial B_{3} + 
    (-1694/27 - 3134Q{}^{2} - 66700 Q{}^{4}/3)$%
$\partial A_{2} A_2 
      \partial{}^{2} B_{2} \partial B_{2} + (5120 Q/9 + 15268 Q{}^{3}/3)
     $%
$\partial A_{2} A_2 \partial{}^{2} B_{2} B_3 + 
    (3800/81 - 4810 Q{}^{2}/27 - 37684 Q{}^{4}/9)$%
$\partial A_{2} A_2 
      \partial{}^{3} B_{2} B_2 + (-5852/9 - 17776 Q{}^{2}/3)
     $%
$\partial A_{2} A_2 \partial B_{3} B_3 + 
    (976/9 + 31832 Q{}^{2}/9 + 80332 Q{}^{4}/3)$%
$\partial A_{2} \partial A_{2} 
      \partial T \partial T + (4444/27 + 39008 Q{}^{2}/9 + 29440Q{}^{4})
     $%
$\partial A_{2} \partial A_{2} \partial{}^{2} T T + 
    (-160/3 - 2348 Q{}^{2}/3 + 4956Q{}^{4})$%
$\partial A_{2} \partial A_{2} A_2 
      \partial{}^{2} T + ((-1028Q)/9 - 3292 Q{}^{3}/3)$%
$\partial A_{2} \partial A_{2} 
      A_2 \partial A_{3} + (-40/9 + 56 Q{}^{2}/9 + 15964 Q{}^{4}/3)
     $%
$\partial A_{2} \partial A_{2} \partial A_{2} \partial T + 
    (566/27 + 9548 Q{}^{2}/9 + 23264 Q{}^{4}/3)$%
$\partial A_{2} \partial A_{2} 
      \partial A_{2} \partial A_{2} + ((-1136Q)/9 - 2788 Q{}^{3}/3)
     $%
$\partial A_{2} \partial A_{2} \partial A_{2} A_3 + 
    ((-280Q)/9 - 7220 Q{}^{3}/3)$%
$\partial A_{2} \partial A_{2} A_3 
      \partial T + (1136/9 + 2860 Q{}^{2}/3)$%
$\partial A_{2} \partial A_{2} 
      A_3 A_3 + (28 Q/9 - 5080 Q{}^{3}/3)
     $%
$\partial A_{2} \partial A_{2} \partial A_{3} T + 
    ((-2440Q)/9 - 6680 Q{}^{3}/3)$%
$\partial A_{2} \partial A_{2} B_2 
      \partial B_{3} + (-836/27 - 11816 Q{}^{2}/9 - 26288 Q{}^{4}/3)
     $%
$\partial A_{2} \partial A_{2} \partial B_{2} \partial B_{2} + 
    (1040 Q/3 + 2788Q{}^{3})$%
$\partial A_{2} \partial A_{2} \partial B_{2} 
      B_3 + (10/3 - 5776 Q{}^{2}/9 - 16628 Q{}^{4}/3)
     $%
$\partial A_{2} \partial A_{2} \partial{}^{2} B_{2} B_2 + 
    (-2140/9 - 6056 Q{}^{2}/3)$%
$\partial A_{2} \partial A_{2} B_3 
      B_3 + ((-8480Q)/9 - 28624 Q{}^{3}/3)$%
$\partial A_{2} A_3 
      \partial T \partial T + (-500Q - 3236Q{}^{3})$%
$\partial A_{2} A_3 
      \partial{}^{2} T T + (-592/9 + 712 Q{}^{2}/3)$%
$\partial A_{2} A_3 
      A_3 \partial T - 48Q$%
$\partial A_{2} A_3 A_3 
      A_3 + (1304/9 + 4036 Q{}^{2}/3)$%
$\partial A_{2} A_3 
      B_2 \partial B_{3} + ((-424Q)/9 - 920 Q{}^{3}/3)
     $%
$\partial A_{2} A_3 \partial B_{2} \partial B_{2} + 
    (944/9 + 3028 Q{}^{2}/3)$%
$\partial A_{2} A_3 \partial B_{2} 
      B_3 + ((-496Q)/9 - 1352 Q{}^{3}/3)$%
$\partial A_{2} A_3 
      \partial{}^{2} B_{2} B_2 + 96Q$%
$\partial A_{2} A_3 B_3 
      B_3 + ((-11192Q)/9 - 42904 Q{}^{3}/3)$%
$\partial A_{2} \partial A_{3} 
      \partial T T + (-380/3 - 1384Q{}^{2})$%
$\partial A_{2} \partial A_{3} 
      A_3 T + (1520/9 + 3892 Q{}^{2}/3)
     $%
$\partial A_{2} \partial A_{3} B_2 B_3 + 
    (208 Q/3 + 1024Q{}^{3})$%
$\partial A_{2} \partial A_{3} \partial B_{2} 
      B_2 + ((-620Q)/3 - 2412Q{}^{3})$%
$\partial A_{2} \partial{}^{2} A_{3} 
      T{}^{2}+ (1064 Q/9 + 2572 Q{}^{3}/3)
     $%
$\partial A_{2} \partial{}^{2} A_{3} B_2{}^{2}+ 
    (1276/27 + 7340 Q{}^{2}/9 + 10424 Q{}^{4}/3)$%
$\partial A_{2} B_2{}^{2}\partial{}^{3} T + (368 Q/3 + 2308Q{}^{3})$%
$\partial A_{2} B_2 
      B_3 \partial{}^{2} T + ((-2864Q)/9 - 3232 Q{}^{3}/3)
     $%
$\partial A_{2} B_2 \partial B_{3} \partial T + 
    ((-1832Q)/3 - 3480Q{}^{3})$%
$\partial A_{2} B_2 \partial{}^{2} B_{3} T + 
    (352/3 + 27388 Q{}^{2}/9 + 49028 Q{}^{4}/3)$%
$\partial A_{2} \partial B_{2} 
      B_2 \partial{}^{2} T + (1304/27 + 4520 Q{}^{2}/3 + 18172 Q{}^{4}/3)
     $%
$\partial A_{2} \partial B_{2} \partial B_{2} \partial T + 
    (296Q + 4596Q{}^{3})$%
$\partial A_{2} \partial B_{2} B_3 \partial T + 
    ((-4352Q)/9 - 6928 Q{}^{3}/3)$%
$\partial A_{2} \partial B_{2} \partial B_{3} 
      T + (4808/27 + 29068 Q{}^{2}/9 + 13700Q{}^{4})
     $%
$\partial A_{2} \partial{}^{2} B_{2} B_2 \partial T + 
    (2182/27 + 16936 Q{}^{2}/9 + 14716 Q{}^{4}/3)$%
$\partial A_{2} \partial{}^{2} B_{2} 
      \partial B_{2} T + (1000 Q/9 + 3128 Q{}^{3}/3)
     $%
$\partial A_{2} \partial{}^{2} B_{2} B_3 T + 
    (296/9 + 42868 Q{}^{2}/27 + 53540 Q{}^{4}/9)$%
$\partial A_{2} \partial{}^{3} B_{2} 
      B_2 T + (1000/9 + 140 Q{}^{2}/3)$%
$\partial A_{2} B_3 
      B_3 \partial T + (1324/9 + 5432 Q{}^{2}/3)
     $%
$\partial A_{2} \partial B_{3} B_3 T + 
    (112 + 15392 Q{}^{2}/3 + 40528Q{}^{4})$%
$\partial{}^{2} A_{2} \partial T \partial T 
      T + (2224/27 + 11408 Q{}^{2}/3 + 81824 Q{}^{4}/3)
     $%
$\partial{}^{2} A_{2} \partial{}^{2} T T{}^{2}+ 
    (2528/27 + 14368 Q{}^{2}/3 + 118000 Q{}^{4}/3)$%
$\partial{}^{2} A_{2} A_2 
      \partial T \partial T + (1844/9 + 6184Q{}^{2} + 46484Q{}^{4})
     $%
$\partial{}^{2} A_{2} A_2 \partial{}^{2} T T + 
    (-1142/27 - 260 Q{}^{2}/3 + 21836 Q{}^{4}/3)$%
$\partial{}^{2} A_{2} A_2{}^{2}\partial{}^{2} T + (-96Q - 752Q{}^{3})$%
$\partial{}^{2} A_{2} A_2{}^{2}\partial A_{3} + ((-724Q)/9 - 2504 Q{}^{3}/3)
     $%
$\partial{}^{2} A_{2} A_2 A_3 \partial T + 
    (1598/9 + 3388 Q{}^{2}/3)$%
$\partial{}^{2} A_{2} A_2 A_3 
      A_3 + (664 Q/9 - 3724 Q{}^{3}/3)$%
$\partial{}^{2} A_{2} A_2 
      \partial A_{3} T + ((-2384Q)/9 - 7396 Q{}^{3}/3)
     $%
$\partial{}^{2} A_{2} A_2 B_2 \partial B_{3} + 
    (-424/9 - 13378 Q{}^{2}/9 - 30152 Q{}^{4}/3)$%
$\partial{}^{2} A_{2} A_2 
      \partial B_{2} \partial B_{2} + (3584 Q/9 + 10216 Q{}^{3}/3)
     $%
$\partial{}^{2} A_{2} A_2 \partial B_{2} B_3 + 
    (1316/27 - 688Q{}^{2} - 19268 Q{}^{4}/3)$%
$\partial{}^{2} A_{2} A_2 
      \partial{}^{2} B_{2} B_2 + (-2866/9 - 7400 Q{}^{2}/3)
     $%
$\partial{}^{2} A_{2} A_2 B_3 B_3 + 
    (6376/27 + 82424 Q{}^{2}/9 + 71560Q{}^{4})$%
$\partial{}^{2} A_{2} \partial A_{2} \partial T 
      T + (-116/9 + 11462 Q{}^{2}/9 + 83044 Q{}^{4}/3)
     $%
$\partial{}^{2} A_{2} \partial A_{2} A_2 \partial T + 
    ((-2524Q)/9 - 6680 Q{}^{3}/3)$%
$\partial{}^{2} A_{2} \partial A_{2} A_2 
      A_3 + (-56 + 7006 Q{}^{2}/9 + 63332 Q{}^{4}/3)
     $%
$\partial{}^{2} A_{2} \partial A_{2} \partial A_{2} T + 
    (586/3 + 71648 Q{}^{2}/9 + 170416 Q{}^{4}/3)$%
$\partial{}^{2} A_{2} \partial A_{2} 
      \partial A_{2} A_2 + (3172 Q/3 + 5216Q{}^{3})
     $%
$\partial{}^{2} A_{2} \partial A_{2} A_3 T + 
    ((-2536Q)/9 - 6308 Q{}^{3}/3)$%
$\partial{}^{2} A_{2} \partial A_{2} B_2 
      B_3 + (-956/27 - 13184 Q{}^{2}/9 - 29192 Q{}^{4}/3)
     $%
$\partial{}^{2} A_{2} \partial A_{2} \partial B_{2} B_2 + 
    (418/27 + 1292Q{}^{2} + 37892 Q{}^{4}/3)$%
$\partial{}^{2} A_{2} \partial{}^{2} A_{2} T{}^{2}+ (-928/27 + 3422 Q{}^{2}/3 + 54460 Q{}^{4}/3)
     $%
$\partial{}^{2} A_{2} \partial{}^{2} A_{2} A_2 T + 
    (1609/27 + 7688 Q{}^{2}/3 + 57104 Q{}^{4}/3)$%
$\partial{}^{2} A_{2} \partial{}^{2} A_{2} 
      A_2{}^{2}+ (-697/27 - 1366 Q{}^{2}/3 - 6032 Q{}^{4}/3)
     $%
$\partial{}^{2} A_{2} \partial{}^{2} A_{2} B_2{}^{2}+ 
    ((-172Q)/3 - 2288Q{}^{3})$%
$\partial{}^{2} A_{2} A_3 \partial T T + 
    (-358/3 - 620Q{}^{2})$%
$\partial{}^{2} A_{2} A_3 A_3 T + 
    (944/9 + 2560 Q{}^{2}/3)$%
$\partial{}^{2} A_{2} A_3 B_2 
      B_3 + ((-40Q)/9 + 2272 Q{}^{3}/3)$%
$\partial{}^{2} A_{2} A_3 
      \partial B_{2} B_2 + ((-484Q)/3 - 2556Q{}^{3})
     $%
$\partial{}^{2} A_{2} \partial A_{3} T{}^{2}+ (1652 Q/9 + 4024 Q{}^{3}/3)
     $%
$\partial{}^{2} A_{2} \partial A_{3} B_2{}^{2}+ 
    (1390/27 + 3188 Q{}^{2}/3 + 17168 Q{}^{4}/3)$%
$\partial{}^{2} A_{2} B_2{}^{2}\partial{}^{2} T + (728 Q/3 + 1540Q{}^{3})$%
$\partial{}^{2} A_{2} B_2 
      B_3 \partial T + ((-1280Q)/3 - 2700Q{}^{3})
     $%
$\partial{}^{2} A_{2} B_2 \partial B_{3} T + 
    (784/9 + 27376 Q{}^{2}/9 + 51272 Q{}^{4}/3)$%
$\partial{}^{2} A_{2} \partial B_{2} 
      B_2 \partial T + (382/27 + 1990 Q{}^{2}/3 + 5720 Q{}^{4}/3)
     $%
$\partial{}^{2} A_{2} \partial B_{2} \partial B_{2} T + 
    ((-464Q)/9 + 5240 Q{}^{3}/3)$%
$\partial{}^{2} A_{2} \partial B_{2} B_3 
      T + (-772/27 + 4504 Q{}^{2}/3 + 20140 Q{}^{4}/3)
     $%
$\partial{}^{2} A_{2} \partial{}^{2} B_{2} B_2 T + 
    (686/9 + 136 Q{}^{2}/3)$%
$\partial{}^{2} A_{2} B_3 B_3 T + 
    (1840/27 + 82448 Q{}^{2}/27 + 208072 Q{}^{4}/9)$%
$\partial{}^{3} A_{2} \partial T 
      T{}^{2}+ (4720/27 + 53812 Q{}^{2}/9 + 129568 Q{}^{4}/3)
     $%
$\partial{}^{3} A_{2} A_2 \partial T T + 
    (-92/9 + 19780 Q{}^{2}/27 + 95312 Q{}^{4}/9)$%
$\partial{}^{3} A_{2} A_2{}^{2}\partial T + ((-380Q)/9 - 348Q{}^{3})$%
$\partial{}^{3} A_{2} A_2{}^{3}3 + (2872 Q/9 + 1536Q{}^{3})
     $%
$\partial{}^{3} A_{2} A_2 A_3 T + 
    ((-316Q)/3 - 740Q{}^{3})$%
$\partial{}^{3} A_{2} A_2 B_2 
      B_3 + (94/81 - 2020 Q{}^{2}/3 - 40072 Q{}^{4}/9)
     $%
$\partial{}^{3} A_{2} A_2 \partial B_{2} B_2 + 
    (4592/81 + 62096 Q{}^{2}/27 + 167120 Q{}^{4}/9)$%
$\partial{}^{3} A_{2} \partial A_{2} 
      T{}^{2}+ (-5024/81 + 14752 Q{}^{2}/9 + 231308 Q{}^{4}/9)
     $%
$\partial{}^{3} A_{2} \partial A_{2} A_2 T + 
    (11174/81 + 111700 Q{}^{2}/27 + 27268Q{}^{4})$%
$\partial{}^{3} A_{2} \partial A_{2} 
      A_2{}^{2}+ (-2702/81 - 16750 Q{}^{2}/27 - 8176 Q{}^{4}/3)
     $%
$\partial{}^{3} A_{2} \partial A_{2} B_2{}^{2}+ 
    (1364 Q/3 + 1956Q{}^{3})$%
$\partial{}^{3} A_{2} A_3 T{}^{2}+ 
    (620 Q/9 + 520Q{}^{3})$%
$\partial{}^{3} A_{2} A_3 B_2{}^{2}+ (3688/81 + 28634 Q{}^{2}/27 + 13952 Q{}^{4}/3)
     $%
$\partial{}^{3} A_{2} B_2{}^{2}\partial T + 
    (568 Q/3 + 940Q{}^{3})$%
$\partial{}^{3} A_{2} B_2 B_3 T + 
    (454/81 + 3752 Q{}^{2}/3 + 49208 Q{}^{4}/9)$%
$\partial{}^{3} A_{2} \partial B_{2} 
      B_2 T + (424/81 + 3428 Q{}^{2}/9 + 28472 Q{}^{4}/9)
     $%
$\partial{}^{4} A_{2} T{}^{3}+ 
    (842/27 + 12668 Q{}^{2}/9 + 30652 Q{}^{4}/3)$%
$\partial{}^{4} A_{2} A_2 
      T{}^{2}+ (-383/27 + 1217 Q{}^{2}/3 + 5536Q{}^{4})
     $%
$\partial{}^{4} A_{2} A_2{}^{2}T + 
    (1759/81 + 587Q{}^{2} + 34268 Q{}^{4}/9)$%
$\partial{}^{4} A_{2} A_2{}^{3}+ (-467/81 - 2183 Q{}^{2}/9 - 11392 Q{}^{4}/9)
     $%
$\partial{}^{4} A_{2} A_2 B_2{}^{2}+ 
    (415/81 + 4841 Q{}^{2}/9 + 25016 Q{}^{4}/9)$%
$\partial{}^{4} A_{2} B_2{}^{2}T + ((-2464Q)/3 - 7056Q{}^{3})
     $%
$A_3 \partial T \partial T \partial T + ((-8528Q)/3 - 20256Q{}^{3})
     $%
$A_3 \partial{}^{2} T \partial T T + ((-2176Q)/3 - 3184Q{}^{3})
     $%
$A_3 \partial{}^{3} T T{}^{2}+ (-88 + 412Q{}^{2})
     $%
$A_3 A_3 \partial T \partial T + 
    (-184 - 764Q{}^{2})$%
$A_3 A_3 \partial{}^{2} T T - 
    144Q$%
$A_3 A_3 A_3 \partial T + 
    24$%
$A_3 A_3 A_3 A_3 - 
    48Q$%
$A_3 A_3 B_2 \partial B_{3} + 
    (-212/3 - 752Q{}^{2})$%
$A_3 A_3 \partial B_{2} \partial B_{2} + 
    240Q$%
$A_3 A_3 \partial B_{2} B_3 + 
    (-224/3 - 560Q{}^{2})$%
$A_3 A_3 \partial{}^{2} B_{2} B_2 - 
    96$%
$A_3 A_3 B_3 B_3 + 
    ((-2168Q)/9 - 1856Q{}^{3})$%
$A_3 B_2{}^{2}\partial{}^{3} T + 
    (-568/9 - 1760 Q{}^{2}/3)$%
$A_3 B_2 B_3 \partial{}^{2} T + 
    (8/9 - 2300 Q{}^{2}/3)$%
$A_3 B_2 \partial B_{3} \partial T + 
    (-316 - 332Q{}^{2})$%
$A_3 B_2 \partial{}^{2} B_{3} T + 
    ((-9472Q)/9 - 25520 Q{}^{3}/3)$%
$A_3 \partial B_{2} B_2 
      \partial{}^{2} T + ((-4528Q)/9 - 14168 Q{}^{3}/3)$%
$A_3 \partial B_{2} 
      \partial B_{2} \partial T + (-2152/9 - 5180 Q{}^{2}/3)
     $%
$A_3 \partial B_{2} B_3 \partial T + 
    (-116 - 776Q{}^{2})$%
$A_3 \partial B_{2} \partial B_{3} T + 
    ((-10216Q)/9 - 19208 Q{}^{3}/3)$%
$A_3 \partial{}^{2} B_{2} B_2 
      \partial T + ((-2540Q)/3 - 4152Q{}^{3})$%
$A_3 \partial{}^{2} B_{2} 
      \partial B_{2} T + (-356/3 - 748Q{}^{2})$%
$A_3 \partial{}^{2} B_{2} 
      B_3 T + ((-1928Q)/9 - 5512 Q{}^{3}/3)
     $%
$A_3 \partial{}^{3} B_{2} B_2 T + 
    240Q$%
$A_3 B_3 B_3 \partial T - 
    288Q$%
$A_3 \partial B_{3} B_3 T + 
    (-2992Q - 23216Q{}^{3})$%
$\partial A_{3} \partial T \partial T T + 
    ((-5968Q)/3 - 14512Q{}^{3})$%
$\partial A_{3} \partial{}^{2} T T{}^{2}+ 
    (-2000/3 - 2104Q{}^{2})$%
$\partial A_{3} A_3 \partial T T + 
    432Q$%
$\partial A_{3} A_3 A_3 T - 
    48Q$%
$\partial A_{3} A_3 B_2 B_3 + 
    (-488/3 - 1760Q{}^{2})$%
$\partial A_{3} A_3 \partial B_{2} B_2 + 
    (-80/3 - 376Q{}^{2})$%
$\partial A_{3} \partial A_{3} T{}^{2}+ 
    (-164/3 - 512Q{}^{2})$%
$\partial A_{3} \partial A_{3} B_2{}^{2}+ 
    ((-3088Q)/9 - 9392 Q{}^{3}/3)$%
$\partial A_{3} B_2{}^{2}\partial{}^{2} T + (-2152/9 - 3740 Q{}^{2}/3)$%
$\partial A_{3} B_2 
      B_3 \partial T + (-280/3 - 632Q{}^{2})$%
$\partial A_{3} B_2 
      \partial B_{3} T + ((-3328Q)/3 - 8704Q{}^{3})
     $%
$\partial A_{3} \partial B_{2} B_2 \partial T + 
    (-140Q - 1064Q{}^{3})$%
$\partial A_{3} \partial B_{2} \partial B_{2} T + 
    (-316/3 - 2024Q{}^{2})$%
$\partial A_{3} \partial B_{2} B_3 T + 
    ((-580Q)/3 - 1880Q{}^{3})$%
$\partial A_{3} \partial{}^{2} B_{2} B_2 T + 
    240Q$%
$\partial A_{3} B_3 B_3 T + 
    ((-4880Q)/3 - 13592Q{}^{3})$%
$\partial{}^{2} A_{3} \partial T T{}^{2}+ 
    (-952/3 - 1272Q{}^{2})$%
$\partial{}^{2} A_{3} A_3 T{}^{2}+ 
    (-212/3 - 512Q{}^{2})$%
$\partial{}^{2} A_{3} A_3 B_2{}^{2}+ 
    ((-2296Q)/9 - 4748 Q{}^{3}/3)$%
$\partial{}^{2} A_{3} B_2{}^{2}\partial T + (-308/3 - 1036Q{}^{2})$%
$\partial{}^{2} A_{3} B_2 B_3 
      T + (236 Q/3 + 8Q{}^{3})$%
$\partial{}^{2} A_{3} \partial B_{2} B_2 
      T + ((-464Q)/3 - 1880Q{}^{3})$%
$\partial{}^{3} A_{3} T{}^{3}+ (86 Q/9 + 132Q{}^{3})$%
$\partial{}^{3} A_{3} B_2{}^{2}T + (974/27 + 1276 Q{}^{2}/3 - 848 Q{}^{4}/3)
     $%
$B_2{}^{2}\partial{}^{2} T \partial{}^{2} T + 
    (136 + 12976 Q{}^{2}/9 + 7832 Q{}^{4}/3)$%
$B_2{}^{2}\partial{}^{3} T 
      \partial T + (8338/81 + 11744 Q{}^{2}/9 + 42920 Q{}^{4}/9)
     $%
$B_2{}^{2}\partial{}^{4} T T - 
    8Q{}^{3}$%
$B_2{}^{3}\partial{}^{3} B_{3} + 
    (24 + 24Q{}^{2})$%
$B_2{}^{2}\partial B_{3} \partial B_{3} + 
    8$%
$B_2{}^{2}\partial{}^{2} B_{3} B_3 + 
    ((-15200Q)/9 - 39460 Q{}^{3}/3)$%
$B_2 B_3 \partial{}^{2} T 
      \partial T + ((-2812Q)/3 - 4948Q{}^{3})$%
$B_2 B_3 \partial{}^{3} T 
      T + ((-14320Q)/9 - 36152 Q{}^{3}/3)$%
$B_2 \partial B_{3} 
      \partial T \partial T + (-2244Q - 14236Q{}^{3})$%
$B_2 \partial B_{3} 
      \partial{}^{2} T T + 96Q$%
$B_2 \partial B_{3} B_3 
      B_3 + ((-3004Q)/3 - 11368Q{}^{3})$%
$B_2 \partial{}^{2} B_{3} 
      \partial T T + ((-236Q)/3 - 2244Q{}^{3})$%
$B_2 \partial{}^{3} B_{3} 
      T{}^{2}+ (11528/27 + 50212 Q{}^{2}/9 + 14412Q{}^{4})
     $%
$\partial B_{2} B_2 \partial{}^{2} T \partial T + 
    (32752/81 + 5820Q{}^{2} + 171812 Q{}^{4}/9)$%
$\partial B_{2} B_2 \partial{}^{3} T 
      T + (-40Q - 120Q{}^{3})$%
$\partial B_{2} B_2{}^{2}\partial{}^{2} B_{3} + (-64 - 336Q{}^{2})$%
$\partial B_{2} B_2 \partial B_{3} 
      B_3 + (3760/27 + 24296 Q{}^{2}/9 + 9972Q{}^{4})
     $%
$\partial B_{2} \partial B_{2} \partial T \partial T + 
    (6004/27 + 4560Q{}^{2} + 58736 Q{}^{4}/3)$%
$\partial B_{2} \partial B_{2} \partial{}^{2} T 
      T + (72Q + 144Q{}^{3})$%
$\partial B_{2} \partial B_{2} B_2 
      \partial B_{3} + (10 + 1076 Q{}^{2}/3 + 2288Q{}^{4})$%
$\partial B_{2} \partial B_{2} 
      \partial B_{2} \partial B_{2} + (-88Q - 528Q{}^{3})$%
$\partial B_{2} \partial B_{2} 
      \partial B_{2} B_3 + (56 + 552Q{}^{2})$%
$\partial B_{2} \partial B_{2} 
      B_3 B_3 + ((-5312Q)/3 - 12304Q{}^{3})
     $%
$\partial B_{2} B_3 \partial T \partial T + 
    ((-16964Q)/9 - 46300 Q{}^{3}/3)$%
$\partial B_{2} B_3 \partial{}^{2} T 
      T - 288Q$%
$\partial B_{2} B_3 B_3 B_3 + 
    ((-33176Q)/9 - 85048 Q{}^{3}/3)$%
$\partial B_{2} \partial B_{3} \partial T 
      T + (-604Q - 7268Q{}^{3})$%
$\partial B_{2} \partial{}^{2} B_{3} T{}^{2}+ (5840/27 + 32744 Q{}^{2}/9 + 39728 Q{}^{4}/3)
     $%
$\partial{}^{2} B_{2} B_2 \partial T \partial T + 
    (11276/27 + 21512 Q{}^{2}/3 + 79612 Q{}^{4}/3)$%
$\partial{}^{2} B_{2} B_2 
      \partial{}^{2} T T + (-36Q - 24Q{}^{3})$%
$\partial{}^{2} B_{2} B_2{}^{2}\partial B_{3} + (40 + 48Q{}^{2})$%
$\partial{}^{2} B_{2} B_2 
      B_3 B_3 + (11848/27 + 93856 Q{}^{2}/9 + 147496 Q{}^{4}/3)
     $%
$\partial{}^{2} B_{2} \partial B_{2} \partial T T + 
    (-32Q - 144Q{}^{3})$%
$\partial{}^{2} B_{2} \partial B_{2} B_2 B_3 + 
    (542/9 + 1868Q{}^{2} + 11872Q{}^{4})$%
$\partial{}^{2} B_{2} \partial B_{2} \partial B_{2} 
      B_2 + (1054/27 + 1628Q{}^{2} + 26564 Q{}^{4}/3)
     $%
$\partial{}^{2} B_{2} \partial{}^{2} B_{2} T{}^{2}+ 
    (197/9 + 1582 Q{}^{2}/3 + 2984Q{}^{4})$%
$\partial{}^{2} B_{2} \partial{}^{2} B_{2} B_2{}^{2}+ ((-25124Q)/9 - 52864 Q{}^{3}/3)$%
$\partial{}^{2} B_{2} B_3 
      \partial T T + ((-3524Q)/3 - 9020Q{}^{3})$%
$\partial{}^{2} B_{2} \partial B_{3} 
      T{}^{2}+ (2632/9 + 128032 Q{}^{2}/27 + 181952 Q{}^{4}/9)
     $%
$\partial{}^{3} B_{2} B_2 \partial T T + 
    (8Q - 8Q{}^{3})$%
$\partial{}^{3} B_{2} B_2{}^{3}3 + 
    (7672/81 + 20144 Q{}^{2}/9 + 108392 Q{}^{4}/9)$%
$\partial{}^{3} B_{2} \partial B_{2} 
      T{}^{2}+ (874/27 + 6340 Q{}^{2}/9 + 3872Q{}^{4})
     $%
$\partial{}^{3} B_{2} \partial B_{2} B_2{}^{2}+ 
    (-692Q - 4252Q{}^{3})$%
$\partial{}^{3} B_{2} B_3 T{}^{2}+ 
    (3766/81 + 2764 Q{}^{2}/3 + 39356 Q{}^{4}/9)$%
$\partial{}^{4} B_{2} B_2 
      T{}^{2}+ (128/27 + 614 Q{}^{2}/9 + 320Q{}^{4})
     $%
$\partial{}^{4} B_{2} B_2{}^{3}+ 
    (-1288/9 - 6332 Q{}^{2}/3)$%
$B_3 B_3 \partial T \partial T + 
    (-3616/9 - 6020 Q{}^{2}/3)$%
$B_3 B_3 \partial{}^{2} T T + 
    72$%
$B_3 B_3 B_3 B_3 + 
    (-5840/9 - 16456 Q{}^{2}/3)$%
$\partial B_{3} B_3 \partial T T + 
    (-464/3 - 1144Q{}^{2})$%
$\partial B_{3} \partial B_{3} T{}^{2}+ 
    (-352/3 - 920Q{}^{2})$%
$\partial{}^{2} B_{3} B_3 T{}^{2}+ 
    ((416\sqrt{2/3}Q)/3 + 1664\sqrt{2/3}Q{}^{3})$%
$J \partial T \partial T 
      \partial T T + (160\sqrt{2/3}Q + 640\sqrt{6}Q{}^{3})
     $%
$J \partial{}^{2} T \partial T T{}^{2}+ 
    ((64\sqrt{2/3}Q)/3 + 256\sqrt{2/3}Q{}^{3})$%
$J \partial{}^{3} T T{}^{3}+ (1048/27 + 5440 Q{}^{2}/3 + 48512 Q{}^{4}/3)
     $%
$J{}^{2}\partial{}^{2} T \partial T \partial T + 
    (1880/27 + 6352 Q{}^{2}/3 + 46144 Q{}^{4}/3)$%
$J{}^{2}\partial{}^{2} T 
      \partial{}^{2} T T + (8992/81 + 3104Q{}^{2} + 191360 Q{}^{4}/9)
     $%
$J{}^{2}\partial{}^{3} T \partial T T + 
    (340/9 + 8504 Q{}^{2}/9 + 17696 Q{}^{4}/3)$%
$J{}^{2}\partial{}^{4} T 
      T{}^{2}+ ((4984\sqrt{2/3}Q)/27 + (20296\sqrt{2/3}Q{}^{3})/9 + 
      160\sqrt{6}Q{}^{5})$%
$J{}^{3}\partial{}^{3} T \partial{}^{2} T + 
    ((9920\sqrt{2/3}Q)/81 + (5672\sqrt{2/3}Q{}^{3})/3 + 
      (45472\sqrt{2/3}Q{}^{5})/9)$%
$J{}^{3}\partial{}^{4} T 
      \partial T + ((5584\sqrt{2/3}Q)/405 + (808\sqrt{2/3}Q{}^{3})/5 - 
      (416\sqrt{2/3}Q{}^{5})/9)$%
$J{}^{3}\partial{}^{5} T 
      T + (5651/7290 - 8624 Q{}^{2}/135 - 695032 Q{}^{4}/405 - 
      90496 Q{}^{6}/9)$%
$J{}^{4}\partial{}^{6} T + 
    (-443/972 + 8726 Q{}^{2}/405 + 3652 Q{}^{4}/135 - 23096 Q{}^{6}/9)
     $%
$J{}^{4}\partial{}^{6} A_{2} + 
    ((-11077Q)/270 - 10747 Q{}^{3}/15 - 10760 Q{}^{5}/3)
     $%
$J{}^{4}\partial{}^{5} A_{3} + 
    ((1312\sqrt{2/3}Q)/9 + (188194\sqrt{2/3}Q{}^{3})/135 + 
      (26308\sqrt{2/3}Q{}^{5})/9)$%
$J{}^{3}A_2 \partial{}^{5} T + ((629\sqrt{2/3})/81 - (1069\sqrt{2/3}Q{}^{2})/3 - 
      (24812\sqrt{2/3}Q{}^{4})/9)$%
$J{}^{3}A_2 \partial{}^{4} A_{3} + ((32828\sqrt{2/3}Q)/81 + 1222\sqrt{6}Q{}^{3} + 
      (55540\sqrt{2/3}Q{}^{5})/9)$%
$J{}^{3}\partial A_{2} \partial{}^{4} T + ((22\sqrt{2/3})/3 - (7046\sqrt{2/3}Q{}^{2})/9 - 
      (13408\sqrt{2/3}Q{}^{4})/3)$%
$J{}^{3}\partial A_{2} \partial{}^{3} A_{3} + ((15082\sqrt{2/3}Q)/27 + 
      (66682\sqrt{2/3}Q{}^{3})/9 + (71240\sqrt{2/3}Q{}^{5})/3)
     $%
$J{}^{3}\partial{}^{2} A_{2} \partial{}^{3} T + 
    ((191\sqrt{2/3})/27 - (3422\sqrt{2/3}Q{}^{2})/9 - 3460\sqrt{2/3}Q{}^{4})
     $%
$J{}^{3}\partial{}^{2} A_{2} \partial{}^{2} A_{3} + 
    ((40874\sqrt{2/3}Q)/81 + (246868\sqrt{2/3}Q{}^{3})/27 + 
      42380\sqrt{2/3}Q{}^{5})$%
$J{}^{3}\partial{}^{3} A_{2} 
      \partial{}^{2} T + ((-9229\sqrt{2/3}Q)/81 - (41441\sqrt{2/3}Q{}^{3})/27 - 
      (5264\sqrt{2/3}Q{}^{5})/3)$%
$J{}^{3}\partial{}^{3} A_{2} 
      \partial{}^{2} A_{2} + ((478\sqrt{2/3})/27 + (4826\sqrt{2/3}Q{}^{2})/9 + 
      (8600\sqrt{2/3}Q{}^{4})/3)$%
$J{}^{3}\partial{}^{3} A_{2} 
      \partial A_{3} + ((20836\sqrt{2/3}Q)/81 + (148456\sqrt{2/3}Q{}^{3})/27 + 
      (263620\sqrt{2/3}Q{}^{5})/9)$%
$J{}^{3}\partial{}^{4} A_{2} \partial T + ((-11719Q)/(81\sqrt{6}) - 
      (17771\sqrt{2/3}Q{}^{3})/27 - (14300\sqrt{2/3}Q{}^{5})/9)
     $%
$J{}^{3}\partial{}^{4} A_{2} \partial A_{2} + 
    (2623/(81\sqrt{6}) + (722\sqrt{2/3}Q{}^{2})/3 + (20104\sqrt{2/3}Q{}^{4})/9)
     $%
$J{}^{3}\partial{}^{4} A_{2} A_3 + 
    ((17854\sqrt{2/3}Q)/405 + (9994\sqrt{2/3}Q{}^{3})/9 + 
      (54352\sqrt{2/3}Q{}^{5})/9)$%
$J{}^{3}\partial{}^{5} A_{2} T + ((-4621Q)/(135\sqrt{6}) + 
      (27112\sqrt{2/3}Q{}^{3})/135 + (17740\sqrt{2/3}Q{}^{5})/9)
     $%
$J{}^{3}\partial{}^{5} A_{2} A_2 + 
    ((-5194\sqrt{2/3})/81 - (1672\sqrt{2/3}Q{}^{2})/9 - (8168\sqrt{2/3}Q{}^{4})/9)
     $%
$J{}^{3}A_3 \partial{}^{4} T + 
    ((-1144\sqrt{2/3})/9 - 2216\sqrt{2/3}Q{}^{2} - (45572\sqrt{2/3}Q{}^{4})/3)
     $%
$J{}^{3}\partial A_{3} \partial{}^{3} T + 
    ((-2728\sqrt{2/3})/27 - (41456\sqrt{2/3}Q{}^{2})/9 - 11120\sqrt{6}Q{}^{4})
     $%
$J{}^{3}\partial{}^{2} A_{3} \partial{}^{2} T + 
    (124\sqrt{6}Q + 344\sqrt{6}Q{}^{3})$%
$J{}^{3}\partial{}^{2} A_{3} \partial A_{3} + ((-836\sqrt{2/3})/27 - 
      (27158\sqrt{2/3}Q{}^{2})/9 - (69956\sqrt{2/3}Q{}^{4})/3)
     $%
$J{}^{3}\partial{}^{3} A_{3} \partial T + 
    (124\sqrt{6}Q + 280\sqrt{6}Q{}^{3})$%
$J{}^{3}\partial{}^{3} A_{3} A_3 + ((-1168\sqrt{2/3})/81 - 
      (2864\sqrt{2/3}Q{}^{2})/3 - (67808\sqrt{2/3}Q{}^{4})/9)
     $%
$J{}^{3}\partial{}^{4} A_{3} T + 
    (-331/(9\sqrt{6}) + 187\sqrt{2/3}Q{}^{2} - 328\sqrt{2/3}Q{}^{4})
     $%
$J{}^{3}B_2 \partial{}^{4} B_{3} + 
    (-10\sqrt{2/3} - 536\sqrt{2/3}Q{}^{2} - 2512\sqrt{6}Q{}^{4})
     $%
$J{}^{3}\partial B_{2} \partial{}^{3} B_{3} + 
    ((-41\sqrt{2/3})/9 - 662\sqrt{6}Q{}^{2} - 17248\sqrt{2/3}Q{}^{4})
     $%
$J{}^{3}\partial{}^{2} B_{2} \partial{}^{2} B_{3} + 
    ((-8011\sqrt{2/3}Q)/27 - 688\sqrt{6}Q{}^{3} + (7168\sqrt{2/3}Q{}^{5})/3)
     $%
$J{}^{3}\partial{}^{3} B_{2} \partial{}^{2} B_{2} + 
    ((518\sqrt{2/3})/9 - 2056\sqrt{2/3}Q{}^{2} - 15856\sqrt{2/3}Q{}^{4})
     $%
$J{}^{3}\partial{}^{3} B_{2} \partial B_{3} + 
    ((-12349Q)/(27\sqrt{6}) - (6928\sqrt{2/3}Q{}^{3})/3 - 
      (10252\sqrt{2/3}Q{}^{5})/3)$%
$J{}^{3}\partial{}^{4} B_{2} \partial B_{2} + (703/(9\sqrt{6}) - (1739\sqrt{2/3}Q{}^{2})/3 - 
      1736\sqrt{6}Q{}^{4})$%
$J{}^{3}\partial{}^{4} B_{2} 
      B_3 + ((-8321\sqrt{2/3}Q)/135 - (17512\sqrt{2/3}Q{}^{3})/15 - 
      (10100\sqrt{2/3}Q{}^{5})/3)$%
$J{}^{3}\partial{}^{5} B_{2} B_2 + (692\sqrt{2/3}Q + 1876\sqrt{6}Q{}^{3})
     $%
$J{}^{3}\partial{}^{2} B_{3} \partial B_{3} + 
    (140\sqrt{2/3}Q + 700\sqrt{6}Q{}^{3})$%
$J{}^{3}\partial{}^{3} B_{3} B_3 + (1472/27 + 10712 Q{}^{2}/3 + 76744 Q{}^{4}/3)
     $%
$J{}^{2}A_2 \partial{}^{2} T \partial{}^{2} T + 
    (5056/27 + 155216 Q{}^{2}/27 + 313624 Q{}^{4}/9)$%
$J{}^{2}A_2 \partial{}^{3} T \partial T + (12784/81 + 9440 Q{}^{2}/3 + 
      148496 Q{}^{4}/9)$%
$J{}^{2}A_2 \partial{}^{4} T T + 
    (253/27 + 2800 Q{}^{2}/3 + 6436Q{}^{4})$%
$J{}^{2}A_2{}^{2}\partial{}^{4} T + ((-65Q)/9 + 14Q{}^{3})$%
$J{}^{2}A_2{}^{2}\partial{}^{3} A_{3} + ((-2530Q)/9 - 1978Q{}^{3})
     $%
$J{}^{2}A_2 A_3 \partial{}^{3} T + 
    ((-2074Q)/9 - 9458 Q{}^{3}/3)$%
$J{}^{2}A_2 
      \partial A_{3} \partial{}^{2} T + (-78 - 640Q{}^{2})$%
$J{}^{2}A_2 \partial A_{3} \partial A_{3} + ((-3130Q)/9 - 11564 Q{}^{3}/3)
     $%
$J{}^{2}A_2 \partial{}^{2} A_{3} \partial T + 
    (-34 - 552Q{}^{2})$%
$J{}^{2}A_2 \partial{}^{2} A_{3} 
      A_3 + ((-692Q)/9 - 3844 Q{}^{3}/3)$%
$J{}^{2}A_2 \partial{}^{3} A_{3} T + (132Q + 732Q{}^{3})
     $%
$J{}^{2}A_2 B_2 \partial{}^{3} B_{3} + 
    (208Q + 826Q{}^{3})$%
$J{}^{2}A_2 \partial B_{2} 
      \partial{}^{2} B_{3} + (802/27 + 691 Q{}^{2}/3 + 1568 Q{}^{4}/3)
     $%
$J{}^{2}A_2 \partial{}^{2} B_{2} \partial{}^{2} B_{2} + 
    ((-302Q)/3 - 1128Q{}^{3})$%
$J{}^{2}A_2 \partial{}^{2} B_{2} 
      \partial B_{3} + (2347/81 - 1046 Q{}^{2}/9 - 8152 Q{}^{4}/9)
     $%
$J{}^{2}A_2 \partial{}^{3} B_{2} \partial B_{2} + 
    ((-10Q)/3 - 422Q{}^{3})$%
$J{}^{2}A_2 \partial{}^{3} B_{2} 
      B_3 + (1396/81 - 617 Q{}^{2}/9 - 5872 Q{}^{4}/9)
     $%
$J{}^{2}A_2 \partial{}^{4} B_{2} B_2 + 
    (286/3 + 948Q{}^{2})$%
$J{}^{2}A_2 \partial B_{3} 
      \partial B_{3} + (296/3 + 748Q{}^{2})$%
$J{}^{2}A_2 
      \partial{}^{2} B_{3} B_3 + (9112/27 + 125960 Q{}^{2}/9 + 92408Q{}^{4})
     $%
$J{}^{2}\partial A_{2} \partial{}^{2} T \partial T + 
    (27152/81 + 24160 Q{}^{2}/3 + 405040 Q{}^{4}/9)$%
$J{}^{2}\partial A_{2} \partial{}^{3} T T + (5944/81 + 132698 Q{}^{2}/27 + 
      99230 Q{}^{4}/3)$%
$J{}^{2}\partial A_{2} A_2 
      \partial{}^{3} T + (680 Q/9 + 4282 Q{}^{3}/3)$%
$J{}^{2}\partial A_{2} A_2 \partial{}^{2} A_{3} + 
    (1822/27 + 39320 Q{}^{2}/9 + 29808Q{}^{4})$%
$J{}^{2}\partial A_{2} \partial A_{2} \partial{}^{2} T + (314 Q/3 + 2156Q{}^{3})
     $%
$J{}^{2}\partial A_{2} \partial A_{2} \partial A_{3} + 
    ((-6074Q)/9 - 8950 Q{}^{3}/3)$%
$J{}^{2}\partial A_{2} 
      A_3 \partial{}^{2} T + (548 Q/9 + 3700 Q{}^{3}/3)
     $%
$J{}^{2}\partial A_{2} \partial A_{3} \partial T + 
    (-190/3 - 500Q{}^{2})$%
$J{}^{2}\partial A_{2} \partial A_{3} 
      A_3 + (404 Q/9 + 1564 Q{}^{3}/3)$%
$J{}^{2}\partial A_{2} \partial{}^{2} A_{3} T + (3556 Q/9 + 4220 Q{}^{3}/3)
     $%
$J{}^{2}\partial A_{2} B_2 \partial{}^{2} B_{3} + 
    (2176 Q/9 + 1976 Q{}^{3}/3)$%
$J{}^{2}\partial A_{2} 
      \partial B_{2} \partial B_{3} + (1091/27 - 3724 Q{}^{2}/9 - 6178 Q{}^{4}/3)
     $%
$J{}^{2}\partial A_{2} \partial{}^{2} B_{2} \partial B_{2} + 
    (284 Q/3 - 428Q{}^{3})$%
$J{}^{2}\partial A_{2} \partial{}^{2} B_{2} 
      B_3 + (148/9 - 10150 Q{}^{2}/27 - 13646 Q{}^{4}/9)
     $%
$J{}^{2}\partial A_{2} \partial{}^{3} B_{2} B_2 + 
    (662/9 + 2716 Q{}^{2}/3)$%
$J{}^{2}\partial A_{2} \partial B_{3} 
      B_3 + (1048/9 + 17392 Q{}^{2}/3 + 40520Q{}^{4})
     $%
$J{}^{2}\partial{}^{2} A_{2} \partial T \partial T + 
    (5768/27 + 82472 Q{}^{2}/9 + 186800 Q{}^{4}/3)$%
$J{}^{2}\partial{}^{2} A_{2} \partial{}^{2} T T + (1622/27 + 47260 Q{}^{2}/9 + 40442Q{}^{4})
     $%
$J{}^{2}\partial{}^{2} A_{2} A_2 \partial{}^{2} T + 
    (1412 Q/9 + 7258 Q{}^{3}/3)$%
$J{}^{2}\partial{}^{2} A_{2} 
      A_2 \partial A_{3} + (3188/27 + 5852Q{}^{2} + 111580 Q{}^{4}/3)
     $%
$J{}^{2}\partial{}^{2} A_{2} \partial A_{2} \partial T + 
    (-28 - 187 Q{}^{2}/3 + 4598Q{}^{4})$%
$J{}^{2}\partial{}^{2} A_{2} 
      \partial A_{2} \partial A_{2} + ((-490Q)/9 - 704 Q{}^{3}/3)
     $%
$J{}^{2}\partial{}^{2} A_{2} \partial A_{2} A_3 + 
    (418/27 + 8252 Q{}^{2}/9 + 7260Q{}^{4})$%
$J{}^{2}\partial{}^{2} A_{2} 
      \partial{}^{2} A_{2} T + (-464/27 - 1283 Q{}^{2}/9 + 6382 Q{}^{4}/3)
     $%
$J{}^{2}\partial{}^{2} A_{2} \partial{}^{2} A_{2} A_2 + 
    ((-2978Q)/9 - 2680 Q{}^{3}/3)$%
$J{}^{2}\partial{}^{2} A_{2} 
      A_3 \partial T + (-179/3 - 118Q{}^{2})$%
$J{}^{2}\partial{}^{2} A_{2} A_3 A_3 + (1292 Q/9 + 9580 Q{}^{3}/3)
     $%
$J{}^{2}\partial{}^{2} A_{2} \partial A_{3} T + 
    (1696 Q/9 + 3566 Q{}^{3}/3)$%
$J{}^{2}\partial{}^{2} A_{2} 
      B_2 \partial B_{3} + (191/27 + 137 Q{}^{2}/9 - 2260 Q{}^{4}/3)
     $%
$J{}^{2}\partial{}^{2} A_{2} \partial B_{2} \partial B_{2} + 
    (328 Q/3 + 404Q{}^{3})$%
$J{}^{2}\partial{}^{2} A_{2} \partial B_{2} 
      B_3 + (-386/27 - 796 Q{}^{2}/9 - 974Q{}^{4})
     $%
$J{}^{2}\partial{}^{2} A_{2} \partial{}^{2} B_{2} B_2 + 
    (343/9 + 644 Q{}^{2}/3)$%
$J{}^{2}\partial{}^{2} A_{2} B_3 
      B_3 + (5104/27 + 173984 Q{}^{2}/27 + 374056 Q{}^{4}/9)
     $%
$J{}^{2}\partial{}^{3} A_{2} \partial T T + 
    (2360/27 + 31082 Q{}^{2}/9 + 73552 Q{}^{4}/3)$%
$J{}^{2}\partial{}^{3} A_{2} A_2 \partial T + ((-188Q)/9 + 48Q{}^{3})
     $%
$J{}^{2}\partial{}^{3} A_{2} A_2 A_3 + 
    (4592/81 + 52208 Q{}^{2}/27 + 129440 Q{}^{4}/9)$%
$J{}^{2}\partial{}^{3} A_{2} \partial A_{2} T + 
    (-2512/81 + 848 Q{}^{2}/9 + 57142 Q{}^{4}/9)$%
$J{}^{2}\partial{}^{3} A_{2} \partial A_{2} A_2 + (2852 Q/9 + 2388Q{}^{3})
     $%
$J{}^{2}\partial{}^{3} A_{2} A_3 T + 
    (92 Q/3 + 150Q{}^{3})$%
$J{}^{2}\partial{}^{3} A_{2} B_2 
      B_3 + (227/81 + 5740 Q{}^{2}/9 + 26908 Q{}^{4}/9)
     $%
$J{}^{2}\partial{}^{3} A_{2} \partial B_{2} B_2 + 
    (2336/81 + 9398 Q{}^{2}/9 + 69508 Q{}^{4}/9)$%
$J{}^{2}\partial{}^{4} A_{2} T{}^{2}+ (842/27 + 11276 Q{}^{2}/9 + 27836 Q{}^{4}/3)
     $%
$J{}^{2}\partial{}^{4} A_{2} A_2 T + 
    (-383/54 + 275 Q{}^{2}/18 + 4064 Q{}^{4}/3)$%
$J{}^{2}\partial{}^{4} A_{2} A_2{}^{2}+ 
    (415/162 + 4121 Q{}^{2}/18 + 13372 Q{}^{4}/9)$%
$J{}^{2}\partial{}^{4} A_{2} B_2{}^{2}+ ((-3112Q)/3 - 14736Q{}^{3})
     $%
$J{}^{2}A_3 \partial{}^{2} T \partial T + 
    ((-1792Q)/3 - 5488Q{}^{3})$%
$J{}^{2}A_3 \partial{}^{3} T 
      T + (4 + 194Q{}^{2})$%
$J{}^{2}A_3 
      A_3 \partial{}^{2} T + (-158 - 358Q{}^{2})$%
$J{}^{2}A_3 B_2 \partial{}^{2} B_{3} + (-58 - 388Q{}^{2})
     $%
$J{}^{2}A_3 \partial B_{2} \partial B_{3} + 
    (1514 Q/3 + 2340Q{}^{3})$%
$J{}^{2}A_3 \partial{}^{2} B_{2} 
      \partial B_{2} + (-178/3 - 182Q{}^{2})$%
$J{}^{2}A_3 
      \partial{}^{2} B_{2} B_3 + (2780 Q/9 + 2812 Q{}^{3}/3)
     $%
$J{}^{2}A_3 \partial{}^{3} B_{2} B_2 + 
    48Q$%
$J{}^{2}A_3 \partial B_{3} B_3 + 
    ((-5368Q)/9 - 25592 Q{}^{3}/3)$%
$J{}^{2}\partial A_{3} 
      \partial T \partial T + ((-9520Q)/9 - 38240 Q{}^{3}/3)
     $%
$J{}^{2}\partial A_{3} \partial{}^{2} T T + 
    (-1000/3 - 1244Q{}^{2})$%
$J{}^{2}\partial A_{3} A_3 
      \partial T - 168Q$%
$J{}^{2}\partial A_{3} A_3 
      A_3 + (-80/3 - 1336Q{}^{2})$%
$J{}^{2}\partial A_{3} 
      \partial A_{3} T + (-140/3 - 316Q{}^{2})$%
$J{}^{2}\partial A_{3} B_2 \partial B_{3} + (622 Q/3 + 876Q{}^{3})
     $%
$J{}^{2}\partial A_{3} \partial B_{2} \partial B_{2} + 
    (-158/3 + 140Q{}^{2})$%
$J{}^{2}\partial A_{3} \partial B_{2} 
      B_3 + (926 Q/3 + 1236Q{}^{3})$%
$J{}^{2}\partial A_{3} \partial{}^{2} B_{2} B_2 - 
    72Q$%
$J{}^{2}\partial A_{3} B_3 B_3 + 
    ((-6448Q)/9 - 31256 Q{}^{3}/3)$%
$J{}^{2}\partial{}^{2} A_{3} 
      \partial T T + (-952/3 - 1848Q{}^{2})$%
$J{}^{2}\partial{}^{2} A_{3} A_3 T + (-154/3 - 134Q{}^{2})
     $%
$J{}^{2}\partial{}^{2} A_{3} B_2 B_3 + 
    (82Q - 60Q{}^{3})$%
$J{}^{2}\partial{}^{2} A_{3} \partial B_{2} 
      B_2 + (728 Q/9 - 5596 Q{}^{3}/3)$%
$J{}^{2}\partial{}^{3} A_{3} T{}^{2}+ (283 Q/9 - 62Q{}^{3})
     $%
$J{}^{2}\partial{}^{3} A_{3} B_2{}^{2}+ 
    (1577/81 + 4808 Q{}^{2}/9 + 22036 Q{}^{4}/9)$%
$J{}^{2}B_2{}^{2}\partial{}^{4} T + ((-1262Q)/3 - 2474Q{}^{3})
     $%
$J{}^{2}B_2 B_3 \partial{}^{3} T + 
    ((-12002Q)/9 - 27754 Q{}^{3}/3)$%
$J{}^{2}B_2 
      \partial B_{3} \partial{}^{2} T + ((-10394Q)/9 - 31612 Q{}^{3}/3)
     $%
$J{}^{2}B_2 \partial{}^{2} B_{3} \partial T + 
    ((-676Q)/3 - 4132Q{}^{3})$%
$J{}^{2}B_2 \partial{}^{3} B_{3} 
      T + (9368/81 + 9610 Q{}^{2}/3 + 142354 Q{}^{4}/9)
     $%
$J{}^{2}\partial B_{2} B_2 \partial{}^{3} T + 
    (1322/27 + 30584 Q{}^{2}/9 + 72248 Q{}^{4}/3)$%
$J{}^{2}\partial B_{2} \partial B_{2} \partial{}^{2} T + (698 Q/3 + 646Q{}^{3})
     $%
$J{}^{2}\partial B_{2} B_3 \partial{}^{2} T + 
    ((-5708Q)/9 - 20188 Q{}^{3}/3)$%
$J{}^{2}\partial B_{2} 
      \partial B_{3} \partial T + ((-508Q)/3 - 4676Q{}^{3})
     $%
$J{}^{2}\partial B_{2} \partial{}^{2} B_{3} T + 
    (3814/27 + 46556 Q{}^{2}/9 + 31658Q{}^{4})$%
$J{}^{2}\partial{}^{2} B_{2} B_2 \partial{}^{2} T + 
    (5924/27 + 90320 Q{}^{2}/9 + 197300 Q{}^{4}/3)$%
$J{}^{2}\partial{}^{2} B_{2} \partial B_{2} \partial T + (1054/27 + 2844Q{}^{2} + 56612 Q{}^{4}/3)
     $%
$J{}^{2}\partial{}^{2} B_{2} \partial{}^{2} B_{2} T + 
    (1402 Q/3 + 3488Q{}^{3})$%
$J{}^{2}\partial{}^{2} B_{2} B_3 
      \partial T + ((-188Q)/3 + 292Q{}^{3})$%
$J{}^{2}\partial{}^{2} B_{2} 
      \partial B_{3} T + (1316/9 + 127760 Q{}^{2}/27 + 261376 Q{}^{4}/9)
     $%
$J{}^{2}\partial{}^{3} B_{2} B_2 \partial T + 
    (7672/81 + 35696 Q{}^{2}/9 + 243368 Q{}^{4}/9)$%
$J{}^{2}\partial{}^{3} B_{2} \partial B_{2} T + (68Q + 2308Q{}^{3})
     $%
$J{}^{2}\partial{}^{3} B_{2} B_3 T + 
    (3766/81 + 14108 Q{}^{2}/9 + 90908 Q{}^{4}/9)$%
$J{}^{2}\partial{}^{4} B_{2} B_2 T + (-368/9 - 7042 Q{}^{2}/3)
     $%
$J{}^{2}B_3 B_3 \partial{}^{2} T + 
    (-2920/9 - 19748 Q{}^{2}/3)$%
$J{}^{2}\partial B_{3} 
      B_3 \partial T + (-464/3 - 2520Q{}^{2})$%
$J{}^{2}\partial B_{3} \partial B_{3} T + (-352/3 - 2648Q{}^{2})
     $%
$J{}^{2}\partial{}^{2} B_{3} B_3 T + 
    ((4768\sqrt{2/3}Q)/9 + (13088\sqrt{2/3}Q{}^{3})/3)
     $%
$J A_2 \partial T \partial T \partial T + 
    ((17648\sqrt{2/3}Q)/9 + (44608\sqrt{2/3}Q{}^{3})/3)
     $%
$J A_2 \partial{}^{2} T \partial T T + 
    (512\sqrt{2/3}Q + 1120\sqrt{6}Q{}^{3})$%
$J A_2 \partial{}^{3} T 
      T{}^{2}+ (368\sqrt{6}Q + 3324\sqrt{6}Q{}^{3})
     $%
$J A_2{}^{2}\partial{}^{2} T \partial T + 
    ((4772\sqrt{2/3}Q)/9 + 1220\sqrt{6}Q{}^{3})$%
$J A_2{}^{2}\partial{}^{3} T T + 
    ((1340\sqrt{2/3}Q)/9 + 1228\sqrt{2/3}Q{}^{3})$%
$J A_2{}^{3}\partial{}^{3} T + (-16\sqrt{2/3} - 20\sqrt{2/3}Q{}^{2})
     $%
$J A_2{}^{3}3 \partial{}^{2} T + 
    (-32\sqrt{2/3} + 24\sqrt{6}Q{}^{2})$%
$J A_2{}^{2}\partial A_{3} \partial T - 224\sqrt{2/3}Q$%
$J A_2{}^{2}\partial A_{3} A_3 + 
    ((-80\sqrt{2/3})/3 + 4\sqrt{6}Q{}^{2})$%
$J A_2{}^{2}\partial{}^{2} A_{3} T + 112\sqrt{2/3}Q{}^{2}$%
$J A_2{}^{2}B_2 \partial{}^{2} B_{3} - 16\sqrt{6}Q{}^{2}
     $%
$J A_2{}^{2}\partial B_{2} \partial B_{3} + 
    (256\sqrt{2/3}Q + 776\sqrt{6}Q{}^{3})$%
$J A_2{}^{2}\partial{}^{2} B_{2} \partial B_{2} - 112\sqrt{6}Q{}^{2}$%
$J A_2{}^{2}\partial{}^{2} B_{2} B_3 + 
    (56\sqrt{2/3}Q + (2008\sqrt{2/3}Q{}^{3})/3)$%
$J A_2{}^{2}\partial{}^{3} B_{2} B_2 + 128\sqrt{6}Q
     $%
$J A_2{}^{2}\partial B_{3} B_3 + 
    ((2392\sqrt{2/3})/9 + (9164\sqrt{2/3}Q{}^{2})/3)$%
$J A_2 
      A_3 \partial T \partial T + 
    ((5656\sqrt{2/3})/9 + (14852\sqrt{2/3}Q{}^{2})/3)$%
$J A_2 
      A_3 \partial{}^{2} T T + 144\sqrt{6}Q$%
$J A_2 
      A_3 A_3 \partial T - 272\sqrt{2/3}Q
     $%
$J A_2 A_3 B_2 \partial B_{3} - 
    32\sqrt{6}Q$%
$J A_2 A_3 \partial B_{2} 
      B_3 + 352\sqrt{2/3}Q{}^{2}$%
$J A_2 A_3 
      \partial{}^{2} B_{2} B_2 + ((6232\sqrt{2/3})/9 + 
      (18884\sqrt{2/3}Q{}^{2})/3)$%
$J A_2 \partial A_{3} \partial T 
      T - 160\sqrt{2/3}Q$%
$J A_2 \partial A_{3} 
      A_3 T - 32\sqrt{6}Q$%
$J A_2 
      \partial A_{3} B_2 B_3 - 32\sqrt{6}Q{}^{2}
     $%
$J A_2 \partial A_{3} \partial B_{2} B_2 + 
    (328\sqrt{2/3} + 652\sqrt{6}Q{}^{2})$%
$J A_2 \partial{}^{2} A_{3} 
      T{}^{2}- 32\sqrt{6}Q{}^{2}$%
$J A_2 \partial{}^{2} A_{3} 
      B_2{}^{2}+ ((604\sqrt{2/3}Q)/9 + 196\sqrt{6}Q{}^{3})
     $%
$J A_2 B_2{}^{2}\partial{}^{3} T + 
    ((400\sqrt{2/3})/3 + 764\sqrt{2/3}Q{}^{2})$%
$J A_2 
      B_2 B_3 \partial{}^{2} T + 
    ((80\sqrt{2/3})/3 + 344\sqrt{2/3}Q{}^{2})$%
$J A_2 
      B_2 \partial B_{3} \partial T + (320\sqrt{2/3} + 164\sqrt{6}Q{}^{2})
     $%
$J A_2 B_2 \partial{}^{2} B_{3} T + 
    ((5464\sqrt{2/3}Q)/9 + (18428\sqrt{2/3}Q{}^{3})/3)
     $%
$J A_2 \partial B_{2} B_2 \partial{}^{2} T + 
    ((3640\sqrt{2/3}Q)/9 + (14216\sqrt{2/3}Q{}^{3})/3)
     $%
$J A_2 \partial B_{2} \partial B_{2} \partial T + 
    ((800\sqrt{2/3})/3 + 2008\sqrt{2/3}Q{}^{2})$%
$J A_2 
      \partial B_{2} B_3 \partial T + (16\sqrt{6} + 1208\sqrt{2/3}Q{}^{2})
     $%
$J A_2 \partial B_{2} \partial B_{3} T + 
    ((9112\sqrt{2/3}Q)/9 + (21320\sqrt{2/3}Q{}^{3})/3)
     $%
$J A_2 \partial{}^{2} B_{2} B_2 \partial T + 
    ((2672\sqrt{2/3}Q)/3 + 6964\sqrt{2/3}Q{}^{3})$%
$J A_2 
      \partial{}^{2} B_{2} \partial B_{2} T + 
    ((352\sqrt{2/3})/3 + 1292\sqrt{2/3}Q{}^{2})$%
$J A_2 
      \partial{}^{2} B_{2} B_3 T + 
    ((1936\sqrt{2/3}Q)/9 + (8308\sqrt{2/3}Q{}^{3})/3)
     $%
$J A_2 \partial{}^{3} B_{2} B_2 T - 
    96\sqrt{6}Q$%
$J A_2 B_3 B_3 \partial T - 
    32\sqrt{6}Q$%
$J A_2 \partial B_{3} B_3 T + 
    ((10672\sqrt{2/3}Q)/9 + (26336\sqrt{2/3}Q{}^{3})/3)
     $%
$J \partial A_{2} \partial T \partial T T + 
    ((1184\sqrt{2/3}Q)/3 + 4544\sqrt{2/3}Q{}^{3})$%
$J \partial A_{2} 
      \partial{}^{2} T T{}^{2}+ (1576\sqrt{2/3}Q + 12628\sqrt{2/3}Q{}^{3})
     $%
$J \partial A_{2} A_2 \partial T \partial T + 
    (464\sqrt{6}Q + 11060\sqrt{2/3}Q{}^{3})$%
$J \partial A_{2} 
      A_2 \partial{}^{2} T T + 
    ((4616\sqrt{2/3}Q)/9 + (11800\sqrt{2/3}Q{}^{3})/3)
     $%
$J \partial A_{2} A_2{}^{2}\partial{}^{2} T + 
    ((664\sqrt{2/3})/9 + (1892\sqrt{2/3}Q{}^{2})/3)$%
$J \partial A_{2} 
      A_2 A_3 \partial T - 32\sqrt{6}Q
     $%
$J \partial A_{2} A_2 A_3 A_3 + 
    ((856\sqrt{2/3})/9 + (3044\sqrt{2/3}Q{}^{2})/3)$%
$J \partial A_{2} 
      A_2 \partial A_{3} T + 368\sqrt{2/3}Q{}^{2}
     $%
$J \partial A_{2} A_2 B_2 \partial B_{3} + 
    ((3256\sqrt{2/3}Q)/9 + (9944\sqrt{2/3}Q{}^{3})/3)
     $%
$J \partial A_{2} A_2 \partial B_{2} \partial B_{2} - 
    64\sqrt{6}Q{}^{2}$%
$J \partial A_{2} A_2 \partial B_{2} 
      B_3 + ((2272\sqrt{2/3}Q)/9 + (8024\sqrt{2/3}Q{}^{3})/3)
     $%
$J \partial A_{2} A_2 \partial{}^{2} B_{2} B_2 + 
    64\sqrt{6}Q$%
$J \partial A_{2} A_2 B_3 
      B_3 + ((656\sqrt{2/3}Q)/3 + 1044\sqrt{6}Q{}^{3})
     $%
$J \partial A_{2} \partial A_{2} \partial T T + 
    (320\sqrt{2/3}Q + 476\sqrt{6}Q{}^{3})$%
$J \partial A_{2} \partial A_{2} 
      A_2 \partial T + 32\sqrt{6}Q{}^{2}$%
$J \partial A_{2} 
      \partial A_{2} A_2 A_3 + 
    ((-784\sqrt{2/3}Q)/9 - (6620\sqrt{2/3}Q{}^{3})/3)
     $%
$J \partial A_{2} \partial A_{2} \partial A_{2} T + 
    ((-4328\sqrt{2/3}Q)/9 - (12160\sqrt{2/3}Q{}^{3})/3)
     $%
$J \partial A_{2} \partial A_{2} \partial A_{2} A_2 + 
    ((520\sqrt{2/3})/9 + (4844\sqrt{2/3}Q{}^{2})/3)$%
$J \partial A_{2} 
      \partial A_{2} A_3 T + 32\sqrt{6}Q{}^{2}
     $%
$J \partial A_{2} \partial A_{2} B_2 B_3 + 
    ((2608\sqrt{2/3}Q)/9 + (9512\sqrt{2/3}Q{}^{3})/3)
     $%
$J \partial A_{2} \partial A_{2} \partial B_{2} B_2 + 
    ((12664\sqrt{2/3})/9 + (35444\sqrt{2/3}Q{}^{2})/3)
     $%
$J \partial A_{2} A_3 \partial T T + 
    240\sqrt{6}Q$%
$J \partial A_{2} A_3 A_3 T - 
    32\sqrt{6}Q$%
$J \partial A_{2} A_3 B_2 
      B_3 - 128\sqrt{6}Q{}^{2}$%
$J \partial A_{2} A_3 
      \partial B_{2} B_2 + (224\sqrt{2/3} + 1120\sqrt{6}Q{}^{2})
     $%
$J \partial A_{2} \partial A_{3} T{}^{2}- 
    64\sqrt{6}Q{}^{2}$%
$J \partial A_{2} \partial A_{3} B_2{}^{2}+ ((352\sqrt{2/3}Q)/3 + 60\sqrt{6}Q{}^{3})
     $%
$J \partial A_{2} B_2{}^{2}\partial{}^{2} T + 
    ((1928\sqrt{2/3})/9 + (3628\sqrt{2/3}Q{}^{2})/3)$%
$J \partial A_{2} 
      B_2 B_3 \partial T + 
    ((2216\sqrt{2/3})/9 + (4156\sqrt{2/3}Q{}^{2})/3)$%
$J \partial A_{2} 
      B_2 \partial B_{3} T + 
    ((1688\sqrt{2/3}Q)/3 + 5468\sqrt{2/3}Q{}^{3})$%
$J \partial A_{2} 
      \partial B_{2} B_2 \partial T + 
    ((1280\sqrt{2/3}Q)/9 + (7684\sqrt{2/3}Q{}^{3})/3)
     $%
$J \partial A_{2} \partial B_{2} \partial B_{2} T + 
    ((1640\sqrt{2/3})/9 + (3196\sqrt{2/3}Q{}^{2})/3)$%
$J \partial A_{2} 
      \partial B_{2} B_3 T + 
    ((8\sqrt{2/3}Q)/9 + (5908\sqrt{2/3}Q{}^{3})/3)$%
$J \partial A_{2} 
      \partial{}^{2} B_{2} B_2 T - 80\sqrt{6}Q
     $%
$J \partial A_{2} B_3 B_3 T + 
    ((416\sqrt{2/3}Q)/3 + 3296\sqrt{2/3}Q{}^{3})$%
$J \partial{}^{2} A_{2} 
      \partial T T{}^{2}+ ((3088\sqrt{2/3}Q)/3 + 9580\sqrt{2/3}Q{}^{3})
     $%
$J \partial{}^{2} A_{2} A_2 \partial T T + 
    ((3824\sqrt{2/3}Q)/9 + (10120\sqrt{2/3}Q{}^{3})/3)
     $%
$J \partial{}^{2} A_{2} A_2{}^{2}\partial T + 
    112\sqrt{2/3}Q{}^{2}$%
$J \partial{}^{2} A_{2} A_2{}^{3}3 + ((616\sqrt{2/3})/9 + (6932\sqrt{2/3}Q{}^{2})/3)
     $%
$J \partial{}^{2} A_{2} A_2 A_3 T + 
    32\sqrt{6}Q{}^{2}$%
$J \partial{}^{2} A_{2} A_2 B_2 
      B_3 + ((2656\sqrt{2/3}Q)/9 + (10424\sqrt{2/3}Q{}^{3})/3)
     $%
$J \partial{}^{2} A_{2} A_2 \partial B_{2} B_2 + 
    ((-176\sqrt{2/3}Q)/3 - 1828\sqrt{2/3}Q{}^{3})$%
$J \partial{}^{2} A_{2} 
      \partial A_{2} T{}^{2}+ ((-4784\sqrt{2/3}Q)/9 - 
      (20104\sqrt{2/3}Q{}^{3})/3)$%
$J \partial{}^{2} A_{2} \partial A_{2} 
      A_2 T + ((-6752\sqrt{2/3}Q)/9 - (19288\sqrt{2/3}Q{}^{3})/3)
     $%
$J \partial{}^{2} A_{2} \partial A_{2} A_2{}^{2}+ 
    (-8\sqrt{6}Q + 64\sqrt{6}Q{}^{3})$%
$J \partial{}^{2} A_{2} \partial A_{2} 
      B_2{}^{2}+ (136\sqrt{6} + 5068\sqrt{2/3}Q{}^{2})
     $%
$J \partial{}^{2} A_{2} A_3 T{}^{2}- 
    48\sqrt{6}Q{}^{2}$%
$J \partial{}^{2} A_{2} A_3 B_2{}^{2}+ (-152\sqrt{2/3}Q - 784\sqrt{2/3}Q{}^{3})
     $%
$J \partial{}^{2} A_{2} B_2{}^{2}\partial T + 
    ((992\sqrt{2/3})/9 + (3664\sqrt{2/3}Q{}^{2})/3)$%
$J \partial{}^{2} A_{2} 
      B_2 B_3 T + 
    ((-280\sqrt{2/3}Q)/9 + (3496\sqrt{2/3}Q{}^{3})/3)
     $%
$J \partial{}^{2} A_{2} \partial B_{2} B_2 T + 
    ((-296\sqrt{2/3}Q)/9 - 32\sqrt{2/3}Q{}^{3})$%
$J \partial{}^{3} A_{2} 
      T{}^{3}+ ((-448\sqrt{2/3}Q)/3 - 
      (1156\sqrt{2/3}Q{}^{3})/3)$%
$J \partial{}^{3} A_{2} A_2 T{}^{2}+ ((-892\sqrt{2/3}Q)/9 - (2252\sqrt{2/3}Q{}^{3})/3)
     $%
$J \partial{}^{3} A_{2} A_2{}^{2}T + 
    ((-268\sqrt{2/3}Q)/3 - (2200\sqrt{2/3}Q{}^{3})/3)
     $%
$J \partial{}^{3} A_{2} A_2{}^{3}+ 
    ((-52\sqrt{2/3}Q)/3 + 32\sqrt{6}Q{}^{3})$%
$J \partial{}^{3} A_{2} 
      A_2 B_2{}^{2}+ 
    ((-2572\sqrt{2/3}Q)/9 - 1168\sqrt{2/3}Q{}^{3})$%
$J \partial{}^{3} A_{2} 
      B_2{}^{2}T + 
    ((3136\sqrt{2/3})/3 + 3008\sqrt{6}Q{}^{2})$%
$J A_3 \partial T 
      \partial T T + (432\sqrt{6} + 6736\sqrt{2/3}Q{}^{2})
     $%
$J A_3 \partial{}^{2} T T{}^{2}- 
    528\sqrt{6}Q$%
$J A_3 A_3 \partial T T - 
    48\sqrt{6}$%
$J A_3 A_3 A_3 T + 
    64\sqrt{6}Q$%
$J A_3 A_3 \partial B_{2} 
      B_2 + ((112\sqrt{2/3})/3 + 784\sqrt{2/3}Q{}^{2})
     $%
$J A_3 B_2{}^{2}\partial{}^{2} T - 
    144\sqrt{6}Q$%
$J A_3 B_2 B_3 \partial T - 
    992\sqrt{2/3}Q$%
$J A_3 B_2 \partial B_{3} 
      T + ((1168\sqrt{2/3})/3 + 2608\sqrt{2/3}Q{}^{2})
     $%
$J A_3 \partial B_{2} B_2 \partial T + 
    ((704\sqrt{2/3})/3 + 848\sqrt{2/3}Q{}^{2})$%
$J A_3 
      \partial B_{2} \partial B_{2} T + 64\sqrt{6}Q
     $%
$J A_3 \partial B_{2} B_3 T + 
    ((1400\sqrt{2/3})/3 + 768\sqrt{6}Q{}^{2})$%
$J A_3 
      \partial{}^{2} B_{2} B_2 T - 16\sqrt{6}$%
$J A_3 
      B_3 B_3 T + 
    ((3760\sqrt{2/3})/3 + 2808\sqrt{6}Q{}^{2})$%
$J \partial A_{3} \partial T 
      T{}^{2}- 1280\sqrt{2/3}Q$%
$J \partial A_{3} 
      A_3 T{}^{2}+ 32\sqrt{6}Q$%
$J \partial A_{3} 
      A_3 B_2{}^{2}+ 
    ((944\sqrt{2/3})/3 + 1616\sqrt{2/3}Q{}^{2})$%
$J \partial A_{3} 
      B_2{}^{2}\partial T + 32\sqrt{6}Q
     $%
$J \partial A_{3} B_2 B_3 T + 
    ((544\sqrt{2/3})/3 + 2080\sqrt{2/3}Q{}^{2})$%
$J \partial A_{3} 
      \partial B_{2} B_2 T + ((464\sqrt{2/3})/3 + 552\sqrt{6}Q{}^{2})
     $%
$J \partial{}^{2} A_{3} T{}^{3}+ 
    ((424\sqrt{2/3})/3 + 1024\sqrt{2/3}Q{}^{2})$%
$J \partial{}^{2} A_{3} 
      B_2{}^{2}T + 
    ((4672\sqrt{2/3}Q)/3 + 12500\sqrt{2/3}Q{}^{3})$%
$J B_2{}^{2}\partial{}^{2} T \partial T + 
    ((7988\sqrt{2/3}Q)/9 + 5164\sqrt{2/3}Q{}^{3})$%
$J B_2{}^{2}\partial{}^{3} T T + 16\sqrt{6}Q{}^{2}$%
$J B_2{}^{3}\partial{}^{2} B_{3} + 
    16\sqrt{6}Q$%
$J B_2{}^{2}\partial B_{3} 
      B_3 + ((4952\sqrt{2/3})/9 + (14428\sqrt{2/3}Q{}^{2})/3)
     $%
$J B_2 B_3 \partial T \partial T + 
    ((12200\sqrt{2/3})/9 + (21076\sqrt{2/3}Q{}^{2})/3)
     $%
$J B_2 B_3 \partial{}^{2} T T + 
    ((14840\sqrt{2/3})/9 + (33028\sqrt{2/3}Q{}^{2})/3)
     $%
$J B_2 \partial B_{3} \partial T T + 
    (56\sqrt{2/3} + 3212\sqrt{2/3}Q{}^{2})$%
$J B_2 \partial{}^{2} B_{3} 
      T{}^{2}+ ((23960\sqrt{2/3}Q)/9 + (66844\sqrt{2/3}Q{}^{3})/3)
     $%
$J \partial B_{2} B_2 \partial T \partial T + 
    ((25616\sqrt{2/3}Q)/9 + (71740\sqrt{2/3}Q{}^{3})/3)
     $%
$J \partial B_{2} B_2 \partial{}^{2} T T + 
    32\sqrt{6}Q{}^{2}$%
$J \partial B_{2} B_2{}^{2}\partial B_{3} - 32\sqrt{6}Q$%
$J \partial B_{2} B_2 
      B_3 B_3 + ((16016\sqrt{2/3}Q)/9 + 
      (52372\sqrt{2/3}Q{}^{3})/3)$%
$J \partial B_{2} \partial B_{2} \partial T 
      T + 128\sqrt{6}Q{}^{2}$%
$J \partial B_{2} \partial B_{2} 
      B_2 B_3 + (-32\sqrt{2/3}Q - 256\sqrt{6}Q{}^{3})
     $%
$J \partial B_{2} \partial B_{2} \partial B_{2} B_2 + 
    ((14840\sqrt{2/3})/9 + (37732\sqrt{2/3}Q{}^{2})/3)
     $%
$J \partial B_{2} B_3 \partial T T + 
    (592\sqrt{2/3} + 6680\sqrt{2/3}Q{}^{2})$%
$J \partial B_{2} 
      \partial B_{3} T{}^{2}+ ((19040\sqrt{2/3}Q)/9 + 
      (60772\sqrt{2/3}Q{}^{3})/3)$%
$J \partial{}^{2} B_{2} B_2 \partial T 
      T + 16\sqrt{6}Q{}^{2}$%
$J \partial{}^{2} B_{2} B_2{}^{3}3 + ((2048\sqrt{2/3}Q)/3 + 7132\sqrt{2/3}Q{}^{3})
     $%
$J \partial{}^{2} B_{2} \partial B_{2} T{}^{2}+ 
    (40\sqrt{2/3}Q - 192\sqrt{6}Q{}^{3})$%
$J \partial{}^{2} B_{2} \partial B_{2} 
      B_2{}^{2}+ ((2440\sqrt{2/3})/3 + 5132\sqrt{2/3}Q{}^{2})
     $%
$J \partial{}^{2} B_{2} B_3 T{}^{2}+ 
    ((4384\sqrt{2/3}Q)/9 + 1020\sqrt{6}Q{}^{3})$%
$J \partial{}^{3} B_{2} 
      B_2 T{}^{2}+ ((56\sqrt{2/3}Q)/3 - 32\sqrt{2/3}Q{}^{3})
     $%
$J \partial{}^{3} B_{2} B_2{}^{3}- 
    208\sqrt{6}Q$%
$J B_3 B_3 \partial T T - 
    448\sqrt{6}Q$%
$J \partial B_{3} B_3 T{}^{2}+ 
    (128\sqrt{6}Q + 1536\sqrt{6}Q{}^{3})$%
$\partial J \partial T \partial T T{}^{2}+ (64\sqrt{6}Q + 768\sqrt{6}Q{}^{3})$%
$\partial J \partial{}^{2} T 
      T{}^{3}+ (1184/27 + 1696Q{}^{2} + 42112 Q{}^{4}/3)
     $%
$\partial J J \partial T \partial T \partial T + 
    (11216/27 + 32288 Q{}^{2}/3 + 208000 Q{}^{4}/3)$%
$\partial J J 
      \partial{}^{2} T \partial T T + (12176/81 + 32720 Q{}^{2}/9 + 197824 Q{}^{4}/9)
     $%
$\partial J J \partial{}^{3} T T{}^{2}+ 
    ((532\sqrt{2/3}Q)/9 - 11456\sqrt{2/3}Q{}^{3} - 145984\sqrt{2/3}Q{}^{5})
     $%
$\partial J J{}^{2}\partial{}^{2} T \partial{}^{2} T + 
    ((-640\sqrt{2/3}Q)/27 - (50240\sqrt{2/3}Q{}^{3})/3 - 
      (592640\sqrt{2/3}Q{}^{5})/3)$%
$\partial J J{}^{2}\partial{}^{3} T 
      \partial T + ((-6356\sqrt{2/3}Q)/27 - (29024\sqrt{2/3}Q{}^{3})/3 - 
      (246592\sqrt{2/3}Q{}^{5})/3)$%
$\partial J J{}^{2}\partial{}^{4} T 
      T + (6376/1215 - 430856 Q{}^{2}/405 - 663752 Q{}^{4}/27 - 
      1194656 Q{}^{6}/9)$%
$\partial J J{}^{3}\partial{}^{5} T + (-1804/243 + 163199 Q{}^{2}/405 + 185674 Q{}^{4}/135 - 
      311072 Q{}^{6}/9)$%
$\partial J J{}^{3}\partial{}^{5} A_{2} + ((-48155Q)/81 - 91240 Q{}^{3}/9 - 388720 Q{}^{5}/9)
     $%
$\partial J J{}^{3}\partial{}^{4} A_{3} + 
    ((1978\sqrt{2/3}Q)/3 - (23206\sqrt{2/3}Q{}^{3})/9 - 
      (204836\sqrt{2/3}Q{}^{5})/3)$%
$\partial J J{}^{2}A_2 \partial{}^{4} T + ((1982\sqrt{2/3})/27 - 1058\sqrt{6}Q{}^{2} - 
      (80582\sqrt{2/3}Q{}^{4})/3)$%
$\partial J J{}^{2}A_2 \partial{}^{3} A_{3} + (556\sqrt{6}Q - (47284\sqrt{2/3}Q{}^{3})/9 - 
      (479468\sqrt{2/3}Q{}^{5})/3)$%
$\partial J J{}^{2}\partial A_{2} \partial{}^{3} T + ((286\sqrt{2/3})/3 - (17372\sqrt{2/3}Q{}^{2})/3 - 
      48094\sqrt{2/3}Q{}^{4})$%
$\partial J J{}^{2}\partial A_{2} 
      \partial{}^{2} A_{3} + (616\sqrt{6}Q + (12920\sqrt{2/3}Q{}^{3})/3 - 
      119372\sqrt{2/3}Q{}^{5})$%
$\partial J J{}^{2}\partial{}^{2} A_{2} 
      \partial{}^{2} T + ((-4666\sqrt{2/3}Q)/9 - (35830\sqrt{2/3}Q{}^{3})/3 - 
      79810\sqrt{2/3}Q{}^{5})$%
$\partial J J{}^{2}\partial{}^{2} A_{2} 
      \partial{}^{2} A_{2} + ((406\sqrt{2/3})/9 - (11602\sqrt{2/3}Q{}^{2})/3 - 
      11518\sqrt{6}Q{}^{4})$%
$\partial J J{}^{2}\partial{}^{2} A_{2} 
      \partial A_{3} + ((31852\sqrt{2/3}Q)/27 + (81164\sqrt{2/3}Q{}^{3})/9 - 
      22960\sqrt{2/3}Q{}^{5})$%
$\partial J J{}^{2}\partial{}^{3} A_{2} 
      \partial T + ((-31132\sqrt{2/3}Q)/27 - 22948\sqrt{2/3}Q{}^{3} - 
      (433808\sqrt{2/3}Q{}^{5})/3)$%
$\partial J J{}^{2}\partial{}^{3} A_{2} \partial A_{2} + (118\sqrt{2/3} + (2936\sqrt{2/3}Q{}^{2})/3 + 
      1438\sqrt{6}Q{}^{4})$%
$\partial J J{}^{2}\partial{}^{3} A_{2} 
      A_3 + ((6794\sqrt{2/3}Q)/27 + (6086\sqrt{2/3}Q{}^{3})/9 - 
      (63796\sqrt{2/3}Q{}^{5})/3)$%
$\partial J J{}^{2}\partial{}^{4} A_{2} T + ((-11836\sqrt{2/3}Q)/27 - 
      (75880\sqrt{2/3}Q{}^{3})/9 - (177754\sqrt{2/3}Q{}^{5})/3)
     $%
$\partial J J{}^{2}\partial{}^{4} A_{2} A_2 + 
    ((-13016\sqrt{2/3})/27 - 2696\sqrt{2/3}Q{}^{2} - (19780\sqrt{2/3}Q{}^{4})/3)
     $%
$\partial J J{}^{2}A_3 \partial{}^{3} T + 
    ((-6712\sqrt{2/3})/9 - (45292\sqrt{2/3}Q{}^{2})/3 - 87688\sqrt{2/3}Q{}^{4})
     $%
$\partial J J{}^{2}\partial A_{3} \partial{}^{2} T + 
    ((4916\sqrt{2/3}Q)/3 + 6652\sqrt{2/3}Q{}^{3})$%
$\partial J J{}^{2}\partial A_{3} \partial A_{3} + 
    ((-5924\sqrt{2/3})/9 - (61502\sqrt{2/3}Q{}^{2})/3 - 117032\sqrt{2/3}Q{}^{4})
     $%
$\partial J J{}^{2}\partial{}^{2} A_{3} \partial T + 
    ((6152\sqrt{2/3}Q)/3 + 7420\sqrt{2/3}Q{}^{3})$%
$\partial J J{}^{2}\partial{}^{2} A_{3} A_3 + 
    ((-848\sqrt{2/3})/3 - 9620\sqrt{2/3}Q{}^{2} - 54308\sqrt{2/3}Q{}^{4})
     $%
$\partial J J{}^{2}\partial{}^{3} A_{3} T + 
    ((74\sqrt{2/3})/3 - (646\sqrt{2/3}Q{}^{2})/3 - 7334\sqrt{2/3}Q{}^{4})
     $%
$\partial J J{}^{2}B_2 \partial{}^{3} B_{3} + 
    (74\sqrt{2/3} - 7466\sqrt{2/3}Q{}^{2} - 20706\sqrt{6}Q{}^{4})
     $%
$\partial J J{}^{2}\partial B_{2} \partial{}^{2} B_{3} + 
    ((-4364\sqrt{2/3}Q)/3 - (38350\sqrt{2/3}Q{}^{3})/3 - 16694\sqrt{2/3}Q{}^{5})
     $%
$\partial J J{}^{2}\partial{}^{2} B_{2} \partial{}^{2} B_{2} + 
    ((986\sqrt{2/3})/3 - 10718\sqrt{2/3}Q{}^{2} - 29886\sqrt{6}Q{}^{4})
     $%
$\partial J J{}^{2}\partial{}^{2} B_{2} \partial B_{3} + 
    ((-60544\sqrt{2/3}Q)/27 - (201236\sqrt{2/3}Q{}^{3})/9 - 14796\sqrt{6}Q{}^{5})
     $%
$\partial J J{}^{2}\partial{}^{3} B_{2} \partial B_{2} + 
    ((658\sqrt{2/3})/3 - (14578\sqrt{2/3}Q{}^{2})/3 - 41402\sqrt{2/3}Q{}^{4})
     $%
$\partial J J{}^{2}\partial{}^{3} B_{2} B_3 + 
    ((-22186\sqrt{2/3}Q)/27 - (97922\sqrt{2/3}Q{}^{3})/9 - 10826\sqrt{6}Q{}^{5})
     $%
$\partial J J{}^{2}\partial{}^{4} B_{2} B_2 + 
    ((4340\sqrt{2/3}Q)/3 + 12868\sqrt{2/3}Q{}^{3})$%
$\partial J J{}^{2}\partial B_{3} \partial B_{3} + 
    ((5348\sqrt{2/3}Q)/3 + 15220\sqrt{2/3}Q{}^{3})$%
$\partial J J{}^{2}\partial{}^{2} B_{3} B_3 + 
    (18224/27 + 156544 Q{}^{2}/9 + 98160Q{}^{4})$%
$\partial J J 
      A_2 \partial{}^{2} T \partial T + (54304/81 + 106928 Q{}^{2}/9 + 
      530144 Q{}^{4}/9)$%
$\partial J J A_2 \partial{}^{3} T T + 
    (1528/27 + 93500 Q{}^{2}/27 + 212500 Q{}^{4}/9)$%
$\partial J J 
      A_2{}^{2}\partial{}^{3} T + (20Q + 184Q{}^{3})
     $%
$\partial J J A_2{}^{2}\partial{}^{2} A_{3} + 
    ((-14120Q)/9 - 40192 Q{}^{3}/3)$%
$\partial J J A_2 
      A_3 \partial{}^{2} T + ((-8360Q)/3 - 25164Q{}^{3})
     $%
$\partial J J A_2 \partial A_{3} \partial T + 
    (-560/9 + 1880 Q{}^{2}/3)$%
$\partial J J A_2 \partial A_{3} 
      A_3 + ((-13604Q)/9 - 41608 Q{}^{3}/3)$%
$\partial J J 
      A_2 \partial{}^{2} A_{3} T + (2308 Q/9 + 3428 Q{}^{3}/3)
     $%
$\partial J J A_2 B_2 \partial{}^{2} B_{3} + 
    (592 Q/3 - 124Q{}^{3})$%
$\partial J J A_2 \partial B_{2} 
      \partial B_{3} + (2668/27 - 7210 Q{}^{2}/3 - 63388 Q{}^{4}/3)
     $%
$\partial J J A_2 \partial{}^{2} B_{2} \partial B_{2} + 
    (2204 Q/9 + 40 Q{}^{3}/3)$%
$\partial J J A_2 
      \partial{}^{2} B_{2} B_3 + (7220/81 - 18214 Q{}^{2}/27 - 71764 Q{}^{4}/9)
     $%
$\partial J J A_2 \partial{}^{3} B_{2} B_2 + 
    (592/9 + 512 Q{}^{2}/3)$%
$\partial J J A_2 \partial B_{3} 
      B_3 + (17536/27 + 177440 Q{}^{2}/9 + 115856Q{}^{4})
     $%
$\partial J J \partial A_{2} \partial T \partial T + 
    (10480/9 + 253984 Q{}^{2}/9 + 468032 Q{}^{4}/3)$%
$\partial J J 
      \partial A_{2} \partial{}^{2} T T + (248 + 131996 Q{}^{2}/9 + 297136 Q{}^{4}/3)
     $%
$\partial J J \partial A_{2} A_2 \partial{}^{2} T + 
    (1928 Q/9 + 10456 Q{}^{3}/3)$%
$\partial J J \partial A_{2} 
      A_2 \partial A_{3} + (6656/27 + 31648 Q{}^{2}/3 + 199636 Q{}^{4}/3)
     $%
$\partial J J \partial A_{2} \partial A_{2} \partial T + 
    (-40/9 + 8024 Q{}^{2}/9 + 33064 Q{}^{4}/3)$%
$\partial J J 
      \partial A_{2} \partial A_{2} \partial A_{2} + (352 Q/3 + 1408Q{}^{3})
     $%
$\partial J J \partial A_{2} \partial A_{2} A_3 + 
    ((-7112Q)/3 - 19340Q{}^{3})$%
$\partial J J \partial A_{2} 
      A_3 \partial T + (-592/9 + 280 Q{}^{2}/3)$%
$\partial J J 
      \partial A_{2} A_3 A_3 + ((-12472Q)/9 - 47528 Q{}^{3}/3)
     $%
$\partial J J \partial A_{2} \partial A_{3} T + 
    (824 Q/3 + 876Q{}^{3})$%
$\partial J J \partial A_{2} B_2 
      \partial B_{3} + (1304/27 - 19544 Q{}^{2}/9 - 18728Q{}^{4})
     $%
$\partial J J \partial A_{2} \partial B_{2} \partial B_{2} + 
    (2432 Q/9 + 2080 Q{}^{3}/3)$%
$\partial J J \partial A_{2} 
      \partial B_{2} B_3 + (4808/27 - 10492 Q{}^{2}/9 - 49144 Q{}^{4}/3)
     $%
$\partial J J \partial A_{2} \partial{}^{2} B_{2} B_2 + 
    (1000/9 + 860 Q{}^{2}/3)$%
$\partial J J \partial A_{2} B_3 
      B_3 + (18224/27 + 246560 Q{}^{2}/9 + 518912 Q{}^{4}/3)
     $%
$\partial J J \partial{}^{2} A_{2} \partial T T + 
    (272 + 119488 Q{}^{2}/9 + 279356 Q{}^{4}/3)$%
$\partial J J 
      \partial{}^{2} A_{2} A_2 \partial T + (820 Q/9 + 3044 Q{}^{3}/3)
     $%
$\partial J J \partial{}^{2} A_{2} A_2 A_3 + 
    (6376/27 + 100376 Q{}^{2}/9 + 82920Q{}^{4})$%
$\partial J J 
      \partial{}^{2} A_{2} \partial A_{2} T + 
    (-116/9 + 35686 Q{}^{2}/9 + 146636 Q{}^{4}/3)$%
$\partial J J 
      \partial{}^{2} A_{2} \partial A_{2} A_2 + ((-5596Q)/3 - 18320Q{}^{3})
     $%
$\partial J J \partial{}^{2} A_{2} A_3 T + 
    (824 Q/3 + 1492Q{}^{3})$%
$\partial J J \partial{}^{2} A_{2} B_2 
      B_3 + (784/9 - 13880 Q{}^{2}/9 - 56752 Q{}^{4}/3)
     $%
$\partial J J \partial{}^{2} A_{2} \partial B_{2} B_2 + 
    (5104/27 + 183128 Q{}^{2}/27 + 390712 Q{}^{4}/9)$%
$\partial J J 
      \partial{}^{3} A_{2} T{}^{2}+ (4720/27 + 20284 Q{}^{2}/3 + 48160Q{}^{4})
     $%
$\partial J J \partial{}^{3} A_{2} A_2 T + 
    (-92/9 + 25312 Q{}^{2}/27 + 108980 Q{}^{4}/9)$%
$\partial J J 
      \partial{}^{3} A_{2} A_2{}^{2}+ 
    (3688/81 + 16502 Q{}^{2}/27 + 6704 Q{}^{4}/3)$%
$\partial J J 
      \partial{}^{3} A_{2} B_2{}^{2}+ ((-19808Q)/3 - 56080Q{}^{3})
     $%
$\partial J J A_3 \partial T \partial T + 
    (-8304Q - 66720Q{}^{3})$%
$\partial J J A_3 \partial{}^{2} T 
      T + (-368 - 472Q{}^{2})$%
$\partial J J A_3 
      A_3 \partial T + (8/9 - 188 Q{}^{2}/3)$%
$\partial J J 
      A_3 B_2 \partial B_{3} + (7952 Q/9 + 15160 Q{}^{3}/3)
     $%
$\partial J J A_3 \partial B_{2} \partial B_{2} + 
    (-2152/9 - 2876 Q{}^{2}/3)$%
$\partial J J A_3 \partial B_{2} 
      B_3 + (4784 Q/9 + 8632 Q{}^{3}/3)$%
$\partial J J 
      A_3 \partial{}^{2} B_{2} B_2 - 
    96Q$%
$\partial J J A_3 B_3 B_3 + 
    ((-118784Q)/9 - 320752 Q{}^{3}/3)$%
$\partial J J \partial A_{3} 
      \partial T T + (-2000/3 - 2232Q{}^{2})$%
$\partial J J 
      \partial A_{3} A_3 T + (-2152/9 - 3452 Q{}^{2}/3)
     $%
$\partial J J \partial A_{3} B_2 B_3 + 
    (3808 Q/3 + 6688Q{}^{3})$%
$\partial J J \partial A_{3} \partial B_{2} 
      B_2 + ((-8384Q)/3 - 27216Q{}^{3})$%
$\partial J J 
      \partial{}^{2} A_{3} T{}^{2}+ (2096 Q/9 + 2452 Q{}^{3}/3)
     $%
$\partial J J \partial{}^{2} A_{3} B_2{}^{2}+ 
    (1192/9 + 21700 Q{}^{2}/9 + 27068 Q{}^{4}/3)$%
$\partial J J 
      B_2{}^{2}\partial{}^{3} T + ((-21224Q)/9 - 55792 Q{}^{3}/3)
     $%
$\partial J J B_2 B_3 \partial{}^{2} T + 
    ((-17128Q)/3 - 40908Q{}^{3})$%
$\partial J J B_2 
      \partial B_{3} \partial T + ((-7564Q)/3 - 23240Q{}^{3})
     $%
$\partial J J B_2 \partial{}^{2} B_{3} T + 
    (11528/27 + 100708 Q{}^{2}/9 + 53120Q{}^{4})$%
$\partial J J 
      \partial B_{2} B_2 \partial{}^{2} T + 
    (10880/27 + 109712 Q{}^{2}/9 + 200468 Q{}^{4}/3)$%
$\partial J J 
      \partial B_{2} \partial B_{2} \partial T + ((-20248Q)/9 - 47924 Q{}^{3}/3)
     $%
$\partial J J \partial B_{2} B_3 \partial T + 
    ((-29432Q)/9 - 74872 Q{}^{3}/3)$%
$\partial J J \partial B_{2} 
      \partial B_{3} T + (15328/27 + 140768 Q{}^{2}/9 + 83684Q{}^{4})
     $%
$\partial J J \partial{}^{2} B_{2} B_2 \partial T + 
    (11848/27 + 139744 Q{}^{2}/9 + 261880 Q{}^{4}/3)$%
$\partial J J 
      \partial{}^{2} B_{2} \partial B_{2} T + ((-13844Q)/9 - 19360 Q{}^{3}/3)
     $%
$\partial J J \partial{}^{2} B_{2} B_3 T + 
    (2632/9 + 200416 Q{}^{2}/27 + 368528 Q{}^{4}/9)$%
$\partial J J 
      \partial{}^{3} B_{2} B_2 T + (-5456/9 - 19432 Q{}^{2}/3)
     $%
$\partial J J B_3 B_3 \partial T + 
    (-5840/9 - 21064 Q{}^{2}/3)$%
$\partial J J \partial B_{3} 
      B_3 T + (112 + 2816 Q{}^{2}/3 - 4864Q{}^{4})
     $%
$(\partial J){}^{2} \partial T \partial T T + 
    (2224/27 + 464Q{}^{2} - 18880 Q{}^{4}/3)$%
$(\partial J){}^{2} \partial{}^{2} T 
      T{}^{2}+ ((-5384\sqrt{2/3}Q)/3 - 86224\sqrt{2/3}Q{}^{3} - 
      258752\sqrt{6}Q{}^{5})$%
$(\partial J){}^{2} J \partial{}^{2} T 
      \partial T + ((-38416\sqrt{2/3}Q)/27 - 49904\sqrt{2/3}Q{}^{3} - 
      (1181888\sqrt{2/3}Q{}^{5})/3)$%
$(\partial J){}^{2} J \partial{}^{3} T 
      T + (2330/243 - 348398 Q{}^{2}/81 - 2137412 Q{}^{4}/27 - 
      2826160 Q{}^{6}/9)$%
$(\partial J){}^{2} J{}^{2}\partial{}^{4} T + (-6118/243 + 269029 Q{}^{2}/81 + 1121536 Q{}^{4}/27 + 
      657554 Q{}^{6}/9)$%
$(\partial J){}^{2} J{}^{2}\partial{}^{4} A_{2} + (-3318Q - 131830 Q{}^{3}/3 - 144986Q{}^{5})
     $%
$(\partial J){}^{2} J{}^{2}\partial{}^{3} A_{3} + 
    ((6904\sqrt{2/3}Q)/27 - (317696\sqrt{2/3}Q{}^{3})/9 - 
      (1012448\sqrt{2/3}Q{}^{5})/3)$%
$(\partial J){}^{2} J 
      A_2 \partial{}^{3} T + ((566\sqrt{2/3})/9 - (24394\sqrt{2/3}Q{}^{2})/3 - 
      67618\sqrt{2/3}Q{}^{4})$%
$(\partial J){}^{2} J A_2 
      \partial{}^{2} A_{3} + ((7856\sqrt{2/3}Q)/9 - 63836\sqrt{2/3}Q{}^{3} - 
      630536\sqrt{2/3}Q{}^{5})$%
$(\partial J){}^{2} J \partial A_{2} 
      \partial{}^{2} T + ((572\sqrt{2/3})/3 - (26336\sqrt{2/3}Q{}^{2})/3 - 
      80116\sqrt{2/3}Q{}^{4})$%
$(\partial J){}^{2} J \partial A_{2} 
      \partial A_{3} + ((2312\sqrt{2/3}Q)/9 - 61124\sqrt{2/3}Q{}^{3} - 
      592880\sqrt{2/3}Q{}^{5})$%
$(\partial J){}^{2} J \partial{}^{2} A_{2} 
      \partial T + ((-31814\sqrt{2/3}Q)/9 - (240364\sqrt{2/3}Q{}^{3})/3 - 
      184082\sqrt{6}Q{}^{5})$%
$(\partial J){}^{2} J \partial{}^{2} A_{2} 
      \partial A_{2} + ((1822\sqrt{2/3})/9 - 610\sqrt{6}Q{}^{2} - 
      23386\sqrt{2/3}Q{}^{4})$%
$(\partial J){}^{2} J \partial{}^{2} A_{2} 
      A_3 + ((-8540\sqrt{2/3}Q)/27 - (289316\sqrt{2/3}Q{}^{3})/9 - 
      (878792\sqrt{2/3}Q{}^{5})/3)$%
$(\partial J){}^{2} J 
      \partial{}^{3} A_{2} T + ((-16582\sqrt{2/3}Q)/9 - 43052\sqrt{2/3}Q{}^{3} - 
      307574\sqrt{2/3}Q{}^{5})$%
$(\partial J){}^{2} J \partial{}^{3} A_{2} 
      A_2 + (-848\sqrt{2/3} + 580\sqrt{2/3}Q{}^{2} + 17040\sqrt{6}Q{}^{4})
     $%
$(\partial J){}^{2} J A_3 \partial{}^{2} T + 
    ((-11528\sqrt{2/3})/9 - 6596\sqrt{2/3}Q{}^{2} + 35876\sqrt{2/3}Q{}^{4})
     $%
$(\partial J){}^{2} J \partial A_{3} \partial T + 
    ((10072\sqrt{2/3}Q)/3 + 10460\sqrt{2/3}Q{}^{3})$%
$(\partial J){}^{2} 
      J \partial A_{3} A_3 + 
    ((-6392\sqrt{2/3})/9 - (32468\sqrt{2/3}Q{}^{2})/3 - 24116\sqrt{2/3}Q{}^{4})
     $%
$(\partial J){}^{2} J \partial{}^{2} A_{3} T + 
    ((874\sqrt{2/3})/3 - 1726\sqrt{2/3}Q{}^{2} - 7282\sqrt{6}Q{}^{4})
     $%
$(\partial J){}^{2} J B_2 \partial{}^{2} B_{3} + 
    ((1028\sqrt{2/3})/3 - (45740\sqrt{2/3}Q{}^{2})/3 - 115696\sqrt{2/3}Q{}^{4})
     $%
$(\partial J){}^{2} J \partial B_{2} \partial B_{3} + 
    ((-53446\sqrt{2/3}Q)/9 - (170332\sqrt{2/3}Q{}^{3})/3 - 94846\sqrt{2/3}Q{}^{5})
     $%
$(\partial J){}^{2} J \partial{}^{2} B_{2} \partial B_{2} + 
    (122\sqrt{6} - (29906\sqrt{2/3}Q{}^{2})/3 - 83866\sqrt{2/3}Q{}^{4})
     $%
$(\partial J){}^{2} J \partial{}^{2} B_{2} B_3 + 
    ((-72310\sqrt{2/3}Q)/27 - 10204\sqrt{6}Q{}^{3} - (239618\sqrt{2/3}Q{}^{5})/3)
     $%
$(\partial J){}^{2} J \partial{}^{3} B_{2} B_2 + 
    ((10136\sqrt{2/3}Q)/3 + 30052\sqrt{2/3}Q{}^{3})$%
$(\partial J){}^{2} 
      J \partial B_{3} B_3 + 
    ((-6344\sqrt{2/3}Q)/9 - (74000\sqrt{2/3}Q{}^{3})/3 - 64832\sqrt{6}Q{}^{5})
     $%
$(\partial J){}^{3} \partial T \partial T + 
    ((-32480\sqrt{2/3}Q)/27 - (108992\sqrt{2/3}Q{}^{3})/3 - 
      (788224\sqrt{2/3}Q{}^{5})/3)$%
$(\partial J){}^{3} \partial{}^{2} T 
      T + (-184/243 - 135536 Q{}^{2}/27 - 1011896 Q{}^{4}/27 + 
      815456 Q{}^{6}/3)$%
$(\partial J){}^{3} J 
      \partial{}^{3} T + (-2872/81 + 271756 Q{}^{2}/27 + 1611160 Q{}^{4}/9 + 
      2474804 Q{}^{6}/3)$%
$(\partial J){}^{3} J 
      \partial{}^{3} A_{2} + (-5960Q - 199492 Q{}^{3}/3 - 151520Q{}^{5})
     $%
$(\partial J){}^{3} J \partial{}^{2} A_{3} + 
    ((-21632Q{}^{2})/27 + 128776 Q{}^{4}/3 + 1891424 Q{}^{6}/3)
     $%
$(\partial J){}^{4} \partial{}^{2} T + 
    (209116 Q{}^{2}/27 + 482804 Q{}^{4}/3 + 2692172 Q{}^{6}/3)
     $%
$(\partial J){}^{4} \partial{}^{2} A_{2} + 
    ((-74548Q)/27 - 28128Q{}^{3} - 135392 Q{}^{5}/3)$%
$(\partial J){}^{4} \partial A_{3} + 
    ((-1184\sqrt{2/3}Q)/9 - (193288\sqrt{2/3}Q{}^{3})/9 - 
      (518048\sqrt{2/3}Q{}^{5})/3)$%
$(\partial J){}^{3} 
      A_2 \partial{}^{2} T + ((796\sqrt{2/3})/27 - 4318\sqrt{2/3}Q{}^{2} - 
      (107386\sqrt{2/3}Q{}^{4})/3)$%
$(\partial J){}^{3} 
      A_2 \partial A_{3} + ((-32992\sqrt{2/3}Q)/27 - 
      (452548\sqrt{2/3}Q{}^{3})/9 - (1035916\sqrt{2/3}Q{}^{5})/3)
     $%
$(\partial J){}^{3} \partial A_{2} \partial T + 
    ((-30502\sqrt{2/3}Q)/27 - (213934\sqrt{2/3}Q{}^{3})/9 - 
      (456958\sqrt{2/3}Q{}^{5})/3)$%
$(\partial J){}^{3} 
      \partial A_{2} \partial A_{2} + (76\sqrt{2/3} - (6118\sqrt{2/3}Q{}^{2})/3 - 
      23222\sqrt{2/3}Q{}^{4})$%
$(\partial J){}^{3} \partial A_{2} 
      A_3 + ((-14456\sqrt{2/3}Q)/27 - (94180\sqrt{2/3}Q{}^{3})/3 - 
      (718948\sqrt{2/3}Q{}^{5})/3)$%
$(\partial J){}^{3} 
      \partial{}^{2} A_{2} T + ((-11354\sqrt{2/3}Q)/9 - 
      (251026\sqrt{2/3}Q{}^{3})/9 - (565484\sqrt{2/3}Q{}^{5})/3)
     $%
$(\partial J){}^{3} \partial{}^{2} A_{2} A_2 + 
    ((-1352\sqrt{2/3})/9 + 4668\sqrt{6}Q{}^{2} + 118532\sqrt{2/3}Q{}^{4})
     $%
$(\partial J){}^{3} A_3 \partial T + 
    (580\sqrt{2/3}Q + 1506\sqrt{6}Q{}^{3})$%
$(\partial J){}^{3} 
      A_3 A_3 + ((-9632\sqrt{2/3})/27 + 7552\sqrt{2/3}Q{}^{2} + 
      (253304\sqrt{2/3}Q{}^{4})/3)$%
$(\partial J){}^{3} 
      \partial A_{3} T + ((380\sqrt{2/3})/3 - (6130\sqrt{2/3}Q{}^{2})/3 - 
      17054\sqrt{2/3}Q{}^{4})$%
$(\partial J){}^{3} B_2 
      \partial B_{3} + ((-17150\sqrt{2/3}Q)/9 - (56234\sqrt{2/3}Q{}^{3})/3 - 
      31222\sqrt{2/3}Q{}^{5})$%
$(\partial J){}^{3} \partial B_{2} 
      \partial B_{2} + ((620\sqrt{2/3})/3 - 3830\sqrt{2/3}Q{}^{2} - 
      9950\sqrt{6}Q{}^{4})$%
$(\partial J){}^{3} \partial B_{2} 
      B_3 + ((-18442\sqrt{2/3}Q)/9 - (65474\sqrt{2/3}Q{}^{3})/3 - 
      44132\sqrt{2/3}Q{}^{5})$%
$(\partial J){}^{3} \partial{}^{2} B_{2} 
      B_2 + ((1688\sqrt{2/3}Q)/3 + 2506\sqrt{2/3}Q{}^{3})
     $%
$(\partial J){}^{3} B_3 B_3 + 
    (2096/9 + 464 Q{}^{2}/3 - 23728Q{}^{4})$%
$(\partial J){}^{2} A_2 
      \partial T \partial T + (11536/27 + 1984 Q{}^{2}/9 - 97424 Q{}^{4}/3)
     $%
$(\partial J){}^{2} A_2 \partial{}^{2} T T + 
    (332/9 + 1280 Q{}^{2}/3 - 6144Q{}^{4})$%
$(\partial J){}^{2} A_2{}^{2}\partial{}^{2} T + (116 Q/3 + 92Q{}^{3})$%
$(\partial J){}^{2} 
      A_2{}^{2}\partial A_{3} + ((-15848Q)/9 - 36520 Q{}^{3}/3)
     $%
$(\partial J){}^{2} A_2 A_3 \partial T + 
    (-856/9 - 5588 Q{}^{2}/3)$%
$(\partial J){}^{2} A_2 A_3 
      A_3 + ((-15140Q)/9 - 38116 Q{}^{3}/3)$%
$(\partial J){}^{2} 
      A_2 \partial A_{3} T + (1756 Q/9 + 4964 Q{}^{3}/3)
     $%
$(\partial J){}^{2} A_2 B_2 \partial B_{3} + 
    (48 - 6484 Q{}^{2}/9 - 15584 Q{}^{4}/3)$%
$(\partial J){}^{2} A_2 
      \partial B_{2} \partial B_{2} + ((-556Q)/9 - 16244 Q{}^{3}/3)
     $%
$(\partial J){}^{2} A_2 \partial B_{2} B_3 + 
    (1520/27 - 3484 Q{}^{2}/3 - 29012 Q{}^{4}/3)$%
$(\partial J){}^{2} 
      A_2 \partial{}^{2} B_{2} B_2 + (968/9 + 8476 Q{}^{2}/3)
     $%
$(\partial J){}^{2} A_2 B_3 B_3 + 
    (18224/27 + 70880 Q{}^{2}/9 + 2528 Q{}^{4}/3)$%
$(\partial J){}^{2} 
      \partial A_{2} \partial T T + (248 + 2844Q{}^{2} - 3820Q{}^{4})
     $%
$(\partial J){}^{2} \partial A_{2} A_2 \partial T + 
    (1324 Q/9 + 9692 Q{}^{3}/3)$%
$(\partial J){}^{2} \partial A_{2} 
      A_2 A_3 + (3644/27 + 22900 Q{}^{2}/9 + 7044Q{}^{4})
     $%
$(\partial J){}^{2} \partial A_{2} \partial A_{2} T + 
    (-160/3 - 6280 Q{}^{2}/3 - 18732Q{}^{4})$%
$(\partial J){}^{2} \partial A_{2} 
      \partial A_{2} A_2 + ((-6316Q)/3 - 20852Q{}^{3})
     $%
$(\partial J){}^{2} \partial A_{2} A_3 T + 
    (344 Q/3 + 664Q{}^{3})$%
$(\partial J){}^{2} \partial A_{2} B_2 
      B_3 + (352/3 - 2828 Q{}^{2}/3 - 12476Q{}^{4})
     $%
$(\partial J){}^{2} \partial A_{2} \partial B_{2} B_2 + 
    (2944/27 + 3176Q{}^{2} + 17672 Q{}^{4}/3)$%
$(\partial J){}^{2} \partial{}^{2} A_{2} 
      T{}^{2}+ (3244/27 + 12076 Q{}^{2}/9 - 25496 Q{}^{4}/3)
     $%
$(\partial J){}^{2} \partial{}^{2} A_{2} A_2 T + 
    (-1142/27 - 1804Q{}^{2} - 48046 Q{}^{4}/3)$%
$(\partial J){}^{2} 
      \partial{}^{2} A_{2} A_2{}^{2}+ 
    (1390/27 + 4072 Q{}^{2}/9 + 682 Q{}^{4}/3)$%
$(\partial J){}^{2} 
      \partial{}^{2} A_{2} B_2{}^{2}+ ((-28784Q)/3 - 64496Q{}^{3})
     $%
$(\partial J){}^{2} A_3 \partial T T + 
    (8 + 1924Q{}^{2})$%
$(\partial J){}^{2} A_3 A_3 T + 
    (-568/9 - 2984 Q{}^{2}/3)$%
$(\partial J){}^{2} A_3 B_2 
      B_3 + (5840 Q/9 + 16456 Q{}^{3}/3)$%
$(\partial J){}^{2} 
      A_3 \partial B_{2} B_2 + ((-14128Q)/3 - 31872Q{}^{3})
     $%
$(\partial J){}^{2} \partial A_{3} T{}^{2}+ 
    (1664 Q/9 + 1648 Q{}^{3}/3)$%
$(\partial J){}^{2} \partial A_{3} 
      B_2{}^{2}+ (1948/27 + 1912 Q{}^{2}/9 - 17480 Q{}^{4}/3)
     $%
$(\partial J){}^{2} B_2{}^{2}\partial{}^{2} T + 
    ((-25880Q)/9 - 53224 Q{}^{3}/3)$%
$(\partial J){}^{2} B_2 
      B_3 \partial T + (-3348Q - 19260Q{}^{3})$%
$(\partial J){}^{2} 
      B_2 \partial B_{3} T + 
    (11528/27 + 9020 Q{}^{2}/3 - 25676 Q{}^{4}/3)$%
$(\partial J){}^{2} 
      \partial B_{2} B_2 \partial T + 
    (2644/27 + 4292 Q{}^{2}/3 - 3796 Q{}^{4}/3)$%
$(\partial J){}^{2} 
      \partial B_{2} \partial B_{2} T + ((-22484Q)/9 - 43660 Q{}^{3}/3)
     $%
$(\partial J){}^{2} \partial B_{2} B_3 T + 
    (7628/27 + 3660Q{}^{2} + 15808 Q{}^{4}/3)$%
$(\partial J){}^{2} \partial{}^{2} B_{2} 
      B_2 T + (-736/9 - 2492 Q{}^{2}/3)
     $%
$(\partial J){}^{2} B_3 B_3 T + 
    ((28240\sqrt{2/3}Q)/9 + (84512\sqrt{2/3}Q{}^{3})/3)
     $%
$\partial J A_2 \partial T \partial T T + 
    ((7504\sqrt{2/3}Q)/3 + 6896\sqrt{6}Q{}^{3})$%
$\partial J A_2 
      \partial{}^{2} T T{}^{2}+ ((4672\sqrt{2/3}Q)/3 + 13952\sqrt{2/3}Q{}^{3})
     $%
$\partial J A_2{}^{2}\partial T \partial T + 
    ((5588\sqrt{2/3}Q)/3 + 15140\sqrt{2/3}Q{}^{3})$%
$\partial J A_2{}^{2}\partial{}^{2} T T + 
    ((2104\sqrt{2/3}Q)/9 + (8462\sqrt{2/3}Q{}^{3})/3)
     $%
$\partial J A_2{}^{3}\partial{}^{2} T + 
    28\sqrt{2/3}Q{}^{2}$%
$\partial J A_2{}^{3}\partial A_{3} + (-32\sqrt{2/3} - 176\sqrt{6}Q{}^{2})
     $%
$\partial J A_2{}^{3}3 \partial T + 
    332\sqrt{2/3}Q$%
$\partial J A_2{}^{3}3 
      A_3 + (-32\sqrt{2/3} - 184\sqrt{6}Q{}^{2})
     $%
$\partial J A_2{}^{2}\partial A_{3} T - 
    220\sqrt{2/3}Q{}^{2}$%
$\partial J A_2{}^{2}B_2 
      \partial B_{3} + ((-328\sqrt{2/3}Q)/3 - 1376\sqrt{2/3}Q{}^{3})
     $%
$\partial J A_2{}^{2}\partial B_{2} \partial B_{2} + 
    472\sqrt{6}Q{}^{2}$%
$\partial J A_2{}^{2}\partial B_{2} 
      B_3 + ((71\sqrt{2/3}Q)/3 - 446\sqrt{2/3}Q{}^{3})
     $%
$\partial J A_2{}^{2}\partial{}^{2} B_{2} B_2 - 
    246\sqrt{6}Q$%
$\partial J A_2{}^{2}B_3 
      B_3 + ((6232\sqrt{2/3})/9 + (6308\sqrt{2/3}Q{}^{2})/3)
     $%
$\partial J A_2 A_3 \partial T T - 
    496\sqrt{2/3}Q$%
$\partial J A_2 A_3 A_3 
      T + 188\sqrt{6}Q$%
$\partial J A_2 A_3 
      B_2 B_3 - 268\sqrt{6}Q{}^{2}$%
$\partial J A_2 
      A_3 \partial B_{2} B_2 + 
    ((832\sqrt{2/3})/3 + 200\sqrt{2/3}Q{}^{2})$%
$\partial J A_2 
      \partial A_{3} T{}^{2}+ 212\sqrt{2/3}Q{}^{2}
     $%
$\partial J A_2 \partial A_{3} B_2{}^{2}+ 
    ((3536\sqrt{2/3}Q)/9 + (10378\sqrt{2/3}Q{}^{3})/3)
     $%
$\partial J A_2 B_2{}^{2}\partial{}^{2} T + 
    ((800\sqrt{2/3})/3 + 904\sqrt{2/3}Q{}^{2})$%
$\partial J A_2 
      B_2 B_3 \partial T + 
    ((80\sqrt{2/3})/3 - 1640\sqrt{2/3}Q{}^{2})$%
$\partial J A_2 
      B_2 \partial B_{3} T + 
    ((19736\sqrt{2/3}Q)/9 + (54052\sqrt{2/3}Q{}^{3})/3)
     $%
$\partial J A_2 \partial B_{2} B_2 \partial T + 
    ((2632\sqrt{2/3}Q)/3 + 8176\sqrt{2/3}Q{}^{3})$%
$\partial J A_2 
      \partial B_{2} \partial B_{2} T + 
    ((800\sqrt{2/3})/3 + 1240\sqrt{2/3}Q{}^{2})$%
$\partial J A_2 
      \partial B_{2} B_3 T + 
    ((4216\sqrt{2/3}Q)/3 + 11104\sqrt{2/3}Q{}^{3})$%
$\partial J A_2 
      \partial{}^{2} B_{2} B_2 T + 16\sqrt{6}Q
     $%
$\partial J A_2 B_3 B_3 T + 
    ((8000\sqrt{2/3}Q)/3 + 26152\sqrt{2/3}Q{}^{3})$%
$\partial J \partial A_{2} 
      \partial T T{}^{2}+ ((17320\sqrt{2/3}Q)/3 + 16364\sqrt{6}Q{}^{3})
     $%
$\partial J \partial A_{2} A_2 \partial T T + 
    ((4340\sqrt{2/3}Q)/9 + (17662\sqrt{2/3}Q{}^{3})/3)
     $%
$\partial J \partial A_{2} A_2{}^{2}\partial T - 
    388\sqrt{2/3}Q{}^{2}$%
$\partial J \partial A_{2} A_2{}^{3}3 + ((664\sqrt{2/3})/9 + (9092\sqrt{2/3}Q{}^{2})/3)
     $%
$\partial J \partial A_{2} A_2 A_3 T - 
    172\sqrt{6}Q{}^{2}$%
$\partial J \partial A_{2} A_2 B_2 
      B_3 + ((-1864\sqrt{2/3}Q)/9 - (5744\sqrt{2/3}Q{}^{3})/3)
     $%
$\partial J \partial A_{2} A_2 \partial B_{2} B_2 + 
    ((2180\sqrt{2/3}Q)/3 + 7340\sqrt{2/3}Q{}^{3})$%
$\partial J \partial A_{2} 
      \partial A_{2} T{}^{2}+ ((-2224\sqrt{2/3}Q)/9 + 
      (52\sqrt{2/3}Q{}^{3})/3)$%
$\partial J \partial A_{2} \partial A_{2} A_2 
      T + ((4844\sqrt{2/3}Q)/9 + (20128\sqrt{2/3}Q{}^{3})/3)
     $%
$\partial J \partial A_{2} \partial A_{2} A_2{}^{2}+ 
    ((-1432\sqrt{2/3}Q)/9 - (3542\sqrt{2/3}Q{}^{3})/3)
     $%
$\partial J \partial A_{2} \partial A_{2} B_2{}^{2}+ 
    ((1808\sqrt{2/3})/3 + 6304\sqrt{2/3}Q{}^{2})$%
$\partial J \partial A_{2} 
      A_3 T{}^{2}+ 272\sqrt{2/3}Q{}^{2}
     $%
$\partial J \partial A_{2} A_3 B_2{}^{2}+ 
    ((8420\sqrt{2/3}Q)/9 + (20926\sqrt{2/3}Q{}^{3})/3)
     $%
$\partial J \partial A_{2} B_2{}^{2}\partial T + 
    ((1928\sqrt{2/3})/9 + (3340\sqrt{2/3}Q{}^{2})/3)$%
$\partial J \partial A_{2} 
      B_2 B_3 T + 
    ((14072\sqrt{2/3}Q)/9 + (40996\sqrt{2/3}Q{}^{3})/3)
     $%
$\partial J \partial A_{2} \partial B_{2} B_2 T + 
    ((952\sqrt{2/3}Q)/3 + 1736\sqrt{6}Q{}^{3})$%
$\partial J \partial{}^{2} A_{2} 
      T{}^{3}+ ((4364\sqrt{2/3}Q)/3 + 5540\sqrt{6}Q{}^{3})
     $%
$\partial J \partial{}^{2} A_{2} A_2 T{}^{2}+ 
    ((196\sqrt{2/3}Q)/3 + 1084\sqrt{6}Q{}^{3})$%
$\partial J \partial{}^{2} A_{2} 
      A_2{}^{2}T + 
    ((2623\sqrt{2/3}Q)/9 + (11120\sqrt{2/3}Q{}^{3})/3)
     $%
$\partial J \partial{}^{2} A_{2} A_2{}^{3}+ 
    ((-2023\sqrt{2/3}Q)/9 - (5876\sqrt{2/3}Q{}^{3})/3)
     $%
$\partial J \partial{}^{2} A_{2} A_2 B_2{}^{2}+ 
    (428\sqrt{2/3}Q + 1668\sqrt{6}Q{}^{3})$%
$\partial J \partial{}^{2} A_{2} 
      B_2{}^{2}T + 
    ((3760\sqrt{2/3})/3 + 4712\sqrt{2/3}Q{}^{2})$%
$\partial J A_3 
      \partial T T{}^{2}- 352\sqrt{6}Q$%
$\partial J A_3 
      A_3 T{}^{2}- 154\sqrt{2/3}Q
     $%
$\partial J A_3 A_3 B_2{}^{2}+ 
    ((224\sqrt{2/3})/3 + 40\sqrt{6}Q{}^{2})$%
$\partial J A_3 
      B_2{}^{2}\partial T - 144\sqrt{6}Q
     $%
$\partial J A_3 B_2 B_3 T + 
    ((1168\sqrt{2/3})/3 + 2000\sqrt{2/3}Q{}^{2})$%
$\partial J A_3 
      \partial B_{2} B_2 T + ((928\sqrt{2/3})/3 + 272\sqrt{6}Q{}^{2})
     $%
$\partial J \partial A_{3} T{}^{3}+ 
    ((944\sqrt{2/3})/3 + 1984\sqrt{2/3}Q{}^{2})$%
$\partial J \partial A_{3} 
      B_2{}^{2}T + 
    (1840\sqrt{2/3}Q + 15248\sqrt{2/3}Q{}^{3})$%
$\partial J B_2{}^{2}\partial T \partial T + 
    ((9268\sqrt{2/3}Q)/3 + 22756\sqrt{2/3}Q{}^{3})$%
$\partial J B_2{}^{2}\partial{}^{2} T T + 4\sqrt{6}Q{}^{2}$%
$\partial J B_2{}^{3}\partial B_{3} - 
    12\sqrt{6}Q$%
$\partial J B_2{}^{3}3 
      B_3 + ((14840\sqrt{2/3})/9 + (21604\sqrt{2/3}Q{}^{2})/3)
     $%
$\partial J B_2 B_3 \partial T T + 
    ((2912\sqrt{2/3})/3 + 3800\sqrt{2/3}Q{}^{2})$%
$\partial J B_2 
      \partial B_{3} T{}^{2}+ ((70232\sqrt{2/3}Q)/9 + 
      (189196\sqrt{2/3}Q{}^{3})/3)$%
$\partial J \partial B_{2} B_2 \partial T 
      T + 60\sqrt{6}Q{}^{2}$%
$\partial J \partial B_{2} B_2{}^{3}3 + ((4732\sqrt{2/3}Q)/3 + 14308\sqrt{2/3}Q{}^{3})
     $%
$\partial J \partial B_{2} \partial B_{2} T{}^{2}+ 
    ((964\sqrt{2/3}Q)/3 + 726\sqrt{6}Q{}^{3})$%
$\partial J \partial B_{2} 
      \partial B_{2} B_2{}^{2}+ 
    ((1472\sqrt{2/3})/3 + 3688\sqrt{2/3}Q{}^{2})$%
$\partial J \partial B_{2} 
      B_3 T{}^{2}+ 
    ((4756\sqrt{2/3}Q)/3 + 15740\sqrt{2/3}Q{}^{3})$%
$\partial J \partial{}^{2} B_{2} 
      B_2 T{}^{2}+ (91\sqrt{2/3}Q + 238\sqrt{6}Q{}^{3})
     $%
$\partial J \partial{}^{2} B_{2} B_2{}^{3}- 
    312\sqrt{6}Q$%
$\partial J B_3 B_3 T{}^{2}+ 
    ((496\sqrt{2/3}Q)/3 + 1984\sqrt{2/3}Q{}^{3})$%
$\partial{}^{2} J \partial T T{}^{3}+ (6224/27 + 44288 Q{}^{2}/9 + 25856Q{}^{4})
     $%
$\partial{}^{2} J J \partial T \partial T T + 
    (6560/27 + 49856 Q{}^{2}/9 + 31488Q{}^{4})$%
$\partial{}^{2} J J \partial{}^{2} T 
      T{}^{2}+ ((-220\sqrt{2/3}Q)/3 - (91360\sqrt{2/3}Q{}^{3})/3 - 
      354880\sqrt{2/3}Q{}^{5})$%
$\partial{}^{2} J J{}^{2}\partial{}^{2} T 
      \partial T + ((-4840\sqrt{2/3}Q)/9 - (66488\sqrt{2/3}Q{}^{3})/3 - 
      188512\sqrt{2/3}Q{}^{5})$%
$\partial{}^{2} J J{}^{2}\partial{}^{3} T 
      T + (2458/243 - 522940 Q{}^{2}/243 - 1269608 Q{}^{4}/27 - 
      6396320 Q{}^{6}/27)$%
$\partial{}^{2} J J{}^{3}\partial{}^{4} T + (-4532/243 + 209054 Q{}^{2}/243 + 82705 Q{}^{4}/27 - 
      1917080 Q{}^{6}/27)$%
$\partial{}^{2} J J{}^{3}\partial{}^{4} A_{2} + ((-44602Q)/27 - 204898 Q{}^{3}/9 - 248824 Q{}^{5}/3)
     $%
$\partial{}^{2} J J{}^{3}\partial{}^{3} A_{3} + 
    ((25360\sqrt{2/3}Q)/27 - (90148\sqrt{2/3}Q{}^{3})/9 - 
      (436700\sqrt{2/3}Q{}^{5})/3)$%
$\partial{}^{2} J J{}^{2}A_2 \partial{}^{3} T + ((1021\sqrt{2/3})/9 - (13540\sqrt{2/3}Q{}^{2})/3 - 
      14500\sqrt{6}Q{}^{4})$%
$\partial{}^{2} J J{}^{2}A_2 
      \partial{}^{2} A_{3} + ((16858\sqrt{2/3}Q)/9 - (183160\sqrt{2/3}Q{}^{3})/9 - 
      (883484\sqrt{2/3}Q{}^{5})/3)$%
$\partial{}^{2} J J{}^{2}\partial A_{2} \partial{}^{2} T + ((454\sqrt{2/3})/3 - 6056\sqrt{2/3}Q{}^{2} - 
      20838\sqrt{6}Q{}^{4})$%
$\partial{}^{2} J J{}^{2}\partial A_{2} 
      \partial A_{3} + ((14986\sqrt{2/3}Q)/9 - (24640\sqrt{2/3}Q{}^{3})/9 - 
      (422900\sqrt{2/3}Q{}^{5})/3)$%
$\partial{}^{2} J J{}^{2}\partial{}^{2} A_{2} \partial T + ((-6787\sqrt{2/3}Q)/3 - 
      (406163\sqrt{2/3}Q{}^{3})/9 - (861544\sqrt{2/3}Q{}^{5})/3)
     $%
$\partial{}^{2} J J{}^{2}\partial{}^{2} A_{2} \partial A_{2} + 
    ((1829\sqrt{2/3})/9 + 782\sqrt{2/3}Q{}^{2} - 1574\sqrt{2/3}Q{}^{4})
     $%
$\partial{}^{2} J J{}^{2}\partial{}^{2} A_{2} A_3 + 
    ((8944\sqrt{2/3}Q)/27 - (64024\sqrt{2/3}Q{}^{3})/9 - 
      (297788\sqrt{2/3}Q{}^{5})/3)$%
$\partial{}^{2} J J{}^{2}\partial{}^{3} A_{2} T + ((-29893\sqrt{2/3}Q)/27 - 
      (202157\sqrt{2/3}Q{}^{3})/9 - (456094\sqrt{2/3}Q{}^{5})/3)
     $%
$\partial{}^{2} J J{}^{2}\partial{}^{3} A_{2} A_2 + 
    ((-2008\sqrt{2/3})/3 - (16378\sqrt{2/3}Q{}^{2})/3 - 3336\sqrt{6}Q{}^{4})
     $%
$\partial{}^{2} J J{}^{2}A_3 \partial{}^{2} T + 
    ((-10264\sqrt{2/3})/9 - (62720\sqrt{2/3}Q{}^{2})/3 - 33388\sqrt{6}Q{}^{4})
     $%
$\partial{}^{2} J J{}^{2}\partial A_{3} \partial T + 
    ((8000\sqrt{2/3}Q)/3 + 8638\sqrt{2/3}Q{}^{3})$%
$\partial{}^{2} J J{}^{2}\partial A_{3} A_3 + 
    ((-5812\sqrt{2/3})/9 - (46076\sqrt{2/3}Q{}^{2})/3 - 74300\sqrt{2/3}Q{}^{4})
     $%
$\partial{}^{2} J J{}^{2}\partial{}^{2} A_{3} T + 
    ((479\sqrt{2/3})/3 - 590\sqrt{2/3}Q{}^{2} - 6824\sqrt{2/3}Q{}^{4})
     $%
$\partial{}^{2} J J{}^{2}B_2 \partial{}^{2} B_{3} + 
    ((958\sqrt{2/3})/3 - (24196\sqrt{2/3}Q{}^{2})/3 - 55234\sqrt{2/3}Q{}^{4})
     $%
$\partial{}^{2} J J{}^{2}\partial B_{2} \partial B_{3} + 
    ((-37105\sqrt{2/3}Q)/9 - (129061\sqrt{2/3}Q{}^{3})/3 - 40112\sqrt{6}Q{}^{5})
     $%
$\partial{}^{2} J J{}^{2}\partial{}^{2} B_{2} \partial B_{2} + 
    ((1159\sqrt{2/3})/3 - 1674\sqrt{6}Q{}^{2} - 13950\sqrt{6}Q{}^{4})
     $%
$\partial{}^{2} J J{}^{2}\partial{}^{2} B_{2} B_3 + 
    ((-49225\sqrt{2/3}Q)/27 - (196709\sqrt{2/3}Q{}^{3})/9 - 
      69142\sqrt{2/3}Q{}^{5})$%
$\partial{}^{2} J J{}^{2}\partial{}^{3} B_{2} 
      B_2 + ((6160\sqrt{2/3}Q)/3 + 15938\sqrt{2/3}Q{}^{3})
     $%
$\partial{}^{2} J J{}^{2}\partial B_{3} B_3 + 
    (12752/27 + 211648 Q{}^{2}/27 + 288800 Q{}^{4}/9)$%
$\partial{}^{2} J J 
      A_2 \partial T \partial T + (9280/9 + 449968 Q{}^{2}/27 + 
      708320 Q{}^{4}/9)$%
$\partial{}^{2} J J A_2 \partial{}^{2} T T + 
    (1984/27 + 36508 Q{}^{2}/9 + 82172 Q{}^{4}/3)$%
$\partial{}^{2} J J 
      A_2{}^{2}\partial{}^{2} T + (148 Q/3 + 2020 Q{}^{3}/3)
     $%
$\partial{}^{2} J J A_2{}^{2}\partial A_{3} + 
    (-2800Q - 24652Q{}^{3})$%
$\partial{}^{2} J J A_2 A_3 
      \partial T + (-664/9 + 76 Q{}^{2}/3)$%
$\partial{}^{2} J J A_2 
      A_3 A_3 + (-2220Q - 66448 Q{}^{3}/3)
     $%
$\partial{}^{2} J J A_2 \partial A_{3} T + 
    (1948 Q/9 + 780Q{}^{3})$%
$\partial{}^{2} J J A_2 B_2 
      \partial B_{3} + (644/9 - 13088 Q{}^{2}/9 - 43240 Q{}^{4}/3)
     $%
$\partial{}^{2} J J A_2 \partial B_{2} \partial B_{2} + 
    (1460 Q/9 + 2380 Q{}^{3}/3)$%
$\partial{}^{2} J J A_2 
      \partial B_{2} B_3 + (3956/27 - 8008 Q{}^{2}/9 - 13460Q{}^{4})
     $%
$\partial{}^{2} J J A_2 \partial{}^{2} B_{2} B_2 + 
    (680/9 - 20 Q{}^{2}/3)$%
$\partial{}^{2} J J A_2 B_3 
      B_3 + (13024/9 + 844144 Q{}^{2}/27 + 1474880 Q{}^{4}/9)
     $%
$\partial{}^{2} J J \partial A_{2} \partial T T + 
    (14896/27 + 389104 Q{}^{2}/27 + 774332 Q{}^{4}/9)$%
$\partial{}^{2} J J 
      \partial A_{2} A_2 \partial T + (940 Q/9 + 2492 Q{}^{3}/3)
     $%
$\partial{}^{2} J J \partial A_{2} A_2 A_3 + 
    (5416/27 + 212620 Q{}^{2}/27 + 510284 Q{}^{4}/9)$%
$\partial{}^{2} J J 
      \partial A_{2} \partial A_{2} T + (-916/27 + 26488 Q{}^{2}/9 + 33528Q{}^{4})
     $%
$\partial{}^{2} J J \partial A_{2} \partial A_{2} A_2 + 
    ((-32428Q)/9 - 115304 Q{}^{3}/3)$%
$\partial{}^{2} J J \partial A_{2} 
      A_3 T + (560 Q/3 + 416Q{}^{3})$%
$\partial{}^{2} J J 
      \partial A_{2} B_2 B_3 + 
    (1736/9 - 21056 Q{}^{2}/9 - 95728 Q{}^{4}/3)$%
$\partial{}^{2} J J 
      \partial A_{2} \partial B_{2} B_2 + 
    (7136/27 + 93008 Q{}^{2}/9 + 63776Q{}^{4})$%
$\partial{}^{2} J J 
      \partial{}^{2} A_{2} T{}^{2}+ (2312/9 + 260452 Q{}^{2}/27 + 
      611744 Q{}^{4}/9)$%
$\partial{}^{2} J J \partial{}^{2} A_{2} A_2 
      T + (-850/27 + 9422 Q{}^{2}/9 + 14812Q{}^{4})
     $%
$\partial{}^{2} J J \partial{}^{2} A_{2} A_2{}^{2}+ 
    (2110/27 + 6410 Q{}^{2}/9 + 1024Q{}^{4})$%
$\partial{}^{2} J J \partial{}^{2} A_{2} 
      B_2{}^{2}+ ((-41552Q)/3 - 105664Q{}^{3})
     $%
$\partial{}^{2} J J A_3 \partial T T + 
    (-376 - 1004Q{}^{2})$%
$\partial{}^{2} J J A_3 A_3 
      T + (-1360/9 - 2192 Q{}^{2}/3)$%
$\partial{}^{2} J J A_3 
      B_2 B_3 + (8408 Q/9 + 15808 Q{}^{3}/3)
     $%
$\partial{}^{2} J J A_3 \partial B_{2} B_2 + 
    ((-17248Q)/3 - 126776 Q{}^{3}/3)$%
$\partial{}^{2} J J \partial A_{3} 
      T{}^{2}+ (3296 Q/9 + 5536 Q{}^{3}/3)
     $%
$\partial{}^{2} J J \partial A_{3} B_2{}^{2}+ 
    (5152/27 + 3620Q{}^{2} + 44284 Q{}^{4}/3)$%
$\partial{}^{2} J J B_2{}^{2}\partial{}^{2} T + ((-35104Q)/9 - 60644 Q{}^{3}/3)
     $%
$\partial{}^{2} J J B_2 B_3 \partial T + 
    ((-39932Q)/9 - 24128Q{}^{3})$%
$\partial{}^{2} J J B_2 
      \partial B_{3} T + (24176/27 + 128032 Q{}^{2}/9 + 137588 Q{}^{4}/3)
     $%
$\partial{}^{2} J J \partial B_{2} B_2 \partial T + 
    (7336/27 + 19996 Q{}^{2}/3 + 26764Q{}^{4})$%
$\partial{}^{2} J J 
      \partial B_{2} \partial B_{2} T + ((-16588Q)/9 - 22376 Q{}^{3}/3)
     $%
$\partial{}^{2} J J \partial B_{2} B_3 T + 
    (14120/27 + 32236 Q{}^{2}/3 + 42752Q{}^{4})$%
$\partial{}^{2} J J 
      \partial{}^{2} B_{2} B_2 T + (-3400/9 - 8660 Q{}^{2}/3)
     $%
$\partial{}^{2} J J B_3 B_3 T + 
    (1840/9 + 2528 Q{}^{2}/3 - 19328Q{}^{4})$%
$\partial{}^{2} J \partial J \partial T 
      T{}^{2}+ ((-22984\sqrt{2/3}Q)/9 - 34432\sqrt{6}Q{}^{3} - 
      871808\sqrt{2/3}Q{}^{5})$%
$\partial{}^{2} J \partial J J \partial T 
      \partial T + ((-41248\sqrt{2/3}Q)/9 - 154064\sqrt{2/3}Q{}^{3} - 
      1188800\sqrt{2/3}Q{}^{5})$%
$\partial{}^{2} J \partial J J \partial{}^{2} T 
      T + (884/81 - 1085432 Q{}^{2}/81 - 1798168 Q{}^{4}/9 - 
      4041376 Q{}^{6}/9)$%
$\partial{}^{2} J \partial J J{}^{2}\partial{}^{3} T + (-4388/27 + 337828 Q{}^{2}/27 + 178896Q{}^{4} + 1409596 Q{}^{6}/3)
     $%
$\partial{}^{2} J \partial J J{}^{2}\partial{}^{3} A_{2} + 
    ((-117166Q)/9 - 1360420 Q{}^{3}/9 - 1179980 Q{}^{5}/3)
     $%
$\partial{}^{2} J \partial J J{}^{2}\partial{}^{2} A_{3} + 
    ((-1112\sqrt{2/3}Q)/9 - (981964\sqrt{2/3}Q{}^{3})/9 - 
      (2863112\sqrt{2/3}Q{}^{5})/3)$%
$\partial{}^{2} J \partial J J 
      A_2 \partial{}^{2} T + ((2044\sqrt{2/3})/9 - (130246\sqrt{2/3}Q{}^{2})/9 - 
      (396272\sqrt{2/3}Q{}^{4})/3)$%
$\partial{}^{2} J \partial J J 
      A_2 \partial A_{3} + ((-22960\sqrt{2/3}Q)/9 - 
      (588760\sqrt{2/3}Q{}^{3})/3 - 512616\sqrt{6}Q{}^{5})
     $%
$\partial{}^{2} J \partial J J \partial A_{2} \partial T + 
    ((-47176\sqrt{2/3}Q)/9 - 116342\sqrt{2/3}Q{}^{3} - 767602\sqrt{2/3}Q{}^{5})
     $%
$\partial{}^{2} J \partial J J \partial A_{2} \partial A_{2} + 
    ((1588\sqrt{2/3})/3 - (34990\sqrt{2/3}Q{}^{2})/9 - (182708\sqrt{2/3}Q{}^{4})/3)
     $%
$\partial{}^{2} J \partial J J \partial A_{2} A_3 + 
    ((-10996\sqrt{2/3}Q)/9 - 128192\sqrt{2/3}Q{}^{3} - 1104848\sqrt{2/3}Q{}^{5})
     $%
$\partial{}^{2} J \partial J J \partial{}^{2} A_{2} T + 
    ((-18148\sqrt{2/3}Q)/3 - (1348414\sqrt{2/3}Q{}^{3})/9 - 
      (3219368\sqrt{2/3}Q{}^{5})/3)$%
$\partial{}^{2} J \partial J J 
      \partial{}^{2} A_{2} A_2 + ((-20528\sqrt{2/3})/9 + 
      (42488\sqrt{2/3}Q{}^{2})/3 + 249512\sqrt{2/3}Q{}^{4})
     $%
$\partial{}^{2} J \partial J J A_3 \partial T + 
    (1436\sqrt{2/3}Q - 890\sqrt{6}Q{}^{3})$%
$\partial{}^{2} J \partial J J 
      A_3 A_3 + ((-23248\sqrt{2/3})/9 - 
      (15880\sqrt{2/3}Q{}^{2})/3 + 141352\sqrt{2/3}Q{}^{4})
     $%
$\partial{}^{2} J \partial J J \partial A_{3} T + 
    ((2596\sqrt{2/3})/3 - (17654\sqrt{2/3}Q{}^{2})/9 - (69448\sqrt{2/3}Q{}^{4})/3)
     $%
$\partial{}^{2} J \partial J J B_2 \partial B_{3} + 
    ((-22952\sqrt{2/3}Q)/3 - (692366\sqrt{2/3}Q{}^{3})/9 - 
      (437926\sqrt{2/3}Q{}^{5})/3)$%
$\partial{}^{2} J \partial J J 
      \partial B_{2} \partial B_{2} + ((2596\sqrt{2/3})/3 - 
      (140270\sqrt{2/3}Q{}^{2})/9 - (398284\sqrt{2/3}Q{}^{4})/3)
     $%
$\partial{}^{2} J \partial J J \partial B_{2} B_3 + 
    ((-26552\sqrt{2/3}Q)/3 - (887558\sqrt{2/3}Q{}^{3})/9 - 
      (740512\sqrt{2/3}Q{}^{5})/3)$%
$\partial{}^{2} J \partial J J 
      \partial{}^{2} B_{2} B_2 + (3428\sqrt{2/3}Q + 10614\sqrt{6}Q{}^{3})
     $%
$\partial{}^{2} J \partial J J B_3 B_3 + 
    ((-41416\sqrt{2/3}Q)/9 - (379144\sqrt{2/3}Q{}^{3})/3 - 284640\sqrt{6}Q{}^{5})
     $%
$\partial{}^{2} J (\partial J){}^{2} \partial T T + 
    (-128/27 - 129256 Q{}^{2}/9 + 105136 Q{}^{4}/3 + 2480448Q{}^{6})
     $%
$\partial{}^{2} J (\partial J){}^{2} J \partial{}^{2} T + 
    (-1448/9 + 1213444 Q{}^{2}/27 + 2607548 Q{}^{4}/3 + 13147328 Q{}^{6}/3)
     $%
$\partial{}^{2} J (\partial J){}^{2} J \partial{}^{2} A_{2} + 
    ((-72728Q)/3 - 2406404 Q{}^{3}/9 - 1849036 Q{}^{5}/3)
     $%
$\partial{}^{2} J (\partial J){}^{2} J \partial A_{3} + 
    (-184/81 + 38936 Q{}^{2}/9 + 3104416 Q{}^{4}/9 + 3512320Q{}^{6})
     $%
$\partial{}^{2} J (\partial J){}^{3} \partial T + 
    (-2872/27 + 35860Q{}^{2} + 793176Q{}^{4} + 4606164Q{}^{6})
     $%
$\partial{}^{2} J (\partial J){}^{3} \partial A_{2} + 
    (-7336Q - 85496Q{}^{3} - 217192Q{}^{5})$%
$\partial{}^{2} J (\partial J){}^{3} A_3 + ((-31736\sqrt{2/3}Q)/9 - 
      (846100\sqrt{2/3}Q{}^{3})/9 - (1712732\sqrt{2/3}Q{}^{5})/3)
     $%
$\partial{}^{2} J (\partial J){}^{2} A_2 \partial T + 
    ((1022\sqrt{2/3})/9 - (44032\sqrt{2/3}Q{}^{2})/9 - (138332\sqrt{2/3}Q{}^{4})/3)
     $%
$\partial{}^{2} J (\partial J){}^{2} A_2 A_3 + 
    ((-38212\sqrt{2/3}Q)/9 - (384868\sqrt{2/3}Q{}^{3})/3 - 
      819664\sqrt{2/3}Q{}^{5})$%
$\partial{}^{2} J (\partial J){}^{2} \partial A_{2} 
      T + ((-51514\sqrt{2/3}Q)/9 - (1057438\sqrt{2/3}Q{}^{3})/9 - 
      (2211350\sqrt{2/3}Q{}^{5})/3)$%
$\partial{}^{2} J (\partial J){}^{2} 
      \partial A_{2} A_2 + ((-8480\sqrt{2/3})/9 + 
      (73468\sqrt{2/3}Q{}^{2})/3 + 238604\sqrt{2/3}Q{}^{4})
     $%
$\partial{}^{2} J (\partial J){}^{2} A_3 T + 
    ((1298\sqrt{2/3})/3 - (5900\sqrt{2/3}Q{}^{2})/9 - (43792\sqrt{2/3}Q{}^{4})/3)
     $%
$\partial{}^{2} J (\partial J){}^{2} B_2 B_3 + 
    ((-24826\sqrt{2/3}Q)/3 - (852794\sqrt{2/3}Q{}^{3})/9 - 
      (706630\sqrt{2/3}Q{}^{5})/3)$%
$\partial{}^{2} J (\partial J){}^{2} 
      \partial B_{2} B_2 + (10208/9 - 92528 Q{}^{2}/27 - 1059856 Q{}^{4}/9)
     $%
$\partial{}^{2} J \partial J A_2 \partial T T + 
    (1528/9 - 10988 Q{}^{2}/9 - 86608 Q{}^{4}/3)$%
$\partial{}^{2} J \partial J 
      A_2{}^{2}\partial T + (60Q + 1480 Q{}^{3}/3)
     $%
$\partial{}^{2} J \partial J A_2{}^{3}3 + 
    ((-30244Q)/9 - 80368 Q{}^{3}/3)$%
$\partial{}^{2} J \partial J A_2 
      A_3 T + (92Q - 160Q{}^{3})$%
$\partial{}^{2} J \partial J 
      A_2 B_2 B_3 + 
    (4936/27 - 5300 Q{}^{2}/9 - 31868 Q{}^{4}/3)$%
$\partial{}^{2} J \partial J 
      A_2 \partial B_{2} B_2 + 
    (5104/9 + 37096 Q{}^{2}/9 - 58984 Q{}^{4}/3)$%
$\partial{}^{2} J \partial J 
      \partial A_{2} T{}^{2}+ (11888/27 + 47044 Q{}^{2}/27 - 
      311488 Q{}^{4}/9)$%
$\partial{}^{2} J \partial J \partial A_{2} A_2 
      T + (-3244/27 - 50848 Q{}^{2}/9 - 146318 Q{}^{4}/3)
     $%
$\partial{}^{2} J \partial J \partial A_{2} A_2{}^{2}+ 
    (1276/9 + 10172 Q{}^{2}/9 + 806Q{}^{4})$%
$\partial{}^{2} J \partial J \partial A_{2} 
      B_2{}^{2}+ ((-25904Q)/3 - 54592Q{}^{3})
     $%
$\partial{}^{2} J \partial J A_3 T{}^{2}+ 
    (1024 Q/3 + 6472 Q{}^{3}/3)$%
$\partial{}^{2} J \partial J A_3 
      B_2{}^{2}+ (1192/3 - 652Q{}^{2} - 88144 Q{}^{4}/3)
     $%
$\partial{}^{2} J \partial J B_2{}^{2}\partial T + 
    ((-46484Q)/9 - 78016 Q{}^{3}/3)$%
$\partial{}^{2} J \partial J B_2 
      B_3 T + (18736/27 + 6092 Q{}^{2}/9 - 47568Q{}^{4})
     $%
$\partial{}^{2} J \partial J \partial B_{2} B_2 T + 
    (808/27 + 1792 Q{}^{2}/9 - 1920Q{}^{4})$%
$\partial{}^{2} J \partial{}^{2} J T{}^{3}+ ((-8768\sqrt{2/3}Q)/3 - 88256\sqrt{2/3}Q{}^{3} - 
      212736\sqrt{6}Q{}^{5})$%
$\partial{}^{2} J \partial{}^{2} J J \partial T 
      T + (1264/81 - 197152 Q{}^{2}/27 - 717212 Q{}^{4}/9 + 122160Q{}^{6})
     $%
$\partial{}^{2} J \partial{}^{2} J J{}^{2}\partial{}^{2} T + 
    (-9272/81 + 239884 Q{}^{2}/27 + 429770 Q{}^{4}/3 + 1494484 Q{}^{6}/3)
     $%
$\partial{}^{2} J \partial{}^{2} J J{}^{2}\partial{}^{2} A_{2} + 
    ((-79138Q)/9 - 853462 Q{}^{3}/9 - 622436 Q{}^{5}/3)
     $%
$\partial{}^{2} J \partial{}^{2} J J{}^{2}\partial A_{3} + 
    ((-16352\sqrt{2/3}Q)/9 - (209692\sqrt{2/3}Q{}^{3})/3 - 
      488936\sqrt{2/3}Q{}^{5})$%
$\partial{}^{2} J \partial{}^{2} J J A_2 
      \partial T + ((1471\sqrt{2/3})/9 - (13636\sqrt{2/3}Q{}^{2})/9 - 
      (67424\sqrt{2/3}Q{}^{4})/3)$%
$\partial{}^{2} J \partial{}^{2} J J 
      A_2 A_3 + ((-22360\sqrt{2/3}Q)/9 - 
      (897284\sqrt{2/3}Q{}^{3})/9 - (2192632\sqrt{2/3}Q{}^{5})/3)
     $%
$\partial{}^{2} J \partial{}^{2} J J \partial A_{2} T + 
    ((-39923\sqrt{2/3}Q)/9 - (926932\sqrt{2/3}Q{}^{3})/9 - 
      (2094308\sqrt{2/3}Q{}^{5})/3)$%
$\partial{}^{2} J \partial{}^{2} J J 
      \partial A_{2} A_2 + ((-10204\sqrt{2/3})/9 + 
      (28528\sqrt{2/3}Q{}^{2})/3 + 166384\sqrt{2/3}Q{}^{4})
     $%
$\partial{}^{2} J \partial{}^{2} J J A_3 T + 
    ((1325\sqrt{2/3})/3 + (13036\sqrt{2/3}Q{}^{2})/9 - (4840\sqrt{2/3}Q{}^{4})/3)
     $%
$\partial{}^{2} J \partial{}^{2} J J B_2 B_3 + 
    ((-16897\sqrt{2/3}Q)/3 - (518972\sqrt{2/3}Q{}^{3})/9 - 
      (287476\sqrt{2/3}Q{}^{5})/3)$%
$\partial{}^{2} J \partial{}^{2} J J 
      \partial B_{2} B_2 + ((-17644\sqrt{2/3}Q)/9 - 
      (145568\sqrt{2/3}Q{}^{3})/3 - 299968\sqrt{2/3}Q{}^{5})
     $%
$\partial{}^{2} J \partial{}^{2} J \partial J T{}^{2}+ 
    (-3436/81 - 71056 Q{}^{2}/27 + 318648Q{}^{4} + 4129440Q{}^{6})
     $%
$\partial{}^{2} J \partial{}^{2} J \partial J J \partial T + 
    (-45580/81 + 1202188 Q{}^{2}/27 + 26492584 Q{}^{4}/27 + 47831024 Q{}^{6}/9)
     $%
$\partial{}^{2} J \partial{}^{2} J \partial J J \partial A_{2} + 
    ((-135712Q)/9 - 509404 Q{}^{3}/3 - 405256Q{}^{5})$%
$\partial{}^{2} J \partial{}^{2} J 
      \partial J J A_3 + 
    (9740/81 + 518636 Q{}^{2}/27 + 5752208 Q{}^{4}/9 + 15334016 Q{}^{6}/3)
     $%
$\partial{}^{2} J \partial{}^{2} J (\partial J){}^{2} T + 
    ((62056\sqrt{2/3}Q)/27 + (504470\sqrt{2/3}Q{}^{3})/9 + 83664\sqrt{6}Q{}^{5} - 
      1088032\sqrt{2/3}Q{}^{7})$%
$\partial{}^{2} J \partial{}^{2} J (\partial J){}^{3} + (-11332/81 + 301132 Q{}^{2}/9 + 23288080 Q{}^{4}/27 + 
      50204852 Q{}^{6}/9)$%
$\partial{}^{2} J \partial{}^{2} J (\partial J){}^{2} 
      A_2 + ((-39772\sqrt{2/3}Q)/9 - (796484\sqrt{2/3}Q{}^{3})/9 - 
      156896\sqrt{6}Q{}^{5})$%
$\partial{}^{2} J \partial{}^{2} J \partial J A_2 
      T + ((-6574\sqrt{2/3}Q)/3 - (139766\sqrt{2/3}Q{}^{3})/3 - 
      (870584\sqrt{2/3}Q{}^{5})/3)$%
$\partial{}^{2} J \partial{}^{2} J \partial J 
      A_2{}^{2}+ ((-24568\sqrt{2/3}Q)/9 - 
      33398\sqrt{2/3}Q{}^{3} - (269432\sqrt{2/3}Q{}^{5})/3)
     $%
$\partial{}^{2} J \partial{}^{2} J \partial J B_2{}^{2}+ 
    (4244/81 + 15136 Q{}^{2}/3 + 1522112 Q{}^{4}/9 + 4180480 Q{}^{6}/3)
     $%
$\partial{}^{2} J \partial{}^{2} J \partial{}^{2} J J T + 
    (-13672/81 + 200228 Q{}^{2}/27 + 5910148 Q{}^{4}/27 + 12613976 Q{}^{6}/9)
     $%
$\partial{}^{2} J \partial{}^{2} J \partial{}^{2} J J A_2 + 
    ((122944\sqrt{2/3}Q)/27 + (566672\sqrt{2/3}Q{}^{3})/3 + 
      (8929616\sqrt{2/3}Q{}^{5})/3 + 5462208\sqrt{6}Q{}^{7})
     $%
$\partial{}^{2} J \partial{}^{2} J \partial{}^{2} J \partial J J + 
    (6832/27 - 16624 Q{}^{2}/9 - 331336 Q{}^{4}/9)$%
$\partial{}^{2} J \partial{}^{2} J 
      A_2 T{}^{2}+ (286/3 + 484 Q{}^{2}/9 - 69920 Q{}^{4}/9)
     $%
$\partial{}^{2} J \partial{}^{2} J A_2{}^{2}T + 
    (-271/9 - 12880 Q{}^{2}/9 - 109138 Q{}^{4}/9)$%
$\partial{}^{2} J \partial{}^{2} J 
      A_2{}^{3}+ 
    (1621/27 + 8920 Q{}^{2}/9 + 53434 Q{}^{4}/9)$%
$\partial{}^{2} J \partial{}^{2} J 
      A_2 B_2{}^{2}+ 
    (4258/27 - 21004 Q{}^{2}/9 - 288896 Q{}^{4}/9)$%
$\partial{}^{2} J \partial{}^{2} J 
      B_2{}^{2}T + 
    (3104\sqrt{2/3}Q + (83992\sqrt{2/3}Q{}^{3})/3)$%
$\partial{}^{2} J A_2 
      \partial T T{}^{2}+ ((22004\sqrt{2/3}Q)/9 + 6240\sqrt{6}Q{}^{3})
     $%
$\partial{}^{2} J A_2{}^{2}\partial T T + 
    (268\sqrt{2/3}Q + (9400\sqrt{2/3}Q{}^{3})/3)$%
$\partial{}^{2} J A_2{}^{3}\partial T + 
    (28\sqrt{2/3}Q{}^{2}$%
$\partial{}^{2} J A_2{}^{4}3)/3 + ((-32\sqrt{2/3})/3 + (140\sqrt{2/3}Q{}^{2})/3)
     $%
$\partial{}^{2} J A_2{}^{3}3 T - 
    20\sqrt{2/3}Q{}^{2}$%
$\partial{}^{2} J A_2{}^{2}B_2 
      B_3 + ((-1756\sqrt{2/3}Q)/9 - (5074\sqrt{2/3}Q{}^{3})/3)
     $%
$\partial{}^{2} J A_2{}^{2}\partial B_{2} B_2 + 
    ((1280\sqrt{2/3})/3 + (5644\sqrt{2/3}Q{}^{2})/3)$%
$\partial{}^{2} J A_2 
      A_3 T{}^{2}+ 
    (116\sqrt{2/3}Q{}^{2}$%
$\partial{}^{2} J A_2 A_3 B_2{}^{2})/3 + ((2404\sqrt{2/3}Q)/3 + (16928\sqrt{2/3}Q{}^{3})/3)
     $%
$\partial{}^{2} J A_2 B_2{}^{2}\partial T + 
    ((664\sqrt{2/3})/3 + 844\sqrt{2/3}Q{}^{2})$%
$\partial{}^{2} J A_2 
      B_2 B_3 T + 
    ((16784\sqrt{2/3}Q)/9 + (43064\sqrt{2/3}Q{}^{3})/3)
     $%
$\partial{}^{2} J A_2 \partial B_{2} B_2 T + 
    (632\sqrt{2/3}Q + 2256\sqrt{6}Q{}^{3})$%
$\partial{}^{2} J \partial A_{2} T{}^{3}+ ((21080\sqrt{2/3}Q)/9 + (61952\sqrt{2/3}Q{}^{3})/3)
     $%
$\partial{}^{2} J \partial A_{2} A_2 T{}^{2}+ 
    ((-1016\sqrt{2/3}Q)/3 + (470\sqrt{2/3}Q{}^{3})/3)
     $%
$\partial{}^{2} J \partial A_{2} A_2{}^{2}T + 
    ((5300\sqrt{2/3}Q)/9 + (17984\sqrt{2/3}Q{}^{3})/3)
     $%
$\partial{}^{2} J \partial A_{2} A_2{}^{3}+ 
    ((-3932\sqrt{2/3}Q)/9 - 1232\sqrt{6}Q{}^{3})$%
$\partial{}^{2} J \partial A_{2} 
      A_2 B_2{}^{2}+ 
    ((10264\sqrt{2/3}Q)/9 + 3538\sqrt{6}Q{}^{3})$%
$\partial{}^{2} J \partial A_{2} 
      B_2{}^{2}T + 
    ((1312\sqrt{2/3})/3 + 1688\sqrt{2/3}Q{}^{2})$%
$\partial{}^{2} J A_3 
      T{}^{3}+ ((680\sqrt{2/3})/3 + (3128\sqrt{2/3}Q{}^{2})/3)
     $%
$\partial{}^{2} J A_3 B_2{}^{2}T + 
    ((12124\sqrt{2/3}Q)/3 + (87104\sqrt{2/3}Q{}^{3})/3)
     $%
$\partial{}^{2} J B_2{}^{2}\partial T T + 
    4\sqrt{2/3}Q{}^{2}$%
$\partial{}^{2} J B_2{}^{4}3 + ((2704\sqrt{2/3})/3 + 1220\sqrt{6}Q{}^{2})
     $%
$\partial{}^{2} J B_2 B_3 T{}^{2}+ 
    (2816\sqrt{2/3}Q + 20624\sqrt{2/3}Q{}^{3})$%
$\partial{}^{2} J \partial B_{2} 
      B_2 T{}^{2}+ 
    ((1876\sqrt{2/3}Q)/9 + (3458\sqrt{2/3}Q{}^{3})/3)
     $%
$\partial{}^{2} J \partial B_{2} B_2{}^{3}+ 
    ((64\sqrt{2/3}Q)/3 + 256\sqrt{2/3}Q{}^{3})$%
$\partial{}^{3} J T{}^{4}+ (14528/81 + 11360 Q{}^{2}/3 + 176512 Q{}^{4}/9)
     $%
$\partial{}^{3} J J \partial T T{}^{2}+ 
    ((-1252\sqrt{2/3}Q)/9 - (38296\sqrt{2/3}Q{}^{3})/3 - 44384\sqrt{6}Q{}^{5})
     $%
$\partial{}^{3} J J{}^{2}\partial T \partial T + 
    ((-21112\sqrt{2/3}Q)/27 - (89896\sqrt{2/3}Q{}^{3})/3 - 
      (740960\sqrt{2/3}Q{}^{5})/3)$%
$\partial{}^{3} J J{}^{2}\partial{}^{2} T 
      T + (1852/243 - 232072 Q{}^{2}/81 - 1576552 Q{}^{4}/27 - 
      2474528 Q{}^{6}/9)$%
$\partial{}^{3} J J{}^{3}\partial{}^{3} T + (-10292/243 + 68062 Q{}^{2}/81 + 1408Q{}^{4} - 107784Q{}^{6})
     $%
$\partial{}^{3} J J{}^{3}\partial{}^{3} A_{2} + 
    ((-187898Q)/81 - 758110 Q{}^{3}/27 - 245804 Q{}^{5}/3)
     $%
$\partial{}^{3} J J{}^{3}\partial{}^{2} A_{3} + 
    ((15898\sqrt{2/3}Q)/27 - (63098\sqrt{2/3}Q{}^{3})/3 - 
      (658432\sqrt{2/3}Q{}^{5})/3)$%
$\partial{}^{3} J J{}^{2}A_2 \partial{}^{2} T + ((860\sqrt{2/3})/9 - (25408\sqrt{2/3}Q{}^{2})/9 - 
      (97079\sqrt{2/3}Q{}^{4})/3)$%
$\partial{}^{3} J J{}^{2}A_2 \partial A_{3} + ((4688\sqrt{2/3}Q)/9 - 
      (269848\sqrt{2/3}Q{}^{3})/9 - (857696\sqrt{2/3}Q{}^{5})/3)
     $%
$\partial{}^{3} J J{}^{2}\partial A_{2} \partial T + 
    ((-34199\sqrt{2/3}Q)/27 - (245767\sqrt{2/3}Q{}^{3})/9 - 
      180176\sqrt{2/3}Q{}^{5})$%
$\partial{}^{3} J J{}^{2}\partial A_{2} 
      \partial A_{2} + ((5548\sqrt{2/3})/27 + (10040\sqrt{2/3}Q{}^{2})/9 + 
      749\sqrt{2/3}Q{}^{4})$%
$\partial{}^{3} J J{}^{2}\partial A_{2} 
      A_3 + (96\sqrt{6}Q - 17852\sqrt{2/3}Q{}^{3} - 61424\sqrt{6}Q{}^{5})
     $%
$\partial{}^{3} J J{}^{2}\partial{}^{2} A_{2} T + 
    ((-12175\sqrt{2/3}Q)/9 - 30533\sqrt{2/3}Q{}^{3} - 211601\sqrt{2/3}Q{}^{5})
     $%
$\partial{}^{3} J J{}^{2}\partial{}^{2} A_{2} A_2 + 
    ((-20696\sqrt{2/3})/27 - (16532\sqrt{2/3}Q{}^{2})/3 + 
      (3692\sqrt{2/3}Q{}^{4})/3)$%
$\partial{}^{3} J J{}^{2}A_3 
      \partial T + (728\sqrt{2/3}Q + 449\sqrt{6}Q{}^{3})
     $%
$\partial{}^{3} J J{}^{2}A_3 A_3 + 
    ((-7888\sqrt{2/3})/9 - 4180\sqrt{6}Q{}^{2} - 44824\sqrt{2/3}Q{}^{4})
     $%
$\partial{}^{3} J J{}^{2}\partial A_{3} T + 
    ((7604\sqrt{2/3})/27 + (928\sqrt{2/3}Q{}^{2})/9 - 979\sqrt{6}Q{}^{4})
     $%
$\partial{}^{3} J J{}^{2}B_2 \partial B_{3} + 
    ((-53009\sqrt{2/3}Q)/27 - (197891\sqrt{2/3}Q{}^{3})/9 - 
      (178604\sqrt{2/3}Q{}^{5})/3)$%
$\partial{}^{3} J J{}^{2}\partial B_{2} \partial B_{2} + ((7604\sqrt{2/3})/27 - 
      (22580\sqrt{2/3}Q{}^{2})/9 - 22571\sqrt{2/3}Q{}^{4})
     $%
$\partial{}^{3} J J{}^{2}\partial B_{2} B_3 + 
    ((-61667\sqrt{2/3}Q)/27 - (242519\sqrt{2/3}Q{}^{3})/9 - 
      (239171\sqrt{2/3}Q{}^{5})/3)$%
$\partial{}^{3} J J{}^{2}\partial{}^{2} B_{2} B_2 + ((1474\sqrt{2/3}Q)/3 + 4811\sqrt{2/3}Q{}^{3})
     $%
$\partial{}^{3} J J{}^{2}B_3 B_3 + 
    (72320/81 + 100832 Q{}^{2}/9 + 380224 Q{}^{4}/9)$%
$\partial{}^{3} J J 
      A_2 \partial T T + (11176/81 + 51280 Q{}^{2}/27 + 
      81220 Q{}^{4}/9)$%
$\partial{}^{3} J J A_2{}^{2}\partial T + (28Q + 656 Q{}^{3}/3)$%
$\partial{}^{3} J J A_2{}^{3}3 + ((-6356Q)/3 - 63424 Q{}^{3}/3)
     $%
$\partial{}^{3} J J A_2 A_3 T + 
    (148 Q/3 + 16Q{}^{3})$%
$\partial{}^{3} J J A_2 B_2 
      B_3 + (3184/27 - 31040 Q{}^{2}/27 - 141952 Q{}^{4}/9)
     $%
$\partial{}^{3} J J A_2 \partial B_{2} B_2 + 
    (36160/81 + 248368 Q{}^{2}/27 + 426160 Q{}^{4}/9)$%
$\partial{}^{3} J J 
      \partial A_{2} T{}^{2}+ (28288/81 + 25804 Q{}^{2}/3 + 
      499616 Q{}^{4}/9)$%
$\partial{}^{3} J J \partial A_{2} A_2 
      T + (-2768/81 + 13462 Q{}^{2}/9 + 153620 Q{}^{4}/9)
     $%
$\partial{}^{3} J J \partial A_{2} A_2{}^{2}+ 
    (7744/81 + 20630 Q{}^{2}/27 + 1060Q{}^{4})$%
$\partial{}^{3} J J 
      \partial A_{2} B_2{}^{2}+ ((-47824Q)/9 - 109312 Q{}^{3}/3)
     $%
$\partial{}^{3} J J A_3 T{}^{2}+ 
    (1472 Q/9 + 880Q{}^{3})$%
$\partial{}^{3} J J A_3 B_2{}^{2}+ (2776/9 + 10336 Q{}^{2}/3 + 31244 Q{}^{4}/3)
     $%
$\partial{}^{3} J J B_2{}^{2}\partial T + 
    ((-27740Q)/9 - 41200 Q{}^{3}/3)$%
$\partial{}^{3} J J B_2 
      B_3 T + (44032/81 + 20972 Q{}^{2}/3 + 133760 Q{}^{4}/9)
     $%
$\partial{}^{3} J J \partial B_{2} B_2 T + 
    (2464/81 - 2032 Q{}^{2}/9 - 63808 Q{}^{4}/9)$%
$\partial{}^{3} J \partial J T{}^{3}+ (-4472\sqrt{2/3}Q - 45856\sqrt{6}Q{}^{3} - 
      335616\sqrt{6}Q{}^{5})$%
$\partial{}^{3} J \partial J J \partial T 
      T + (3412/243 - 939748 Q{}^{2}/81 - 3982840 Q{}^{4}/27 - 
      677024 Q{}^{6}/9)$%
$\partial{}^{3} J \partial J J{}^{2}\partial{}^{2} T + (-44780/243 + 1017572 Q{}^{2}/81 + 5689856 Q{}^{4}/27 + 
      6839404 Q{}^{6}/9)$%
$\partial{}^{3} J \partial J J{}^{2}\partial{}^{2} A_{2} + ((-343636Q)/27 - 1284140 Q{}^{3}/9 - 344464Q{}^{5})
     $%
$\partial{}^{3} J \partial J J{}^{2}\partial A_{3} + 
    ((-87272\sqrt{2/3}Q)/27 - (373712\sqrt{2/3}Q{}^{3})/3 - 
      (2710744\sqrt{2/3}Q{}^{5})/3)$%
$\partial{}^{3} J \partial J J 
      A_2 \partial T + ((1720\sqrt{2/3})/9 - (8276\sqrt{2/3}Q{}^{2})/3 - 
      35318\sqrt{2/3}Q{}^{4})$%
$\partial{}^{3} J \partial J J A_2 
      A_3 + ((-100432\sqrt{2/3}Q)/27 - (1328024\sqrt{2/3}Q{}^{3})/9 - 
      1072400\sqrt{2/3}Q{}^{5})$%
$\partial{}^{3} J \partial J J \partial A_{2} 
      T + ((-175660\sqrt{2/3}Q)/27 - (1379498\sqrt{2/3}Q{}^{3})/9 - 
      1044802\sqrt{2/3}Q{}^{5})$%
$\partial{}^{3} J \partial J J \partial A_{2} 
      A_2 + ((-14528\sqrt{2/3})/9 + 8576\sqrt{2/3}Q{}^{2} + 
      165872\sqrt{2/3}Q{}^{4})$%
$\partial{}^{3} J \partial J J A_3 
      T + ((15208\sqrt{2/3})/27 + (3820\sqrt{2/3}Q{}^{2})/3 - 
      (16522\sqrt{2/3}Q{}^{4})/3)$%
$\partial{}^{3} J \partial J J 
      B_2 B_3 + ((-73820\sqrt{2/3}Q)/9 - 
      (807182\sqrt{2/3}Q{}^{3})/9 - (577006\sqrt{2/3}Q{}^{5})/3)
     $%
$\partial{}^{3} J \partial J J \partial B_{2} B_2 + 
    ((-34568\sqrt{2/3}Q)/27 - (92632\sqrt{2/3}Q{}^{3})/3 - 
      (558496\sqrt{2/3}Q{}^{5})/3)$%
$\partial{}^{3} J (\partial J){}^{2} T{}^{2}+ (-6592/243 - 216440 Q{}^{2}/81 + 5739472 Q{}^{4}/27 + 
      25999040 Q{}^{6}/9)$%
$\partial{}^{3} J (\partial J){}^{2} J 
      \partial T + (-102976/243 + 2574472 Q{}^{2}/81 + 19494508 Q{}^{4}/27 + 
      36065168 Q{}^{6}/9)$%
$\partial{}^{3} J (\partial J){}^{2} J 
      \partial A_{2} + ((-298648Q)/27 - 127108Q{}^{3} - 936632 Q{}^{5}/3)
     $%
$\partial{}^{3} J (\partial J){}^{2} J A_3 + 
    (536/9 + 699464 Q{}^{2}/81 + 2582872 Q{}^{4}/9 + 20729248 Q{}^{6}/9)
     $%
$\partial{}^{3} J (\partial J){}^{3} T + 
    ((71948\sqrt{2/3}Q)/81 + (636904\sqrt{2/3}Q{}^{3})/27 + 
      (1277860\sqrt{2/3}Q{}^{5})/9 - (474352\sqrt{2/3}Q{}^{7})/3)
     $%
$\partial{}^{3} J (\partial J){}^{4} + 
    (-17656/243 + 1234004 Q{}^{2}/81 + 10613080 Q{}^{4}/27 + 22806724 Q{}^{6}/9)
     $%
$\partial{}^{3} J (\partial J){}^{3} A_2 + 
    ((-8444\sqrt{2/3}Q)/3 - (494744\sqrt{2/3}Q{}^{3})/9 - 
      (857668\sqrt{2/3}Q{}^{5})/3)$%
$\partial{}^{3} J (\partial J){}^{2} 
      A_2 T + ((-38540\sqrt{2/3}Q)/27 - 
      (257516\sqrt{2/3}Q{}^{3})/9 - (516674\sqrt{2/3}Q{}^{5})/3)
     $%
$\partial{}^{3} J (\partial J){}^{2} A_2{}^{2}+ 
    ((-53308\sqrt{2/3}Q)/27 - (219472\sqrt{2/3}Q{}^{3})/9 - 
      (193738\sqrt{2/3}Q{}^{5})/3)$%
$\partial{}^{3} J (\partial J){}^{2} 
      B_2{}^{2}+ (8384/27 - 106184 Q{}^{2}/27 - 527368 Q{}^{4}/9)
     $%
$\partial{}^{3} J \partial J A_2 T{}^{2}+ 
    (7976/81 - 42164 Q{}^{2}/27 - 75884 Q{}^{4}/3)$%
$\partial{}^{3} J \partial J 
      A_2{}^{2}T + 
    (-3280/81 - 56558 Q{}^{2}/27 - 51958 Q{}^{4}/3)$%
$\partial{}^{3} J \partial J 
      A_2{}^{3}+ 
    (5744/81 + 29006 Q{}^{2}/27 + 46226 Q{}^{4}/9)$%
$\partial{}^{3} J \partial J 
      A_2 B_2{}^{2}+ 
    (17176/81 - 25388 Q{}^{2}/9 - 324436 Q{}^{4}/9)$%
$\partial{}^{3} J \partial J 
      B_2{}^{2}T + 
    ((-6688\sqrt{2/3}Q)/3 - (559136\sqrt{2/3}Q{}^{3})/9 - 
      (1273472\sqrt{2/3}Q{}^{5})/3)$%
$\partial{}^{3} J \partial{}^{2} J J T{}^{2}+ (-5692/243 - 680164 Q{}^{2}/81 - 1609448 Q{}^{4}/27 + 
      4080544 Q{}^{6}/9)$%
$\partial{}^{3} J \partial{}^{2} J J{}^{2}\partial T + (-114988/243 + 737414 Q{}^{2}/81 + 712156 Q{}^{4}/3 + 
      10697528 Q{}^{6}/9)$%
$\partial{}^{3} J \partial{}^{2} J J{}^{2}\partial A_{2} + ((-71188Q)/9 - 279850 Q{}^{3}/3 - 263708Q{}^{5})
     $%
$\partial{}^{3} J \partial{}^{2} J J{}^{2}A_3 + 
    ((-45088\sqrt{2/3}Q)/9 - (1166872\sqrt{2/3}Q{}^{3})/9 - 
      (2506288\sqrt{2/3}Q{}^{5})/3)$%
$\partial{}^{3} J \partial{}^{2} J J 
      A_2 T + ((-63328\sqrt{2/3}Q)/27 - 
      (526886\sqrt{2/3}Q{}^{3})/9 - (1216238\sqrt{2/3}Q{}^{5})/3)
     $%
$\partial{}^{3} J \partial{}^{2} J J A_2{}^{2}+ 
    ((-25424\sqrt{2/3}Q)/9 - (103934\sqrt{2/3}Q{}^{3})/3 - 
      (291470\sqrt{2/3}Q{}^{5})/3)$%
$\partial{}^{3} J \partial{}^{2} J J 
      B_2{}^{2}+ (17288/81 + 577136 Q{}^{2}/27 + 
      6611584 Q{}^{4}/9 + 18318592 Q{}^{6}/3)$%
$\partial{}^{3} J \partial{}^{2} J \partial J 
      J T + (-55576/81 + 925528 Q{}^{2}/27 + 9148844 Q{}^{4}/9 + 
      19919752 Q{}^{6}/3)$%
$\partial{}^{3} J \partial{}^{2} J \partial J J 
      A_2 + ((829388\sqrt{2/3}Q)/81 + (11462900\sqrt{2/3}Q{}^{3})/27 + 
      (58676584\sqrt{2/3}Q{}^{5})/9 + (104380768\sqrt{2/3}Q{}^{7})/3)
     $%
$\partial{}^{3} J \partial{}^{2} J (\partial J){}^{2} J + 
    ((327050\sqrt{2/3}Q)/81 + (4953934\sqrt{2/3}Q{}^{3})/27 + 
      (27403372\sqrt{2/3}Q{}^{5})/9 + (51281744\sqrt{2/3}Q{}^{7})/3)
     $%
$\partial{}^{3} J \partial{}^{2} J \partial{}^{2} J J{}^{2}+ 
    (6832/243 + 73804 Q{}^{2}/81 + 999824 Q{}^{4}/27 + 3255680 Q{}^{6}/9)
     $%
$\partial{}^{3} J \partial{}^{3} J J{}^{2}T + 
    (-34720/243 + 40474 Q{}^{2}/81 + 476390 Q{}^{4}/9 + 1157812 Q{}^{6}/3)
     $%
$\partial{}^{3} J \partial{}^{3} J J{}^{2}A_2 + 
    ((81244\sqrt{2/3}Q)/27 + (3714200\sqrt{2/3}Q{}^{3})/27 + 
      (6825572\sqrt{2/3}Q{}^{5})/3 + (38078512\sqrt{2/3}Q{}^{7})/3)
     $%
$\partial{}^{3} J \partial{}^{3} J \partial J J{}^{2}+ 
    ((6088\sqrt{2/3}Q)/9 + (18632\sqrt{2/3}Q{}^{3})/3)
     $%
$\partial{}^{3} J A_2 T{}^{3}+ 
    ((7856\sqrt{2/3}Q)/9 + 6640\sqrt{2/3}Q{}^{3})$%
$\partial{}^{3} J A_2{}^{2}T{}^{2}+ ((-1036\sqrt{2/3}Q)/9 - 10\sqrt{6}Q{}^{3})
     $%
$\partial{}^{3} J A_2{}^{3}T + 
    ((1477\sqrt{2/3}Q)/9 + (4316\sqrt{2/3}Q{}^{3})/3)
     $%
$\partial{}^{3} J A_2{}^{4}+ 
    (-66\sqrt{6}Q - 1486\sqrt{2/3}Q{}^{3})$%
$\partial{}^{3} J A_2{}^{2}B_2{}^{2}+ 
    ((7124\sqrt{2/3}Q)/9 + (18082\sqrt{2/3}Q{}^{3})/3)
     $%
$\partial{}^{3} J A_2 B_2{}^{2}T + 
    ((12464\sqrt{2/3}Q)/9 + (27808\sqrt{2/3}Q{}^{3})/3)
     $%
$\partial{}^{3} J B_2{}^{2}T{}^{2}+ 
    ((497\sqrt{2/3}Q)/9 + (782\sqrt{2/3}Q{}^{3})/3)$%
$\partial{}^{3} J B_2{}^{4}+ 
    (2344/81 + 5888 Q{}^{2}/9 + 33152 Q{}^{4}/9)$%
$\partial{}^{4} J J T{}^{3}+ ((-18884\sqrt{2/3}Q)/27 - (67532\sqrt{2/3}Q{}^{3})/3 - 
      (508240\sqrt{2/3}Q{}^{5})/3)$%
$\partial{}^{4} J J{}^{2}\partial T 
      T + (614/81 - 62512 Q{}^{2}/27 - 148640 Q{}^{4}/3 - 248064Q{}^{6})
     $%
$\partial{}^{4} J J{}^{3}\partial{}^{2} T + 
    (-8738/243 + 5138 Q{}^{2}/9 + 75464 Q{}^{4}/27 - 67644Q{}^{6})
     $%
$\partial{}^{4} J J{}^{3}\partial{}^{2} A_{2} + 
    ((-151472Q)/81 - 658114 Q{}^{3}/27 - 238088 Q{}^{5}/3)
     $%
$\partial{}^{4} J J{}^{3}\partial A_{3} + 
    ((-6842\sqrt{2/3}Q)/27 - (173950\sqrt{2/3}Q{}^{3})/9 - 53772\sqrt{6}Q{}^{5})
     $%
$\partial{}^{4} J J{}^{2}A_2 \partial T + 
    (1241/(9\sqrt{6}) + (3809\sqrt{2/3}Q{}^{2})/9 - (131\sqrt{2/3}Q{}^{4})/3)
     $%
$\partial{}^{4} J J{}^{2}A_2 A_3 + 
    ((-10862\sqrt{2/3}Q)/27 - (194758\sqrt{2/3}Q{}^{3})/9 - 57208\sqrt{6}Q{}^{5})
     $%
$\partial{}^{4} J J{}^{2}\partial A_{2} T + 
    ((-20411Q)/(9\sqrt{6}) - (236873\sqrt{2/3}Q{}^{3})/9 - 
      (542809\sqrt{2/3}Q{}^{5})/3)$%
$\partial{}^{4} J J{}^{2}\partial A_{2} A_2 + ((-13342\sqrt{2/3})/27 - 2200\sqrt{6}Q{}^{2} - 
      (69176\sqrt{2/3}Q{}^{4})/3)$%
$\partial{}^{4} J J{}^{2}A_3 T + (8659/(27\sqrt{6}) + (9793\sqrt{2/3}Q{}^{2})/9 + 
      2063\sqrt{2/3}Q{}^{4})$%
$\partial{}^{4} J J{}^{2}B_2 
      B_3 + ((-97633Q)/(27\sqrt{6}) - (213943\sqrt{2/3}Q{}^{3})/9 - 
      75823\sqrt{2/3}Q{}^{5})$%
$\partial{}^{4} J J{}^{2}\partial B_{2} 
      B_2 + (2144/9 + 93904 Q{}^{2}/27 + 133712 Q{}^{4}/9)
     $%
$\partial{}^{4} J J A_2 T{}^{2}+ 
    (6434/81 + 37864 Q{}^{2}/27 + 7692Q{}^{4})$%
$\partial{}^{4} J J 
      A_2{}^{2}T + (-781/81 + 1000 Q{}^{2}/27 + 1300Q{}^{4})
     $%
$\partial{}^{4} J J A_2{}^{3}+ 
    (3125/81 + 8528 Q{}^{2}/27 + 4532 Q{}^{4}/9)$%
$\partial{}^{4} J J 
      A_2 B_2{}^{2}+ 
    (12862/81 + 17984 Q{}^{2}/9 + 58700 Q{}^{4}/9)$%
$\partial{}^{4} J J 
      B_2{}^{2}T + 
    ((-34844\sqrt{2/3}Q)/27 - (110516\sqrt{2/3}Q{}^{3})/3 - 
      (768688\sqrt{2/3}Q{}^{5})/3)$%
$\partial{}^{4} J \partial J J T{}^{2}+ (-1670/81 - 53576 Q{}^{2}/9 - 568516 Q{}^{4}/9 + 63568Q{}^{6})
     $%
$\partial{}^{4} J \partial J J{}^{2}\partial T + 
    (-7610/27 + 125206 Q{}^{2}/27 + 1198472 Q{}^{4}/9 + 679368Q{}^{6})
     $%
$\partial{}^{4} J \partial J J{}^{2}\partial A_{2} + 
    ((-125402Q)/27 - 59246Q{}^{3} - 545392 Q{}^{5}/3)$%
$\partial{}^{4} J \partial J 
      J{}^{2}A_3 + 
    ((-77920\sqrt{2/3}Q)/27 - (712684\sqrt{2/3}Q{}^{3})/9 - 
      (1589248\sqrt{2/3}Q{}^{5})/3)$%
$\partial{}^{4} J \partial J J 
      A_2 T + ((-12425\sqrt{2/3}Q)/9 - 
      (322573\sqrt{2/3}Q{}^{3})/9 - (760154\sqrt{2/3}Q{}^{5})/3)
     $%
$\partial{}^{4} J \partial J J A_2{}^{2}+ 
    ((-44525\sqrt{2/3}Q)/27 - (65783\sqrt{2/3}Q{}^{3})/3 - 
      (202570\sqrt{2/3}Q{}^{5})/3)$%
$\partial{}^{4} J \partial J J 
      B_2{}^{2}+ (14296/243 + 465836 Q{}^{2}/81 + 
      5570216 Q{}^{4}/27 + 15742432 Q{}^{6}/9)$%
$\partial{}^{4} J (\partial J){}^{2} J T + (-49184/243 + 782036 Q{}^{2}/81 + 
      8149664 Q{}^{4}/27 + 18260728 Q{}^{6}/9)$%
$\partial{}^{4} J (\partial J){}^{2} J A_2 + 
    ((167546\sqrt{2/3}Q)/81 + (2344840\sqrt{2/3}Q{}^{3})/27 + 
      (11810506\sqrt{2/3}Q{}^{5})/9 + (20447528\sqrt{2/3}Q{}^{7})/3)
     $%
$\partial{}^{4} J (\partial J){}^{3} J + 
    (10340/243 + 29716 Q{}^{2}/27 + 1356724 Q{}^{4}/27 + 518032Q{}^{6})
     $%
$\partial{}^{4} J \partial{}^{2} J J{}^{2}T + 
    (-58546/243 + 9448 Q{}^{2}/81 + 688850 Q{}^{4}/9 + 1766132 Q{}^{6}/3)
     $%
$\partial{}^{4} J \partial{}^{2} J J{}^{2}A_2 + 
    ((395201\sqrt{2/3}Q)/81 + (676154\sqrt{2/3}Q{}^{3})/3 + 
      (33658726\sqrt{2/3}Q{}^{5})/9 + 6951288\sqrt{6}Q{}^{7})
     $%
$\partial{}^{4} J \partial{}^{2} J \partial J J{}^{2}+ 
    ((70655\sqrt{2/3}Q)/81 + (1150406\sqrt{2/3}Q{}^{3})/27 + 
      (6672520\sqrt{2/3}Q{}^{5})/9 + (12805504\sqrt{2/3}Q{}^{7})/3)
     $%
$\partial{}^{4} J \partial{}^{3} J J{}^{3}+ 
    ((-28994\sqrt{2/3}Q)/135 - (274666\sqrt{2/3}Q{}^{3})/45 - 
      (126952\sqrt{2/3}Q{}^{5})/3)$%
$\partial{}^{5} J J{}^{2}T{}^{2}+ (-1426/1215 - 99496 Q{}^{2}/81 - 3526784 Q{}^{4}/135 - 
      1247744 Q{}^{6}/9)$%
$\partial{}^{5} J J{}^{3}\partial T + (-53186/1215 - 35312 Q{}^{2}/81 - 615484 Q{}^{4}/135 - 
      381040 Q{}^{6}/9)$%
$\partial{}^{5} J J{}^{3}\partial A_{2} + ((-43232Q)/81 - 111844 Q{}^{3}/15 - 250684 Q{}^{5}/9)
     $%
$\partial{}^{5} J J{}^{3}A_3 + 
    ((-61078\sqrt{2/3}Q)/135 - (117590\sqrt{2/3}Q{}^{3})/9 - 
      (265784\sqrt{2/3}Q{}^{5})/3)$%
$\partial{}^{5} J J{}^{2}A_2 T + ((-23813\sqrt{2/3}Q)/135 - 
      417239 Q{}^{3}/(45\sqrt{6}) - (101788\sqrt{2/3}Q{}^{5})/3)
     $%
$\partial{}^{5} J J{}^{2}A_2{}^{2}+ 
    ((-40856\sqrt{2/3}Q)/135 - 414731 Q{}^{3}/(45\sqrt{6}) - 
      (54724\sqrt{2/3}Q{}^{5})/3)$%
$\partial{}^{5} J J{}^{2}B_2{}^{2}+ (18872/1215 + 41644 Q{}^{2}/405 + 
      1719604 Q{}^{4}/135 + 1483984 Q{}^{6}/9)$%
$\partial{}^{5} J \partial J 
      J{}^{2}T + (-136408/1215 - 61906 Q{}^{2}/405 + 
      861950 Q{}^{4}/27 + 2315108 Q{}^{6}/9)$%
$\partial{}^{5} J \partial J J{}^{2}A_2 + ((54197\sqrt{2/3}Q)/45 + 
      (516460\sqrt{2/3}Q{}^{3})/9 + (1621061\sqrt{6}Q{}^{5})/5 + 
      5489444\sqrt{2/3}Q{}^{7})$%
$\partial{}^{5} J (\partial J){}^{2} J{}^{2}+ ((232511\sqrt{2/3}Q)/405 + (11536078\sqrt{2/3}Q{}^{3})/405 + 
      (7515376\sqrt{2/3}Q{}^{5})/15 + (26123680\sqrt{2/3}Q{}^{7})/9)
     $%
$\partial{}^{5} J \partial{}^{2} J J{}^{3}+ 
    (772/405 - 154412 Q{}^{2}/1215 - 19176 Q{}^{4}/5 - 659552 Q{}^{6}/27)
     $%
$\partial{}^{6} J J{}^{3}T + 
    (-18874/1215 - 491348 Q{}^{2}/1215 - 708338 Q{}^{4}/135 - 738140 Q{}^{6}/27)
     $%
$\partial{}^{6} J J{}^{3}A_2 + 
    ((95401\sqrt{2/3}Q)/405 + (4888426\sqrt{2/3}Q{}^{3})/405 + 
      (9778232\sqrt{2/3}Q{}^{5})/45 + (11488192\sqrt{2/3}Q{}^{7})/9)
     $%
$\partial{}^{6} J \partial J J{}^{3}+ 
    (29473 Q/(1134\sqrt{6}) + (640582\sqrt{2/3}Q{}^{3})/945 + 
      (3895873\sqrt{2/3}Q{}^{5})/315 + (219772\sqrt{2/3}Q{}^{7})/3)
     $%
$\partial{}^{7} J J{}^{4}+ 
    (-64 - 2432 Q{}^{2}/3)$%
$A_2 \partial T \partial T T{}^{2}+ 
    (-128/3 - 384Q{}^{2})$%
$A_2 \partial{}^{2} T T{}^{3}+ 
    (-1696/27 - 5504 Q{}^{2}/9)$%
$A_2{}^{2}\partial T \partial T 
      T + (-2840/27 - 3784 Q{}^{2}/9)$%
$A_2{}^{2}\partial{}^{2} T 
      T{}^{2}+ (56/27 - 1076 Q{}^{2}/9)$%
$A_2{}^{3}\partial T \partial T + (-184/27 + 148 Q{}^{2}/9)
     $%
$A_2{}^{3}\partial{}^{2} T T + 
    (298/27 + 296 Q{}^{2}/9)$%
$A_2{}^{4}\partial{}^{2} T + (56Q$%
$A_2{}^{4}3 
       \partial T)/3 - (56$%
$A_2{}^{4}3 
       A_3)/3 + (56Q$%
$A_2{}^{3}\partial A_{3} T)/3 + (56Q$%
$A_2{}^{3}B_2 \partial B_{3})/3 + (64/3 + 584 Q{}^{2}/3)
     $%
$A_2{}^{3}\partial B_{2} \partial B_{2} - 
    120Q$%
$A_2{}^{3}\partial B_{2} B_3 + 
    (26/3 + 116Q{}^{2})$%
$A_2{}^{3}\partial{}^{2} B_{2} 
      B_2 + 60$%
$A_2{}^{3}B_3 
      B_3 + (400Q$%
$A_2{}^{3}3 \partial T 
       T)/3 + (152$%
$A_2{}^{3}3 A_3 
       T)/3 - 64$%
$A_2{}^{3}3 B_2 
      B_3 + (248Q$%
$A_2{}^{3}3 \partial B_{2} 
       B_2)/3 + (64Q$%
$A_2{}^{2}\partial A_{3} T{}^{2})/3 - (56Q$%
$A_2{}^{2}\partial A_{3} B_2{}^{2})/3 + (-356/27 - 148 Q{}^{2}/9)$%
$A_2{}^{2}B_2{}^{2}\partial{}^{2} T - 
    40Q$%
$A_2{}^{2}B_2 B_3 \partial T + 
    (520Q$%
$A_2{}^{2}B_2 \partial B_{3} T)/3 + 
    (8/27 - 1604 Q{}^{2}/9)$%
$A_2{}^{2}\partial B_{2} B_2 
      \partial T + (-224/9 - 256 Q{}^{2}/3)$%
$A_2{}^{2}\partial B_{2} 
      \partial B_{2} T + 112Q$%
$A_2{}^{2}\partial B_{2} 
      B_3 T + (-50/9 - 212Q{}^{2})$%
$A_2{}^{2}\partial{}^{2} B_{2} B_2 T - 
    76$%
$A_2{}^{2}B_3 B_3 T + 
    (1376Q$%
$A_2 A_3 \partial T T{}^{2})/3 - 
    (128$%
$A_2 A_3 A_3 T{}^{2})/3 + 
    (92$%
$A_2 A_3 A_3 B_2{}^{2})/3 + 
    (232Q$%
$A_2 A_3 B_2{}^{2}\partial T)/3 + 
    88$%
$A_2 A_3 B_2 B_3 T - 
    (424Q$%
$A_2 A_3 \partial B_{2} B_2 T)/3 + 
    (704Q$%
$A_2 \partial A_{3} T{}^{3})/3 - 
    (472Q$%
$A_2 \partial A_{3} B_2{}^{2}T)/3 + 
    (-1592/27 - 3436 Q{}^{2}/9)$%
$A_2 B_2{}^{2}\partial T 
      \partial T + (-3656/27 - 7924 Q{}^{2}/9)$%
$A_2 B_2{}^{2}\partial{}^{2} T T - 8Q$%
$A_2 B_2{}^{3}\partial B_{3} - 
    16$%
$A_2 B_2{}^{3}3 B_3 + 
    160Q$%
$A_2 B_2 B_3 \partial T T + 
    (896Q$%
$A_2 B_2 \partial B_{3} T{}^{2})/3 + 
    (-9232/27 - 20408 Q{}^{2}/9)$%
$A_2 \partial B_{2} B_2 \partial T 
      T + 32Q$%
$A_2 \partial B_{2} B_2{}^{3}3 + (-424/9 - 448Q{}^{2})$%
$A_2 \partial B_{2} \partial B_{2} 
      T{}^{2}+ (-248/9 - 260Q{}^{2})$%
$A_2 \partial B_{2} 
      \partial B_{2} B_2{}^{2}+ 
    128Q$%
$A_2 \partial B_{2} B_3 T{}^{2}+ 
    (-1624/9 - 928Q{}^{2})$%
$A_2 \partial{}^{2} B_{2} B_2 T{}^{2}+ (-26/3 - 84Q{}^{2})$%
$A_2 \partial{}^{2} B_{2} B_2{}^{3}+ 64$%
$A_2 B_3 B_3 
      T{}^{2}+ (-32 - 384Q{}^{2})$%
$\partial A_{2} \partial T T{}^{3}+ (-6736/27 - 19016 Q{}^{2}/9)$%
$\partial A_{2} A_2 
      \partial T T{}^{2}+ (-952/27 - 4196 Q{}^{2}/9)
     $%
$\partial A_{2} A_2{}^{2}\partial T T + 
    (64/3 + 832 Q{}^{2}/3)$%
$\partial A_{2} A_2{}^{3}\partial T + (56Q$%
$\partial A_{2} A_2{}^{4}3)/3 - (488Q$%
$\partial A_{2} A_2{}^{3}3 T)/3 + 64Q$%
$\partial A_{2} A_2{}^{2}B_2 B_3 + (1160/27 + 4300 Q{}^{2}/9)
     $%
$\partial A_{2} A_2{}^{2}\partial B_{2} B_2 - 
    (256Q$%
$\partial A_{2} A_2 A_3 T{}^{2})/3 - 
    (160Q$%
$\partial A_{2} A_2 A_3 B_2{}^{2})/
     3 + (-256/9 - 368Q{}^{2})$%
$\partial A_{2} A_2 B_2{}^{2}\partial T + 8Q$%
$\partial A_{2} A_2 B_2 
      B_3 T + (-1744/27 - 5744 Q{}^{2}/9)
     $%
$\partial A_{2} A_2 \partial B_{2} B_2 T + 
    (-24 - 712 Q{}^{2}/3)$%
$\partial A_{2} \partial A_{2} T{}^{3}+ 
    (-1048/27 - 4280 Q{}^{2}/9)$%
$\partial A_{2} \partial A_{2} A_2 T{}^{2}+ (728/9 + 1616 Q{}^{2}/3)$%
$\partial A_{2} \partial A_{2} A_2{}^{2}T + (-1784/27 - 6136 Q{}^{2}/9)
     $%
$\partial A_{2} \partial A_{2} A_2{}^{3}+ 
    (1184/27 + 3244 Q{}^{2}/9)$%
$\partial A_{2} \partial A_{2} A_2 
      B_2{}^{2}+ (-1040/27 - 4012 Q{}^{2}/9)
     $%
$\partial A_{2} \partial A_{2} B_2{}^{2}T - 
    192Q$%
$\partial A_{2} A_3 T{}^{3}- 
    (64Q$%
$\partial A_{2} A_3 B_2{}^{2}T)/3 + 
    (-7832/27 - 19204 Q{}^{2}/9)$%
$\partial A_{2} B_2{}^{2}\partial T 
      T - 8Q$%
$\partial A_{2} B_2{}^{4}3 - 272Q$%
$\partial A_{2} B_2 B_3 T{}^{2}+ (-1360/9 - 4168 Q{}^{2}/3)$%
$\partial A_{2} \partial B_{2} 
      B_2 T{}^{2}+ (-392/27 - 820 Q{}^{2}/9)
     $%
$\partial A_{2} \partial B_{2} B_2{}^{3}- 
    64Q{}^{2}$%
$\partial{}^{2} A_{2} T{}^{4}+ 
    (-104/3 - 1832 Q{}^{2}/3)$%
$\partial{}^{2} A_{2} A_2 T{}^{3}+ (-584/27 - 3796 Q{}^{2}/9)$%
$\partial{}^{2} A_{2} A_2{}^{2}T{}^{2}+ (998/27 + 1600 Q{}^{2}/9)
     $%
$\partial{}^{2} A_{2} A_2{}^{3}T + 
    (-668/27 - 2296 Q{}^{2}/9)$%
$\partial{}^{2} A_{2} A_2{}^{4}+ (862/27 + 2300 Q{}^{2}/9)
     $%
$\partial{}^{2} A_{2} A_2{}^{2}B_2{}^{2}+ 
    (-998/27 - 3496 Q{}^{2}/9)$%
$\partial{}^{2} A_{2} A_2 B_2{}^{2}T + (-160/3 - 2660 Q{}^{2}/3)
     $%
$\partial{}^{2} A_{2} B_2{}^{2}T{}^{2}+ 
    (-194/27 - 292 Q{}^{2}/9)$%
$\partial{}^{2} A_{2} B_2{}^{4}+ 128Q$%
$A_3 \partial T T{}^{3}+ 128$%
$A_3 A_3 T{}^{3}- 
    (148$%
$A_3 A_3 B_2{}^{2}T)/3 + 
    (1552Q$%
$A_3 B_2{}^{2}\partial T T)/3 + 
    8$%
$A_3 B_2{}^{4}3 + 
    176$%
$A_3 B_2 B_3 T{}^{2}+ 
    304Q$%
$A_3 \partial B_{2} B_2 T{}^{2}- 
    16Q$%
$A_3 \partial B_{2} B_2{}^{3}+ 
    64Q$%
$\partial A_{3} T{}^{4}+ 
    (832Q$%
$\partial A_{3} B_2{}^{2}T{}^{2})/3 + 
    8Q$%
$\partial A_{3} B_2{}^{4}+ 
    (-1568/9 - 4288 Q{}^{2}/3)$%
$B_2{}^{2}\partial T \partial T 
      T + (-216 - 3656 Q{}^{2}/3)$%
$B_2{}^{2}\partial{}^{2} T 
      T{}^{2}+ (58/27 - 148 Q{}^{2}/9)$%
$B_2{}^{4}\partial{}^{2} T + 
    8Q$%
$B_2{}^{4}3 \partial T + 
    72Q$%
$B_2{}^{3}\partial B_{3} T + 
    8$%
$B_2{}^{3}3 B_3 T + 
    608Q$%
$B_2 B_3 \partial T T{}^{2}+ 
    192Q$%
$B_2 \partial B_{3} T{}^{3}+ 
    (-3760/9 - 9032 Q{}^{2}/3)$%
$\partial B_{2} B_2 \partial T T{}^{2}+ (-680/27 - 460 Q{}^{2}/9)$%
$\partial B_{2} B_2{}^{3}\partial T + 
    184Q$%
$\partial B_{2} B_2{}^{3}3 T + 
    (-296/9 - 536Q{}^{2})$%
$\partial B_{2} \partial B_{2} T{}^{3}+ 
    (-152/9 - 180Q{}^{2})$%
$\partial B_{2} \partial B_{2} B_2{}^{2}T + 352Q$%
$\partial B_{2} B_3 T{}^{3}+ 
    (-632/9 - 680Q{}^{2})$%
$\partial{}^{2} B_{2} B_2 T{}^{3}+ 
    (-30 - 68Q{}^{2})$%
$\partial{}^{2} B_{2} B_2{}^{3}T - 16$%
$B_3 B_3 T{}^{3}+ 
    (-32 - 384Q{}^{2})$%
$J{}^{2}\partial T \partial T T{}^{2}+ (-64/3 - 256Q{}^{2})$%
$J{}^{2}\partial{}^{2} T T{}^{3}+ ((-656\sqrt{2/3}Q)/9 - (2624\sqrt{2/3}Q{}^{3})/3)
     $%
$J{}^{3}\partial T \partial T \partial T + 
    ((-640\sqrt{2/3}Q)/9 - (2560\sqrt{2/3}Q{}^{3})/3)
     $%
$J{}^{3}\partial{}^{2} T \partial T T + 
    ((32\sqrt{2/3}Q)/9 + (128\sqrt{2/3}Q{}^{3})/3)$%
$J{}^{3}\partial{}^{3} T T{}^{2}+ 
    (130/81 + 1124 Q{}^{2}/3 + 38384 Q{}^{4}/9)$%
$J{}^{4}\partial{}^{2} T \partial{}^{2} T + 
    (40/27 + 11480 Q{}^{2}/27 + 44000 Q{}^{4}/9)$%
$J{}^{4}\partial{}^{3} T \partial T + 
    (1250/243 + 716 Q{}^{2}/3 + 57328 Q{}^{4}/27)$%
$J{}^{4}\partial{}^{4} T T + 
    ((8648\sqrt{2/3}Q)/405 + (5608\sqrt{2/3}Q{}^{3})/15 + 
      (12704\sqrt{2/3}Q{}^{5})/9)$%
$J{}^{5}\partial{}^{5} T + ((-2249Q)/(81\sqrt{6}) - 
      34999 Q{}^{3}/(135\sqrt{6}) - (1804\sqrt{2/3}Q{}^{5})/9)
     $%
$J{}^{5}\partial{}^{5} A_{2} + 
    ((292\sqrt{2/3})/81 - 32\sqrt{2/3}Q{}^{2} - (4972\sqrt{2/3}Q{}^{4})/9)
     $%
$J{}^{5}\partial{}^{4} A_{3} + 
    (-1766/243 + 12344 Q{}^{2}/27 + 105548 Q{}^{4}/27)$%
$J{}^{4}A_2 \partial{}^{4} T + (1421 Q/9 + 2459 Q{}^{3}/3)
     $%
$J{}^{4}A_2 \partial{}^{3} A_{3} + 
    (-860/81 + 31736 Q{}^{2}/27 + 84880 Q{}^{4}/9)$%
$J{}^{4}\partial A_{2} \partial{}^{3} T + 
    (6289 Q/27 + 9725 Q{}^{3}/9)$%
$J{}^{4}\partial A_{2} \partial{}^{2} A_{3} + 
    (-1016/81 + 22840 Q{}^{2}/27 + 79312 Q{}^{4}/9)$%
$J{}^{4}\partial{}^{2} A_{2} \partial{}^{2} T + 
    (209/54 + 1033 Q{}^{2}/27 + 3995 Q{}^{4}/9)$%
$J{}^{4}\partial{}^{2} A_{2} \partial{}^{2} A_{2} + 
    (3475 Q/27 + 7577 Q{}^{3}/9)$%
$J{}^{4}\partial{}^{2} A_{2} \partial A_{3} + 
    (-28/9 + 11020 Q{}^{2}/27 + 35746 Q{}^{4}/9)$%
$J{}^{4}\partial{}^{3} A_{2} \partial T + 
    (1148/81 + 692 Q{}^{2}/3 + 19804 Q{}^{4}/9)$%
$J{}^{4}\partial{}^{3} A_{2} \partial A_{2} + 
    (211 Q/9 - 1085 Q{}^{3}/3)$%
$J{}^{4}\partial{}^{3} A_{2} A_3 + (-746/243 + 169 Q{}^{2}/3 + 20366 Q{}^{4}/27)
     $%
$J{}^{4}\partial{}^{4} A_{2} T + 
    (421/54 + 727 Q{}^{2}/9 + 2735 Q{}^{4}/3)$%
$J{}^{4}\partial{}^{4} A_{2} A_2 + ((-800Q)/3 - 412Q{}^{3})
     $%
$J{}^{4}A_3 \partial{}^{3} T + 
    (748 Q/27 + 18044 Q{}^{3}/9)$%
$J{}^{4}\partial A_{3} \partial{}^{2} T + (-20/3 + 322Q{}^{2})
     $%
$J{}^{4}\partial A_{3} \partial A_{3} + 
    (5020 Q/27 + 15602 Q{}^{3}/9)$%
$J{}^{4}\partial{}^{2} A_{3} \partial T + (-238/3 + 162Q{}^{2})
     $%
$J{}^{4}\partial{}^{2} A_{3} A_3 + 
    (196 Q/3 + 2378 Q{}^{3}/3)$%
$J{}^{4}\partial{}^{3} A_{3} T + (59 Q/9 + 2525 Q{}^{3}/3)
     $%
$J{}^{4}B_2 \partial{}^{3} B_{3} + 
    (2773 Q/9 + 9061 Q{}^{3}/3)$%
$J{}^{4}\partial B_{2} \partial{}^{2} B_{3} + 
    (527/54 - 4561 Q{}^{2}/9 - 12767 Q{}^{4}/3)$%
$J{}^{4}\partial{}^{2} B_{2} \partial{}^{2} B_{2} + 
    (5561 Q/9 + 10403 Q{}^{3}/3)$%
$J{}^{4}\partial{}^{2} B_{2} \partial B_{3} + 
    (1918/81 - 596Q{}^{2} - 52102 Q{}^{4}/9)$%
$J{}^{4}\partial{}^{3} B_{2} \partial B_{2} + (3271 Q/9 + 5243 Q{}^{3}/3)
     $%
$J{}^{4}\partial{}^{3} B_{2} B_3 + 
    (1883/162 + 245 Q{}^{2}/9 - 9313 Q{}^{4}/9)$%
$J{}^{4}\partial{}^{4} B_{2} B_2 + 
    (-116/3 - 206Q{}^{2})$%
$J{}^{4}\partial B_{3} \partial B_{3} + (-88/3 - 326Q{}^{2})$%
$J{}^{4}\partial{}^{2} B_{3} B_3 + 
    ((-8440\sqrt{2/3}Q)/27 - (9680\sqrt{2/3}Q{}^{3})/9)
     $%
$J{}^{3}A_2 \partial{}^{2} T \partial T + 
    ((-656\sqrt{2/3}Q)/27 + (112\sqrt{2/3}Q{}^{3})/3)
     $%
$J{}^{3}A_2 \partial{}^{3} T T + 
    ((-4730\sqrt{2/3}Q)/27 - (1342\sqrt{2/3}Q{}^{3})/3)
     $%
$J{}^{3}A_2{}^{2}\partial{}^{3} T + 
    ((40\sqrt{2/3})/9 + 94\sqrt{2/3}Q{}^{2})$%
$J{}^{3}A_2{}^{2}\partial{}^{2} A_{3} + 
    ((1012\sqrt{2/3})/27 + (2486\sqrt{2/3}Q{}^{2})/9)$%
$J{}^{3}A_2 A_3 \partial{}^{2} T + 
    ((1876\sqrt{2/3})/27 + (13454\sqrt{2/3}Q{}^{2})/9)
     $%
$J{}^{3}A_2 \partial A_{3} \partial T + 
    112\sqrt{2/3}Q$%
$J{}^{3}A_2 
      \partial A_{3} A_3 + ((328\sqrt{2/3})/3 + 1292\sqrt{2/3}Q{}^{2})
     $%
$J{}^{3}A_2 \partial{}^{2} A_{3} T + 
    ((-160\sqrt{2/3})/3 - (374\sqrt{2/3}Q{}^{2})/3)$%
$J{}^{3}A_2 B_2 \partial{}^{2} B_{3} + 
    (-8\sqrt{2/3} + (772\sqrt{2/3}Q{}^{2})/3)$%
$J{}^{3}A_2 \partial B_{2} \partial B_{3} + 
    ((3544\sqrt{2/3}Q)/9 + 1954\sqrt{2/3}Q{}^{3})$%
$J{}^{3}A_2 \partial{}^{2} B_{2} \partial B_{2} + 
    ((-176\sqrt{2/3})/9 + (506\sqrt{2/3}Q{}^{2})/3)$%
$J{}^{3}A_2 \partial{}^{2} B_{2} B_3 + 
    ((7192\sqrt{2/3}Q)/27 + (9350\sqrt{2/3}Q{}^{3})/9)
     $%
$J{}^{3}A_2 \partial{}^{3} B_{2} 
      B_2 - 48\sqrt{6}Q$%
$J{}^{3}A_2 \partial B_{3} B_3 + 
    ((-12344\sqrt{2/3}Q)/27 - (31312\sqrt{2/3}Q{}^{3})/9)
     $%
$J{}^{3}\partial A_{2} \partial T \partial T + 
    (-464\sqrt{2/3}Q - (9680\sqrt{2/3}Q{}^{3})/3)$%
$J{}^{3}\partial A_{2} \partial{}^{2} T T + 
    ((-15448\sqrt{2/3}Q)/27 - (15542\sqrt{2/3}Q{}^{3})/9)
     $%
$J{}^{3}\partial A_{2} A_2 \partial{}^{2} T + 
    ((-428\sqrt{2/3})/27 - (3826\sqrt{2/3}Q{}^{2})/9)$%
$J{}^{3}\partial A_{2} A_2 \partial A_{3} + 
    ((-7912\sqrt{2/3}Q)/27 - (4394\sqrt{2/3}Q{}^{3})/9)
     $%
$J{}^{3}\partial A_{2} \partial A_{2} \partial T + 
    ((1240\sqrt{2/3}Q)/27 + (5726\sqrt{2/3}Q{}^{3})/9)
     $%
$J{}^{3}\partial A_{2} \partial A_{2} 
      \partial A_{2} + ((-260\sqrt{2/3})/27 - (4150\sqrt{2/3}Q{}^{2})/9)
     $%
$J{}^{3}\partial A_{2} \partial A_{2} 
      A_3 + ((4516\sqrt{2/3})/27 + (2438\sqrt{2/3}Q{}^{2})/9)
     $%
$J{}^{3}\partial A_{2} A_3 \partial T + 
    24\sqrt{6}Q$%
$J{}^{3}\partial A_{2} A_3 
      A_3 + ((224\sqrt{2/3})/3 + 544\sqrt{2/3}Q{}^{2})
     $%
$J{}^{3}\partial A_{2} \partial A_{3} T + 
    ((-1108\sqrt{2/3})/27 - (1022\sqrt{2/3}Q{}^{2})/9)
     $%
$J{}^{3}\partial A_{2} B_2 
      \partial B_{3} + ((7280\sqrt{2/3}Q)/27 + (12238\sqrt{2/3}Q{}^{3})/9)
     $%
$J{}^{3}\partial A_{2} \partial B_{2} 
      \partial B_{2} + ((-820\sqrt{2/3})/27 + (1282\sqrt{2/3}Q{}^{2})/9)
     $%
$J{}^{3}\partial A_{2} \partial B_{2} 
      B_3 + ((11132\sqrt{2/3}Q)/27 + (13894\sqrt{2/3}Q{}^{3})/9)
     $%
$J{}^{3}\partial A_{2} \partial{}^{2} B_{2} 
      B_2 - 88\sqrt{2/3}Q$%
$J{}^{3}\partial A_{2} B_3 B_3 + 
    ((-5488\sqrt{2/3}Q)/9 - (14912\sqrt{2/3}Q{}^{3})/3)
     $%
$J{}^{3}\partial{}^{2} A_{2} \partial T T + 
    ((-13336\sqrt{2/3}Q)/27 - (25154\sqrt{2/3}Q{}^{3})/9)
     $%
$J{}^{3}\partial{}^{2} A_{2} A_2 \partial T + 
    ((-308\sqrt{2/3})/27 - (3850\sqrt{2/3}Q{}^{2})/9)$%
$J{}^{3}\partial{}^{2} A_{2} A_2 A_3 + 
    ((-464\sqrt{2/3}Q)/3 + 28\sqrt{6}Q{}^{3})$%
$J{}^{3}\partial{}^{2} A_{2} \partial A_{2} T + 
    ((2120\sqrt{2/3}Q)/27 + (7684\sqrt{2/3}Q{}^{3})/9)
     $%
$J{}^{3}\partial{}^{2} A_{2} \partial A_{2} 
      A_2 + (136\sqrt{2/3} + 580\sqrt{2/3}Q{}^{2})
     $%
$J{}^{3}\partial{}^{2} A_{2} A_3 T + 
    ((-496\sqrt{2/3})/27 - (104\sqrt{2/3}Q{}^{2})/9)$%
$J{}^{3}\partial{}^{2} A_{2} B_2 B_3 + 
    ((4748\sqrt{2/3}Q)/27 + (11164\sqrt{2/3}Q{}^{3})/9)
     $%
$J{}^{3}\partial{}^{2} A_{2} \partial B_{2} 
      B_2 + ((-1324\sqrt{2/3}Q)/9 - (3776\sqrt{2/3}Q{}^{3})/3)
     $%
$J{}^{3}\partial{}^{3} A_{2} T{}^{2}+ 
    ((-3008\sqrt{2/3}Q)/27 - (2636\sqrt{2/3}Q{}^{3})/3)
     $%
$J{}^{3}\partial{}^{3} A_{2} A_2 T + 
    ((-74\sqrt{2/3}Q)/27 - 74\sqrt{2/3}Q{}^{3})$%
$J{}^{3}\partial{}^{3} A_{2} A_2{}^{2}+ 
    ((-82\sqrt{2/3}Q)/27 - (440\sqrt{2/3}Q{}^{3})/3)$%
$J{}^{3}\partial{}^{3} A_{2} B_2{}^{2}+ 
    ((1856\sqrt{2/3})/9 + 3824\sqrt{2/3}Q{}^{2})$%
$J{}^{3}A_3 \partial T \partial T + 
    ((4208\sqrt{2/3})/9 + 5632\sqrt{2/3}Q{}^{2})$%
$J{}^{3}A_3 \partial{}^{2} T T - 
    104\sqrt{6}Q$%
$J{}^{3}A_3 A_3 
      \partial T + 8\sqrt{6}$%
$J{}^{3}A_3 
      A_3 A_3 - 48\sqrt{6}Q$%
$J{}^{3}A_3 B_2 \partial B_{3} + 
    ((-352\sqrt{2/3})/9 + (728\sqrt{2/3}Q{}^{2})/3)$%
$J{}^{3}A_3 \partial B_{2} \partial B_{2} - 
    224\sqrt{2/3}Q$%
$J{}^{3}A_3 
      \partial B_{2} B_3 + ((-700\sqrt{2/3})/9 + (704\sqrt{2/3}Q{}^{2})/3)
     $%
$J{}^{3}A_3 \partial{}^{2} B_{2} 
      B_2 + 8\sqrt{2/3}$%
$J{}^{3}A_3 B_3 B_3 + 
    (208\sqrt{6} + 7768\sqrt{2/3}Q{}^{2})$%
$J{}^{3}\partial A_{3} \partial T T + 64\sqrt{2/3}Q$%
$J{}^{3}\partial A_{3} A_3 T - 
    208\sqrt{2/3}Q$%
$J{}^{3}\partial A_{3} 
      B_2 B_3 + ((-272\sqrt{2/3})/9 + (1840\sqrt{2/3}Q{}^{2})/3)
     $%
$J{}^{3}\partial A_{3} \partial B_{2} 
      B_2 + ((1160\sqrt{2/3})/9 + 2084\sqrt{2/3}Q{}^{2})
     $%
$J{}^{3}\partial{}^{2} A_{3} T{}^{2}+ 
    ((-212\sqrt{2/3})/9 + (352\sqrt{2/3}Q{}^{2})/3)$%
$J{}^{3}\partial{}^{2} A_{3} B_2{}^{2}+ 
    ((-1874\sqrt{2/3}Q)/27 - (254\sqrt{2/3}Q{}^{3})/3)
     $%
$J{}^{3}B_2{}^{2}\partial{}^{3} T + 
    ((3788\sqrt{2/3})/27 + (9598\sqrt{2/3}Q{}^{2})/9)$%
$J{}^{3}B_2 B_3 \partial{}^{2} T + 
    ((10052\sqrt{2/3})/27 + (19534\sqrt{2/3}Q{}^{2})/9)
     $%
$J{}^{3}B_2 \partial B_{3} \partial T + 
    ((56\sqrt{2/3})/3 + (3340\sqrt{2/3}Q{}^{2})/3)$%
$J{}^{3}B_2 \partial{}^{2} B_{3} T + 
    ((-5944\sqrt{2/3}Q)/27 - (5990\sqrt{2/3}Q{}^{3})/9)
     $%
$J{}^{3}\partial B_{2} B_2 \partial{}^{2} T + 
    ((-12376\sqrt{2/3}Q)/27 - (31898\sqrt{2/3}Q{}^{3})/9)
     $%
$J{}^{3}\partial B_{2} \partial B_{2} \partial T + 
    ((1412\sqrt{2/3})/27 + (9886\sqrt{2/3}Q{}^{2})/9)$%
$J{}^{3}\partial B_{2} B_3 \partial T + 
    ((592\sqrt{2/3})/3 + (5656\sqrt{2/3}Q{}^{2})/3)$%
$J{}^{3}\partial B_{2} \partial B_{3} T + 
    ((-12448\sqrt{2/3}Q)/27 - (28226\sqrt{2/3}Q{}^{3})/9)
     $%
$J{}^{3}\partial{}^{2} B_{2} B_2 \partial T + 
    ((-7616\sqrt{2/3}Q)/9 - 5788\sqrt{2/3}Q{}^{3})$%
$J{}^{3}\partial{}^{2} B_{2} \partial B_{2} T + 
    ((2440\sqrt{2/3})/9 + (3980\sqrt{2/3}Q{}^{2})/3)$%
$J{}^{3}\partial{}^{2} B_{2} B_3 T + 
    ((-4640\sqrt{2/3}Q)/27 - (15316\sqrt{2/3}Q{}^{3})/9)
     $%
$J{}^{3}\partial{}^{3} B_{2} B_2 T + 
    872\sqrt{2/3}Q$%
$J{}^{3}B_3 
      B_3 \partial T + 320\sqrt{6}Q$%
$J{}^{3}\partial B_{3} B_3 T + (-6400/27 - 20096 Q{}^{2}/9)
     $%
$J{}^{2}A_2 \partial T \partial T T + 
    (-8672/27 - 21088 Q{}^{2}/9)$%
$J{}^{2}A_2 \partial{}^{2} T 
      T{}^{2}+ (-1904/27 - 8800 Q{}^{2}/9)$%
$J{}^{2}A_2{}^{2}\partial T \partial T + (-4760/27 - 13960 Q{}^{2}/9)
     $%
$J{}^{2}A_2{}^{2}\partial{}^{2} T T + 
    (-188/27 - 1462 Q{}^{2}/9)$%
$J{}^{2}A_2{}^{3}\partial{}^{2} T + (28Q$%
$J{}^{2}A_2{}^{3}\partial A_{3})/3 + 
    (200Q$%
$J{}^{2}A_2{}^{3}3 
       \partial T)/3 + (76$%
$J{}^{2}A_2{}^{3}3 A_3)/3 + 
    (224Q$%
$J{}^{2}A_2{}^{2}\partial A_{3} 
       T)/3 - (188Q$%
$J{}^{2}A_2{}^{2}B_2 \partial B_{3})/3 + (-112/9 - 224 Q{}^{2}/3)
     $%
$J{}^{2}A_2{}^{2}\partial B_{2} 
      \partial B_{2} + 24Q$%
$J{}^{2}A_2{}^{2}\partial B_{2} B_3 + (-25/9 - 62 Q{}^{2}/3)
     $%
$J{}^{2}A_2{}^{2}\partial{}^{2} B_{2} 
      B_2 - 38$%
$J{}^{2}A_2{}^{2}B_3 B_3 + (2240Q$%
$J{}^{2}A_2 
       A_3 \partial T T)/3 - 
    (128$%
$J{}^{2}A_2 A_3 A_3 
       T)/3 + 44$%
$J{}^{2}A_2 A_3 
      B_2 B_3 + 4Q$%
$J{}^{2}A_2 
      A_3 \partial B_{2} B_2 + 
    (896Q$%
$J{}^{2}A_2 \partial A_{3} T{}^{2})/
     3 + (148Q$%
$J{}^{2}A_2 \partial A_{3} B_2{}^{2})/3 + (-724/27 - 3962 Q{}^{2}/9)$%
$J{}^{2}A_2 B_2{}^{2}\partial{}^{2} T + 
    80Q$%
$J{}^{2}A_2 B_2 B_3 
      \partial T + (896Q$%
$J{}^{2}A_2 B_2 
       \partial B_{3} T)/3 + (-4616/27 - 14428 Q{}^{2}/9)
     $%
$J{}^{2}A_2 \partial B_{2} B_2 \partial T + 
    (-424/9 - 2080 Q{}^{2}/3)$%
$J{}^{2}A_2 \partial B_{2} 
      \partial B_{2} T - 224Q$%
$J{}^{2}A_2 
      \partial B_{2} B_3 T + (-1624/9 - 3872 Q{}^{2}/3)
     $%
$J{}^{2}A_2 \partial{}^{2} B_{2} B_2 T + 
    64$%
$J{}^{2}A_2 B_3 B_3 T + 
    (-9008/27 - 30352 Q{}^{2}/9)$%
$J{}^{2}\partial A_{2} \partial T 
      T{}^{2}+ (-16400/27 - 50632 Q{}^{2}/9)
     $%
$J{}^{2}\partial A_{2} A_2 \partial T T + 
    (-476/27 - 2002 Q{}^{2}/9)$%
$J{}^{2}\partial A_{2} A_2{}^{2}\partial T - (20Q$%
$J{}^{2}\partial A_{2} 
       A_2{}^{3}3)/3 - 
    128Q$%
$J{}^{2}\partial A_{2} A_2 A_3 
      T - 60Q$%
$J{}^{2}\partial A_{2} A_2 
      B_2 B_3 + (-872/27 - 4696 Q{}^{2}/9)
     $%
$J{}^{2}\partial A_{2} A_2 \partial B_{2} 
      B_2 + (-1772/27 - 8308 Q{}^{2}/9)$%
$J{}^{2}\partial A_{2} \partial A_{2} T{}^{2}+ (-1048/27 - 248 Q{}^{2}/9)
     $%
$J{}^{2}\partial A_{2} \partial A_{2} A_2 T + 
    (364/9 + 1000 Q{}^{2}/3)$%
$J{}^{2}\partial A_{2} \partial A_{2} 
      A_2{}^{2}+ (-520/27 - 2582 Q{}^{2}/9)
     $%
$J{}^{2}\partial A_{2} \partial A_{2} B_2{}^{2}- (1600Q$%
$J{}^{2}\partial A_{2} A_3 
       T{}^{2})/3 + (256Q$%
$J{}^{2}\partial A_{2} 
       A_3 B_2{}^{2})/3 + (-3916/27 - 10466 Q{}^{2}/9)
     $%
$J{}^{2}\partial A_{2} B_2{}^{2}\partial T - 
    272Q$%
$J{}^{2}\partial A_{2} B_2 B_3 
      T + (-1360/9 - 1656Q{}^{2})$%
$J{}^{2}\partial A_{2} 
      \partial B_{2} B_2 T + (-304/9 - 2320 Q{}^{2}/3)
     $%
$J{}^{2}\partial{}^{2} A_{2} T{}^{3}+ 
    (-3692/27 - 17284 Q{}^{2}/9)$%
$J{}^{2}\partial{}^{2} A_{2} 
      A_2 T{}^{2}+ (-584/27 - 2548 Q{}^{2}/9)
     $%
$J{}^{2}\partial{}^{2} A_{2} A_2{}^{2}T + 
    (499/27 + 1472 Q{}^{2}/9)$%
$J{}^{2}\partial{}^{2} A_{2} A_2{}^{3}+ (-499/27 - 2612 Q{}^{2}/9)
     $%
$J{}^{2}\partial{}^{2} A_{2} A_2 B_2{}^{2}+ (-160/3 - 2372 Q{}^{2}/3)$%
$J{}^{2}\partial{}^{2} A_{2} B_2{}^{2}T + 
    960Q$%
$J{}^{2}A_3 \partial T T{}^{2}+ 
    384$%
$J{}^{2}A_3 A_3 T{}^{2}- 
    (74$%
$J{}^{2}A_3 A_3 B_2{}^{2})/3 + (776Q$%
$J{}^{2}A_3 
       B_2{}^{2}\partial T)/3 + 
    176$%
$J{}^{2}A_3 B_2 B_3 
      T + (1520Q$%
$J{}^{2}A_3 \partial B_{2} 
       B_2 T)/3 + (832Q$%
$J{}^{2}\partial A_{3} 
       T{}^{3})/3 + 
    (352Q$%
$J{}^{2}\partial A_{3} B_2{}^{2}T)/3 + (-1520/9 - 1824Q{}^{2})$%
$J{}^{2}B_2{}^{2}\partial T \partial T + (-3848/9 - 2808Q{}^{2})
     $%
$J{}^{2}B_2{}^{2}\partial{}^{2} T T - 
    28Q$%
$J{}^{2}B_2{}^{3}\partial B_{3} + 4$%
$J{}^{2}B_2{}^{3}3 B_3 + 896Q$%
$J{}^{2}B_2 
      B_3 \partial T T + 
    (1408Q$%
$J{}^{2}B_2 \partial B_{3} T{}^{2})/
     3 + (-8432/9 - 25000 Q{}^{2}/3)$%
$J{}^{2}\partial B_{2} 
      B_2 \partial T T - 100Q$%
$J{}^{2}\partial B_{2} B_2{}^{3}3 + 
    (-1564/9 - 6092 Q{}^{2}/3)$%
$J{}^{2}\partial B_{2} \partial B_{2} 
      T{}^{2}+ (-76/9 + 230Q{}^{2})$%
$J{}^{2}\partial B_{2} \partial B_{2} B_2{}^{2}- 
    144Q$%
$J{}^{2}\partial B_{2} B_3 T{}^{2}+ 
    (-2164/9 - 8212 Q{}^{2}/3)$%
$J{}^{2}\partial{}^{2} B_{2} B_2 
      T{}^{2}+ (-15 + 30Q{}^{2})$%
$J{}^{2}\partial{}^{2} B_{2} 
      B_2{}^{3}+ 
    296$%
$J{}^{2}B_3 B_3 T{}^{2}- 
    (256\sqrt{2/3}Q$%
$J A_2 \partial T T{}^{3})/3 - (1376\sqrt{2/3}Q$%
$J A_2{}^{2}\partial T T{}^{2})/3 - 
    (208\sqrt{2/3}Q$%
$J A_2{}^{3}\partial T T)/3 - 56\sqrt{2/3}Q$%
$J A_2{}^{4}\partial T + 
    (56\sqrt{2/3}$%
$J A_2{}^{4}3 T)/3 - (112\sqrt{2/3}Q$%
$J A_2{}^{3}\partial B_{2} B_2)/3 - 
    (32\sqrt{2/3}$%
$J A_2{}^{3}3 T{}^{2})/3 + 16\sqrt{2/3}Q$%
$J A_2{}^{2}B_2{}^{2}\partial T - 88\sqrt{2/3}
     $%
$J A_2{}^{2}B_2 B_3 T - 
    8\sqrt{6}Q$%
$J A_2{}^{2}\partial B_{2} B_2 
      T - (640\sqrt{2/3}$%
$J A_2 A_3 T{}^{3})/3 + (280\sqrt{2/3}$%
$J A_2 
       A_3 B_2{}^{2}T)/3 - 
    (2096\sqrt{2/3}Q$%
$J A_2 B_2{}^{2}\partial T T)/3 - 176\sqrt{2/3}$%
$J A_2 
      B_2 B_3 T{}^{2}- 
    (1616\sqrt{2/3}Q$%
$J A_2 \partial B_{2} B_2 
       T{}^{2})/3 + 16\sqrt{2/3}Q$%
$J A_2 
      \partial B_{2} B_2{}^{3}- 
    (320\sqrt{2/3}Q$%
$J \partial A_{2} A_2 T{}^{3})/3 + 160\sqrt{2/3}Q$%
$J \partial A_{2} A_2{}^{2}T{}^{2}- 8\sqrt{2/3}Q$%
$J \partial A_{2} 
      A_2{}^{3}T + 
    (112\sqrt{2/3}Q$%
$J \partial A_{2} A_2{}^{4})/3 - 16\sqrt{2/3}Q$%
$J \partial A_{2} 
      A_2{}^{2}B_2{}^{2}+ 
    (296\sqrt{2/3}Q$%
$J \partial A_{2} A_2 B_2{}^{2}T)/3 + (272\sqrt{2/3}Q$%
$J \partial A_{2} 
       B_2{}^{2}T{}^{2})/3 - 
    64\sqrt{2/3}$%
$J A_3 T{}^{4}- 
    (1136\sqrt{2/3}$%
$J A_3 B_2{}^{2}T{}^{2})/3 - (1184\sqrt{2/3}Q$%
$J B_2{}^{2}\partial T T{}^{2})/3 - 8\sqrt{6}Q$%
$J B_2{}^{4}\partial T - 
    88\sqrt{2/3}$%
$J B_2{}^{4}3 T - 352\sqrt{2/3}$%
$J B_2 
      B_3 T{}^{3}- 
    32\sqrt{6}Q$%
$J \partial B_{2} B_2 T{}^{3}- 88\sqrt{2/3}Q$%
$J \partial B_{2} B_2{}^{3}T + (-32 - 384Q{}^{2})
     $%
$\partial J J \partial T T{}^{3}+ 
    ((1120\sqrt{2/3}Q)/3 + 4480\sqrt{2/3}Q{}^{3})$%
$\partial J J{}^{2}\partial T \partial T T + 
    (416\sqrt{2/3}Q + 1664\sqrt{6}Q{}^{3})$%
$\partial J J{}^{2}\partial{}^{2} T T{}^{2}+ (4280/81 + 46960 Q{}^{2}/9 + 495040 Q{}^{4}/9)
     $%
$\partial J J{}^{3}\partial{}^{2} T \partial T + 
    (15952/243 + 24368 Q{}^{2}/9 + 622016 Q{}^{4}/27)$%
$\partial J J{}^{3}\partial{}^{3} T T + 
    ((18179\sqrt{2/3}Q)/81 + (21058\sqrt{2/3}Q{}^{3})/9 - 
      (38168\sqrt{2/3}Q{}^{5})/9)$%
$\partial J J{}^{4}\partial{}^{4} T + ((-16087Q)/(27\sqrt{6}) - 
      213523 Q{}^{3}/(27\sqrt{6}) - (164533\sqrt{2/3}Q{}^{5})/9)
     $%
$\partial J J{}^{4}\partial{}^{4} A_{2} + 
    ((2326\sqrt{2/3})/27 - (10892\sqrt{2/3}Q{}^{2})/9 - (38909\sqrt{2/3}Q{}^{4})/3)
     $%
$\partial J J{}^{4}\partial{}^{3} A_{3} + 
    (-4880/243 + 123160 Q{}^{2}/27 + 957920 Q{}^{4}/27)
     $%
$\partial J J{}^{3}A_2 \partial{}^{3} T + 
    (10366 Q/9 + 21172 Q{}^{3}/3)$%
$\partial J J{}^{3}A_2 \partial{}^{2} A_{3} + 
    (-1256/27 + 234856 Q{}^{2}/27 + 613280 Q{}^{4}/9)$%
$\partial J J{}^{3}\partial A_{2} \partial{}^{2} T + 
    (50668 Q/27 + 112100 Q{}^{3}/9)$%
$\partial J J{}^{3}\partial A_{2} \partial A_{3} + 
    (1544/81 + 173696 Q{}^{2}/27 + 457184 Q{}^{4}/9)$%
$\partial J J{}^{3}\partial{}^{2} A_{2} \partial T + 
    (3188/27 - 2644 Q{}^{2}/9 - 1580 Q{}^{4}/3)$%
$\partial J J{}^{3}\partial{}^{2} A_{2} \partial A_{2} + 
    (2434 Q/3 + 4160Q{}^{3})$%
$\partial J J{}^{3}\partial{}^{2} A_{2} A_3 + (-112/9 + 43208 Q{}^{2}/27 + 131000 Q{}^{4}/9)
     $%
$\partial J J{}^{3}\partial{}^{3} A_{2} T + 
    (2360/27 + 19774 Q{}^{2}/27 + 6712Q{}^{4})$%
$\partial J J{}^{3}\partial{}^{3} A_{2} A_2 + ((-3848Q)/3 - 13664 Q{}^{3}/3)
     $%
$\partial J J{}^{3}A_3 \partial{}^{2} T + 
    ((-25952Q)/27 - 36856 Q{}^{3}/9)$%
$\partial J J{}^{3}\partial A_{3} \partial T + (-1000/3 + 932Q{}^{2})
     $%
$\partial J J{}^{3}\partial A_{3} A_3 + 
    (-64Q - 4352 Q{}^{3}/3)$%
$\partial J J{}^{3}\partial{}^{2} A_{3} T + (10526 Q/9 + 26012 Q{}^{3}/3)
     $%
$\partial J J{}^{3}B_2 \partial{}^{2} B_{3} + 
    (30884 Q/9 + 62804 Q{}^{3}/3)$%
$\partial J J{}^{3}\partial B_{2} \partial B_{3} + 
    (5924/27 - 57616 Q{}^{2}/9 - 55556Q{}^{4})$%
$\partial J J{}^{3}\partial{}^{2} B_{2} \partial B_{2} + (22310 Q/9 + 40696 Q{}^{3}/3)
     $%
$\partial J J{}^{3}\partial{}^{2} B_{2} B_3 + 
    (1316/9 - 26192 Q{}^{2}/27 - 136384 Q{}^{4}/9)$%
$\partial J J{}^{3}\partial{}^{3} B_{2} B_2 + 
    (-2920/9 - 6308 Q{}^{2}/3)$%
$\partial J J{}^{3}\partial B_{3} B_3 + ((10504\sqrt{2/3}Q)/9 + 
      (43088\sqrt{2/3}Q{}^{3})/3)$%
$\partial J J{}^{2}A_2 \partial T \partial T + (720\sqrt{6}Q + 21152\sqrt{2/3}Q{}^{3})
     $%
$\partial J J{}^{2}A_2 \partial{}^{2} T T + 
    ((-2194\sqrt{2/3}Q)/9 + (11842\sqrt{2/3}Q{}^{3})/3)
     $%
$\partial J J{}^{2}A_2{}^{2}\partial{}^{2} T + 
    (16\sqrt{2/3} + 308\sqrt{2/3}Q{}^{2})$%
$\partial J J{}^{2}A_2{}^{2}\partial A_{3} + 
    ((3116\sqrt{2/3})/9 + (3586\sqrt{2/3}Q{}^{2})/3)$%
$\partial J J{}^{2}A_2 A_3 \partial T - 
    104\sqrt{2/3}Q$%
$\partial J J{}^{2}A_2 
      A_3 A_3 + ((3736\sqrt{2/3})/9 + 
      (11852\sqrt{2/3}Q{}^{2})/3)$%
$\partial J J{}^{2}A_2 \partial A_{3} T + ((-40\sqrt{2/3})/3 + 124\sqrt{6}Q{}^{2})
     $%
$\partial J J{}^{2}A_2 B_2 
      \partial B_{3} + ((10804\sqrt{2/3}Q)/9 + (20000\sqrt{2/3}Q{}^{3})/3)
     $%
$\partial J J{}^{2}A_2 \partial B_{2} 
      \partial B_{2} + ((-400\sqrt{2/3})/3 + 580\sqrt{2/3}Q{}^{2})
     $%
$\partial J J{}^{2}A_2 \partial B_{2} 
      B_3 + ((12124\sqrt{2/3}Q)/9 + (21752\sqrt{2/3}Q{}^{3})/3)
     $%
$\partial J J{}^{2}A_2 \partial{}^{2} B_{2} 
      B_2 - 24\sqrt{6}Q$%
$\partial J J{}^{2}A_2 B_3 B_3 + 
    ((3808\sqrt{2/3}Q)/3 + 5864\sqrt{6}Q{}^{3})$%
$\partial J J{}^{2}\partial A_{2} \partial T T + 
    ((-1988\sqrt{2/3}Q)/9 + (31286\sqrt{2/3}Q{}^{3})/3)
     $%
$\partial J J{}^{2}\partial A_{2} A_2 \partial T + 
    ((-332\sqrt{2/3})/9 - (2338\sqrt{2/3}Q{}^{2})/3)$%
$\partial J J{}^{2}\partial A_{2} A_2 A_3 + 
    ((-2132\sqrt{2/3}Q)/9 + (7976\sqrt{2/3}Q{}^{3})/3)
     $%
$\partial J J{}^{2}\partial A_{2} \partial A_{2} T + 
    ((632\sqrt{2/3}Q)/3 + 3226\sqrt{2/3}Q{}^{3})$%
$\partial J J{}^{2}\partial A_{2} \partial A_{2} A_2 + 
    ((7240\sqrt{2/3})/9 + (22004\sqrt{2/3}Q{}^{2})/3)$%
$\partial J J{}^{2}\partial A_{2} A_3 T + 
    ((-964\sqrt{2/3})/9 - (230\sqrt{2/3}Q{}^{2})/3)$%
$\partial J J{}^{2}\partial A_{2} B_2 B_3 + 
    ((19828\sqrt{2/3}Q)/9 + (40622\sqrt{2/3}Q{}^{3})/3)
     $%
$\partial J J{}^{2}\partial A_{2} \partial B_{2} 
      B_2 + ((-388\sqrt{2/3}Q)/3 + 3580\sqrt{2/3}Q{}^{3})
     $%
$\partial J J{}^{2}\partial{}^{2} A_{2} T{}^{2}+ 
    ((2500\sqrt{2/3}Q)/9 + (23816\sqrt{2/3}Q{}^{3})/3)
     $%
$\partial J J{}^{2}\partial{}^{2} A_{2} A_2 T + 
    ((1558\sqrt{2/3}Q)/9 + (6758\sqrt{2/3}Q{}^{3})/3)
     $%
$\partial J J{}^{2}\partial{}^{2} A_{2} A_2{}^{2}+ ((5246\sqrt{2/3}Q)/9 + (11506\sqrt{2/3}Q{}^{3})/3)
     $%
$\partial J J{}^{2}\partial{}^{2} A_{2} B_2{}^{2}+ ((7760\sqrt{2/3})/3 + 19448\sqrt{2/3}Q{}^{2})
     $%
$\partial J J{}^{2}A_3 \partial T T - 
    304\sqrt{6}Q$%
$\partial J J{}^{2}A_3 A_3 
      T - 120\sqrt{6}Q$%
$\partial J J{}^{2}A_3 
      B_2 B_3 + ((-584\sqrt{2/3})/3 + 488\sqrt{2/3}Q{}^{2})
     $%
$\partial J J{}^{2}A_3 \partial B_{2} 
      B_2 + ((3296\sqrt{2/3})/3 + 8368\sqrt{2/3}Q{}^{2})
     $%
$\partial J J{}^{2}\partial A_{3} T{}^{2}+ 
    ((-472\sqrt{2/3})/3 - 32\sqrt{6}Q{}^{2})$%
$\partial J J{}^{2}\partial A_{3} B_2{}^{2}+ 
    ((1718\sqrt{2/3}Q)/9 + (11938\sqrt{2/3}Q{}^{3})/3)
     $%
$\partial J J{}^{2}B_2{}^{2}\partial{}^{2} T + 
    ((7420\sqrt{2/3})/9 + (11234\sqrt{2/3}Q{}^{2})/3)$%
$\partial J J{}^{2}B_2 B_3 \partial T + 
    ((6104\sqrt{2/3})/9 + (9052\sqrt{2/3}Q{}^{2})/3)$%
$\partial J J{}^{2}B_2 \partial B_{3} T + 
    ((5420\sqrt{2/3}Q)/9 + (22438\sqrt{2/3}Q{}^{3})/3)
     $%
$\partial J J{}^{2}\partial B_{2} B_2 \partial T + 
    ((-6796\sqrt{2/3}Q)/9 - (10784\sqrt{2/3}Q{}^{3})/3)
     $%
$\partial J J{}^{2}\partial B_{2} \partial B_{2} T + 
    ((10424\sqrt{2/3})/9 + (21484\sqrt{2/3}Q{}^{2})/3)
     $%
$\partial J J{}^{2}\partial B_{2} B_3 T + 
    ((-1588\sqrt{2/3}Q)/9 + (1528\sqrt{2/3}Q{}^{3})/3)
     $%
$\partial J J{}^{2}\partial{}^{2} B_{2} B_2 T + 
    120\sqrt{6}Q$%
$\partial J J{}^{2}B_3 B_3 
      T + (-18016/27 - 40352 Q{}^{2}/9)$%
$\partial J J 
      A_2 \partial T T{}^{2}+ (-13184/27 - 18304 Q{}^{2}/9)
     $%
$\partial J J A_2{}^{2}\partial T T + 
    (304/27 + 968 Q{}^{2}/9)$%
$\partial J J A_2{}^{3}\partial T + (176Q$%
$\partial J J A_2{}^{3}3 T)/3 - 
    16Q$%
$\partial J J A_2{}^{2}B_2 
      B_3 + (8/27 + 2164 Q{}^{2}/9)$%
$\partial J J 
      A_2{}^{2}\partial B_{2} B_2 + 
    (4928Q$%
$\partial J J A_2 A_3 T{}^{2})/
     3 + (16Q$%
$\partial J J A_2 A_3 B_2{}^{2})/3 + (-5392/27 - 8456 Q{}^{2}/9)$%
$\partial J J 
      A_2 B_2{}^{2}\partial T + 
    864Q$%
$\partial J J A_2 B_2 B_3 
      T + (-9232/27 - 18296 Q{}^{2}/9)$%
$\partial J J 
      A_2 \partial B_{2} B_2 T + 
    (-1696/9 - 6656 Q{}^{2}/3)$%
$\partial J J \partial A_{2} T{}^{3}+ (-16400/27 - 42376 Q{}^{2}/9)
     $%
$\partial J J \partial A_{2} A_2 T{}^{2}+ 
    (-952/27 - 6020 Q{}^{2}/9)$%
$\partial J J \partial A_{2} A_2{}^{2}T + (64/3 - 328 Q{}^{2}/3)$%
$\partial J J 
      \partial A_{2} A_2{}^{3}+ 
    (-256/9 - 264Q{}^{2})$%
$\partial J J \partial A_{2} A_2 
      B_2{}^{2}+ (-7832/27 - 21604 Q{}^{2}/9)
     $%
$\partial J J \partial A_{2} B_2{}^{2}T + 
    1728Q$%
$\partial J J A_3 T{}^{3}+ 
    592Q$%
$\partial J J A_3 B_2{}^{2}T + (-1056 - 19456 Q{}^{2}/3)$%
$\partial J J B_2{}^{2}\partial T T - 32Q$%
$\partial J J B_2{}^{4}3 + 
    2656Q$%
$\partial J J B_2 B_3 T{}^{2}+ 
    (-8432/9 - 21640 Q{}^{2}/3)$%
$\partial J J \partial B_{2} 
      B_2 T{}^{2}+ (-680/27 + 44 Q{}^{2}/9)
     $%
$\partial J J \partial B_{2} B_2{}^{3}+ ((2992\sqrt{2/3}Q)/3 + 11968\sqrt{2/3}Q{}^{3})
     $%
$(\partial J){}^{2} J \partial T T{}^{2}+ 
    (2312/27 + 14608 Q{}^{2}/3 + 138304 Q{}^{4}/3)$%
$(\partial J){}^{2} 
      J{}^{2}\partial T \partial T + 
    (4208/27 + 17296 Q{}^{2}/3 + 140224 Q{}^{4}/3)$%
$(\partial J){}^{2} 
      J{}^{2}\partial{}^{2} T T + 
    ((57544\sqrt{2/3}Q)/81 - (36008\sqrt{2/3}Q{}^{3})/9 - 
      (1352800\sqrt{2/3}Q{}^{5})/9)$%
$(\partial J){}^{2} J{}^{3}\partial{}^{3} T + ((-170290\sqrt{2/3}Q)/81 - 
      (828278\sqrt{2/3}Q{}^{3})/27 - (451876\sqrt{2/3}Q{}^{5})/3)
     $%
$(\partial J){}^{2} J{}^{3}\partial{}^{3} A_{2} + 
    ((15044\sqrt{2/3})/27 - (76562\sqrt{2/3}Q{}^{2})/9 - 29350\sqrt{6}Q{}^{4})
     $%
$(\partial J){}^{2} J{}^{3}\partial{}^{2} A_{3} + 
    (-1160/27 + 82688 Q{}^{2}/9 + 57224Q{}^{4})$%
$(\partial J){}^{2} 
      J{}^{2}A_2 \partial{}^{2} T + (8450 Q/3 + 18918Q{}^{3})
     $%
$(\partial J){}^{2} J{}^{2}A_2 \partial A_{3} + 
    (2152/27 + 130264 Q{}^{2}/9 + 257896 Q{}^{4}/3)$%
$(\partial J){}^{2} 
      J{}^{2}\partial A_{2} \partial T + 
    (5150/27 - 11414 Q{}^{2}/9 - 14366Q{}^{4})$%
$(\partial J){}^{2} J{}^{2}\partial A_{2} \partial A_{2} + (2310Q + 14874Q{}^{3})
     $%
$(\partial J){}^{2} J{}^{2}\partial A_{2} A_3 + 
    (-280/9 + 42952 Q{}^{2}/9 + 89504 Q{}^{4}/3)$%
$(\partial J){}^{2} 
      J{}^{2}\partial{}^{2} A_{2} T + 
    (5294/27 - 4702 Q{}^{2}/3 - 45608 Q{}^{4}/3)$%
$(\partial J){}^{2} 
      J{}^{2}\partial{}^{2} A_{2} A_2 + 
    ((-14264Q)/3 - 28280Q{}^{3})$%
$(\partial J){}^{2} J{}^{2}A_3 \partial T + (-180 + 942Q{}^{2})$%
$(\partial J){}^{2} 
      J{}^{2}A_3 A_3 + 
    ((-10976Q)/3 - 26096Q{}^{3})$%
$(\partial J){}^{2} J{}^{2}\partial A_{3} T + (36398 Q/9 + 68314 Q{}^{3}/3)
     $%
$(\partial J){}^{2} J{}^{2}B_2 \partial B_{3} + 
    (2254/9 - 79906 Q{}^{2}/9 - 227138 Q{}^{4}/3)$%
$(\partial J){}^{2} 
      J{}^{2}\partial B_{2} \partial B_{2} + 
    (48782 Q/9 + 93718 Q{}^{3}/3)$%
$(\partial J){}^{2} J{}^{2}\partial B_{2} B_3 + 
    (3826/9 - 56722 Q{}^{2}/9 - 199316 Q{}^{4}/3)$%
$(\partial J){}^{2} 
      J{}^{2}\partial{}^{2} B_{2} B_2 + 
    (-344 - 2382Q{}^{2})$%
$(\partial J){}^{2} J{}^{2}B_3 B_3 + ((18416\sqrt{2/3}Q)/3 + 52832\sqrt{2/3}Q{}^{3})
     $%
$(\partial J){}^{2} J A_2 \partial T T + 
    ((6640\sqrt{2/3}Q)/9 + (29756\sqrt{2/3}Q{}^{3})/3)
     $%
$(\partial J){}^{2} J A_2{}^{2}\partial T + 
    (16\sqrt{2/3} + 236\sqrt{2/3}Q{}^{2})$%
$(\partial J){}^{2} J 
      A_2{}^{3}3 + 
    ((4208\sqrt{2/3})/9 + (3352\sqrt{2/3}Q{}^{2})/3)$%
$(\partial J){}^{2} 
      J A_2 A_3 T + 
    ((-400\sqrt{2/3})/3 - 68\sqrt{2/3}Q{}^{2})$%
$(\partial J){}^{2} 
      J A_2 B_2 B_3 + 
    ((7088\sqrt{2/3}Q)/3 + 13432\sqrt{2/3}Q{}^{3})$%
$(\partial J){}^{2} 
      J A_2 \partial B_{2} B_2 + 
    ((7424\sqrt{2/3}Q)/3 + 26056\sqrt{2/3}Q{}^{3})$%
$(\partial J){}^{2} 
      J \partial A_{2} T{}^{2}+ 
    ((18536\sqrt{2/3}Q)/9 + (78712\sqrt{2/3}Q{}^{3})/3)
     $%
$(\partial J){}^{2} J \partial A_{2} A_2 T + 
    ((4700\sqrt{2/3}Q)/9 + (21430\sqrt{2/3}Q{}^{3})/3)
     $%
$(\partial J){}^{2} J \partial A_{2} A_2{}^{2}+ (332\sqrt{6}Q + 2194\sqrt{6}Q{}^{3})
     $%
$(\partial J){}^{2} J \partial A_{2} B_2{}^{2}+ (800\sqrt{2/3} - 2056\sqrt{2/3}Q{}^{2})
     $%
$(\partial J){}^{2} J A_3 T{}^{2}+ 
    ((-112\sqrt{2/3})/3 + 48\sqrt{6}Q{}^{2})$%
$(\partial J){}^{2} J 
      A_3 B_2{}^{2}+ 
    ((15088\sqrt{2/3}Q)/9 + (33788\sqrt{2/3}Q{}^{3})/3)
     $%
$(\partial J){}^{2} J B_2{}^{2}\partial T + 
    ((7264\sqrt{2/3})/9 - (1480\sqrt{2/3}Q{}^{2})/3)$%
$(\partial J){}^{2} 
      J B_2 B_3 T + 
    ((17992\sqrt{2/3}Q)/9 + (47240\sqrt{2/3}Q{}^{3})/3)
     $%
$(\partial J){}^{2} J \partial B_{2} B_2 T + 
    ((320\sqrt{2/3}Q)/3 + 1280\sqrt{2/3}Q{}^{3})$%
$(\partial J){}^{3} T{}^{3}+ 
    (7280/81 + 912Q{}^{2} - 17984 Q{}^{4}/9)$%
$(\partial J){}^{3} 
      J \partial T T + ((19456\sqrt{2/3}Q)/27 - 
      (95552\sqrt{2/3}Q{}^{3})/3 - (1457920\sqrt{2/3}Q{}^{5})/3)
     $%
$(\partial J){}^{3} J{}^{2}\partial{}^{2} T + 
    ((-126344\sqrt{2/3}Q)/27 - (219526\sqrt{2/3}Q{}^{3})/3 - 
      (1085998\sqrt{2/3}Q{}^{5})/3)$%
$(\partial J){}^{3} 
      J{}^{2}\partial{}^{2} A_{2} + 
    ((29768\sqrt{2/3})/27 - (54604\sqrt{2/3}Q{}^{2})/3 - 
      (555416\sqrt{2/3}Q{}^{4})/3)$%
$(\partial J){}^{3} 
      J{}^{2}\partial A_{3} + 
    (1808/81 + 12560 Q{}^{2}/9 - 142208 Q{}^{4}/9)$%
$(\partial J){}^{3} J A_2 \partial T + (4736 Q/3 + 10900Q{}^{3})
     $%
$(\partial J){}^{3} J A_2 A_3 + 
    (-2512/27 - 1120Q{}^{2} - 85904 Q{}^{4}/3)$%
$(\partial J){}^{3} 
      J \partial A_{2} T + (248 - 13756 Q{}^{2}/3 - 49256Q{}^{4})
     $%
$(\partial J){}^{3} J \partial A_{2} A_2 + 
    (-4432Q - 25296Q{}^{3})$%
$(\partial J){}^{3} J 
      A_3 T + (22816 Q/9 + 43484 Q{}^{3}/3)
     $%
$(\partial J){}^{3} J B_2 B_3 + 
    (11528/27 - 27716 Q{}^{2}/3 - 271232 Q{}^{4}/3)$%
$(\partial J){}^{3} J \partial B_{2} B_2 + 
    (200/81 - 6944 Q{}^{2}/9 - 86528 Q{}^{4}/9)$%
$(\partial J){}^{4} T{}^{2}+ 
    ((248\sqrt{2/3}Q)/9 - (103016\sqrt{2/3}Q{}^{3})/3 - 416032\sqrt{2/3}Q{}^{5})
     $%
$(\partial J){}^{4} J \partial T + 
    ((-14072\sqrt{2/3}Q)/3 - (624052\sqrt{2/3}Q{}^{3})/9 - 
      (905540\sqrt{2/3}Q{}^{5})/3)$%
$(\partial J){}^{4} J \partial A_{2} + 
    ((4492\sqrt{2/3})/9 - 2804\sqrt{6}Q{}^{2} - 89956\sqrt{2/3}Q{}^{4})
     $%
$(\partial J){}^{4} J A_3 + 
    ((232\sqrt{2/3}Q)/27 - (11072\sqrt{2/3}Q{}^{3})/3 - 
      (136576\sqrt{2/3}Q{}^{5})/3)$%
$(\partial J){}^{5} T + (33044 Q{}^{2}/27 + 113200 Q{}^{4}/3 + 
      829696 Q{}^{6}/3)$%
$(\partial J){}^{6} + ((-21452\sqrt{2/3}Q)/27 - 
      (102256\sqrt{2/3}Q{}^{3})/9 - 13440\sqrt{6}Q{}^{5})
     $%
$(\partial J){}^{5} A_2 + 
    (-4384/81 - 3360Q{}^{2} - 281360 Q{}^{4}/9)$%
$(\partial J){}^{4} A_2 T + 
    (166/9 - 10492 Q{}^{2}/9 - 11928Q{}^{4})$%
$(\partial J){}^{4} A_2{}^{2}+ (974/27 - 1228Q{}^{2} - 11904Q{}^{4})
     $%
$(\partial J){}^{4} B_2{}^{2}+ 
    ((11872\sqrt{2/3}Q)/9 + (25856\sqrt{2/3}Q{}^{3})/3)
     $%
$(\partial J){}^{3} A_2 T{}^{2}+ 
    ((2764\sqrt{2/3}Q)/9 + (7420\sqrt{2/3}Q{}^{3})/3)
     $%
$(\partial J){}^{3} A_2{}^{2}T + 
    ((328\sqrt{2/3}Q)/9 + (670\sqrt{2/3}Q{}^{3})/3)$%
$(\partial J){}^{3} A_2{}^{3}+ 
    ((1040\sqrt{2/3}Q)/3 + (7450\sqrt{2/3}Q{}^{3})/3)
     $%
$(\partial J){}^{3} A_2 B_2{}^{2}+ ((7180\sqrt{2/3}Q)/9 + 4052\sqrt{2/3}Q{}^{3})
     $%
$(\partial J){}^{3} B_2{}^{2}T + 
    (-608/9 + 352 Q{}^{2}/3)$%
$(\partial J){}^{2} A_2 T{}^{3}+ (-1976/27 + 6008 Q{}^{2}/9)$%
$(\partial J){}^{2} 
      A_2{}^{2}T{}^{2}+ (-376/27 - 1196 Q{}^{2}/9)
     $%
$(\partial J){}^{2} A_2{}^{3}T + 
    (298/27 + 1844 Q{}^{2}/9)$%
$(\partial J){}^{2} A_2{}^{4}+ (-356/27 - 2188 Q{}^{2}/9)
     $%
$(\partial J){}^{2} A_2{}^{2}B_2{}^{2}+ (-1448/27 + 4172 Q{}^{2}/9)$%
$(\partial J){}^{2} 
      A_2 B_2{}^{2}T + 
    (-712/3 - 616 Q{}^{2}/3)$%
$(\partial J){}^{2} B_2{}^{2}T{}^{2}+ (58/27 + 728 Q{}^{2}/9)$%
$(\partial J){}^{2} 
      B_2{}^{4}- 
    64\sqrt{2/3}Q$%
$\partial J A_2 T{}^{4}- (896\sqrt{2/3}Q$%
$\partial J A_2{}^{2}T{}^{3})/3 - 
    (64\sqrt{2/3}Q$%
$\partial J A_2{}^{3}T{}^{2})/3 + (160\sqrt{2/3}Q$%
$\partial J A_2{}^{4}T)/3 - 28\sqrt{2/3}Q
     $%
$\partial J A_2{}^{5}+ (152\sqrt{2/3}Q$%
$\partial J A_2{}^{3}B_2{}^{2})/3 - 
    (224\sqrt{2/3}Q$%
$\partial J A_2{}^{2}B_2{}^{2}T)/3 - (1984\sqrt{2/3}Q$%
$\partial J A_2 
       B_2{}^{2}T{}^{2})/3 - 
    (68\sqrt{2/3}Q$%
$\partial J A_2 B_2{}^{4})/3 - 128\sqrt{6}Q$%
$\partial J B_2{}^{2}T{}^{3}- 
    (128\sqrt{2/3}Q$%
$\partial J B_2{}^{4}T)/3 + (-64/9 - 256 Q{}^{2}/3)
     $%
$\partial{}^{2} J J T{}^{4}+ 
    (632\sqrt{2/3}Q + 2528\sqrt{6}Q{}^{3})$%
$\partial{}^{2} J J{}^{2}\partial T T{}^{2}+ (728/27 + 77824 Q{}^{2}/27 + 276352 Q{}^{4}/9)
     $%
$\partial{}^{2} J J{}^{3}\partial T \partial T + 
    (736/9 + 100384 Q{}^{2}/27 + 295552 Q{}^{4}/9)$%
$\partial{}^{2} J J{}^{3}\partial{}^{2} T T + 
    ((3382\sqrt{2/3}Q)/9 + (7418\sqrt{2/3}Q{}^{3})/3 - 24440\sqrt{2/3}Q{}^{5})
     $%
$\partial{}^{2} J J{}^{4}\partial{}^{3} T + 
    ((-59218\sqrt{2/3}Q)/81 - (274562\sqrt{2/3}Q{}^{3})/27 - 
      (139433\sqrt{2/3}Q{}^{5})/3)$%
$\partial{}^{2} J J{}^{4}\partial{}^{3} A_{2} + 
    ((6593\sqrt{2/3})/27 - (11926\sqrt{2/3}Q{}^{2})/9 - (58063\sqrt{2/3}Q{}^{4})/3)
     $%
$\partial{}^{2} J J{}^{4}\partial{}^{2} A_{3} + 
    (-4000/81 + 473368 Q{}^{2}/81 + 1167296 Q{}^{4}/27)
     $%
$\partial{}^{2} J J{}^{3}A_2 \partial{}^{2} T + 
    (40766 Q/27 + 97312 Q{}^{3}/9)$%
$\partial{}^{2} J J{}^{3}A_2 \partial A_{3} + 
    (784/27 + 712744 Q{}^{2}/81 + 1579472 Q{}^{4}/27)$%
$\partial{}^{2} J J{}^{3}\partial A_{2} \partial T + 
    (2708/27 + 13666 Q{}^{2}/81 - 9670 Q{}^{4}/27)$%
$\partial{}^{2} J J{}^{3}\partial A_{2} \partial A_{2} + 
    (32846 Q/27 + 77596 Q{}^{3}/9)$%
$\partial{}^{2} J J{}^{3}\partial A_{2} A_3 + 
    (-3152/81 + 64256 Q{}^{2}/27 + 19792Q{}^{4})$%
$\partial{}^{2} J J{}^{3}\partial{}^{2} A_{2} T + 
    (1156/9 + 73798 Q{}^{2}/81 + 181928 Q{}^{4}/27)$%
$\partial{}^{2} J J{}^{3}\partial{}^{2} A_{2} A_2 + 
    ((-15800Q)/9 - 30560 Q{}^{3}/3)$%
$\partial{}^{2} J J{}^{3}A_3 \partial T + (-188 - 342Q{}^{2})
     $%
$\partial{}^{2} J J{}^{3}A_3 A_3 + 
    ((-7264Q)/9 - 16888 Q{}^{3}/3)$%
$\partial{}^{2} J J{}^{3}\partial A_{3} T + (52726 Q/27 + 85736 Q{}^{3}/9)
     $%
$\partial{}^{2} J J{}^{3}B_2 \partial B_{3} + 
    (3668/27 - 9026 Q{}^{2}/3 - 214850 Q{}^{4}/9)$%
$\partial{}^{2} J J{}^{3}\partial B_{2} \partial B_{2} + 
    (75454 Q/27 + 130988 Q{}^{3}/9)$%
$\partial{}^{2} J J{}^{3}\partial B_{2} B_3 + 
    (7060/27 - 10798 Q{}^{2}/9 - 157976 Q{}^{4}/9)$%
$\partial{}^{2} J J{}^{3}\partial{}^{2} B_{2} B_2 + 
    (-1700/9 - 3146 Q{}^{2}/3)$%
$\partial{}^{2} J J{}^{3}B_3 B_3 + ((29696\sqrt{2/3}Q)/9 + 
      (88184\sqrt{2/3}Q{}^{3})/3)$%
$\partial{}^{2} J J{}^{2}A_2 \partial T T + 
    ((-98\sqrt{2/3}Q)/3 + (9652\sqrt{2/3}Q{}^{3})/3)$%
$\partial{}^{2} J J{}^{2}A_2{}^{2}\partial T + 
    ((16\sqrt{2/3})/3 + (130\sqrt{2/3}Q{}^{2})/3)$%
$\partial{}^{2} J J{}^{2}A_2{}^{3}3 + 
    ((4696\sqrt{2/3})/9 + 1224\sqrt{6}Q{}^{2})$%
$\partial{}^{2} J J{}^{2}A_2 A_3 T + 
    ((-332\sqrt{2/3})/3 - 114\sqrt{6}Q{}^{2})$%
$\partial{}^{2} J J{}^{2}A_2 B_2 B_3 + 
    ((16376\sqrt{2/3}Q)/9 + (34480\sqrt{2/3}Q{}^{3})/3)
     $%
$\partial{}^{2} J J{}^{2}A_2 \partial B_{2} 
      B_2 + ((8116\sqrt{2/3}Q)/9 + 9952\sqrt{2/3}Q{}^{3})
     $%
$\partial{}^{2} J J{}^{2}\partial A_{2} T{}^{2}+ 
    ((2552\sqrt{2/3}Q)/3 + 4276\sqrt{6}Q{}^{3})$%
$\partial{}^{2} J J{}^{2}\partial A_{2} A_2 T + 
    ((532\sqrt{2/3}Q)/9 + (4579\sqrt{2/3}Q{}^{3})/3)$%
$\partial{}^{2} J J{}^{2}\partial A_{2} A_2{}^{2}+ 
    ((8068\sqrt{2/3}Q)/9 + (19093\sqrt{2/3}Q{}^{3})/3)
     $%
$\partial{}^{2} J J{}^{2}\partial A_{2} B_2{}^{2}+ ((3808\sqrt{2/3})/3 + 7556\sqrt{2/3}Q{}^{2})
     $%
$\partial{}^{2} J J{}^{2}A_3 T{}^{2}+ 
    ((-340\sqrt{2/3})/3 - (788\sqrt{2/3}Q{}^{2})/3)$%
$\partial{}^{2} J J{}^{2}A_3 B_2{}^{2}+ 
    ((8810\sqrt{2/3}Q)/9 + (25172\sqrt{2/3}Q{}^{3})/3)
     $%
$\partial{}^{2} J J{}^{2}B_2{}^{2}\partial T + 
    ((8840\sqrt{2/3})/9 + (12352\sqrt{2/3}Q{}^{2})/3)$%
$\partial{}^{2} J J{}^{2}B_2 B_3 T + 
    ((5072\sqrt{2/3}Q)/9 + (23252\sqrt{2/3}Q{}^{3})/3)
     $%
$\partial{}^{2} J J{}^{2}\partial B_{2} B_2 T + 
    (-2048/9 - 13216 Q{}^{2}/9)$%
$\partial{}^{2} J J A_2 T{}^{3}+ (-7928/27 - 4216 Q{}^{2}/3)$%
$\partial{}^{2} J J 
      A_2{}^{2}T{}^{2}+ (-184/27 - 92Q{}^{2})
     $%
$\partial{}^{2} J J A_2{}^{3}T + 
    (262/27 + 128 Q{}^{2}/9)$%
$\partial{}^{2} J J A_2{}^{4}+ (-284/27 - 100 Q{}^{2}/9)
     $%
$\partial{}^{2} J J A_2{}^{2}B_2{}^{2}+ (-5960/27 - 3340 Q{}^{2}/3)$%
$\partial{}^{2} J J 
      A_2 B_2{}^{2}T + 
    (-4952/9 - 25304 Q{}^{2}/9)$%
$\partial{}^{2} J J B_2{}^{2}T{}^{2}+ (-170/27 - 220 Q{}^{2}/9)
     $%
$\partial{}^{2} J J B_2{}^{4}+ (544\sqrt{2/3}Q + 2176\sqrt{6}Q{}^{3})
     $%
$\partial{}^{2} J \partial J J T{}^{3}+ 
    (9616/27 + 91616 Q{}^{2}/9 + 212608 Q{}^{4}/3)$%
$\partial{}^{2} J \partial J 
      J{}^{2}\partial T T + 
    ((13984\sqrt{2/3}Q)/9 - (257864\sqrt{2/3}Q{}^{3})/9 - 
      (1702688\sqrt{2/3}Q{}^{5})/3)$%
$\partial{}^{2} J \partial J J{}^{3}\partial{}^{2} T + ((-180182\sqrt{2/3}Q)/27 - 
      (318164\sqrt{2/3}Q{}^{3})/3 - (1572832\sqrt{2/3}Q{}^{5})/3)
     $%
$\partial{}^{2} J \partial J J{}^{3}\partial{}^{2} A_{2} + 
    ((18488\sqrt{2/3})/9 - (122204\sqrt{2/3}Q{}^{2})/9 - 
      (581788\sqrt{2/3}Q{}^{4})/3)$%
$\partial{}^{2} J \partial J J{}^{3}\partial A_{3} + 
    (3568/27 + 308648 Q{}^{2}/27 + 334216 Q{}^{4}/9)$%
$\partial{}^{2} J \partial J 
      J{}^{2}A_2 \partial T + (30710 Q/9 + 24344Q{}^{3})
     $%
$\partial{}^{2} J \partial J J{}^{2}A_2 A_3 + 
    (-208/3 + 249736 Q{}^{2}/27 + 436760 Q{}^{4}/9)$%
$\partial{}^{2} J \partial J 
      J{}^{2}\partial A_{2} T + 
    (20840/27 + 19714 Q{}^{2}/27 - 156976 Q{}^{4}/9)$%
$\partial{}^{2} J \partial J 
      J{}^{2}\partial A_{2} A_2 + 
    (-10880Q - 81904Q{}^{3})$%
$\partial{}^{2} J \partial J J{}^{2}A_3 T + (47822 Q/9 + 81376 Q{}^{3}/3)
     $%
$\partial{}^{2} J \partial J J{}^{2}B_2 B_3 + 
    (33544/27 - 31726 Q{}^{2}/3 - 405064 Q{}^{4}/3)$%
$\partial{}^{2} J \partial J 
      J{}^{2}\partial B_{2} B_2 + 
    ((56000\sqrt{2/3}Q)/9 + (150536\sqrt{2/3}Q{}^{3})/3)
     $%
$\partial{}^{2} J \partial J J A_2 T{}^{2}+ 
    ((17036\sqrt{2/3}Q)/9 + (54152\sqrt{2/3}Q{}^{3})/3)
     $%
$\partial{}^{2} J \partial J J A_2{}^{2}T + 
    ((3304\sqrt{2/3}Q)/9 + (14570\sqrt{2/3}Q{}^{3})/3)
     $%
$\partial{}^{2} J \partial J J A_2{}^{3}+ ((12752\sqrt{2/3}Q)/9 + 8426\sqrt{2/3}Q{}^{3})
     $%
$\partial{}^{2} J \partial J J A_2 B_2{}^{2}+ (3980\sqrt{2/3}Q + 9736\sqrt{6}Q{}^{3})
     $%
$\partial{}^{2} J \partial J J B_2{}^{2}T + 
    (928/9 - 14384 Q{}^{2}/9 - 102080 Q{}^{4}/3)$%
$\partial{}^{2} J (\partial J){}^{2} J T{}^{2}+ 
    (260\sqrt{6}Q - (378292\sqrt{2/3}Q{}^{3})/3 - 1625488\sqrt{2/3}Q{}^{5})
     $%
$\partial{}^{2} J (\partial J){}^{2} J{}^{2}\partial T + 
    ((-62150\sqrt{2/3}Q)/3 - 331250\sqrt{2/3}Q{}^{3} - 526476\sqrt{6}Q{}^{5})
     $%
$\partial{}^{2} J (\partial J){}^{2} J{}^{2}\partial A_{2} + 
    ((27520\sqrt{2/3})/9 - 11326\sqrt{2/3}Q{}^{2} - 237466\sqrt{2/3}Q{}^{4})
     $%
$\partial{}^{2} J (\partial J){}^{2} J{}^{2}A_3 + 
    (-5920/27 - 406432 Q{}^{2}/27 - 1427360 Q{}^{4}/9)$%
$\partial{}^{2} J (\partial J){}^{2} J A_2 T + 
    (6568/27 - 43436 Q{}^{2}/9 - 56504Q{}^{4})$%
$\partial{}^{2} J (\partial J){}^{2} 
      J A_2{}^{2}+ 
    (15880/27 - 29692 Q{}^{2}/9 - 170128 Q{}^{4}/3)$%
$\partial{}^{2} J (\partial J){}^{2} J B_2{}^{2}+ 
    ((-25552\sqrt{2/3}Q)/27 - (280000\sqrt{2/3}Q{}^{3})/3 - 
      (2951168\sqrt{2/3}Q{}^{5})/3)$%
$\partial{}^{2} J (\partial J){}^{3} J T + ((-339320\sqrt{2/3}Q)/27 - 
      (2019580\sqrt{2/3}Q{}^{3})/9 - 376348\sqrt{6}Q{}^{5})
     $%
$\partial{}^{2} J (\partial J){}^{3} J A_2 + 
    (-6608/81 + 480872 Q{}^{2}/27 + 5380616 Q{}^{4}/9 + 13405600 Q{}^{6}/3)
     $%
$\partial{}^{2} J (\partial J){}^{4} J + 
    (596/9 + 848Q{}^{2} + 640Q{}^{4})$%
$\partial{}^{2} J \partial{}^{2} J J{}^{2}T{}^{2}+ ((5008\sqrt{2/3}Q)/9 - 
      (183592\sqrt{2/3}Q{}^{3})/9 - (974752\sqrt{2/3}Q{}^{5})/3)
     $%
$\partial{}^{2} J \partial{}^{2} J J{}^{3}\partial T + 
    ((-44464\sqrt{2/3}Q)/9 - (2126846\sqrt{2/3}Q{}^{3})/27 - 
      (3379564\sqrt{2/3}Q{}^{5})/9)$%
$\partial{}^{2} J \partial{}^{2} J J{}^{3}\partial A_{2} + 
    ((28574\sqrt{2/3})/27 + (8824\sqrt{2/3}Q{}^{2})/3 - (94112\sqrt{2/3}Q{}^{4})/3)
     $%
$\partial{}^{2} J \partial{}^{2} J J{}^{3}A_3 + 
    (-1432/27 - 33008 Q{}^{2}/27 - 180632 Q{}^{4}/9)$%
$\partial{}^{2} J \partial{}^{2} J 
      J{}^{2}A_2 T + 
    (4261/27 + 874Q{}^{2} + 7436 Q{}^{4}/9)$%
$\partial{}^{2} J \partial{}^{2} J J{}^{2}A_2{}^{2}+ 
    (8419/27 + 3986 Q{}^{2}/9 - 130804 Q{}^{4}/9)$%
$\partial{}^{2} J \partial{}^{2} J 
      J{}^{2}B_2{}^{2}+ 
    ((-14092\sqrt{2/3}Q)/9 - 38536\sqrt{6}Q{}^{3} - 1161824\sqrt{2/3}Q{}^{5})
     $%
$\partial{}^{2} J \partial{}^{2} J \partial J J{}^{2}T + 
    ((-40510\sqrt{2/3}Q)/3 - (2275982\sqrt{2/3}Q{}^{3})/9 - 
      (4031720\sqrt{2/3}Q{}^{5})/3)$%
$\partial{}^{2} J \partial{}^{2} J \partial J 
      J{}^{2}A_2 + 
    (-30736/81 + 483950 Q{}^{2}/27 + 6958432 Q{}^{4}/9 + 18123424 Q{}^{6}/3)
     $%
$\partial{}^{2} J \partial{}^{2} J (\partial J){}^{2} J{}^{2}+ 
    (-2672/27 - 28072 Q{}^{2}/81 + 146240 Q{}^{4}/3 + 4174720 Q{}^{6}/9)
     $%
$\partial{}^{2} J \partial{}^{2} J \partial{}^{2} J J{}^{3}+ 
    ((1504\sqrt{2/3}Q)/9 + (6016\sqrt{2/3}Q{}^{3})/3)
     $%
$\partial{}^{3} J J{}^{2}T{}^{3}+ 
    (15136/243 + 71776 Q{}^{2}/27 + 619136 Q{}^{4}/27)$%
$\partial{}^{3} J J{}^{3}\partial T T + 
    ((25426\sqrt{2/3}Q)/81 + (1390\sqrt{2/3}Q{}^{3})/9 - 
      (390136\sqrt{2/3}Q{}^{5})/9)$%
$\partial{}^{3} J J{}^{4}\partial{}^{2} T + 
    ((-7622\sqrt{2/3}Q)/9 - (362255\sqrt{2/3}Q{}^{3})/27 - 
      (584110\sqrt{2/3}Q{}^{5})/9)$%
$\partial{}^{3} J J{}^{4}\partial{}^{2} A_{2} + 
    ((1004\sqrt{2/3})/3 - (343\sqrt{2/3}Q{}^{2})/9 - 4774\sqrt{6}Q{}^{4})
     $%
$\partial{}^{3} J J{}^{4}\partial A_{3} + 
    (256/9 + 359984 Q{}^{2}/81 + 781168 Q{}^{4}/27)$%
$\partial{}^{3} J J{}^{3}A_2 \partial T + 
    (16534 Q/27 + 42928 Q{}^{3}/9)$%
$\partial{}^{3} J J{}^{3}A_2 A_3 + 
    (-4960/243 + 226960 Q{}^{2}/81 + 174608 Q{}^{4}/9)$%
$\partial{}^{3} J J{}^{3}\partial A_{2} T + 
    (14144/81 + 135250 Q{}^{2}/81 + 258848 Q{}^{4}/27)$%
$\partial{}^{3} J J{}^{3}\partial A_{2} A_2 + 
    ((-14128Q)/9 - 33232 Q{}^{3}/3)$%
$\partial{}^{3} J J{}^{3}A_3 T + (25414 Q/27 + 37208 Q{}^{3}/9)
     $%
$\partial{}^{3} J J{}^{3}B_2 B_3 + 
    (22016/81 - 218 Q{}^{2}/3 - 96704 Q{}^{4}/9)$%
$\partial{}^{3} J J{}^{3}\partial B_{2} B_2 + 
    ((14764\sqrt{2/3}Q)/9 + (38380\sqrt{2/3}Q{}^{3})/3)
     $%
$\partial{}^{3} J J{}^{2}A_2 T{}^{2}+ 
    ((4652\sqrt{2/3}Q)/9 + (15052\sqrt{2/3}Q{}^{3})/3)
     $%
$\partial{}^{3} J J{}^{2}A_2{}^{2}T + 
    ((-14\sqrt{2/3}Q)/3 + 103\sqrt{6}Q{}^{3})$%
$\partial{}^{3} J J{}^{2}A_2{}^{3}+ 
    ((4174\sqrt{2/3}Q)/9 + (9629\sqrt{2/3}Q{}^{3})/3)
     $%
$\partial{}^{3} J J{}^{2}A_2 B_2{}^{2}+ ((9220\sqrt{2/3}Q)/9 + (22588\sqrt{2/3}Q{}^{3})/3)
     $%
$\partial{}^{3} J J{}^{2}B_2{}^{2}T + 
    (8000/81 + 4696 Q{}^{2}/3 + 41056 Q{}^{4}/9)$%
$\partial{}^{3} J \partial J 
      J{}^{2}T{}^{2}+ 
    ((60724\sqrt{2/3}Q)/81 - 28528\sqrt{2/3}Q{}^{3} - (4052608\sqrt{2/3}Q{}^{5})/
       9)$%
$\partial{}^{3} J \partial J J{}^{3}\partial T + 
    ((-579712\sqrt{2/3}Q)/81 - (3217160\sqrt{2/3}Q{}^{3})/27 - 
      197184\sqrt{6}Q{}^{5})$%
$\partial{}^{3} J \partial J J{}^{3}\partial A_{2} + ((112496\sqrt{2/3})/81 + 
      (37016\sqrt{2/3}Q{}^{2})/9 - (329792\sqrt{2/3}Q{}^{4})/9)
     $%
$\partial{}^{3} J \partial J J{}^{3}A_3 + 
    (-4096/81 - 33352 Q{}^{2}/27 - 201448 Q{}^{4}/9)$%
$\partial{}^{3} J \partial J 
      J{}^{2}A_2 T + 
    (15164/81 + 2654 Q{}^{2}/9 - 52394 Q{}^{4}/9)$%
$\partial{}^{3} J \partial J 
      J{}^{2}A_2{}^{2}+ 
    (33572/81 + 2522 Q{}^{2}/3 - 139538 Q{}^{4}/9)$%
$\partial{}^{3} J \partial J 
      J{}^{2}B_2{}^{2}+ 
    ((-25096\sqrt{2/3}Q)/27 - (231832\sqrt{2/3}Q{}^{3})/3 - 
      (2380448\sqrt{2/3}Q{}^{5})/3)$%
$\partial{}^{3} J (\partial J){}^{2} 
      J{}^{2}T + ((-264566\sqrt{2/3}Q)/27 - 
      (559148\sqrt{2/3}Q{}^{3})/3 - (3010162\sqrt{2/3}Q{}^{5})/3)
     $%
$\partial{}^{3} J (\partial J){}^{2} J{}^{2}A_2 + 
    (-15388/81 + 221164 Q{}^{2}/27 + 3319972 Q{}^{4}/9 + 8756432 Q{}^{6}/3)
     $%
$\partial{}^{3} J (\partial J){}^{3} J{}^{2}+ 
    ((-26192\sqrt{2/3}Q)/81 - (300080\sqrt{2/3}Q{}^{3})/9 - 
      (3181888\sqrt{2/3}Q{}^{5})/9)$%
$\partial{}^{3} J \partial{}^{2} J J{}^{3}T + ((-369152\sqrt{2/3}Q)/81 - 
      (800980\sqrt{2/3}Q{}^{3})/9 - (1487200\sqrt{2/3}Q{}^{5})/3)
     $%
$\partial{}^{3} J \partial{}^{2} J J{}^{3}A_2 + 
    (-11188/27 - 235480 Q{}^{2}/243 + 2029808 Q{}^{4}/9 + 57507904 Q{}^{6}/27)
     $%
$\partial{}^{3} J \partial{}^{2} J \partial J J{}^{3}+ 
    (-26543/729 - 68113 Q{}^{2}/81 - 227308 Q{}^{4}/81 + 220480 Q{}^{6}/9)
     $%
$\partial{}^{3} J \partial{}^{3} J J{}^{4}+ 
    (4180/243 + 14480 Q{}^{2}/27 + 106880 Q{}^{4}/27)$%
$\partial{}^{4} J J{}^{3}T{}^{2}+ 
    ((4783\sqrt{2/3}Q)/27 + (3695\sqrt{2/3}Q{}^{3})/9 - 
      (61748\sqrt{2/3}Q{}^{5})/3)$%
$\partial{}^{4} J J{}^{4}\partial T + ((-119651Q)/(81\sqrt{6}) - 
      700771 Q{}^{3}/(27\sqrt{6}) - (199694\sqrt{2/3}Q{}^{5})/3)
     $%
$\partial{}^{4} J J{}^{4}\partial A_{2} + 
    (31915/(81\sqrt{6}) + (17384\sqrt{2/3}Q{}^{2})/9 + (47938\sqrt{2/3}Q{}^{4})/9)
     $%
$\partial{}^{4} J J{}^{4}A_3 + 
    (-56/9 + 67984 Q{}^{2}/81 + 198752 Q{}^{4}/27)$%
$\partial{}^{4} J J{}^{3}A_2 T + 
    (3217/81 + 14644 Q{}^{2}/27 + 27782 Q{}^{4}/9)$%
$\partial{}^{4} J J{}^{3}A_2{}^{2}+ 
    (6431/81 + 7376 Q{}^{2}/9 + 21350 Q{}^{4}/9)$%
$\partial{}^{4} J J{}^{3}B_2{}^{2}+ 
    ((-4652\sqrt{2/3}Q)/27 - 18764\sqrt{2/3}Q{}^{3} - (601072\sqrt{2/3}Q{}^{5})/3)
     $%
$\partial{}^{4} J \partial J J{}^{3}T + 
    ((-216592\sqrt{2/3}Q)/81 - (492742\sqrt{2/3}Q{}^{3})/9 - 
      (2823296\sqrt{2/3}Q{}^{5})/9)$%
$\partial{}^{4} J \partial J J{}^{3}A_2 + 
    (-29750/243 - 42250 Q{}^{2}/81 + 1706300 Q{}^{4}/27 + 5597200 Q{}^{6}/9)
     $%
$\partial{}^{4} J (\partial J){}^{2} J{}^{3}+ 
    (-29521/486 - 366301 Q{}^{2}/243 - 196207 Q{}^{4}/27 + 672316 Q{}^{6}/27)
     $%
$\partial{}^{4} J \partial{}^{2} J J{}^{4}+ 
    ((689\sqrt{2/3}Q)/81 - (7817\sqrt{2/3}Q{}^{3})/9 - 
      (104828\sqrt{2/3}Q{}^{5})/9)$%
$\partial{}^{5} J J{}^{4}T + ((-169757Q)/(405\sqrt{6}) - 
      391319 Q{}^{3}/(45\sqrt{6}) - (227956\sqrt{2/3}Q{}^{5})/9)
     $%
$\partial{}^{5} J J{}^{4}A_2 + 
    (-2321/81 - 316163 Q{}^{2}/405 - 81083 Q{}^{4}/15 - 17708 Q{}^{6}/9)
     $%
$\partial{}^{5} J \partial J J{}^{4}+ 
    (-1733/810 - 107293 Q{}^{2}/1215 - 20138 Q{}^{4}/15 - 191464 Q{}^{6}/27)
     $%
$\partial{}^{6} J J{}^{5}+ 
    (32$%
$A_2{}^{2}T{}^{4})/9 + 
    (8$%
$A_2{}^{4}T{}^{2})/
     3 - (40$%
$A_2{}^{5}T)/9 + (4$%
$A_2{}^{6})/3 - 
    (28$%
$A_2{}^{4}B_2{}^{2})/9 + (16$%
$A_2{}^{3}B_2{}^{2}T)/3 - 
    (16$%
$A_2{}^{2}B_2{}^{2}T{}^{2})/
     3 + (20$%
$A_2{}^{2}B_2{}^{4})/9 + (512$%
$A_2 B_2{}^{2}T{}^{3})/9 - (8$%
$A_2 B_2{}^{4}T)/9 + 
    (32$%
$B_2{}^{2}T{}^{4})/3 + 
    (152$%
$B_2{}^{4}T{}^{2})/
     9 - (4$%
$B_2{}^{6})/9 + (-32/27 - 128 Q{}^{2}/9)$%
$J{}^{4}\partial T \partial T T + 
    (-224/27 - 896 Q{}^{2}/9)$%
$J{}^{4}\partial{}^{2} T T{}^{2}+ ((-680\sqrt{2/3}Q)/9 - 
      (2720\sqrt{2/3}Q{}^{3})/3)$%
$J{}^{5}\partial{}^{2} T \partial T + ((-112\sqrt{2/3}Q)/9 - 
      (448\sqrt{2/3}Q{}^{3})/3)$%
$J{}^{5}\partial{}^{3} T T + (-637/729 + 1106 Q{}^{2}/81 + 
      23464 Q{}^{4}/81)$%
$J{}^{6}\partial{}^{4} T + 
    (902/729 - 533 Q{}^{2}/54 - 2549 Q{}^{4}/81)$%
$J{}^{6}\partial{}^{4} A_{2} + 
    (50Q + 1115 Q{}^{3}/3)$%
$J{}^{6}\partial{}^{3} A_{3} + 
    ((-3784\sqrt{2/3}Q)/27 - (2464\sqrt{2/3}Q{}^{3})/3)
     $%
$J{}^{5}A_2 
      \partial{}^{3} T + ((-82\sqrt{2/3})/3 + 29\sqrt{2/3}Q{}^{2})
     $%
$J{}^{5}A_2 
      \partial{}^{2} A_{3} + ((-5600\sqrt{2/3}Q)/27 - (9640\sqrt{2/3}Q{}^{3})/9)
     $%
$J{}^{5}\partial A_{2} 
      \partial{}^{2} T + ((-56\sqrt{2/3})/3 + 24\sqrt{6}Q{}^{2})
     $%
$J{}^{5}\partial A_{2} 
      \partial A_{3} + ((-3728\sqrt{2/3}Q)/27 - (10552\sqrt{2/3}Q{}^{3})/9)
     $%
$J{}^{5}\partial{}^{2} A_{2} 
      \partial T + ((3620\sqrt{2/3}Q)/27 + (7723\sqrt{2/3}Q{}^{3})/9)
     $%
$J{}^{5}\partial{}^{2} A_{2} 
      \partial A_{2} + (-34\sqrt{2/3} - 27\sqrt{6}Q{}^{2})
     $%
$J{}^{5}\partial{}^{2} A_{2} 
      A_3 + ((-1066\sqrt{2/3}Q)/27 - (728\sqrt{2/3}Q{}^{3})/3)
     $%
$J{}^{5}\partial{}^{3} A_{2} 
      T + ((1360\sqrt{2/3}Q)/27 + 137\sqrt{2/3}Q{}^{3})
     $%
$J{}^{5}\partial{}^{3} A_{2} 
      A_2 + ((340\sqrt{2/3})/9 - 404\sqrt{2/3}Q{}^{2})
     $%
$J{}^{5}A_3 
      \partial{}^{2} T + ((-4\sqrt{2/3})/3 - 398\sqrt{2/3}Q{}^{2})
     $%
$J{}^{5}\partial A_{3} 
      \partial T - 96\sqrt{6}Q$%
$J{}^{5}\partial A_{3} A_3 + 
    ((116\sqrt{2/3})/9 + 74\sqrt{2/3}Q{}^{2})$%
$J{}^{5}\partial{}^{2} A_{3} T + 
    ((-14\sqrt{2/3})/3 - (355\sqrt{2/3}Q{}^{2})/3)$%
$J{}^{5}B_2 \partial{}^{2} B_{3} + 
    ((-148\sqrt{2/3})/3 + (122\sqrt{2/3}Q{}^{2})/3)$%
$J{}^{5}\partial B_{2} \partial B_{3} + 
    ((2912\sqrt{2/3}Q)/9 + 1979\sqrt{2/3}Q{}^{3})$%
$J{}^{5}\partial{}^{2} B_{2} \partial B_{2} + 
    ((-610\sqrt{2/3})/9 + (61\sqrt{2/3}Q{}^{2})/3)$%
$J{}^{5}\partial{}^{2} B_{2} B_3 + 
    ((3032\sqrt{2/3}Q)/27 + (7729\sqrt{2/3}Q{}^{3})/9)
     $%
$J{}^{5}\partial{}^{3} B_{2} 
      B_2 - 80\sqrt{2/3}Q$%
$J{}^{5}\partial B_{3} B_3 + 
    (-1072/81 - 7712 Q{}^{2}/27)$%
$J{}^{4}A_2 \partial T \partial T + (-3328/81 - 15488 Q{}^{2}/27)
     $%
$J{}^{4}A_2 \partial{}^{2} T 
      T + (250/27 - 2834 Q{}^{2}/9)$%
$J{}^{4}A_2{}^{2}\partial{}^{2} T - 
    (160Q$%
$J{}^{4}A_2{}^{2}\partial A_{3})/9 + 
    (200Q$%
$J{}^{4}A_2 
       A_3 \partial T)/3 - (32$%
$J{}^{4}A_2 A_3 A_3)/3 - 
    (944Q$%
$J{}^{4}A_2 
       \partial A_{3} T)/9 - (352Q$%
$J{}^{4}A_2 B_2 \partial B_{3})/9 + 
    (-106/9 + 128Q{}^{2})$%
$J{}^{4}A_2 \partial B_{2} \partial B_{2} - 
    (304Q$%
$J{}^{4}A_2 
       \partial B_{2} B_3)/3 + (-406/9 + 248 Q{}^{2}/9)
     $%
$J{}^{4}A_2 \partial{}^{2} B_{2} 
      B_2 + 16$%
$J{}^{4}A_2 B_3 B_3 + (-2680/81 - 9104 Q{}^{2}/27)
     $%
$J{}^{4}\partial A_{2} \partial T 
      T + (-220/81 - 19958 Q{}^{2}/27)$%
$J{}^{4}\partial A_{2} A_2 \partial T + 
    (128Q$%
$J{}^{4}\partial A_{2} 
       A_2 A_3)/3 + (-658/81 - 6614 Q{}^{2}/27)
     $%
$J{}^{4}\partial A_{2} \partial A_{2} 
      T + (-262/27 - 1742 Q{}^{2}/9)$%
$J{}^{4}\partial A_{2} \partial A_{2} A_2 + 
    (368Q$%
$J{}^{4}\partial A_{2} 
       A_3 T)/3 - (76Q$%
$J{}^{4}\partial A_{2} B_2 B_3)/3 + 
    (-340/9 + 74Q{}^{2})$%
$J{}^{4}\partial A_{2} \partial B_{2} B_2 + (-152/27 - 680 Q{}^{2}/9)
     $%
$J{}^{4}\partial{}^{2} A_{2} T{}^{2}+ (182/81 - 5246 Q{}^{2}/27)$%
$J{}^{4}\partial{}^{2} A_{2} A_2 T + 
    (-146/27 - 581 Q{}^{2}/9)$%
$J{}^{4}\partial{}^{2} A_{2} A_2{}^{2}+ (-40/3 - 137 Q{}^{2}/3)
     $%
$J{}^{4}\partial{}^{2} A_{2} B_2{}^{2}- 544Q$%
$J{}^{4}A_3 \partial T T + 32$%
$J{}^{4}A_3 A_3 T + 
    44$%
$J{}^{4}A_3 B_2 
      B_3 - (748Q$%
$J{}^{4}A_3 \partial B_{2} B_2)/3 - 
    (1984Q$%
$J{}^{4}\partial A_{3} 
       T{}^{2})/9 - (352Q$%
$J{}^{4}\partial A_{3} B_2{}^{2})/3 + 
    (38/27 - 1894 Q{}^{2}/9)$%
$J{}^{4}B_2{}^{2}\partial{}^{2} T + 
    104Q$%
$J{}^{4}B_2 
      B_3 \partial T + (752Q$%
$J{}^{4}B_2 \partial B_{3} T)/9 + 
    (-484/27 - 4678 Q{}^{2}/9)$%
$J{}^{4}\partial B_{2} B_2 \partial T + (454/27 - 2 Q{}^{2}/9)
     $%
$J{}^{4}\partial B_{2} \partial B_{2} 
      T + (56Q$%
$J{}^{4}\partial B_{2} B_3 T)/3 + (-206/27 - 538 Q{}^{2}/3)
     $%
$J{}^{4}\partial{}^{2} B_{2} B_2 
      T - (356$%
$J{}^{4}B_3 B_3 T)/3 - 
    (256\sqrt{2/3}Q$%
$J{}^{3}A_2 \partial T 
       T{}^{2})/9 - (224\sqrt{2/3}Q$%
$J{}^{3}A_2{}^{2}\partial T T)/9 + 
    (104\sqrt{2/3}Q$%
$J{}^{3}A_2{}^{3}\partial T)/9 - 
    (28\sqrt{2/3}$%
$J{}^{3}A_2{}^{4}3)/9 - 
    (32\sqrt{2/3}$%
$J{}^{3}A_2{}^{3}3 T)/9 + (44\sqrt{2/3}$%
$J{}^{3}A_2{}^{2}B_2 B_3)/3 - 
    (124\sqrt{2/3}Q$%
$J{}^{3}A_2{}^{2}\partial B_{2} B_2)/9 - 
    (1600\sqrt{2/3}$%
$J{}^{3}A_2 
       A_3 T{}^{2})/9 - 
    (140\sqrt{2/3}$%
$J{}^{3}A_2 
       A_3 B_2{}^{2})/9 - 
    (488\sqrt{2/3}Q$%
$J{}^{3}A_2 
       B_2{}^{2}\partial T)/9 - 
    (176\sqrt{2/3}$%
$J{}^{3}A_2 
       B_2 B_3 T)/3 + 
    (176\sqrt{2/3}Q$%
$J{}^{3}A_2 
       \partial B_{2} B_2 T)/9 + 
    (320\sqrt{2/3}Q$%
$J{}^{3}\partial A_{2} T{}^{3})/3 + (992\sqrt{2/3}Q$%
$J{}^{3}\partial A_{2} A_2 T{}^{2})/9 - 
    (512\sqrt{2/3}Q$%
$J{}^{3}\partial A_{2} 
       A_2{}^{2}T)/9 - 
    (148\sqrt{2/3}Q$%
$J{}^{3}\partial A_{2} 
       A_2{}^{3})/9 + 
    (524\sqrt{2/3}Q$%
$J{}^{3}\partial A_{2} 
       A_2 B_2{}^{2})/9 + 
    (560\sqrt{2/3}Q$%
$J{}^{3}\partial A_{2} 
       B_2{}^{2}T)/9 - 
    (640\sqrt{2/3}$%
$J{}^{3}A_3 T{}^{3})/3 - (1136\sqrt{2/3}$%
$J{}^{3}A_3 B_2{}^{2}T)/9 - 
    (800\sqrt{2/3}Q$%
$J{}^{3}B_2{}^{2}\partial T T)/9 + 
    (44\sqrt{2/3}$%
$J{}^{3}B_2{}^{4}3)/3 - 
    (880\sqrt{2/3}$%
$J{}^{3}B_2 
       B_3 T{}^{2})/3 + 112\sqrt{2/3}Q
     $%
$J{}^{3}\partial B_{2} B_2 T{}^{2}- (244\sqrt{2/3}Q$%
$J{}^{3}\partial B_{2} B_2{}^{3})/3 + 
    (128$%
$J{}^{2}A_2 T{}^{4})/9 + (704$%
$J{}^{2}A_2{}^{2}T{}^{3})/9 - 
    (64$%
$J{}^{2}A_2{}^{3}T{}^{2})/9 + (8$%
$J{}^{2}A_2{}^{4}T)/3 - 
    (20$%
$J{}^{2}A_2{}^{5})/9 + 
    (8$%
$J{}^{2}A_2{}^{3}B_2{}^{2})/3 - 
    (16$%
$J{}^{2}A_2{}^{2}B_2{}^{2}T)/3 + (1504$%
$J{}^{2}A_2 
       B_2{}^{2}T{}^{2})/9 - 
    (4$%
$J{}^{2}A_2 B_2{}^{4})/9 + 
    (736$%
$J{}^{2}B_2{}^{2}T{}^{3})/9 + (152$%
$J{}^{2}B_2{}^{4}T)/9 + (-496/9 - 1984 Q{}^{2}/3)
     $%
$\partial J J{}^{3}\partial T T{}^{2}+ 
    ((-784\sqrt{2/3}Q)/3 - 3136\sqrt{2/3}Q{}^{3})$%
$\partial J J{}^{4}\partial T \partial T + 
    ((-176\sqrt{2/3}Q)/3 - 704\sqrt{2/3}Q{}^{3})$%
$\partial J J{}^{4}\partial{}^{2} T T + 
    (-1732/243 + 10124 Q{}^{2}/27 + 149200 Q{}^{4}/27)$%
$\partial J J{}^{5}\partial{}^{3} T + 
    (244/9 - 1574 Q{}^{2}/9 - 12434 Q{}^{4}/9)$%
$\partial J J{}^{5}\partial{}^{3} A_{2} + 
    (22864 Q/27 + 49748 Q{}^{3}/9)$%
$\partial J J{}^{5}\partial{}^{2} A_{3} + 
    ((-21548\sqrt{2/3}Q)/27 - (26644\sqrt{2/3}Q{}^{3})/9)
     $%
$\partial J J{}^{4}A_2 
      \partial{}^{2} T + ((-3740\sqrt{2/3})/27 + (2072\sqrt{2/3}Q{}^{2})/9)
     $%
$\partial J J{}^{4}A_2 
      \partial A_{3} + ((-10784\sqrt{2/3}Q)/9 - (15434\sqrt{2/3}Q{}^{3})/3)
     $%
$\partial J J{}^{4}\partial A_{2} 
      \partial T + ((7459\sqrt{2/3}Q)/9 + (17707\sqrt{2/3}Q{}^{3})/3)
     $%
$\partial J J{}^{4}\partial A_{2} 
      \partial A_{2} + ((-7688\sqrt{2/3})/27 - (11170\sqrt{2/3}Q{}^{2})/9)
     $%
$\partial J J{}^{4}\partial A_{2} 
      A_3 + ((-3962\sqrt{2/3}Q)/9 - (5938\sqrt{2/3}Q{}^{3})/3)
     $%
$\partial J J{}^{4}\partial{}^{2} A_{2} 
      T + ((20683\sqrt{2/3}Q)/27 + (45425\sqrt{2/3}Q{}^{3})/9)
     $%
$\partial J J{}^{4}\partial{}^{2} A_{2} 
      A_2 + ((820\sqrt{2/3})/9 - 490\sqrt{6}Q{}^{2})
     $%
$\partial J J{}^{4}A_3 
      \partial T - 160\sqrt{6}Q$%
$\partial J J{}^{4}A_3 A_3 + 
    ((712\sqrt{2/3})/3 + (3892\sqrt{2/3}Q{}^{2})/3)$%
$\partial J J{}^{4}\partial A_{3} T + 
    ((-9604\sqrt{2/3})/27 - (7892\sqrt{2/3}Q{}^{2})/9)
     $%
$\partial J J{}^{4}B_2 
      \partial B_{3} + ((38183\sqrt{2/3}Q)/27 + (78583\sqrt{2/3}Q{}^{3})/9)
     $%
$\partial J J{}^{4}\partial B_{2} 
      \partial B_{2} + ((-8524\sqrt{2/3})/27 + (256\sqrt{2/3}Q{}^{2})/9)
     $%
$\partial J J{}^{4}\partial B_{2} 
      B_3 + ((41909\sqrt{2/3}Q)/27 + (89029\sqrt{2/3}Q{}^{3})/9)
     $%
$\partial J J{}^{4}\partial{}^{2} B_{2} 
      B_2 - 194\sqrt{2/3}Q$%
$\partial J J{}^{4}B_3 B_3 + 
    (-23584/81 - 80864 Q{}^{2}/27)$%
$\partial J J{}^{3}A_2 \partial T T + (-640/27 - 11456 Q{}^{2}/9)
     $%
$\partial J J{}^{3}A_2{}^{2}\partial T - (376Q$%
$\partial J J{}^{3}A_2{}^{3}3)/9 + 
    (4448Q$%
$\partial J J{}^{3}A_2 
       A_3 T)/9 - (656Q$%
$\partial J J{}^{3}A_2 B_2 B_3)/3 + 
    (-4616/27 + 2660 Q{}^{2}/9)$%
$\partial J J{}^{3}A_2 \partial B_{2} B_2 + (-1360/9 - 5296 Q{}^{2}/3)
     $%
$\partial J J{}^{3}\partial A_{2} T{}^{2}+ (-880/81 - 50456 Q{}^{2}/27)$%
$\partial J J{}^{3}\partial A_{2} A_2 T + 
    (-476/27 - 242Q{}^{2})$%
$\partial J J{}^{3}\partial A_{2} A_2{}^{2}+ (-3916/27 - 4882 Q{}^{2}/9)
     $%
$\partial J J{}^{3}\partial A_{2} B_2{}^{2}+ (1888Q$%
$\partial J J{}^{3}A_3 T{}^{2})/3 - 
    (2680Q$%
$\partial J J{}^{3}A_3 
       B_2{}^{2})/9 + (-1136/27 - 1984 Q{}^{2}/3)
     $%
$\partial J J{}^{3}B_2{}^{2}\partial T + 768Q$%
$\partial J J{}^{3}B_2 B_3 T + (-1936/27 - 8120 Q{}^{2}/9)
     $%
$\partial J J{}^{3}\partial B_{2} B_2 
      T - (1024\sqrt{2/3}Q$%
$\partial J J{}^{2}A_2 T{}^{3})/3 - 
    (2144\sqrt{2/3}Q$%
$\partial J J{}^{2}A_2{}^{2}T{}^{2})/3 - 
    (16\sqrt{2/3}Q$%
$\partial J J{}^{2}A_2{}^{3}T)/3 + 8\sqrt{2/3}Q
     $%
$\partial J J{}^{2}A_2{}^{4}- 32\sqrt{2/3}Q$%
$\partial J J{}^{2}A_2{}^{2}B_2{}^{2}- 
    (1552\sqrt{2/3}Q$%
$\partial J J{}^{2}A_2 
       B_2{}^{2}T)/3 - 848\sqrt{2/3}Q
     $%
$\partial J J{}^{2}B_2{}^{2}T{}^{2}- (88\sqrt{2/3}Q$%
$\partial J J{}^{2}B_2{}^{4})/3 + 
    (-224/9 - 896 Q{}^{2}/3)$%
$(\partial J){}^{2} J{}^{2}T{}^{3}+ ((1168\sqrt{2/3}Q)/9 + 
      (4672\sqrt{2/3}Q{}^{3})/3)$%
$(\partial J){}^{2} J{}^{3}\partial T T + (-628/27 + 20260 Q{}^{2}/9 + 91088 Q{}^{4}/3)
     $%
$(\partial J){}^{2} J{}^{4}\partial{}^{2} T + (9428/81 - 6322 Q{}^{2}/3 - 176054 Q{}^{4}/9)
     $%
$(\partial J){}^{2} J{}^{4}\partial{}^{2} A_{2} + (116540 Q/27 + 238120 Q{}^{3}/9)
     $%
$(\partial J){}^{2} J{}^{4}\partial A_{3} + ((-34760\sqrt{2/3}Q)/27 + (6944\sqrt{2/3}Q{}^{3})/9)
     $%
$(\partial J){}^{2} J{}^{3}A_2 
      \partial T + ((-8336\sqrt{2/3})/27 - (6208\sqrt{2/3}Q{}^{2})/9)
     $%
$(\partial J){}^{2} J{}^{3}A_2 
      A_3 + ((-2816\sqrt{2/3}Q)/3 + 1256\sqrt{2/3}Q{}^{3})
     $%
$(\partial J){}^{2} J{}^{3}\partial A_{2} 
      T + ((88036\sqrt{2/3}Q)/27 + (201872\sqrt{2/3}Q{}^{3})/9)
     $%
$(\partial J){}^{2} J{}^{3}\partial A_{2} 
      A_2 + ((4640\sqrt{2/3})/9 + 1144\sqrt{2/3}Q{}^{2})
     $%
$(\partial J){}^{2} J{}^{3}A_3 
      T + ((-18472\sqrt{2/3})/27 - (15824\sqrt{2/3}Q{}^{2})/9)
     $%
$(\partial J){}^{2} J{}^{3}B_2 
      B_3 + ((150740\sqrt{2/3}Q)/27 + (315544\sqrt{2/3}Q{}^{3})/9)
     $%
$(\partial J){}^{2} J{}^{3}\partial B_{2} 
      B_2 + (-1616/9 - 400Q{}^{2})$%
$(\partial J){}^{2} J{}^{2}A_2 T{}^{2}+ (40/3 + 296 Q{}^{2}/3)
     $%
$(\partial J){}^{2} J{}^{2}A_2{}^{2}T + (-4/3 + 2 Q{}^{2}/3)$%
$(\partial J){}^{2} J{}^{2}A_2{}^{3}+ 
    (-380/3 - 290 Q{}^{2}/3)$%
$(\partial J){}^{2} J{}^{2}A_2 B_2{}^{2}+ (-328/9 + 2344 Q{}^{2}/3)
     $%
$(\partial J){}^{2} J{}^{2}B_2{}^{2}T + ((3472\sqrt{2/3}Q)/9 + (13888\sqrt{2/3}Q{}^{3})/3)
     $%
$(\partial J){}^{3} J{}^{2}T{}^{2}+ (-184/27 + 39176 Q{}^{2}/9 + 53216Q{}^{4})
     $%
$(\partial J){}^{3} J{}^{3}\partial T + (3416/9 - 133832 Q{}^{2}/27 - 532960 Q{}^{4}/9)
     $%
$(\partial J){}^{3} J{}^{3}\partial A_{2} + (16232 Q/3 + 103000 Q{}^{3}/3)$%
$(\partial J){}^{3} J{}^{3}A_3 + 
    ((-16\sqrt{2/3}Q)/9 + (22384\sqrt{2/3}Q{}^{3})/3)
     $%
$(\partial J){}^{3} J{}^{2}A_2 
      T + ((14006\sqrt{2/3}Q)/9 + 3834\sqrt{6}Q{}^{3})
     $%
$(\partial J){}^{3} J{}^{2}A_2{}^{2}+ ((22190\sqrt{2/3}Q)/9 + 16294\sqrt{2/3}Q{}^{3})
     $%
$(\partial J){}^{3} J{}^{2}B_2{}^{2}+ (-1096/27 + 2288 Q{}^{2}/9 + 8896Q{}^{4})
     $%
$(\partial J){}^{4} J{}^{2}T + (18536/81 - 3776Q{}^{2} - 465272 Q{}^{4}/9)
     $%
$(\partial J){}^{4} J{}^{2}A_2 + ((-10244\sqrt{2/3}Q)/9 - 31112\sqrt{2/3}Q{}^{3} - 
      209440\sqrt{2/3}Q{}^{5})$%
$(\partial J){}^{5} J{}^{2}+ (-448/27 - 1792 Q{}^{2}/9)
     $%
$\partial{}^{2} J J{}^{3}T{}^{3}+ 
    ((-364\sqrt{2/3}Q)/3 - 1456\sqrt{2/3}Q{}^{3})$%
$\partial{}^{2} J J{}^{4}\partial T T + 
    (-344/27 + 15104 Q{}^{2}/27 + 76928 Q{}^{4}/9)$%
$\partial{}^{2} J J{}^{5}\partial{}^{2} T + 
    (2896/81 - 6620 Q{}^{2}/27 - 20864 Q{}^{4}/9)$%
$\partial{}^{2} J J{}^{5}\partial{}^{2} A_{2} + 
    (34504 Q/27 + 69110 Q{}^{3}/9)$%
$\partial{}^{2} J J{}^{5}\partial A_{3} + 
    (-872\sqrt{2/3}Q - 970\sqrt{6}Q{}^{3})$%
$\partial{}^{2} J J{}^{4}A_2 \partial T + 
    ((-5228\sqrt{2/3})/27 - (8365\sqrt{2/3}Q{}^{2})/9)
     $%
$\partial{}^{2} J J{}^{4}A_2 
      A_3 + ((-15058\sqrt{2/3}Q)/27 - (12548\sqrt{2/3}Q{}^{3})/9)
     $%
$\partial{}^{2} J J{}^{4}\partial A_{2} 
      T + ((31258\sqrt{2/3}Q)/27 + (68606\sqrt{2/3}Q{}^{3})/9)
     $%
$\partial{}^{2} J J{}^{4}\partial A_{2} 
      A_2 + ((2168\sqrt{2/3})/9 + 470\sqrt{6}Q{}^{2})
     $%
$\partial{}^{2} J J{}^{4}A_3 
      T + ((-10504\sqrt{2/3})/27 - (12587\sqrt{2/3}Q{}^{2})/9)
     $%
$\partial{}^{2} J J{}^{4}B_2 
      B_3 + ((56360\sqrt{2/3}Q)/27 + (109594\sqrt{2/3}Q{}^{3})/9)
     $%
$\partial{}^{2} J J{}^{4}\partial B_{2} 
      B_2 + (-4000/27 - 31792 Q{}^{2}/27)$%
$\partial{}^{2} J J{}^{3}A_2 T{}^{2}+ 
    (8/3 - 10040 Q{}^{2}/27)$%
$\partial{}^{2} J J{}^{3}A_2{}^{2}T + (-92/27 - 826 Q{}^{2}/27)
     $%
$\partial{}^{2} J J{}^{3}A_2{}^{3}+ (-2980/27 - 12374 Q{}^{2}/27)$%
$\partial{}^{2} J J{}^{3}A_2 B_2{}^{2}+ 
    (-1256/27 - 13160 Q{}^{2}/27)$%
$\partial{}^{2} J J{}^{3}B_2{}^{2}T + 
    ((4880\sqrt{2/3}Q)/9 + (19520\sqrt{2/3}Q{}^{3})/3)
     $%
$\partial{}^{2} J \partial J J{}^{3}T{}^{2}+ (-1396/81 + 145400 Q{}^{2}/27 + 67104Q{}^{4})
     $%
$\partial{}^{2} J \partial J J{}^{4}\partial T + (13988/27 - 94286 Q{}^{2}/81 - 883954 Q{}^{4}/27)
     $%
$\partial{}^{2} J \partial J J{}^{4}\partial A_{2} + (52228 Q/9 + 111512 Q{}^{3}/3)$%
$\partial{}^{2} J \partial J 
      J{}^{4}A_3 + 
    ((5984\sqrt{2/3}Q)/27 + (135128\sqrt{2/3}Q{}^{3})/9)
     $%
$\partial{}^{2} J \partial J J{}^{3}A_2 
      T + ((18806\sqrt{2/3}Q)/9 + (134488\sqrt{2/3}Q{}^{3})/9)
     $%
$\partial{}^{2} J \partial J J{}^{3}A_2{}^{2}+ ((32246\sqrt{2/3}Q)/9 + (214552\sqrt{2/3}Q{}^{3})/9)
     $%
$\partial{}^{2} J \partial J J{}^{3}B_2{}^{2}+ (-13456/81 + 51376 Q{}^{2}/27 + 420800 Q{}^{4}/9)
     $%
$\partial{}^{2} J (\partial J){}^{2} J{}^{3}T + (75136/81 - 35504 Q{}^{2}/81 - 1864192 Q{}^{4}/27)
     $%
$\partial{}^{2} J (\partial J){}^{2} J{}^{3}A_2 + ((-16352\sqrt{2/3}Q)/3 - (1383632\sqrt{2/3}Q{}^{3})/9 - 
      (3179840\sqrt{2/3}Q{}^{5})/3)$%
$\partial{}^{2} J (\partial J){}^{3} J{}^{3}+ 
    (-3638/81 + 11344 Q{}^{2}/27 + 34528 Q{}^{4}/3)$%
$\partial{}^{2} J \partial{}^{2} J 
      J{}^{4}T + 
    (19520/81 + 159412 Q{}^{2}/81 + 42722 Q{}^{4}/27)$%
$\partial{}^{2} J \partial{}^{2} J 
      J{}^{4}A_2 + 
    ((-75961\sqrt{2/3}Q)/27 - (239620\sqrt{2/3}Q{}^{3})/3 - 
      (1660064\sqrt{2/3}Q{}^{5})/3)$%
$\partial{}^{2} J \partial{}^{2} J \partial J 
      J{}^{4}+ 
    ((848\sqrt{2/3}Q)/27 + (3392\sqrt{2/3}Q{}^{3})/9)
     $%
$\partial{}^{3} J J{}^{4}T{}^{2}+ (-1216/243 + 14264 Q{}^{2}/27 + 190624 Q{}^{4}/27)
     $%
$\partial{}^{3} J J{}^{5}\partial T + (15776/243 + 25444 Q{}^{2}/81 - 164 Q{}^{4}/9)
     $%
$\partial{}^{3} J J{}^{5}\partial A_{2} + (5276 Q/9 + 10712 Q{}^{3}/3)$%
$\partial{}^{3} J J{}^{5}A_3 + 
    ((-730\sqrt{2/3}Q)/9 + (11090\sqrt{2/3}Q{}^{3})/9)
     $%
$\partial{}^{3} J J{}^{4}A_2 
      T + ((2362\sqrt{2/3}Q)/9 + (16070\sqrt{2/3}Q{}^{3})/9)
     $%
$\partial{}^{3} J J{}^{4}A_2{}^{2}+ ((12098\sqrt{2/3}Q)/27 + (8342\sqrt{2/3}Q{}^{3})/3)
     $%
$\partial{}^{3} J J{}^{4}B_2{}^{2}+ (-5144/81 + 1076 Q{}^{2}/3 + 121040 Q{}^{4}/9)
     $%
$\partial{}^{3} J \partial J J{}^{4}T + (25520/81 + 23210 Q{}^{2}/9 + 20494 Q{}^{4}/9)
     $%
$\partial{}^{3} J \partial J J{}^{4}A_2 + ((-165382\sqrt{2/3}Q)/81 - 58498\sqrt{2/3}Q{}^{3} - 
      (3671672\sqrt{2/3}Q{}^{5})/9)$%
$\partial{}^{3} J (\partial J){}^{2} 
      J{}^{4}+ 
    ((-14672\sqrt{2/3}Q)/27 - (139904\sqrt{2/3}Q{}^{3})/9 - 36096\sqrt{6}Q{}^{5})
     $%
$\partial{}^{3} J \partial{}^{2} J J{}^{5}+ (-1862/243 - 784 Q{}^{2}/27 + 20384 Q{}^{4}/27)
     $%
$\partial{}^{4} J J{}^{5}T + (316/9 + 43084 Q{}^{2}/81 + 68084 Q{}^{4}/27)
     $%
$\partial{}^{4} J J{}^{5}A_2 + ((-25685\sqrt{2/3}Q)/81 - (27821\sqrt{2/3}Q{}^{3})/3 - 
      (590596\sqrt{2/3}Q{}^{5})/9)$%
$\partial{}^{4} J \partial J J{}^{5}+ 
    ((-7655Q)/(243\sqrt{6}) - 24985 Q{}^{3}/(27\sqrt{6}) - 
      (88670\sqrt{2/3}Q{}^{5})/27)$%
$\partial{}^{5} J J{}^{6}+ 
    (88/81 + 352 Q{}^{2}/27)$%
$J{}^{6}\partial T \partial T + (112/81 + 448 Q{}^{2}/27)
     $%
$J{}^{6}\partial{}^{2} T T + ((-40\sqrt{2/3}Q)/9 - (160\sqrt{2/3}Q{}^{3})/3)
     $%
$J{}^{7}\partial{}^{3} T + ((349\sqrt{2/3}Q)/27 + (680\sqrt{2/3}Q{}^{3})/9)
     $%
$J{}^{7}\partial{}^{3} A_{2} + ((-58\sqrt{2/3})/9 - 7\sqrt{6}Q{}^{2})
     $%
$J{}^{7}\partial{}^{2} A_{3} + (1576/243 - 2200 Q{}^{2}/81)
     $%
$J{}^{6}A_2 \partial{}^{2} T - (496Q$%
$J{}^{6}A_2 \partial A_{3})/9 + 
    (1324/243 - 4612 Q{}^{2}/81)$%
$J{}^{6}\partial A_{2} \partial T + (-929/243 - 223 Q{}^{2}/81)
     $%
$J{}^{6}\partial A_{2} \partial A_{2} - (128Q$%
$J{}^{6}\partial A_{2} A_3)/3 + 
    (380/81 + 916 Q{}^{2}/27)$%
$J{}^{6}\partial{}^{2} A_{2} T + (-1625/243 - 571 Q{}^{2}/81)
     $%
$J{}^{6}\partial{}^{2} A_{2} A_2 + (304Q$%
$J{}^{6}A_3 \partial T)/3 + 
    (64$%
$J{}^{6}A_3 A_3)/3 + 
    (272Q$%
$J{}^{6}\partial A_{3} T)/9 - 
    (592Q$%
$J{}^{6}B_2 \partial B_{3})/9 + (-613/81 + 503 Q{}^{2}/27)
     $%
$J{}^{6}\partial B_{2} \partial B_{2} - (860Q$%
$J{}^{6}\partial B_{2} B_3)/9 + 
    (-1015/81 - 703 Q{}^{2}/27)$%
$J{}^{6}\partial{}^{2} B_{2} B_2 + 
    (62$%
$J{}^{6}B_3 B_3)/9 + 
    (320\sqrt{2/3}Q$%
$J{}^{5}A_2 \partial T T)/9 + 
    (920\sqrt{2/3}Q$%
$J{}^{5}A_2{}^{2}\partial T)/9 + 
    (8\sqrt{2/3}$%
$J{}^{5}A_2{}^{3}3)/9 - 
    (160\sqrt{2/3}$%
$J{}^{5}A_2 A_3 T)/9 + 
    (44\sqrt{2/3}$%
$J{}^{5}A_2 B_2 B_3)/3 - 
    140\sqrt{2/3}Q$%
$J{}^{5}A_2 \partial B_{2} B_2 + 
    (32\sqrt{2/3}Q$%
$J{}^{5}\partial A_{2} T{}^{2})/9 + 
    (272\sqrt{2/3}Q$%
$J{}^{5}\partial A_{2} A_2 T)/3 + 
    (88\sqrt{2/3}Q$%
$J{}^{5}\partial A_{2} A_2{}^{2})/9 - 
    (500\sqrt{2/3}Q$%
$J{}^{5}\partial A_{2} B_2{}^{2})/9 + 
    32\sqrt{2/3}$%
$J{}^{5}A_3 T{}^{2}+ 
    (284\sqrt{2/3}$%
$J{}^{5}A_3 B_2{}^{2})/9 + 
    (488\sqrt{2/3}Q$%
$J{}^{5}B_2{}^{2}\partial T)/9 - 
    (88\sqrt{2/3}$%
$J{}^{5}B_2 B_3 T)/3 + 
    (632\sqrt{2/3}Q$%
$J{}^{5}\partial B_{2} B_2 T)/9 + 
    (256$%
$J{}^{4}A_2 T{}^{3})/27 + (464$%
$J{}^{4}A_2{}^{2}T{}^{2})/27 + 
    (64$%
$J{}^{4}A_2{}^{3}T)/27 + 
    (2$%
$J{}^{4}A_2{}^{4})/3 - 
    (4$%
$J{}^{4}A_2{}^{2}B_2{}^{2})/3 + 
    (416$%
$J{}^{4}A_2 
       B_2{}^{2}T)/27 + 
    (704$%
$J{}^{4}B_2{}^{2}T{}^{2})/27 + 
    (38$%
$J{}^{4}B_2{}^{4})/9 + (184/27 + 736 Q{}^{2}/9)
     $%
$\partial J J{}^{5}\partial T T + ((-1160\sqrt{2/3}Q)/27 - (4640\sqrt{2/3}Q{}^{3})/9)
     $%
$\partial J J{}^{6}\partial{}^{2} T + (211\sqrt{2/3}Q + 499\sqrt{6}Q{}^{3})
     $%
$\partial J J{}^{6}\partial{}^{2} A_{2} + ((-352\sqrt{2/3})/3 - 232\sqrt{2/3}Q{}^{2})
     $%
$\partial J J{}^{6}\partial A_{3} + (1720/81 - 10168 Q{}^{2}/27)
     $%
$\partial J J{}^{5}A_2 \partial T - (2048Q$%
$\partial J J{}^{5}A_2 A_3)/9 + 
    (4712/81 + 3088 Q{}^{2}/27)$%
$\partial J J{}^{5}\partial A_{2} T + (-7468/81 - 4838 Q{}^{2}/27)
     $%
$\partial J J{}^{5}\partial A_{2} A_2 + (688Q$%
$\partial J J{}^{5}A_3 T)/3 - 
    344Q$%
$\partial J J{}^{5}B_2 B_3 + (-3988/27 - 2870 Q{}^{2}/9)
     $%
$\partial J J{}^{5}\partial B_{2} B_2 - (544\sqrt{2/3}Q$%
$\partial J J{}^{4}A_2 T{}^{2})/9 + 
    (1280\sqrt{2/3}Q$%
$\partial J J{}^{4}A_2{}^{2}T)/9 - 
    8\sqrt{2/3}Q$%
$\partial J J{}^{4}A_2{}^{3}- 
    (2216\sqrt{2/3}Q$%
$\partial J J{}^{4}A_2 B_2{}^{2})/9 - 
    (80\sqrt{2/3}Q$%
$\partial J J{}^{4}B_2{}^{2}T)/3 + (-208/27 - 832 Q{}^{2}/9)
     $%
$(\partial J){}^{2} J{}^{4}T{}^{2}+ ((-1916\sqrt{2/3}Q)/9 - (7664\sqrt{2/3}Q{}^{3})/3)
     $%
$(\partial J){}^{2} J{}^{5}\partial T + (1280\sqrt{2/3}Q + 9250\sqrt{2/3}Q{}^{3})
     $%
$(\partial J){}^{2} J{}^{5}\partial A_{2} + ((-1160\sqrt{2/3})/3 - 1330\sqrt{2/3}Q{}^{2})
     $%
$(\partial J){}^{2} J{}^{5}A_3 + (7672/81 - 184 Q{}^{2}/27)
     $%
$(\partial J){}^{2} J{}^{4}A_2 T + (-2302/27 - 898 Q{}^{2}/9)
     $%
$(\partial J){}^{2} J{}^{4}A_2{}^{2}+ (-5450/27 - 2578 Q{}^{2}/3)
     $%
$(\partial J){}^{2} J{}^{4}B_2{}^{2}+ ((-1024\sqrt{2/3}Q)/9 - 
      (4096\sqrt{2/3}Q{}^{3})/3)$%
$(\partial J){}^{3} J{}^{4}T + 
    ((16672\sqrt{2/3}Q)/9 + (46024\sqrt{2/3}Q{}^{3})/3)
     $%
$(\partial J){}^{3} J{}^{4}A_2 + (1622/27 + 6064 Q{}^{2}/9 - 1696 Q{}^{4}/3)
     $%
$(\partial J){}^{4} J{}^{4}+ ((-1762\sqrt{2/3}Q)/27 - 
      (7048\sqrt{2/3}Q{}^{3})/9)$%
$\partial{}^{2} J J{}^{6}\partial T + 
    ((26095\sqrt{2/3}Q)/81 + (60716\sqrt{2/3}Q{}^{3})/27)
     $%
$\partial{}^{2} J J{}^{6}\partial A_{2} + ((-1280\sqrt{2/3})/9 - (2023\sqrt{2/3}Q{}^{2})/3)
     $%
$\partial{}^{2} J J{}^{6}A_3 + (4256/81 + 1000 Q{}^{2}/9)
     $%
$\partial{}^{2} J J{}^{5}A_2 T + (-394/9 - 5134 Q{}^{2}/27)
     $%
$\partial{}^{2} J J{}^{5}A_2{}^{2}+ (-2354/27 - 11602 Q{}^{2}/27)
     $%
$\partial{}^{2} J J{}^{5}B_2{}^{2}+ ((-440\sqrt{2/3}Q)/3 - 1760\sqrt{2/3}Q{}^{3})
     $%
$\partial{}^{2} J \partial J J{}^{5}T + ((41216\sqrt{2/3}Q)/27 + (37682\sqrt{2/3}Q{}^{3})/3)
     $%
$\partial{}^{2} J \partial J J{}^{5}A_2 + (10592/81 + 22132 Q{}^{2}/9 + 96112 Q{}^{4}/9)
     $%
$\partial{}^{2} J (\partial J){}^{2} J{}^{5}+ (4997/243 + 4196 Q{}^{2}/9 + 71104 Q{}^{4}/27)
     $%
$\partial{}^{2} J \partial{}^{2} J J{}^{6}+ ((-152\sqrt{2/3}Q)/9 - (608\sqrt{2/3}Q{}^{3})/3)
     $%
$\partial{}^{3} J J{}^{6}T + ((9833\sqrt{2/3}Q)/81 + (26249\sqrt{2/3}Q{}^{3})/27)
     $%
$\partial{}^{3} J J{}^{6}A_2 + (19736/729 + 16778 Q{}^{2}/27 + 288232 Q{}^{4}/81)
     $%
$\partial{}^{3} J \partial J J{}^{6}+ (1445/729 + 1396 Q{}^{2}/27 + 27136 Q{}^{4}/81)
     $%
$\partial{}^{4} J J{}^{7}+ (8/81 + 32 Q{}^{2}/27)$%
$J{}^{8}\partial{}^{2} T + (-38/81 + 2 Q{}^{2}/27)$%
$J{}^{8}\partial{}^{2} A_{2} - 
    (116Q$%
$J{}^{8}\partial A_{3})/9 + 
    (128\sqrt{2/3}Q$%
$J{}^{7}A_2 \partial T)/9 + 
    (80\sqrt{2/3}$%
$J{}^{7}A_2 A_3)/9 + 
    (80\sqrt{2/3}Q$%
$J{}^{7}\partial A_{2} T)/9 - 
    8\sqrt{6}Q$%
$J{}^{7}\partial A_{2} A_2 - 
    (32\sqrt{2/3}$%
$J{}^{7}A_3 T)/3 + 
    (44\sqrt{2/3}$%
$J{}^{7}B_2 B_3)/3 - 
    (100\sqrt{2/3}Q$%
$J{}^{7}\partial B_{2} B_2)/3 - 
    (64$%
$J{}^{6}A_2 T{}^{2})/81 - 
    (656$%
$J{}^{6}A_2{}^{2}T)/81 - 
    (16$%
$J{}^{6}A_2{}^{3})/81 + 
    (760$%
$J{}^{6}A_2 B_2{}^{2})/81 - 
    (248$%
$J{}^{6}B_2{}^{2}T)/81 + (44/81 + 176 Q{}^{2}/27)
     $%
$\partial J J{}^{7}\partial T + (-3428/243 - 2884 Q{}^{2}/81)
     $%
$\partial J J{}^{7}\partial A_{2} - 
    (248Q$%
$\partial J J{}^{7}A_3)/3 + 
    (832\sqrt{2/3}Q$%
$\partial J J{}^{6}A_2 T)/27 - 
    (1688\sqrt{2/3}Q$%
$\partial J J{}^{6}A_2{}^{2})/27 - 
    (2012\sqrt{2/3}Q$%
$\partial J J{}^{6}B_2{}^{2})/27 + 
    (952/81 + 3808 Q{}^{2}/27)$%
$(\partial J){}^{2} J{}^{6}T + 
    (-12980/243 - 26764 Q{}^{2}/81)$%
$(\partial J){}^{2} J{}^{6}A_2 + 
    ((2908\sqrt{2/3}Q)/27 + (11632\sqrt{2/3}Q{}^{3})/9)
     $%
$(\partial J){}^{3} J{}^{6}+ (208/81 + 832 Q{}^{2}/27)
     $%
$\partial{}^{2} J J{}^{7}T + (-3944/243 - 9052 Q{}^{2}/81)
     $%
$\partial{}^{2} J J{}^{7}A_2 + 
    ((524\sqrt{2/3}Q)/9 + (2096\sqrt{2/3}Q{}^{3})/3)$%
$\partial{}^{2} J \partial J 
      J{}^{7}+ ((88\sqrt{2/3}Q)/27 + (352\sqrt{2/3}Q{}^{3})/9)
     $%
$\partial{}^{3} J J{}^{8}- 
    (56\sqrt{2/3}Q$%
$J{}^{9}\partial A_{2})/27 + 
    (52\sqrt{2/3}$%
$J{}^{9}A_3)/27 - 
    (64$%
$J{}^{8}A_2 T)/81 + 
    (98$%
$J{}^{8}A_2{}^{2})/81 + 
    (122$%
$J{}^{8}B_2{}^{2})/81 - 
    (220\sqrt{2/3}Q$%
$\partial J J{}^{8}A_2)/
     27 + (-44/27 - 176 Q{}^{2}/9)$%
$(\partial J){}^{2} J{}^{8}+ (-28/81 - 112 Q{}^{2}/27)$%
$\partial{}^{2} J J{}^{9}+ (8$%
$J{}^{10}A_2)/81.
$%

\endgroup

\bibliographystyle{ytphys}
\small\baselineskip=.84\baselineskip
\let\bbb\bibitem\def\bibitem{\itemsep1.5pt\bbb}
\bibliography{ref}

\end{document}